\input{epsf}
\documentclass{aa}
\usepackage{graphicx}
\usepackage{longtable}
\usepackage{natbib}
\bibpunct{(}{)}{;}{a}{}{,} 
\usepackage{epsfig}
\usepackage{rotating}
\usepackage{lscape}

\begin{document}

\newcounter{myFootNote}

\title{A spectroscopic atlas of post-AGB stars and planetary nebulae \\
  selected from the IRAS Point Source Catalogue  \thanks{Based mainly
    on observations collected at the European Southern Observatory (La
    Silla, Chile) and at the Observatorio del Roque de los Muchachos
    (La Palma, Spain) }}

\author{ O. Su\'arez\inst{1}
\and P. Garc\'\i a-Lario\inst{2}
\and A. Manchado\inst{3,4}
\and M. Manteiga\inst{5}
\and A. Ulla\inst{6}
\and S.R. Pottasch \inst{7}
}

\institute{Laboratorio de Astrof\'\i sica Espacial y F\'\i sica 
Fundamental, INTA, Apartado de Correos 50727, E-28080 Madrid, Spain
\and 
Research and Scientific Support Department of ESA, European 
Space Astronomy Centre, Villafranca del Castillo, Apartado de
Correos 50727. E-28080 Madrid, Spain
\and
Instituto de Astrof\'\i sica de Canarias, c/Via Lactea, s/n, E-38200
La Laguna, Tenerife, Spain
\and
Consejo Superior de Investigaciones Cient\'\i ficas (CSIC), Spain
\and
Departamento de Ciencias de la Navegaci\'on y de la Tierra, E.T.S. de
N\'autica y M\'aquinas, Universidade da Coru\~na, E-15011 A Coru\~na, Spain
\and
Departamento de F\'\i sica Aplicada, Facultade de Ciencias do Mar, Campus
Marcosende-Lagoas, Universidade de Vigo, E-36310 Vigo, Pontevedra,
Spain
\and
Kapteyn Astronomical Institute, Postbus 800 NL-9700 AV Groningen, The 
Netherlands
}

\offprints{O. Su\'arez. 
\email{olga@laeff.inta.es}}

\date{Received <date>/Acepted <date>}

\abstract
        {}
   {We study the optical spectral properties of a sample of stars
   showing far infrared colours similar to those of well-known
   planetary nebulae. The large majority of them were unidentified sources or 
  poorly known in the  literature  at the time when this spectroscopic
  survey started, some 15 years ago.}
   {We present low-resolution optical spectroscopy, finding
  charts and improved astrometric coordinates of a sample of 253 IRAS
  sources.}
   {We have identified 103 sources as post-AGB stars, 21 as
   ``transition sources'', and 36 as planetary nebulae, some of them
   strongly reddened. Among the rest of sources in the sample, we were
   also able to identify 38 young stellar objects, 5 peculiar stars,
   and 2 Seyfert galaxies. Up to 49 sources in our spectroscopic
   sample do not show any optical counterpart, and most of them are
   suggested to be heavily obscured post-AGB stars, rapidly evolving
   on their way to becoming planetary nebulae.}
   {An analysis of the galactic distribution of the sources identified
   as evolved stars in the sample is presented together with a study
   of the distribution of these stars in the IRAS two-colour diagram.
   Finally, the spectral type distribution and other properties of the
   sources identified as post-AGB in this spectroscopic survey are
   discussed in the framework of stellar evolution.}

\titlerunning{Spectroscopic atlas of post-AGB and PNe}
\authorrunning{Su\'arez et al.}

\maketitle

\section{Introduction}
\label{intro}

Post-Asymptotic Giant Branch (post-AGB, hereafter) stars are rapidly
evolving low- and intermediate-mass stars (1$-$8 M$_{\odot}$) in 
the transition phase from the AGB to the Planetary Nebula (PNe, hereafter)
stage \citep{kwokan93,habing96,vanwinckel03}. Their precursors, the 
AGB, stars are pulsating stars, very bright in the infrared, which can become 
heavily obscured in the optical by thick circumstellar envelopes formed as 
a consequence of their strong mass loss (up to $10^{-4}\rm{M_{\sun}/yr}$). 
 When the mass loss stops, the AGB star enters the so-called {\it post-AGB}
 stage, which is also accompanied by the cessation of the stellar pulsations. 
 This is followed by a decrease in the optical depth of the circumstellar
envelope as a consequence of the expansion, which implies that the
central star can be seen again in the optical range if it were ever
obscured at the end of the AGB.  During this process, the effective
temperature of the central star increases. This leads to a rapid
change in the spectral type, which migrates from late- to early-type in
very short timescales of just a few thousand years \citep{vassi94}.

The terminology used to define the various stages preceding the
formation of a PN is sometimes confusing. In this work the term
`post-AGB star' will be applied to those sources that have already
left the strong mass-losing AGB phase. When the temperature of the
central star is hot enough (T$\ge$20$\,$000K), the ionization of the
envelope starts, and we consider that the star has entered the PN stage. 
We will adopt the term `transition source' for those stars in an intermediate
stage between the post-AGB and PN stages, whose spectra are characterised 
by the simultaneous detection of a prominent stellar continuum and 
shock-excited emission lines.

 The initial goal of the observations presented here was the discovery
 of new PNe among the 1084 sources included in the so-called `GLMP
 catalogue' \citep{gltesis}. This was a colour-selected sample of IRAS
 sources showing the characteristic far-infrared colours of well known
 PNe. The strategy was based on the fact that PNe are located in a
 well-defined region of the IRAS two-colour diagram [12]$-$[25] $\it
 {vs.}$ [25]$-$[60] almost exclusively populated by PNe and their
 immediate precursors, the post-AGB stars. In this region, only a
 small overlap exists with some young stellar objects and a few
 Seyfert galaxies, while normal stars and galaxies show completely
 different far-infrared colours \citep{pottasch88}.  Not unexpectedly,
 as a byproduct of our search for new PNe, we found that many of the
 observed stars were actually post-AGB stars and transition sources, as
 well as peculiar very young PNe, rather than genuine, classical PNe.

 Indeed, the use of IRAS data proved to be a highly successful method to
identify new candidate sources in the transition from the AGB to the
PN stage. Their strong infrared excess makes them very bright in the
infrared and easily detectable at these wavelengths. Based on their
characteristic colours, several lists with potential candidates were 
compiled in the past by \citet{preite88}, \citet{pottasch88}, 
\citet{vanderveen89}, \citet{ratag90}, \citet{hu93}, and other authors reported 
in the Strasbourg-ESO Catalogue of Galactic PNe \citep{acker92}. More
specific searches for PNe based on other observational criteria were
also made by \cite{vdsteene93} and \cite{jacoby04}. Other searches for
post-AGB stars were based on the detection of infrared excesses around
stars with optically bright counterparts, like those performed by
\citep{oudmaijer92,oudmaijer96} in the last decade.  Unfortunately, in
many cases no spectroscopic confirmation was ever provided in the
literature about the nature of the newly discovered sources. One of
the main reasons for this is the poor astrometric accuracy of the IRAS
data ($\sim$15-30 arcsec), which makes the identification of the
optical counterparts of the selected IRAS sources very difficult,
especially in crowded fields close to the galactic plane and/or in the
direction of the Galactic Centre.

 In this paper we present, for the first time, the whole optical
spectroscopic database of stars resulting from the survey carried
out by our group in the optical from 1988 to 2003, concentrated on the 
stars in the GLMP sample for which indications of the presence of an optical
counterpart were available. At the time when the observations were
performed, priority was given to those sources suspected to be evolved 
stars. The results are presented in the form of
an atlas including the largest set of spectroscopic data on 
post-AGB stars and transition sources 
published before that time. The atlas also contains
improved astrometric coordinates and finding charts for all the
sources observed. The spectral energy distributions (SEDs) from 1 to 
100 $\mu$m of the objects observed were also studied, but they will be 
presented  elsewhere, together with a
more detailed analysis of some of the new sources found. Several papers 
dealing with other individual sources of strong interest found during the 
survey have already been published (e.g.,\cite{manchado89b,gl91,bobrowsky98,gl99})

In Sect.~\ref{criteria} we explain the selection criteria applied to
include an IRAS source in our sample. In Sect.~\ref{observations} we
present the observations performed and the data reduction process.
Section~\ref{results} describes the identification of the sources in the
sample and the way the classification of the optical spectra 
was performed. The main properties of the
sources identified as evolved stars are discussed in
Sect.~\ref{discussion}. Finally, in Sect.~\ref{conclusion} we
present our conclusions. Spectra and finding charts for individual
sources are provided in Appendices A, B, C, D, E, and F (only available 
in electronic format).

\section{Selection criteria} 
\label{criteria}

The selection criteria applied in this work are essentially the same
as those adopted by \citet{pottasch88}, with some modifications:

\begin{itemize}
  
\item [(i)]The source must be well detected at 12, 25, and 60~$\mu$m in the
  IRAS Point Source Catalogue \citep[see][]{beichmancat88}. The flux 
  quality for each band must be: 
  \begin{displaymath}
     FQUAL~(12~\rm{\mu}m)~\ge~2; \\   
  \end{displaymath}

  \begin{displaymath}
     FQUAL~(25~\rm{\mu}m)~=~3; \\
  \end{displaymath}

  \begin{displaymath}
     FQUAL~(60~\rm{\mu}m)~=~3.
  \end{displaymath}

\item [(ii)] The ratios between the IRAS photometric fluxes must satisfy 
  the following conditions:

  \begin{displaymath}
    \frac{F_{\nu}~(12~\rm\rm{\mu}m)}{F_{\nu}~(25~\rm{\mu}m)}\le0.50;   
  \end{displaymath}

  \begin{displaymath}
    \frac{F_{\nu}~(25~\rm{\mu}m)}{F_{\nu}~(60~\rm{\mu}m)}\ge0.35.
  \end{displaymath}

\item [(iii)]When data at 100 $\mu$m are of good quality 
   (FQUAL(100~$\mu$m)~$=$~3), we further impose: 
  \begin{displaymath}
    \frac{F_{\nu}~(60~\rm{\mu}m)}{F_{\nu}~(100~\rm{\mu}m)}\ge0.60.
  \end{displaymath}

\item [(iv)]Sources must show a low IRAS variability index:
   \begin{displaymath}
     VAR\le~60~\%
   \end{displaymath}

\end{itemize}

The choice of criteria (i) and (ii) was mainly a direct consequence of the
range of dust temperatures ($T_d$) expected in the circumstellar
shells of post-AGB stars. If we assume a typical luminosity for a
post-AGB star of $L$=$10^4L_{\sun}$, the 
dust temperature in the shell, following \citet{scoville76},  would be:

\begin{equation}
 T_{\rm d} = 1.658\,f^{-1/5}r^{2/5}L_*^{1/5},
\end{equation}

\noindent where $f$ is the value of the emissivity of the dust grains at 
50$\mu$m, $r$ is the radius of the shell in pc, and $L_*$ is the luminosity 
of the central star in solar units.
 
Using $f$(50\,$\mu$m)~=~0.004 \citep{draine84}, we find a $T_{\rm d}$ 
between 200\,K and 80\,K for $r$ between 0.01\,pc and 0.1\,pc, respectively 
(the expected range of radii of the expanding circumstellar envelope). 

Criterium (iii) was  added to avoid the potential contamination by young
stellar objects, ultra-compact H II regions, and Seyfert galaxies, 
which can sometimes show very similar colours up to 60 $\mu$m.

 Finally, the low IRAS variability was imposed to exclude 
AGB stars from the sample, since they are known to be strongly variable 
stars.

All the sources in the original GLMP sample satisfy criteria (i),
(ii), and (iii), but not necessarily the last one, concerning the
variability. Moreover, in the original GLMP catalogue, about half of
the entries were associated with objects that were well identified in
the literature, most of them PNe, and they were excluded from the
atlas presented in this paper. There was also a significant number of
known young stellar objects ($\sim$20\%) and a few Seyfert galaxies
($\sim$5\%), which were not considered either.

 A slight, more recent modification of the selection criteria allowed
the inclusion of about 100 additional sources in an extended version of the
GLMP catalogue (Garc\'\i a-Lario, private communication). This 
extension also contains those IRAS sources not detected in the 12
$\mu$m IRAS band (FQUAL~(12~$\mu$m)~$=$~1), but satisfying the rest of 
conditions mentioned above. Some of these sources were also included in 
our spectroscopic survey in the latest years and, as such, they are also 
considered in the following analysis.

The sample presented in this paper is thus comprised by a subset of
sources taken from  the two GLMP samples described above for which no
identification was available (or was poorly established) at the time 
when these two catalogues were created, and for which optical spectroscopy 
was obtained.

A total of 253 different IRAS fields were searched, resulting in the 
successful identification of 205 optical counterparts.

 With these selection criteria, our sample of post-AGB stars with
optical counterparts is expected to be essentially complete, limited only 
by the IRAS sensitivity.  Only a few C-rich post-AGB stars showing 
prominent solid state features attributed to SiC at 11.3 $\mu$m, and
thus located in region VII of the IRAS colour-colour diagram, as defined by
\citet{vdveenhabing88}, may have escaped our identification.
 In addition, some post-AGB
stars in the vicinity of the Galactic Centre could have also been
missed due to the IRAS confusion in that region.

%---------------OBSERVATIONS-----------------------------------

\section{Observations and data reduction}
\label{observations}

The spectroscopic observations were conducted during several runs
spanning 15 years from March 1988 to June 2003. The observations from
the Southern Hemisphere were carried out in most cases at the European
Southern Observatory (ESO, La Silla, Chile) with the 1.5\,m ESO telescope,
equipped with a Boller \& Chivens spectrograph. The first and last runs
of observations were carried out at the 3.6\,m ESO telescope, located
at the same site, using the ESO Faint Object Spectrograph and Camera
EFOSC1 in the first run and EFOSC2 in the last one.  The observations
from the Northern Hemisphere were carried out at the
2.5m Isaac Newton Telescope at the Observatorio del Roque de los
Muchachos (La Palma, Spain), using the IDS spectrograph, and at the
2.2\,m telescope at the Observatorio Hispano-Alem\'an (Calar Alto,
Spain), also equipped with a Boller \& Chivens spectrograph.

The full log of the spectroscopic observations is shown in
Table~\ref{catalog}, where we list the telescopes and dates of the
observations, together with the instrumentation used in each run, as
well as the spectral resolution and the spectral range covered in 
each case. Exposure times were in the range from 10 to 60 min.
depending on the brightness  of the source observed, with mean 
exposure times of 20--30 min.

\begin{table*}[htb]
\begin{center}
\caption{\label{catalog}Log of the spectroscopic observations.}
\begin{tabular}{lcccccc}
\hline 
\hline 
Run& Telescope& Instrumentation& Dates& Dispersion& Spectral range& Number of\\
& & & &(\AA/pix) &(\AA) & obs. objects\\
\hline
\#1& 3.6~m ESO La Silla& EFOSC1& 19-21 March 1988 & 3.79& 3406-6975 & 12\\
\#2& 1.5~m ESO La Silla& Boller \& Chivens& 2-4 January 1989 & 2.47& 4272-6812& 24 \\
\#3& 1.5~m ESO La Silla& Boller \& Chivens& 23-25 February 1990& 2.80& 4046-6920& 41 \\
\#4& 1.5~m ESO La Silla& Boller \& Chivens& 25-29 June 1990& 3.47& 4050-6925&34 \\ 
\#5& 2.5~m INT La Palma& IDS &23-25 May 1991&  2.81 & 3570-5226 / 5538-7240 & 4\\
\#6& 2.2~m Calar Alto&Boller \& Chivens & November 1991 & 2.50& 3830-6736&27 \\
\#7& 1.5~m ESO La Silla& Boller \& Chivens& 19-25 August 1992& 2.83& 3590-9425& 34\\
\#8& 1.5~m ESO La Silla& Boller \& Chivens& 10-13 March 1993& 3.74& 3321-11015& 32\\
\#9& 2.5~m INT La Palma& IDS& 1-4 July 1993& 1.48& 5376-7269& 8\\
\#10& 1.5~m ESO La Silla& Boller \& Chivens& 13-17 March 1994& 3.74& 3285-10980&38 \\
\#11& 2.5~m INT La Palma&IDS & 18-24 August 1994& 1.57& 3779-5401 / 5570-7195&32 \\
\#12& 1.5~m ESO La Silla& Boller \& Chivens& 11-15 February 1995& 3.79& 3532-11190& 21\\
\#13& 2.5~m INT La Palma&IDS & 13-16 June 1995& 1.57 & 3700-5455 / 5488-7255& 9\\
\#14& 2.5~m INT La Palma&IDS& 9-10 April 2001& 3.30& 3702-7401& 4\\
\#15& 3.6~m ESO La Silla& EFOSC2& 23-25 June 2003& 5.30& 3600-9200&45 \\
\hline
\end{tabular}
\end{center}
\end{table*}

The optical spectra were reduced using IRAF,
following standard techniques. After subtracting the bias, the
two-dimensional CCD spectra were divided by the normalised
flat-field of the corresponding night. Then, the cosmic ray events 
were identified and removed. 

The sky contribution was determined by averaging two narrow
regions of the CCD located at both sides of the object and then the mean 
sky was subtracted. For the wavelength calibration, several Cu-Ne or 
Cu-Ar lamp exposures were taken.  

Several photometric standards were observed during each observing
night and were used for flux calibration. The spectra of the target 
stars were flux calibrated and corrected for atmospheric extinction using
these standards. We estimate typical errors in the absolute fluxes of
$\sim$ 20\%. These errors depend mainly on the stability of the 
seeing conditions for each night of observation.

In parallel with these spectroscopic observations, we also carried out
a near infrared photometric survey \citep{manchado89,gl90,gl97}. The
data obtained in the near infrared helped in many cases to identify
the right optical counterpart of the IRAS source, or to confirm the
lack of it, in the case of the most obscured sources.

 The counterpart identification strategy was the following: in many cases,
before the observations took place, a single optical counterpart,
close to the IRAS position, was identified in the Digitized Sky Survey
(DSS) plates as the most plausible one. When no other information was
available, this was the source observed at the telescope.

Sometimes, however, several faint sources with similar brightness were
found  located within the elliptical IRAS error box. In this case, we usually
tried to observe the field in the near-infrared, where we identified the
tentative counterpart on the basis of its peculiar near infrared
colours (usually the brightest source or the most heavily reddened in
the near infrared) first. If no near infrared information was available at
the time of the spectroscopic observations, we usually took spectra of
all the optical sources located in the field within the elliptical
IRAS error box, starting from the closest one to the nominal IRAS position
and/or the redder one.

In some cases the identification of the optical counterpart was clear
(e.g., when a PN or a source displaying H$\alpha$
emission was found). Sometimes, however, no special features were detected in
the spectra of any of the stars tried, and it was more difficult to
determine which of them (if any) was the star physically associated
with the IRAS source. It might also be the case that none of the
observed stars is the optical counterpart of the IRAS source, since
stars on their way to becoming new PNe can become so heavily obscured by
the material expelled during the previous AGB phase that they may not be
detectable in the optical range. We will further discuss these
uncertainties in Sect.~\ref{finding}.

The full list of the 253 IRAS fields observed is shown in
Table~3 (only available in electronic format), where we also 
give: the entry number of 
the object in the GLMP catalogue, the IRAS name, the improved
astrometric coordinates taken from the 2MASS Point Source Catalogue \citep{2mass}
for which the estimated errors are of the order of $\sim$0.2$\arcsec$ 
corresponding to the source identified as the right counterpart, 
and the far infrared IRAS
colours [12]$-$[25] and [25]$-$[60], defined in the classical way as:

\begin{equation}
 [12]-[25] = -2.5~ \log \frac{F_{12\rm{\mu}m}}{F_{25\rm{\mu}m}} 
\end{equation}

\begin{equation}
[25]-[60] = -2.5~ \log \frac{F_{25\rm{\mu}m}}{F_{60\rm{\mu}m}}.  
\end{equation}

\noindent In the last columns of this table, we also provide the 
classification assigned to each source and, in the case of the 
non-evolved  objects, the spectral class or type, and we 
identify the spectroscopic observing run when every source was observed,
according to the code given in Table~\ref{catalog}.

In addition to the spectroscopic observations, CCD images of selected
fields were also obtained through various standard broad and narrow
filters covering the whole optical range and using several telescopes
listed in Table~\ref{loginimage}.  These images were bias and
flat-field corrected and cleaned for cosmic rays using standard IRAF
routines. No flux calibration was performed in this case. 

In Appendices A, B, C, D, E, and F (only in electronic format) we show
the spectrum and the finding chart for all the targets in the survey
for which an optical counterpart was identified and a spectrum was
taken. In most cases the finding charts are taken from the
DSS. However, when extended emission was detected in our images, we
used them for the atlas.

 We prefer showing the images obtained through red filters
(usually the R-band) if available, because in most cases the
optical counterparts of the IRAS sources in our sample are expected to
be brighter at longer wavelengths since on many occasions they are
strongly reddened by their circumstellar envelopes. If convenient, we
have also used our H$\alpha$ images, in some cases, to better illustrate the 
morphology of those sources showing extended ionised emission.

\begin{table*}[htb]
\begin{center}
\caption{\label{loginimage}Log of the direct imaging observations.}
\begin{tabular}{lccccccc}
\hline 
\hline 
Run& Telescope& Instrumentation& Dates\\
\hline 
\#a& 3.6m ESO (La Silla)& EFOSC1& June 1990  \\
\#b& JKT (La Palma)& CCD & July 1990  \\
\#c& 0.90m Dutch (La Palma)& CCD & March 1994  \\
\#d& 2.5m NOT (La Palma) &ALFOSC & June 2000 \\
\#e& 2.5m NOT (La Palma) &ALFOSC & May 2001 \\
\#f& 3.6m ESO (La Silla) & EFOSC2 & June 2003 \\
\hline
\end{tabular}
\end{center}
\end{table*}

\section{Results}
\label{results}

\subsection{Classification of the sources in the sample}
\label{finding}

To classify the objects in the sample as a function of their nature
and evolutionary stage, we have studied their optical spectra and the
complementary information coming from the near infrared measurements
made by \citet{gl97} and from other sources in the literature. The
assigned classifications are shown in Table~3\footnote{These tables
are only available in electronic form.}.\setcounter{myFootNote}{\value{footnote}}

The determination of the evolutionary stage of an object can lead to
some uncertainties, especially when trying to discriminate between
post-AGB stars and young stellar objects. If the source is located
within the boundaries or close to a known star-forming region, we have
assumed it is most probably young. The luminosity obtained in the
spectral classification can also give us some hints about its
evolutionary stage, since the young stars are expected to show spectra
corresponding to a low luminosity class (usually V), while post-AGB
stars show extended atmospheres and low gravities usually
corresponding to the high luminosity classes I and III.

To take these uncertainties into account, we have introduced a
two-letter code in Tables\footnotemark[\themyFootNote]~\ref{postagbtabla},
\ref{transtabla}, and \ref{PNtabla}. The letters rang from A to D and
rate our confidence in the data presented, with A representing
the maximum reliability. The first letter marks our confidence in the
correct identification of the optical counterpart, and the second
letter indicates how confident we are in the evolutionary
classification assigned.

The adopted selection criteria turned out to be very efficient in the
detection of evolved stars. Up to 209 sources, representing 81\% of
the stars in the sample, were found to be evolved stars. Among them,
the majority of these sources are considered to be post-AGB stars: 103
with an optical counterpart (see Table~\ref{postagbtabla}\footnotemark[\themyFootNote]), plus
possibly 49 more heavily obscured ones, for which we did not find any
optical counterpart during our observations, although most of which are
believed to be obscured post-AGB stars. They are labeled as ``No
counterpart'' in Table~3. We have also found 21 `transition sources'
(see Table~\ref{transtabla}\footnotemark[\themyFootNote]).
 
Only 38 of the sources observed in the sample ($\sim$17\%) were
identified as young stellar objects and 2 other ones as galaxies (less
than 1\%). Their identifications and the spectral type of the young
objects are shown in Table~3.

\subsection{Classification of the optical spectra}
\label{clas_sp}

\subsubsection{Spectral types in the MK system}

The classification of the optical spectra in the MK system 
was performed taking several spectral libraries as references  
\citep{silva,jacoby,pickles}, and it was applied to all the observed 
sources showing a stellar continuum. By default, we always
assumed a luminosity class I in our first try 
for the sources suspected to be post-AGB 
stars. To perform this classification,
we first normalised both the spectral templates taken from these catalogues 
and our target spectra. This way we concentrate our attention on 
the spectral lines, rather than on the stellar continuum, which may appear
extremely reddened in many stars of our sample.
 
We used Silva \& Cornell's templates for the spectra with the
lowest signal-to-noise ratios and for the young stellar objects.  The
estimated error in this classification is about plus or minus one
subclass, which implies a total error of around five subtypes in the
determination of the spectral type.

For the sources in our sample classified as evolved objects whose
spectra had a good signal-to-noise ratio, we used the other two
catalogues \citep{jacoby,pickles} to derive a more accurate spectral
classification, as they provide one spectrum for each subtype and for
a wide range of spectral types for stars with luminosity class I. We
estimate that the stars classified using these catalogues are affected
by errors that are always of the order of less than two spectral
subtypes.

We have had difficulties classifying with accuracy the stars with the
earliest spectral types showing the Balmer lines in emission. In
many cases these lines show emission over absorption, making
it impossible to use their strength for classification purposes.

In Table~\ref{postagbtabla}\footnotemark[\themyFootNote] we list all
the sources in the sample classified as post-AGB stars together with
their associated spectral classification, and in
Table~\ref{transtabla}\footnotemark[\themyFootNote] we show the
classification of the transition sources\footnote{The rough spectral
types assigned to the young stellar objects identified in our survey
are shown in Table~3. Note that for these sources we did not try to
make any luminosity class determination.}.

 The spectral type in these two tables is given, as usual, by the letter
and number corresponding to the MK classification system. Those
spectra for which Silva's rough classification was used
maintain the nomenclature followed in that catalogue, i.e, one letter
followed by two numbers indicating the range of subtypes. The letter
``e'' appended to the spectral type indicates that the object shows
emission lines. When the spectrum is dominated by emission lines, and
the continuum is too faint to derive its spectral type, it is
classified with the code ``em''.

\subsubsection{Determination of the extinction constant, the
  excitation classes for PNe, and the WR type of the central stars}
\label{excitation}

To determine the extinction corrected line intensities of the PNe, we
calculated the extinction coefficient {\it c}, using the observed
Balmer line decrement (see results in
Table~\ref{PNtabla}\footnotemark[\themyFootNote]).  Then we
dereddened the observed spectra using Whitford's extinction law
\citep{whitford58}, and we used the dereddened lines to calculate
the excitation classes.

The excitation classes were determined using both
\citet{morgan84} and \citet{dopita90} criteria to provide a
more reliable classification.  \citet{morgan84} based this
classification on the value of the ratios [O~{\sc
  iii}]~($\lambda$\,4959\,\AA)/H$\beta$,
H$\beta$/HeII~($\lambda$\,4686\,\AA), and
H$\beta$/[NeIII]~($\lambda$\,3869\,\AA). It establishes 12 different
classes, ranging from low (class 0) to high (class 10) excitation.
The classification proposed by \citet{dopita90} is a continuous one,
based on the ratios [O~{\sc iii}]~($\lambda$\,4959\,\AA)/H$\beta$ and
HeII~($\lambda$\,4686\,\AA)/H$\beta$.

We grouped together all possible excitation classes in only three
main subgroups: low, medium, and high excitation (see
Table~\ref{PNtabla}).

 Note that some PNe were impossible to classify using these criteria, as they
appeared so extinguished in the blue region that they did not show several
of the lines needed for the classification.

In the case of the PNe with H-poor envelopes (i.e., with WR-type
central stars), we used the criteria defined by \citet{acker03} which
is, in turn, also based on the classification previously defined by
\citet{crowther98}. They identify [WO] subtypes ranging from [W01] to
[W04] and [WC] subtypes ranging from [WC4] to [WC11]. This
classification is based on the dereddened strength of several O and C
lines with respect to the continuum, and implies a decrease in
temperature from 100\,000~K for [WO1-3] stars to 20\,000 for [WC10].

 The final classification derived for the PNe found in our spectroscopic
survey and additional information on the morphology of individual sources 
is shown in Table~\ref{PNtabla}.

\section{Discussion}
\label{discussion}

In the following we will concentrate our discussion on the subsample
of evolved stars, e.g., the sources classified either as optically
bright post-AGB stars, transition sources or PNe, plus the subsample
of heavily obscured sources also tentatively classified as post-AGB
stars.

\subsection{Galactic latitude distribution}
\label{galactic}

To study the evolutionary connection between the various
types of evolved stars identified in this work (post-AGB stars
with and without optical counterpart, transition sources, and PNe) we
can compare their relative galactic latitude distribution. For a given class,
this is expected to be an indicator of the average mass of the
progenitor stars. The youngest, and thus more massive, stars are
expected to appear more concentrated towards the galactic plane, while
an older population of stars, with lower masses, should show a higher
dispersion. 

\begin{figure}[t!]
\centering
 \includegraphics[width=7cm]{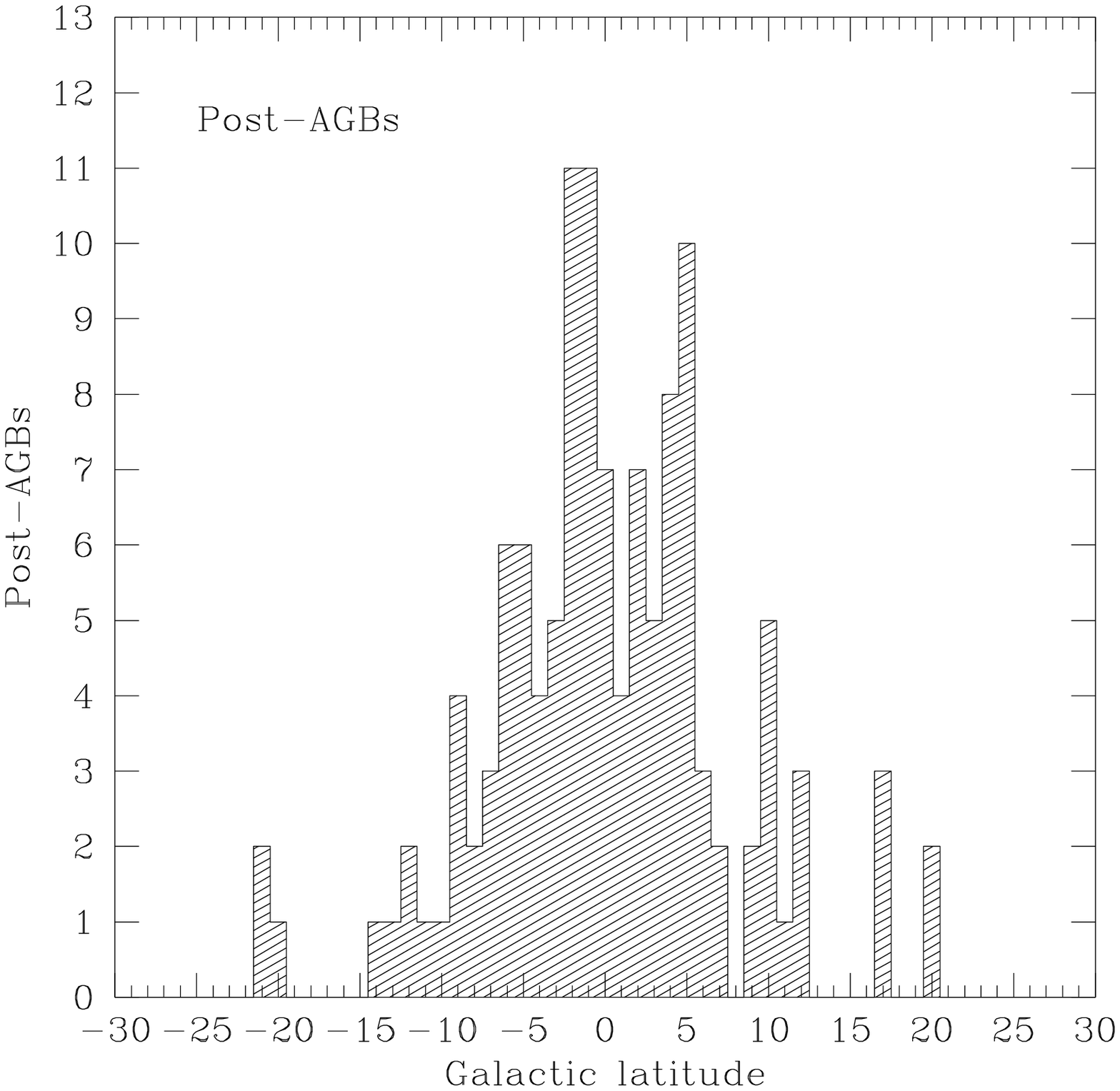}
 \includegraphics[width=7cm]{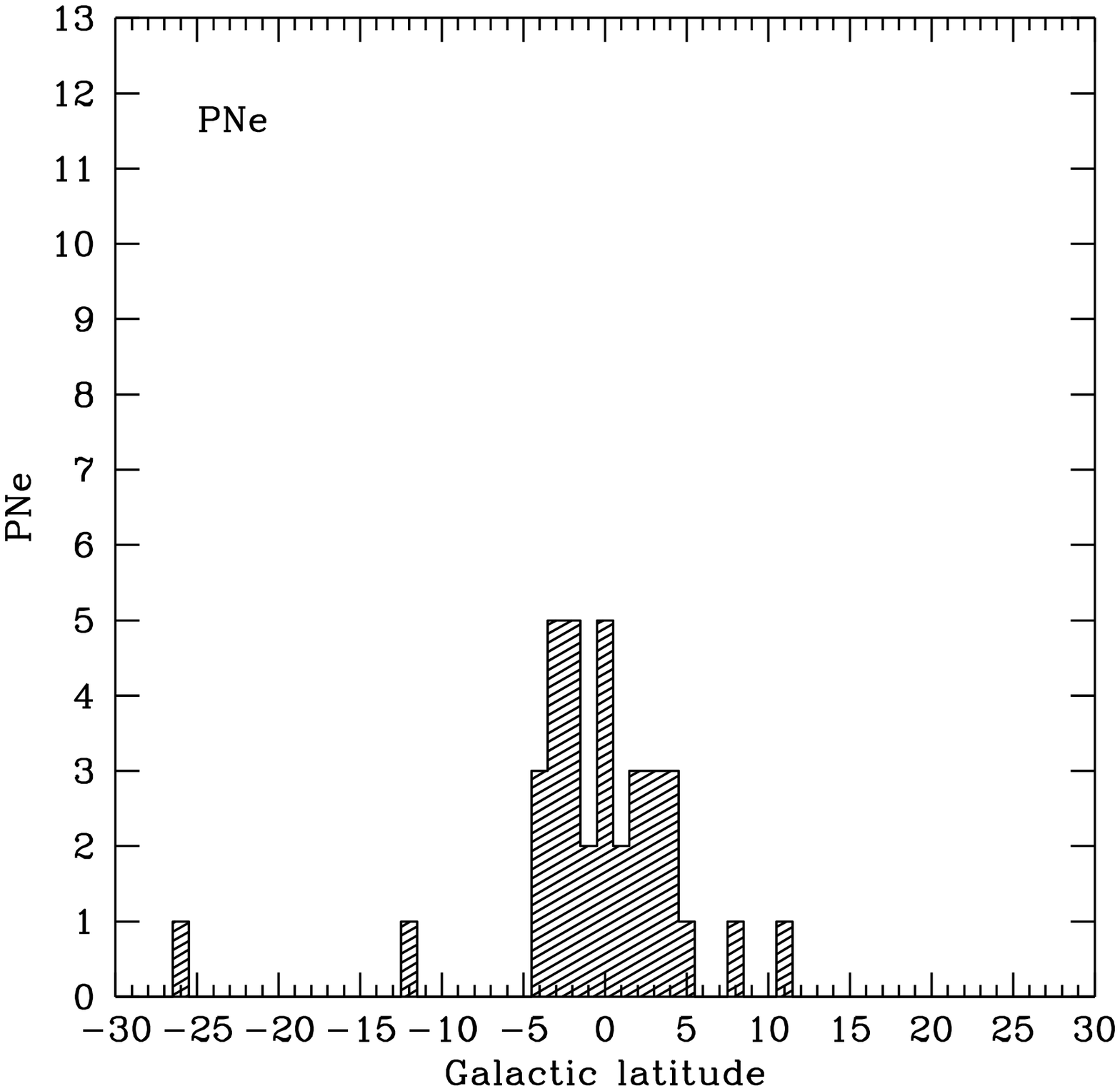}
 \includegraphics[width=7cm]{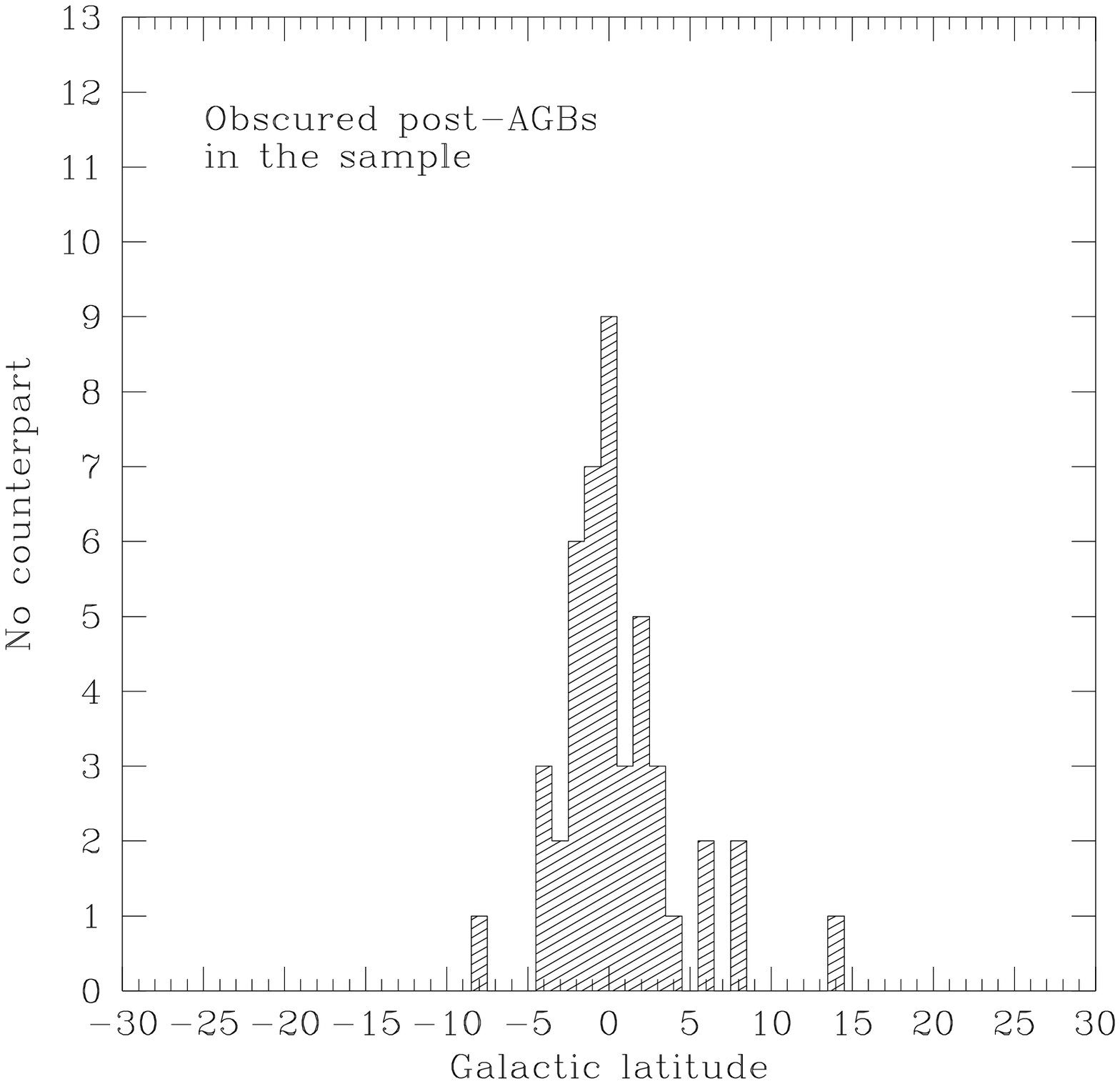}
 \caption{Galactic latitude distribution of optically bright post-AGB
 stars, PNe, and obscured post-AGB stars identified in the sample. }
 \label{histo}
\end{figure}

For this purpose, in Fig. \ref{histo}, we have plotted the galactic
latitude distribution of the optically bright post-AGB stars and
transition sources, PNe, and obscured post-AGB stars identified in our
sample.

To test if the apparent differences among the distributions shown in
Fig.~\ref{histo} are statistically significant, we have used the
F-Snedecor test (or F-variance test). This test determines whether two
samples are drawn from populations with the same variances.

The result of this study shows 
that there are no statistically significant differences between the
galactic distribution of optically bright post-AGB stars (including the
transition sources) and the galactic distribution of PNe 
with a 99\% confidence. This can be interpreted as the confirmation that 
an evolutionary link exists between the two populations.

However, the galactic distribution of post-AGB stars with no optical
counterparts seems to be different from both the post-AGB stars with
optical counterparts and the PNe. The narrow galactic latitude
distribution of the obscured post-AGB stars suggests that
they represent a population of more massive progenitors. These stars
may be evolving very rapidly to the PN stage and have very
recently left the AGB phase, as they still preserve the optically thick
envelopes formed in the mass-losing phase. In contrast, the post-AGB
stars showing a bright optical counterpart may represent a less
massive population of stars that would evolve more slowly.
Alternatively, they may never develop optically thick envelopes,
evolving all the way from the end of the AGB to the PN stage as
optically bright stars.

\subsection{Spectral type distribution of post-AGB stars}
\label{spectral}

\begin{figure}[h!]
 \centering
 \includegraphics[width=7.4cm]{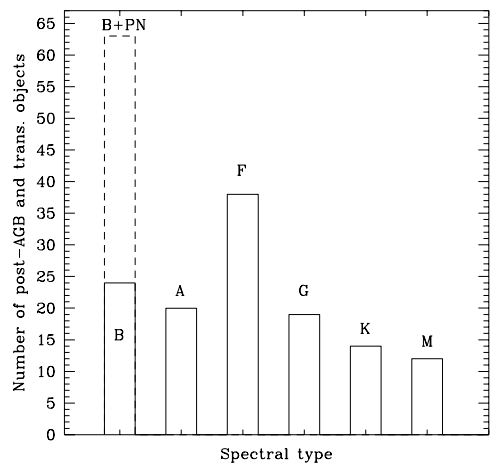}
 \caption{Relative distribution of spectral types for the post-AGB stars in 
our sample.}
 \label{tipos}
\end{figure}

We have found post-AGB stars belonging to all spectral types from M to
B, shown in Fig.~\ref{tipos}, in what can be interpreted as
an evolutionary sequence towards higher effective temperatures on
their way to becoming PNe. If this interpretation is correct, the
distribution of spectral types should be a good indicator of the time
spent by these stars in each range of temperatures. The same analysis
could be applied to high mass stars, but in this case, part of the
evolution would take place while the central star is still heavily
obscured by the circumstellar envelope, and by the time it could
be observed in the optical, it would have already
evolved significantly towards hotter temperatures.

The models developed by ~\citet{vanhoof97} (VH97, hereafter) predict
the relative number of stars that we should expect to find within each
spectral type. 

To test if the distribution of spectral types observed (see
Fig.~\ref{tipos}) is similar to any of the models produced by VH97, we
have applied the Pearson's Chi-square test. This test verifies if the
shape of the distribution of a given variable fits a theoretical
distribution with known parameters reasonably well.

The result of this analysis indicates that the distribution of
spectral types observed in our sample of post-AGB stars and transition
objects, excluding the M- and K- spectral types, is not compatible with
any of the three theoretical models used by VH97 with a 95\%
confidence level.

We suggest that the incompatibility can be due to the fact that our
sample is a combination of objects with different core masses, as
is to be expected from the selection criteria chosen, which do not favour
any particular core mass range.

\subsection{Distribution in the IRAS colour-colour diagram}
\label{IRASc_c}

\subsubsection{Overall distribution}

\begin{figure*}[t!]
 \centering
 \includegraphics[width=11cm]{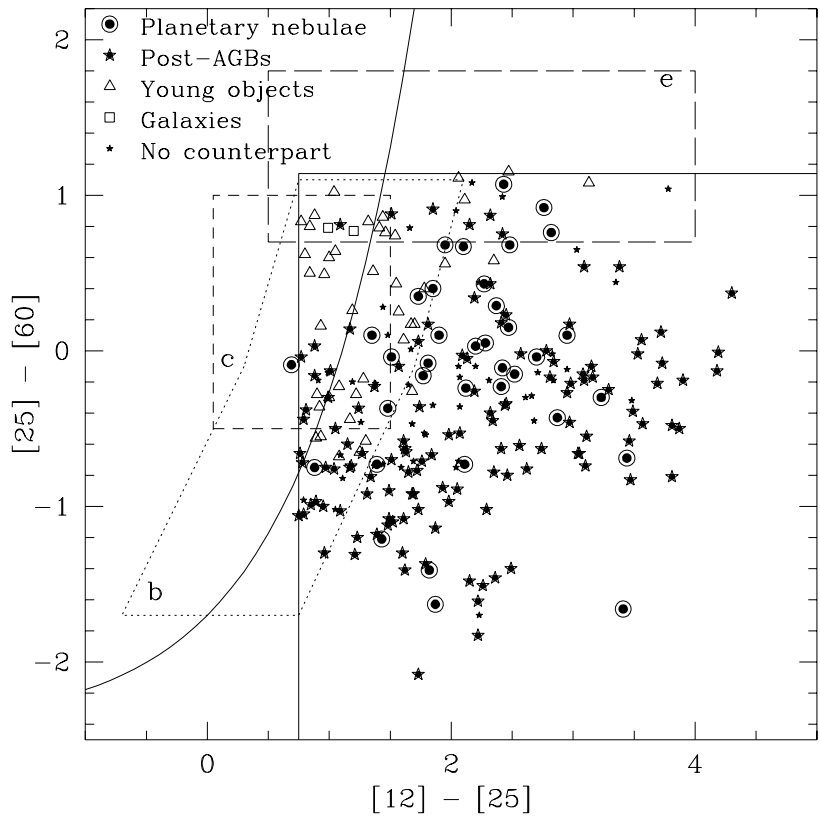}
 \caption{IRAS colour-colour diagram showing the location of all the objects 
included in our spectroscopic survey.}
 \label{completo}
\end{figure*}

In Fig.~\ref{completo} in the IRAS colour-colour diagram, we show the
distribution of all the sources included in our spectroscopic survey,
where special symbols are used to distinguish between the different
types of objects: PNe, optically bright post-AGB stars, galaxies,
young stellar objects, and objects for which no counterpart was found
(transition sources have again been included together with the
post-AGB stars).

The boxes with dashed boundary lines indicate areas mostly populated
by variable OH/IR stars (box $\it{b}$, dotted line), T-Tauri and Herbig Ae/Be
stars (box $\it{c}$, short dash), and compact HII regions (box $\it{e}$, long 
dash) \citep[see][~and references therein]{gl97}. In 
Fig.~\ref{completo} we have maintained the labels used in that paper.

The two solid straight lines (vertical and horizontal) show the limits
of the area defined by our selection criteria (see
Sect.~\ref{criteria}). The solid curve has been modeled by
\citet{bedijn}, and shows the location of O-rich stars as they evolve
along the AGB from M-type Miras to variable OH/IR stars.

 We can see that PNe and post-AGB stars spread almost regularly over the diagram.
However, it is important to remark that there is a high probability ($>$ 95\%)
of finding a post-AGB or a PN in the region that does not overlap with
regions $\it{b}$, $\it{c}$, or $\it{e}$.

Almost all of young sources (marked with open triangles) are located,
as expected, either in box $\it{c}$ (T-Tauri, Herbig Ae/Be, and
Vega-like stars) or $\it{e}$ (usually populated by protostars and
ultra-compact H II regions).

We have also plotted the objects for which we did not find any optical
counterpart as small asterisks. A few of them are located in box $e$,
suggesting that they could probably be heavily obscured young stellar
objects, surrounded by thick protoplanetary discs or by their parent
molecular cloud in which they are embedded. The rest of the sources
with no counterpart are more or less clustered around a value of
[25]$-$[60]$\sim$$-$0.2 and/or relatively close to box $b$, and these
are the ones that have been classified as heavily obscured post-AGB
stars in this paper.

\subsubsection{Post-AGB stars and transition objects}
\label{irasdiagrampagb}

\begin{figure*}[t!]
\centering
 \includegraphics[width=11cm]{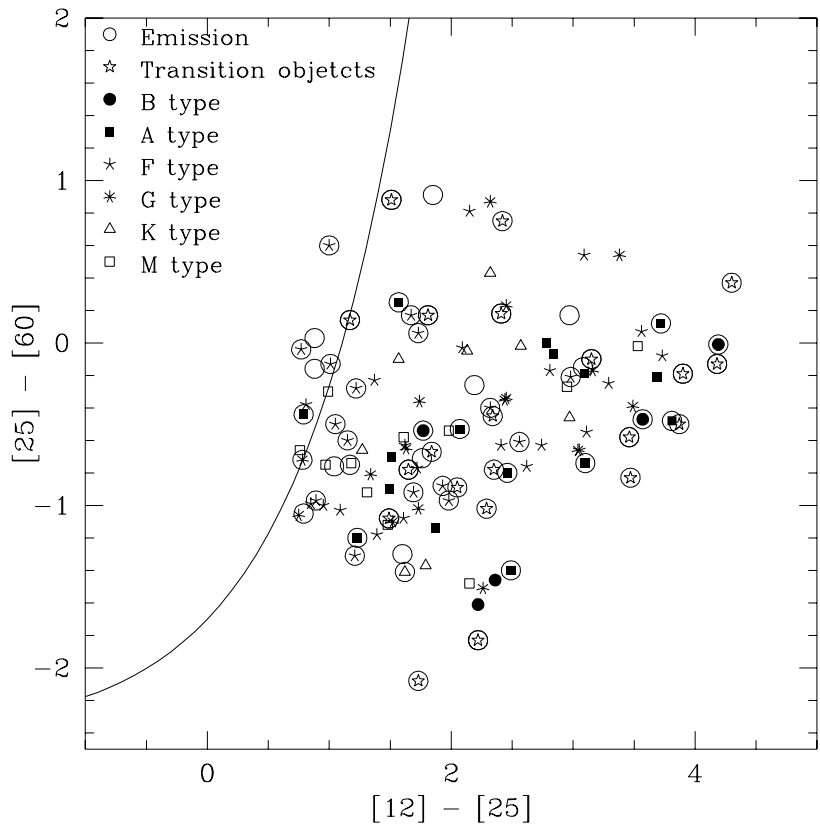}
 \caption{IRAS colour-colour diagram showing the location of the post-AGB stars and transition objects in the survey.}
 \label{cc_pagb}
\end{figure*}

In Fig. \ref{cc_pagb} we show more in detail the distribution of 
post-AGB stars and ``transition sources'' in the IRAS two-colour diagram as 
a function of the spectral type. Post-AGB stars with different spectral 
types are plotted with different colours, and a different
symbol is used to plot the transition objects. Objects showing
H$\alpha$ emission are surrounded with a circle. Sources with emission,
but for which we could not determine the spectral type because of their
faint continuum, are plotted as open circles.

It is already known that there is not a one-to-one correspondence
between the position of a post-AGB star in the IRAS two-colour diagram
and its evolutionary stage (VH97), but this is the first time that a large
sample of spectroscopically classified post-AGB stars show this
evidence.

Neither the C-rich models nor the O-rich models predict an unambiguous
relationship between the evolutionary stage in the post-AGB phase and
the position in the IRAS two colour diagram. Perhaps models for
different core masses and precise temporal marks could be used to create
statistics and check whether there is an adequate correspondance between 
the position of our objects and their evolutionary status.

\section{Conclusions}
\label{conclusion}

We have presented the results of an extensive spectroscopic survey 
carried out on
a sample of 253 IRAS sources showing infrared colour characteristics
of PNe. The selection criteria used turned out to be very efficient in
discovering new post-AGB stars and transition sources evolving from
the AGB to the PN stage, resulting in the identification of 152
post-AGB stars (49 of them  without an
optical counterpart), 21 transition sources, and 36 PNe. This
constitutes the largest catalogue of post-AGB stars built so far.

The results from the survey are presented in an atlas format, 
including  the spectroscopic data and finding charts.
Improved astrometric information and spectral classification is also 
provided.

Our spectroscopic study has substantially increased the knowledge of
the optical properties of galactic post-AGB stars, as can be seen
in Table~\ref{postagbtabla}; only 35\% of our 103 objects had
their spectral types identified in the SIMBAD database.

An analysis of the galactic latitude distribution of the sources in
the sample shows that  the new population of obscured post-AGB stars
identified in our spectroscopic survey  must in general have
higher progenitor masses than the population of post-AGB stars with
optical counterparts for which spectroscopic data is presented in the
atlas. The relatively large number of post-AGB stars showing optically thick
envelopes suggests that many stars may evolve during a significant
part of their post-AGB evolution hidden from detection in the optical.

The similar galactic latitude distribution shown by the 
PNe included in our sample and the population of post-AGB stars
with optical counterparts suggests that many of these post-AGB stars 
will  develop observable PNe, as they seem to belong to the same galactic
population. Unfortunately, we do not know whether the small sample of PNe
here studied is representative of the more general population of
galactic PNe.

The spectral type distribution of the post-AGB stars observed is
not compatible with any of the theoretical models developed by VH97
to predict the distribution expected for 
stars with several specific core masses. We suggest that this is due
to our sample being a combination of sources with a wide variety of core
masses.

The IRAS two-colour diagram has proved to be very useful in finding
new post-AGB stars. The distribution of the post-AGB stars and the PNe
in the diagram is similar, and suggests a direct evolutionary
connection. We have also confirmed that the precise location of a
given post-AGB star or a PNe in this diagram by itself cannot be
directly interpreted as an indication of their evolutionary stage, as
was suggested by previous theoretical works.

\begin{acknowledgements}

IRAF is distributed by the National Optical Astronomy Observatories,
which are operated by the Association of Universities for Research in
Astronomy, Inc., under cooperative agreement with the National Science
Foundation

This research was funded through grant AYA-2003-09499 from the Spanish
Ministerio de Ciencia y Tecnolog\'\i a. AM acknowledges support from
grant AYA2004-03136 funded by the Spanish Ministerio de Educacion y
Ciencia. It made extensive use of the Simbad database, operated at
CDS, Strasbourg (France), as well as of the Space Telescope Science
Institute Digitized Sky Survey (DSS).  The STScI DSS is based on
photographic data obtained using the UK Schmidt Telescope and produced
at the STScI under US Government grant NAG W-2166.  Some images
presented in the Appendices were
obtained at the 2.5 m Nordic Optical Telescope, operated jointly by
Denmark, Finland, Iceland, Norway, and Sweden, also at the Spanish
Observatorio del Roque de los Muchachos of the Instituto de Astrof\'\i
sica de Canarias.
\end{acknowledgements}

%------------------------------------------
%    TABLES AND APPENDIXES
%------------------------------------------

\clearpage
        %   TABLA de FUENTES OBSERVADAS

\clearpage
\onecolumn
{
\begin{landscape}

\setcounter{table}{2}
\begin{longtable}{cccrrrccl}
\caption{List of IRAS fields observed. The  astrometric coordinates are given for all sources but for those without optical counterparts and without precise infrared measurements. IRAS colours are defined in Sect.~\ref{observations}. The run number is referred to Table~\ref{catalog}.  \label{lista}}\\
\hline \hline 
GLMP& IRAS&RA (J2000)&Dec (J2000)&  [12]$-$[25]& [25]$-$[60]& Classification& Spectral type/class&  Run\\
& & & & & & & (not evolved)& \\
\hline
\endfirsthead
\caption[]{List of IRAS fields observed (continued).}\\
\hline\hline
GLMP& IRAS&RA (J2000)&Dec (J2000)&  [12]$-$[25]& [25]$-$[60]& Classification& Spectral type/class&  Run\\
& & & & & & & (not evolved) &\\
\hline
\endhead
\hline
\endlastfoot

  12 &   00509+6623 & 00:54:07.7 &    66:40:12.8 &      1.05 &       $-$1.02 & No counterpart &&                 \#11  \\
  15 &   01005+7910 & 01:04:45.5 &    79:26:46.3 &      1.98 &       $-$0.97 &       Post-AGB &&                  \#6  \\
  19 & 01156$-$5249 & 01:17:43.5 & $-$52:33:30.8 &      1.22 &       $-$0.28 &          Young &F6e&              \#4,\#15 \\
  20 &   01174+6110 & 01:20:44.2 &    61:26:15.9 &      1.54 &          0.74 &          Young &em&                  \#6  \\
  23 &   01259+6823 & 01:29:33.4 &    68:39:16.9 &      1.05 &       $-$0.50 &       Post-AGB &&                  \#6  \\
& & & & & & & \\                                                                                
  26 &   02143+5852 & 02:17:57.8 &    59:05:52.0 &      1.21 &       $-$1.31 &       Post-AGB &&             \#6,\#11  \\
  34 &   02528+4350 & 02:56:11.3 &    44:02:52.2 &      1.57 &          0.25 &          Young &A0e&                  \#6  \\
  53 &   04101+3103 & 04:13:20.0 &    31:10:47.3 &      0.90 &       $-$0.28 &          Young &A5e&                  \#6  \\
  58 &   04137+7016 &   ...      &    ...        &      1.26 &       $-$0.46 & No counterpart &&             \#6,\#11  \\
  63 &   04189+2650 & 04:22:02.2 &    26:57:30.5 &      1.05 &          0.64 &          Young &em&              \#8      \\
& & & & & & & & \\                                                                              
  74 &   04296+3429 & 04:32:57.0 &    34:36:12.4 &      1.39 &       $-$1.18 &       Post-AGB &&                 \#11  \\
  87 &   05089+0459 & 05:11:36.2 &    05:03:26.3 &      1.18 &       $-$0.74 &       Post-AGB &&              \#1      \\
  88 &   05113+1347 & 05:14:07.8 &    13:50:28.3 &      1.52 &       $-$1.10 &       Post-AGB &&              \#2      \\
  91 & 05209$-$0107 & 05:23:31.0 & $-$01:04:23.8 &      1.19 &          0.26 &          Young &F7e&              \#2,\#7  \\
  95 & 05238$-$0626 & 05:26:19.8 & $-$06:23:57.4 &      1.17 &       $-$0.44 &          Young &F4&          \#7,\#6      \\
& & & & & & & & \\                                                                              
  96 &   05245+0022 & 05:27:05.5 &    00:25:07.5 &      1.00 &       $-$0.30 &          Young &A03&              \#7      \\
 100 &   05273+2517 & 05:30:27.5 &    25:19:57.1 &      0.88 &          0.87 &          Young &A5e&                  \#6  \\
 106 &   05341+0852 & 05:36:55.1 &    08:54:08.7 &      0.85 &       $-$0.99 &       Post-AGB &&             \#7,\#6   \\
 109 & 05357$-$0217 & 05:38:14.1 & $-$02:15:59.7 &      1.25 &       $-$0.65 &          Young &F6e&          \#2,\#6,\#7  \\
 117 &   05381+1012 & 05:40:57.1 &    10:14:25.0 &      1.34 &       $-$0.81 &       Post-AGB &&             \#6,\#11  \\
& & & & & & & & \\                                                                              
 131 &   05471+2351 & 05:50:13.9 &    23:52:17.7 &      2.00 &          0.96 &       Peculiar &BQ[~]&             \#6,\#11  \\
 136 &   05573+3156 & 06:00:33.4 &    31:56:44.5 &      1.95 &          0.56 & No counterpart &&                  \#6  \\
 138 & 06013$-$1452 & 06:03:37.1 & $-$14:53:02.5 &      0.92 &       $-$0.36 &          Young &A03&     \#7,\#6,\#11      \\
 155 & 06464$-$1644 & 06:48:41.7 & $-$16:48:05.6 &      0.89 &       $-$0.56 &          Young &A0e&      \#6,\#8          \\
 159 &   06499+0145 & 06:52:28.2 &    01:42:12.0 &      3.03 &          0.65 & No counterpart &&             \#12      \\
& & & & & & & & \\                                                                              
 160 & 06518$-$1041 & 06:54:13.4 & $-$10:45:38.2 &      1.77 &       $-$0.16 &             PN &&            \#2,\#12   \\
 161 & 06530$-$0213 & 06:55:31.8 & $-$02:17:28.3 &      1.63 &       $-$0.65 &       Post-AGB &&             \#3       \\
 162 & 06549$-$2330 & 06:57:06.5 & $-$23:34:09.2 &      1.70 &          0.17 &          Young &F4&             \#3       \\
 163 &   06556+1623 & 06:58:30.4 &    16:19:26.1 &      1.77 &       $-$0.54 &       Peculiar &BeI-BQ[~]&            \#3,\#10   \\
 165 & 06562$-$0337 & 06:58:44.4 & $-$03:41:10.0 &      1.55 &          0.43 &          Young &em&\#6,\#7,\#8,\#10,\#12  \\
& & & & & & & & \\                                                                              
 170 & 07027$-$7934 & 06:59:26.4 & $-$79:38:47.0 &      1.39 &       $-$0.73 &             PN &&             \#3       \\
 175 &   07134+1005 & 07:16:10.3 &    09:59:48.0 &      1.69 &       $-$0.92 &       Post-AGB &&             \#3,\#6   \\
 178 & 07173$-$1733 & 07:19:35.9 & $-$17:39:18.0 &      0.84 &          0.50 &          Young &Be&                  \#2  \\
 182 & 07227$-$1320 & 07:25:03.1 & $-$13:26:19.9 &      0.76 &       $-$0.66 &       Post-AGB &&             \#2       \\
 183 & 07253$-$2001 & 07:27:33.0 & $-$20:07:19.6 &      0.95 &       $-$1.00 &       Post-AGB &&             \#2       \\
& & & & & & & & \\                                                                              
 186 & 07280$-$1829 & 07:30:16.7 & $-$18:35:49.1 &      2.11 &          0.97 &          Young &Ae&         \#2           \\
 188 &   07331+0021 & 07:35:41.2 &    00:14:57.9 &      1.62 &       $-$1.41 &       Post-AGB &&                  \#3  \\
 192 &   07430+1115 & 07:45:51.4 &    11:08:19.6 &      1.48 &       $-$1.12 &       Post-AGB &&             \#2       \\
 197 & 07506$-$0345 & 07:53:07.4 & $-$03:53:32.2 &      0.93 &       $-$0.55 &          Young &B7e&            \#2,\#10   \\
 201 & 07582$-$4059 & 07:59:57.7 & $-$41:07:23.3 &      1.74 &       $-$0.36 &       Post-AGB &&             \#3       \\
& & & & & & & & \\                                                                              
 202 & 08005$-$2356 & 08:02:40.7 & $-$24:04:42.7 &      1.15 &       $-$0.60 &       Post-AGB &&         \#1,\#8,\#12  \\
   - & 08046$-$3844 & 08:06:28.4 & $-$38:53:24.0 & $\ge$2.41 &       $-$0.23 &             PN &&                \#10   \\
 206 & 08143$-$4406 & 08:16:03.0 & $-$44:16:04.6 &      2.97 &       $-$0.46 &       Post-AGB &&              \#1,\#3  \\
 209 & 08187$-$1905 & 08:20:57.1 & $-$19:15:03.4 &      3.49 &       $-$0.39 &       Post-AGB &&              \#2,\#3  \\
 212 & 08213$-$3857 & 08:23:12.1 & $-$39:07:07.4 &      1.73 &          0.06 &       Post-AGB &&              \#3      \\
& & & & & & & & \\                                                                              
 214 & 08242$-$3828 & 08:26:03.8 & $-$38:38:47.5 &      0.91 &       $-$0.19 & No counterpart &&             \#12      \\
 218 & 08281$-$4850 & 08:29:40.6 & $-$49:00:04.3 &      1.61 &       $-$1.08 &       Post-AGB &&             \#2,\#12  \\
 222 & 08351$-$4634 & ...        &      ...      &      0.79 &       $-$0.96 & No counterpart &&             \#12      \\
 223 & 08355$-$4027 & 08:37:24.7 & $-$40:38:04.2 &      2.70 &       $-$0.04 &             PN &&             \#3,\#8   \\
 227 & 08418$-$4843 & 08:43:29.5 & $-$48:54:46.8 &      2.47 &          0.15 &             PN &&                  \#3  \\
& & & & & & & & \\                                                                              
 236 & 08574$-$5011 & 08:59:02.3 & $-$50:23:40.2 &      1.81 &       $-$0.08 &             PN &&          \#8          \\
 251 & 09149$-$6206 & 09:16:09.4 & $-$62:19:29.6 &      0.99 &          0.79 &         Galaxy &&              \#1      \\
 254 & 09362$-$5413 & 09:37:51.8 & $-$54:27:08.7 &      1.43 &       $-$1.21 &             PN &&            \#8        \\
 255 & 09370$-$4826 &   ...      &       ...     &      1.11 &       $-$0.82 & No counterpart &&              \#2      \\
 260 & 09425$-$6040 & 09:44:01.7 & $-$60:54:25.7 &      0.79 &       $-$1.05 &       Post-AGB &&              \#3      \\
& & & & & & & & \\                                                                              
 261 & 09500$-$5236 & 09:51:49.2 & $-$52:50:52.7 &      1.46 &          0.76 & No counterpart &&              \#2      \\
   - & 09517$-$5438 &   ...      &      ...      & $\ge$1.95 &          0.68 &             PN &&             \#10      \\
 262 & 10001$-$5857 & 10:01:48.1 & $-$59:12:12.5 &      1.00 &          0.60 &          Young &F3e&              \#1      \\
 263 & 10029$-$5553 & 10:04:40.1 & $-$56:08:37.2 &      2.11 &       $-$0.73 &             PN &&              \#8      \\
 268 & 10115$-$5640 & 10:13:19.7 & $-$56:55:32.3 &      2.12 &       $-$0.24 &             PN &&            \#8,\#10   \\
& & & & & & & & \\                                                                              
 270 & 10178$-$5958 & 10:19:32.5 & $-$60:13:29.3 &      2.42 &          0.75 &Transition      &&      \#3,\#10,\#12    \\
 272 & 10197$-$5750 & 10:21:33.9 & $-$58:05:47.7 &      1.84 &       $-$0.67 &Transition      &&                  \#10 \\
 273 & 10211$-$5922 & 10:22:53.9 & $-$59:37:28.4 &      1.72 &       $-$0.40 &       Peculiar &LBV&              \#2      \\
 275 & 10215$-$5916 & 10:23:19.5 & $-$59:32:04.7 &      2.35 &       $-$0.78 &Transition      &&    \#3,\#8,\#10,\#12  \\
 277 & 10256$-$5628 & 10:27:35.2 & $-$56:44:19.8 &      2.74 &       $-$0.63 &       Post-AGB &&         \#3,\#15      \\
& & & & & & & & \\                                                                              
 285 & 10594$-$3426 & 11:01:51.9 & $-$34:42:17.0 &      1.36 &          0.51 &          Young &K1e&              \#2      \\
 289 & 11065$-$6026 & 11:08:40.1 & $-$60:42:51.7 &      3.02 &       $-$0.50 &       Peculiar &LBV&          \#1,\#8,\#12 \\
 294 & 11195$-$2430 & 11:22:05.3 & $-$24:46:39.4 &      1.68 &       $-$0.26 &          Young &K5&              \#3      \\
 295 & 11201$-$6545 & 11:22:18.9 & $-$66:01:50.7 &      3.10 &       $-$0.74 &       Post-AGB &&              \#2      \\
 300 & 11307$-$5402 & 11:33:05.6 & $-$54:19:28.5 &      1.67 &          0.17 &          Young &F0e&             \#2,\#3   \\
& & & & & & & & \\                                                                              
 303 & 11339$-$6004 & 11:36:20.7 & $-$60:20:53.3 &      1.59 &       $-$0.75 & No counterpart &&             \#7,\#12  \\
   - & 11353$-$6037 & 11:37:42.9 & $-$60:53:51.4 & $\ge$4.30 &          0.37 &Transition      &&          \#10,\#15    \\
 306 & 11381$-$6401 & 11:40:32.0 & $-$64:18:34.9 &      1.78 &       $-$0.53 & No counterpart &&              \#3      \\
 307 & 11387$-$6113 & 11:41:08.7 & $-$61:30:17.3 &      2.07 &       $-$0.53 &       Post-AGB &&              \#3      \\
 312 & 11472$-$7834 & 11:49:31.8 & $-$78:51:01.1 &      1.28 &       $-$0.18 &          Young &M&             \#10      \\
& & & & & & & & \\                                                                              
   - & 11531$-$6111 & 11:55:38.0 & $-$61:28:16.8 & $\ge$2.34 &       $-$0.45 &Transition      &&          \#10,\#15    \\
 317 & 12002$-$5333 & 12:02:47.6 & $-$53:50:07.8 &      1.20 &          0.77 &         Galaxy &&              \#8      \\
 318 & 12067$-$4508 & 12:09:23.8 & $-$45:25:34.7 &      0.78 &       $-$0.72 &       Post-AGB &&                  \#2  \\
   - & 12145$-$5834 & 12:17:16.1 & $-$58:51:29.5 & $\ge$2.19 &       $-$0.26 &       Post-AGB &&             \#10      \\
 323 & 12262$-$6417 &   ...      &      ...      &      1.38 &       $-$0.21 & No counterpart &&             \#10      \\
& & & & & & & & \\                                                                              
 326 & 12302$-$6317 & 12:33:07.0 & $-$63:33:42.7 &      0.96 &       $-$1.30 &       Post-AGB &&        \#12,\#15      \\
 327 & 12309$-$5928 & 12:33:44.6 & $-$59:45:18.5 &      1.09 &       $-$0.67 & No counterpart &&              \#8      \\
 328 & 12316$-$6401 & 12:34:36.0 & $-$64:18:16.8 &      1.85 &          0.40 &             PN &&              \#7      \\
 335 & 12405$-$6219 & 12:43:31.5 & $-$62:36:13.5 &      2.04 &          0.90 & No counterpart &&              \#2,\#8  \\
 339 & 13010$-$6012 & 13:04:05.5 & $-$60:28:45.6 &      1.61 &       $-$0.58 &       Post-AGB &&         \#3,\#15      \\
& & & & & & & & \\                                                                              
 343 & 13185$-$6922 & 13:22:07.5 & $-$69:38:12.2 &      0.93 &          0.16 &          Young &K0e&              \#2      \\
 345 & 13203$-$5917 & 13:23:32.2 & $-$59:32:49.7 &      2.26 &       $-$1.51 &       Post-AGB &&             \#12      \\
 346 & 13208$-$6020 & 13:24:04.4 & $-$60:36:30.7 &      0.84 &          0.80 &          Young &em&              \#4      \\
 347 & 13245$-$5036 & 13:27:36.1 & $-$50:52:05.7 &      1.23 &       $-$1.20 &       Post-AGB &&              \#2      \\
 348 & 13266$-$5551 & 13:29:51.0 & $-$56:06:53.7 &      4.19 &       $-$0.01 &       Post-AGB &&               \#10    \\
& & & & & & & & \\                                                                              
 349 & 13293$-$6000 & 13:32:39.2 & $-$60:15:39.2 &      1.04 &          1.02 & No counterpart &&              \#4      \\
 352 & 13313$-$5838 & 13:34:37.4 & $-$58:53:32.3 &      1.27 &       $-$0.66 &       Post-AGB &&              \#8      \\
 353 & 13356$-$6249 &   ...      &    ...        &      3.48 &       $-$0.32 & No counterpart &&             \#3,\#12  \\
 360 & 13416$-$6243 & 13:45:07.3 & $-$62:58:16.9 &      1.24 &       $-$0.37 &       Post-AGB &&           \#2,\#15    \\
 362 & 13427$-$6531 & 13:46:25.7 & $-$65:46:24.4 &      2.68 &       $-$0.45 & No counterpart &&                \#3    \\
& & & & & & & & \\                                                                              
 363 & 13428$-$6232 & 13:46:20.5 & $-$62:47:59.6 &      2.97 &          0.17 &       Post-AGB &&               \#15    \\
 366 & 13483$-$6232 & 13:51:55.1 & $-$62:46:56.6 &      2.06 &          1.11 &          Young & -&                \#8    \\
 368 & 13500$-$6106 &    ...     &    ...        &      1.65 &       $-$0.22 & No counterpart &&               \#12    \\
 371 & 13529$-$5934 & 13:56:24.6 & $-$59:48:57.2 &      2.20 &       $-$0.10 & No counterpart &&                \#3    \\
 374 & 14079$-$6402 & 14:11:46.3 & $-$64:16:24.0 &      1.82 &       $-$1.41 &             PN &&                   \#3 \\
& & & & & & & & \\                                                                              
 377 & 14122$-$5947 & 14:15:53.3 & $-$60:01:37.9 &      1.48 &       $-$0.37 &             PN &&\#7,\#8,\#12           \\
 381 & 14177$-$5824 & 14:21:19.9 & $-$58:38:22.5 &      2.37 &          0.29 &             PN &&    \#1,\#7,\#8        \\
 384 & 14325$-$6428 & 14:36:34.4 & $-$64:41:31.1 &      2.41 &       $-$0.63 &       Post-AGB &&              \#1      \\
 385 & 14331$-$6435 & 14:37:10.1 & $-$64:48:04.7 &      3.57 &       $-$0.47 &       Post-AGB &&             \#10      \\
 386 & 14341$-$6211 & 14:38:05.0 & $-$62:24:47.6 &      1.68 &       $-$0.92 &       Post-AGB &&         \#3,\#15      \\
& & & & & & & & \\                                                                              
 387 & 14345$-$5858 & 14:38:20.0 & $-$59:11:46.1 &      2.20 &          0.03 &             PN &&        \#10           \\
 388 & 14346$-$5952 & 14:38:24.6 & $-$60:04:53.0 &      0.88 &          0.03 &       Post-AGB &&            \#8        \\
 391 & 14429$-$4539 & 14:46:13.8 & $-$45:52:05.1 &      0.89 &       $-$0.97 &       Post-AGB &&         \#1,\#15      \\
 393 & 14482$-$5725 & 14:51:57.3 & $-$57:38:19.0 &      1.49 &       $-$0.90 &       Post-AGB &&              \#3      \\
 394 & 14488$-$5405 & 14:52:28.7 & $-$54:17:42.7 &     2.458 &       $-$0.81 &       Post-AGB &&              \#2      \\
& & & & & & & & \\                                                                              
 399 & 14592$-$6311 & 15:03:23.8 & $-$63:22:58.9 &      0.80 &          0.62 &          Young & em&            \#1        \\
   - & 15039$-$4806 & 15:07:27.4 & $-$48:17:53.8 & $\ge$3.09 &       $-$0.19 &       Post-AGB &&            \#7,\#15   \\
 403 & 15066$-$5532 & 15:10:26.0 & $-$55:44:13.3 &      3.90 &       $-$0.19 &Transition      &&              \#3      \\
 404 & 15093$-$5732 &   ...      &    ...        &      1.48 &          0.10 & No counterpart &&        \#10,\#15      \\
 405 & 15103$-$5754 & 15:14:18.5 & $-$58:05:20.3 &      2.43 &          0.24 & No counterpart &&             \#8       \\
& & & & & & & & \\                                                                              
 408 & 15144$-$5812 & 15:18:21.9 & $-$58:23:12.0 &      1.67 &          0.01 & No counterpart &&            \#3        \\
 409 & 15154$-$5258 & 15:19:08.2 & $-$53:09:46.6 &      1.90 &          0.10 &             PN &&               \#1,\#7 \\
 411 & 15210$-$6554 & 15:25:31.7 & $-$66:05:20.0 &      1.79 &       $-$1.37 &       Post-AGB &&         \#3,\#15      \\
 416 & 15310$-$6149 & 15:35:17.1 & $-$61:59:04.1 &      1.51 &       $-$0.70 &       Post-AGB &&              \#3      \\
 422 & 15373$-$4220 & 15:40:46.4 & $-$42:29:53.6 &      1.61 &          0.07 &          Young & A79&              \#3      \\
& & & & & & & & \\                                                                              
 424 & 15406$-$4946 & 15:44:20.5 & $-$49:56:24.1 &      0.99 &       $-$0.30 &       Post-AGB &&              \#3      \\
 431 & 15482$-$5741 & 15:52:19.4 & $-$57:50:53.2 &      2.56 &       $-$0.61 &       Post-AGB &&             \#12      \\
 437 & 15534$-$5422 & 15:57:21.1 & $-$54:30:46.1 &      2.23 &          0.44 & No counterpart &&            \#10       \\
 441 & 15556$-$2248 & 15:58:36.9 & $-$22:57:15.3 &      1.41 &          0.79 &          Young & G8e&    \#1,\#7,\#13       \\
 442 & 15559$-$5546 & 15:59:57.4 & $-$55:55:34.0 &      0.88 &       $-$0.75 &             PN &&                 \#3   \\
& & & & & & & & \\                                                                              
 444 & 15579$-$5445 & 16:01:50.8 & $-$54:53:40.1 &      2.52 &       $-$0.15 &             PN &&       \#3,\#10        \\
 446 & 16053$-$5528 & 16:09:20.2 & $-$55:36:10.0 &      2.76 &          0.92 &             PN &&       \#7,\#12        \\
 456 & 16114$-$4504 & 16:15:03.0 & $-$45:11:54.5 &      2.27 &          0.43 &             PN &&        \#7            \\
 457 & 16115$-$5044 & 16:15:18.0 & $-$50:52:19.7 &      3.13 &          1.08 &          Young & -&        \#4            \\
   - & 16206$-$5956 & 16:25:02.6 & $-$60:03:32.3 & $\ge$3.72 &          0.12 &       Post-AGB &&              \#7      \\
& & & & & & & & \\                                                                              
 464 & 16228$-$5014 & 16:26:31.3 & $-$50:21:26.9 &      3.78 &          1.04 & No counterpart &&        \#4            \\
 466 & 16241$-$2412 & 16:27:10.3 & $-$24:19:18.9 &      2.35 &          0.58 &          Young & G1&             \#3       \\
 469 & 16279$-$4757 & 16:31:38.7 & $-$48:04:05.7 &      1.98 &       $-$0.54 &       Post-AGB &&             \#10      \\
 472 & 16283$-$4424 & 16:31:58.4 & $-$44:31:18.6 &      2.49 &       $-$1.40 &       Post-AGB &&              \#8      \\
 474 & 16289$-$4449 & 16:32:32.2 & $-$44:55:30.7 &      1.32 &          0.83 &          Young & em&            \#7        \\
& & & & & & & & \\                                                                              
 478 & 16333$-$4807 &   ...      &      ...      &      1.66 &          0.79 & No counterpart &&            \#7        \\
 488 & 16476$-$1122 & 16:50:24.3 & $-$11:27:57.8 &      1.31 &       $-$0.92 &       Post-AGB &&             \#11      \\
 490 & 16494$-$3930 & 16:52:55.4 & $-$39:34:56.2 &      1.73 &       $-$1.02 &       Post-AGB &&        \#12,\#15      \\
 494 & 16518$-$3425 &   ...      &      ...      &      2.23 &       $-$1.70 & No counterpart &&           \#10        \\
 495 & 16529$-$4341 & 16:56:34.0 & $-$43:46:14.7 &      2.95 &          0.10 &             PN &&    \#7,\#8,\#15       \\
& & & & & & & & \\                                                                              
 498 & 16552$-$3050 & 16:58:27.8 & $-$30:55:06.2 &      1.57 &       $-$0.10 &       Post-AGB &&                \#3    \\
 500 & 16559$-$2957 & 16:59:08.2 & $-$30:01:40.3 &      1.37 &       $-$0.74 & No counterpart &&        \#7,\#8,\#9    \\
 504 & 16584$-$3710 & 17:01:52.1 & $-$37:14:53.8 &      2.07 &       $-$0.36 & No counterpart &&           \#10        \\
 507 & 16594$-$4656 & 17:03:11.2 & $-$47:00:21.0 &      2.05 &       $-$0.89 &Transition      &&              \#3      \\
 509 & 17009$-$4154 & 17:04:29.6 & $-$41:58:38.7 &      2.66 &       $-$0.29 & No counterpart &&      \#10,\#15        \\
& & & & & & & & \\                                                                              
 510 & 17010$-$3810 & 17:04:27.3 & $-$38:14:41.7 &      2.06 &       $-$0.71 & No counterpart &&           \#10        \\
 522 & 17074$-$1845 & 17:10:24.2 & $-$18:49:00.7 &      3.47 &       $-$0.83 &Transition      &&     \#10,\#13,\#15    \\
 525 & 17086$-$2403 & 17:11:38.9 & $-$24:07:33.1 &      2.22 &       $-$1.83 &Transition      &&     \#4,\#13          \\
 527 & 17088$-$4221 &      ...   &      ...      &      1.19 &       $-$0.20 & No counterpart &&              \#7      \\
 526 & 17088$-$4227 & 17:12:21.8 & $-$42:30:50.3 &      3.23 &       $-$0.30 &             PN &&              \#8      \\
& & & & & & & & \\                                                                              
 531 & 17106$-$3046 & 17:13:51.8 & $-$30:49:40.7 &      2.98 &       $-$0.21 &       Post-AGB &&        \#10,\#15      \\
 535 & 17119$-$5926 & 17:16:21.1 & $-$59:29:23.3 &      3.44 &       $-$0.69 &             PN &&\#7,\#8,\#10,\#12,\#15 \\  
 537 & 17130$-$4029 & ...        &     ...       &      2.83 &       $-$0.02 & No counterpart &&              \#4      \\
 538 & 17131$-$3330 & 17:16:26.2 & $-$33:33:23.8 &      2.43 &          1.07 &             PN &&              \#8      \\
 542 & 17153$-$3814 & 17:18:44.7 & $-$38:17:21.2 &      2.17 &          1.08 & No counterpart &&              \#3      \\
& & & & & & & & \\                                                                              
 546 & 17164$-$3226 & 17:19:40.8 & $-$32:29:52.7 &      2.48 &          0.68 &             PN &&         \#8,\#15      \\
 547 & 17168$-$3736 &      ...   &      ...      &      1.44 &          0.28 & No counterpart &&         \#3,\#15      \\
 554 & 17195$-$2710 & 17:22:43.6 & $-$27:13:36.9 &      1.17 &       $-$0.75 &       Post-AGB &&                  \#11 \\
   - & 17203$-$1534 & 17:23:11.9 & $-$15:37:15.1 &      3.81 &       $-$0.48 &       Post-AGB &&              \#7      \\
 556 & 17208$-$3859 & 17:24:19.5 & $-$39:01:46.0 &      2.84 &       $-$0.07 &       Post-AGB &&              \#4      \\
& & & & & & & & \\                                                                              
 558 & 17223$-$2659 & 17:25:25.2 & $-$27:02:03.3 &      2.15 &       $-$1.48 &       Post-AGB &&         \#3,\#15      \\
 563 & 17234$-$4008 & 17:26:56.1 & $-$40:11:03.7 &      2.31 &       $-$0.19 & No counterpart &&        \#10,\#15      \\
 565 & 17245$-$3951 & 17:28:04.6 & $-$39:53:44.3 &      2.81 &       $-$0.17 &       Post-AGB &&         \#4,\#15      \\
 567 & 17253$-$2831 & 17:28:33.0 & $-$28:33:25.8 &      2.95 &       $-$0.27 &       Post-AGB &&              \#4      \\
 572 & 17269$-$2235 & 17:29:58.7 & $-$22:37:44.5 &      3.35 &          0.44 & No counterpart &&              \#7      \\
& & & & & & & & \\                                                                              
 574 & 17287$-$3443 & 17:32:04.8 & $-$34:45:32.6 &      3.81 &       $-$0.81 &       Post-AGB &&        \#10,\#15      \\
 575 & 17291$-$2402 & 17:32:12.8 & $-$24:04:59.9 &      2.41 &          0.18 &Transition      &&              \#7      \\
 578 & 17300$-$3509 & 17:33:22.6 & $-$35:11:03.7 &      2.32 &          0.87 &       Post-AGB &&              \#4      \\
 580 & 17310$-$3432 & 17:34:20.8 & $-$34:34:54.9 &      2.78 &          0.00 &       Post-AGB &&        \#10,\#15      \\
 581 & 17311$-$4924 & 17:35:02.5 & $-$49:26:26.4 &      2.29 &       $-$1.02 &Transition      &&                \#8    \\
& & & & & & & & \\                                                                              
 584 & 17317$-$2743 & 17:34:53.3 & $-$27:45:11.5 &      3.56 &          0.07 &       Post-AGB &&           \#11,\#15   \\
 588 & 17332$-$2215 & 17:36:17.7 & $-$22:17:24.9 &      2.57 &       $-$0.02 &       Post-AGB &&         \#4,\#15      \\
 591 & 17347$-$3139 & 17:38:00.6 & $-$31:40:55.2 &      1.81 &          0.17 &Transition      &&             \#15      \\
   - & 17364$-$1238 & 17:39:16.9 & $-$12:40:29.7 & $\ge$1.76 &       $-$0.71 &       Post-AGB &&               \#9     \\
 607 & 17370$-$3357 & 17:40:20.2 & $-$33:59:14.4 &      2.45 &          0.23 &       Post-AGB &&             \#15      \\
& & & & & & & & \\                                                                              
 608 & 17371$-$2747 & 17:40:23.3 & $-$27:49:11.7 &      1.73 &          0.35 &             PN &&         \#4,\#15      \\
 611 & 17376$-$2040 & 17:40:38.6 & $-$20:41:52.6 &      0.81 &       $-$0.38 &       Post-AGB &&                \#11   \\
   - & 17381$-$1616 & 17:40:59.8 & $-$16:17:58.9 & $\ge$3.41 &       $-$1.66 &             PN &&                 \#9   \\
 616 & 17388$-$2203 & 17:41:49.0 & $-$22:05:15.9 &      3.16 &       $-$0.17 &       Post-AGB &&         \#7,\#15      \\
 619 & 17392$-$3020 & 17:42:30.4 & $-$30:22:10.7 &      1.09 &          0.81 &       Post-AGB &&             \#15      \\
& & & & & & & & \\                                                                              
 621 & 17395$-$0841 & 17:42:14.4 & $-$08:43:19.5 &      2.82 &          0.76 &             PN &&         \#4,\#15      \\
 631 & 17418$-$3335 & 17:45:08.7 & $-$33:36:06.2 &      2.25 &       $-$0.21 & No counterpart &&               \#15    \\
 632 & 17423$-$1755 & 17:45:14.2 & $-$17:56:46.9 &      1.51 &          0.88 &Transition      &&              \#4      \\
 639 &   17436+5003 & 17:44:55.5 &    50:02:39.5 &      3.69 &       $-$0.21 &       Post-AGB &&                \#13   \\
 640 & 17441$-$2411 & 17:47:13.5 & $-$24:12:51.4 &      1.62 &       $-$0.63 &       Post-AGB &&             \#15      \\
& & & & & & & & \\                                                                              
 641 & 17443$-$2949 &    ...     &      ...      &      0.99 &       $-$0.14 & No counterpart &&              \#4      \\
 646 & 17460$-$3114 & 17:49:16.6 & $-$31:15:18.1 &      1.30 &       $-$0.58 &          Young & B3&                 \#8   \\
 647 & 17466$-$3031 & 17:49:52.5 & $-$30:33:02.5 &      1.17 &          0.14 &Transition      &&             \#15      \\
 651 & 17476$-$4446 & 17:51:16.4 & $-$44:47:28.9 &      2.22 &       $-$1.61 &       Post-AGB &&             \#10      \\
   - & 17488$-$1741 & 17:51:48.9 & $-$17:42:25.8 & $\ge$3.09 &          0.54 &       Post-AGB &&                 \#9   \\
& & & & & & & & \\                                                                              
 661 & 17514$-$1555 & 17:54:21.1 & $-$15:55:52.0 &      1.51 &       $-$0.04 &             PN &&              \#4      \\
 662 & 17516$-$2526 & 17:54:43.4 & $-$25:26:28.0 &      0.88 &       $-$0.16 &       Post-AGB &&             \#11      \\
 669 & 17542$-$0603 & 17:56:56.0 & $-$06:04:09.7 &      1.60 &       $-$1.30 &       Post-AGB &&                \#11   \\
 684 & 17576$-$2653 & 18:00:49.6 & $-$26:53:12.7 &      1.87 &       $-$1.14 &       Post-AGB &&           \#11,\#15   \\
 686 & 17579$-$3121 & 18:01:12.8 & $-$31:21:59.7 &      3.73 &       $-$0.08 &       Post-AGB &&              \#4      \\
& & & & & & & & \\                                                                              
 687 & 17580$-$3111 & 18:01:19.6 & $-$31:11:22.2 &      1.69 &       $-$0.71 & No counterpart &&              \#4      \\
 689 & 17582$-$2619 & 18:01:21.6 & $-$26:19:37.3 &      2.07 &       $-$0.17 & No counterpart &&               \#4     \\
 698 & 17597$-$1442 & 18:02:38.3 & $-$14:42:02.8 &      2.28 &          0.05 &             PN &&     \#8,\#11          \\
 711 & 18019$-$3403 & 18:05:13.7 & $-$34:03:16.4 &      2.36 &       $-$1.46 &       Post-AGB &&             \#10      \\
 713 & 18025$-$3906 & 18:06:03.3 & $-$39:05:56.6 &      2.45 &       $-$0.34 &       Post-AGB &&              \#4      \\
& & & & & & & & \\                                                                              
   - & 18044$-$1303 & 18:07:15.3 & $-$13:03:29.0 & $\ge$2.15 &          0.81 &       Post-AGB &&              \#9      \\
 723 & 18061$-$2505 & 18:09:12.4 & $-$25:04:34.5 &      1.35 &          0.10 &             PN &&           \#10,\#11   \\
 724 &   18062+2410 & 18:08:20.1 &    24:10:43.3 &      1.73 &       $-$2.08 &Transition      &&   \#5,\#8,\#12,\#13   \\
 726 & 18075$-$0924 & 18:10:15.1 & $-$09:23:35.1 &      2.44 &       $-$0.35 &       Post-AGB &&              \#4      \\
 737 & 18096$-$3230 & 18:12:57.5 & $-$32:30:08.9 &      3.38 &          0.54 &       Post-AGB &&              \#4      \\
& & & & & & & & \\                                                                              
 744 & 18108$-$3248 & 18:14:10.5 & $-$32:47:34.4 &      1.44 &          0.86 &          Young & K1e&              \#7      \\
 759 & 18186$-$0833 & 18:21:21.1 & $-$08:31:42.4 &      2.87 &       $-$0.43 &             PN &&              \#4      \\
 771 & 18216$-$0156 & 18:24:14.3 & $-$01:54:24.2 &      2.47 &          1.15 &          Young & em&              \#4      \\
 777 & 18246$-$1032 &   ...      &       ...     &      2.42 &          0.99 & No counterpart &&              \#4      \\
& & & & & & & & \\                                                                              
   - & 18401$-$1109 & 18:42:57.1 & $-$11:06:53.0 &      2.10 &          0.67 &             PN &&             \#10      \\
 822 & 18415$-$2100 & 18:44:32.0 & $-$20:57:12.8 &      1.04 &       $-$0.76 &       Peculiar &RCrB star&               \#10    \\
 823 & 18420$-$0512 & 18:44:41.7 & $-$05:09:17.0 &      3.53 &       $-$0.02 &       Post-AGB &&                 \#11  \\
 829 & 18442$-$1144 & 18:47:04.0 & $-$11:41:12.0 &      3.15 &       $-$0.10 &Transition      &&              \#7      \\
 844 &   18520+0007 & 18:54:34.8 &    00:11:04.4 &      2.42 &       $-$0.11 &             PN &&                \#14   \\
& & & & & & & & \\                                                                              
 845 &   18524+0544 & 18:54:54.1 &    05:48:11.3 &      2.95 &       $-$0.12 & No counterpart &&             \#10      \\
 849 &   18533+0523 & 18:55:46.7 &    05:27:03   &      2.61 &       $-$0.30 & No counterpart &&         \#7,\#14      \\
 855 &   18576+0341 &    ...     &      ...      &      1.68 &       $-$0.47 & No counterpart &&                \#15   \\
 858 &   18582+0001 & 19:00:49.0 &    00:06:14.1 &      2.32 &          0.43 &       Post-AGB &&           \#11,\#15   \\
 869 & 19016$-$2330 & 19:04:43.6 & $-$23:26:09.1 &      1.65 &       $-$0.78 &Transition      && \#4,\#6,\#7,\#15      \\
& & & & & & & & \\                                                                              
 879 &   19024+0044 & 19:05:02.1 &    00:48:50.9 &      3.08 &       $-$0.15 &       Post-AGB &&           \#11,\#15   \\
 875 & 19063$-$3709 & 19:09:45.9 & $-$37:04:26.1 &      0.96 &          0.49 &          Young & K5e&              \#4      \\
 883 &   19083+0119 & 19:10:54.5 &    01:24:45.0 &      2.06 &       $-$0.10 & No counterpart &&              \#4      \\
 890 &   19114+0002 & 19:13:58.6 &    00:07:31.9 &      3.29 &       $-$0.25 &       Post-AGB &&              \#4,\#13 \\
 895 &   19154+0809 & 19:17:50.6 &    08:15:08.5 &      1.87 &       $-$1.63 &             PN &&             \#11      \\
& & & & & & & & \\                                                                              
   - &   19200+3457 & 19:21:55.3 &    35:02:55.1 &      2.32 &  $\ge$$-$0.41 &       Post-AGB &&              \#9      \\
 907 &   19207+2023 & 19:22:55.8 &    20:28:54.8 &      3.05 &       $-$0.66 &       Post-AGB &&              \#4      \\
 908 &   19208+1541 & 19:23:05.9 &    15:47:31.3 &      0.77 &          0.83 & No counterpart &&              \#4      \\
 914 &   19225+3013 & 19:24:26.9 &    30:19:26.7 &      0.97 &       $-$0.75 &       Post-AGB &&             \#11      \\
 923 &   19306+1407 & 19:32:55.1 &    14:13:36.9 &      3.04 &       $-$0.66 &       Post-AGB &&              \#4      \\
& & & & & & & & \\                                                                              
 935 &   19356+0754 & 19:38:02.2 &    08:01:33.8 &      2.13 &       $-$0.05 &       Post-AGB &&              \#4      \\
 941 &   19386+0155 & 19:41:08.3 &    02:02:31.3 &      1.09 &       $-$1.03 &       Post-AGB &&           \#11,\#15   \\
   - &   19422+1438 & 19:44:31.7 &    14:45:24.6 & $\ge$2.09 &       $-$0.03 &       Post-AGB &&                 \#9   \\
 950 &   19454+2920 &    ...     &       ...     &      1.79 &       $-$0.54 & No counterpart &&                \#11   \\
 952 &   19477+2401 & 19:49:54.9 &    24:08:53.3 &      1.72 &       $-$0.77 &       Post-AGB &&                \#11   \\
& & & & & & & & \\                                                                              
 954 & 19500$-$1709 & 19:52:52.7 & $-$17:01:50.4 &      1.93 &       $-$0.88 &       Post-AGB &&           \#14,\#15   \\
 961 &   19589+4020 & 20:00:43.0 &    40:29:09.6 &      2.62 &       $-$0.76 &       Post-AGB &&            \#11       \\
 962 &   19589+3419 & 20:00:52.9 &    34:28:22.2 &      0.69 &       $-$0.09 &             PN &&            \#11       \\
   - & 19590$-$1249 & 20:01:49.8 & $-$12:41:17.8 &      3.87 &       $-$0.50 &Transition      &&             \#7,\#8   \\
 987 &   20160+2734 & 20:18:05.9 &    27:44:03.6 &      0.77 &       $-$0.04 &       Post-AGB &&                \#13   \\
& & & & & & & & \\                                                                              
 988 &   20174+3222 & 20:19:27.8 &    32:32:15.2 &      2.04 &       $-$0.75 & No counterpart &&               \#5,\#6 \\
 998 &   20259+4206 & 20:27:42.3 &    42:16:44.1 &      1.37 &       $-$0.23 &       Post-AGB &&                \#11   \\
1003 &   20406+2953 & 20:42:46.0 &    30:04:06.4 &      1.85 &       $-$0.35 & No counterpart &&                 \#6   \\
1008 &   20462+3416 & 20:48:16.6 &    34:27:24.3 &      4.18 &       $-$0.13 &Transition      &&      \#11,\#13,\#14   \\
1012 &   20490+5934 & 20:50:13.6 &    59:45:51.2 &      1.78 &          0.40 &          Young & A3e&                \#11   \\
& & & & & & & & \\                                                                              
1013 &   20559+6416 & 20:15:58.4 &    47:05:35.9 &      1.08 &       $-$0.23 &          Young & A3e&            \#6,\#11   \\
1015 &   20572+4919 & 20:58:55.6 &    49:31:13.2 &      1.01 &       $-$0.13 &       Post-AGB &&            \#9,\#11   \\
1034 &   21289+5815 & 21:30:22.8 &    58:28:52.0 &      0.79 &       $-$0.44 &       Post-AGB &&                \#6    \\
1044 &   21537+6435 & ...        &        ...    &      1.44 &       $-$0.73 & No counterpart &&                \#6    \\
1047 &   21546+4721 & 21:56:33.0 &    47:36:12.8 &      1.49 &       $-$1.08 &Transition      &&                \#5    \\
& & & & & & & & \\                                                                              
1051 &   22023+5249 & 22:05:30.3 &    53:21:33 &      3.46 &       $-$0.58 &Transition      &&             \#5,\#6   \\
1052 &   22036+5306 & 22:05:30.3 &    53:21:32.8 &      1.85 &          0.91 &       Post-AGB &&                 \#6   \\
1058 &   22223+4327 & 22:24:31.4 &    43:43:10.9 &      3.11 &       $-$0.55 &       Post-AGB &&                \#13   \\
1073 & 23198$-$0230 & 23:22:24.7 & $-$02:13:41.4 &      1.08 &       $-$0.68 &          Young &G2e&      \#6,\#7,\#15     \\

\end{longtable}

\end{landscape}
}

        %   TABLA DE POST-AGBS
\clearpage
\onecolumn
\setcounter{table}{3}
\begin{longtable}{lcccc}
\caption{Main characteristics derived for the IRAS sources identified as post-AGB stars. \label{postagbtabla}}\\
\hline \hline 
IRAS name& Other identifications& Spectral type&Spectral type& Reliability \\
& &our survey &(SIMBAD) & code{$^\dagger$} \\   
\hline
\endfirsthead
\caption[]{Main characteristics derived of the IRAS sources identified as Post AGB stars (continued).}\\
\hline
IRAS name& Other identifications&Spectral type&Spectral type& Reliability \\
& &our survey &(SIMBAD) & code{$^\dagger$}
  \\   
\hline
\endhead
\hline
\endlastfoot

01005$+$7910&            &Fe  	& B2Iab:e & AA \\
01259$+$6823&            &F5Ie  & GIab:   & AA \\
02143$+$5852&            &F7Ie  & A5      & AA\\
04296$+$3429&            &F7I 	& G0Ia    & AA \\
05089$+$0459&            &M3I   & M       & AB\\
& & & & \\
05113$+$1347&  PM 2-4   &G5I  	& G8Ia    & AA \\
05341$+$0852&            &F5I  	& F4Iab:  & AA \\
05381$+$1012&            &G2I  	& G       &AA\\ 
06530$-$0213&  PM 1-24  &G1I  	& F0Iab:  & AA \\
07134$+$1005&            &F7Ie	& F5Iab:  & AA \\
& & & & \\
07227$-$1320&            &M1I  	&         & AA \\
07253$-$2001&            &F2I  	&         &AA \\
07331$+$0021&            &K35I 	& G5Iab   & AA\\
07430$+$1115&            &M2I  	& G5Ia    & AA\\
07582$-$4059&  PM 1-35  &G5I  	&         & AA \\
& & & & \\
08005$-$2356&            &F5Ie  & F5e     & AA \\
08143$-$4406&  PM 1-39  &K12I  &         & AA \\
08187$-$1905&            &G1I   & F6Ib/II &  AA \\
08213$-$3857&            &F2Ie  & F3V     & AA\\
08281$-$4850&  PM 1-40  &F0I   &         & AA \\
& & & & \\
09425$-$6040&            &C     &         &  AA   \\
10256$-$5628&  PM 2-11  &F5I  	&         & AA \\
11201$-$6545&  PM 1-56  &A3Ie 	&         &AA\\
11387$-$6113&            &A3Ie 	&         &AA\\
12067$-$4508&Hen 3-755   &G1Ie  & G2w...  &AA\\
& & & & \\
12145$-$5834&  PM 1-64  &em    &         &AA\\
12302$-$6317&            &*    	&         & AA  \\
13010$-$6012&  PM 1-72  &M2I 	&         & CA\\
13203$-$5917&            &G2I 	&         &AA\\
13245$-$5036&            &A79Ie &         &AA\\
& & & & \\
13266$-$5551&            &B1Ie	& O+...   &AA\\
13313$-$5838&            &K5I   & K1III   &AA\\
13416$-$6243&            &*     &          &AA\\
13428$-$6232&  PM 2-14  &em    &         &AA\\
14325$-$6428&            &F5I   &         &AA\\
& & & & \\
14331$-$6435&Hen 3-1013  &B8Ie  & B3Iab:e &AA\\
14341$-$6211&  PM 1-85  &*     &         & BA \\
14346$-$5952&            &em    	&         &AA    \\
14429$-$4539&            &F4Ie  & M       &AA\\  
14482$-$5725&  PM 1-86  &A2I   &         &AA\\
& & & & \\
14488$-$5405&            &A0Ie  &         &AA\\
15039$-$4806&            &A0I   & A1/A2Ib/&AA\\
15210$-$6554&  PM 2-16  &K2I  	&         &BA\\
15310$-$6149&            &A7I  	& F0V     &AA\\
15406$-$4946&            &M4II 	&         &AA\\
& & & & \\
15482$-$5741&            &F7I  	&         &AA\\
16206$-$5956&            &A3Ie 	& A3Iab:e &AA\\
16279$-$4757&  PM 2-18  &M3II  &         &AA\\
16283$-$4424&  PM 2-19  &A2Ie  &         &AA\\
16476$-$1122&            &M1I 	&         &AA\\
& & & & \\
16494$-$3930&            &G2I  	&         &CA\\
16552$-$3050&  PM 1-120 &K0I 	&         &AA\\
17106$-$3046&  PM 2-23  &F5I 	&         &AA \\
17195$-$2710&  PN G358.7+05.1&em	&         &AA\\
17203$-$1534&            &A0Ie  & B1IIIpe &AA\\
& & & & \\
17208$-$3859&  RPZM 14  &A2I   &         &AA\\
17223$-$2659&  PM 2-26  &M5III &         &AA\\
17245$-$3951&  PM 2-27  &F6I   &         &AA\\
17253$-$2831&            &M4II  &         &AA\\
17287$-$3443&            &*     &         &CA\\
& & & & \\
17300$-$3509&  PM 1-156  &G2I &     &AA\\
17310$-$3432&  PM 1-157 &A2I   &         &AA\\
17317$-$2743&  PM 1-158 &F5I   &         &CA \\
17332$-$2215&  PM 1-160 &K2I   &         &AA\\
17364$-$1238&  PM 1-167 & em   &         &BB\\ 
& & & & \\
17370$-$3357&  PM 1-169 &G3I   &         &AA\\
17376$-$2040&            &F6I   &         &BA\\
17388$-$2203&            &G0I   &         &AA\\
17392$-$3020&            &*     &         &AA\\
17436$+$5003&            &A7I   & F3Ib    &AA\\                 
& & & & \\
17441$-$2411&            &F4I   & F5:     &AA\\
17476$-$4446&  PM 2-29  &B7I   &         &AA \\
17488$-$1741&  PM 1-184 &F7I   &         &BA \\
17516$-$2526&            &em    &         &AB \\ 
17542$-$0603&  PM 2-32  &em    & Ge      &AA\\ 
& & & & \\
17576$-$2653&  PM 1-198 &A7I   &         &BA\\
17579$-$3121&  RPZM 44    PM 2-33&F4I &     &  AA\\
18019$-$3403&            &B8I   &         &  AA \\
18025$-$3906&  PM 2-34  &G1I   & G2      & AA \\
18044$-$1303&  PM 1-212 &F7I   &         & AA\\
& & & & \\
18075$-$0924&            &G2I   &         & AA \\
18096$-$3230&  PM 1-218 &G3I   &         &  AA\\
18420$-$0512&  PM 1-255 &M1I   &         & AA\\
18582$+$0001&  PM 1-272 &K2I   &         & AA\\
19024$+$0044&            &em    &         & AA \\
& & & & \\
19114$+$0002&            &F7I   & G5Ia    & AA \\
19200$+$3457&  PM 1-300 &Fe    & B...    & AA \\
19207$+$2023&            &F6I   &         & AA \\
19225$+$3013&            &M2II  &         & AA \\ 
19306$+$1407&            &G5I   & B0:e    & AA \\
& & & & \\
19356$+$0754&            &K2I   &         & AA \\
19386$+$0155&            &F5I   & F       & AA \\  
19422$+$1438&  PM 1-312 &F5I   &         & AA \\
19477$+$2401&            &F4I-F7I&         & AA \\
19500$-$1709&            &F0Ie  & F2/F3Iab& AA \\
& & & & \\
19589$+$4020&  PM 1-315 &F5I   &         & AA \\
20160$+$2734&            &F3Ie  &         & AA \\
20259$+$4206&            &F3I   &         & AA \\
20572$+$4919&            &F3Ie  & Fe      & AA \\
21289$+$5815&            &A2Ie  &         & AA \\
& & & & \\
22036$+$5306&            &em    &         & AA \\  
22223$+$4327&            &F7I   & G0Ia    & AA \\

\end{longtable}
\noindent $\dagger$ This code rates our confidence in
the data presented, with A representing the maximum reliability. The
first letter marks our confidence in the correct identification of the
optical counterpart, and the second letter indicates how confident we
are in the evolutionary classification assigned.

\noindent * Faint and red continuum.

        \normalsize

\clearpage
\setcounter{table}{4}
\begin{table*}[ht!]
\begin{center}
\caption{ \label{transtabla}Main characteristics derived for the IRAS sources identified as transition objects.}
\begin{tabular}{lccccc}
\hline \hline 
IRAS name& Other names & Spectral type&Spectral type& Reliability&Morphology  \\
& &our survey &(SIMBAD) & code$^\dagger$ & \\   
\hline
10178$-$5958&PN G285.1-02.7  Hen 3-401 &BeI    & Be        &AA & Bipolar [1]\\
10197$-$5750&Hen 3-404   &A0Ie   & A2Iabe    &AA & Bipolar [2]\\
10215$-$5916&         &A7eI   & K0        & AA\\ 
11353$-$6037&PM 1-60  &B5Ie   &           &AA\\
11531$-$6111&PM 1-61  &B8eI   &           &AA\\
15066$-$5532&PM 2-15  &       &           &AA\\
16594$-$4656&PN G340.3-03.2&Ae     & B7        &AA & Multipolar [3]\\
17074$-$1845&Hen 3-1347&B3eI   & B3IIIe    &AA\\
17086$-$2403&PN G359.8+08.9 PM 2-22&       & G5IV-V    & AA\\
17291$-$2402&PN G002.5+05.1 PM 1-155&       &           &AA\\
17311$-$4924&PN G341.4-09.0 Hen 3-1428&B3Ie   &  B1IIe    &AA&\\
17347$-$3139&RPZM 28  &       &          &AA& Bipolar [4]\\
17423$-$1755&PN G009.3+05.7 Hen 3-1475 &       & Be         &AA& Bipolar [5]\\
17466$-$3031& RPZM 42  &       &           &AA\\
18062$+$2410&PN G050.6+19.7  &B3e    & B1IIIpe   &AA\\
18442$-$1144& PN G022.0-04.3 &       & A3V       & AA \\
19016$-$2330&PN G013.1-13.2  &       &           &AA\\
19590$-$1249&PN G029.1-21.2  & B2eI  & B1Ibe      &AA\\
20462$+$3416&PN G076.6-05.7  &       &  B1Iae        & AA\\
21546$+$4721&PN G095.0-05.5  &       &           & AA \\
22023$+$5249&PN G099.3-01.9  &       &  Be        & AA \\
\hline
\end{tabular}
\end{center}
\noindent $\dagger$ This code rates our confidence in
the data presented, with A representing the maximum reliability. The
first letter marks our confidence in the correct identification of the
optical counterpart, and the second letter indicates how confident we
are in the evolutionary classification assigned.

\noindent [1]~\citet{gl99};
[2]~\citet{allen80};
[3]~\citet{gl99b};
[4]~\citet{itzi04};
[5]~\citet{riera95}.

\end{table*}
\normalsize

        \begin{landscape}

\setcounter{table}{5}
\begin{longtable}{llccccccc}
\caption{\label{PNtabla}
Main characteristics derived from the optical images and spectra  of
the IRAS sources identified as  PNe.}\\
\hline \hline 
IRAS name& PN G name& Other names &Morphology& Size (\arcsec)& c &Excitation class & WR subtype&Code$^\dagger$\\
\hline
\endfirsthead
\caption[]{Characteristics of PNe (continued).}\\
\hline \hline
IRAS name& PN G name& Other names &Morphology& Size (\arcsec)& c &Excitation class & WR subtype &Code$^\dagger$\\
\hline
\endhead 
\hline 
\endlastfoot
06518$-$1041&222.8$-$04.2& PM 1-23  & Bipolar  & 20   & 1.1&  &[WC7]     & AA\\
07027$-$7934&291.3$-$26.2&             & Round    & $<$15& 2.3& &[WC10]    & AA \\
08046$-$3844&255.3-03.6  &             & Compact  &  -   &1.5&Medium   &   & AA \\
08355$-$4027&260.7+00.9  & PM 1-41  & Round    &  12 &3.0&Medium   &   & AA \\
08418$-$4843&                &             & -        &  -   &2.5&Medium   &   & AA \\
& & & & & & & &\\		         
08574$-$5011&270.1$-$02.9&             & Round    & 10   &2.7&Medium   &   & AA \\
09362$-$5413&277.1$-$01.5&             & Compact  & $<$2 &2.8&  Low    &   & AA \\
09517$-$5438&279.1-00.4  &             & -        &  -   &2.4&Medium     &    & AA \\
10029$-$5553&281.1$-$00.4& PM 2-10  & Compact  &$<$3.4&High&  -  &    & AB \\
10115$-$5640&282.6$-$00.4& PM 1-51  & Compact  &$<$6.1&High&  -  &    & AB \\
& & & & & & & &\\		         
12316$-$6401&301.1$-$01.4& PM 1-68  & Compact  &$<$2.3&1.1 &Medium&      & AA\\
14079$-$6402&                &             & Compact  &   -  & 2.2&  Low    &  & AA \\
14122$-$5947&313.3+01.1  &             & Elliptical/Bipolar& $<$4.5&High& -& & AA\\
14177$-$5824&314.4+02.2  & PM 1-81  & Bipolar  &  6   &4.9& Medium    &      & AA \\
14345$-$5858&316.2+00.8  &             & -        &$<$9.9&High& -&       & AA\\
& & & & & & & &\\		         
15154$-$5258&324.0+03.5  & PM 1-89  & Round    & 10   & 0.9& &[WC4] & AA \\
15559$-$5546&327.1-02.2  &  Hen 2-142  & Elliptical/Bipolar& 5.3$\times$3.2&1.4&  &[WC9] & AA \\
15579$-$5445&328.0$-$01.6& PM 1-105 &  -       &$<$1.8& High& - & & AA\\
16053$-$5528&328.4$-$02.8& PM 1-106 &  -       &  -   & 2.8& Low & & AA\\
16114$-$4504&336.1+04.1  & PM 2-17  & Elliptical?&  8 & 3.0& High& &AA \\
& & & & & & & &\\		         
16529$-$4341&342.2$-$00.3& PM 1-119 & -        &$<$4.7&High& - & & AA\\
17088$-$4227&344.9$-$01.9& PM 1-131 & -        &$<$2.2&High& - & & AA \\
17119$-$5926&331.3-12.1  &  Hen 3-1357 & Elliptical/Bipolar&2.1$\times$1.9& 0.3& Medium& &AA \\
17131$-$3330&                & PM 1-136 RPZM 10& -        &$<$1.6&High& - & & AA\\
17164$-$3226&                & PM 1-140 RPZM 12 & - &$<$2.1&2.84& Medium& &CA \\
& & & & & & & &\\		         
17371$-$2747&                & OH6      & -        &-     &4.7&  Low& &AA\\
17381$-$1616&010.2+07.5  & PM 2-28  & -         &-      &High& - & & AA \\
17395$-$0841&017.0+11.1  &             & Bipolar  &5$\times$8&1.9 &Low & & AA \\
17514$-$1555&012.2+04.9  & PM 1-188 & Round    &  18  &1.2& &[WC9]  & AA \\
17597$-$1442&014.2+03.8  & PM 1-205 & Round    &  20  &0.8& &[WC9] & AA \\
& & & & & & & &\\		         
18061$-$2505&005.9$-$02.6& PM 1-213 & Bipolar  &15$\times$46&2.1& Low & & AA \\
18186$-$0833&021.9+02.7  & PM 1-226 & Round    & 8   &2.31  &Medium & & AA \\
18401$-$1109&022.0$-$03.1& PM 1-252 & Elliptical&9$\times$11&1.0& High & & AA\\
18520$+$0007&                &             & Compact  &-     & High& - &    & AB\\
19154$+$0809&043.2$-$02.0& PM 2-40  & Compact  &-     &2.7& Low &  & AA \\
19589$+$3419&070.9+02.2  & PM 1-316 & Compact? & -    &-& Low & & AA \\

\end{longtable}
\noindent $\dagger$ This code rates our confidence in
the data presented, with A representing the maximum reliability. The
first letter marks our confidence in the correct identification of the
optical counterpart, and the second letter indicates how confident we
are in the evolutionary classification assigned.

\noindent * Spectroscopic range covered not enough to assign a classification.
\end{landscape}

\clearpage
\appendix 
\section{Atlas of post-AGB stars}
        \begin{figure*}[h!]

\begin{center}
\epsfxsize=13.5cm
\epsfysize=4cm
\epsfbox{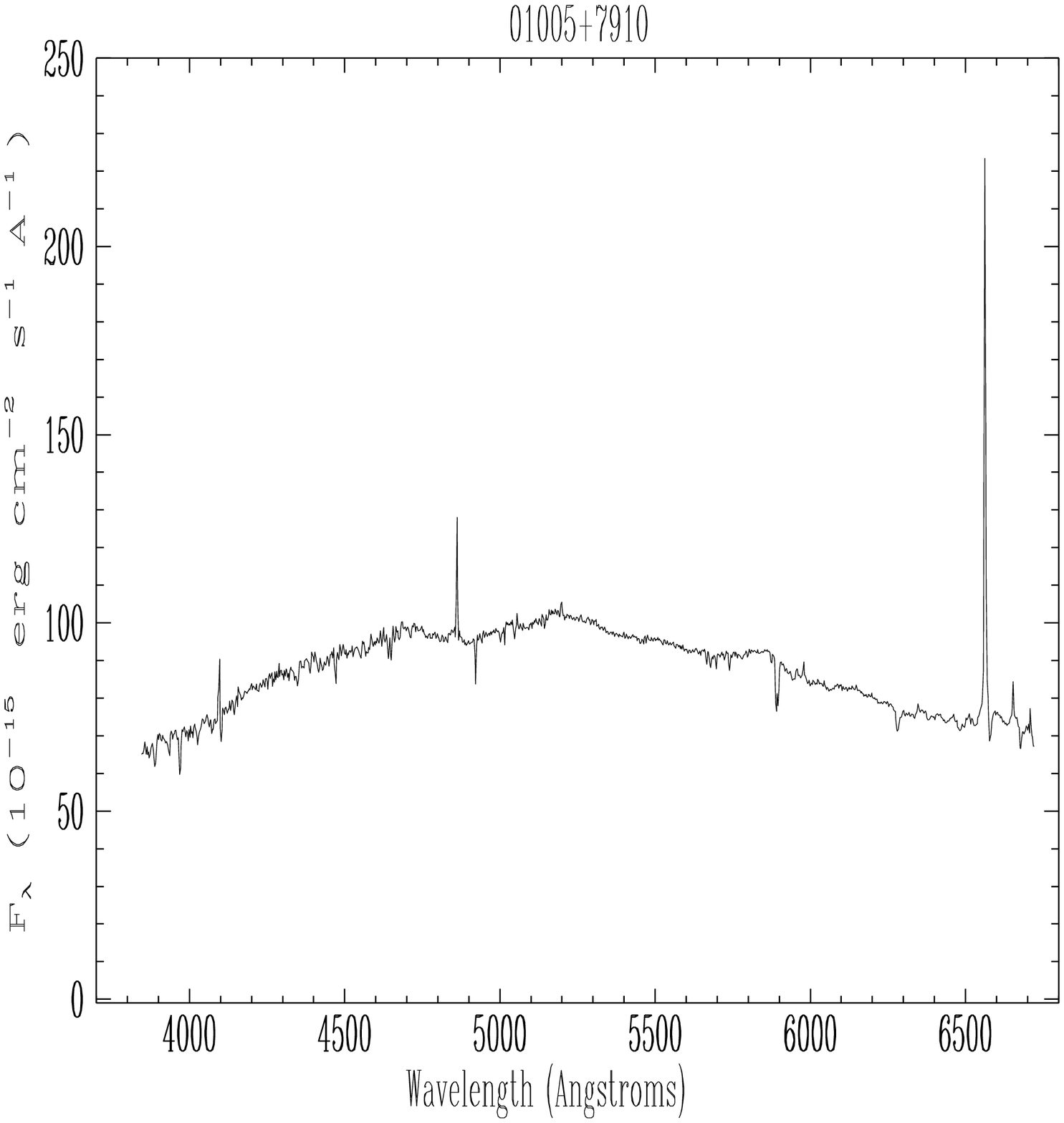}
%\psdraft
\epsfxsize=4cm
\epsfysize=4cm
\epsfbox{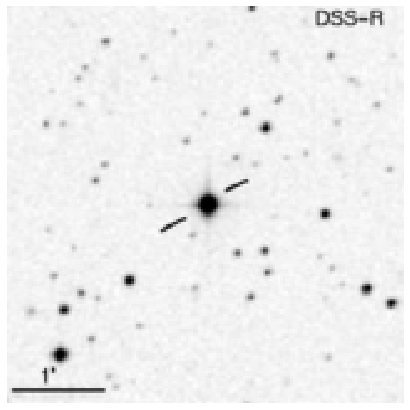}
%\psfull
\end{center}

\begin{center}
\epsfxsize=13.5cm
\epsfysize=4cm
\epsfbox{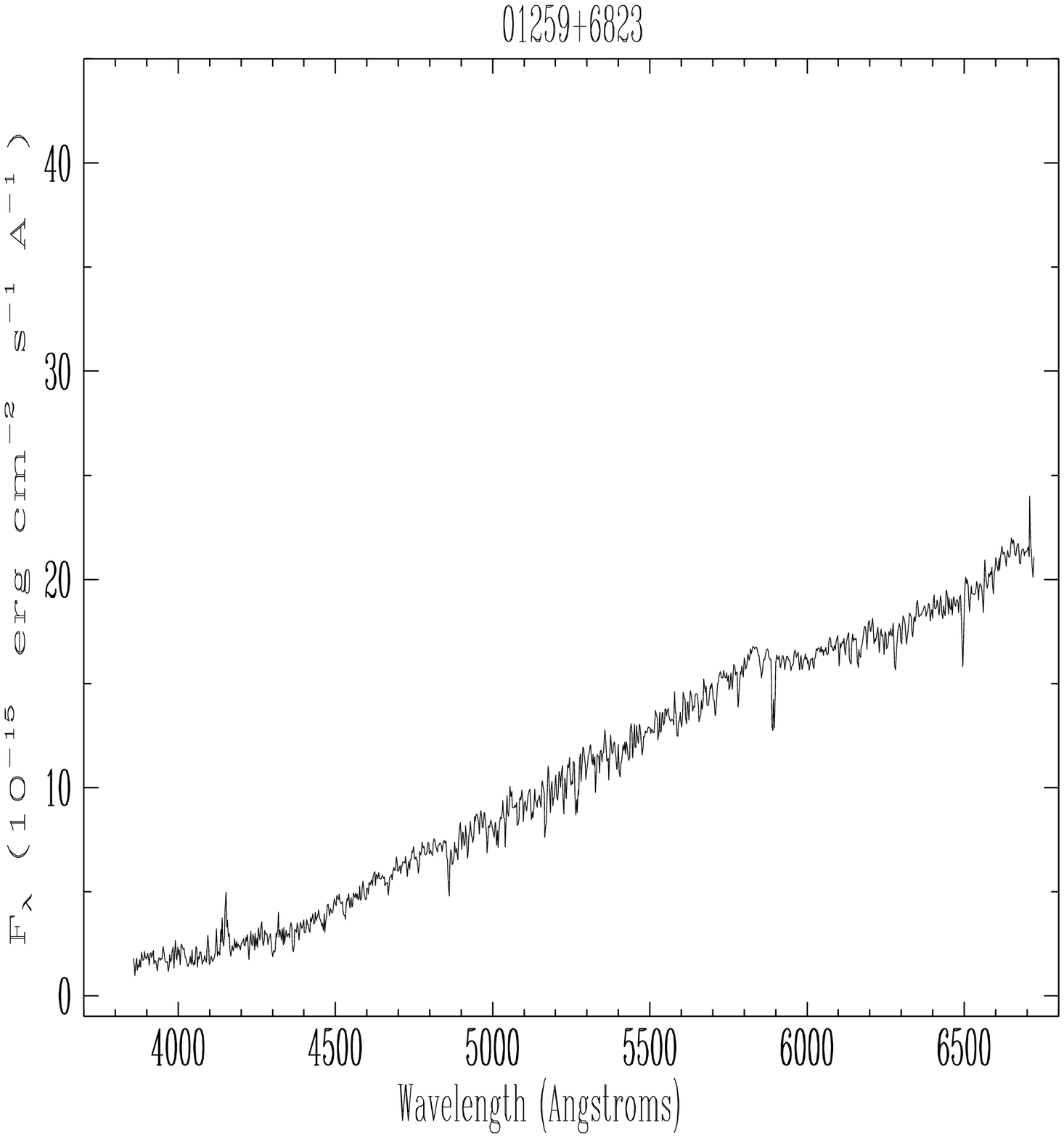}
%\psdraft
\epsfxsize=4cm
\epsfysize=4cm
\epsfbox{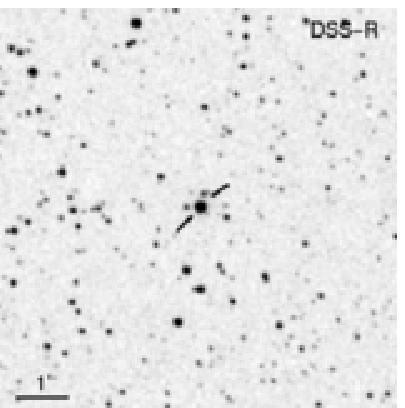}
%\psfull
\end{center}

\begin{center}
\epsfxsize=13.5cm
\epsfysize=4cm
\epsfbox{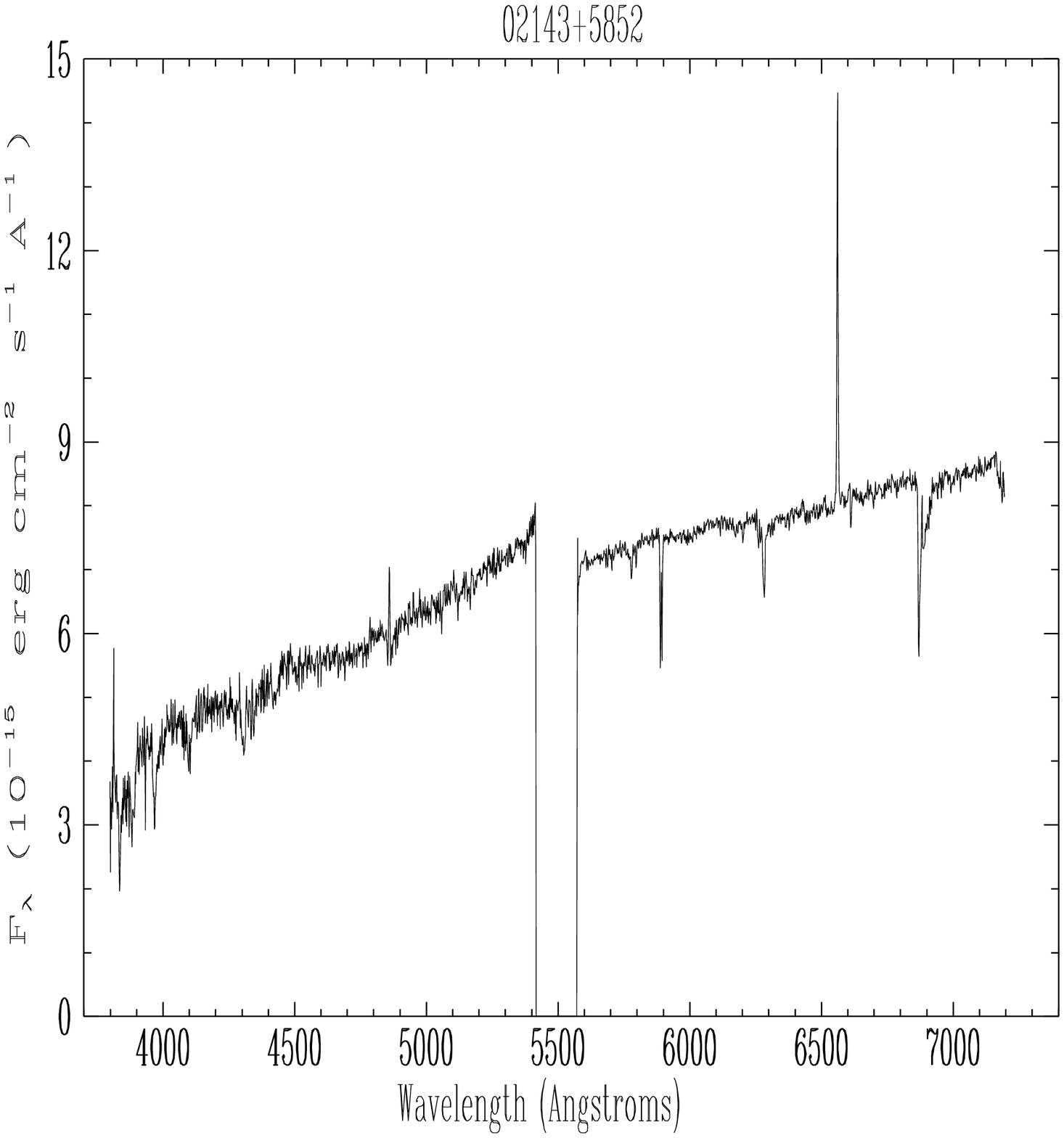}
%\psdraft
\epsfxsize=4cm
\epsfysize=4cm
\epsfbox{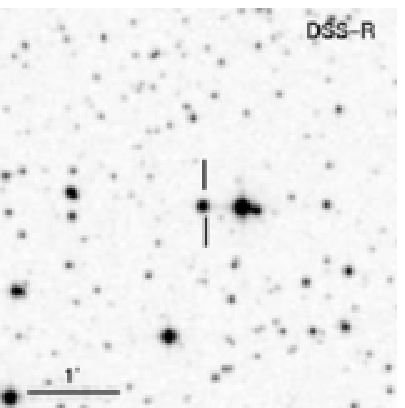}
%\psfull
\end{center}

\begin{center}
\epsfxsize=13.5cm
\epsfysize=4cm
\epsfbox{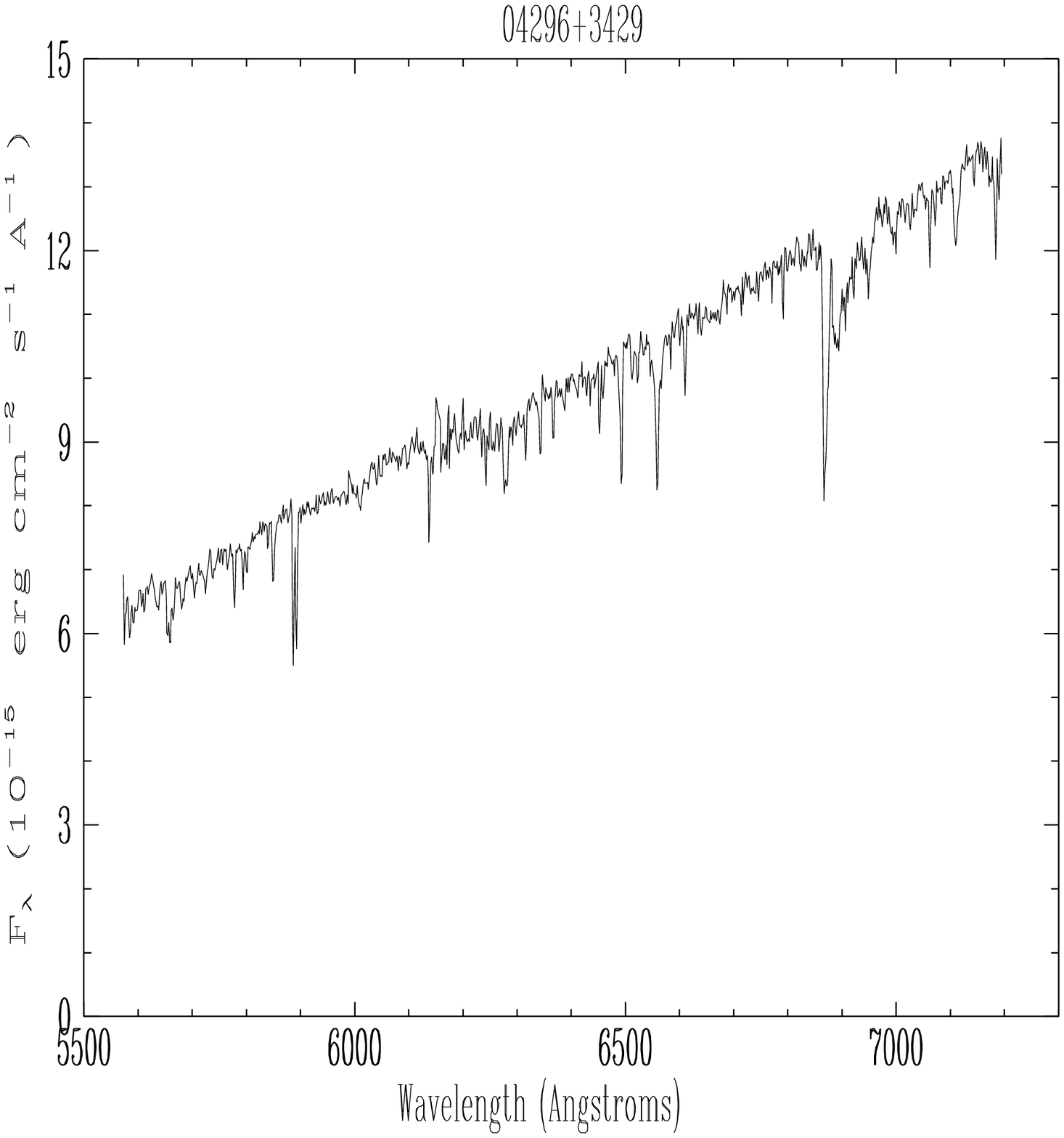}
%\psdraft
\epsfxsize=4cm
\epsfysize=4cm
\epsfbox{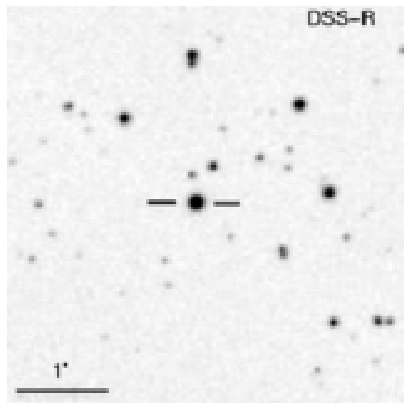}
%\psfull
\end{center}

\caption{Spectra of the objects classified as post-AGB in the sample together with their 
corresponding identification charts. }
\end{figure*}

%-------------------------------------------------------------
%pg2

\clearpage

\setcounter{figure}{0}
\begin{figure*}

\begin{center}
\epsfxsize=13.5cm
\epsfysize=4cm
\epsfbox{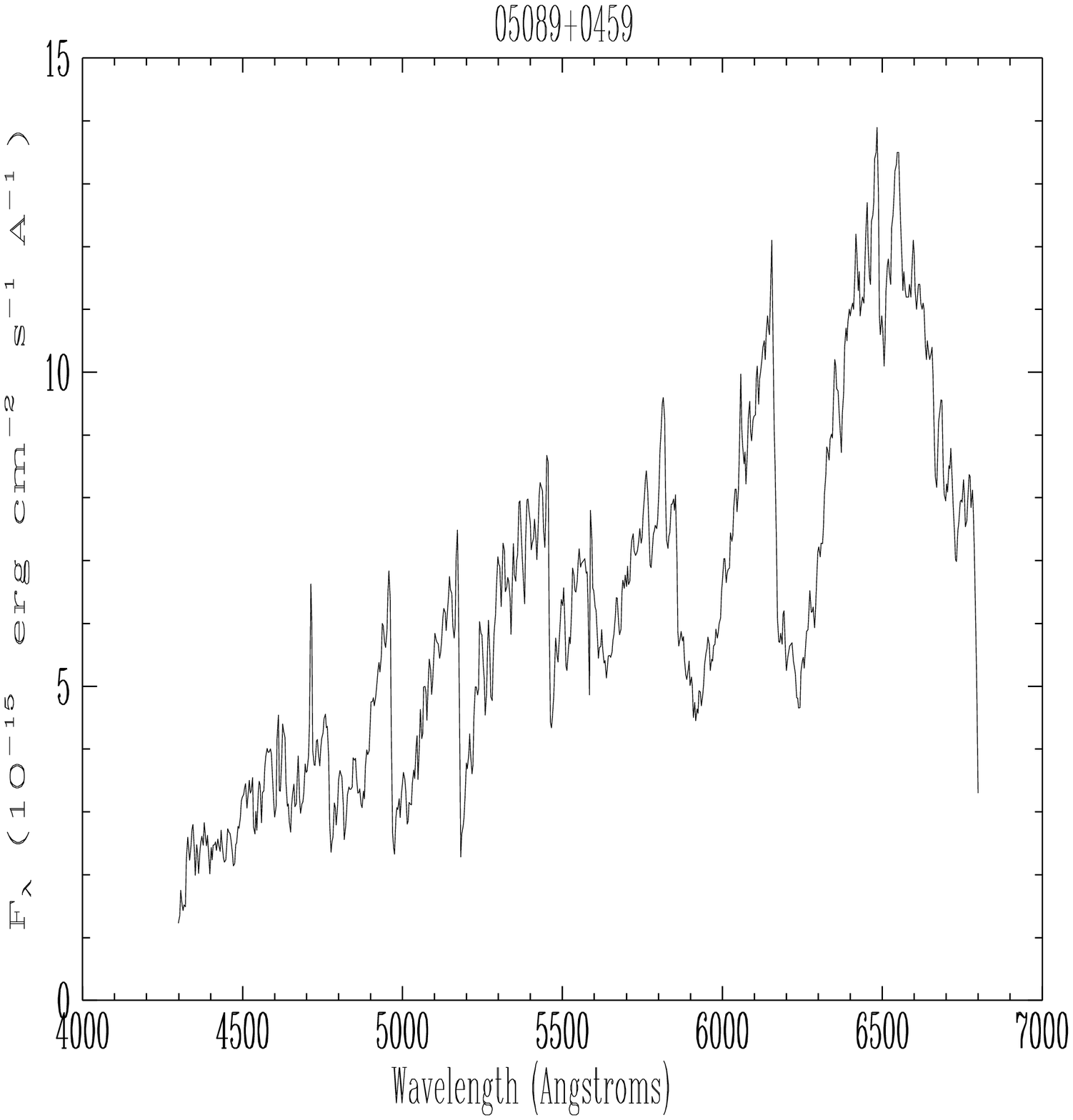}
%\psdraft
\epsfxsize=4cm
\epsfysize=4cm
\epsfbox{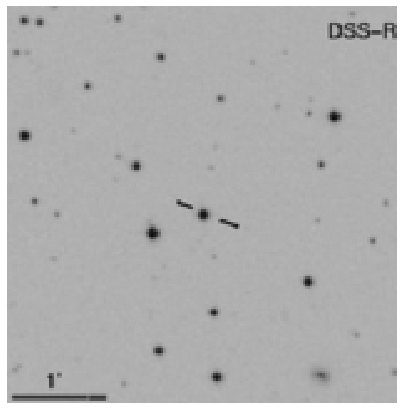}
%\psfull
\end{center}

\begin{center}
\epsfxsize=13.5cm
\epsfysize=4cm
\epsfbox{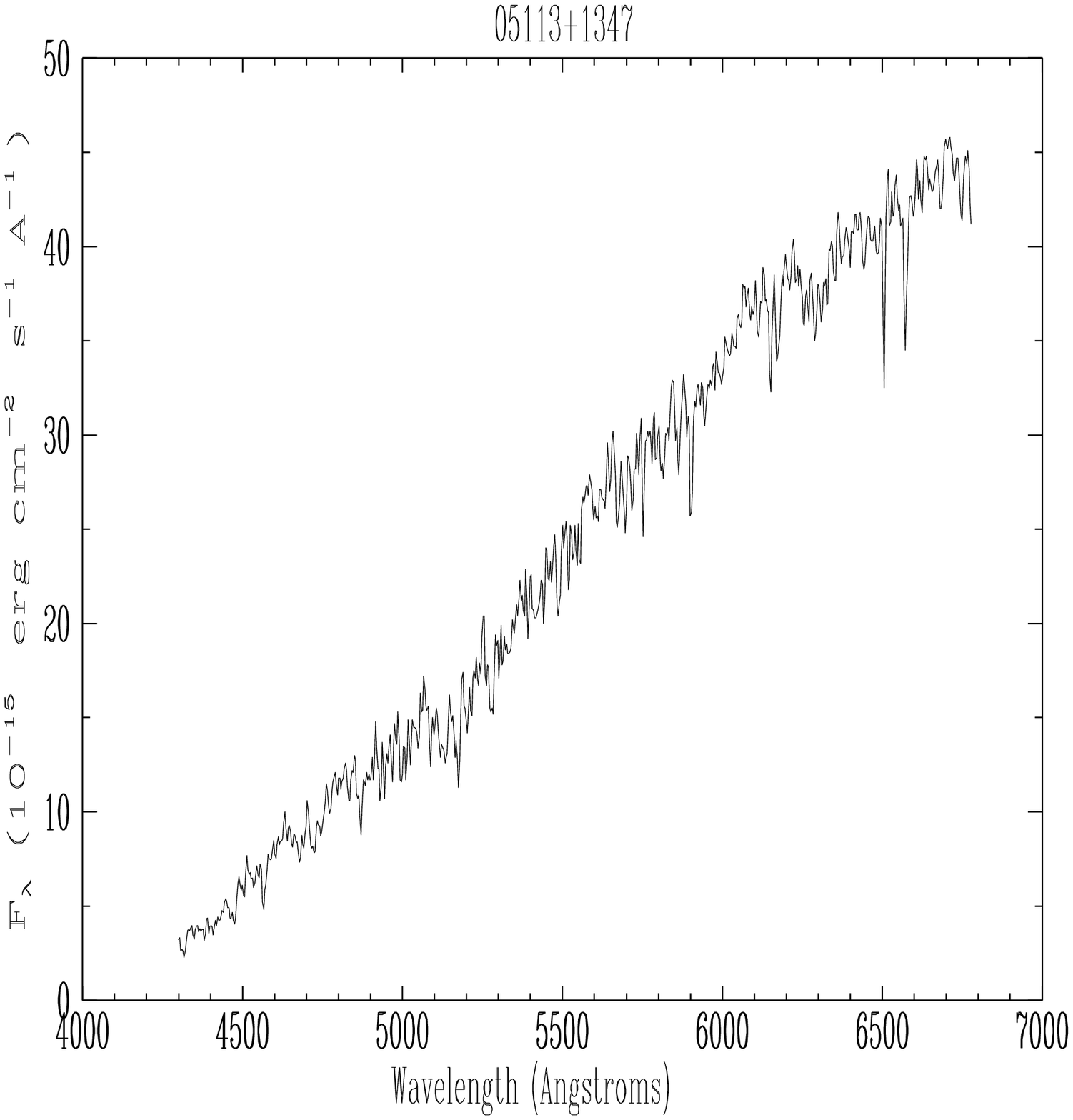}
%\psdraft
\epsfxsize=4cm
\epsfysize=4cm
\epsfbox{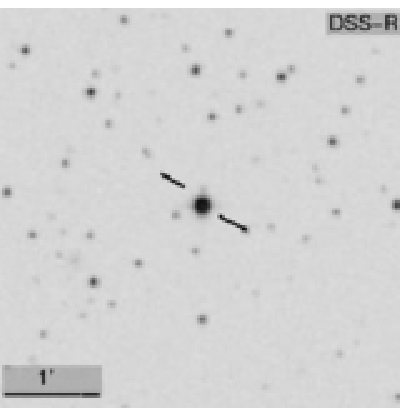}
%\psfull
\end{center}

\begin{center}
\epsfxsize=13.5cm
\epsfysize=4cm
\epsfbox{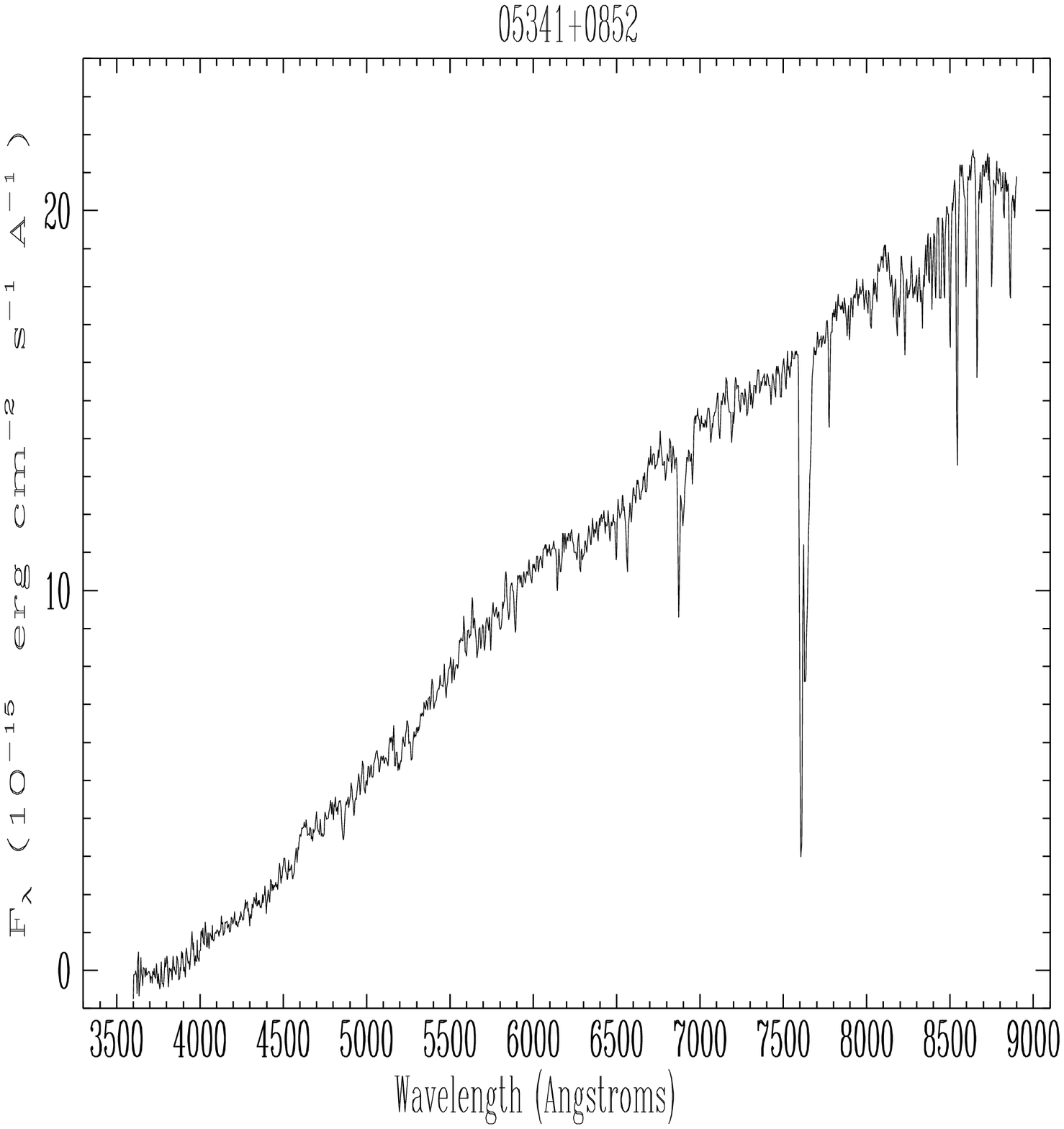}
%\psdraft
\epsfxsize=4cm
\epsfysize=4cm
\epsfbox{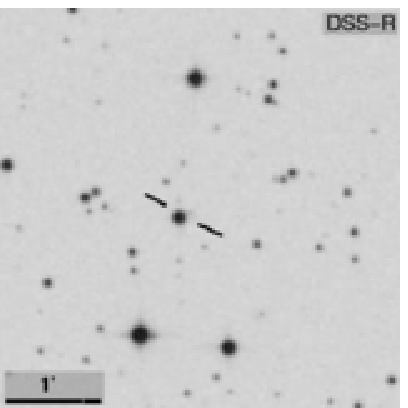}
%\psfull
\end{center}

\begin{center}
\epsfxsize=13.5cm
\epsfysize=4cm
\epsfbox{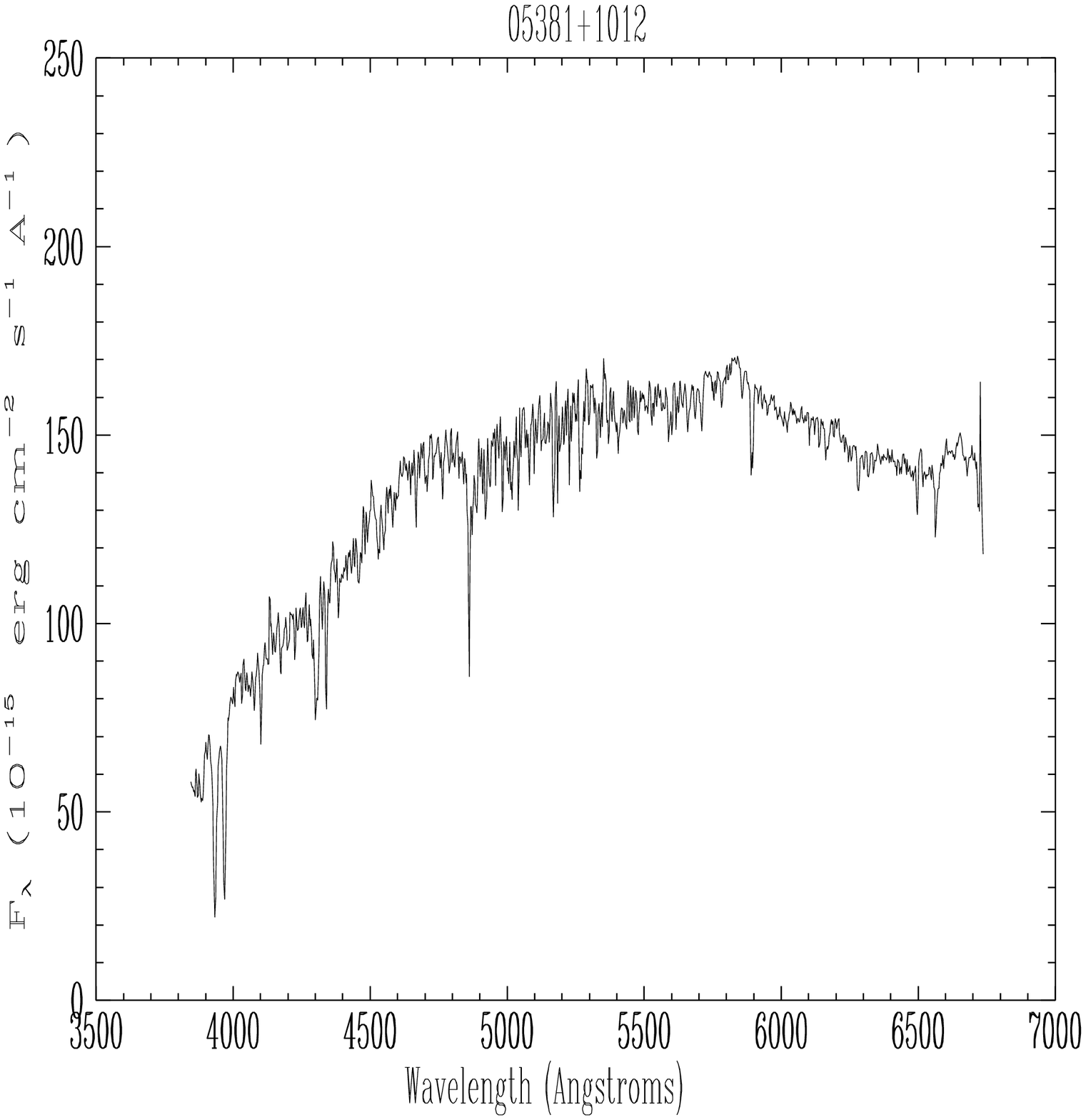}
%\psdraft
\epsfxsize=4cm
\epsfysize=4cm
\epsfbox{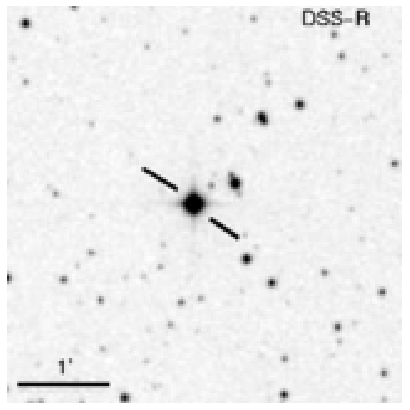}
%\psfull
\end{center}

\begin{center}
\epsfxsize=13.5cm
\epsfysize=4cm
\epsfbox{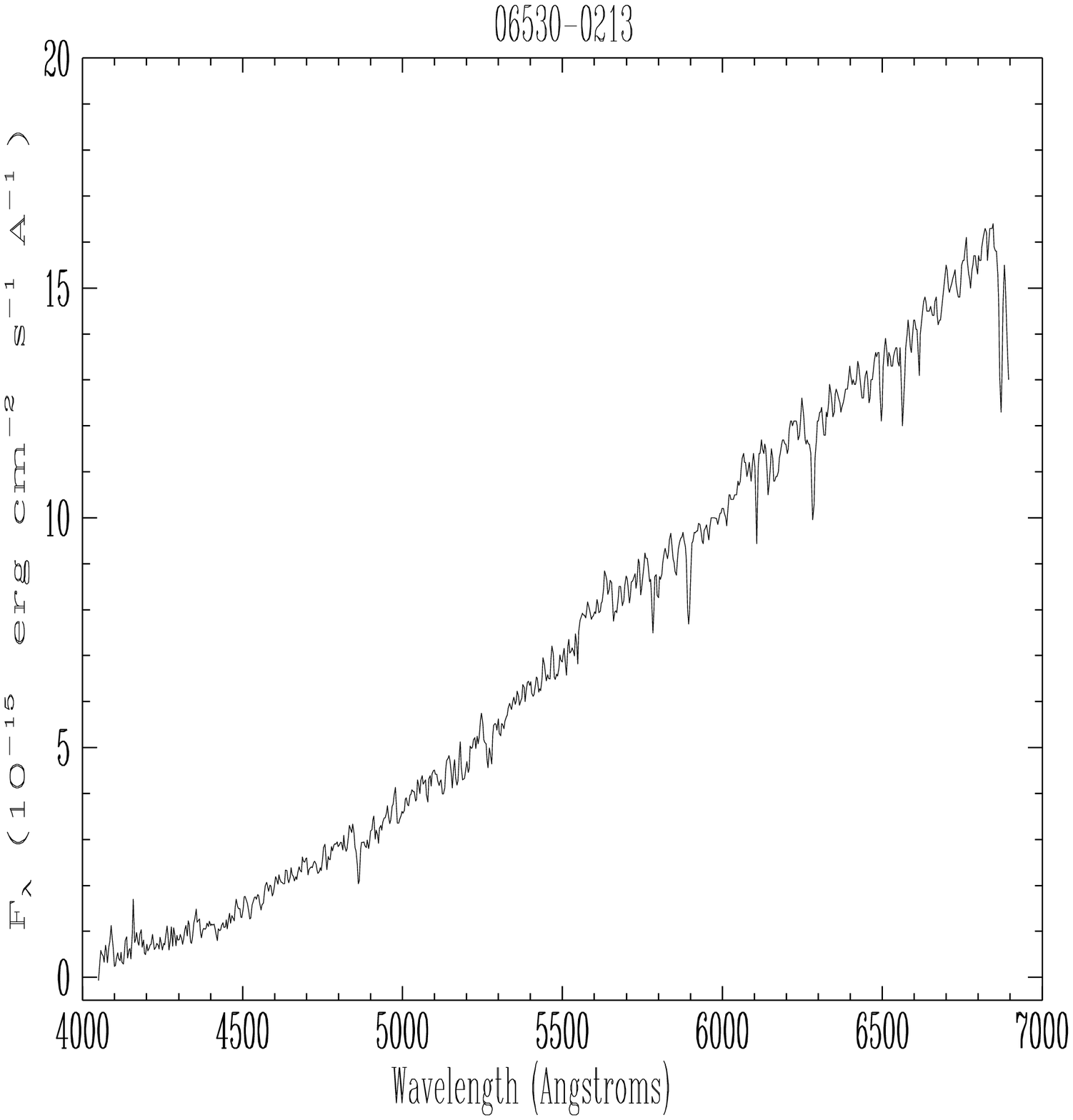}
%\psdraft
\epsfxsize=4cm
\epsfysize=4cm
\epsfbox{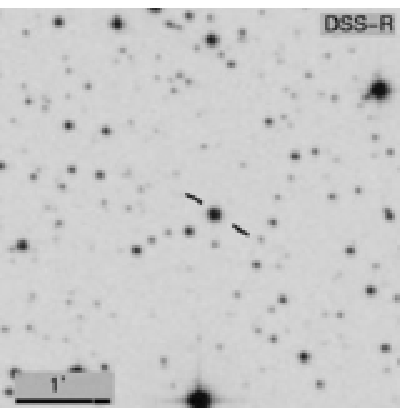}
%\psfull
\end{center}

\caption{Spectra of the objects classified as post-AGB in the sample together with their 
corresponding identification charts (continued). }
\end{figure*}

%-------------------------------------------------------------
%pg3
\setcounter{figure}{0}
\begin{figure*}

\begin{center}
\epsfxsize=13.5cm
\epsfysize=4cm
\epsfbox{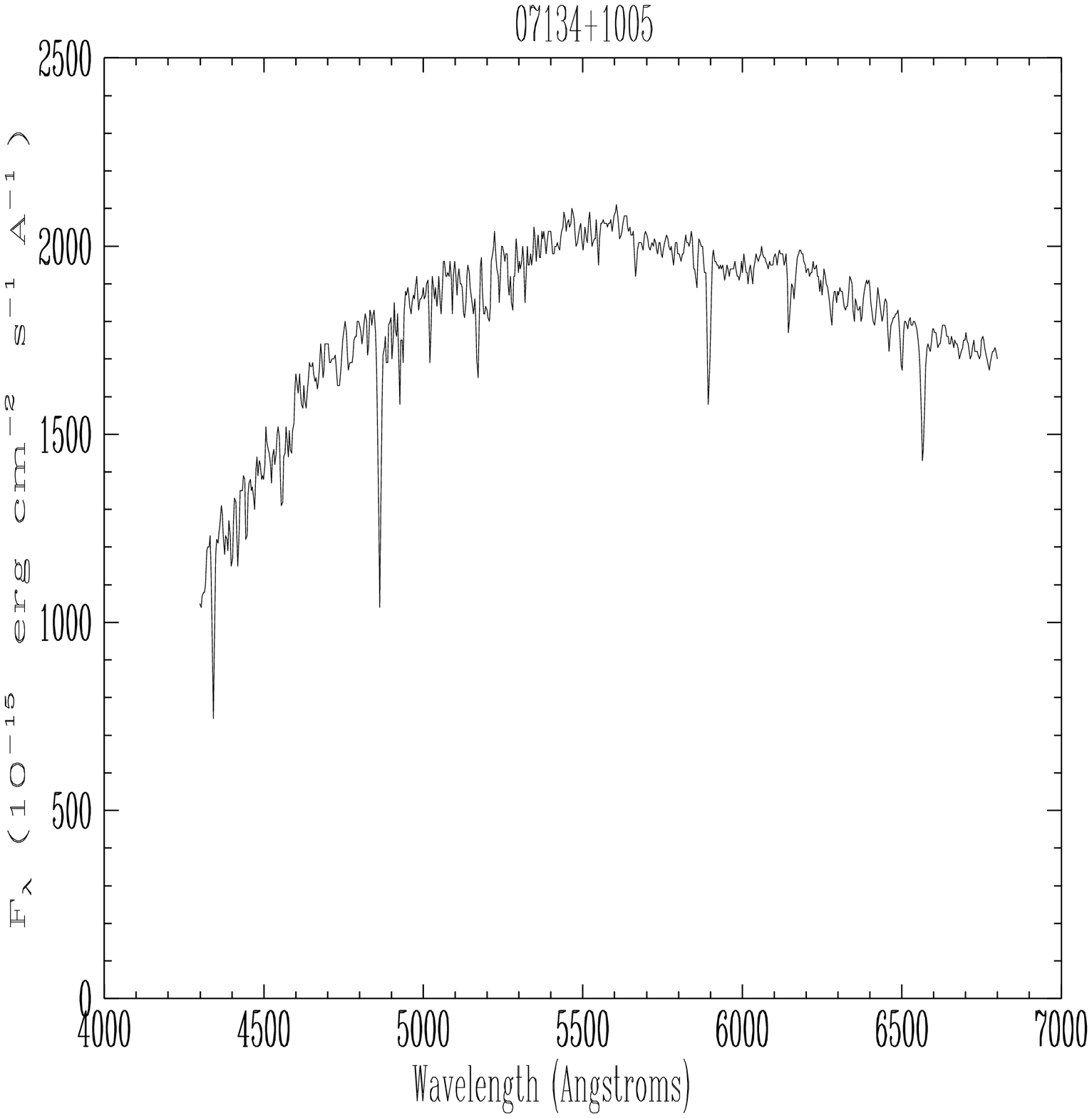}
%\psdraft
\epsfxsize=4cm
\epsfysize=4cm
\epsfbox{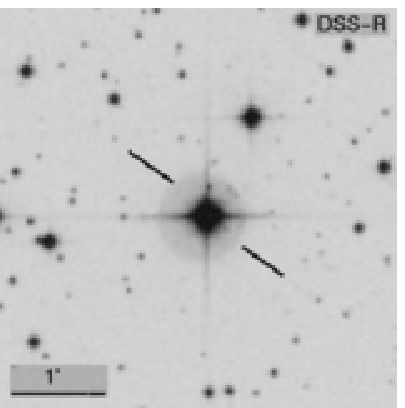}
%\psfull
\end{center}

\begin{center}
\epsfxsize=13.5cm
\epsfysize=4cm
\epsfbox{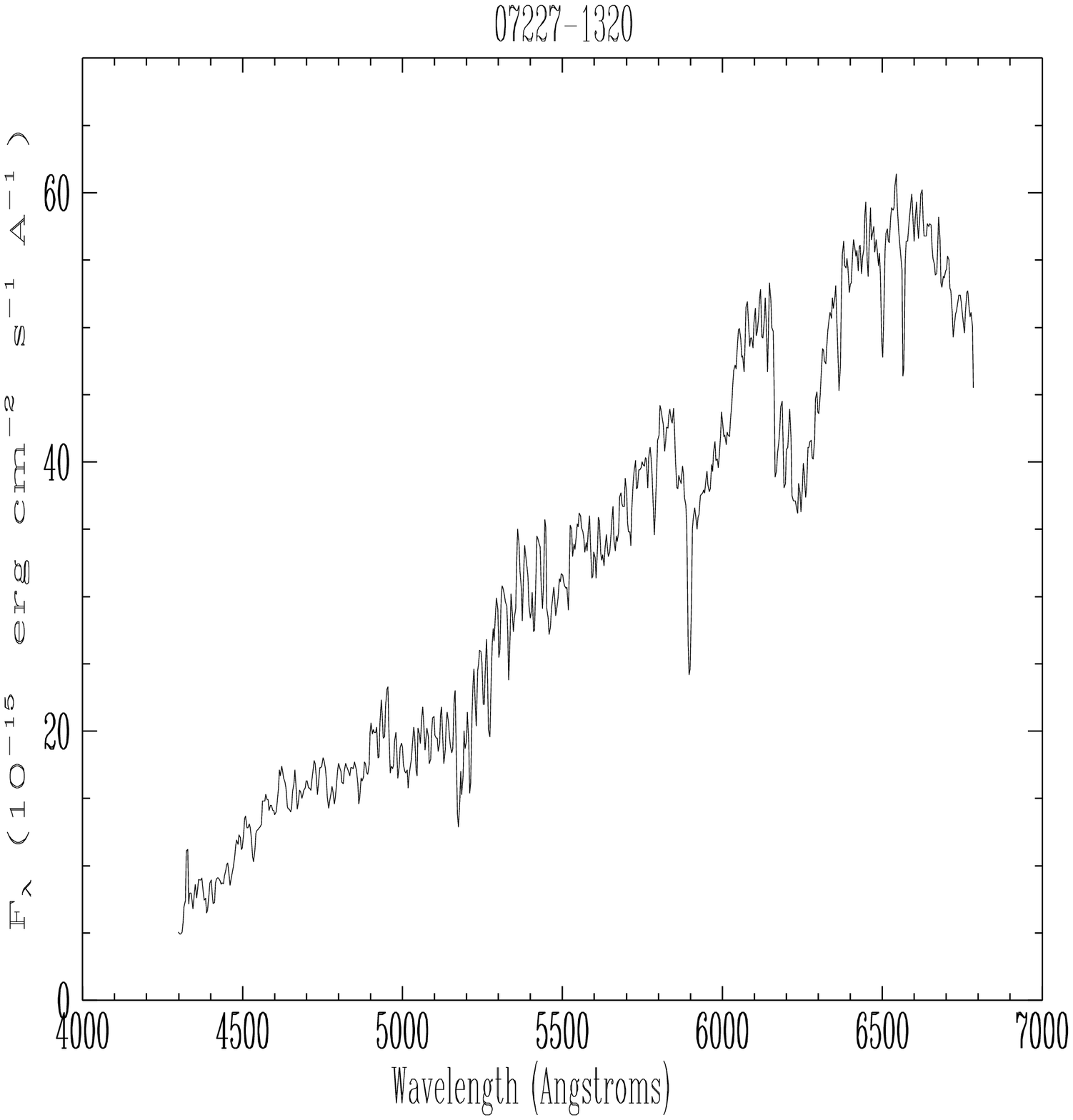}
%\psdraft
\epsfxsize=4cm
\epsfysize=4cm
\epsfbox{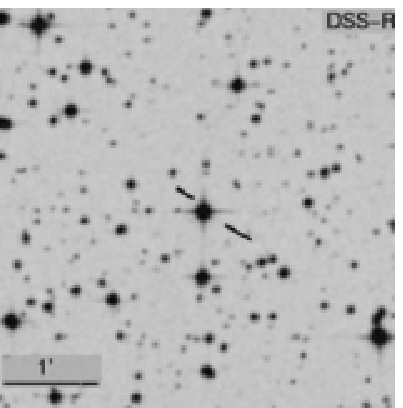}
%\psfull
\end{center}

\begin{center}
\epsfxsize=13.5cm
\epsfysize=4cm
\epsfbox{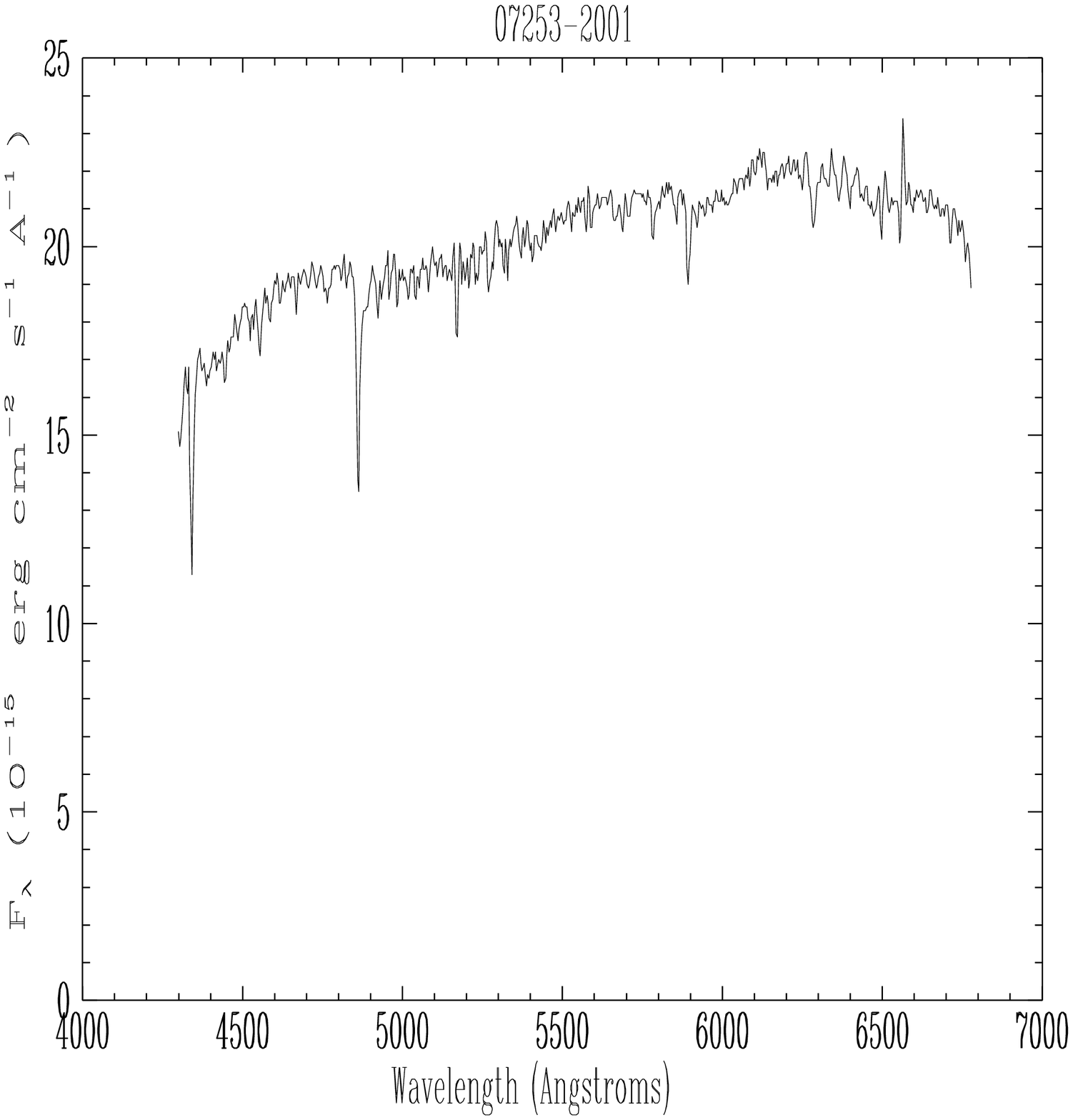}
%\psdraft
\epsfxsize=4cm
\epsfysize=4cm
\epsfbox{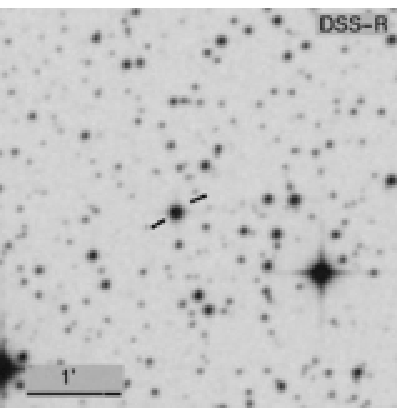}
%\psfull
\end{center}

\begin{center}
\epsfxsize=13.5cm
\epsfysize=4cm
\epsfbox{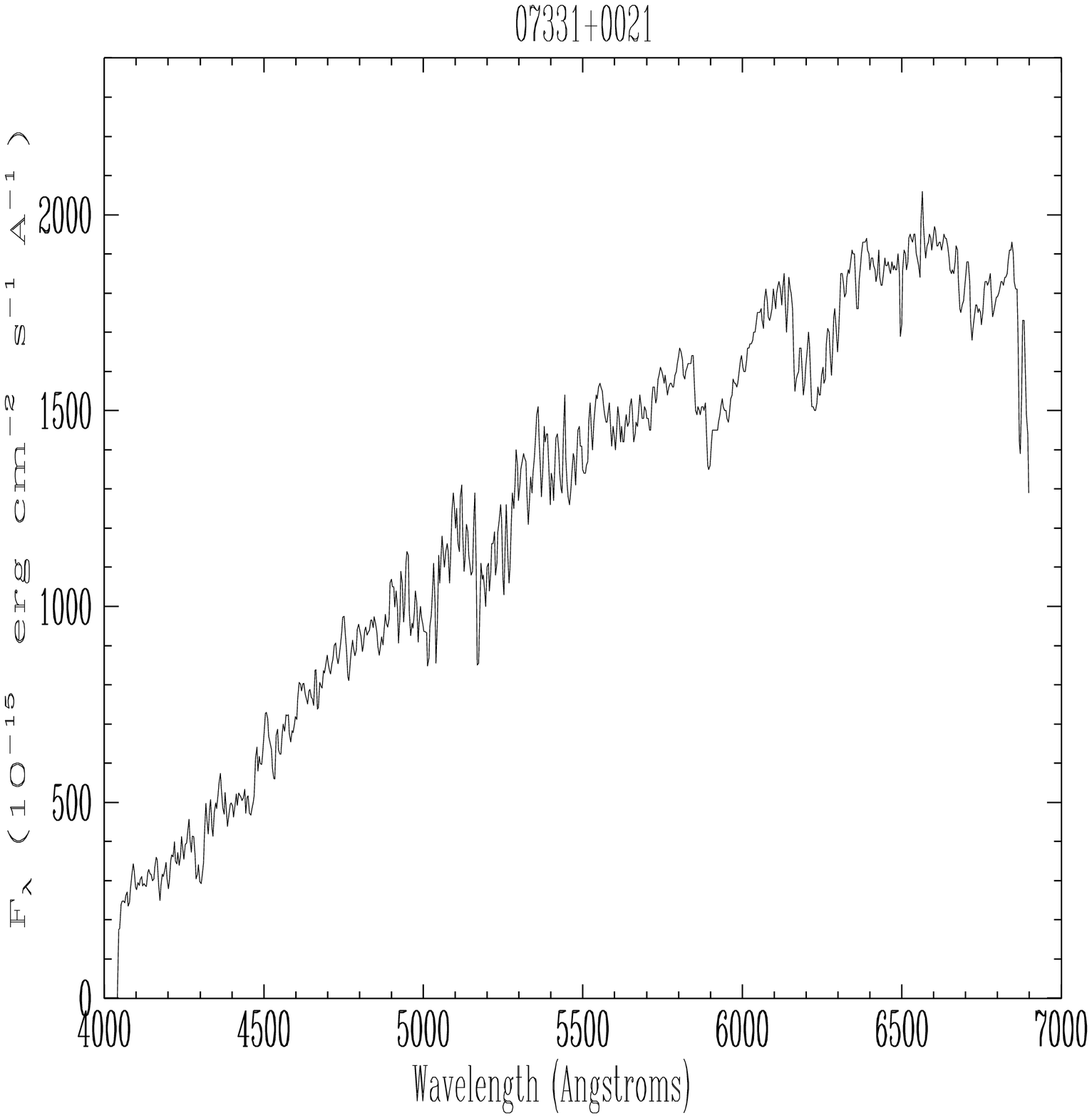}
%\psdraft
\epsfxsize=4cm
\epsfysize=4cm
\epsfbox{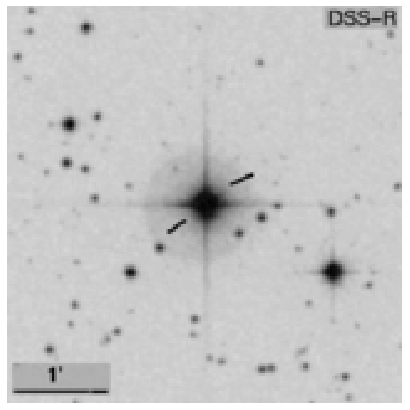}
%\psfull
\end{center}

\begin{center}
\epsfxsize=13.5cm
\epsfysize=4cm
\epsfbox{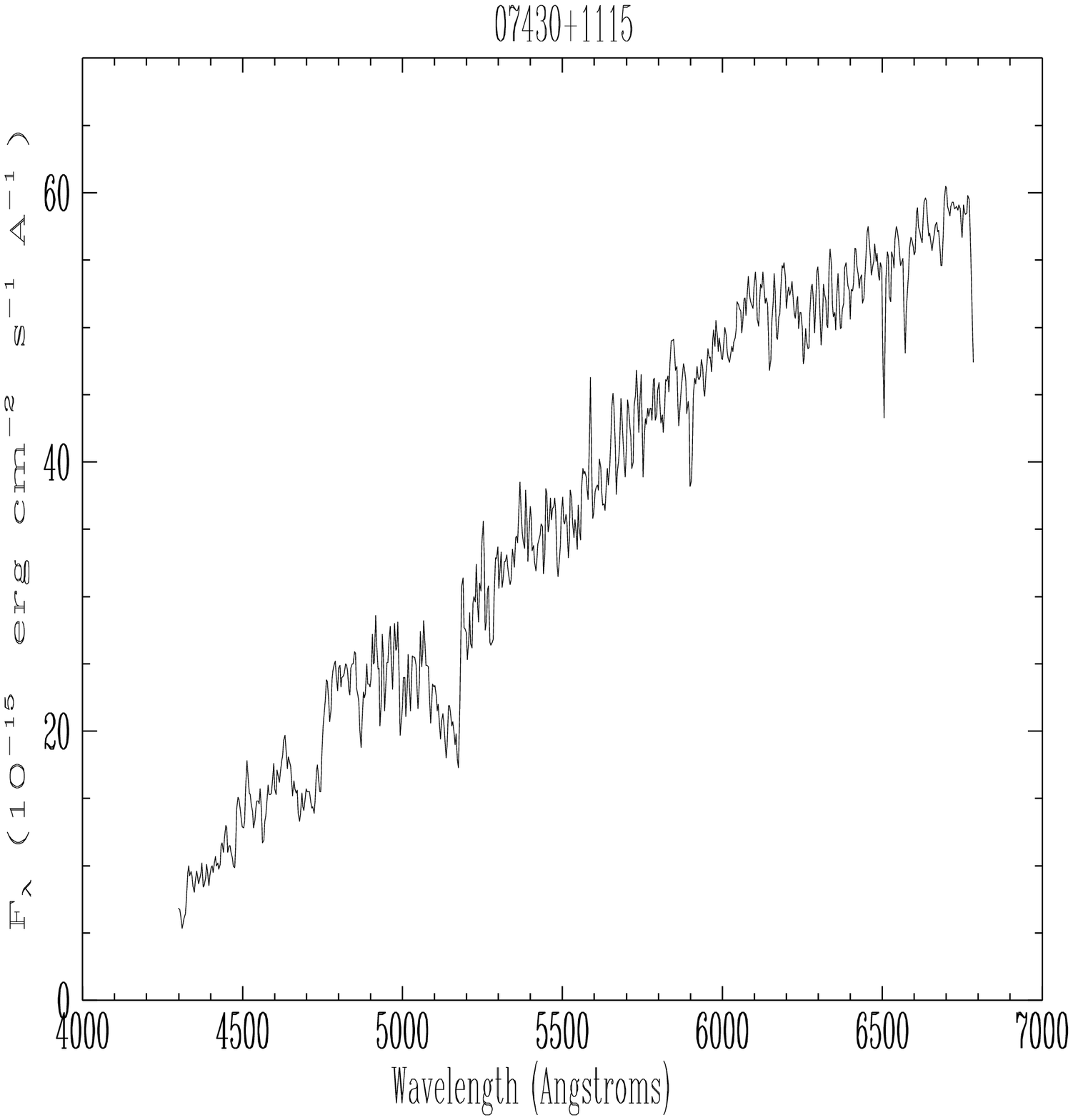}
%\psdraft
\epsfxsize=4cm
\epsfysize=4cm
\epsfbox{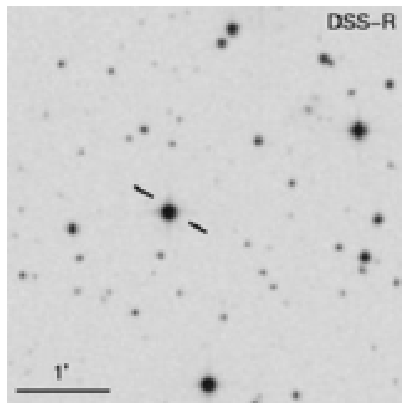}
%\psfull
\end{center}

\caption{Spectra of the objects classified as post-AGB in the sample together with their 
corresponding identification charts (continued). }
\end{figure*}

%-------------------------------------------------------------
%pg4
\setcounter{figure}{0}
\begin{figure*}

\begin{center}
\epsfxsize=13.5cm
\epsfysize=4cm
\epsfbox{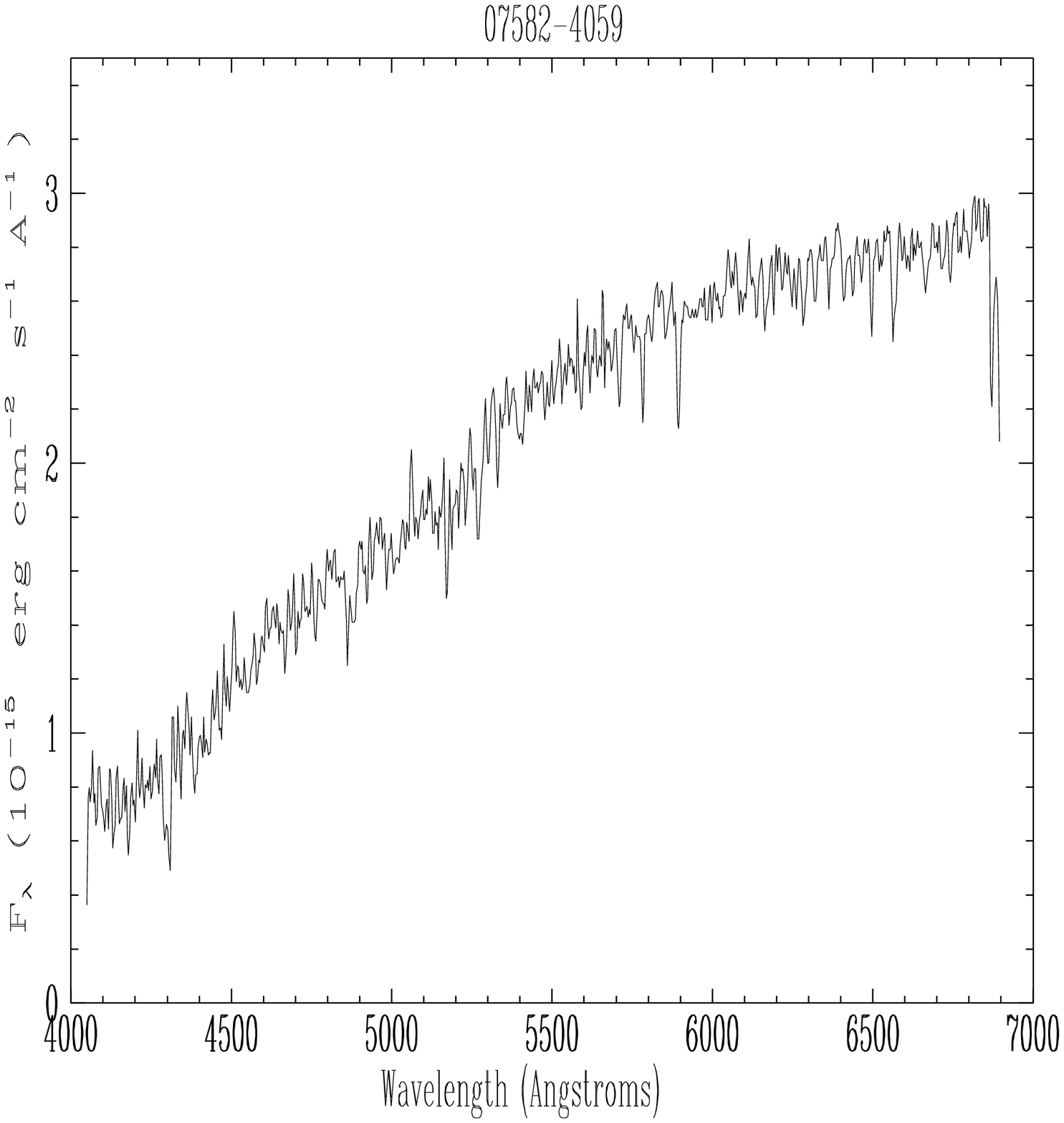}
%\psdraft
\epsfxsize=4cm
\epsfysize=4cm
\epsfbox{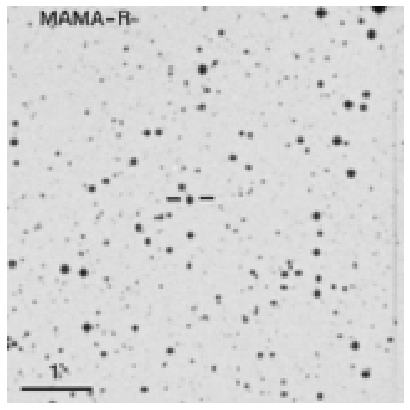}
%\psfull
\end{center}

\begin{center}
\epsfxsize=13.5cm
\epsfysize=4cm
\epsfbox{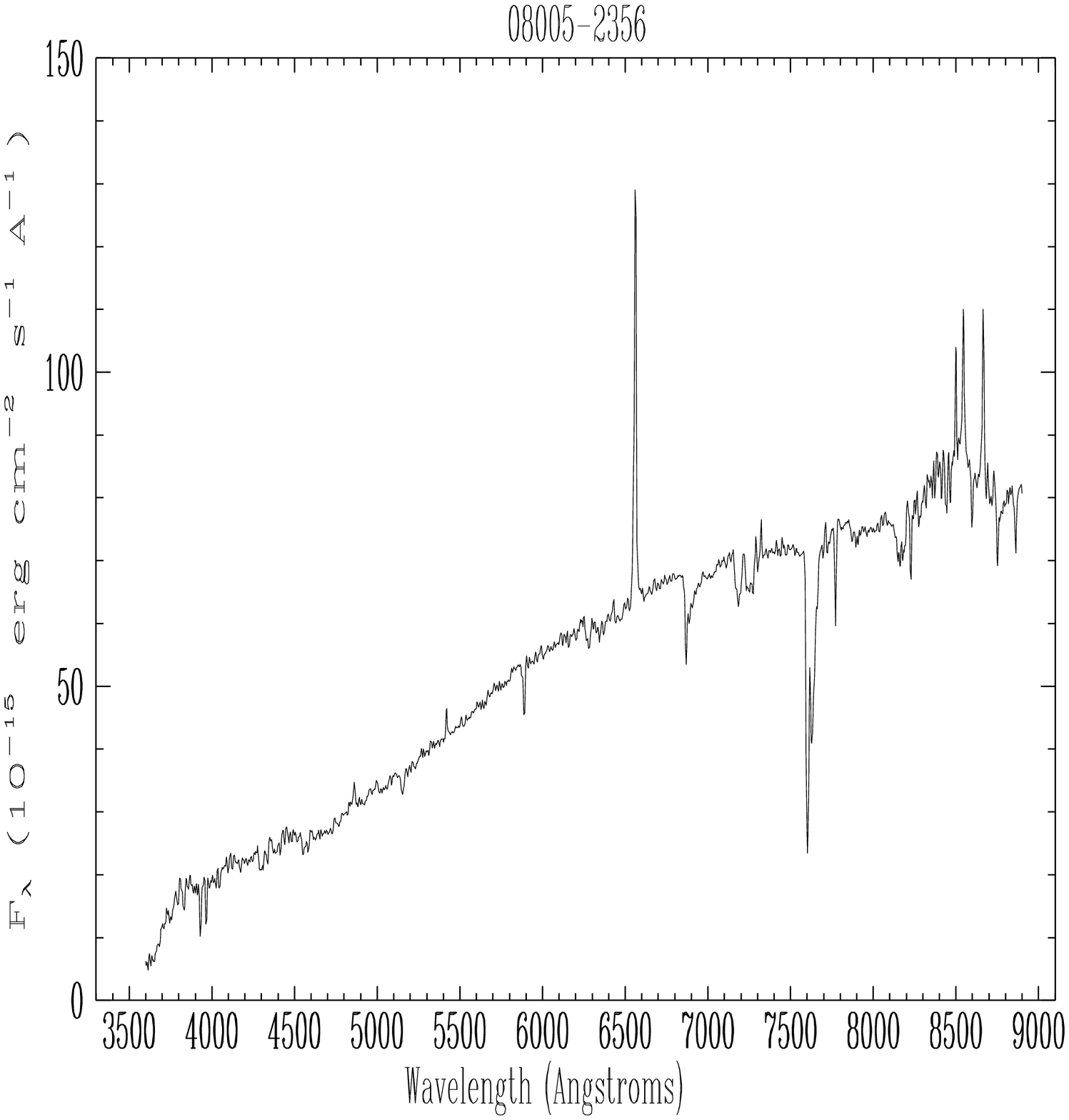}
%\psdraft
\epsfxsize=4cm
\epsfysize=4cm
\epsfbox{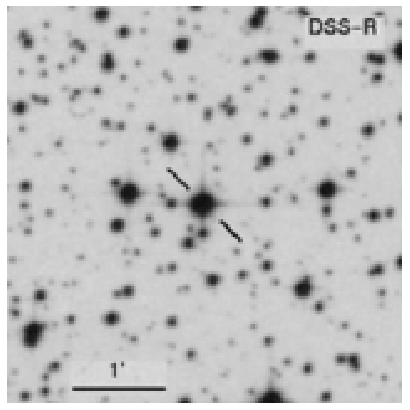}
%\psfull
\end{center}

\begin{center}
\epsfxsize=13.5cm
\epsfysize=4cm
\epsfbox{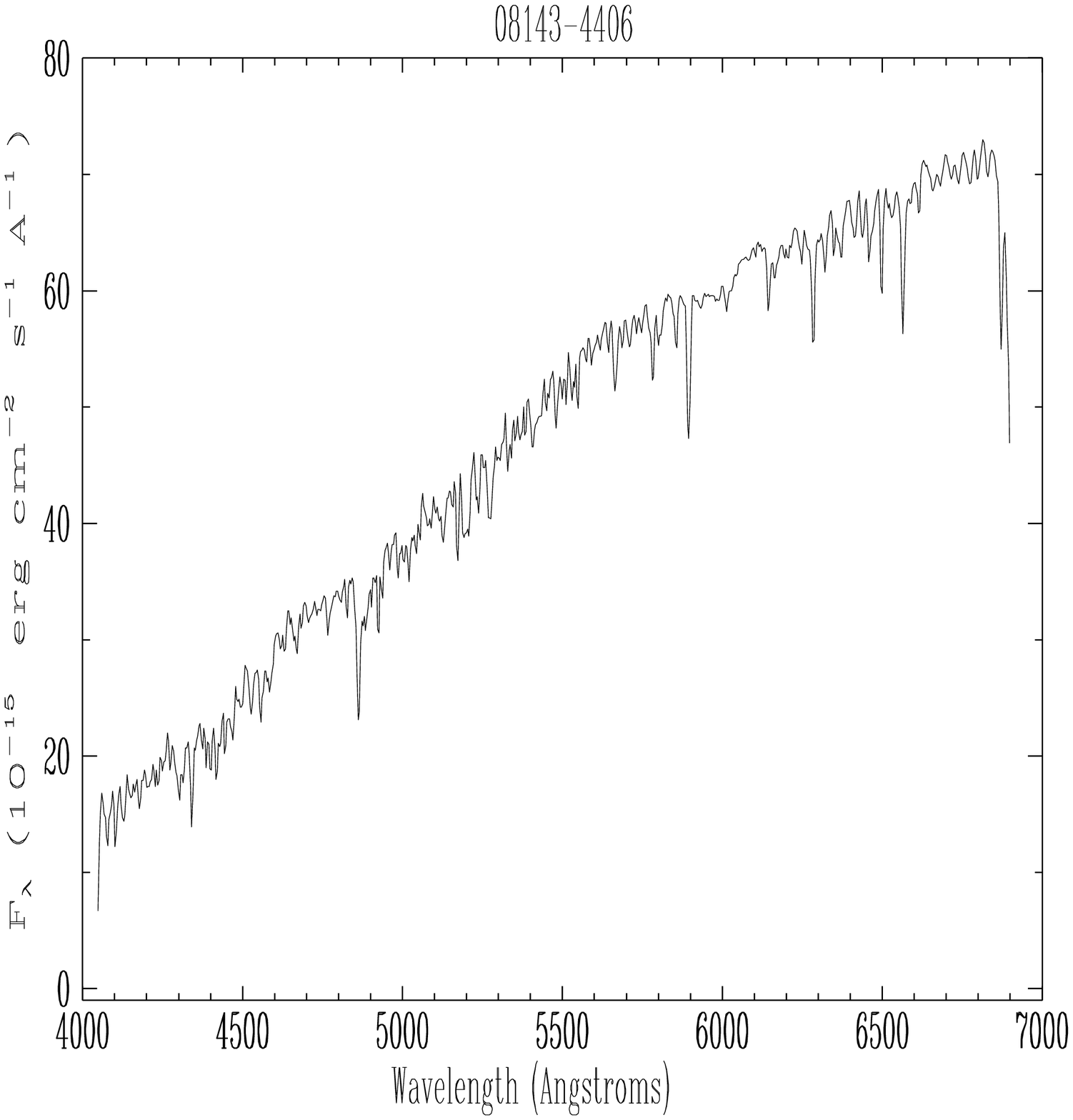}
%\psdraft
\epsfxsize=4cm
\epsfysize=4cm
\epsfbox{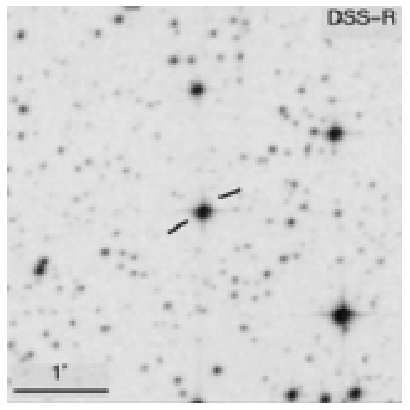}
%\psfull
\end{center}

\begin{center}
\epsfxsize=13.5cm
\epsfysize=4cm
\epsfbox{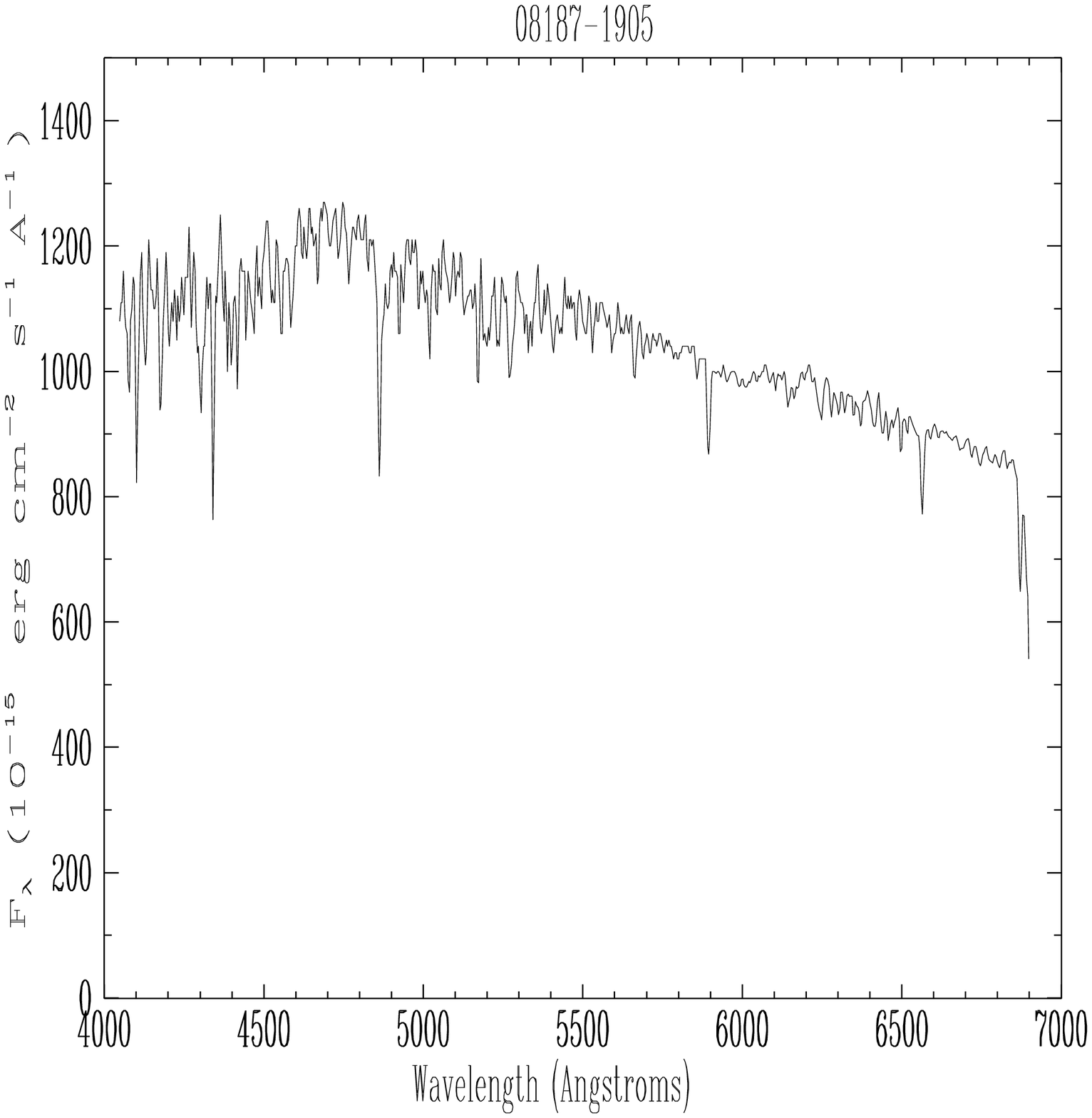}
%\psdraft
\epsfxsize=4cm
\epsfysize=4cm
\epsfbox{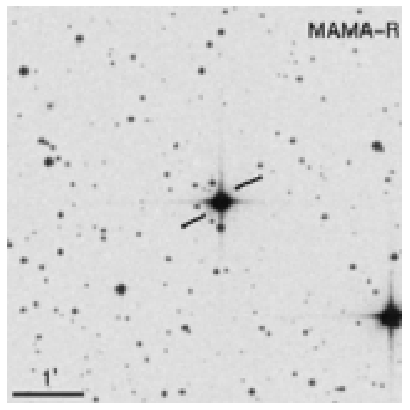}
%\psfull
\end{center}

\begin{center}
\epsfxsize=13.5cm
\epsfysize=4cm
\epsfbox{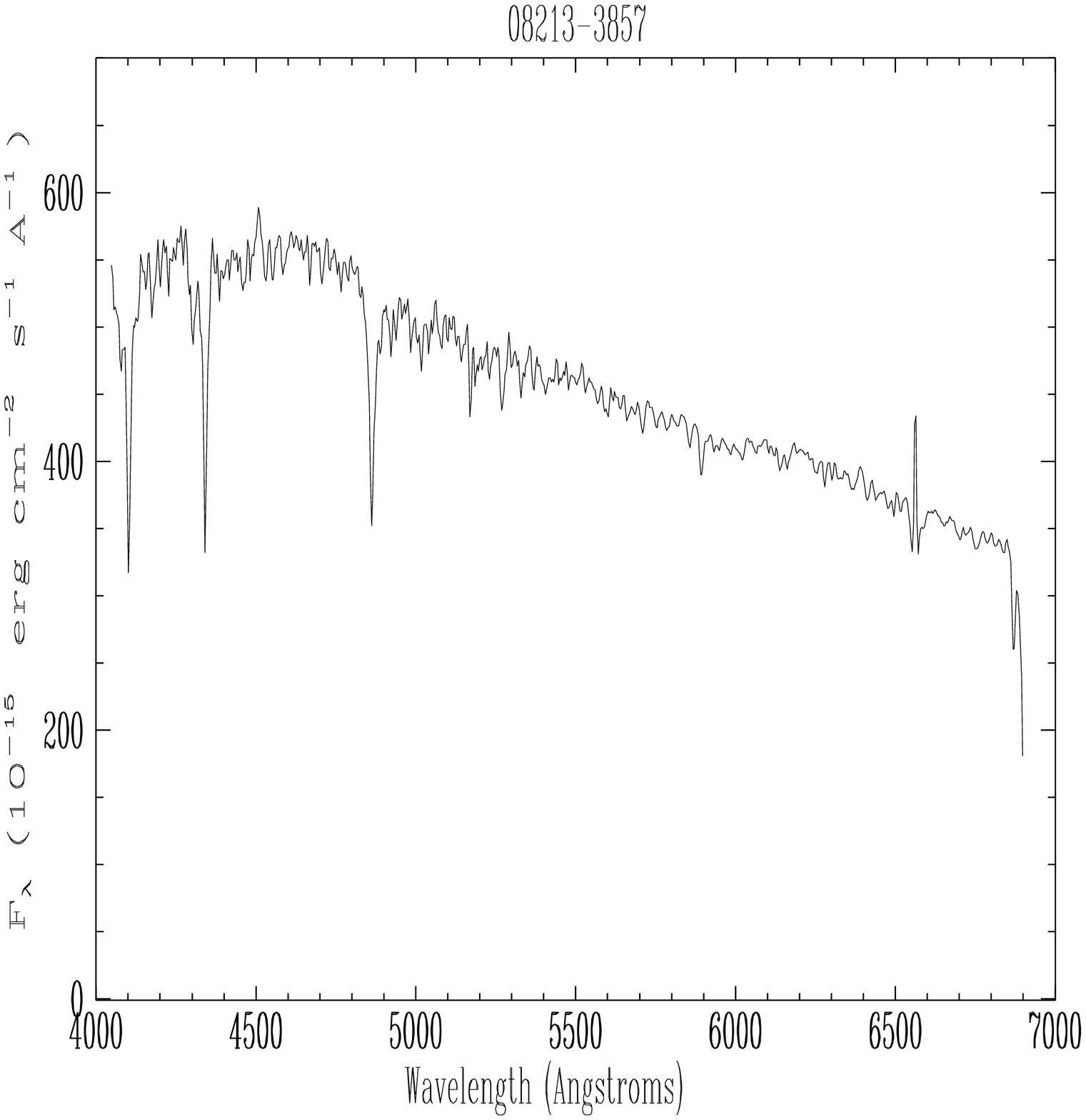}
%\psdraft
\epsfxsize=4cm
\epsfysize=4cm
\epsfbox{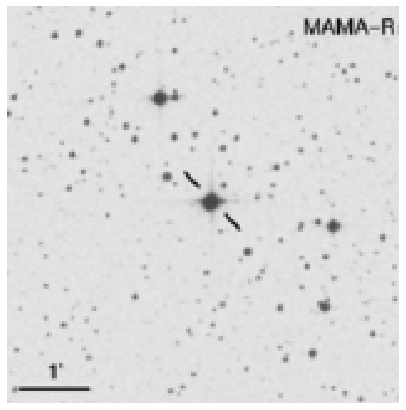}
%\psfull
\end{center}

\caption{Spectra of the objects classified as post-AGB in the sample together with their 
corresponding identification charts (continued). }
\end{figure*}

%-------------------------------------------------------------
%pg5
\setcounter{figure}{0}
\begin{figure*}

\begin{center}
\epsfxsize=13.5cm
\epsfysize=4cm
\epsfbox{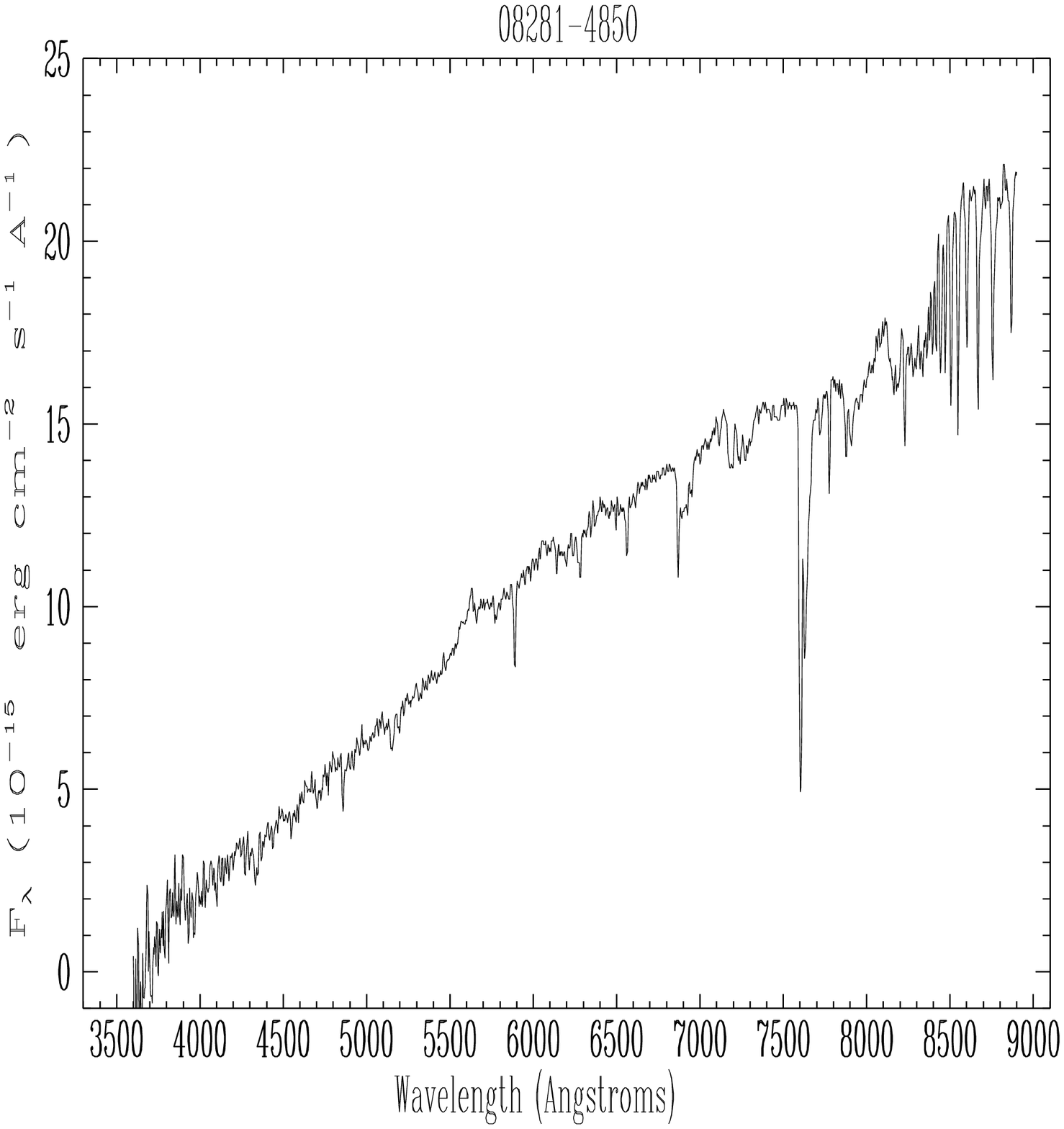}
%\psdraft
\epsfxsize=4cm
\epsfysize=4cm
\epsfbox{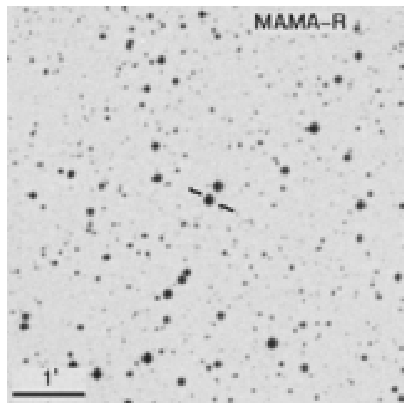}
%\psfull
\end{center}

\begin{center}
\epsfxsize=13.5cm
\epsfysize=4cm
\epsfbox{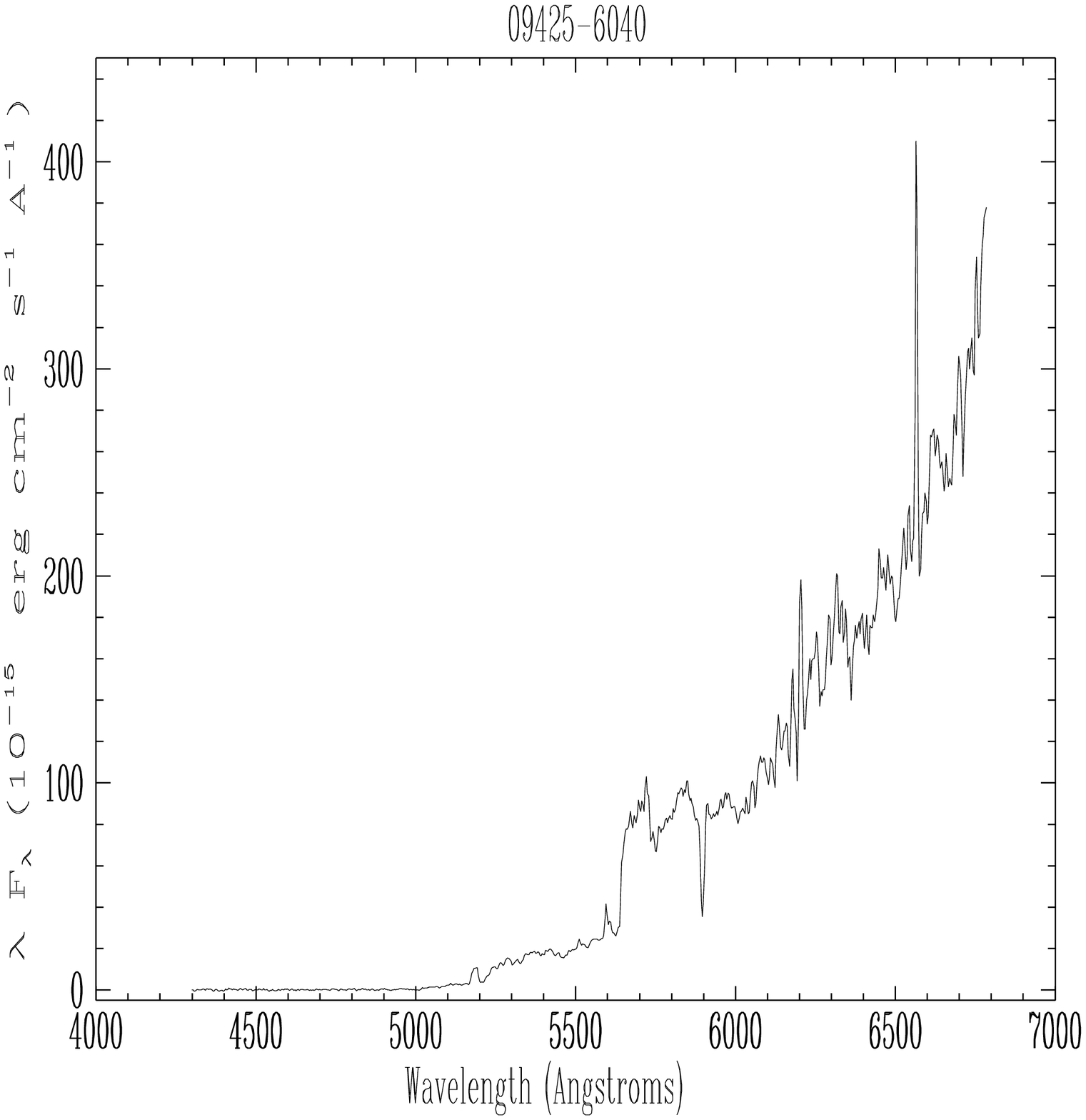}
%\psdraft
\epsfxsize=4cm
\epsfysize=4cm
\epsfbox{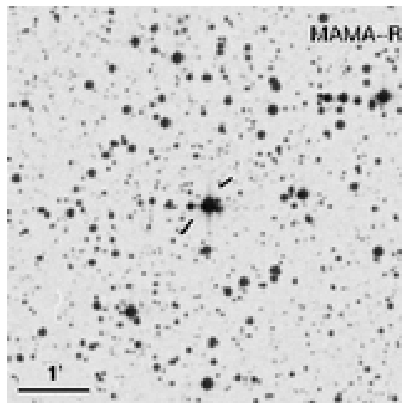}
%\psfull
\end{center}

\begin{center}
\epsfxsize=13.5cm
\epsfysize=4cm
\epsfbox{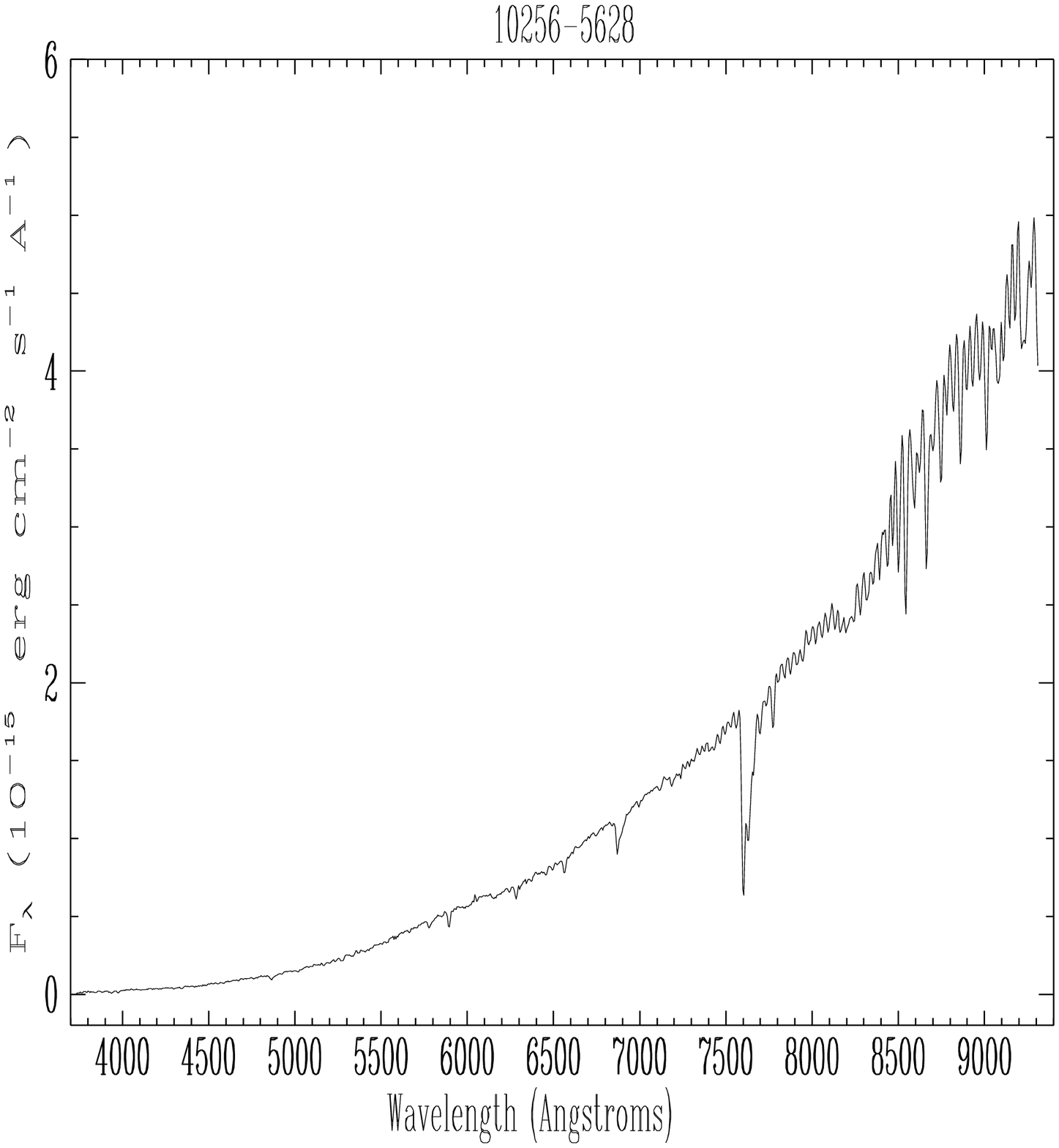}
%\psdraft
\epsfxsize=4cm
\epsfysize=4cm
\epsfbox{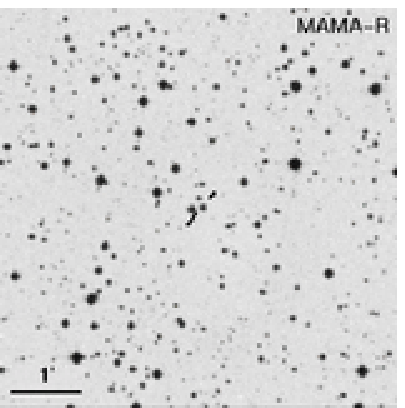}
%\psfull
\end{center}

\begin{center}
\epsfxsize=13.5cm
\epsfysize=4cm
\epsfbox{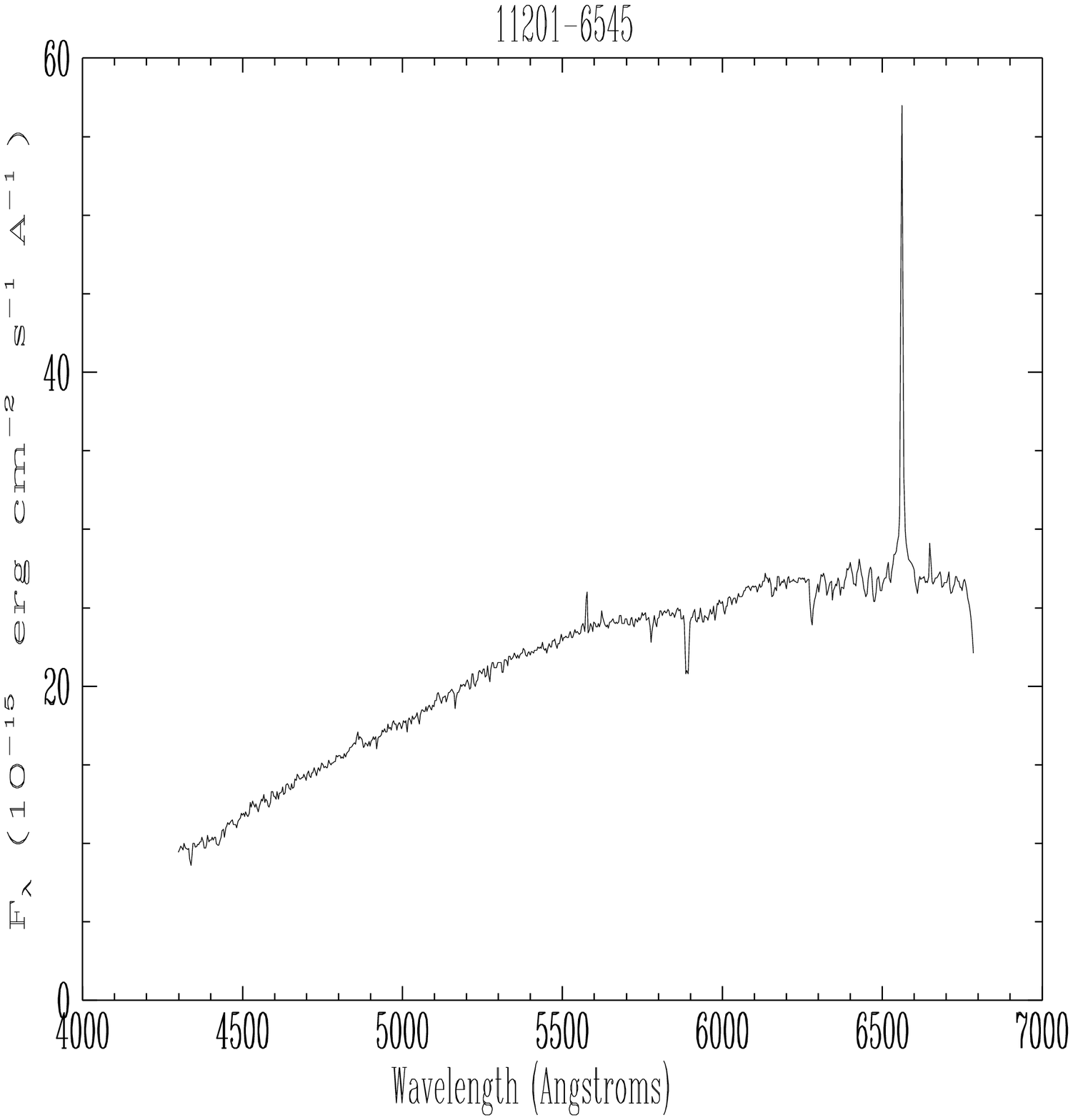}
%\psdraft
\epsfxsize=4cm
\epsfysize=4cm
\epsfbox{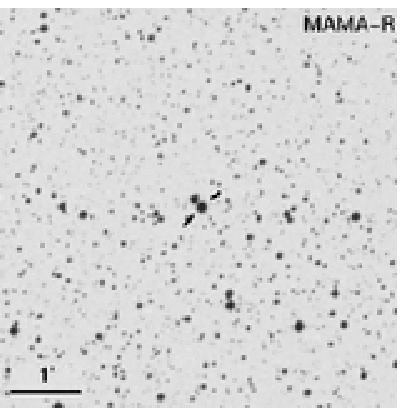}
%\psfull
\end{center}

\begin{center}
\epsfxsize=13.5cm
\epsfysize=4cm
\epsfbox{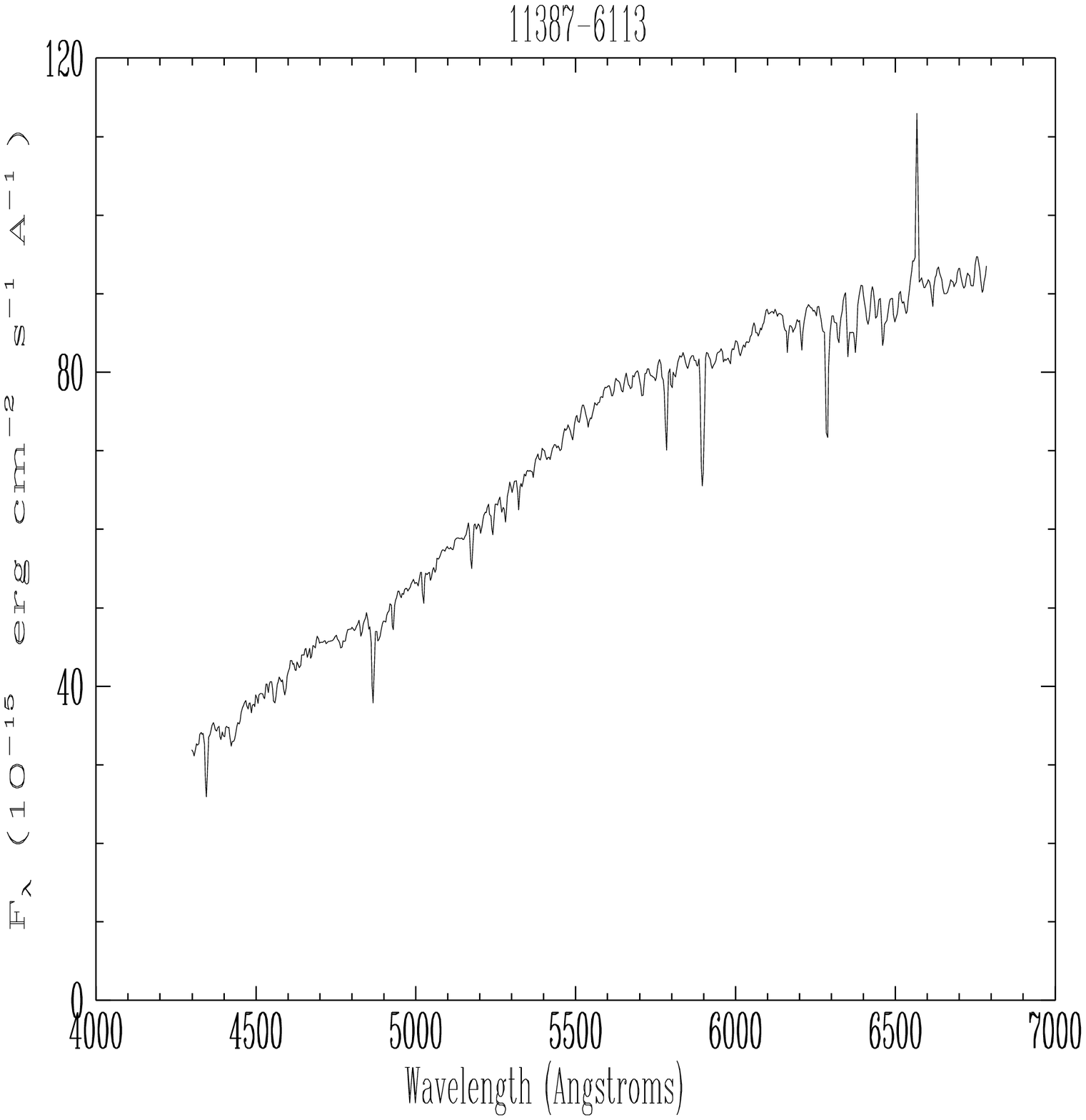}
%\psdraft
\epsfxsize=4cm
\epsfysize=4cm
\epsfbox{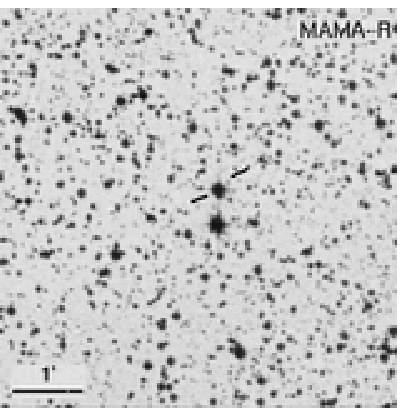}
%\psfull
\end{center}

\caption{Spectra of the objects classified as post-AGB in the sample together with their 
corresponding identification charts (continued). }
\end{figure*}

%-------------------------------------------------------------
%pg7
\setcounter{figure}{0}
\begin{figure*}

\begin{center}
\epsfxsize=13.5cm
\epsfysize=4cm
\epsfbox{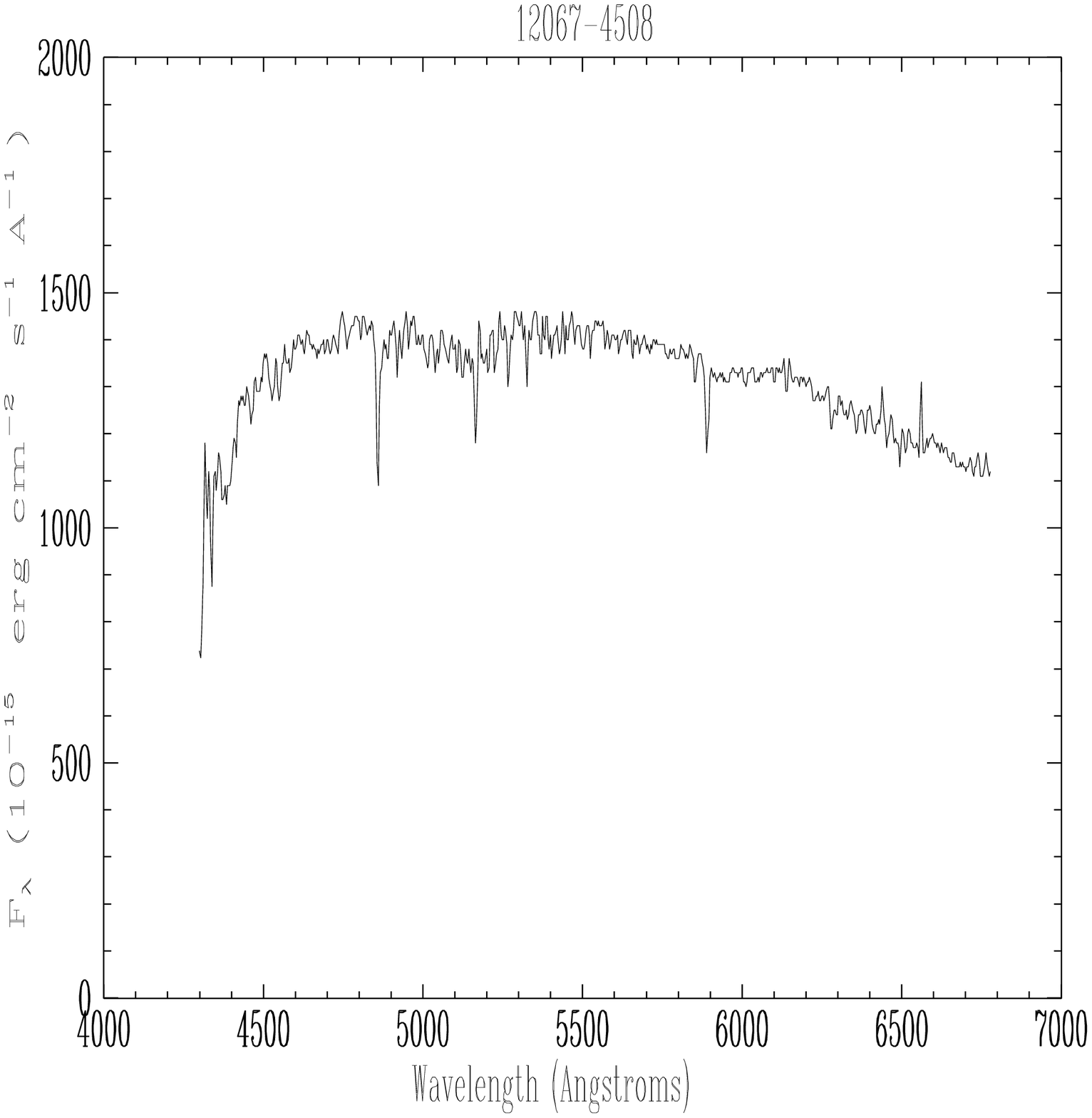}
%\psdraft
\epsfxsize=4cm
\epsfysize=4cm
\epsfbox{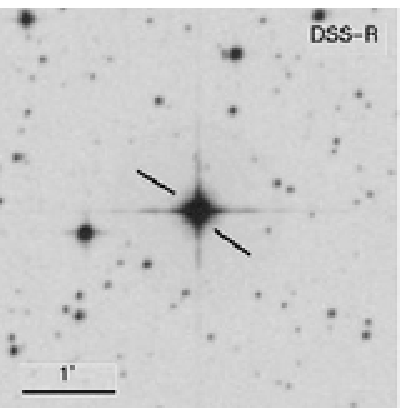}
%\psfull
\end{center}

\begin{center}
\epsfxsize=13.5cm
\epsfysize=4cm
\epsfbox{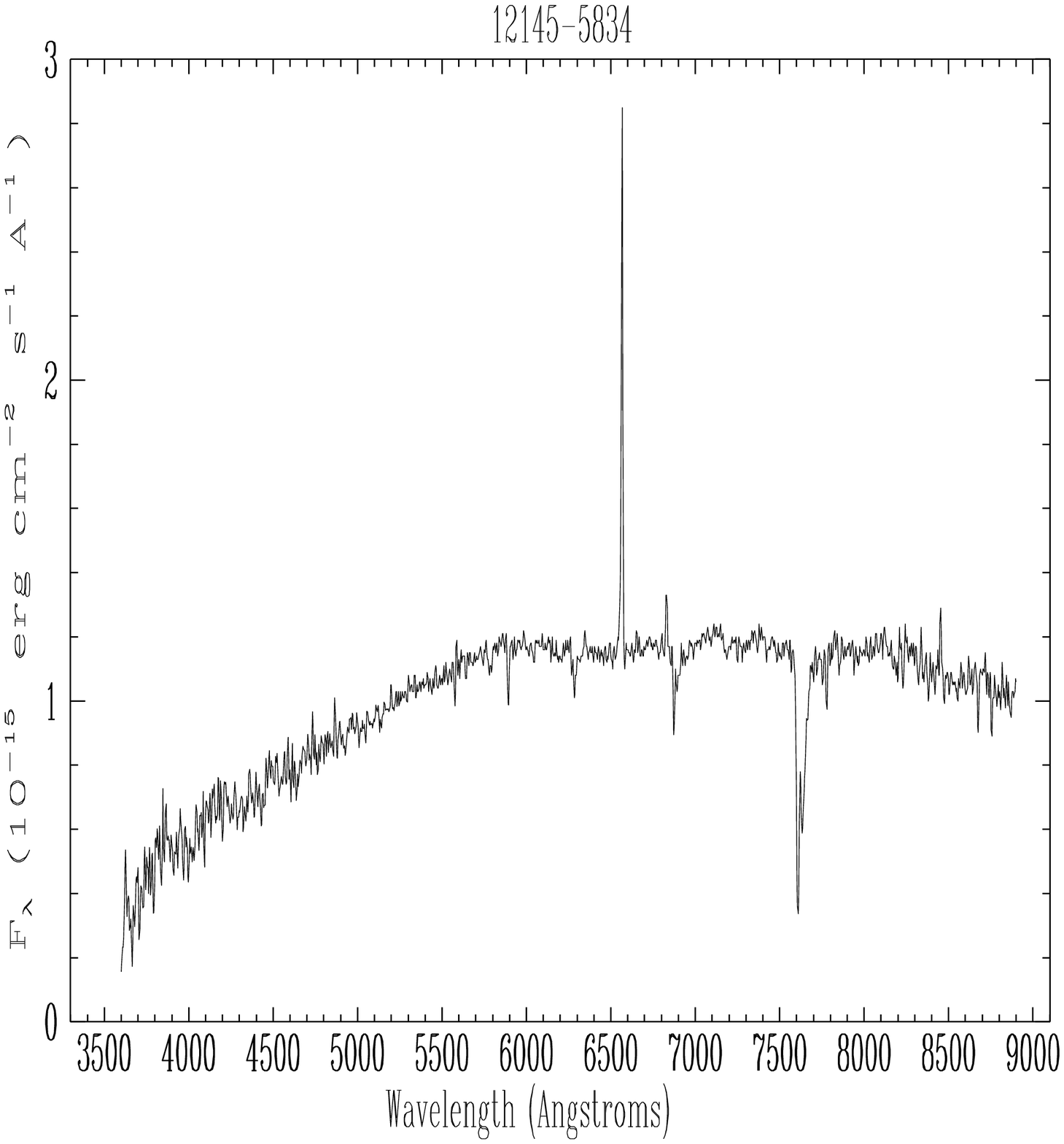}
%\psdraft
\epsfxsize=4cm
\epsfysize=4cm
\epsfbox{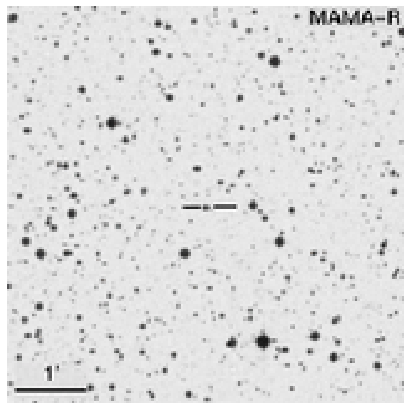}
%\psfull
\end{center}

\begin{center}
\epsfxsize=13.5cm
\epsfysize=4cm
\epsfbox{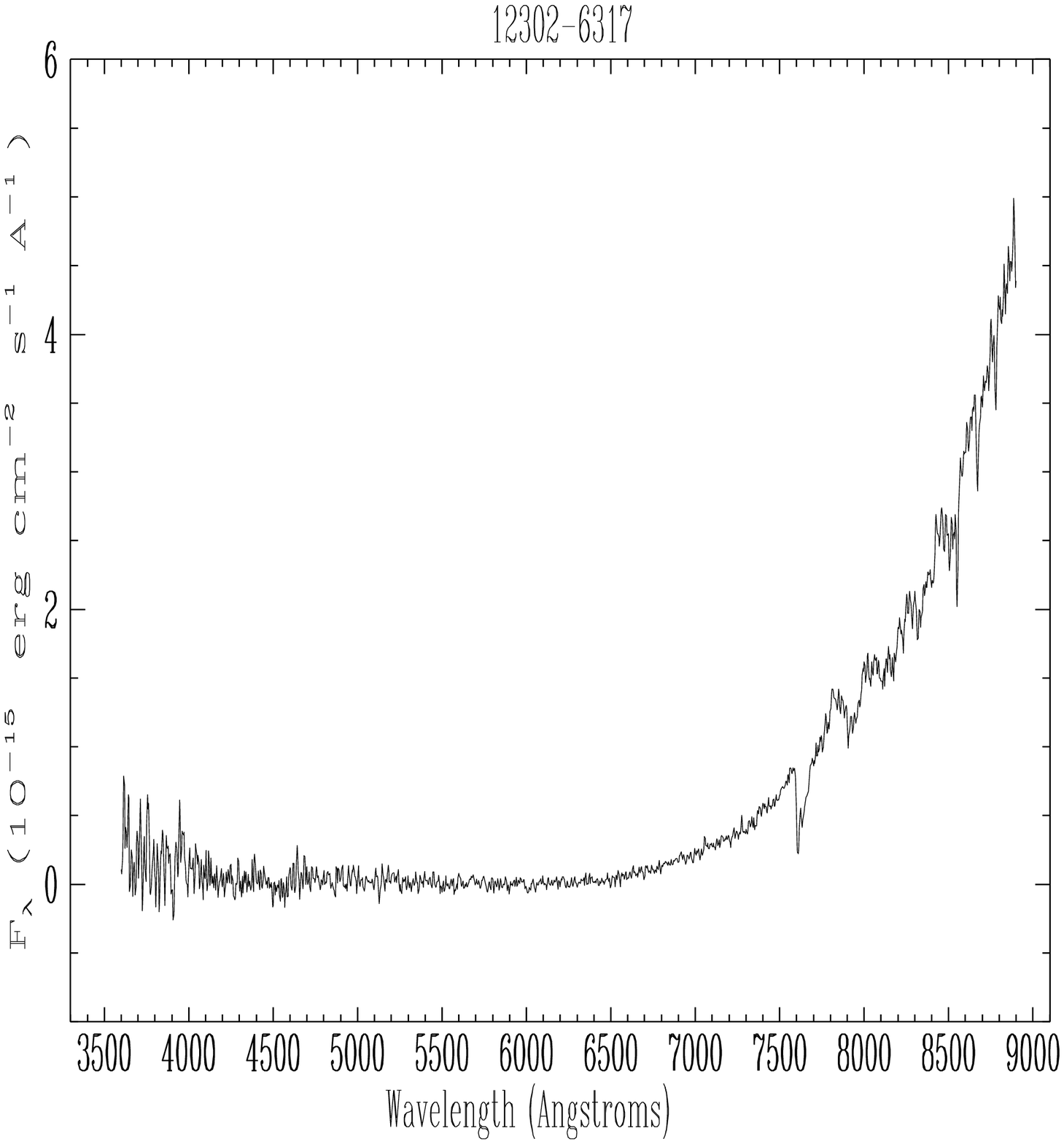}
%\psdraft
\epsfxsize=4cm
\epsfysize=4cm
\epsfbox{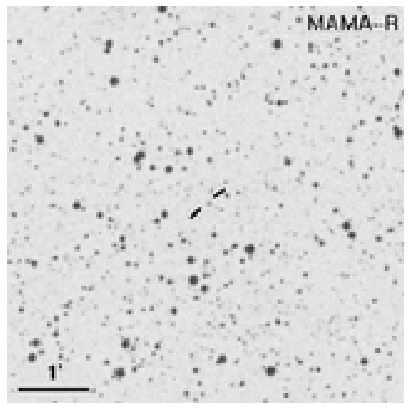}
%\psfull
\end{center}

\begin{center}
\epsfxsize=13.5cm
\epsfysize=4cm
\epsfbox{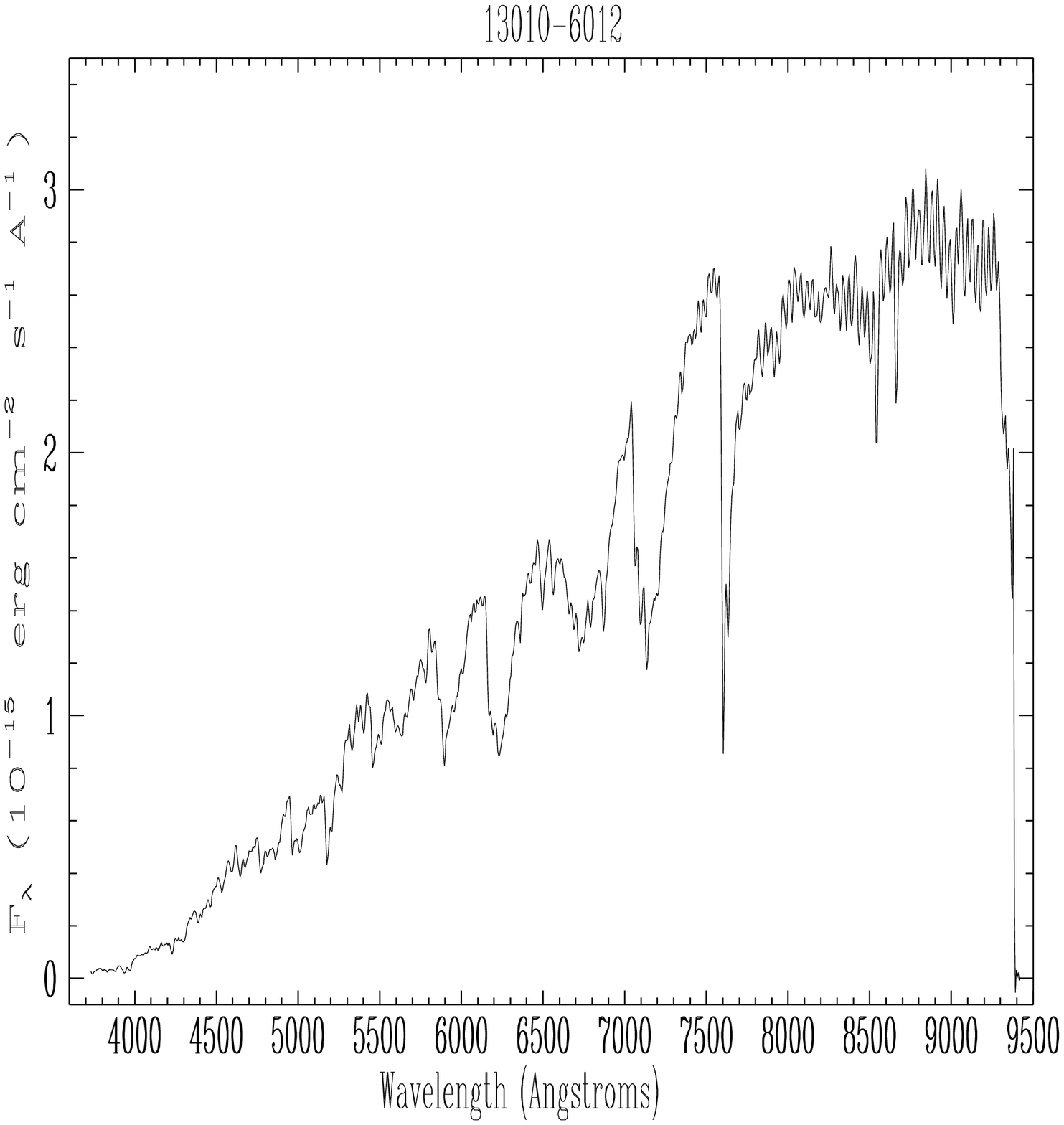}
%\psdraft
\epsfxsize=4cm
\epsfysize=4cm
\epsfbox{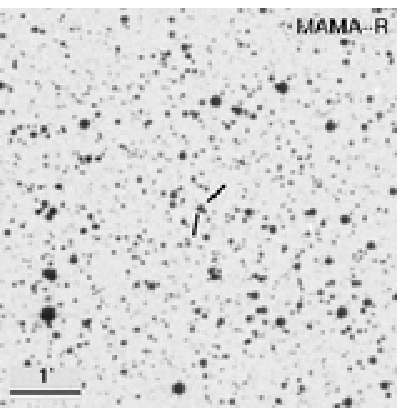}
%\psfull
\end{center}

\begin{center}
\epsfxsize=13.5cm
\epsfysize=4cm
\epsfbox{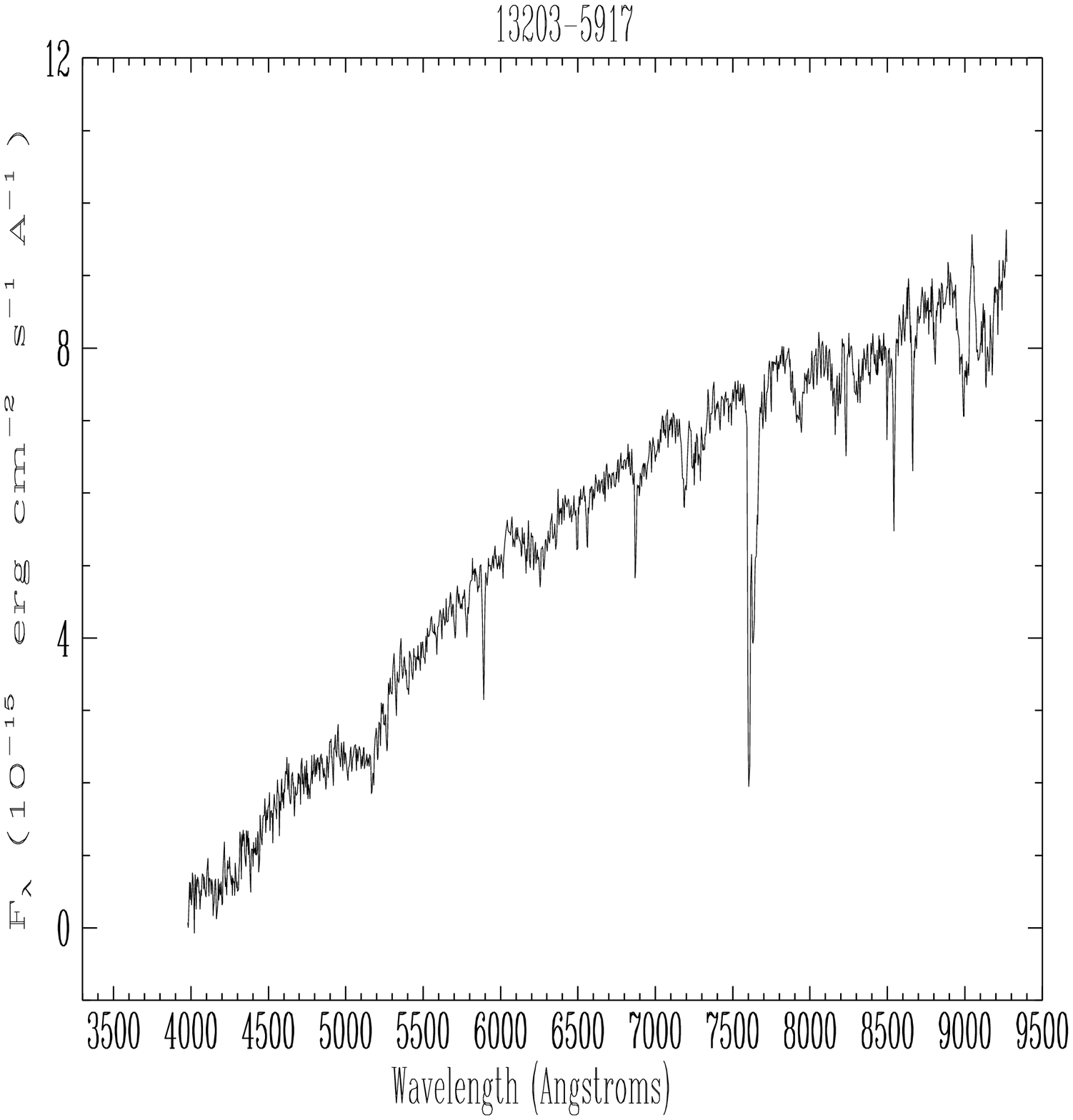}
%\psdraft
\epsfxsize=4cm
\epsfysize=4cm
\epsfbox{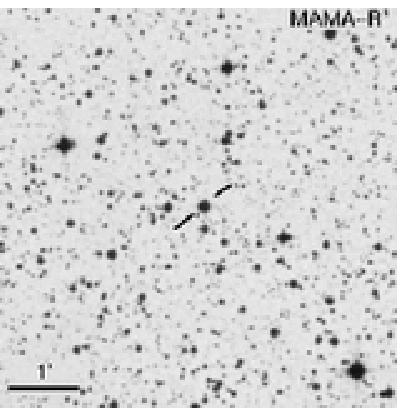}
%\psfull
\end{center}

\caption{Spectra of the objects classified as post-AGB in the sample together with their 
corresponding identification charts (continued). }
\end{figure*}

%-------------------------------------------------------------
%pg8
\setcounter{figure}{0}
\begin{figure*}

\begin{center}
\epsfxsize=13.5cm
\epsfysize=4cm
\epsfbox{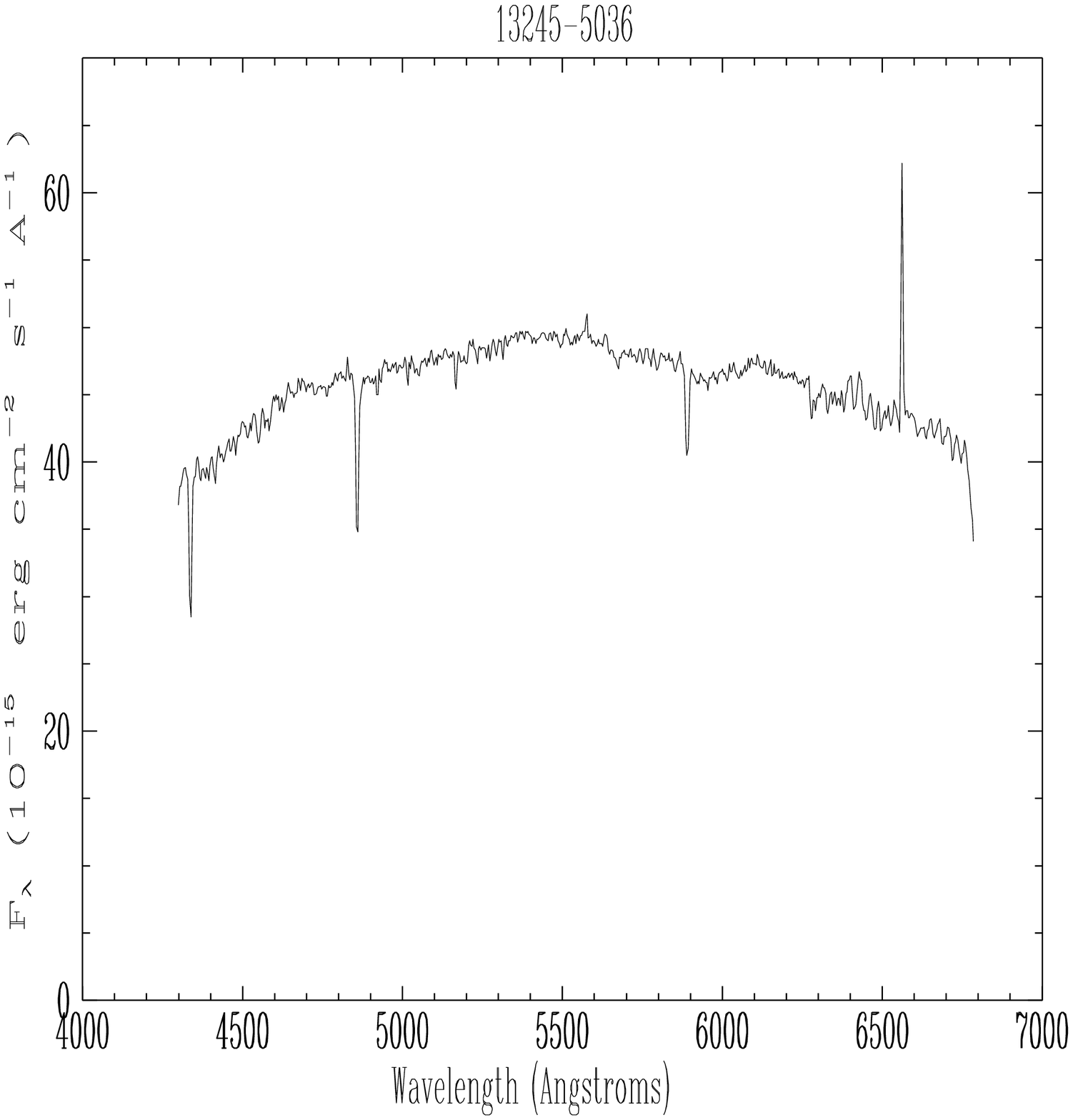}
%\psdraft
\epsfxsize=4cm
\epsfysize=4cm
\epsfbox{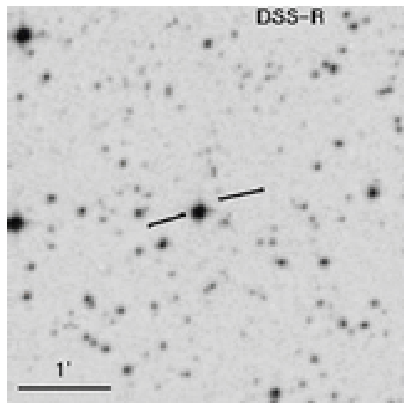}
%\psfull
\end{center}

\begin{center}
\epsfxsize=13.5cm
\epsfysize=4cm
\epsfbox{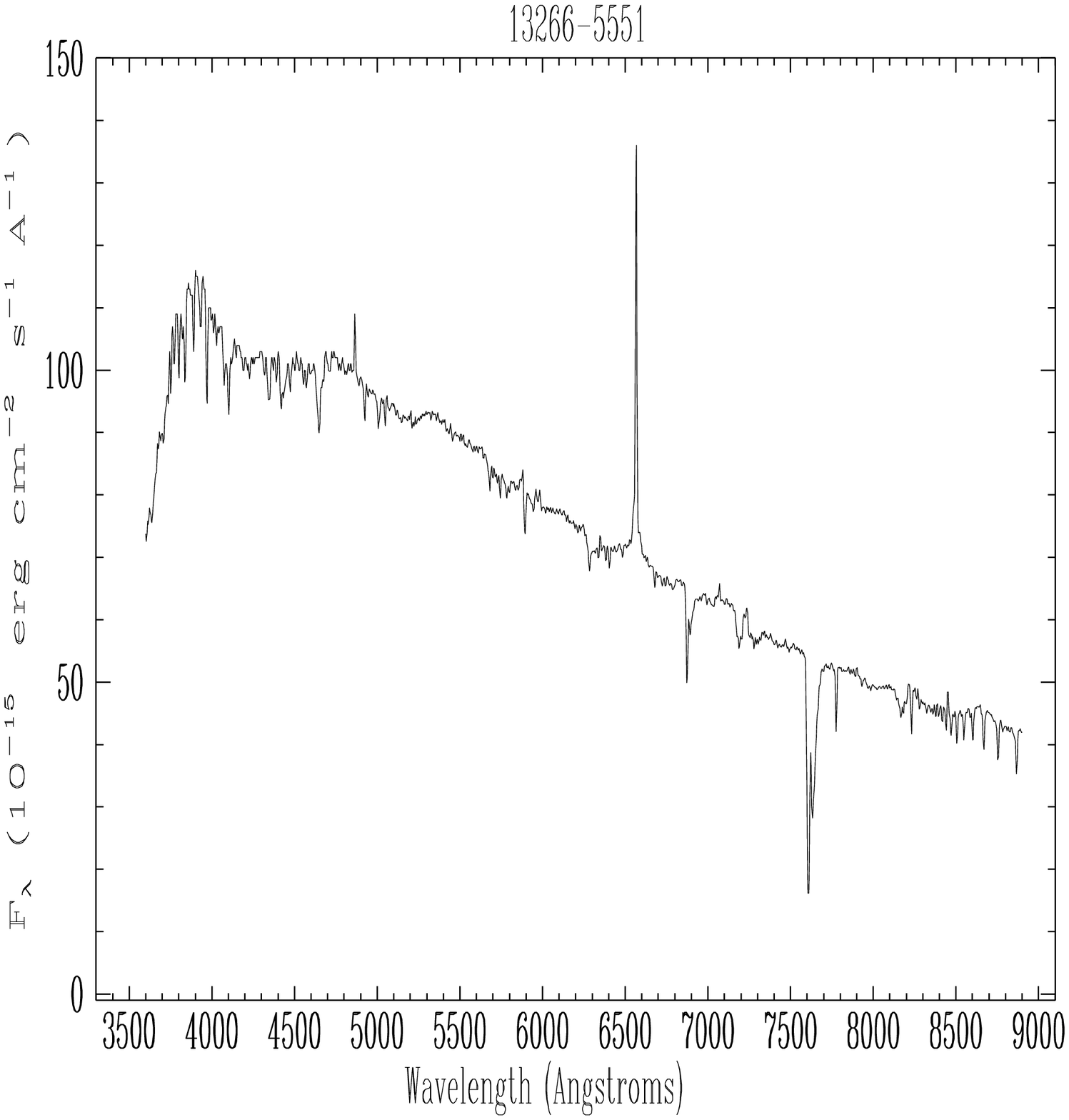}
%\psdraft
\epsfxsize=4cm
\epsfysize=4cm
\epsfbox{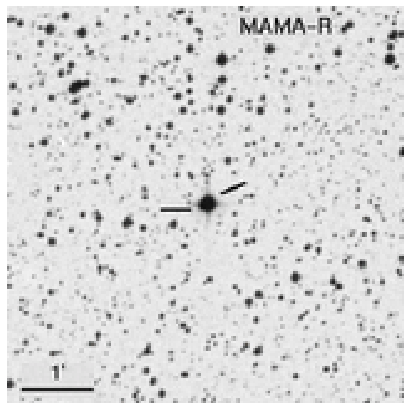}
%\psfull
\end{center}

\begin{center}
\epsfxsize=13.5cm
\epsfysize=4cm
\epsfbox{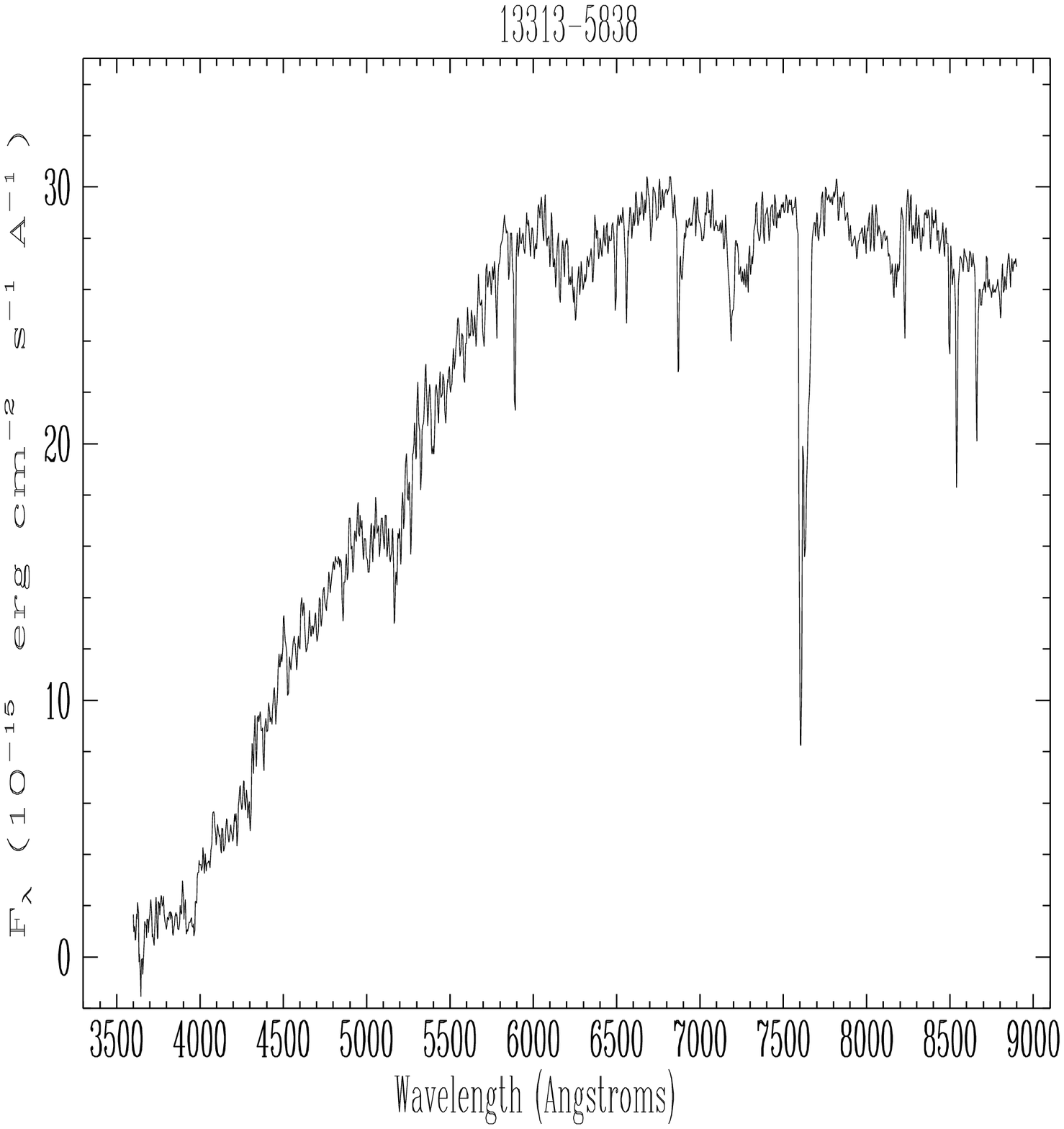}
%\psdraft
\epsfxsize=4cm
\epsfysize=4cm
\epsfbox{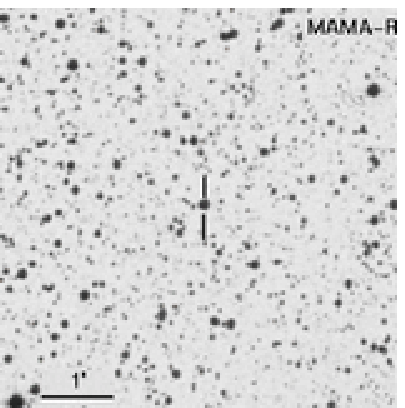}
%\psfull
\end{center}

\begin{center}
\epsfxsize=13.5cm
\epsfysize=4cm
\epsfbox{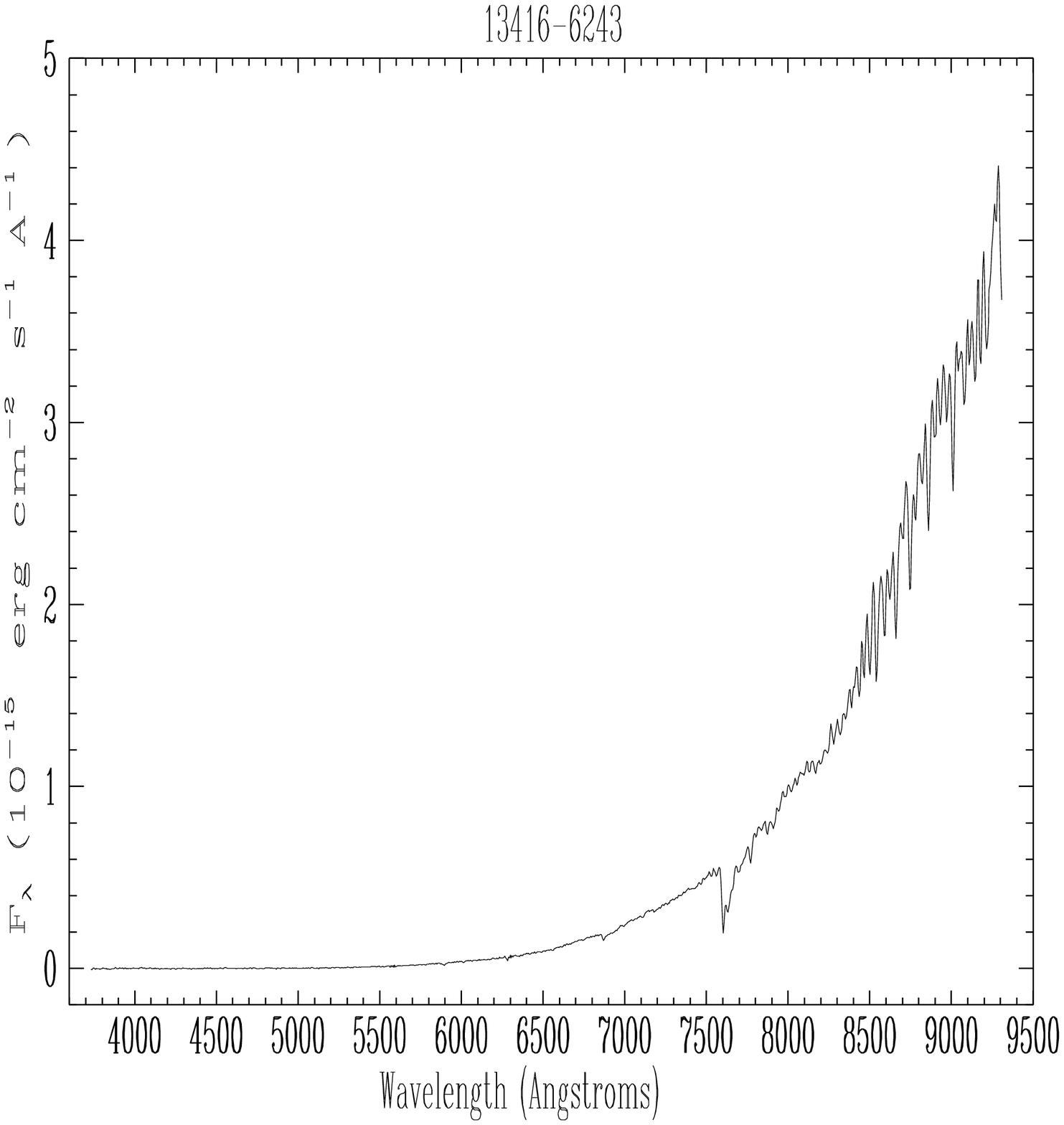}
%\psdraft
\epsfxsize=4cm
\epsfysize=4cm
\epsfbox{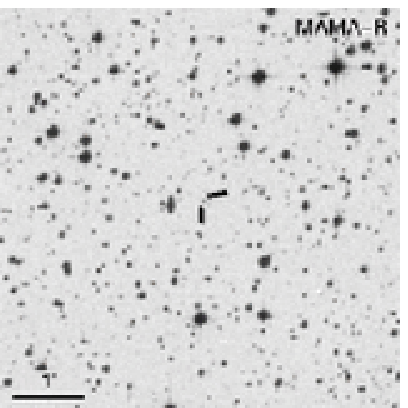}
%\psfull
\end{center}

\begin{center}
\epsfxsize=13.5cm
\epsfysize=4cm
\epsfbox{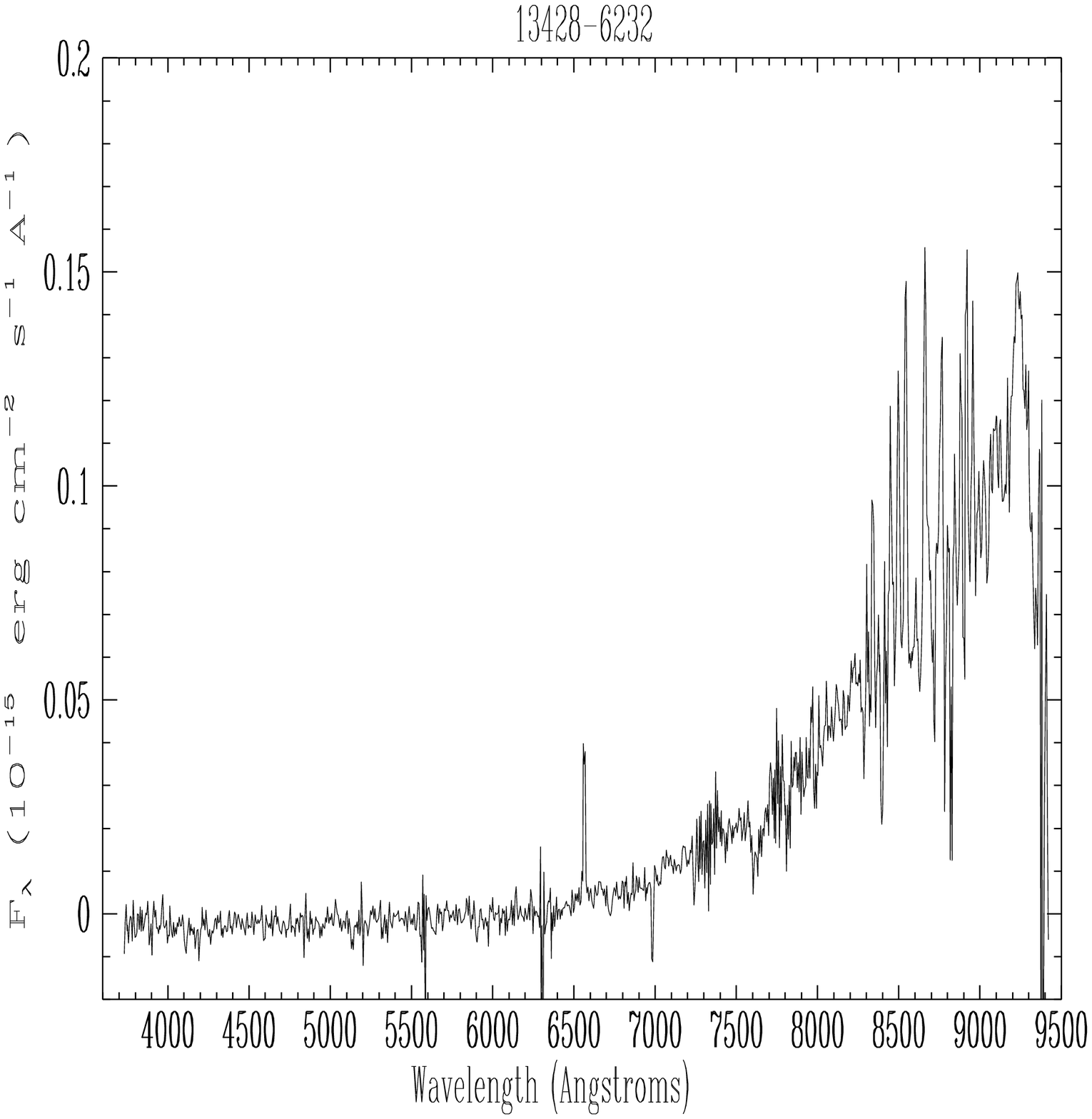}
%\psdraft
\epsfxsize=4cm
\epsfysize=4cm
\epsfbox{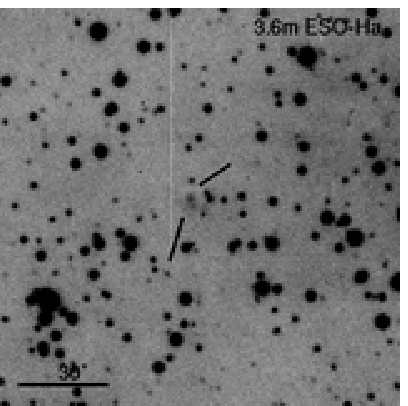}
%\psfull
\end{center}

\caption{Spectra of the objects classified as post-AGB in the sample together with their 
corresponding identification charts (continued). }
\end{figure*}

%-------------------------------------------------------------
%pg9
\setcounter{figure}{0}
\begin{figure*}

\begin{center}
\epsfxsize=13.5cm
\epsfysize=4cm
\epsfbox{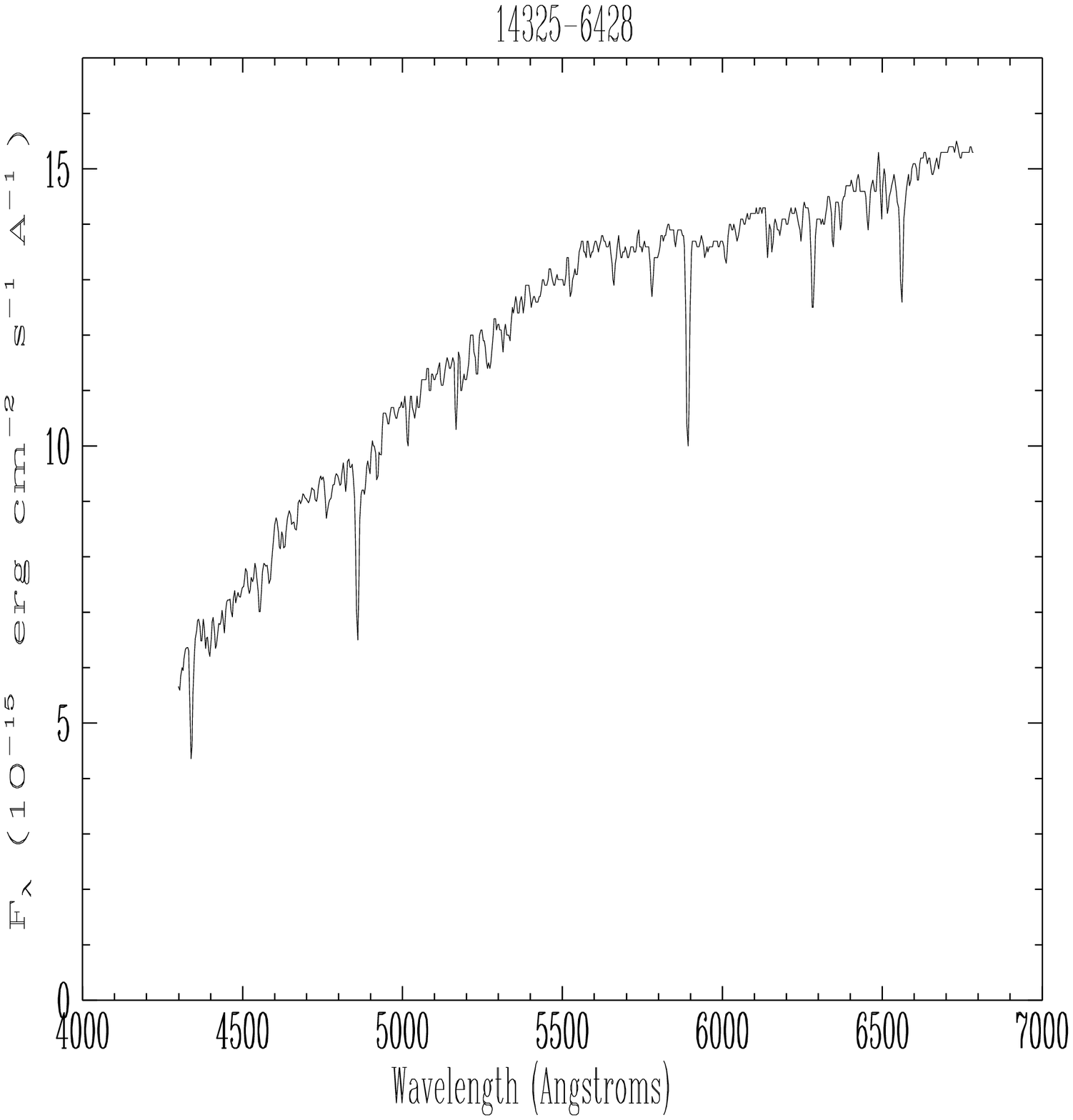}
%\psdraft
\epsfxsize=4cm
\epsfysize=4cm
\epsfbox{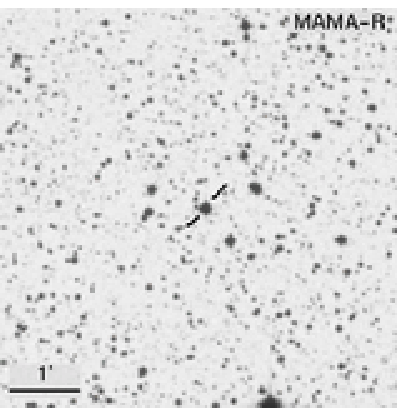}
%\psfull
\end{center}

\begin{center}
\epsfxsize=13.5cm
\epsfysize=4cm
\epsfbox{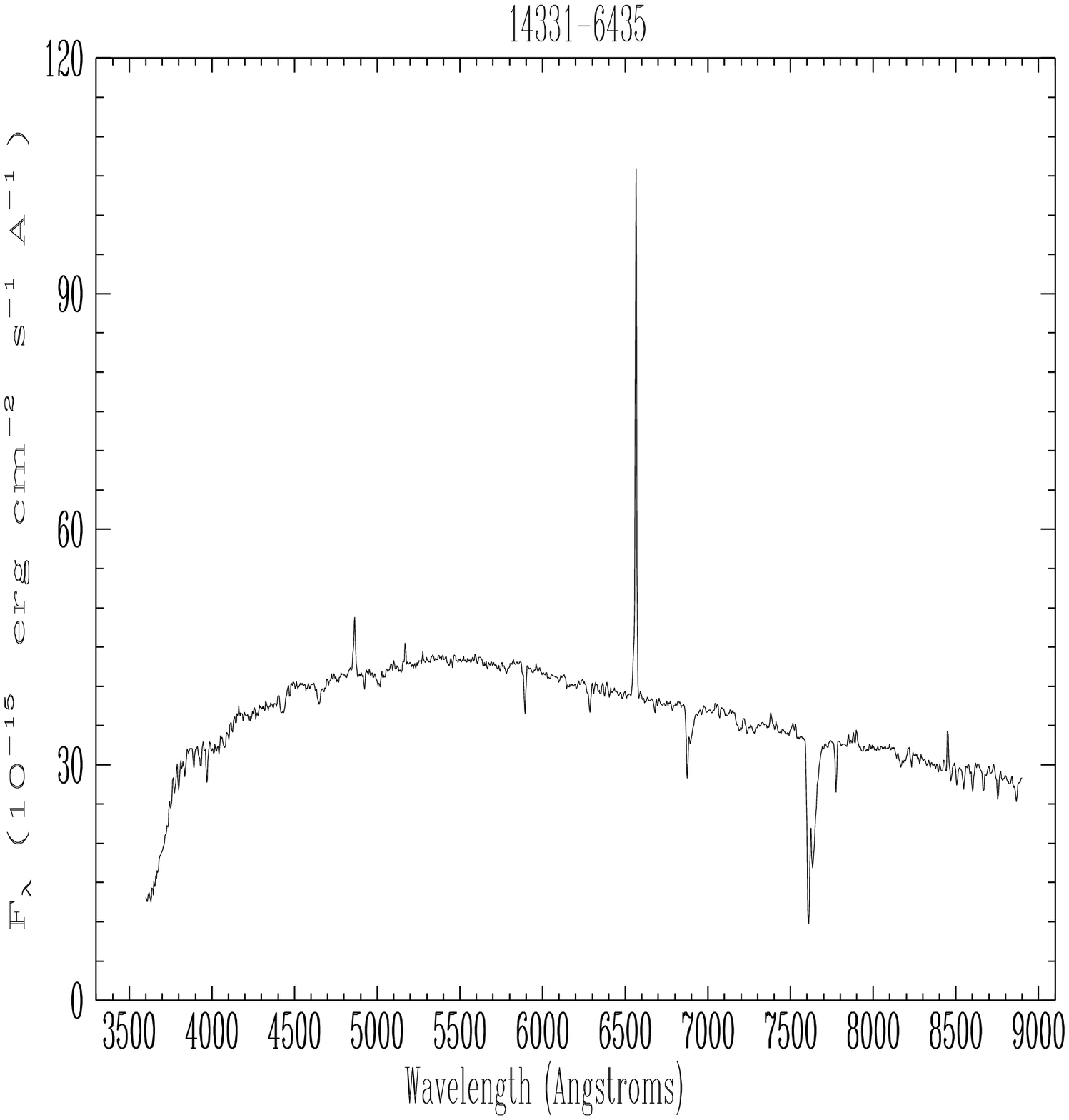}
%\psdraft
\epsfxsize=4cm
\epsfysize=4cm
\epsfbox{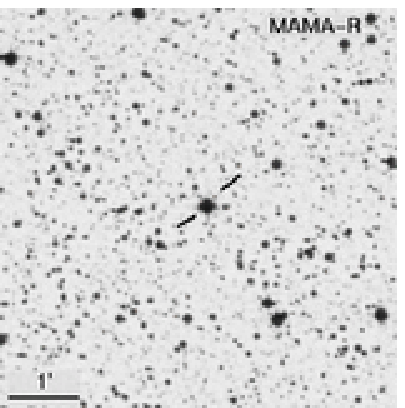}
%\psfull
\end{center}

\begin{center}
\epsfxsize=13.5cm
\epsfysize=4cm
\epsfbox{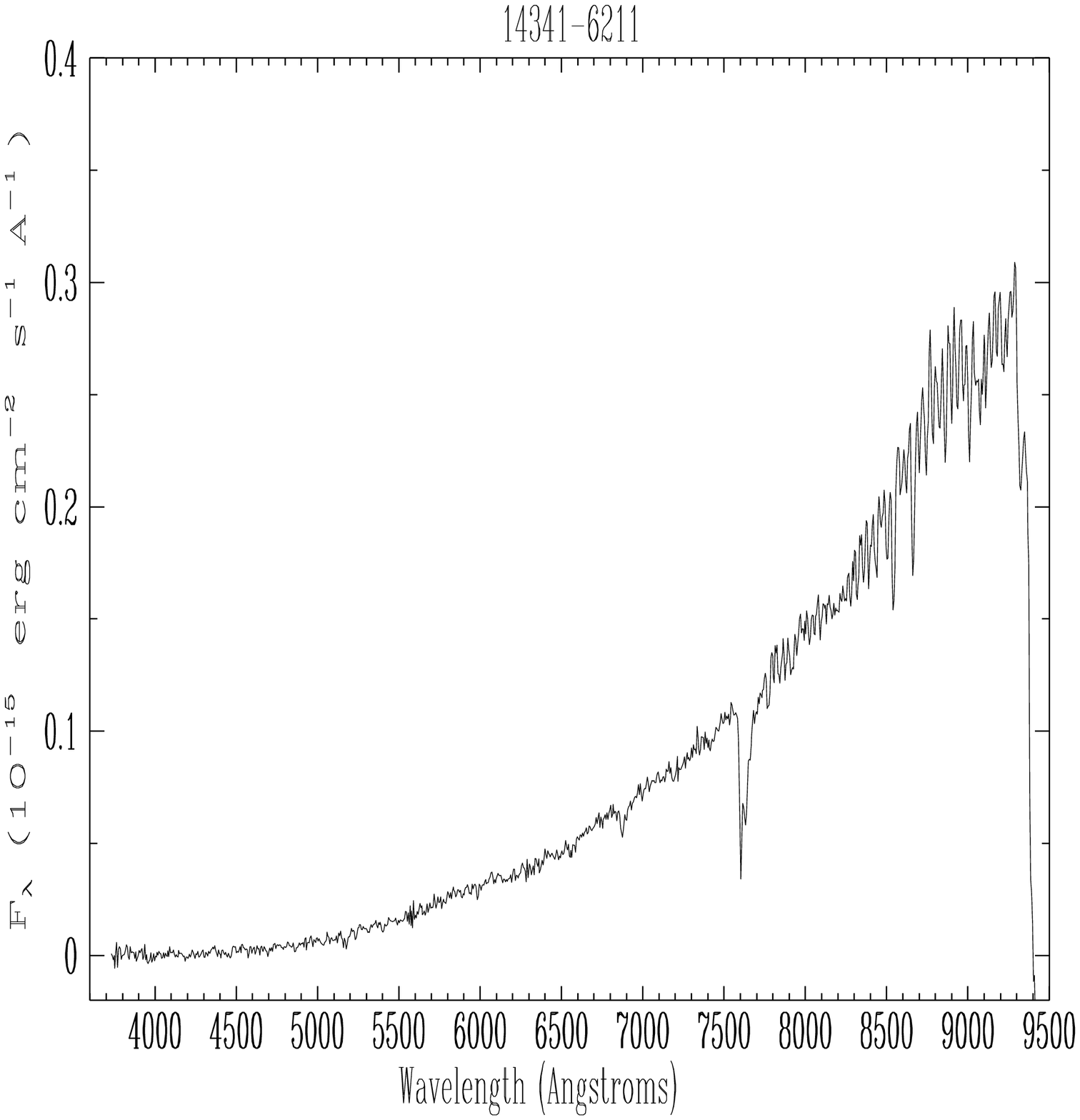}
%\psdraft
\epsfxsize=4cm
\epsfysize=4cm
\epsfbox{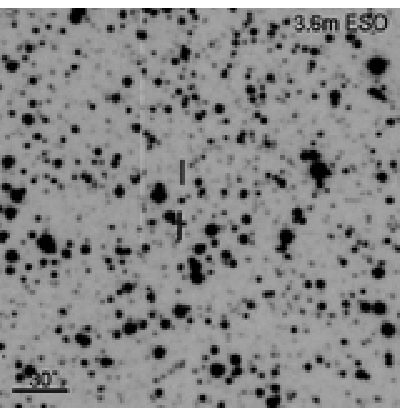}
%\psfull
\end{center}

\begin{center}
\epsfxsize=13.5cm
\epsfysize=4cm
\epsfbox{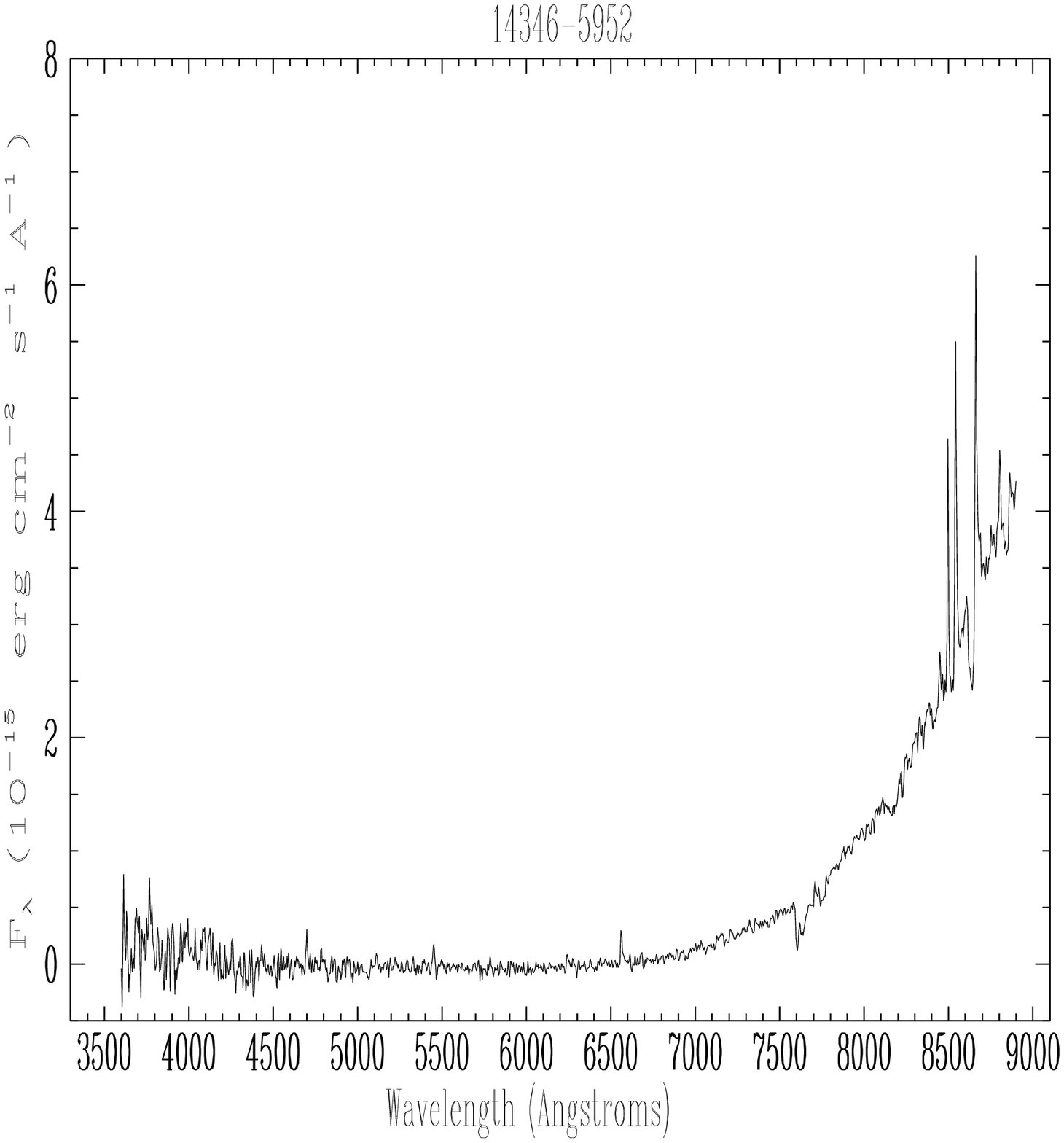}
%\psdraft
\epsfxsize=4cm
\epsfysize=4cm
\epsfbox{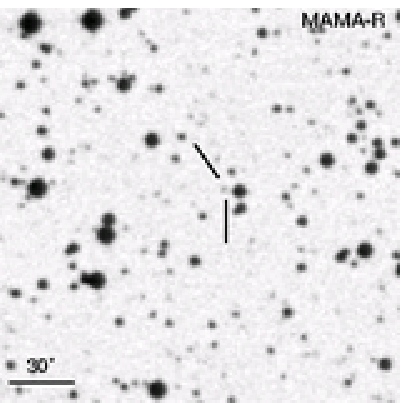}
%\psfull
\end{center}

\begin{center}
\epsfxsize=13.5cm
\epsfysize=4cm
\epsfbox{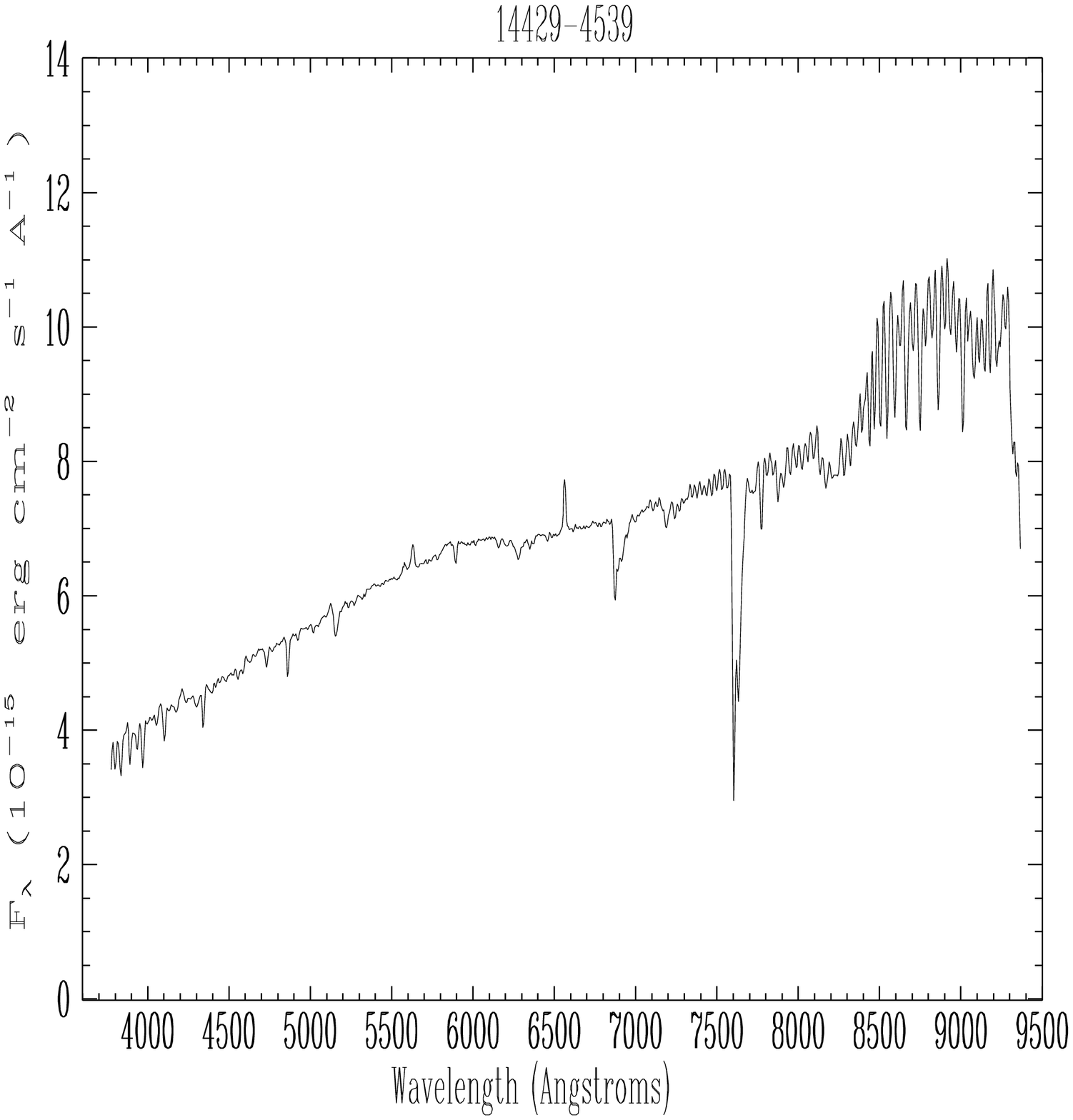}
%\psdraft
\epsfxsize=4cm
\epsfysize=4cm
\epsfbox{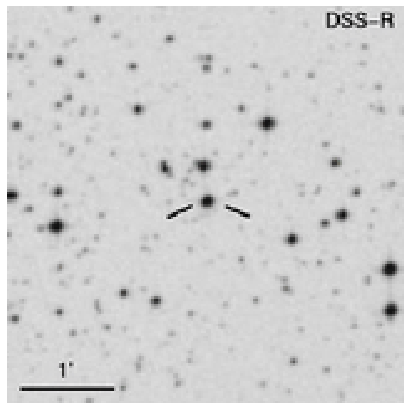}
%\psfull
\end{center}

\caption{Spectra of the objects classified as post-AGB in the sample together with their 
corresponding identification charts (continued). }
\end{figure*}

%%% Local Variables: 
%%% mode: latex
%%% TeX-master: "~/tesis/mitesis/final/tesis"
%%% End: 

        \begin{figure*}
\setcounter{figure}{0}

\begin{center}
\epsfxsize=13.5cm
\epsfysize=4cm
\epsfbox{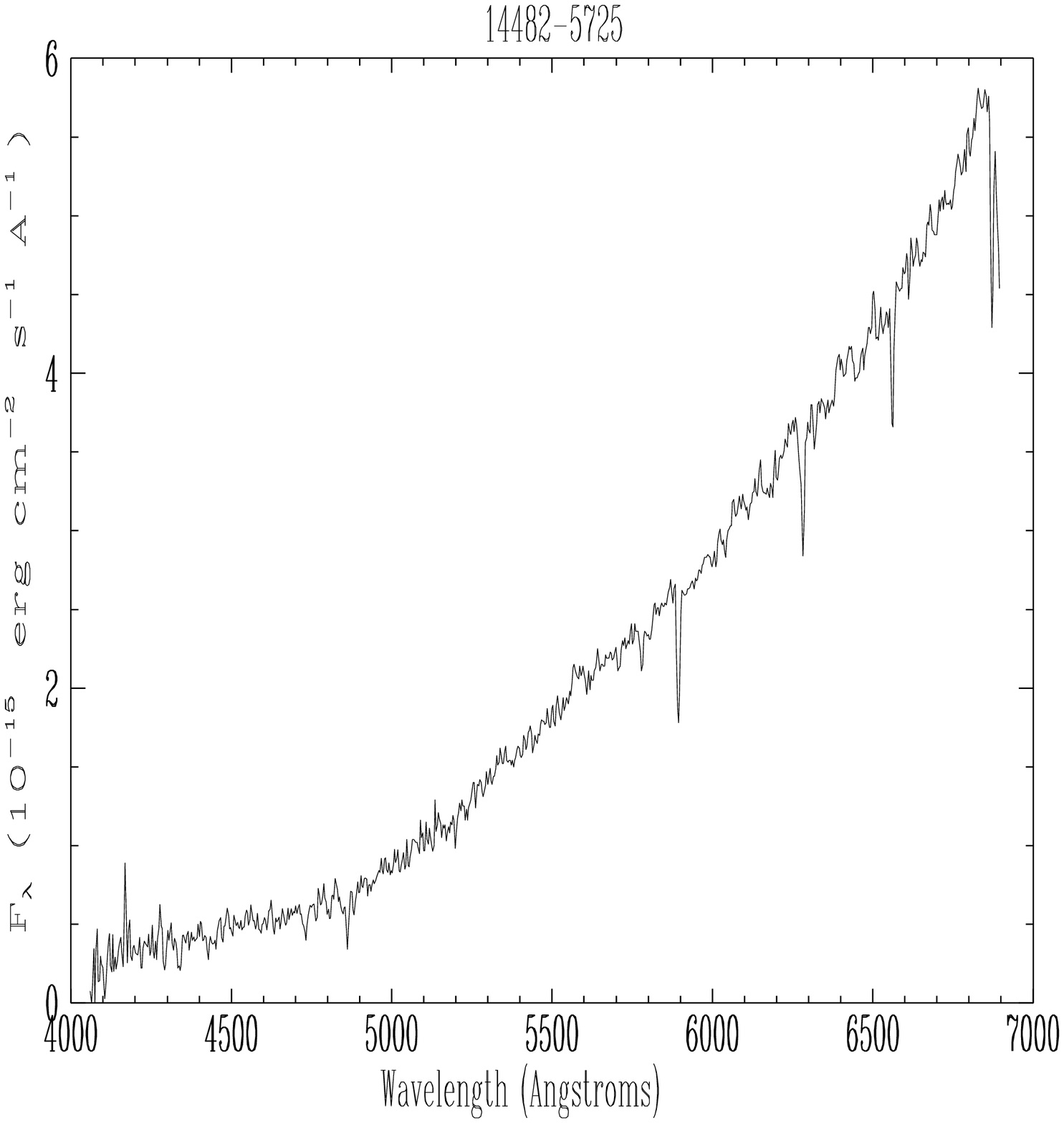}
%\psdraft
\epsfxsize=4cm
\epsfysize=4cm
\epsfbox{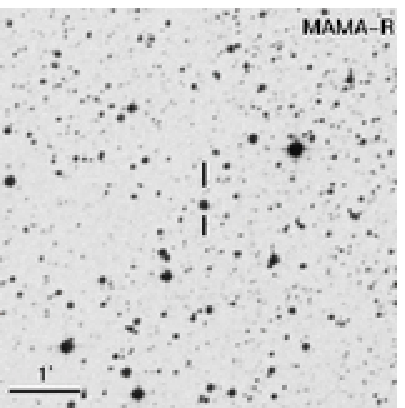}
%\psfull
\end{center}

\begin{center}
\epsfxsize=13.5cm
\epsfysize=4cm
\epsfbox{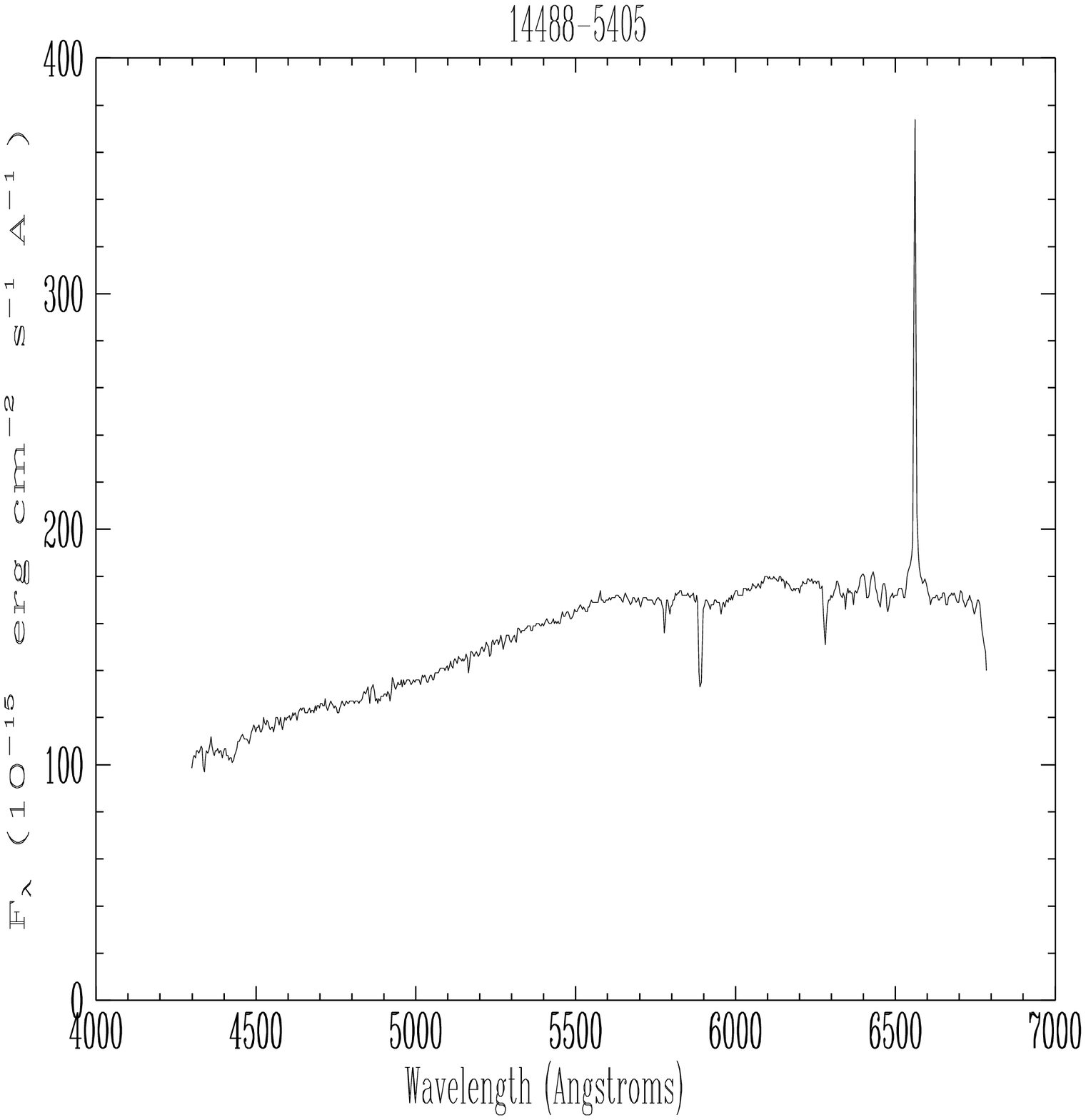}
%\psdraft
\epsfxsize=4cm
\epsfysize=4cm
\epsfbox{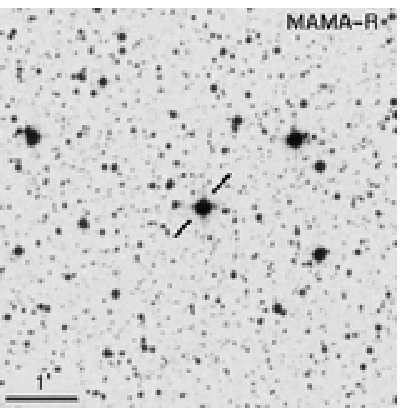}
%\psfull
\end{center}

\begin{center}
\epsfxsize=13.5cm
\epsfysize=4cm
\epsfbox{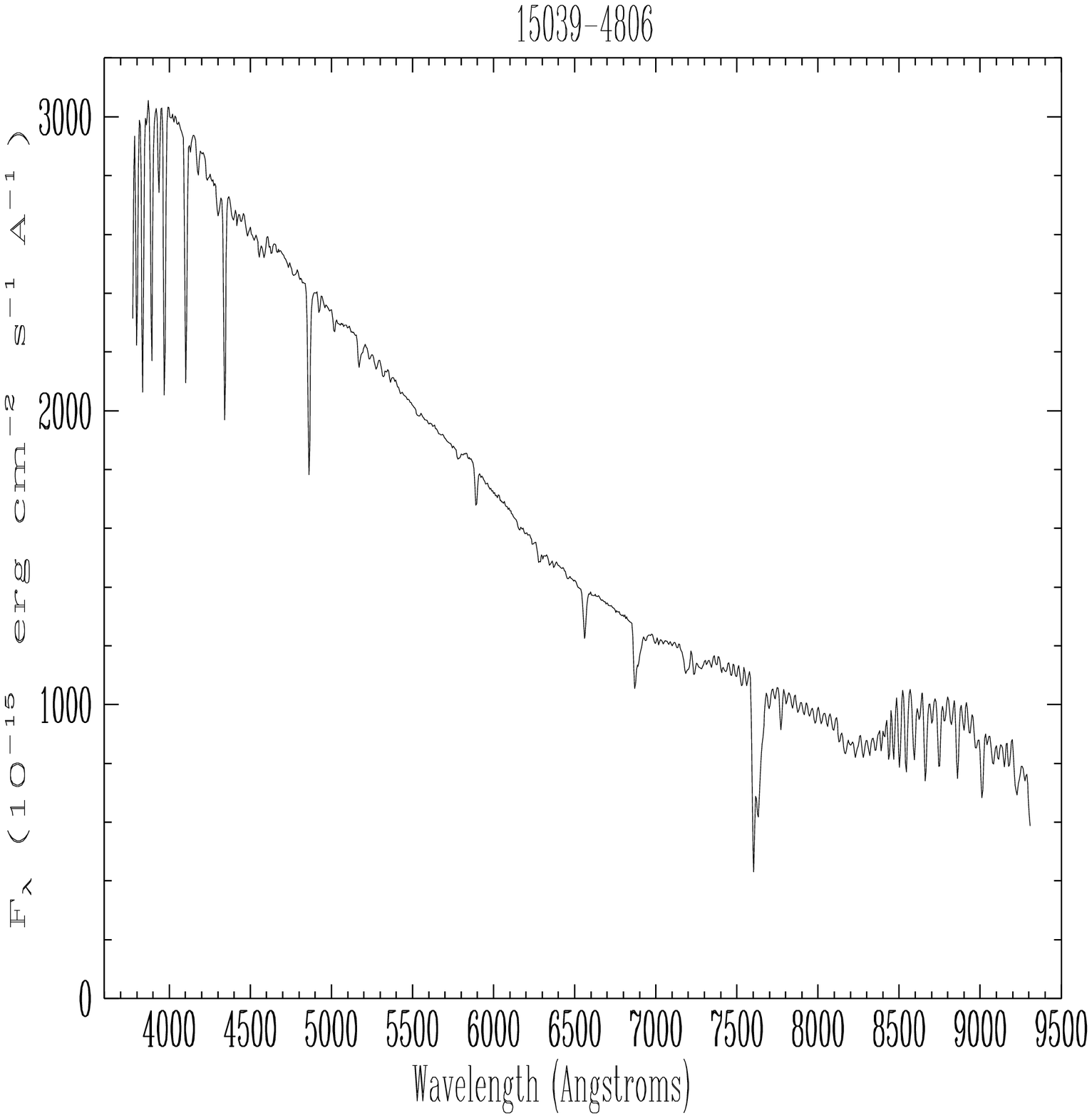}
%\psdraft
\epsfxsize=4cm
\epsfysize=4cm
\epsfbox{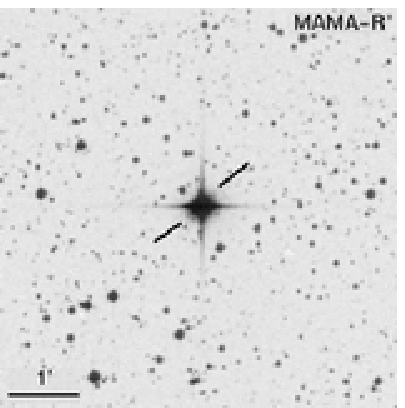}
%\psfull
\end{center}

\begin{center}
\epsfxsize=13.5cm
\epsfysize=4cm
\epsfbox{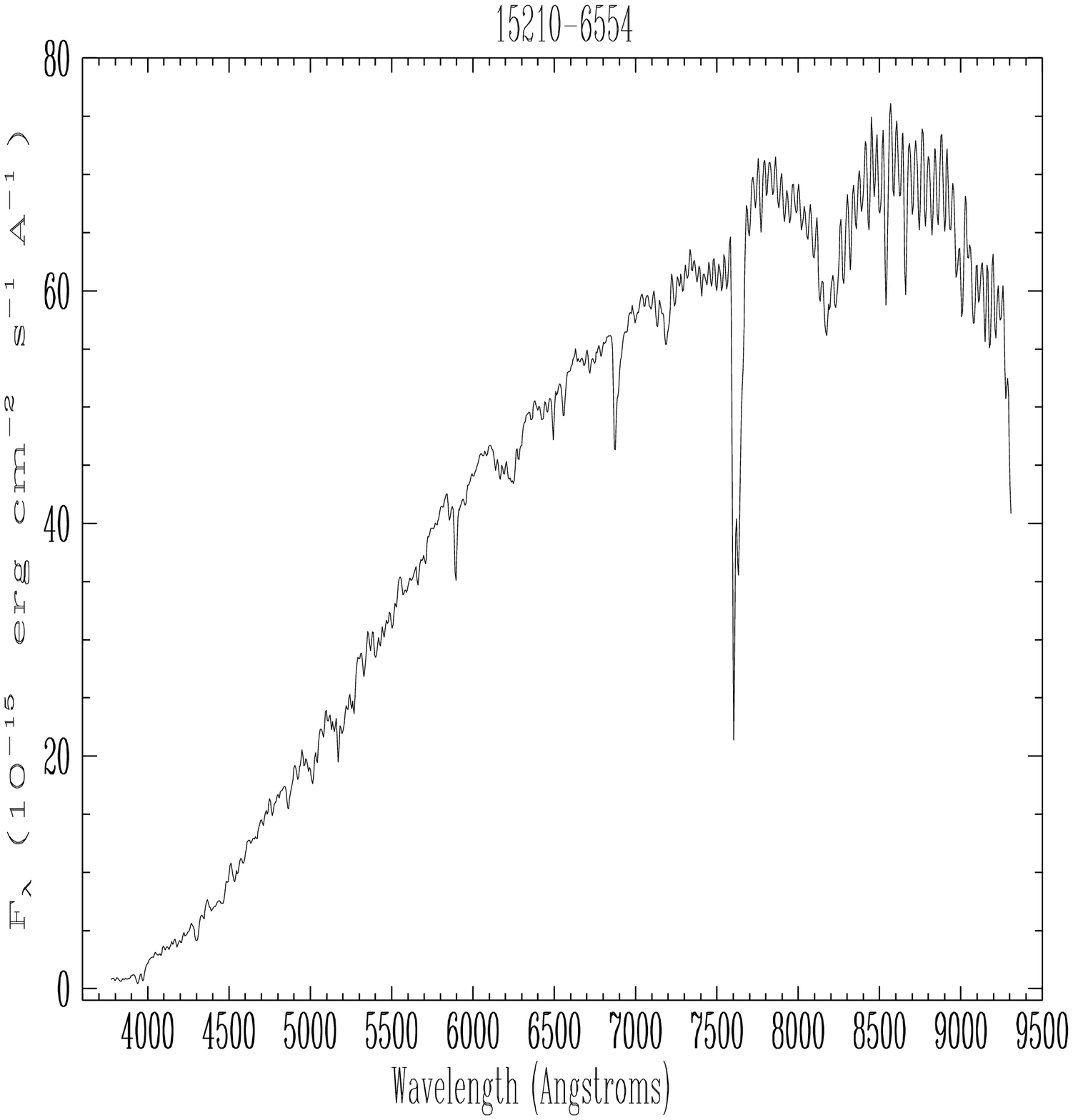}
%\psdraft
\epsfxsize=4cm
\epsfysize=4cm
\epsfbox{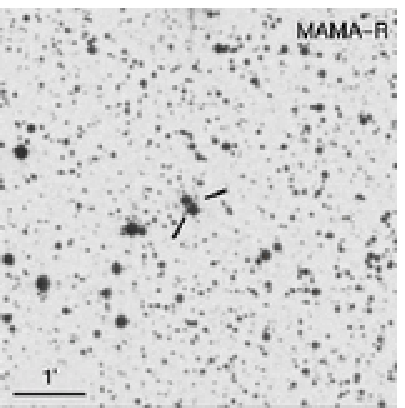}
%\psfull
\end{center}

\begin{center}
\epsfxsize=13.5cm
\epsfysize=4cm
\epsfbox{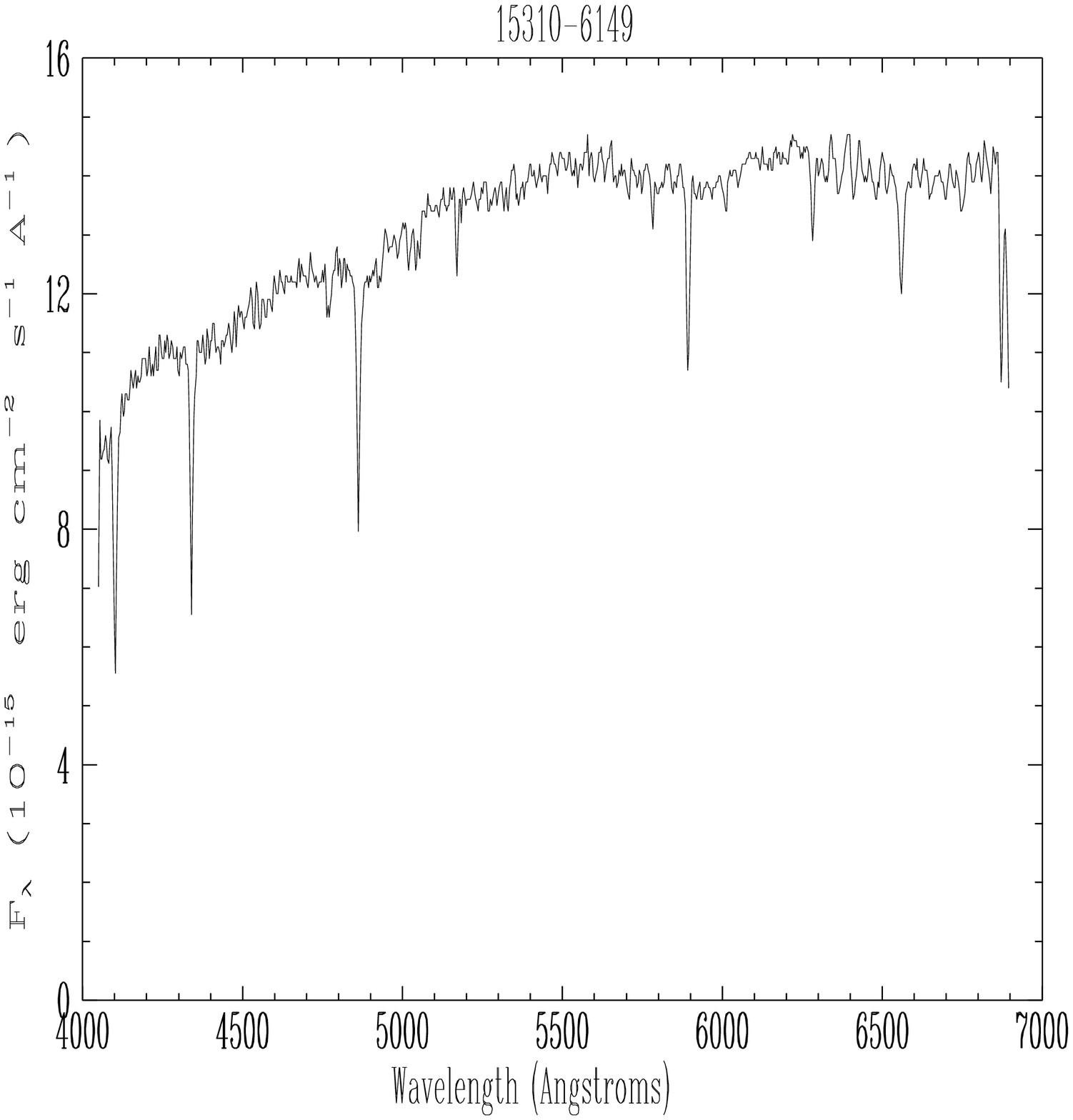}
%\psdraft
\epsfxsize=4cm
\epsfysize=4cm
\epsfbox{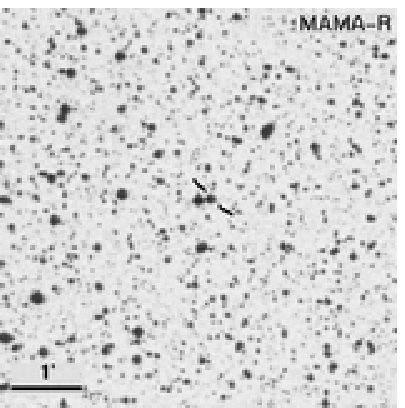}
%\psfull
\end{center}

\caption{Spectra of the objects classified as post-AGB in the sample together with their 
corresponding identification charts (continued). }
\end{figure*}

%-------------------------------------------------------------
%pg11
\clearpage
\setcounter{figure}{0}
\begin{figure*}

\begin{center}
\epsfxsize=13.5cm
\epsfysize=4cm
\epsfbox{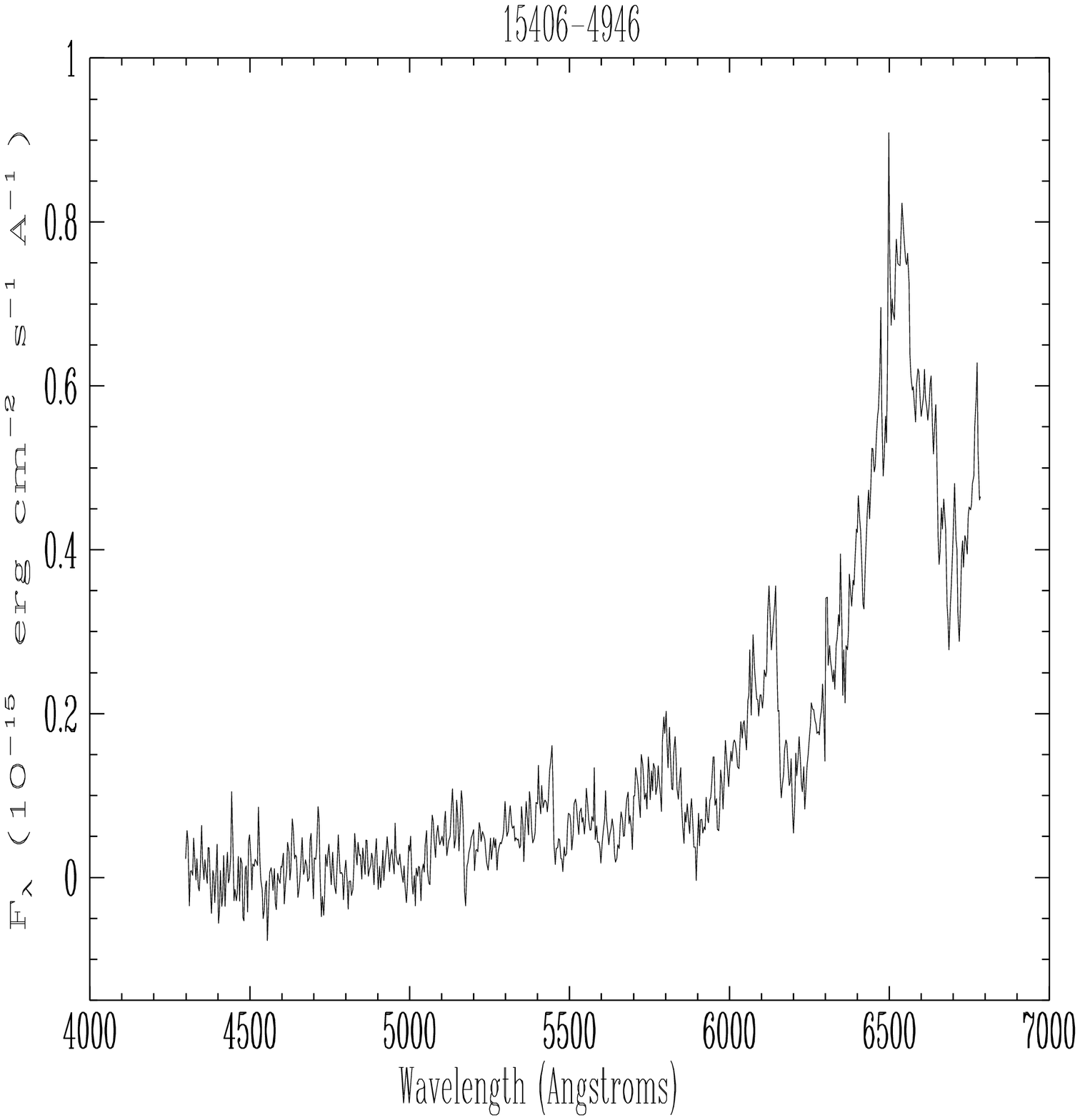}
%\psdraft
\epsfxsize=4cm
\epsfysize=4cm
\epsfbox{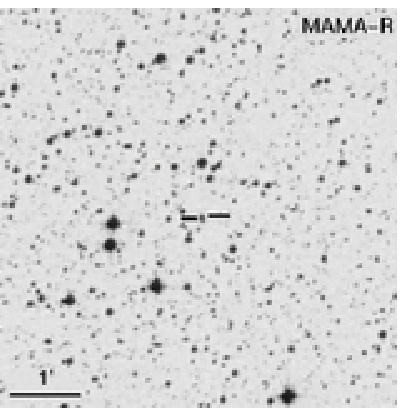}
%\psfull
\end{center}

\begin{center}
\epsfxsize=13.5cm
\epsfysize=4cm
\epsfbox{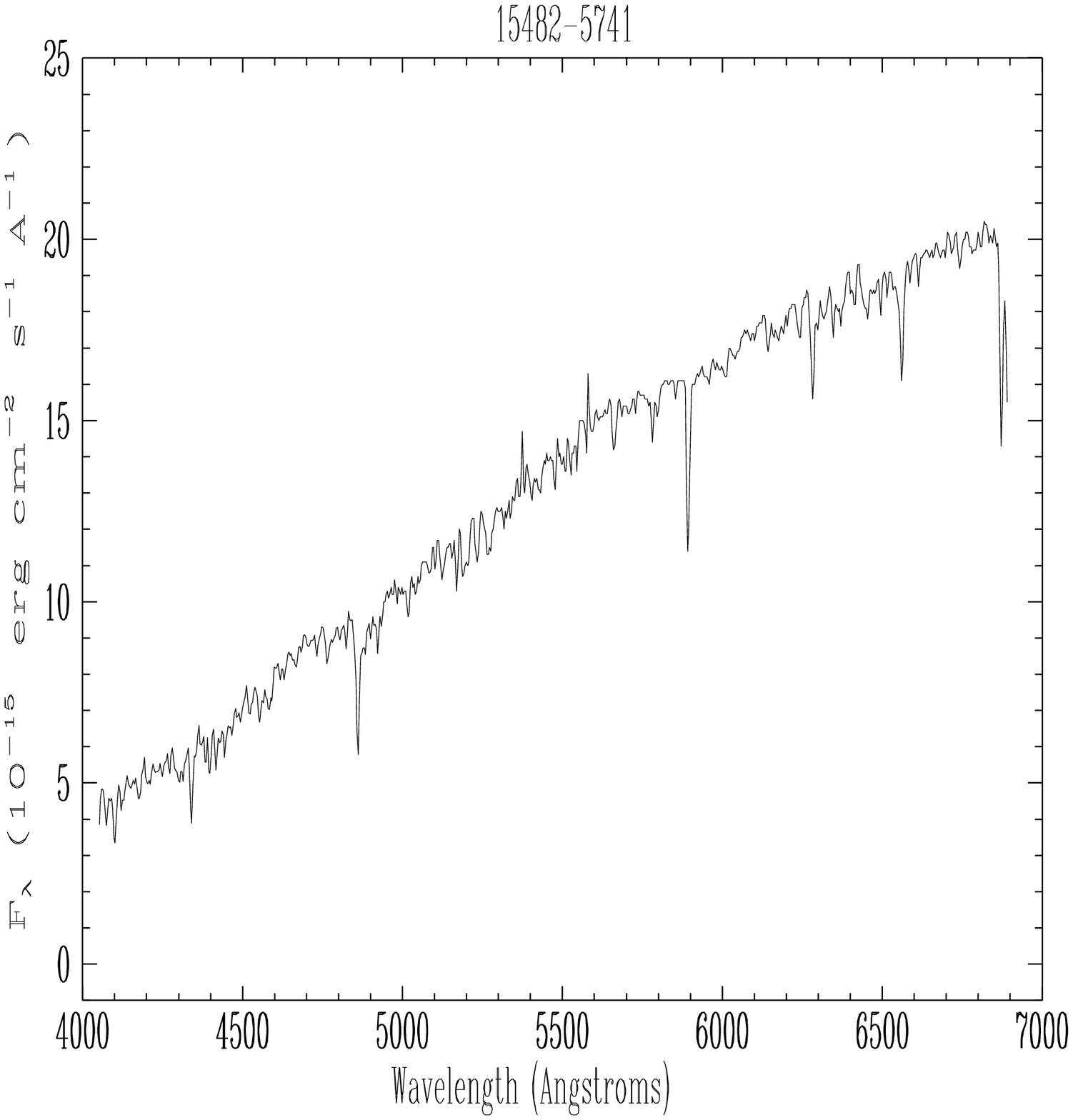}
%\psdraft
\epsfxsize=4cm
\epsfysize=4cm
\epsfbox{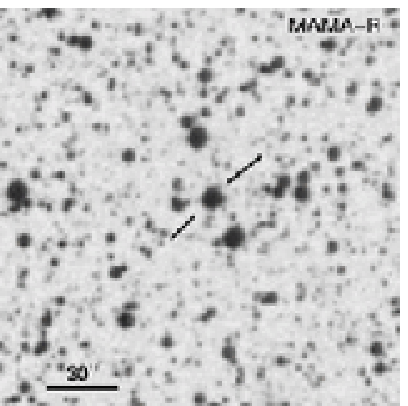}
%\psfull
\end{center}

\begin{center}
\epsfxsize=13.5cm
\epsfysize=4cm
\epsfbox{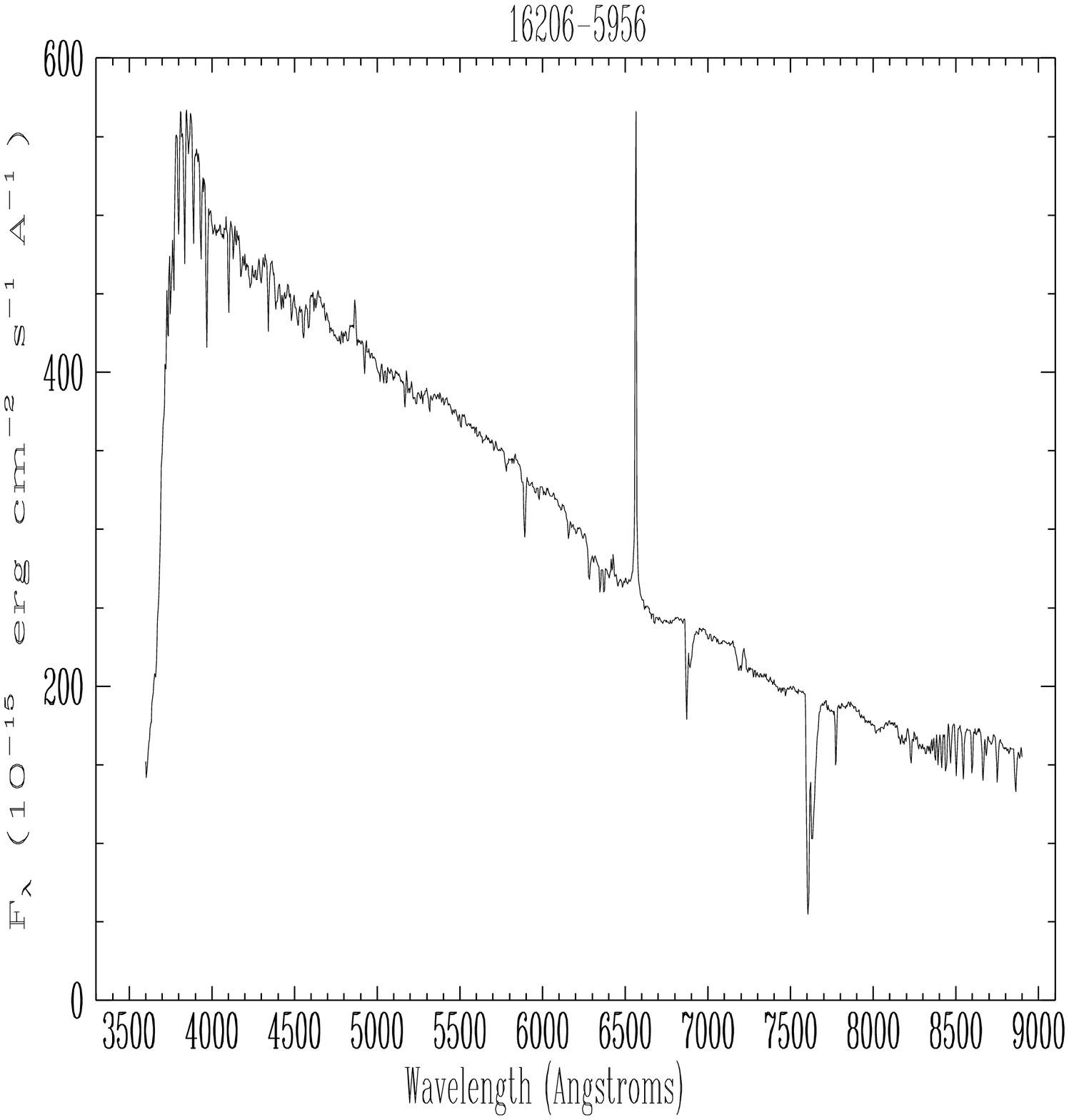}
%\psdraft
\epsfxsize=4cm
\epsfysize=4cm
\epsfbox{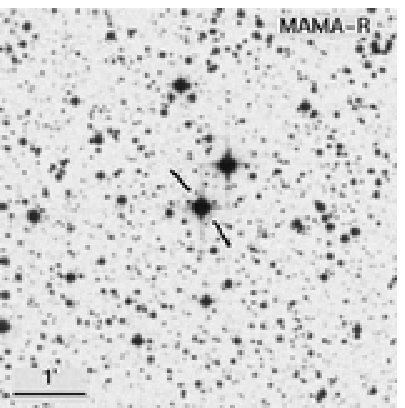}
%\psfull
\end{center}

\begin{center}
\epsfxsize=13.5cm
\epsfysize=4cm
\epsfbox{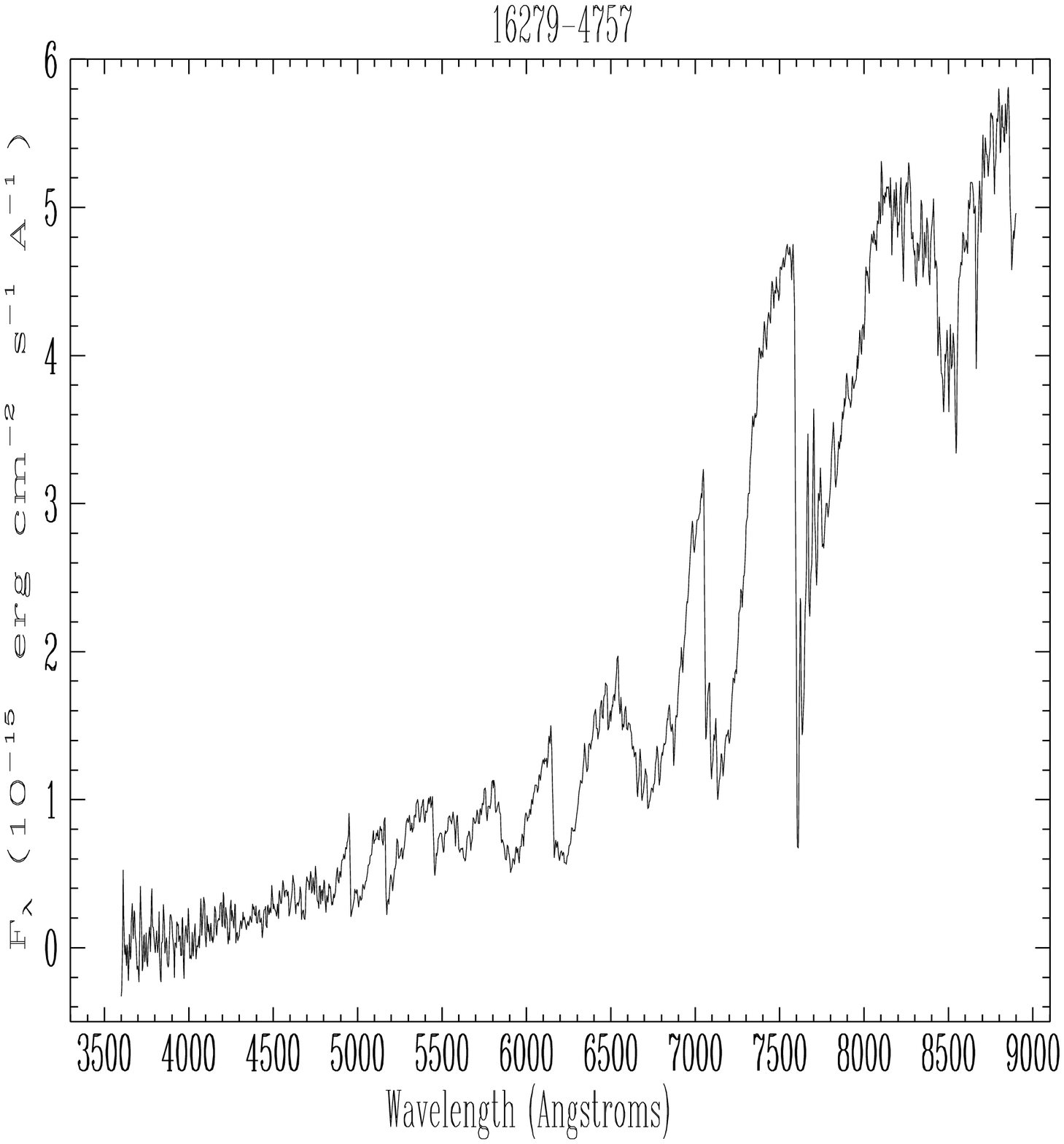}
%\psdraft
\epsfxsize=4cm
\epsfysize=4cm
\epsfbox{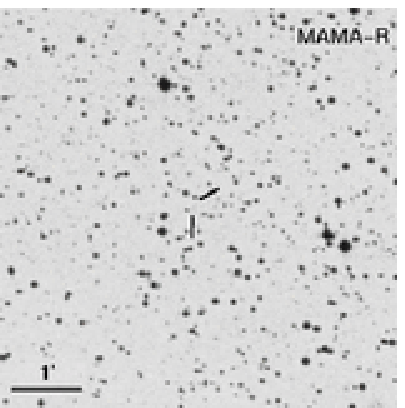}
%\psfull
\end{center}

\begin{center}
\epsfxsize=13.5cm
\epsfysize=4cm
\epsfbox{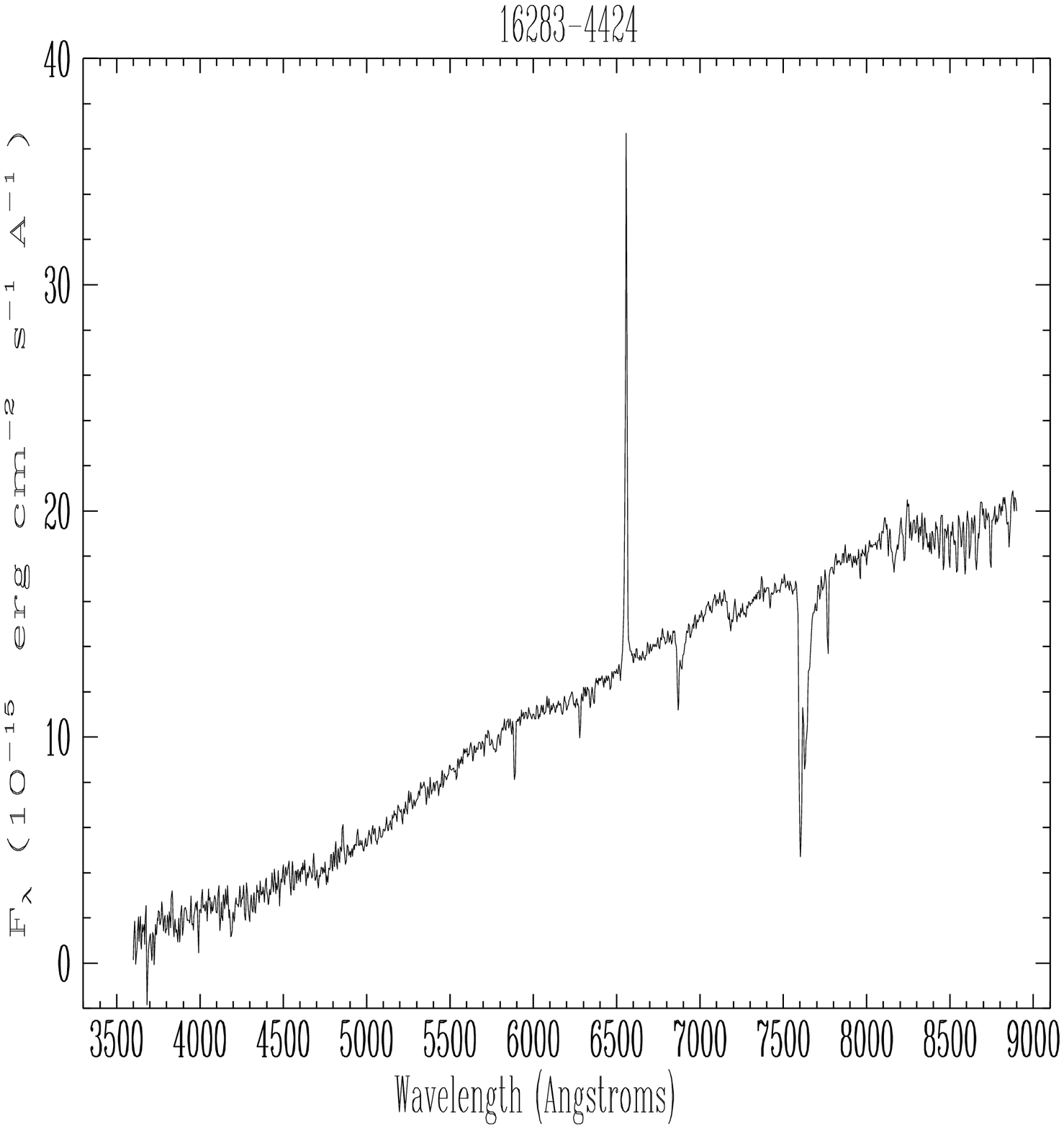}
%\psdraft
\epsfxsize=4cm
\epsfysize=4cm
\epsfbox{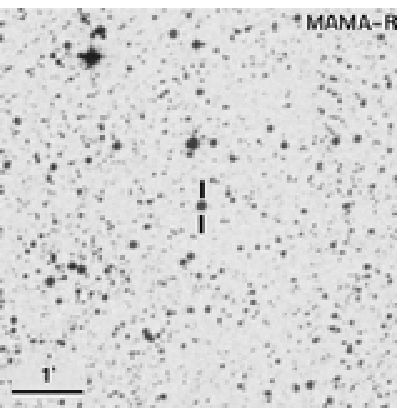}
%\psfull
\end{center}

\caption{Spectra of the objects classified as post-AGB in the sample together with their 
corresponding identification charts (continued). }
\end{figure*}

%%-------------------------------------------------------------
%%pg12
\clearpage
\setcounter{figure}{0}
\begin{figure*}

\begin{center}
\epsfxsize=13.5cm
\epsfysize=4cm
\epsfbox{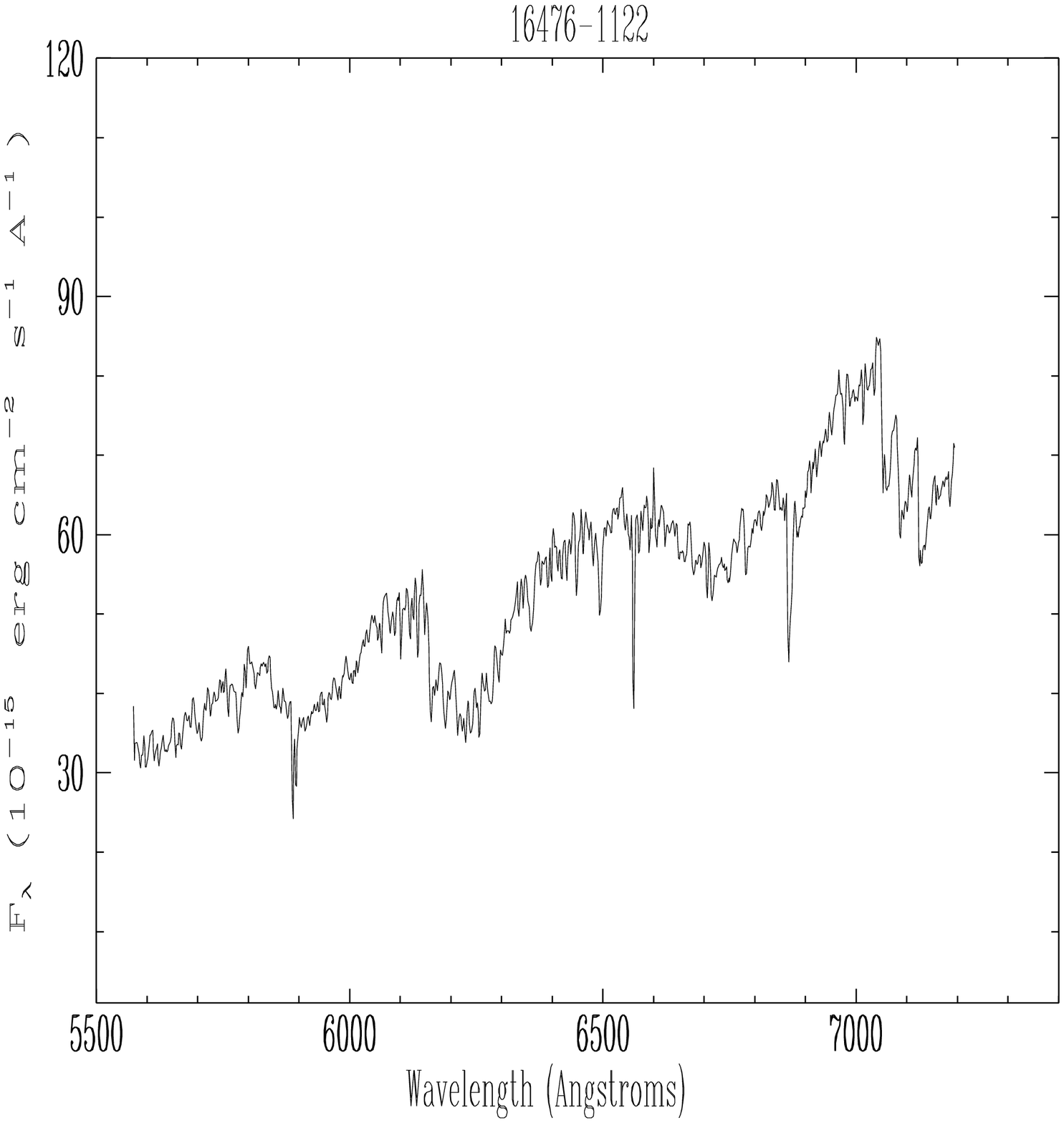}
%\psdraft
\epsfxsize=4cm
\epsfysize=4cm
\epsfbox{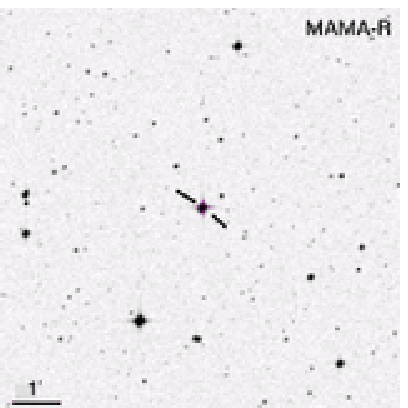}
%\psfull
\end{center}

\begin{center}
\epsfxsize=13.5cm
\epsfysize=4cm
\epsfbox{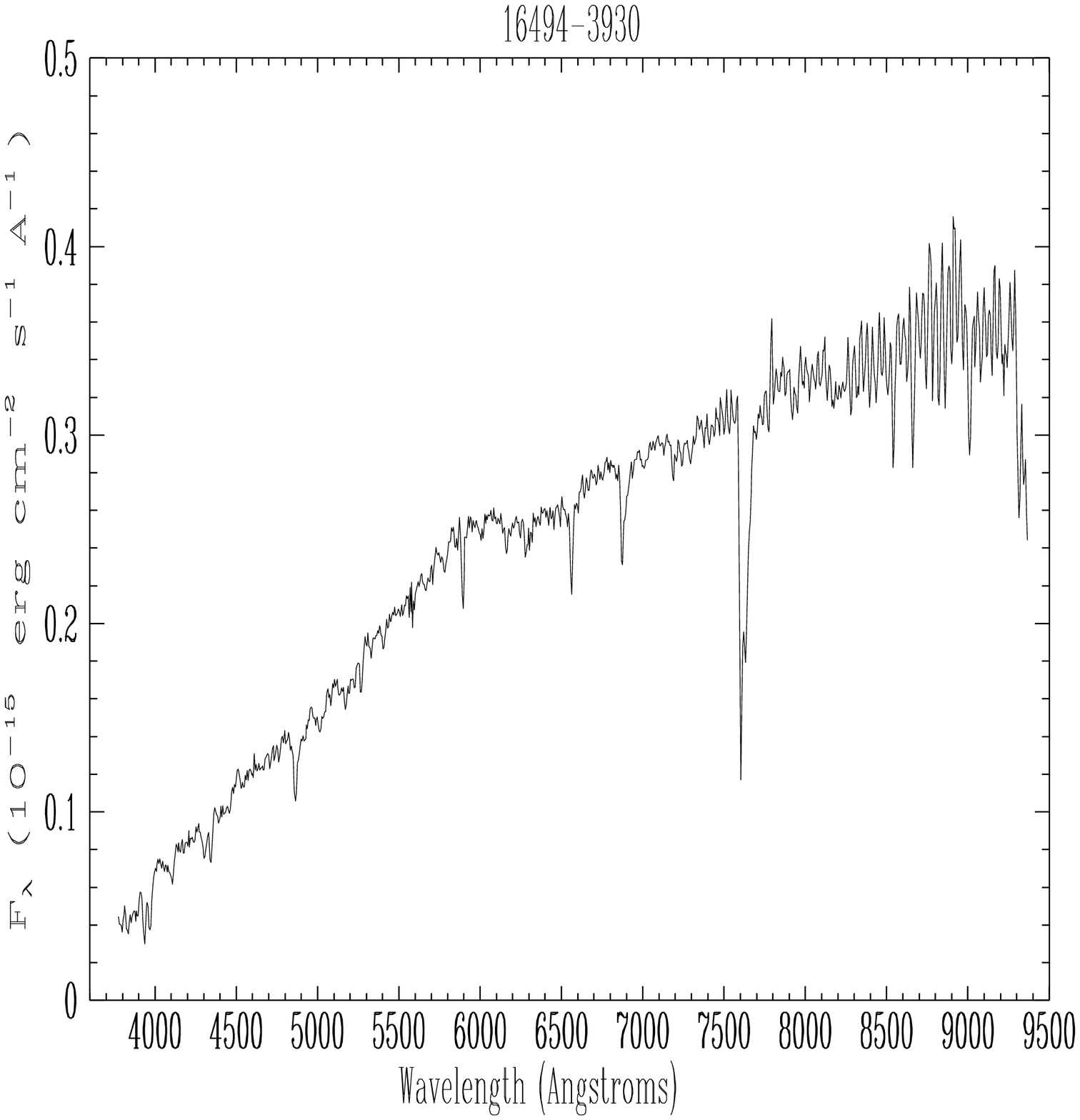}
%\psdraft
\epsfxsize=4cm
\epsfysize=4cm
\epsfbox{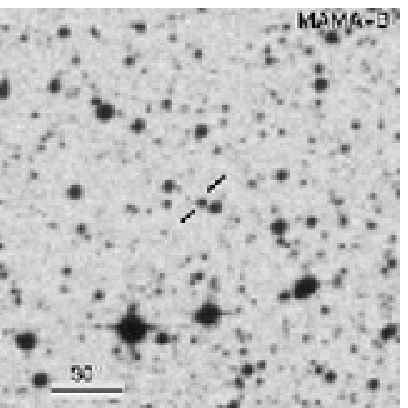}
%\psfull
\end{center}

\begin{center}
\epsfxsize=13.5cm
\epsfysize=4cm
\epsfbox{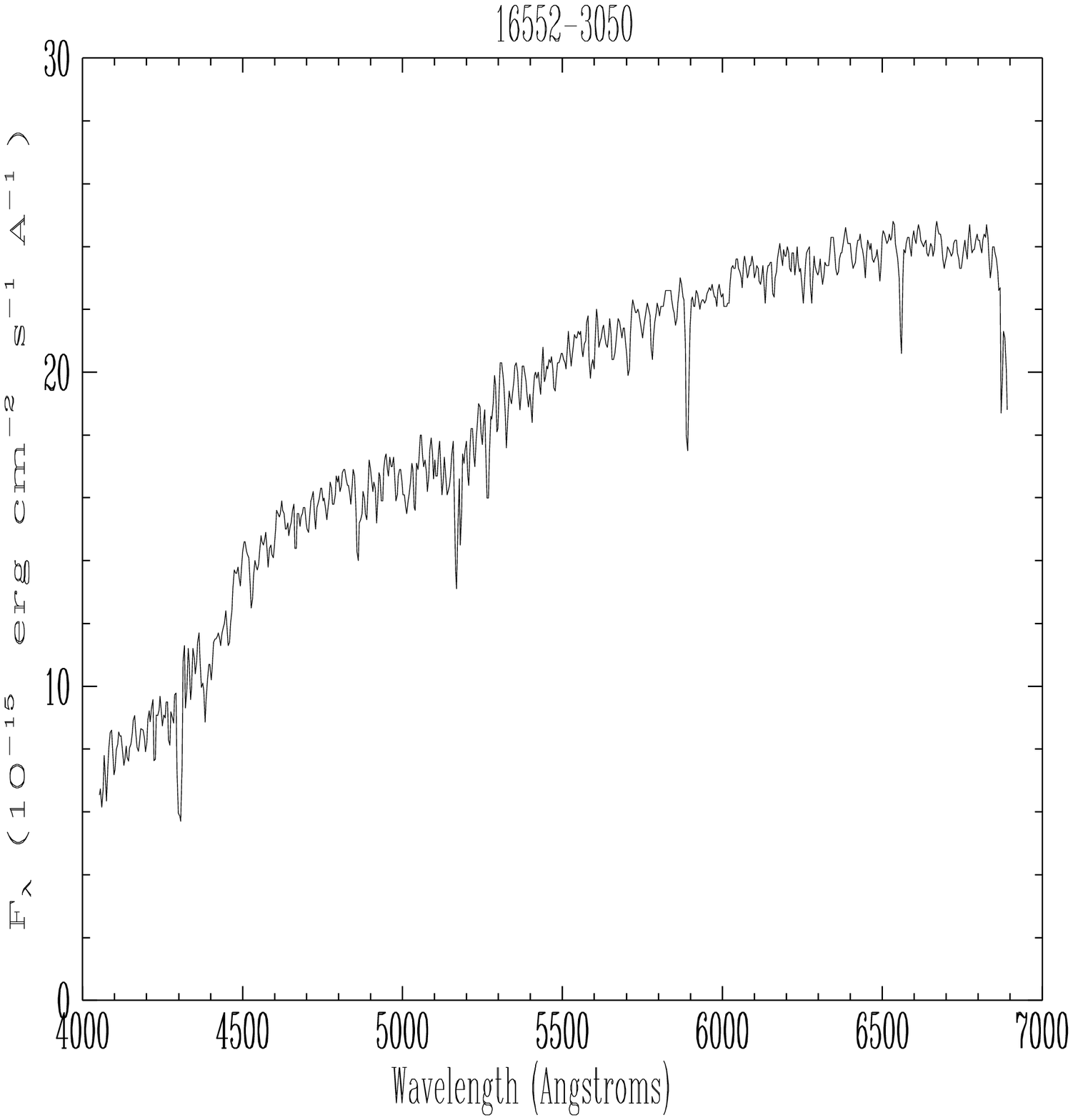}
%\psdraft
\epsfxsize=4cm
\epsfysize=4cm
\epsfbox{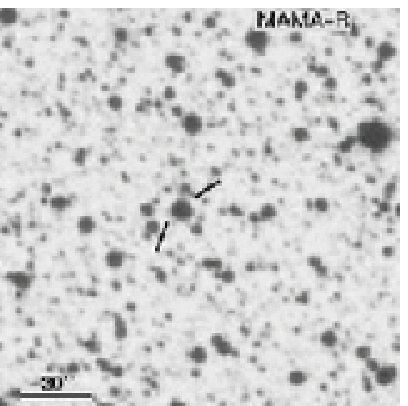}
%\psfull
\end{center}

\begin{center}
\epsfxsize=13.5cm
\epsfysize=4cm
\epsfbox{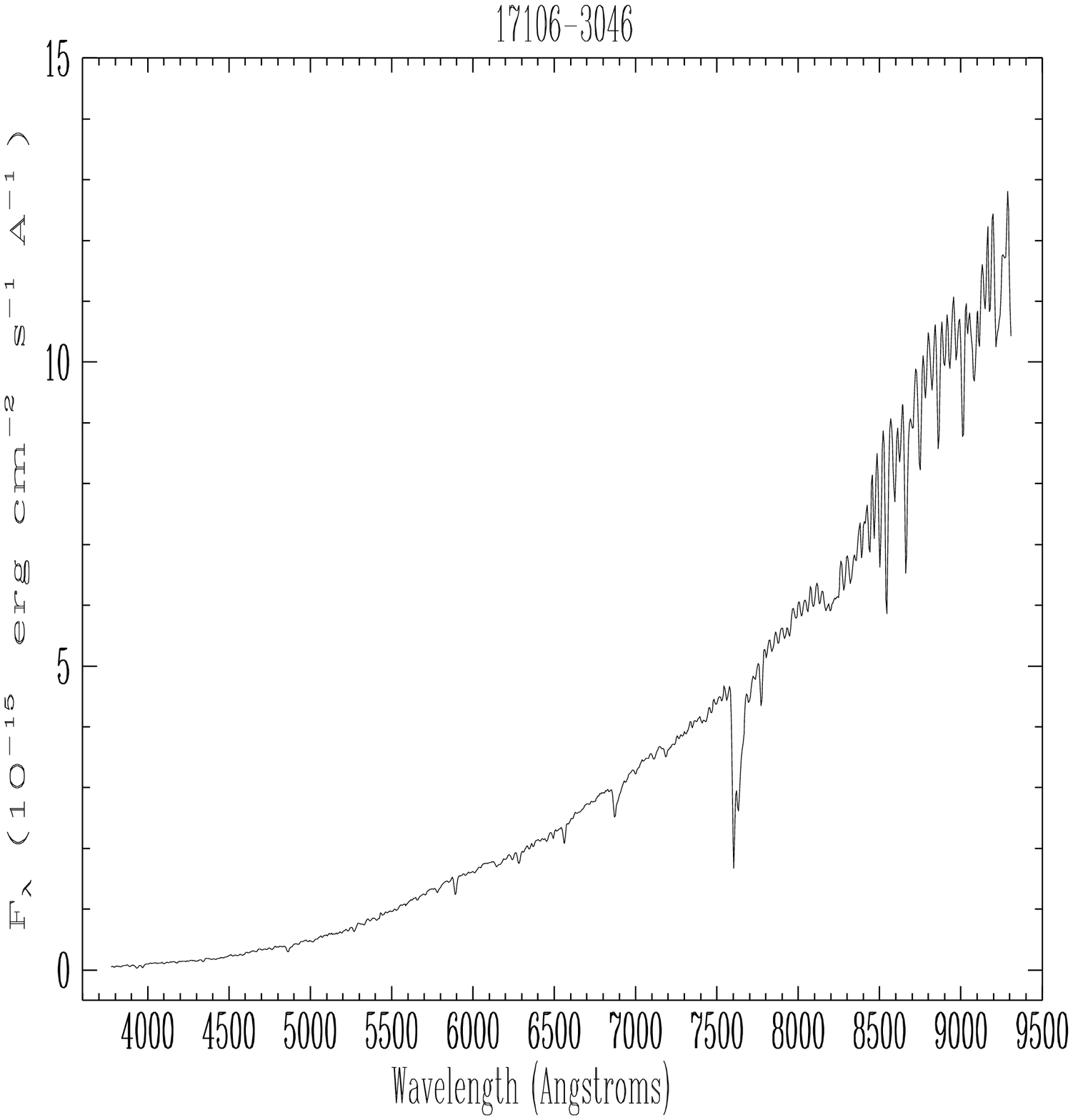}
%\psdraft
\epsfxsize=4cm
\epsfysize=4cm
\epsfbox{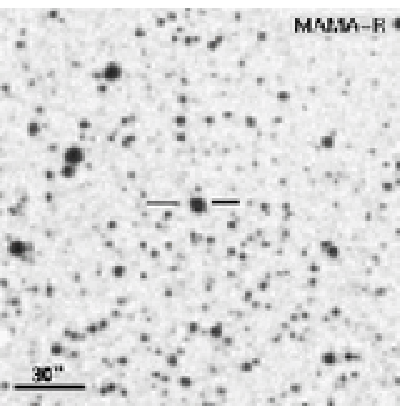}
%\psfull
\end{center}

\begin{center}
\epsfxsize=13.5cm
\epsfysize=4cm
\epsfbox{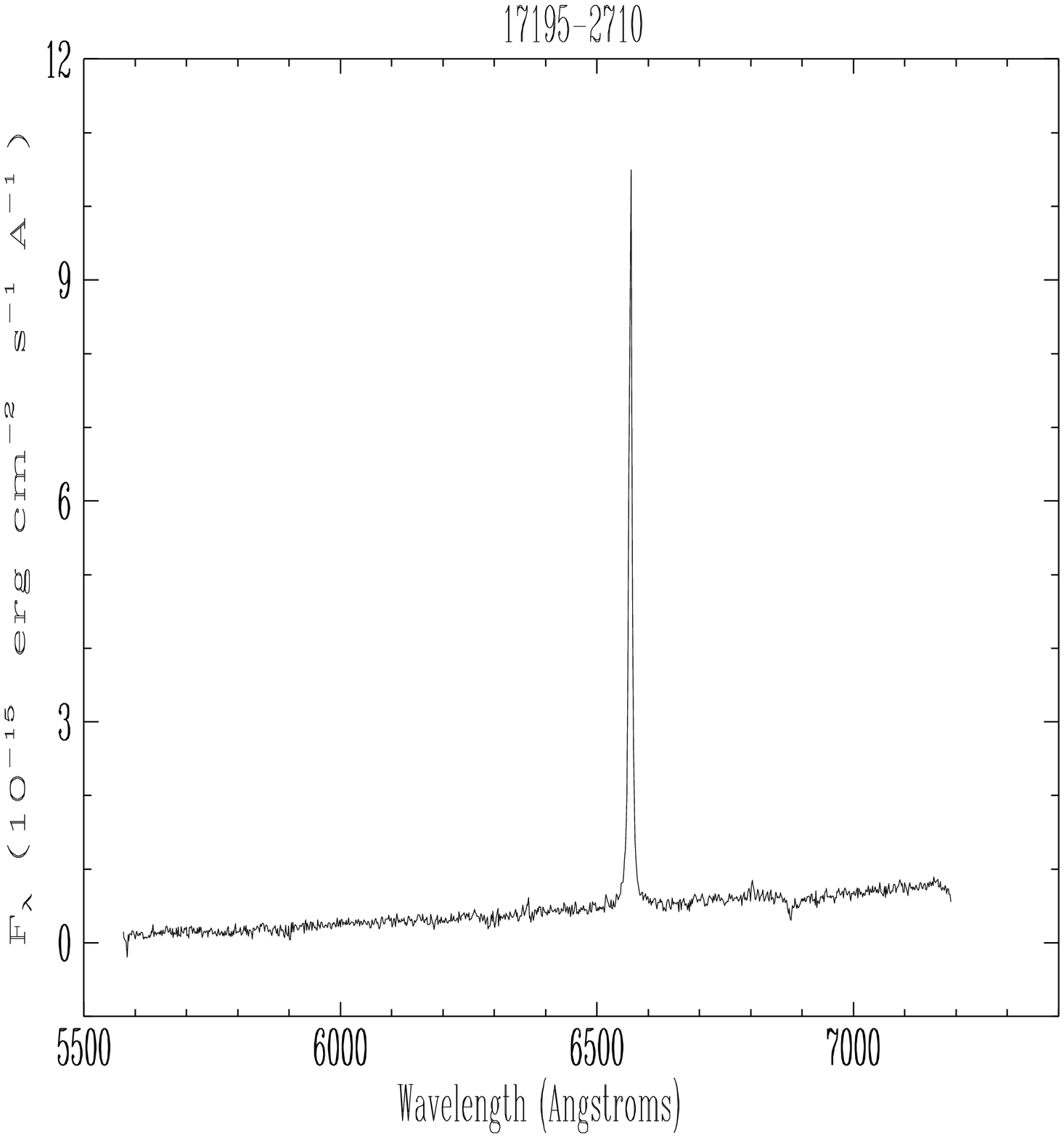}
%\psdraft
\epsfxsize=4cm
\epsfysize=4cm
\epsfbox{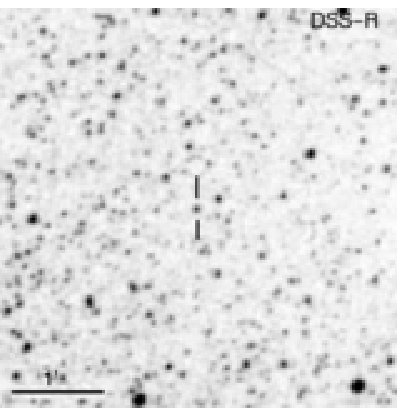}
%\psfull
\end{center}

\caption{Spectra of the objects classified as post-AGB in the sample together with their 
corresponding identification charts (continued). }
\end{figure*}

%%-------------------------------------------------------------
%%pg13
\clearpage
\setcounter{figure}{0}
\begin{figure*}

\begin{center}
\epsfxsize=13.5cm
\epsfysize=4cm
\epsfbox{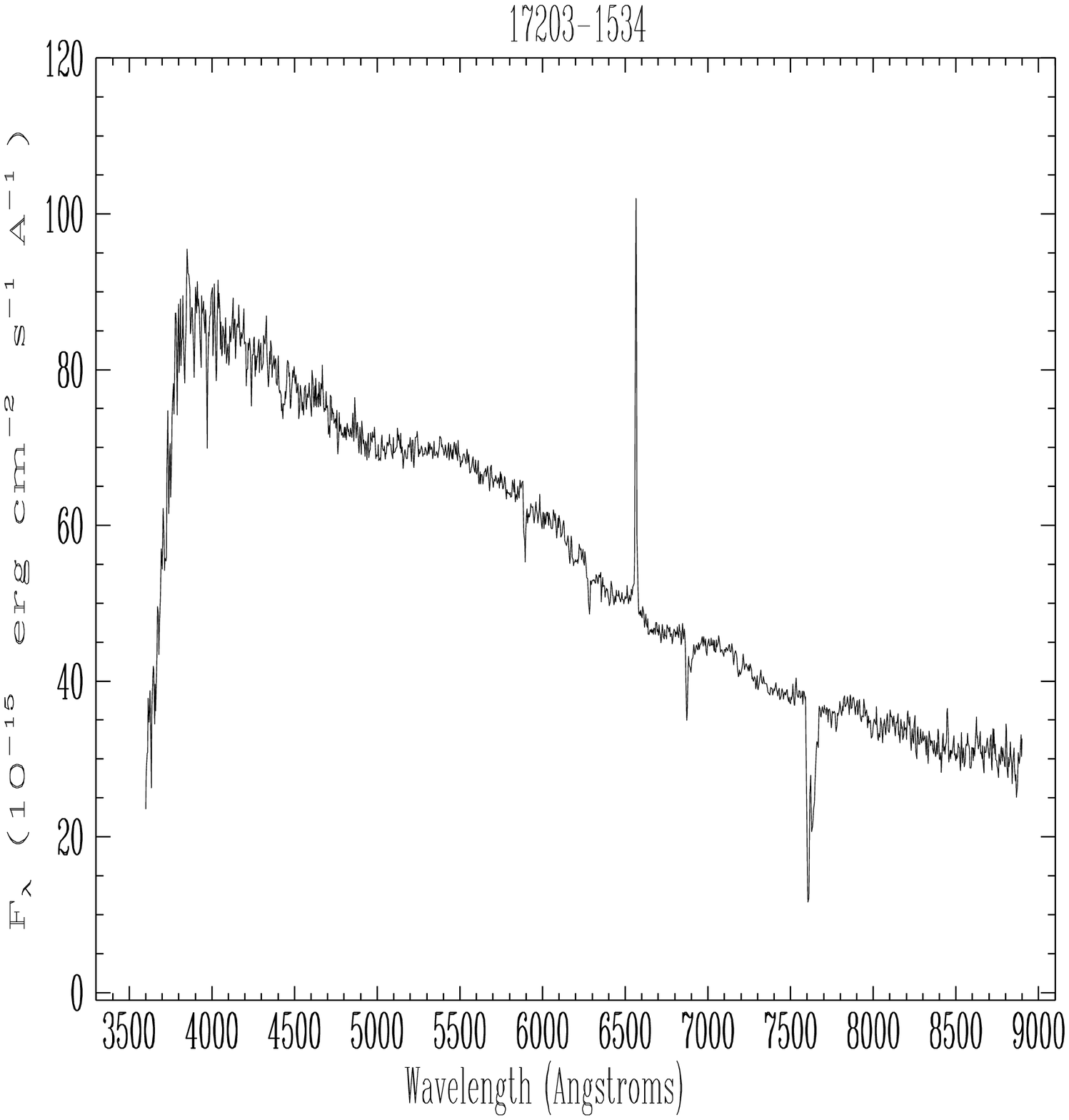}
%\psdraft
\epsfxsize=4cm
\epsfysize=4cm
\epsfbox{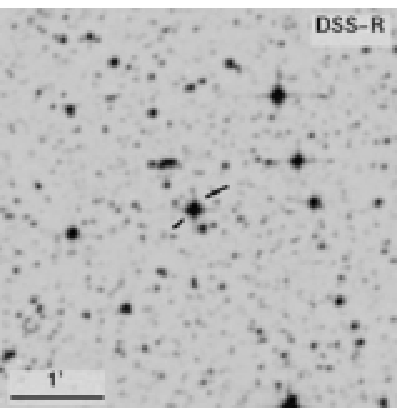}
%\psfull
\end{center}

\begin{center}
\epsfxsize=13.5cm
\epsfysize=4cm
\epsfbox{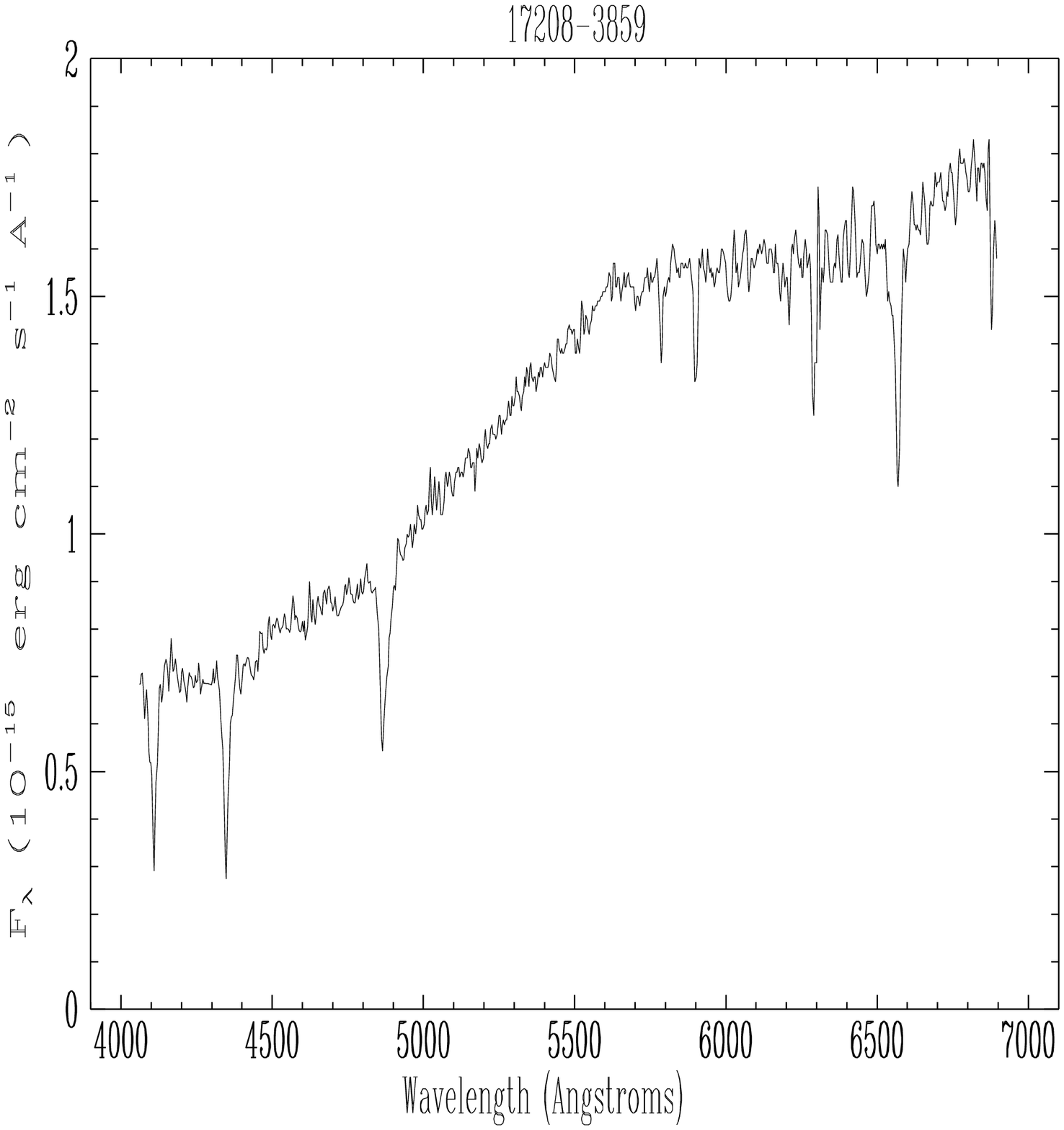}
%\psdraft
\epsfxsize=4cm
\epsfysize=4cm
\epsfbox{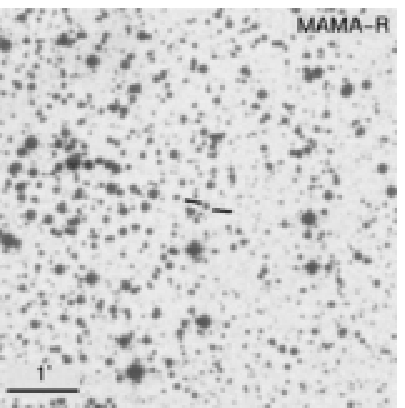}
%\psfull
\end{center}

\begin{center}
\epsfxsize=13.5cm
\epsfysize=4cm
\epsfbox{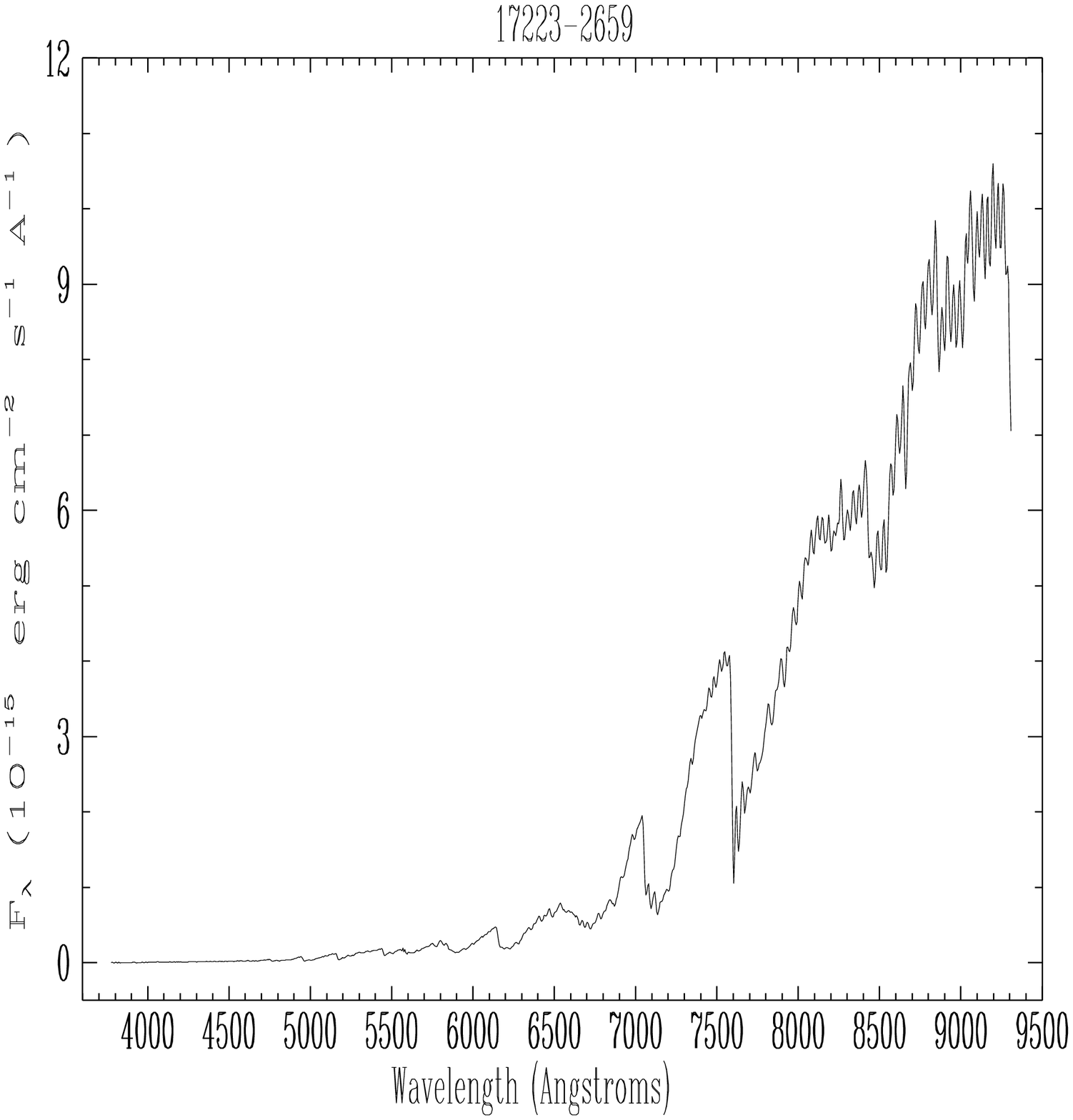}
%\psdraft
\epsfxsize=4cm
\epsfysize=4cm
\epsfbox{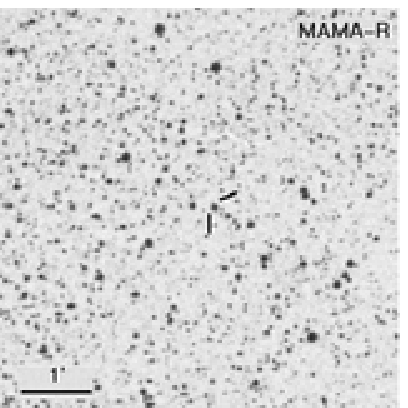}
%\psfull
\end{center}

\begin{center}
\epsfxsize=13.5cm
\epsfysize=4cm
\epsfbox{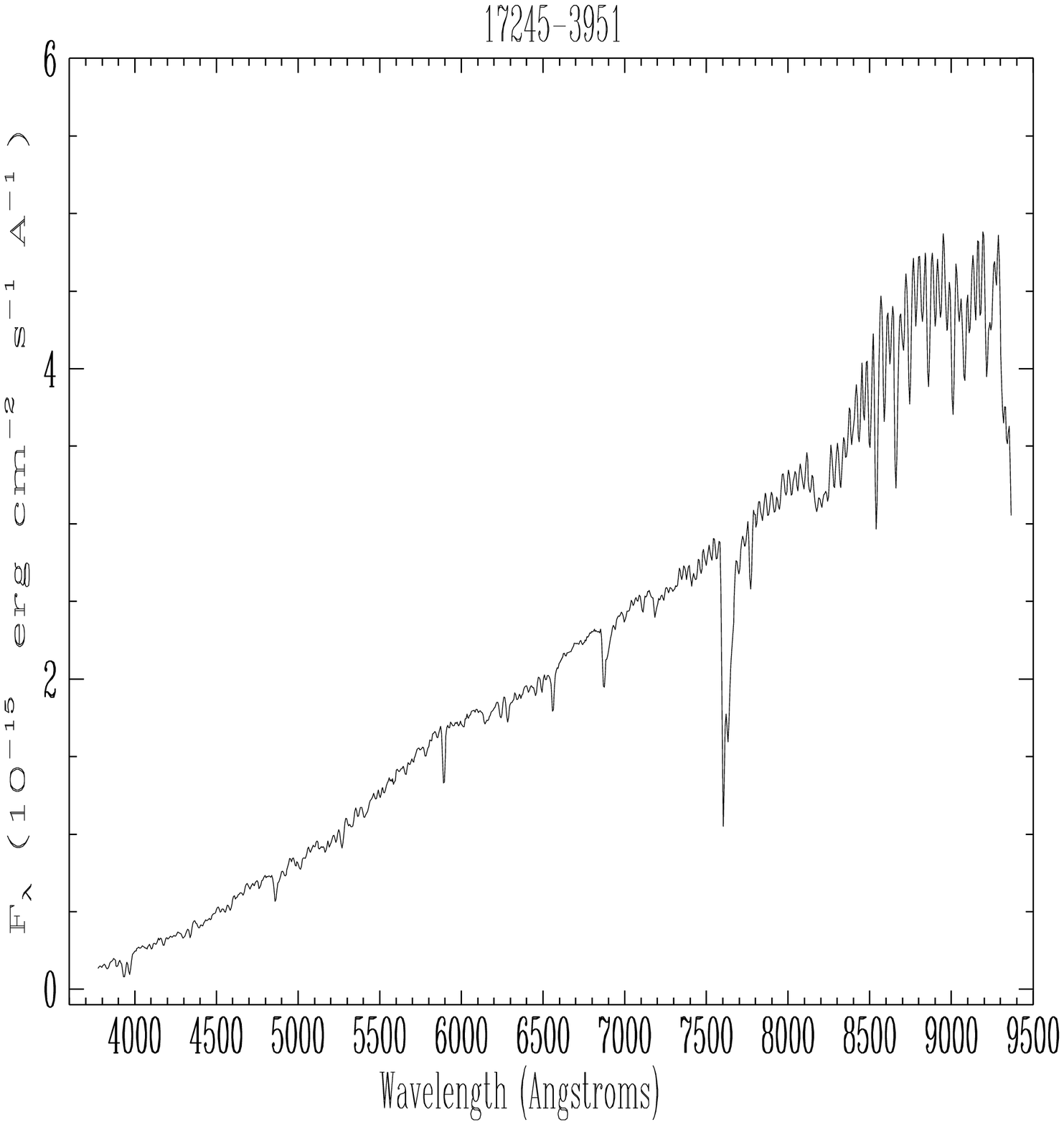}
%\psdraft
\epsfxsize=4cm
\epsfysize=4cm
\epsfbox{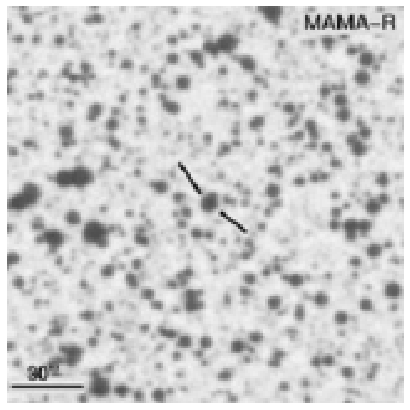}
%\psfull
\end{center}

\begin{center}
\epsfxsize=13.5cm
\epsfysize=4cm
\epsfbox{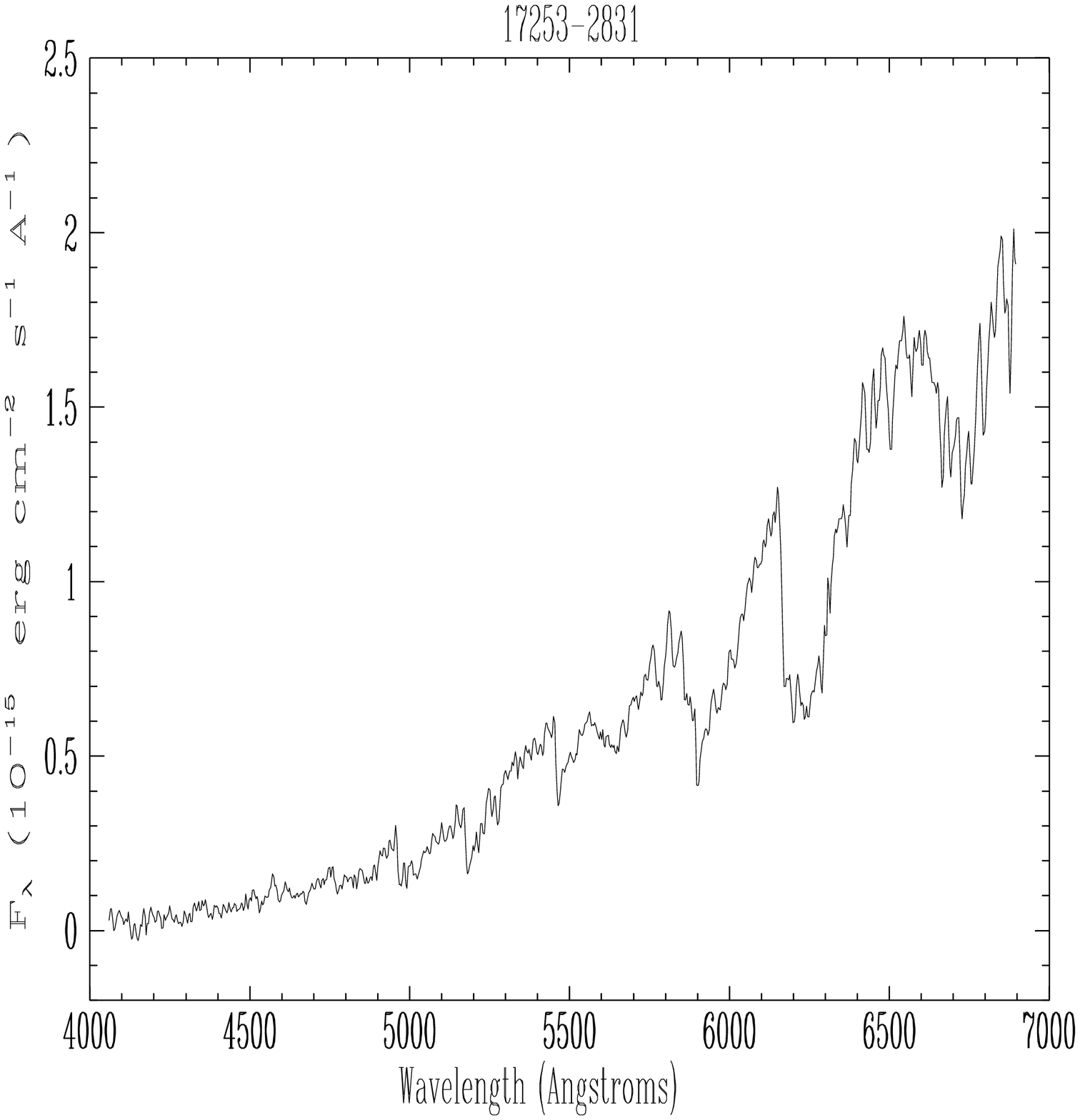}
%\psdraft
\epsfxsize=4cm
\epsfysize=4cm
\epsfbox{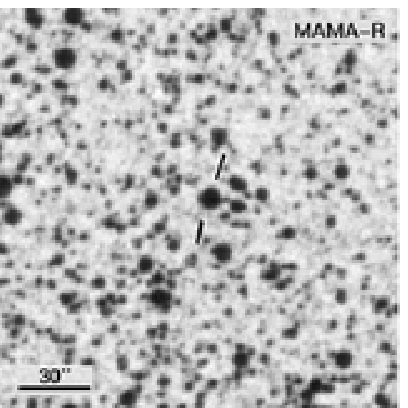}
%\psfull
\end{center}

\caption{Spectra of the objects classified as post-AGB in the sample together with their 
corresponding identification charts (continued). }
\end{figure*}

%%-------------------------------------------------------------
%%pg15
\clearpage
\setcounter{figure}{0}
\begin{figure*}

\begin{center}
\epsfxsize=13.5cm
\epsfysize=4cm
\epsfbox{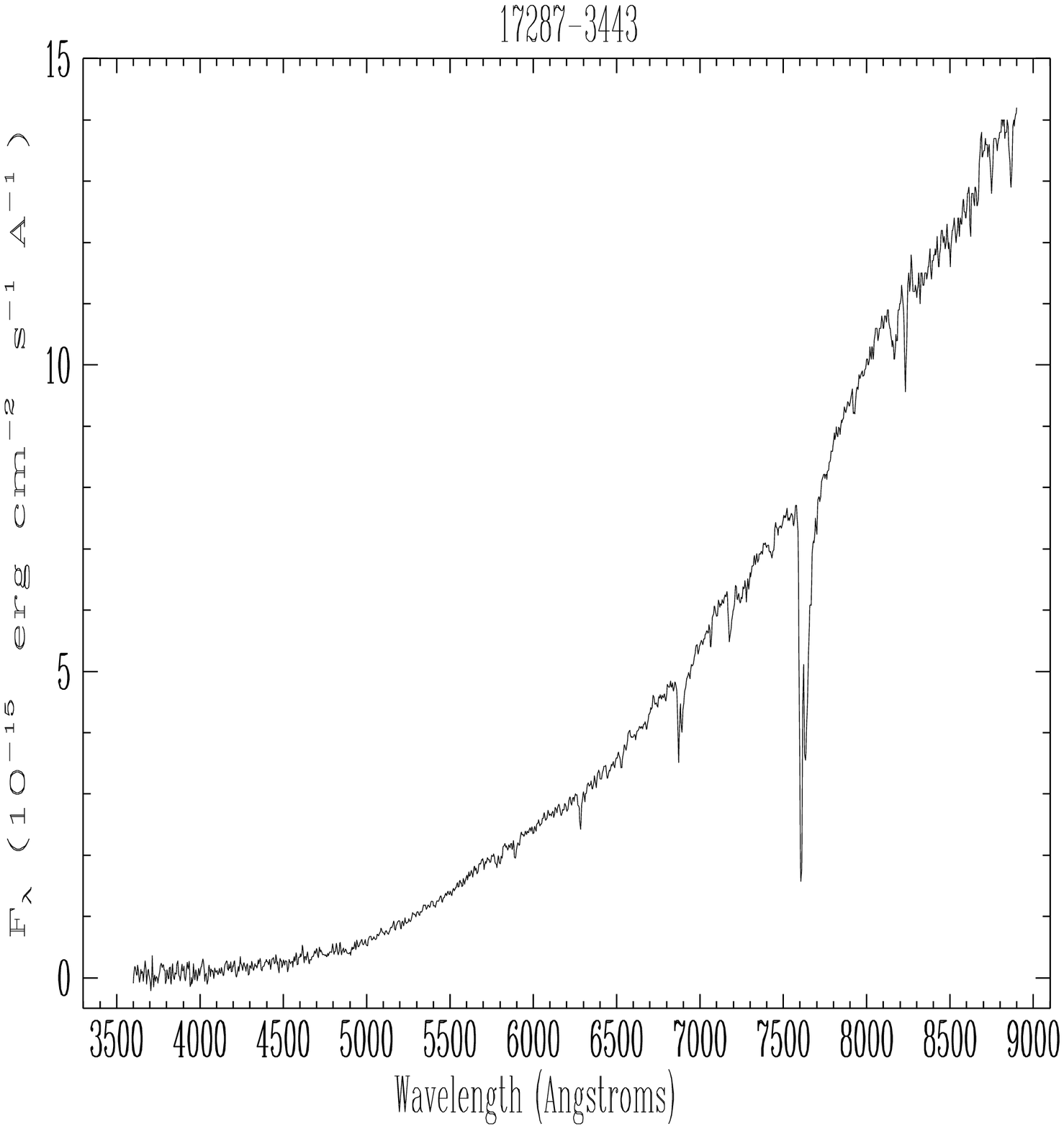}
%\psdraft
\epsfxsize=4cm
\epsfysize=4cm
\epsfbox{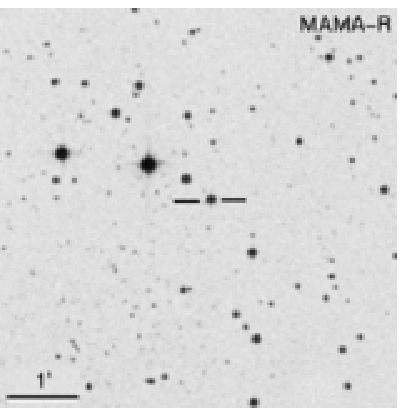}
%\psfull
\end{center}

\begin{center}
\epsfxsize=13.5cm
\epsfysize=4cm
\epsfbox{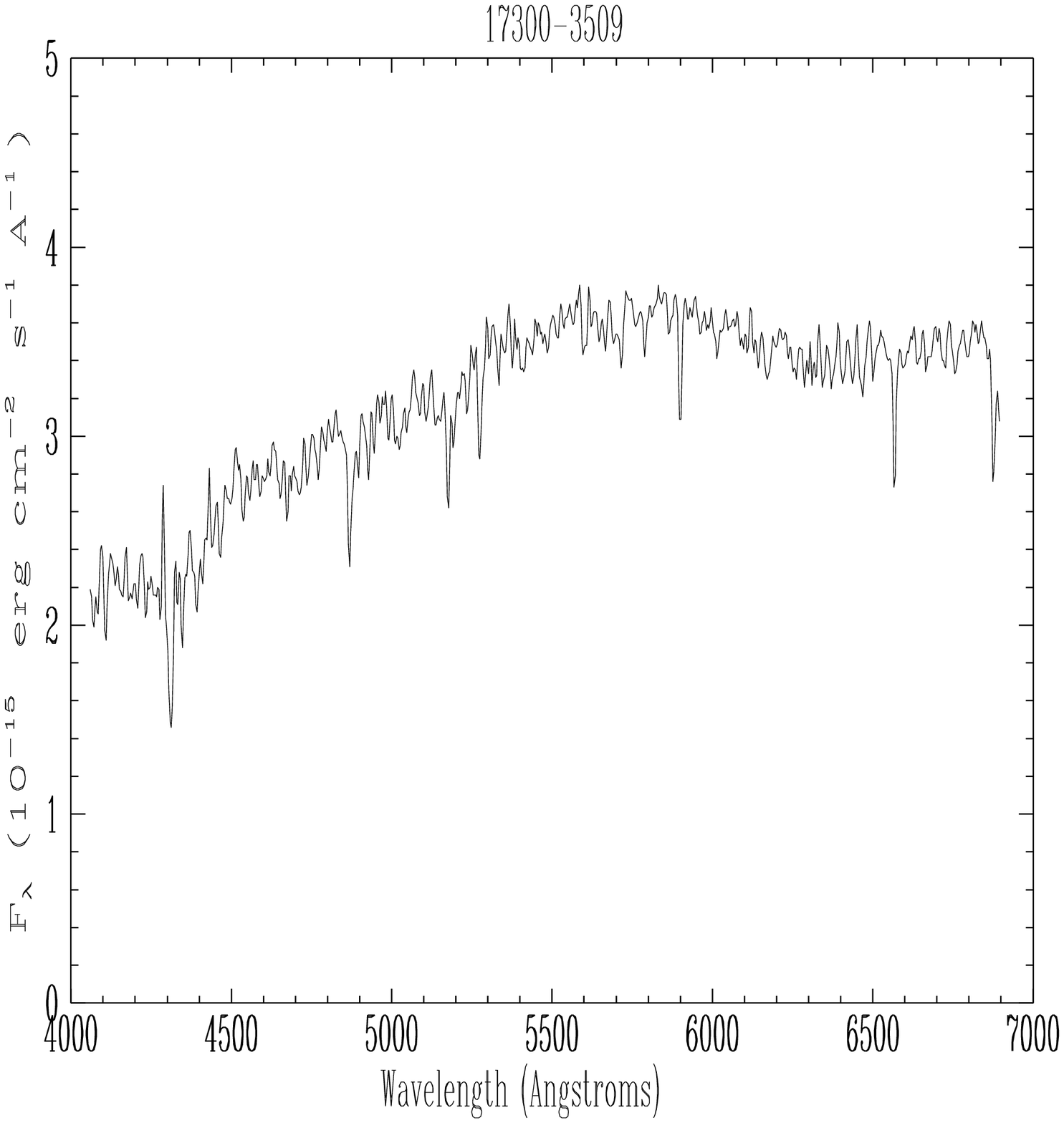}
%\psdraft
\epsfxsize=4cm
\epsfysize=4cm
\epsfbox{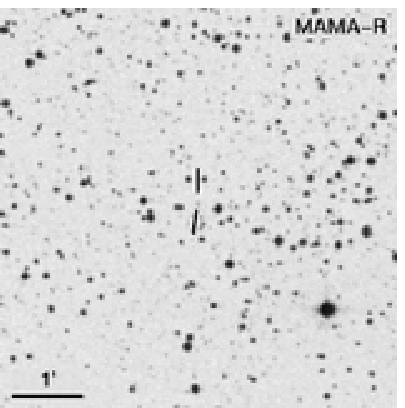}
%\psfull
\end{center}

\begin{center}
\epsfxsize=13.5cm
\epsfysize=4cm
\epsfbox{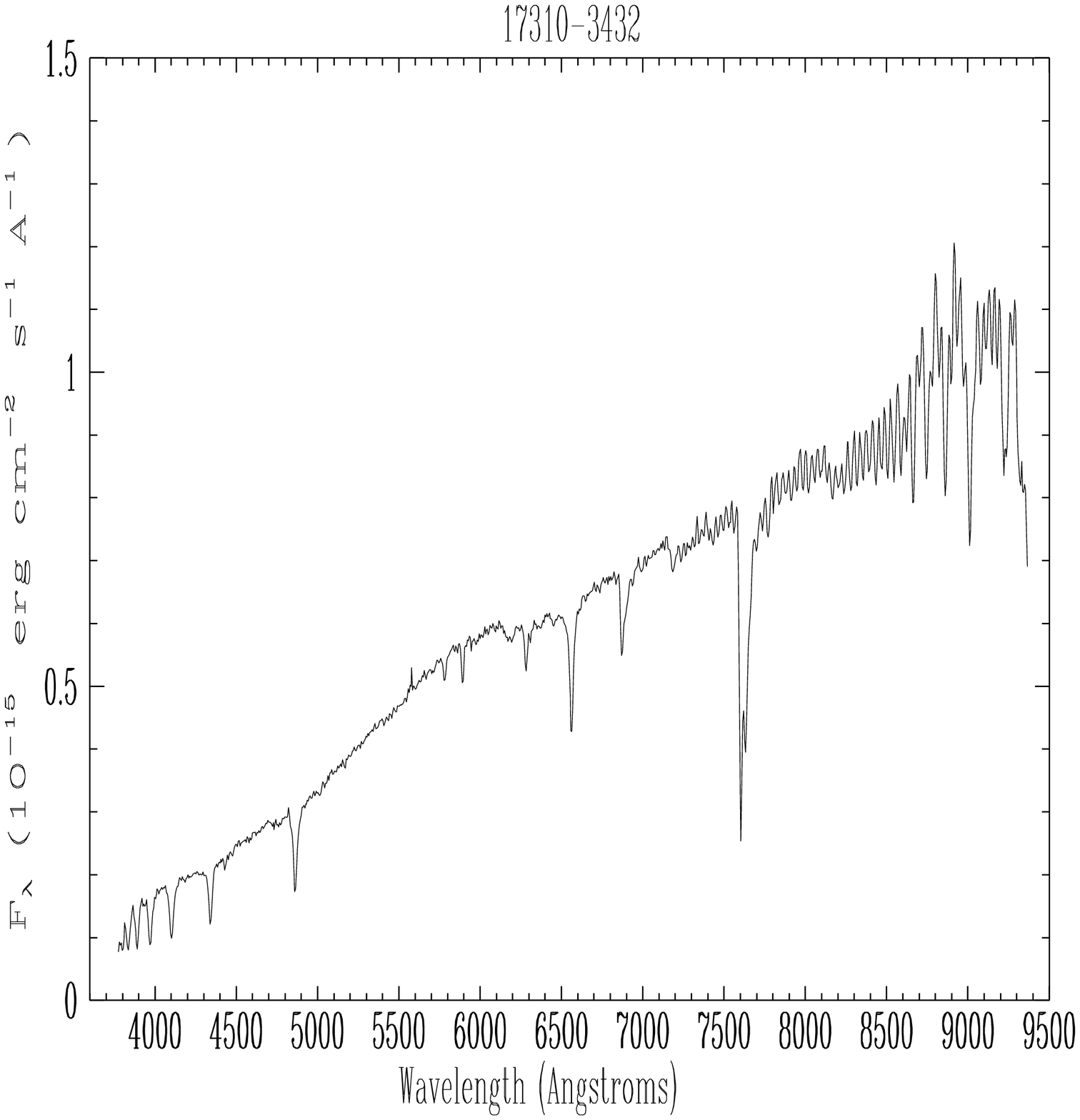}
%\psdraft
\epsfxsize=4cm
\epsfysize=4cm
\epsfbox{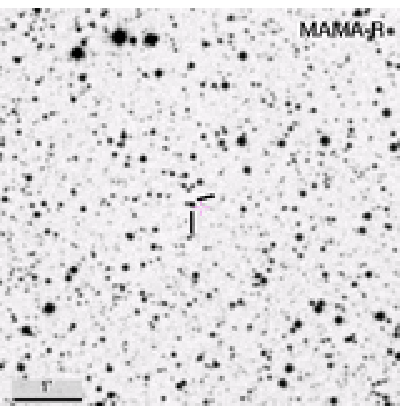}
%\psfull
\end{center}

\begin{center}
\epsfxsize=13.5cm
\epsfysize=4cm
\epsfbox{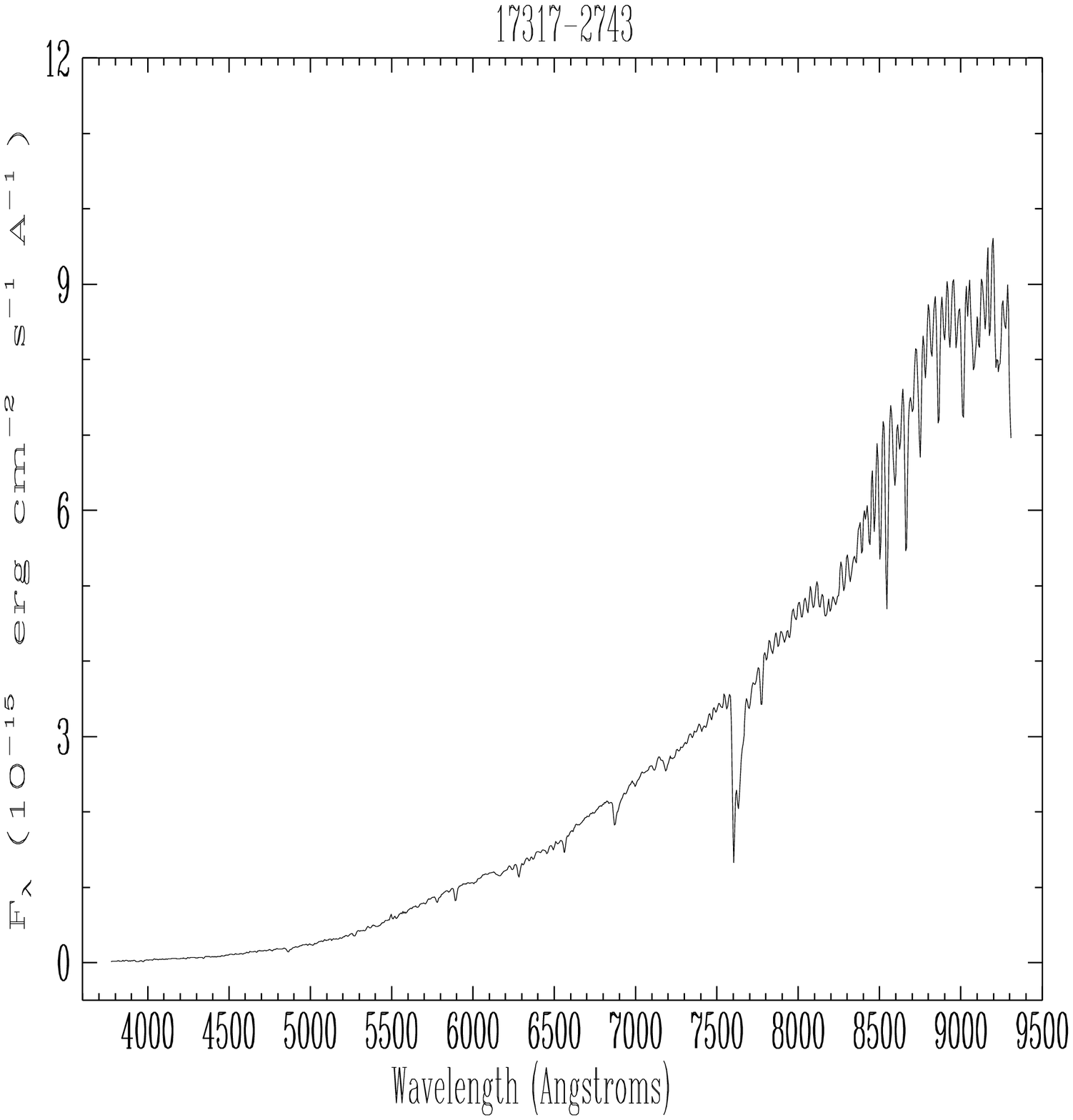}
%\psdraft
\epsfxsize=4cm
\epsfysize=4cm
\epsfbox{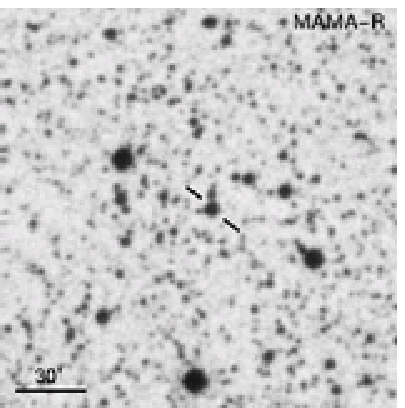}
%\psfull
\end{center}

\begin{center}
\epsfxsize=13.5cm
\epsfysize=4cm
\epsfbox{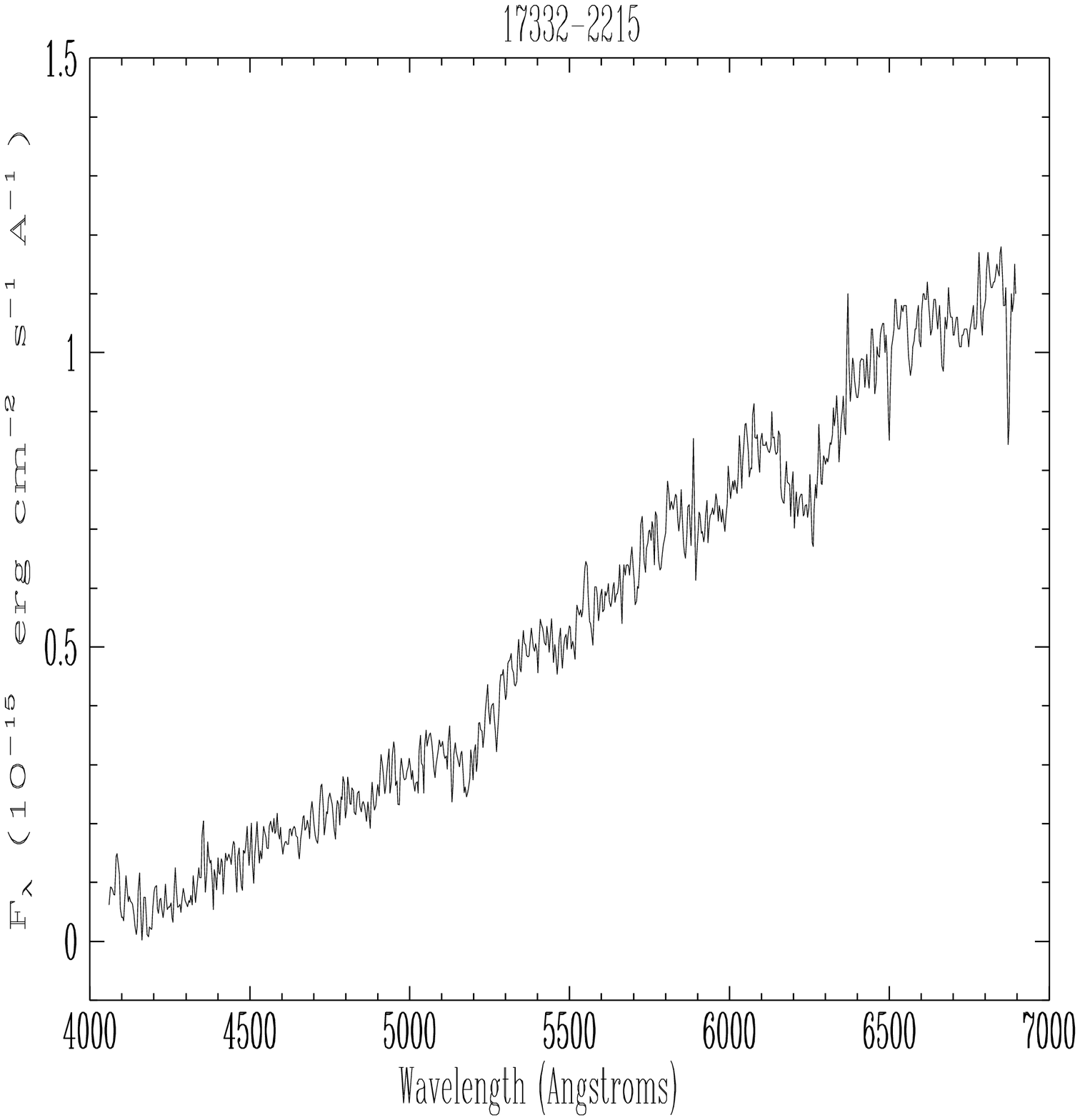}
%\psdraft
\epsfxsize=4cm
\epsfysize=4cm
\epsfbox{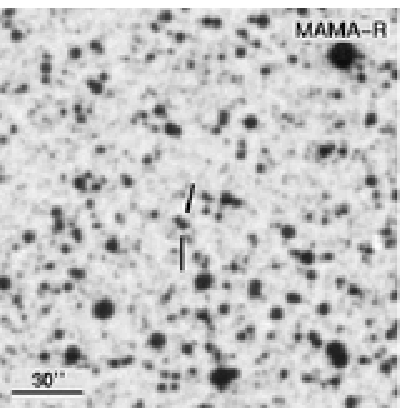}
%\psfull
\end{center}

\caption{Spectra of the objects classified as post-AGB in the sample together with their 
corresponding identification charts (continued). }
\end{figure*}

%pg16
\clearpage
\setcounter{figure}{0}
\begin{figure*}

\begin{center}
\epsfxsize=13.5cm
\epsfysize=4cm
\epsfbox{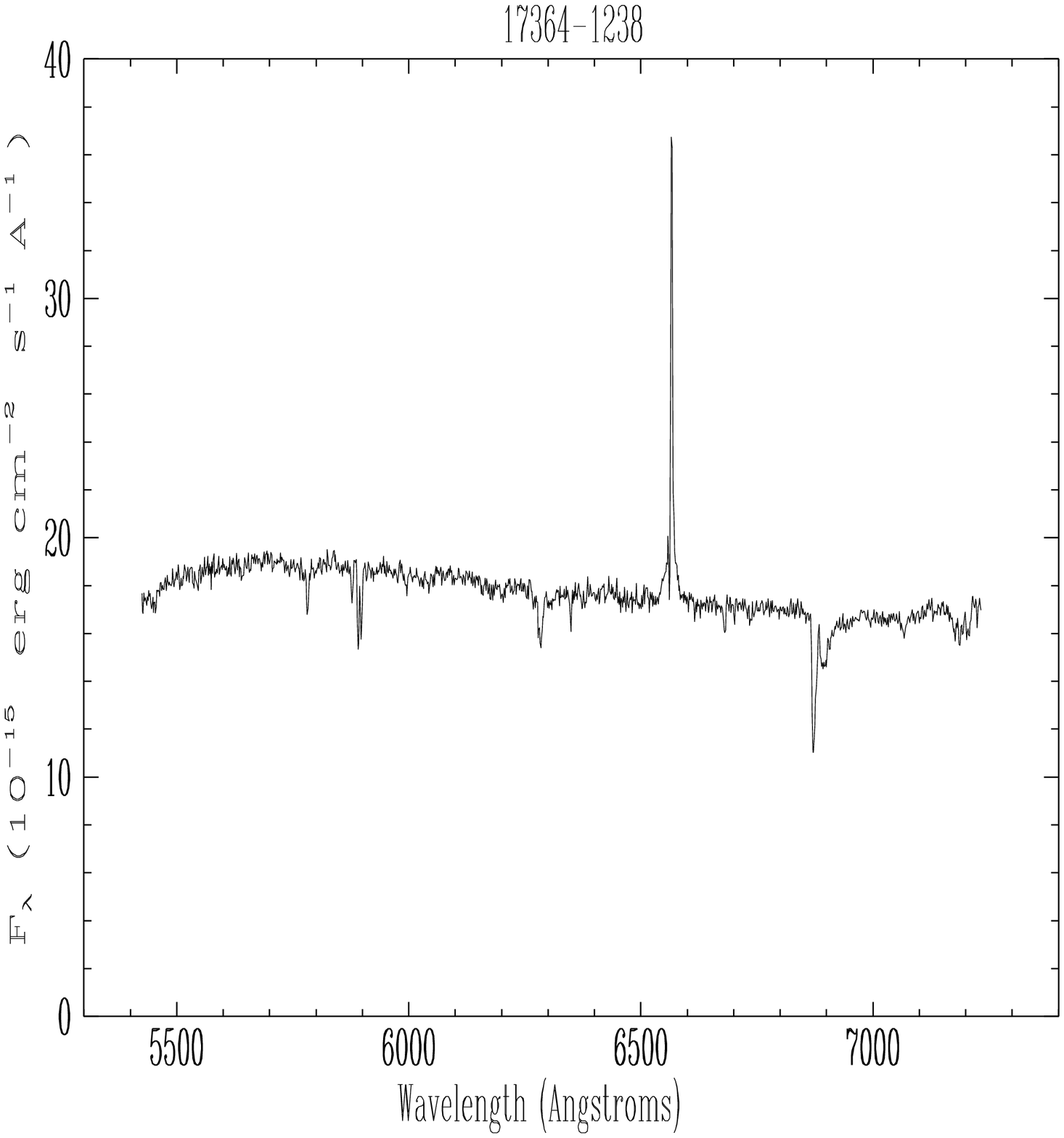}
%\psdraft
\epsfxsize=4cm
\epsfysize=4cm
\epsfbox{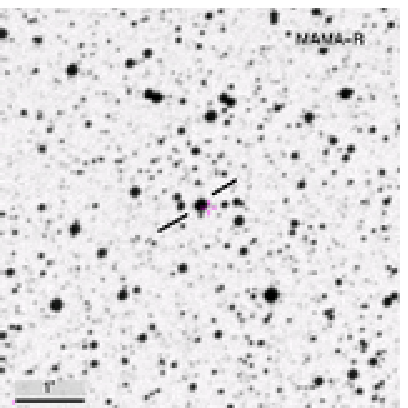}
%\psfull
\end{center}

\begin{center}
\epsfxsize=13.5cm
\epsfysize=4cm
\epsfbox{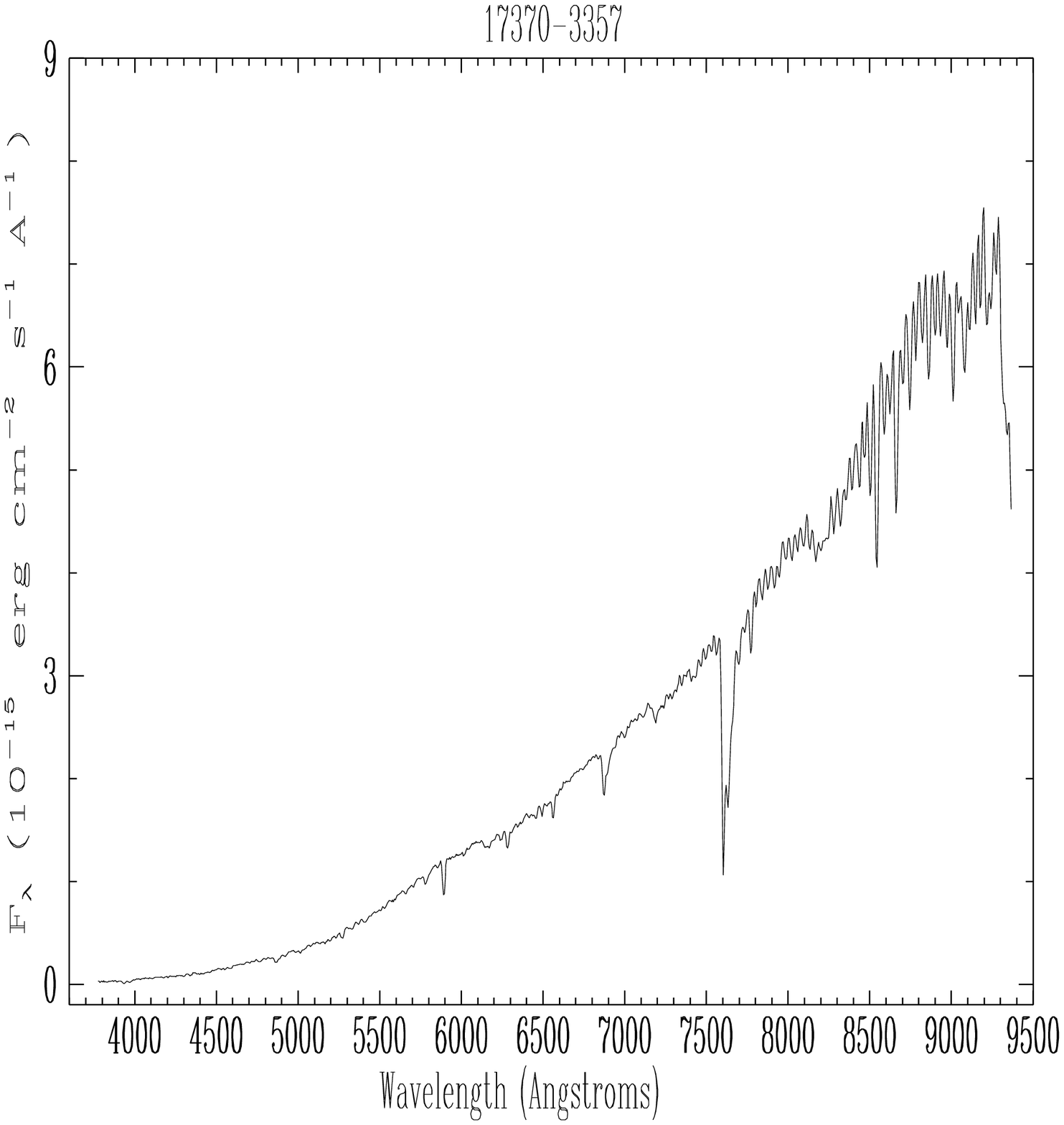}
%\psdraft
\epsfxsize=4cm
\epsfysize=4cm
\epsfbox{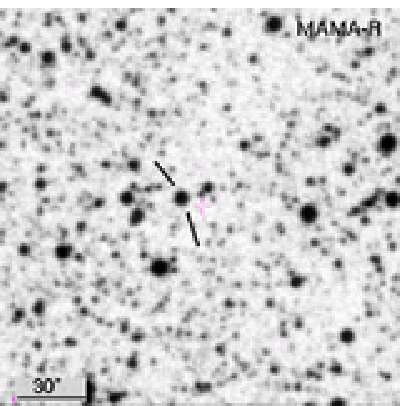}
%\psfull
\end{center}

\begin{center}
\epsfxsize=13.5cm
\epsfysize=4cm
\epsfbox{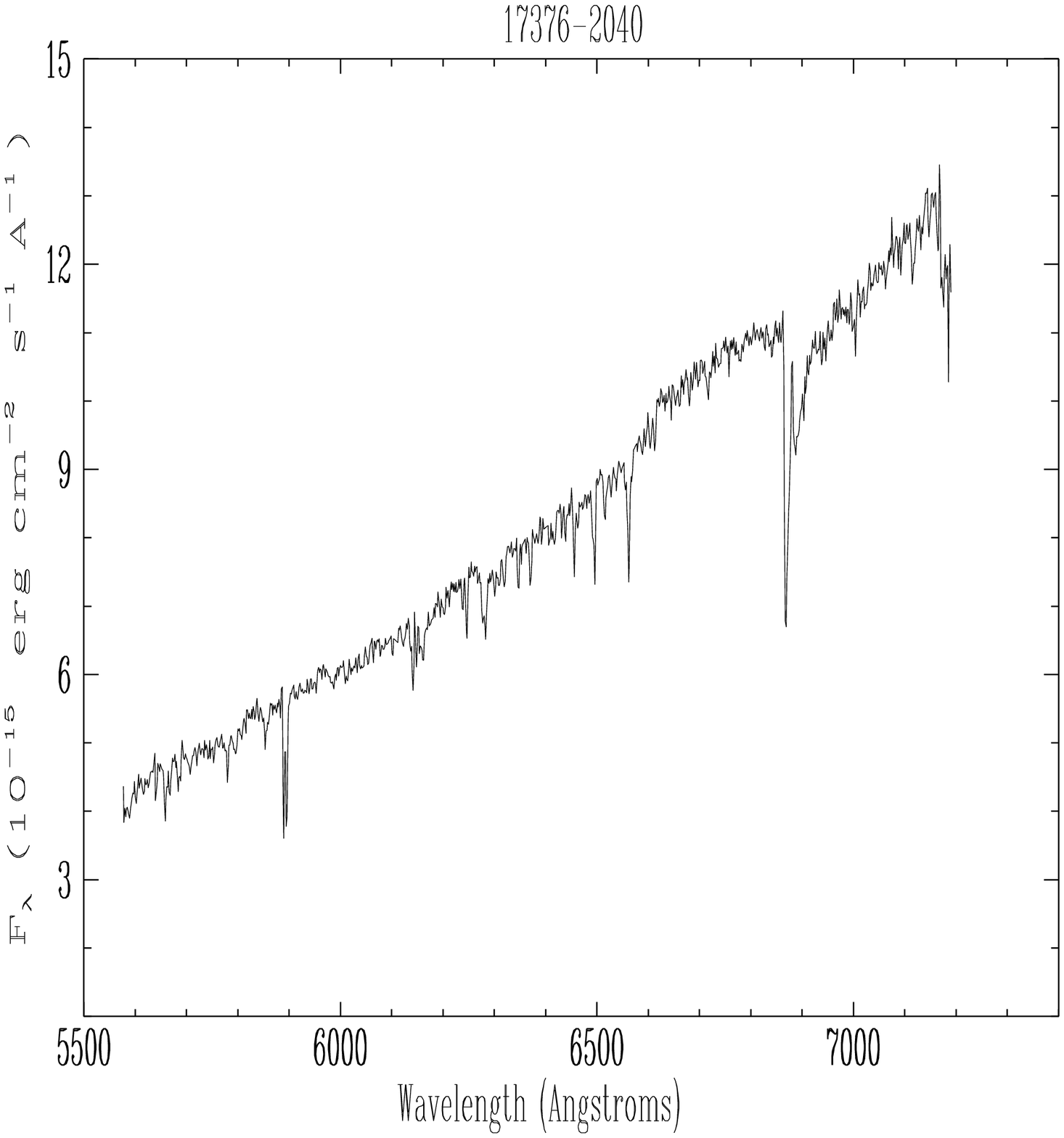}
%\psdraft
\epsfxsize=4cm
\epsfysize=4cm
\epsfbox{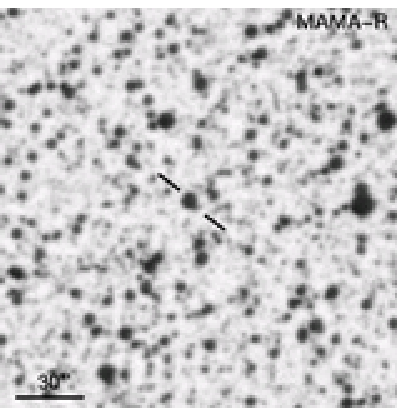}
%\psfull
\end{center}

\begin{center}
\epsfxsize=13.5cm
\epsfysize=4cm
\epsfbox{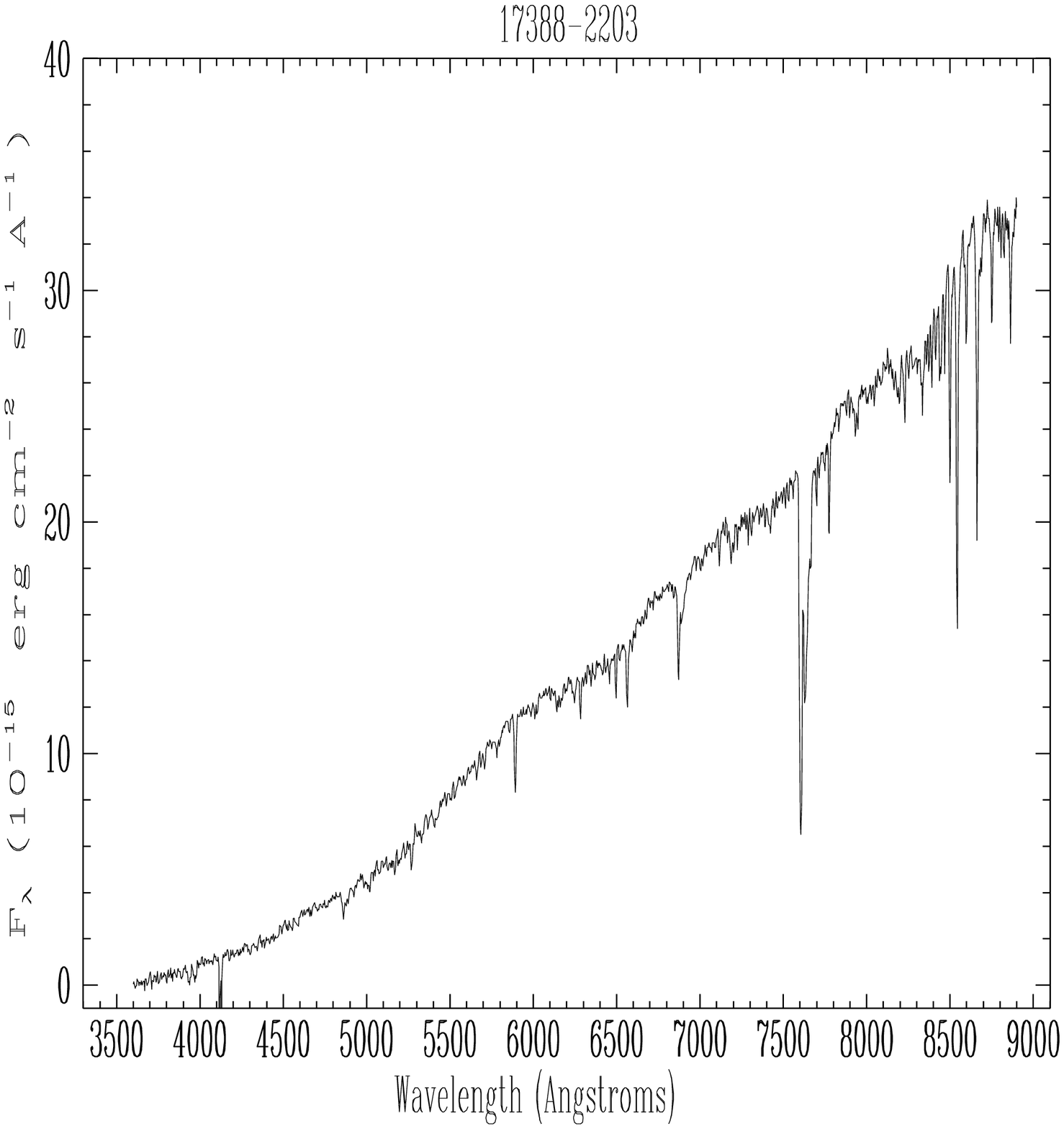}
%\psdraft
\epsfxsize=4cm
\epsfysize=4cm
\epsfbox{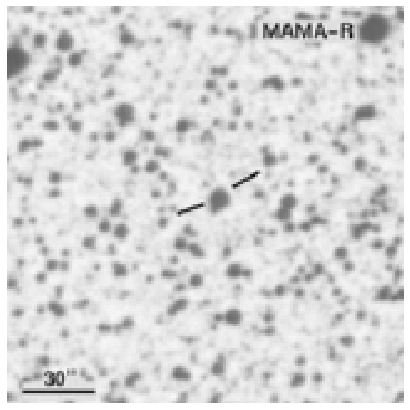}
%\psfull
\end{center}

\begin{center}
\epsfxsize=13.5cm
\epsfysize=4cm
\epsfbox{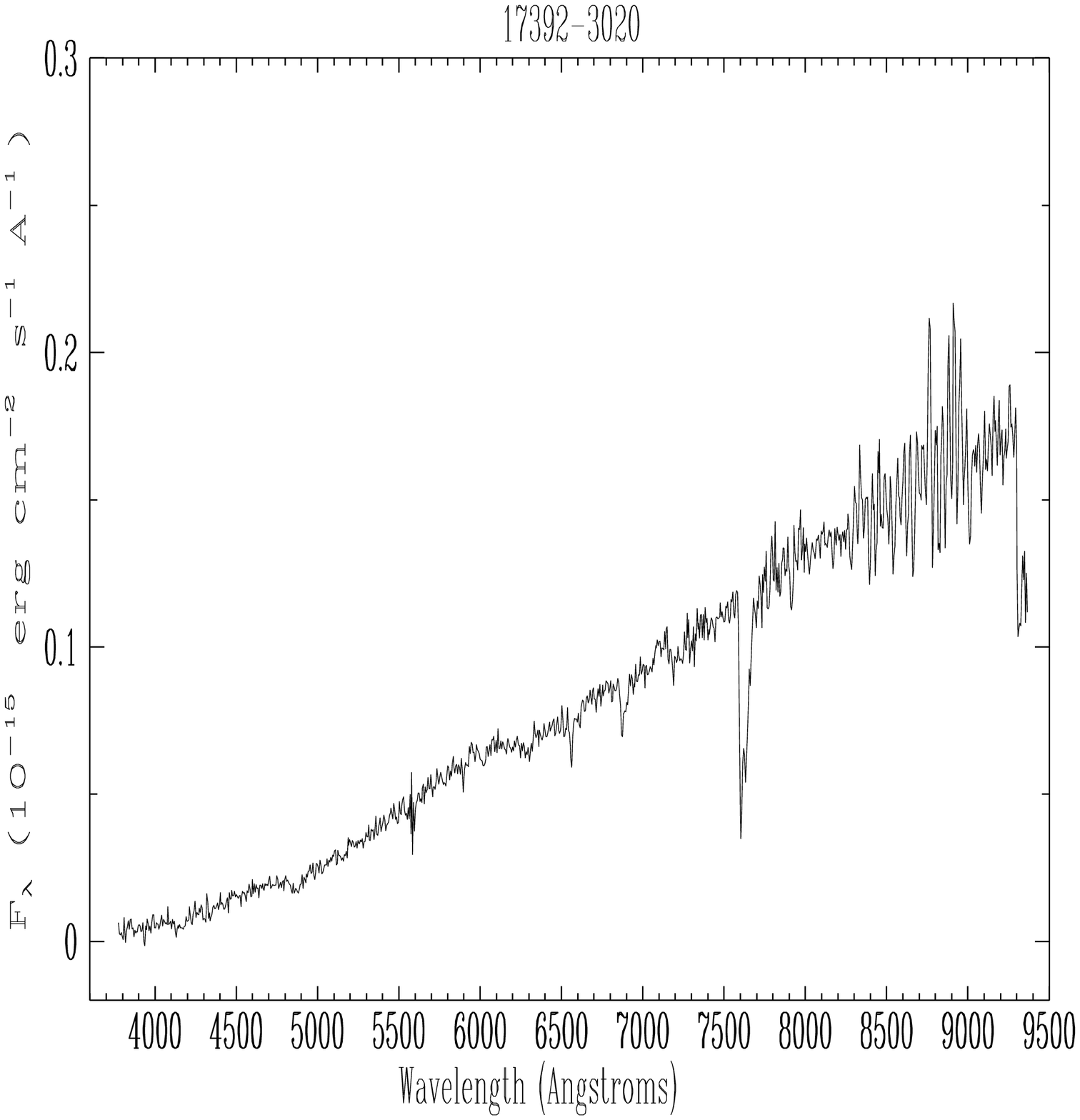}
%\psdraft
\epsfxsize=4cm
\epsfysize=4cm
\epsfbox{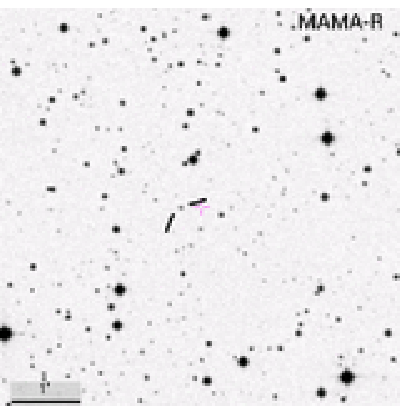}
%\psfull
\end{center}

\caption{Spectra of the objects classified as post-AGB in the sample together with their 
corresponding identification charts (continued). }
\end{figure*}

%%-------------------------------------------------------------
%%pg17
\clearpage
\setcounter{figure}{0}
\begin{figure*}

\begin{center}
\epsfxsize=13.5cm
\epsfysize=4cm
\epsfbox{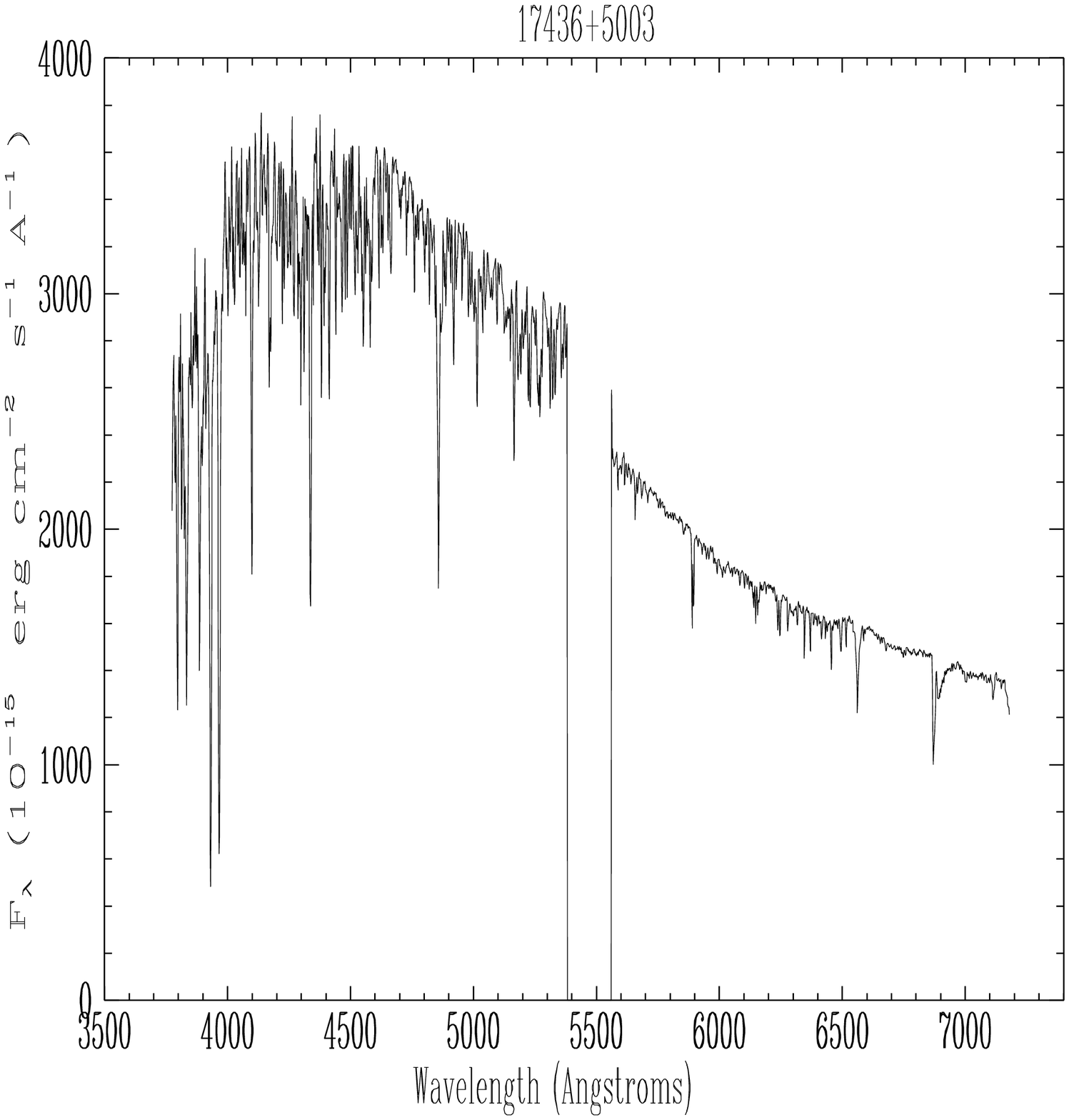}
%\psdraft
\epsfxsize=4cm
\epsfysize=4cm
\epsfbox{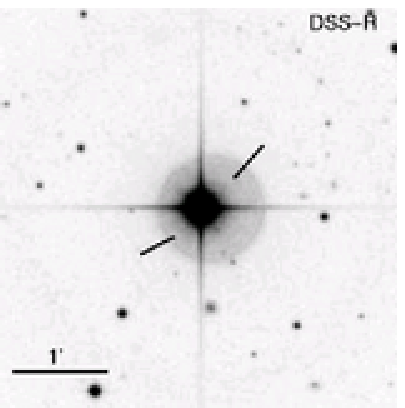}
%\psfull
\end{center}

\begin{center}
\epsfxsize=13.5cm
\epsfysize=4cm
\epsfbox{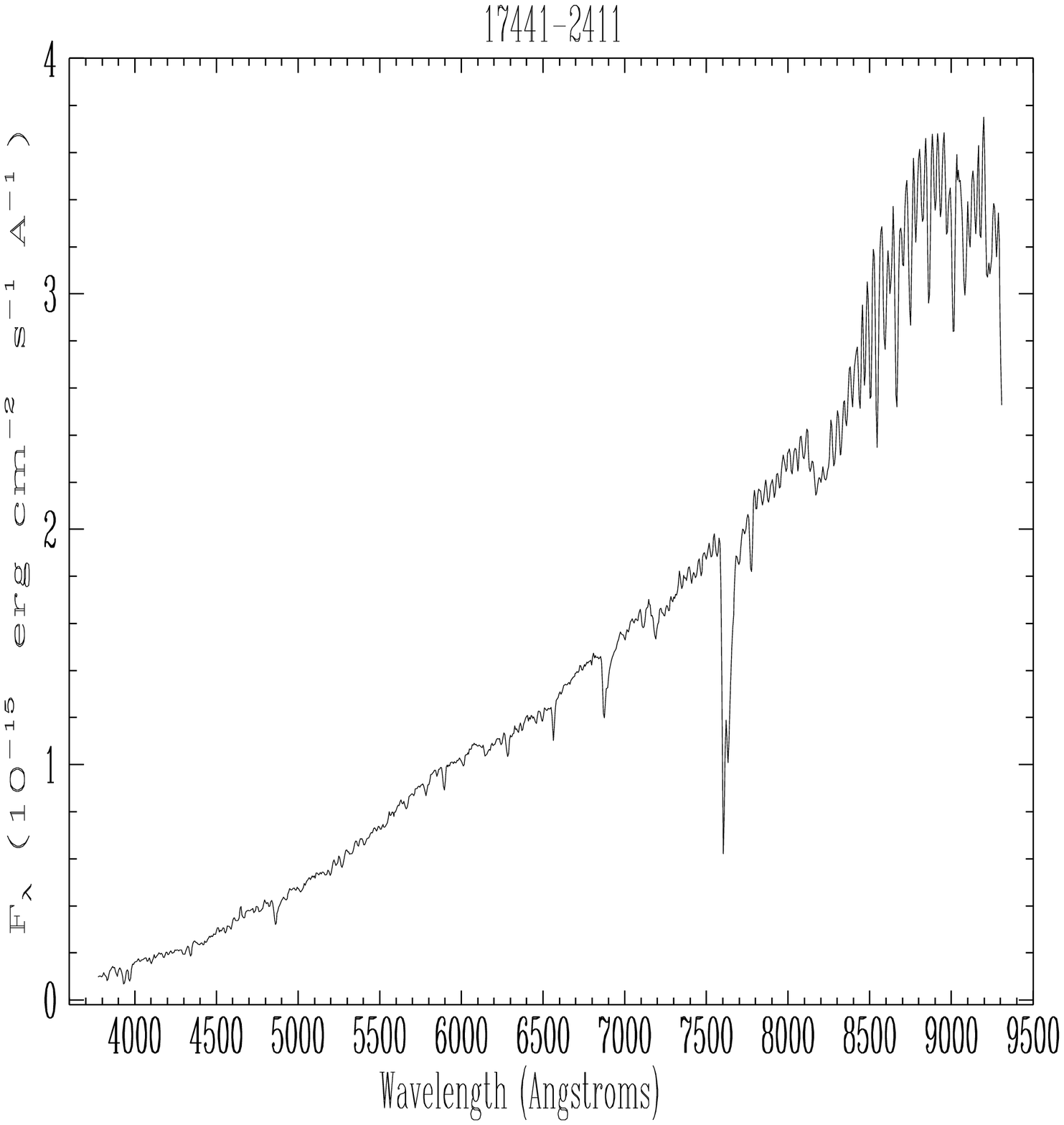}
%\psdraft
\epsfxsize=4cm
\epsfysize=4cm
\epsfbox{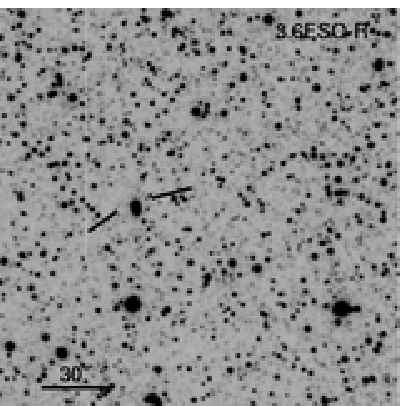}
%\psfull
\end{center}

\begin{center}
\epsfxsize=13.5cm
\epsfysize=4cm
\epsfbox{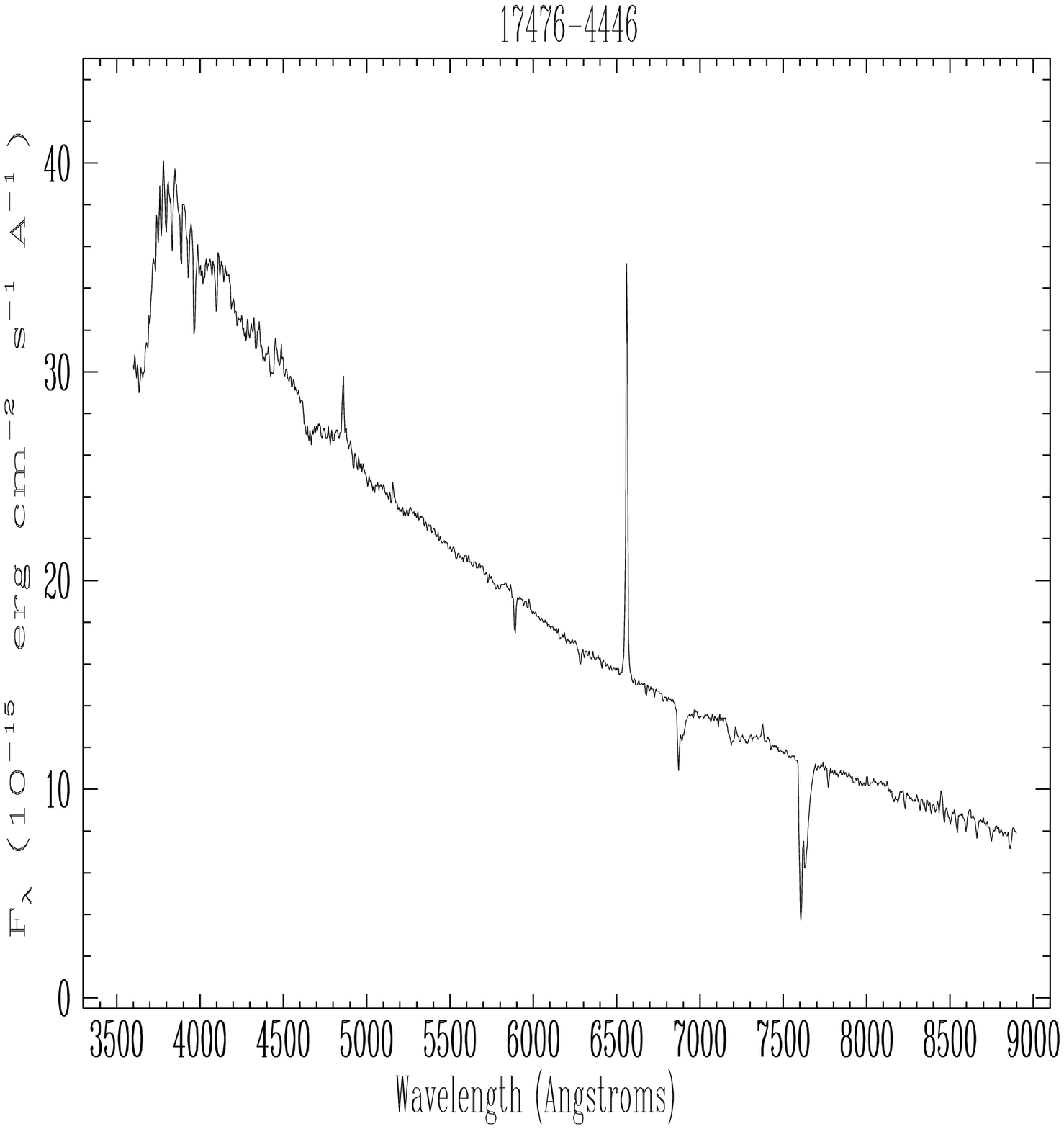}
%\psdraft
\epsfxsize=4cm
\epsfysize=4cm
\epsfbox{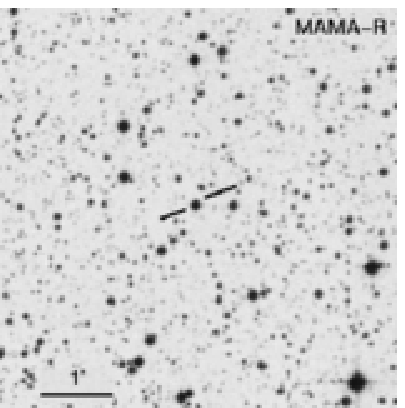}
%\psfull
\end{center}

\begin{center}
\epsfxsize=13.5cm
\epsfysize=4cm
\epsfbox{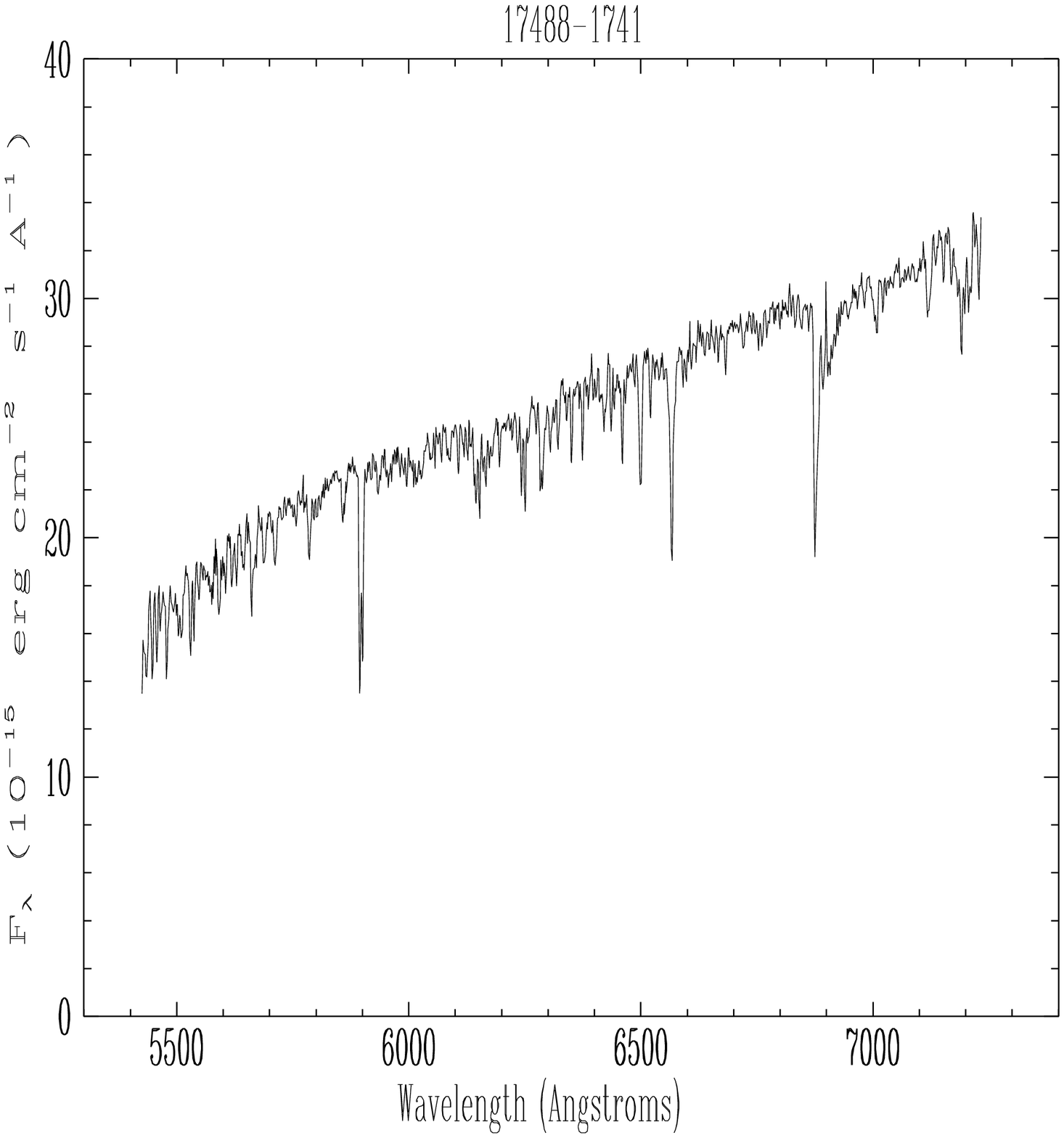}
%\psdraft
\epsfxsize=4cm
\epsfysize=4cm
\epsfbox{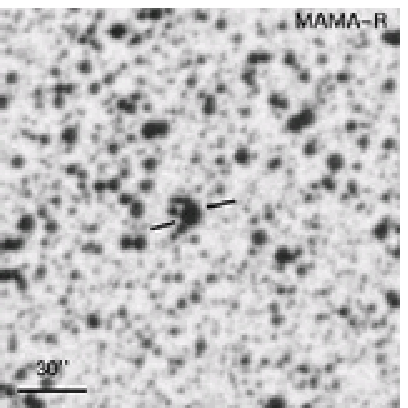}
%\psfull
\end{center}

\begin{center}
\epsfxsize=13.5cm
\epsfysize=4cm
\epsfbox{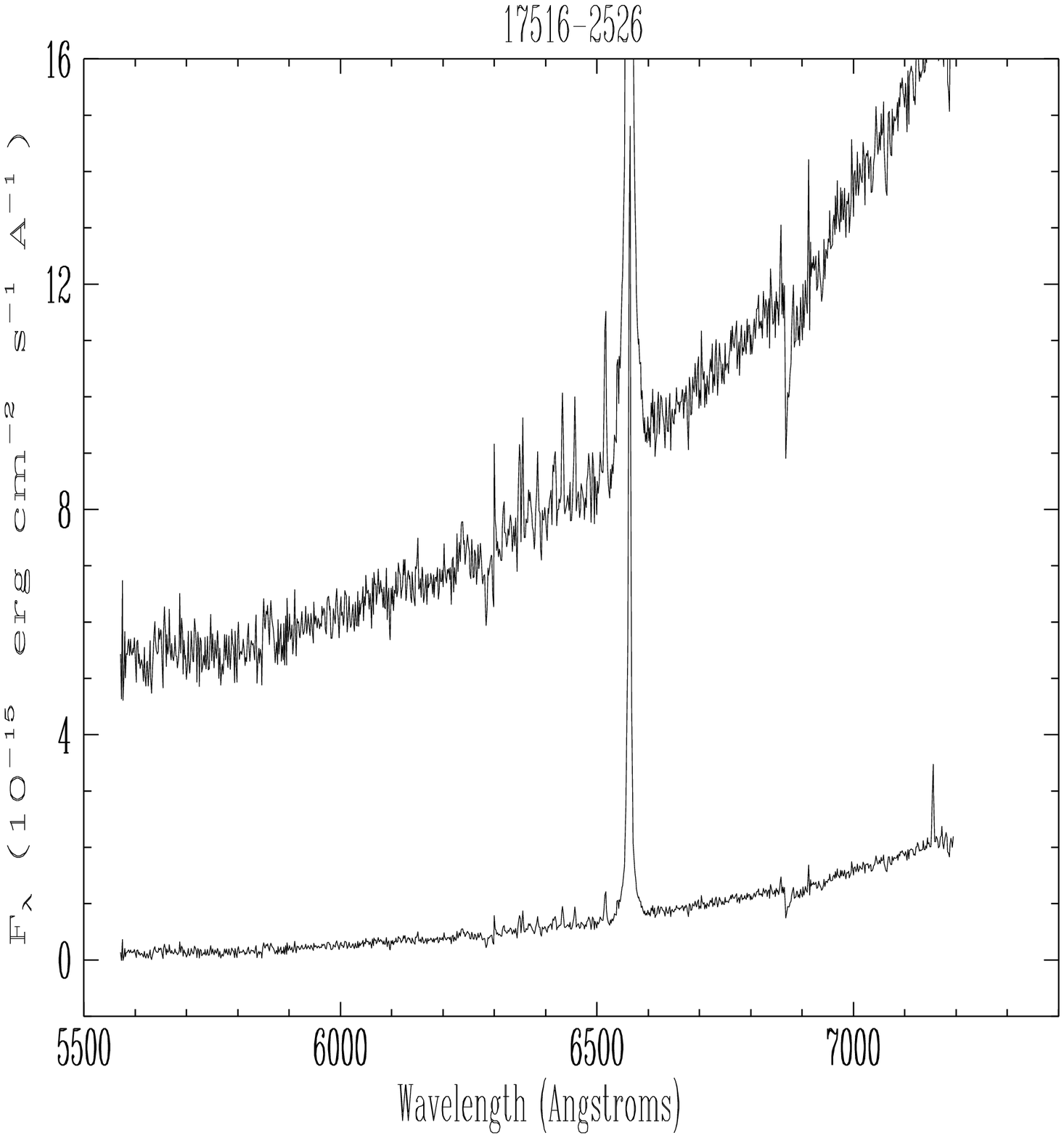}
%\psdraft
\epsfxsize=4cm
\epsfysize=4cm
\epsfbox{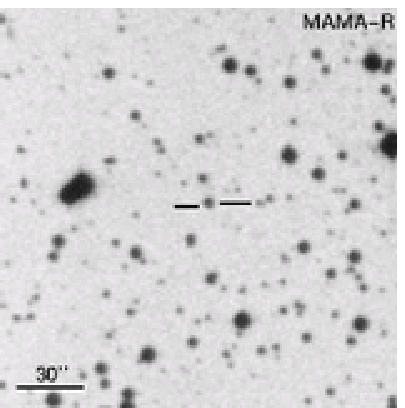}
%\psfull
\end{center}

\caption{Spectra of the objects classified as post-AGB in the sample together with their 
corresponding identification charts (continued). }
\end{figure*}

%%% Local Variables: 
%%% mode: latex
%%% TeX-master: "~/tesis/mitesis/final/tesis"
%%% End: 

        %pg18
\setcounter{figure}{0}
\begin{figure*}

\begin{center}
\epsfxsize=13.5cm
\epsfysize=4cm
\epsfbox{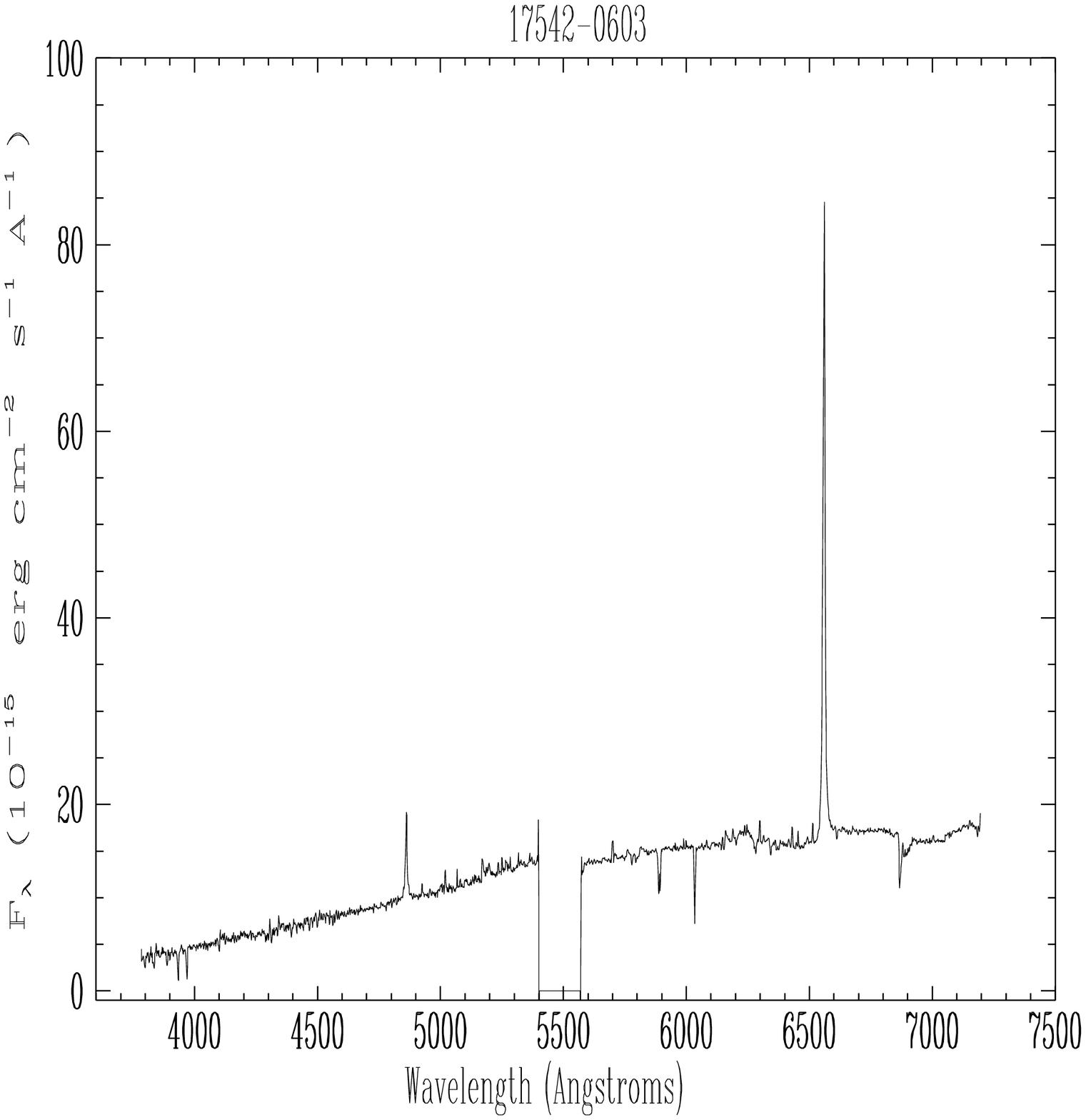}
%\psdraft
\epsfxsize=4cm
\epsfysize=4cm
\epsfbox{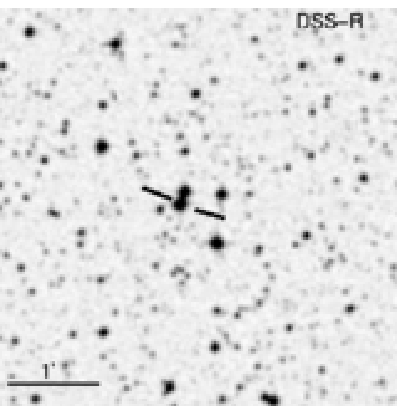}
%\psfull
\end{center}

\begin{center}
\epsfxsize=13.5cm
\epsfysize=4cm
\epsfbox{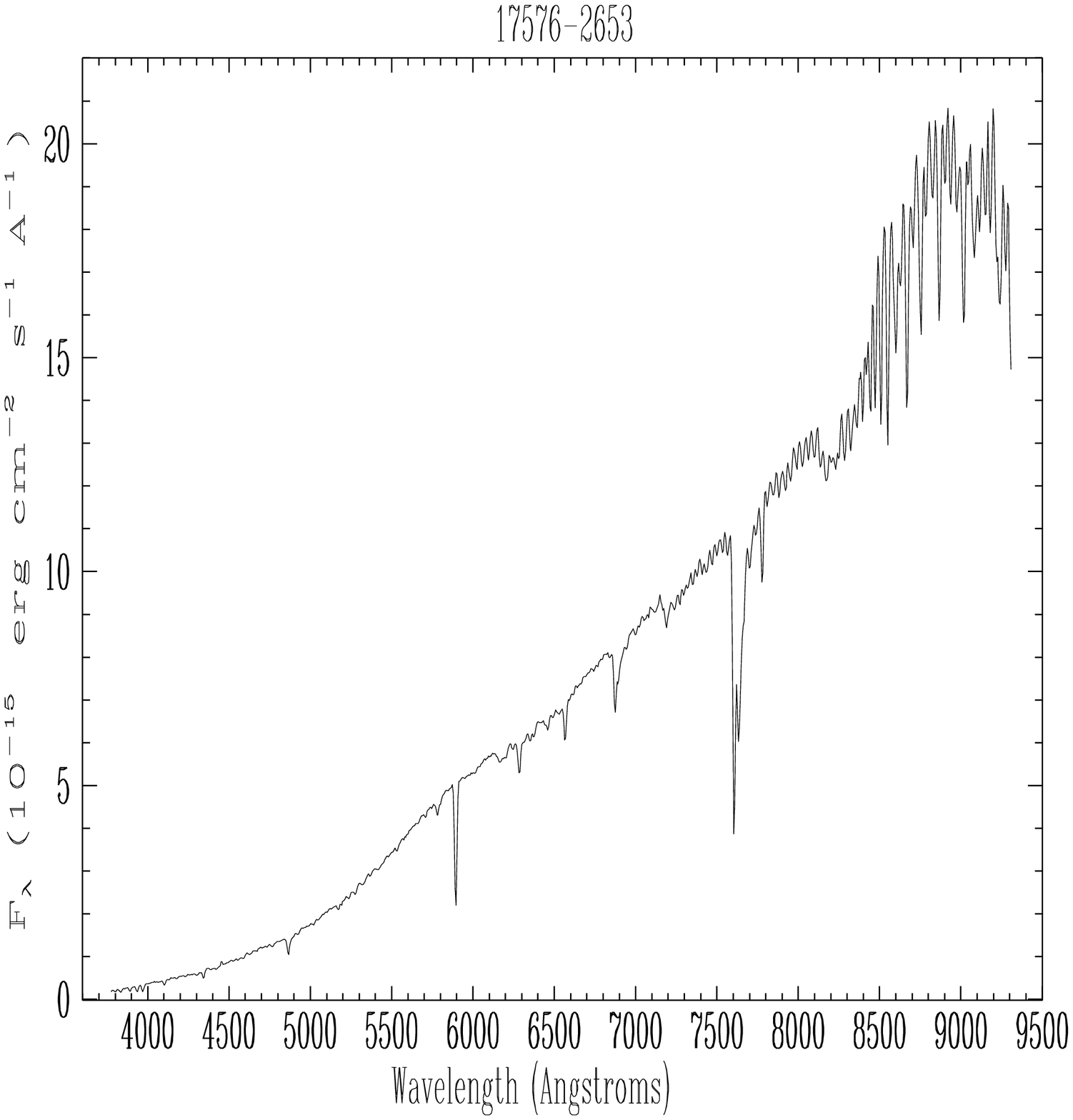}
%\psdraft
\epsfxsize=4cm
\epsfysize=4cm
\epsfbox{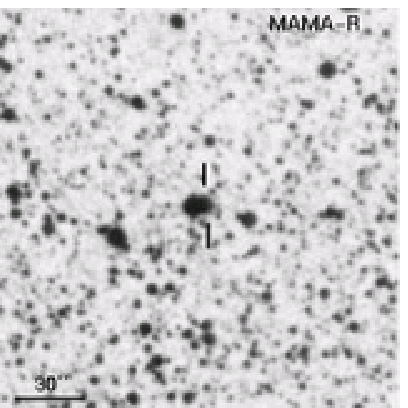}
%\psfull
\end{center}

\begin{center}
\epsfxsize=13.5cm
\epsfysize=4cm
\epsfbox{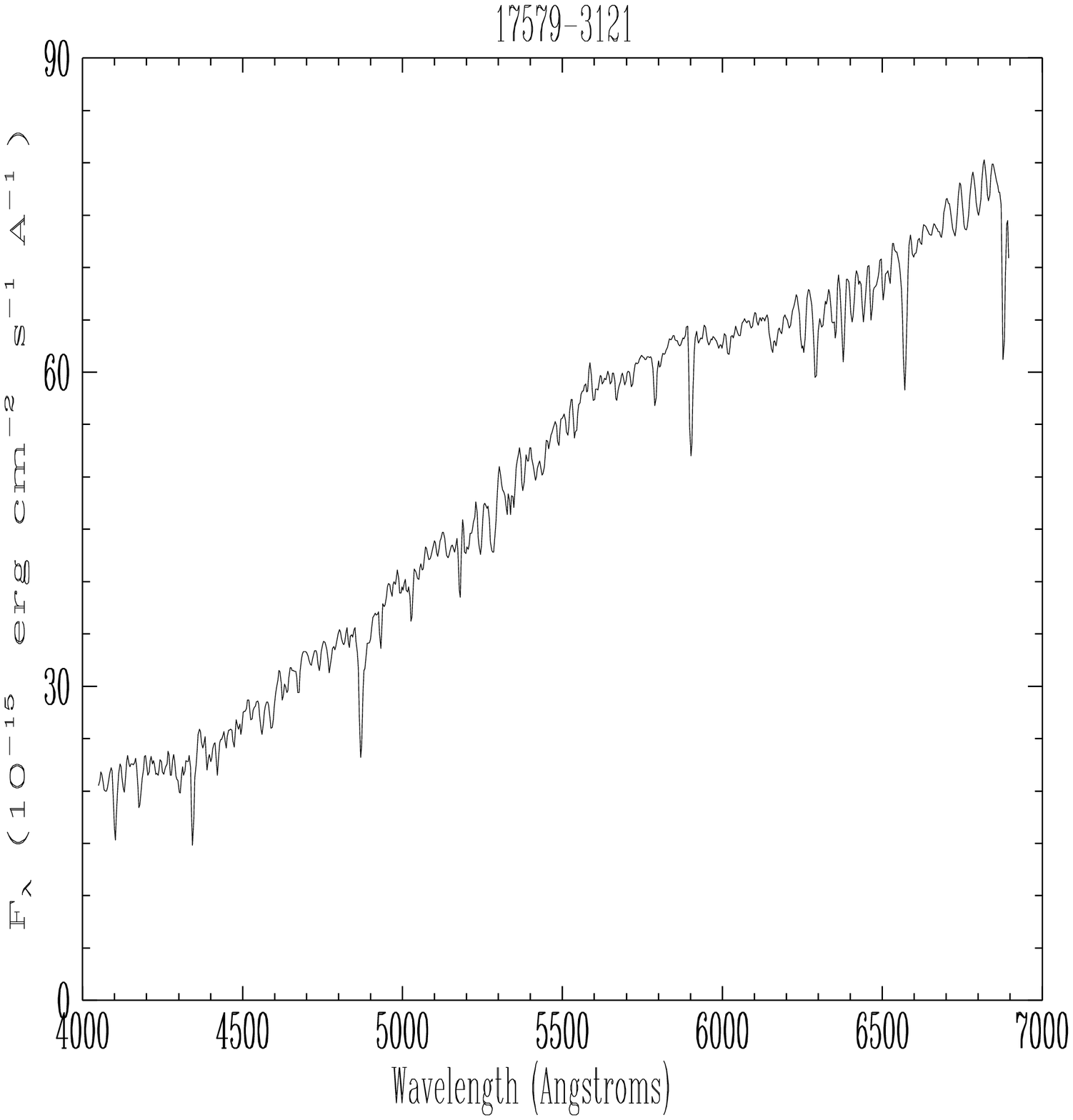}
%\psdraft
\epsfxsize=4cm
\epsfysize=4cm
\epsfbox{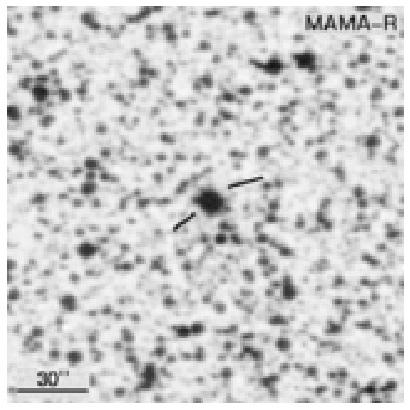}
%\psfull
\end{center}

\begin{center}
\epsfxsize=13.5cm
\epsfysize=4cm
\epsfbox{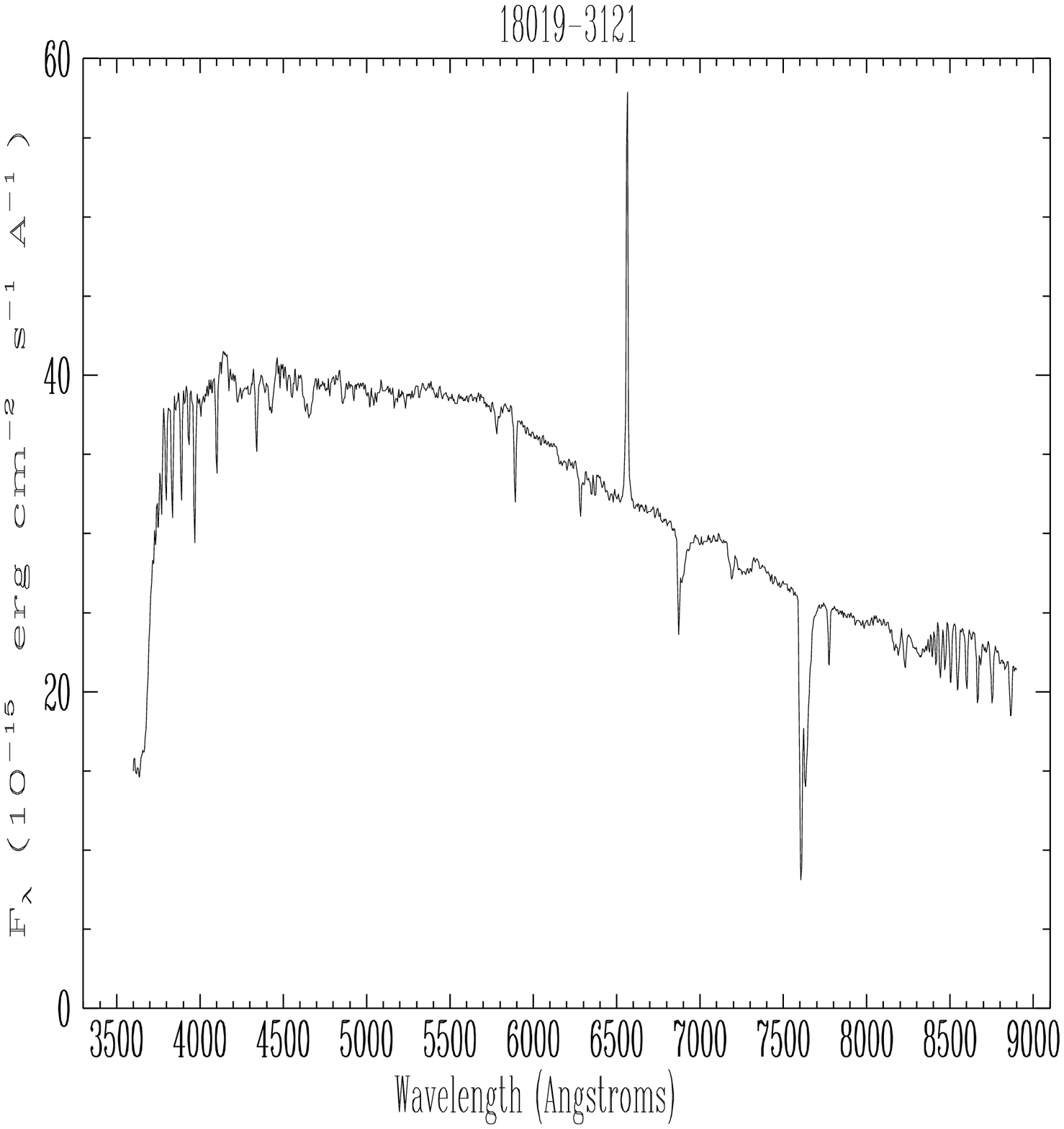}
%\psdraft
\epsfxsize=4cm
\epsfysize=4cm
\epsfbox{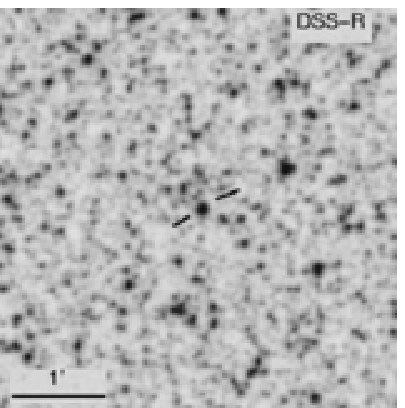}
%\psfull
\end{center}

\begin{center}
\epsfxsize=13.5cm
\epsfysize=4cm
\epsfbox{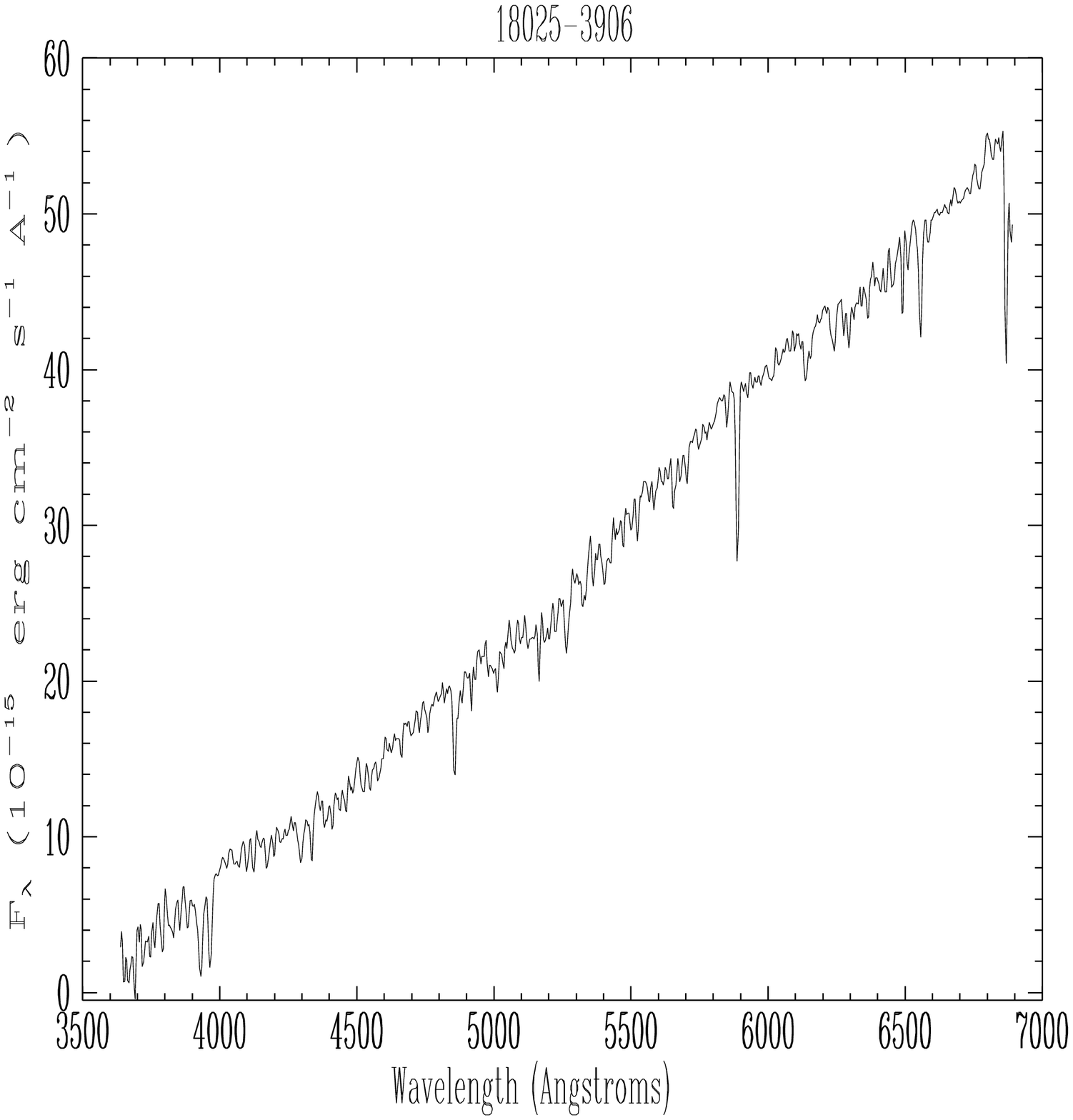}
%\psdraft
\epsfxsize=4cm
\epsfysize=4cm
\epsfbox{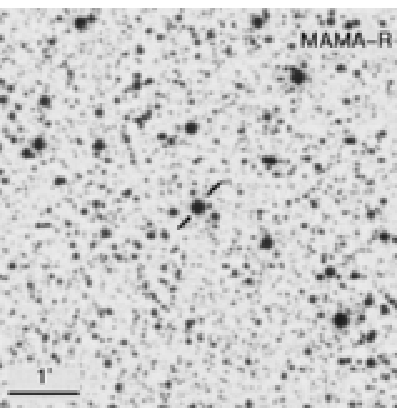}
%\psfull
\end{center}

\caption{Spectra of the objects classified as post-AGB in the sample together with their 
corresponding identification charts (continued). }
\end{figure*}

%-------------------------------------------------------------
%pg19

\setcounter{figure}{0}
\begin{figure*}

\begin{center}
\epsfxsize=13.5cm
\epsfysize=4cm
\epsfbox{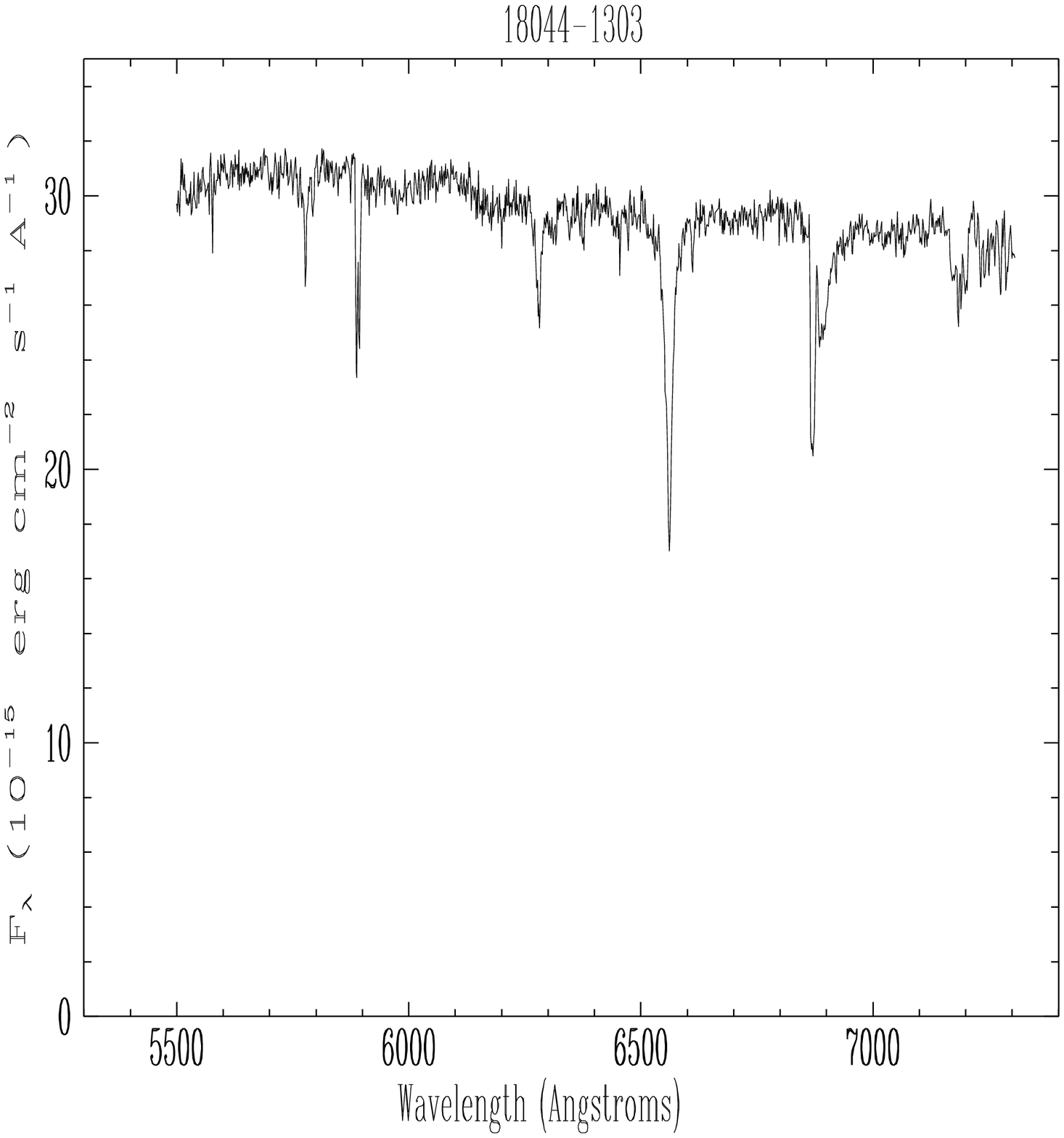}
%\psdraft
\epsfxsize=4cm
\epsfysize=4cm
\epsfbox{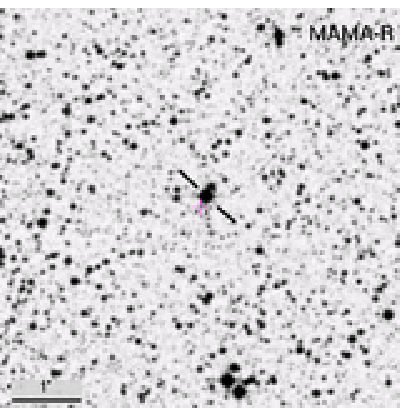}
%\psfull
\end{center}

\begin{center}
\epsfxsize=13.5cm
\epsfysize=4cm
\epsfbox{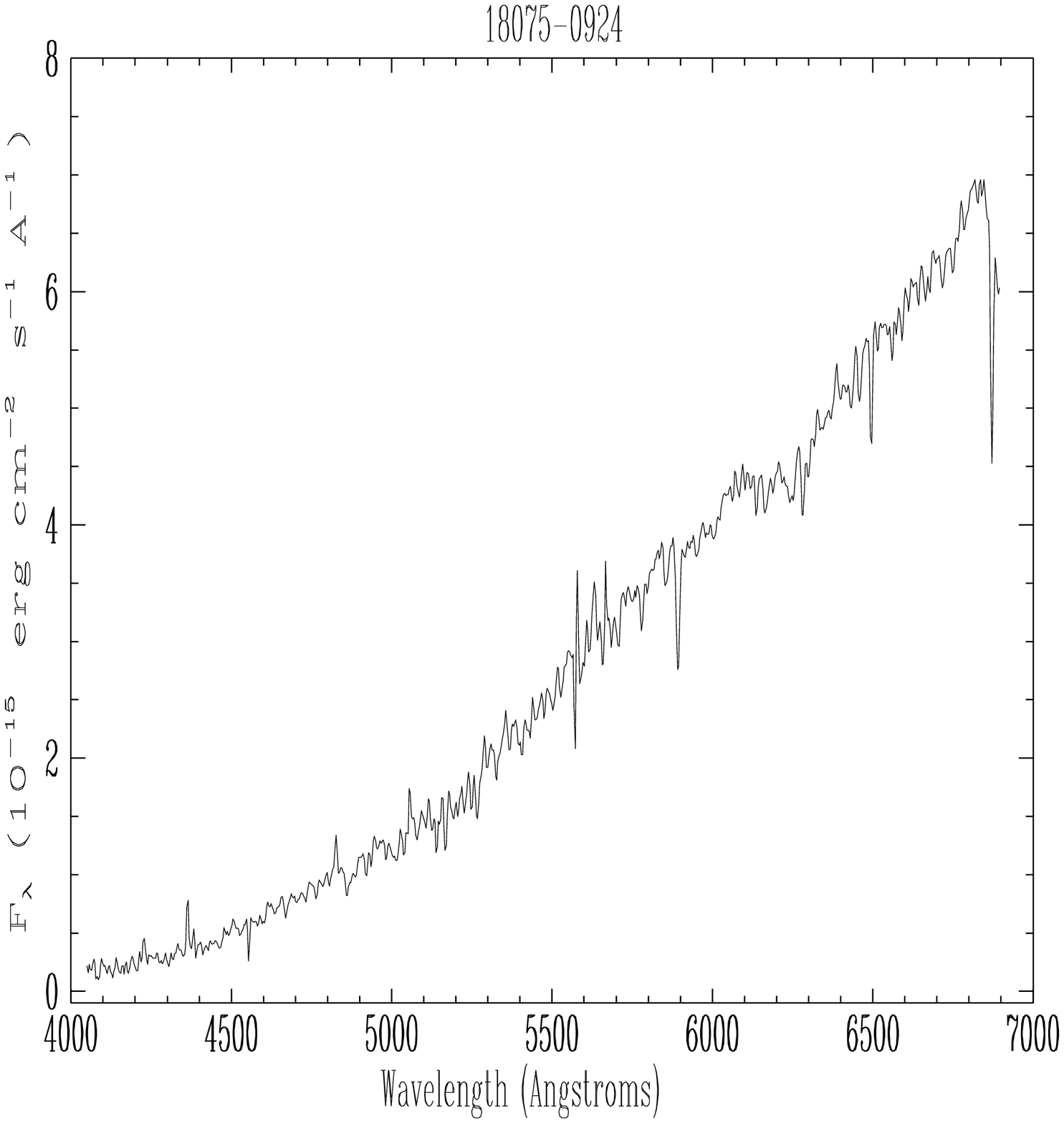}
%\psdraft
\epsfxsize=4cm
\epsfysize=4cm
\epsfbox{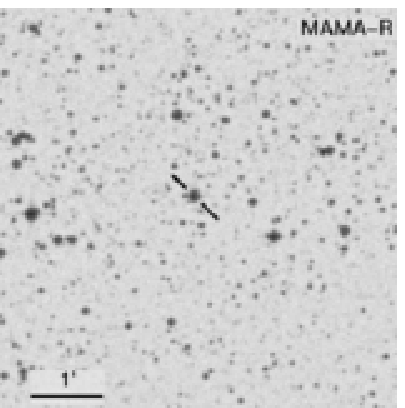}
%\psfull
\end{center}

\begin{center}
\epsfxsize=13.5cm
\epsfysize=4cm
\epsfbox{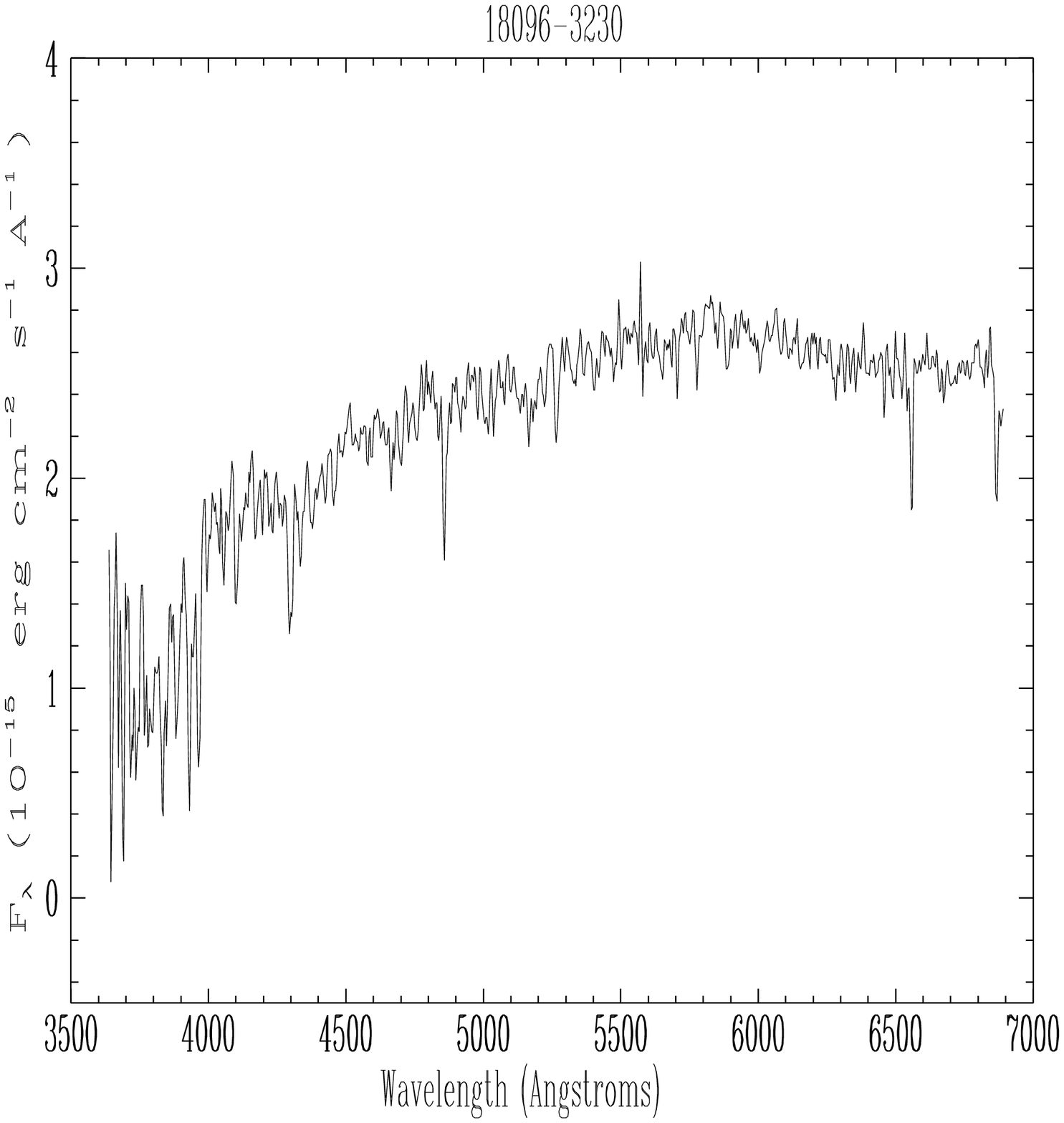}
%\psdraft
\epsfxsize=4cm
\epsfysize=4cm
\epsfbox{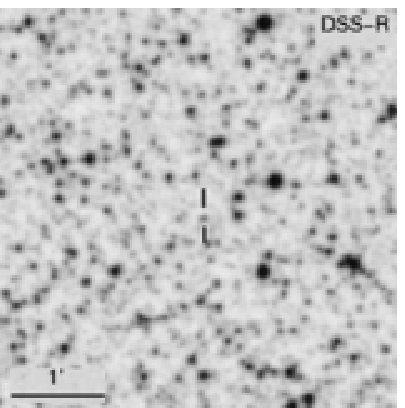}
%\psfull
\end{center}

\begin{center}
\epsfxsize=13.5cm
\epsfysize=4cm
\epsfbox{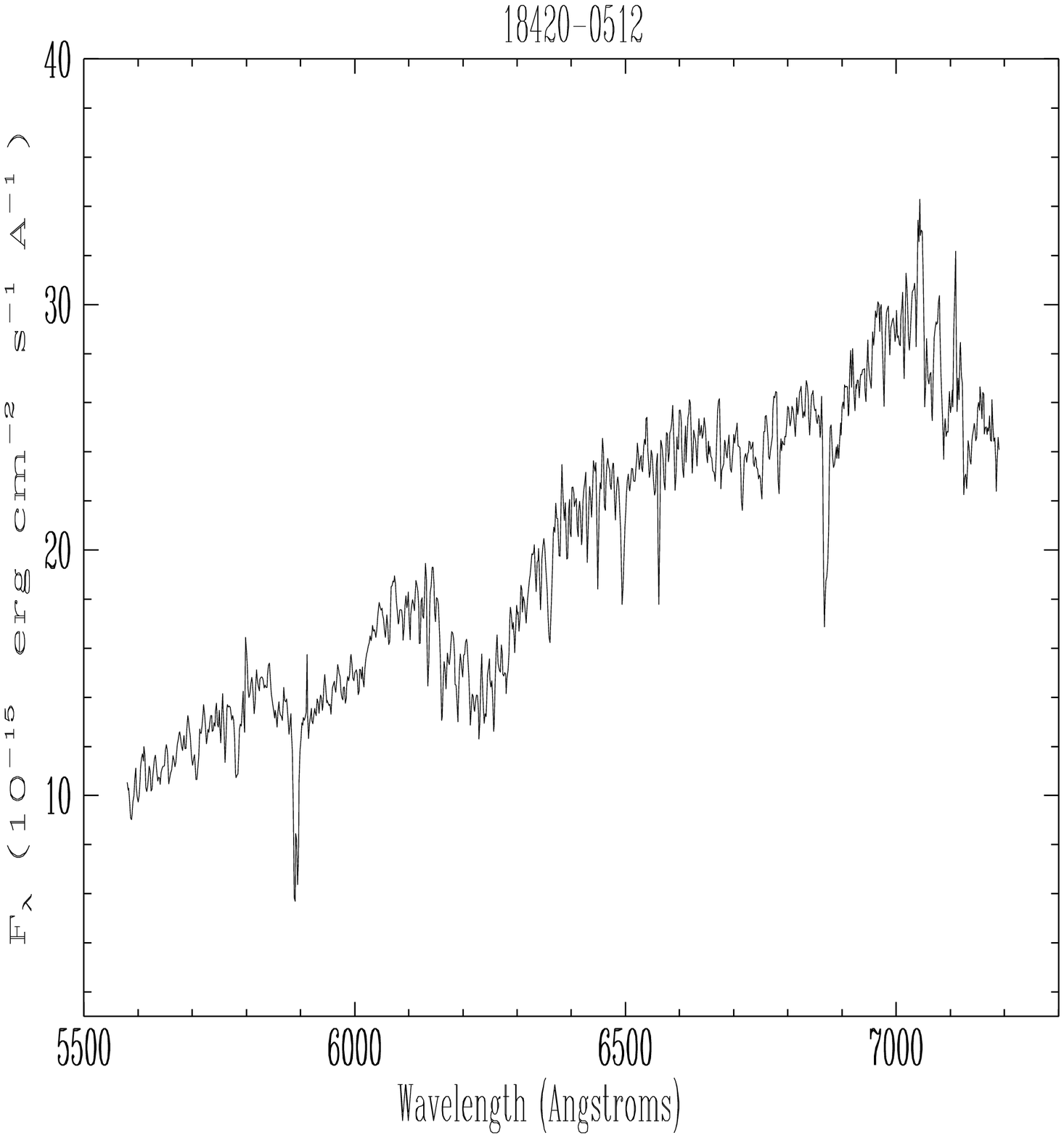}
%\psdraft
\epsfxsize=4cm
\epsfysize=4cm
\epsfbox{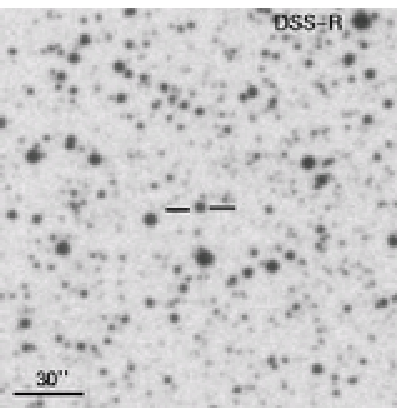}
%\psfull
\end{center}

\begin{center}
\epsfxsize=13.5cm
\epsfysize=4cm
\epsfbox{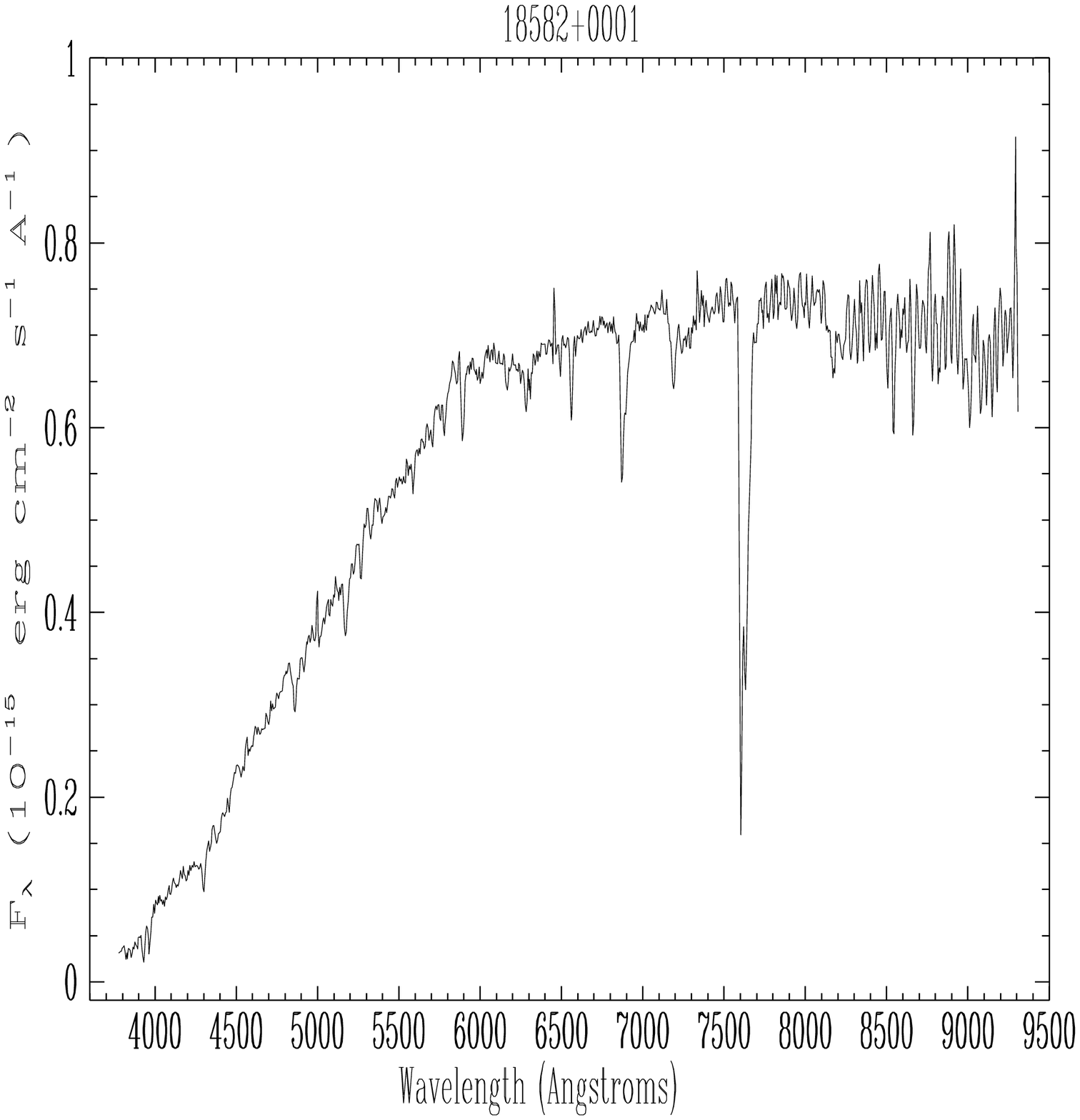}
%\psdraft
\epsfxsize=4cm
\epsfysize=4cm
\epsfbox{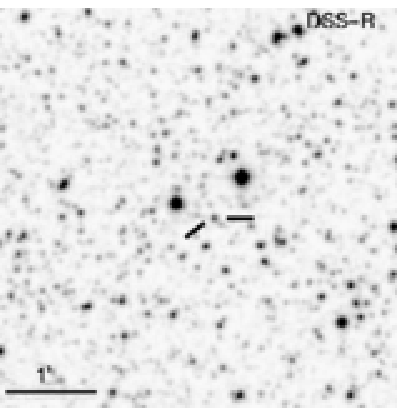}
%\psfull
\end{center}

\caption{Spectra of the objects classified as post-AGB in the sample together with their 
corresponding identification charts (continued). }
\end{figure*}

%-------------------------------------------------------------
%pg20
\setcounter{figure}{0}
\begin{figure*}

\begin{center}
\epsfxsize=13.5cm
\epsfysize=4cm
\epsfbox{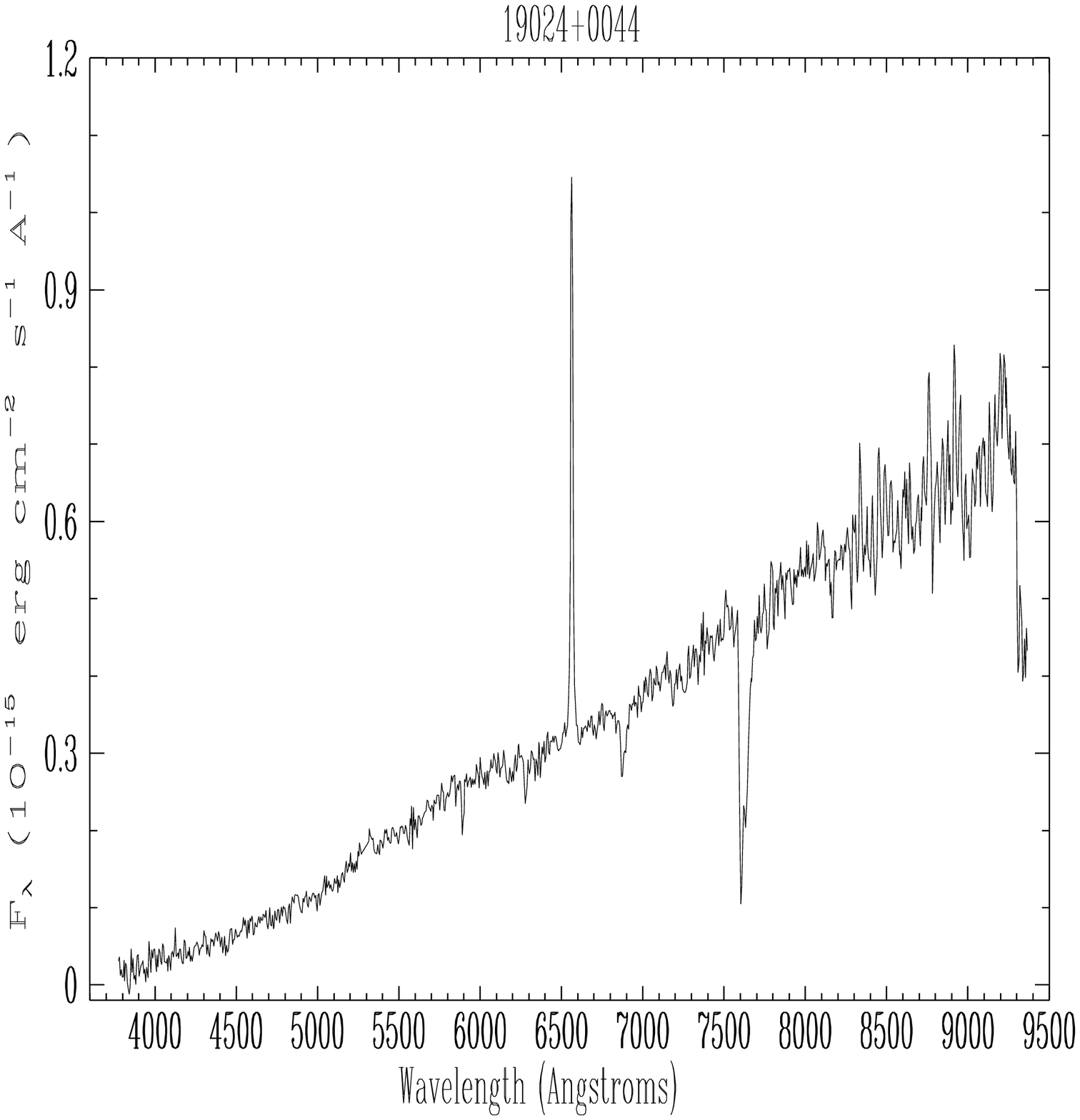}
%\psdraft
\epsfxsize=4cm
\epsfysize=4cm
\epsfbox{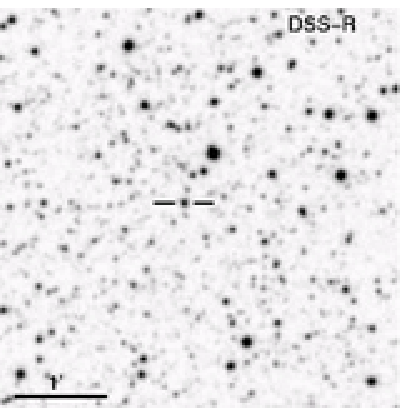}
%\psfull
\end{center}

\begin{center}
\epsfxsize=13.5cm
\epsfysize=4cm
\epsfbox{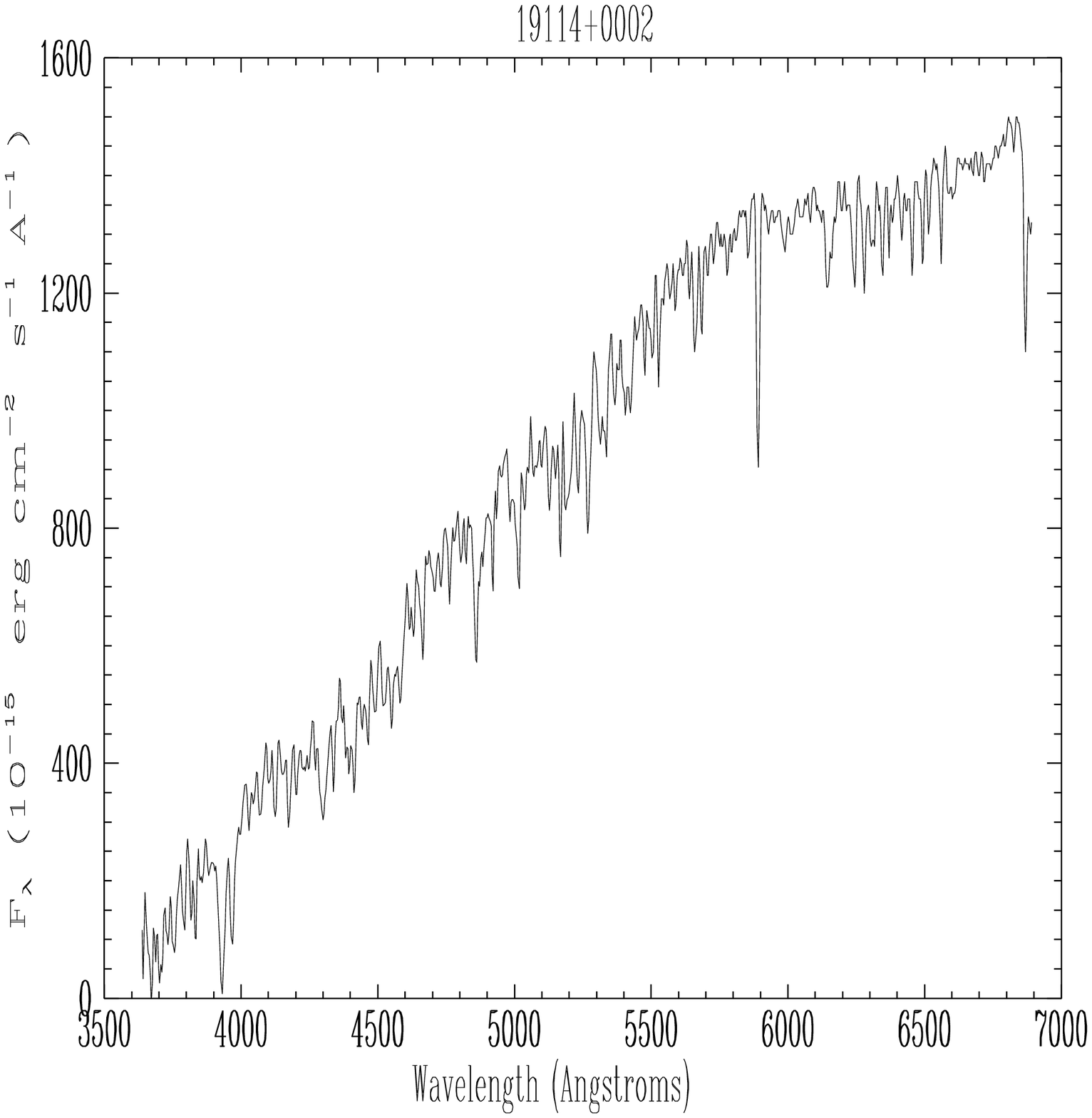}
%\psdraft
\epsfxsize=4cm
\epsfysize=4cm
\epsfbox{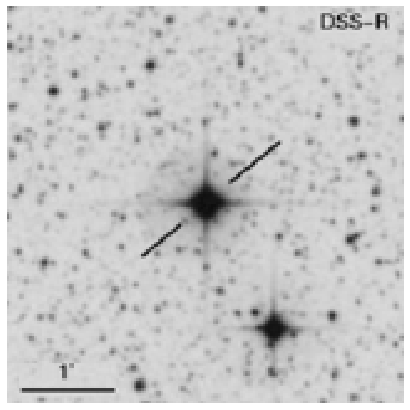}
%\psfull
\end{center}

\begin{center}
\epsfxsize=13.5cm
\epsfysize=4cm
\epsfbox{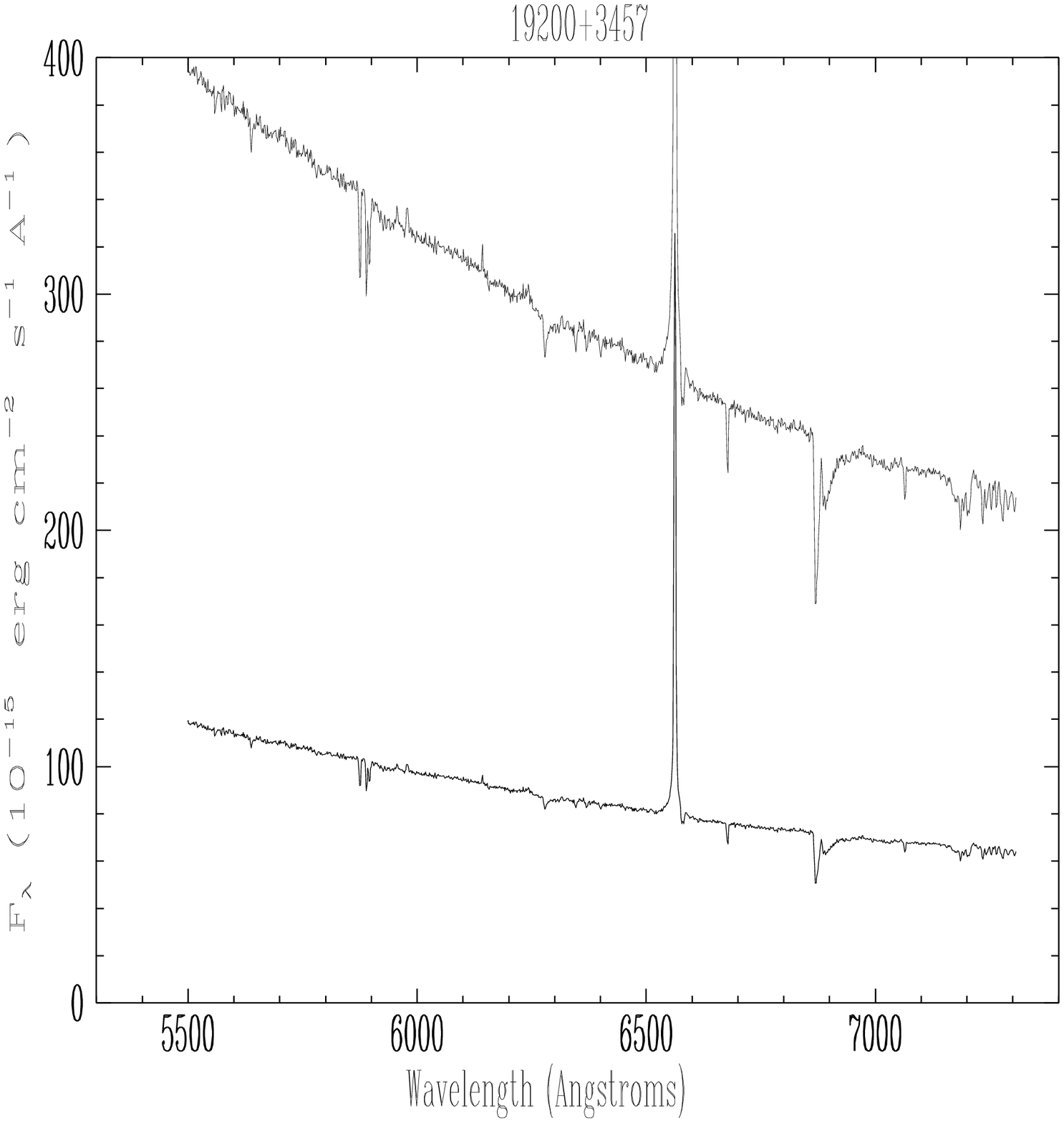}
%\psdraft
\epsfxsize=4cm
\epsfysize=4cm
\epsfbox{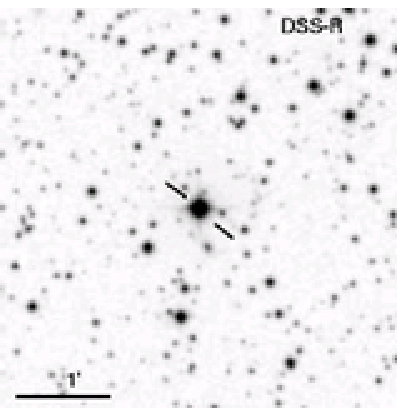}
%\psfull
\end{center}

\begin{center}
\epsfxsize=13.5cm
\epsfysize=4cm
\epsfbox{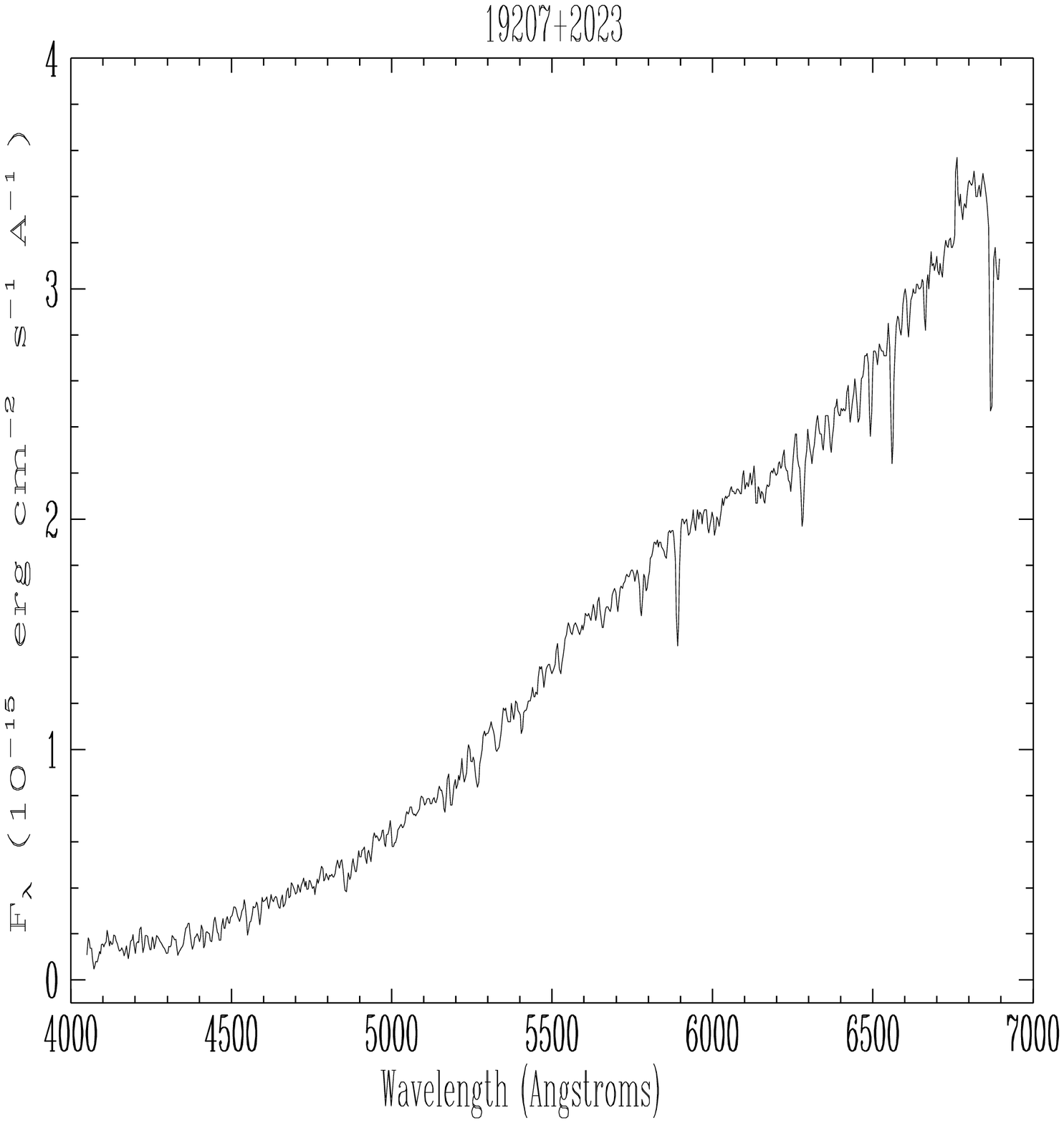}
%\psdraft
\epsfxsize=4cm
\epsfysize=4cm
\epsfbox{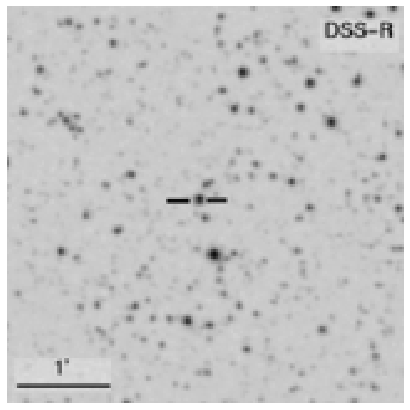}
%\psfull
\end{center}

\begin{center}
\epsfxsize=13.5cm
\epsfysize=4cm
\epsfbox{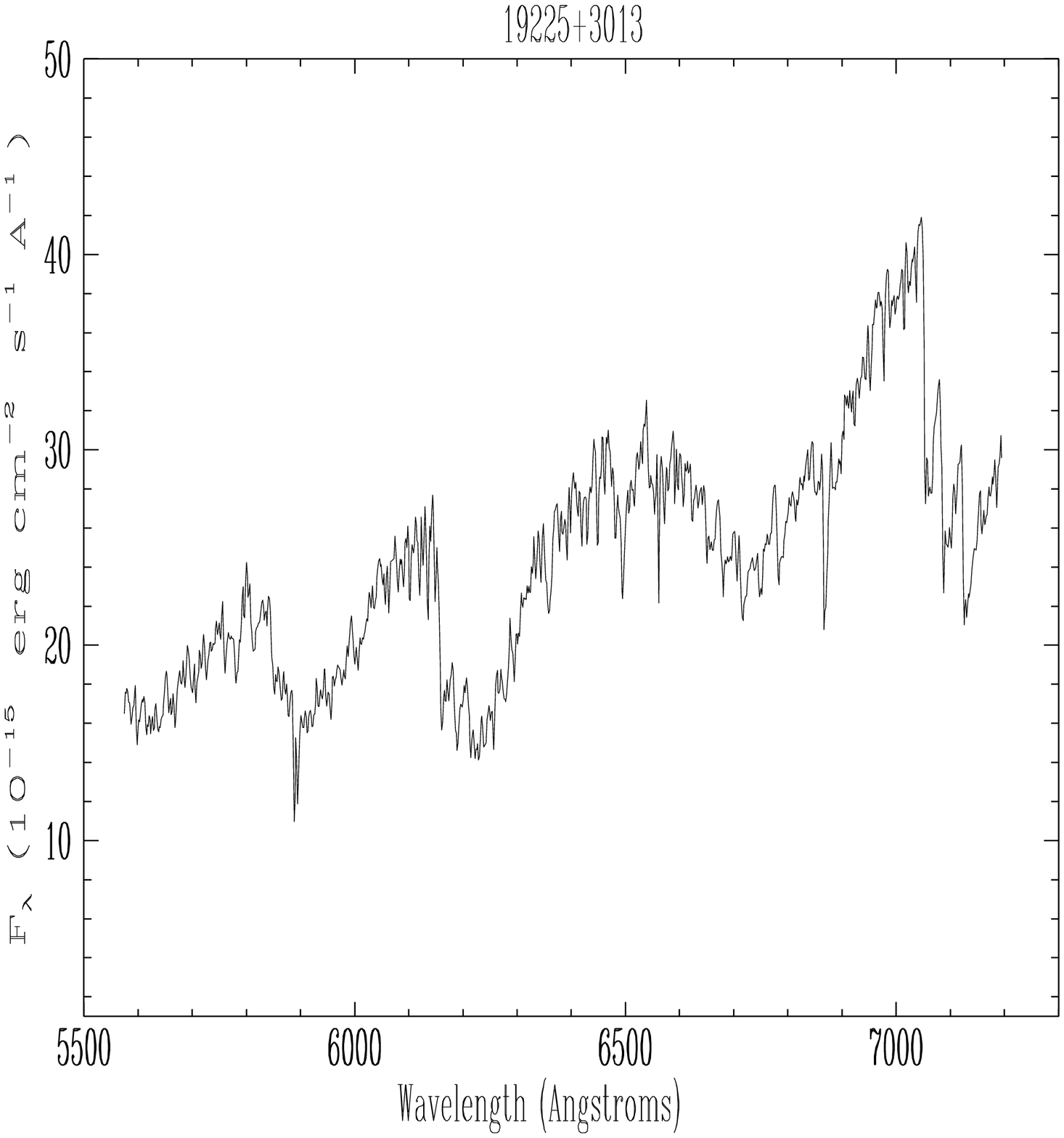}
%\psdraft
\epsfxsize=4cm
\epsfysize=4cm
\epsfbox{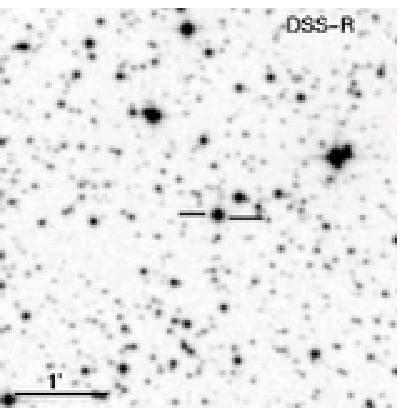}
%\psfull
\end{center}

\caption{Spectra of the objects classified as post-AGB in the sample together with their 
corresponding identification charts (continued). }
\end{figure*}

%-------------------------------------------------------------
%pg20
\setcounter{figure}{0}
\begin{figure*}

\begin{center}
\epsfxsize=13.5cm
\epsfysize=4cm
\epsfbox{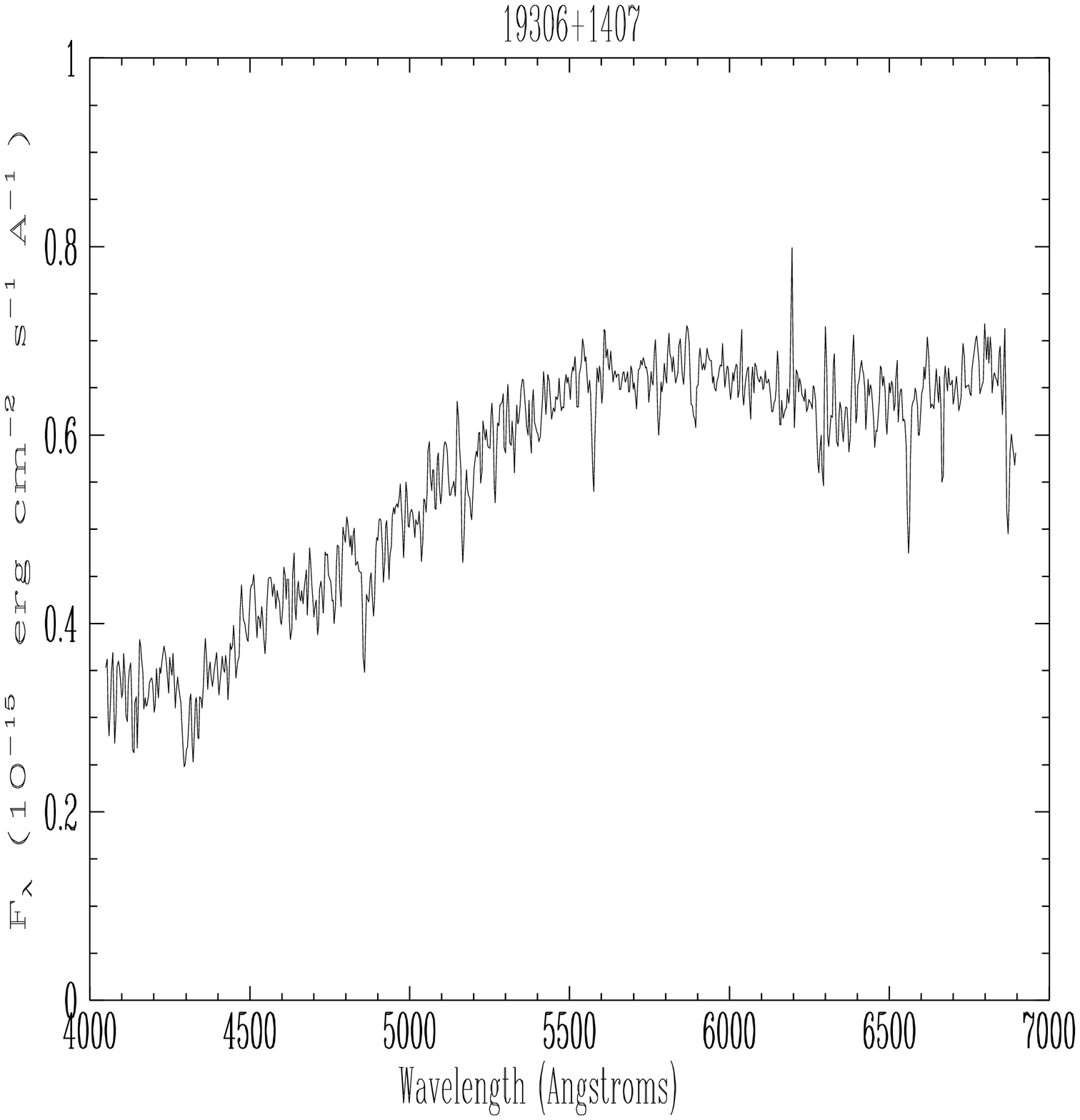}
%\psdraft
\epsfxsize=4cm
\epsfysize=4cm
\epsfbox{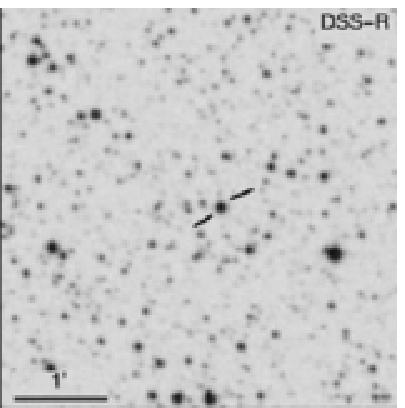}
%\psfull
\end{center}

\begin{center}
\epsfxsize=13.5cm
\epsfysize=4cm
\epsfbox{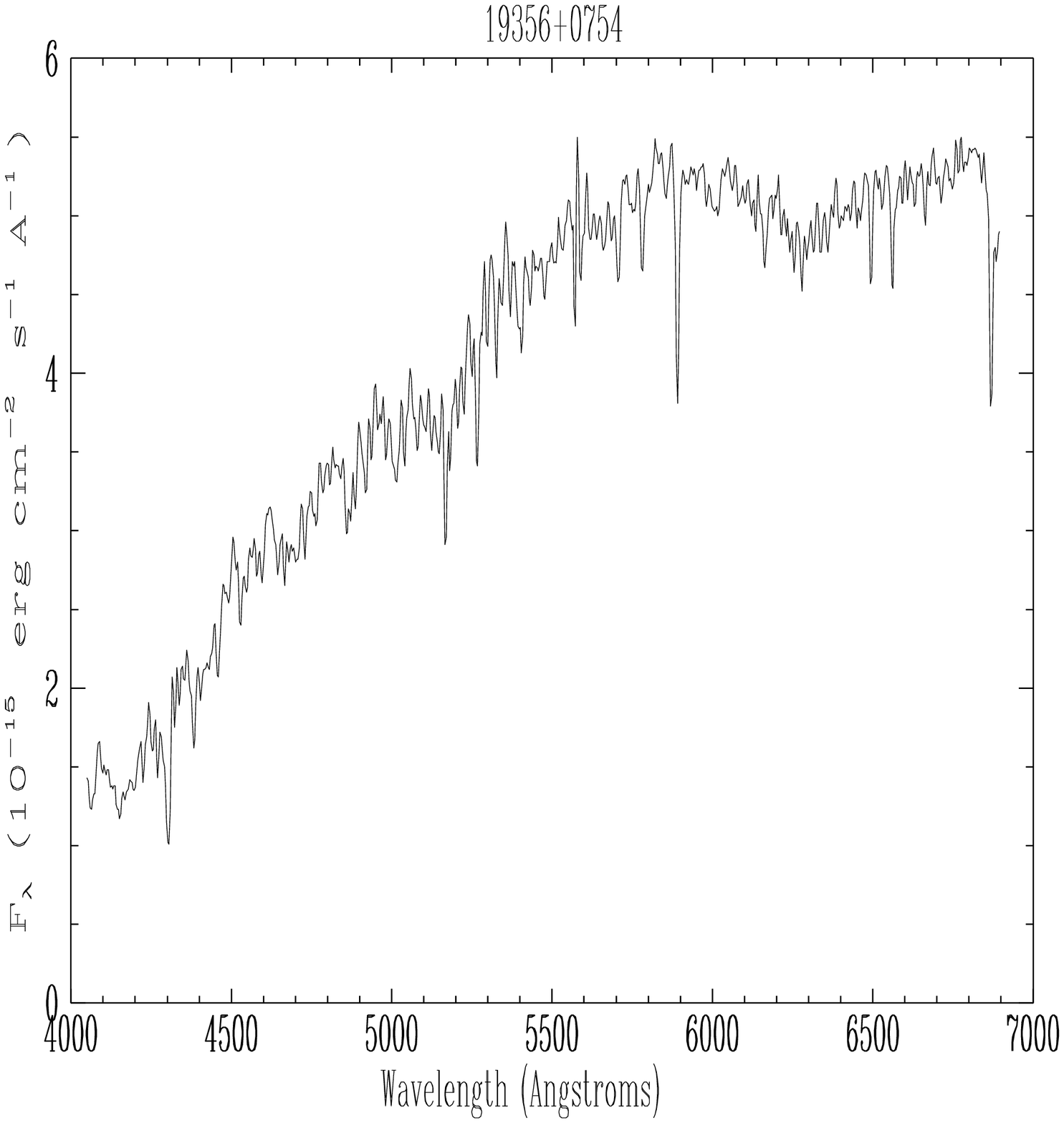}
%\psdraft
\epsfxsize=4cm
\epsfysize=4cm
\epsfbox{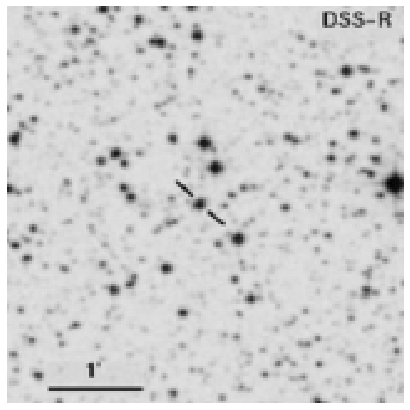}
%\psfull
\end{center}

\begin{center}
\epsfxsize=13.5cm
\epsfysize=4cm
\epsfbox{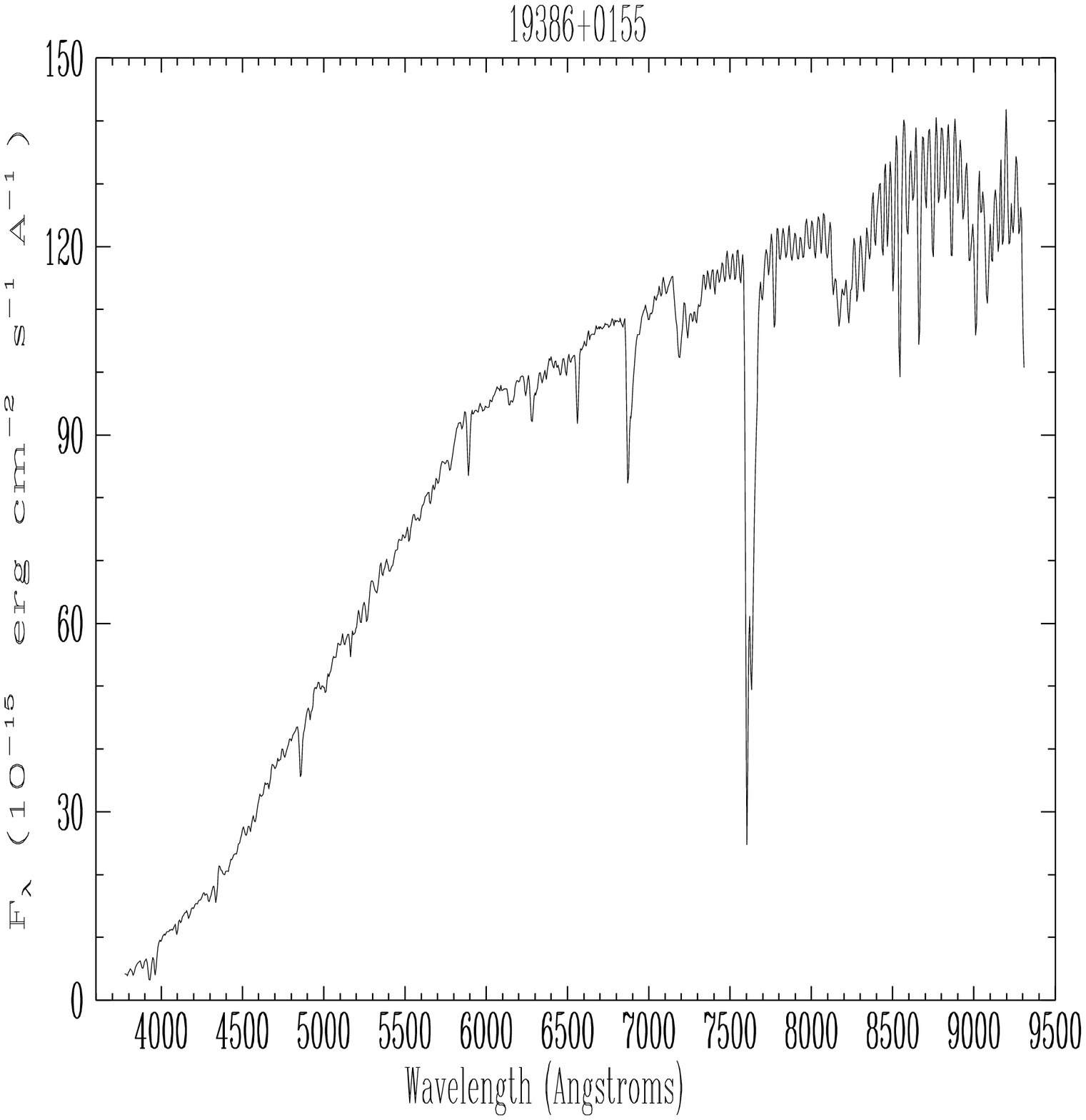}
%\psdraft
\epsfxsize=4cm
\epsfysize=4cm
\epsfbox{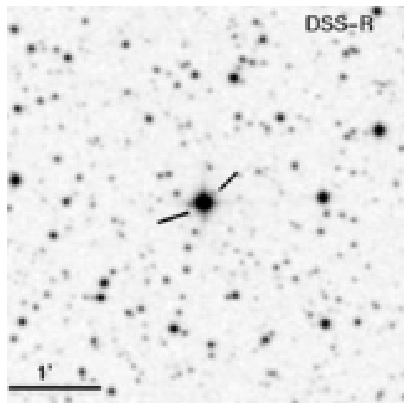}
%\psfull
\end{center}

\begin{center}
\epsfxsize=13.5cm
\epsfysize=4cm
\epsfbox{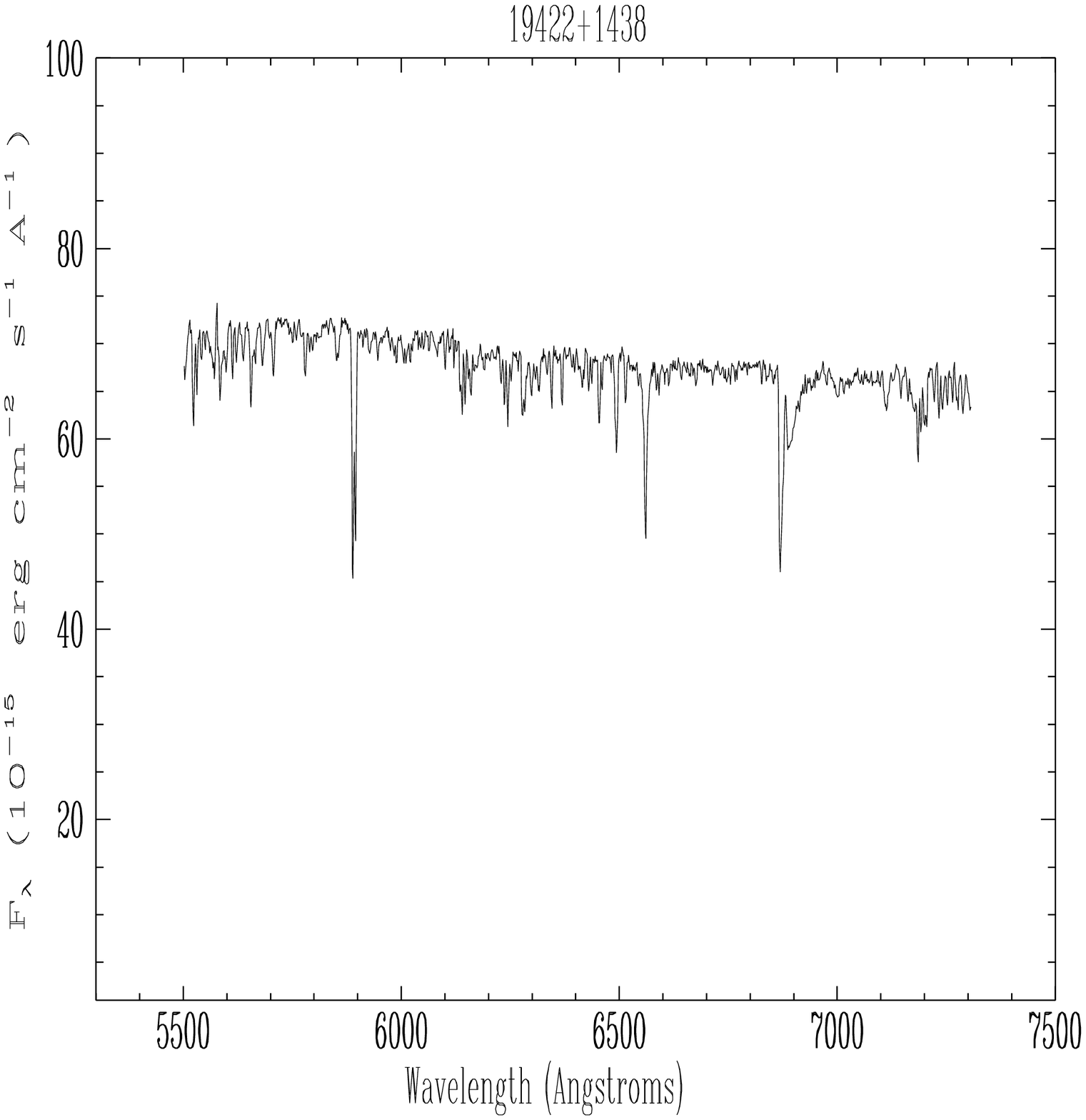}
%\psdraft
\epsfxsize=4cm
\epsfysize=4cm
\epsfbox{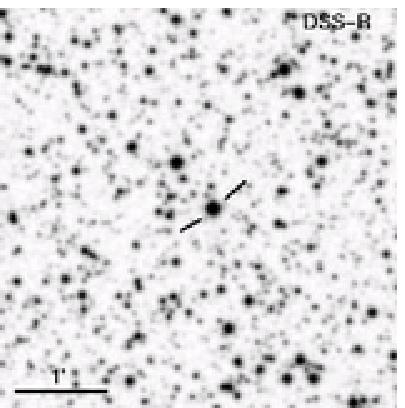}
%\psfull
\end{center}

\begin{center}
\epsfxsize=13.5cm
\epsfysize=4cm
\epsfbox{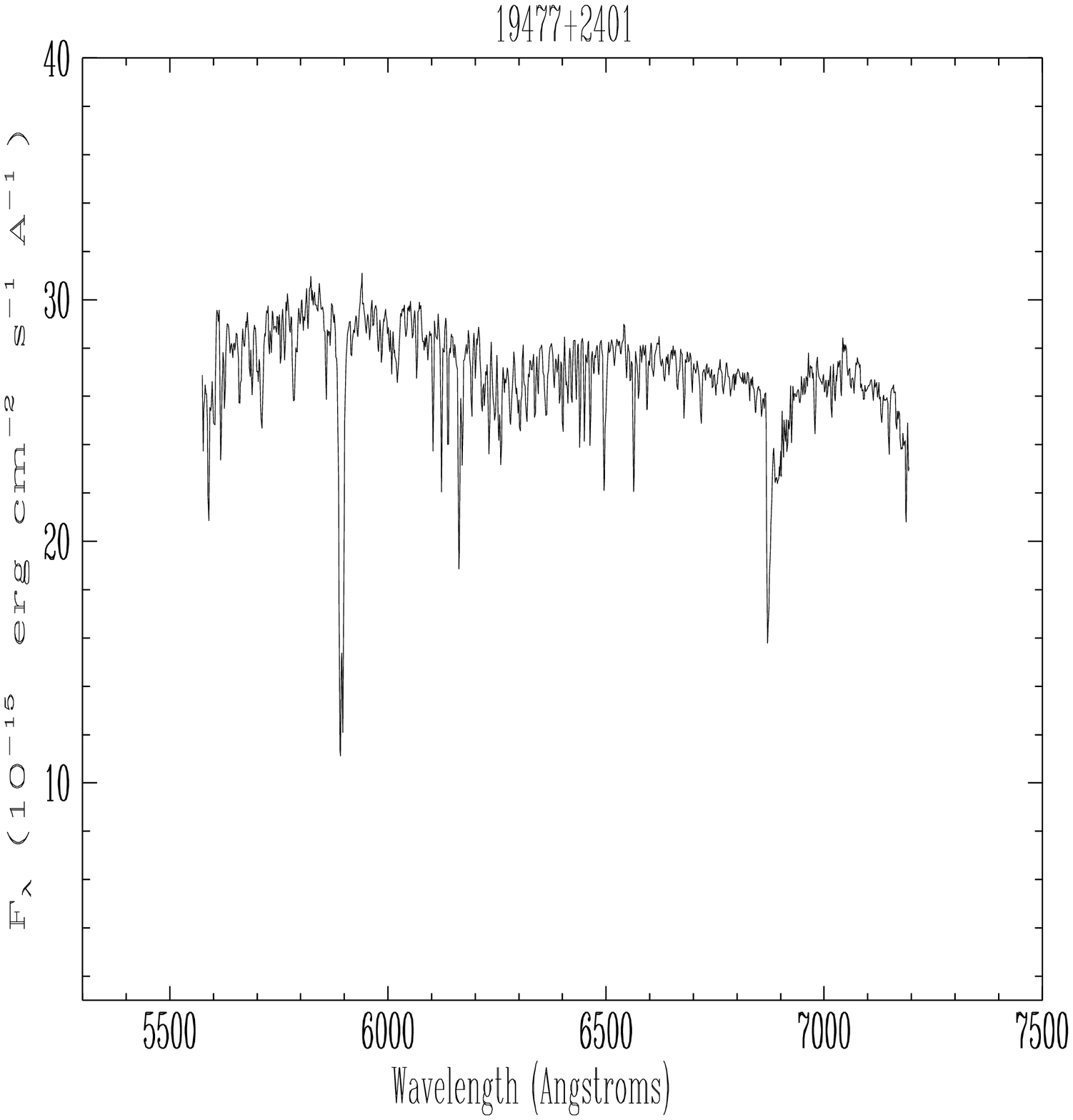}
%\psdraft
\epsfxsize=4cm
\epsfysize=4cm
\epsfbox{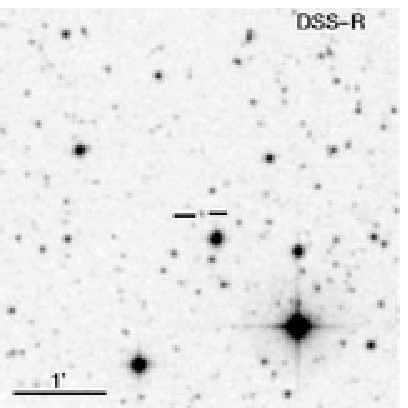}
%\psfull
\end{center}

\caption{Spectra of the objects classified as post-AGB in the sample together with their 
corresponding identification charts (continued). }
\end{figure*}

%-------------------------------------------------------------
%pg21
\setcounter{figure}{0}
\begin{figure*}

\begin{center}
\epsfxsize=13.5cm
\epsfysize=4cm
\epsfbox{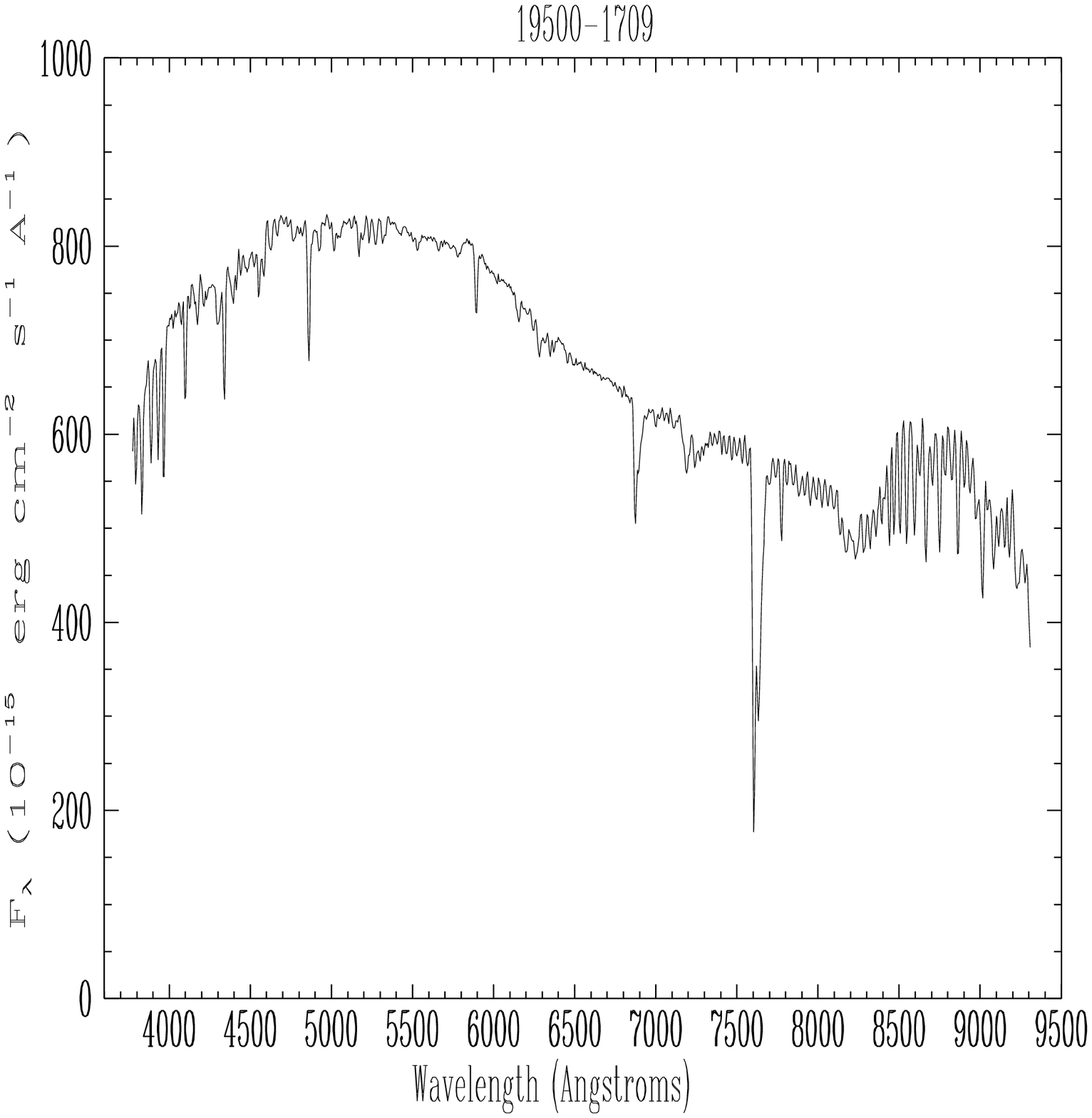}
%\psdraft
\epsfxsize=4cm
\epsfysize=4cm
\epsfbox{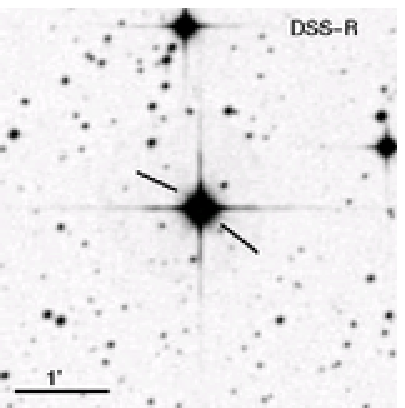}
%\psfull
\end{center}

\begin{center}
\epsfxsize=13.5cm
\epsfysize=4cm
\epsfbox{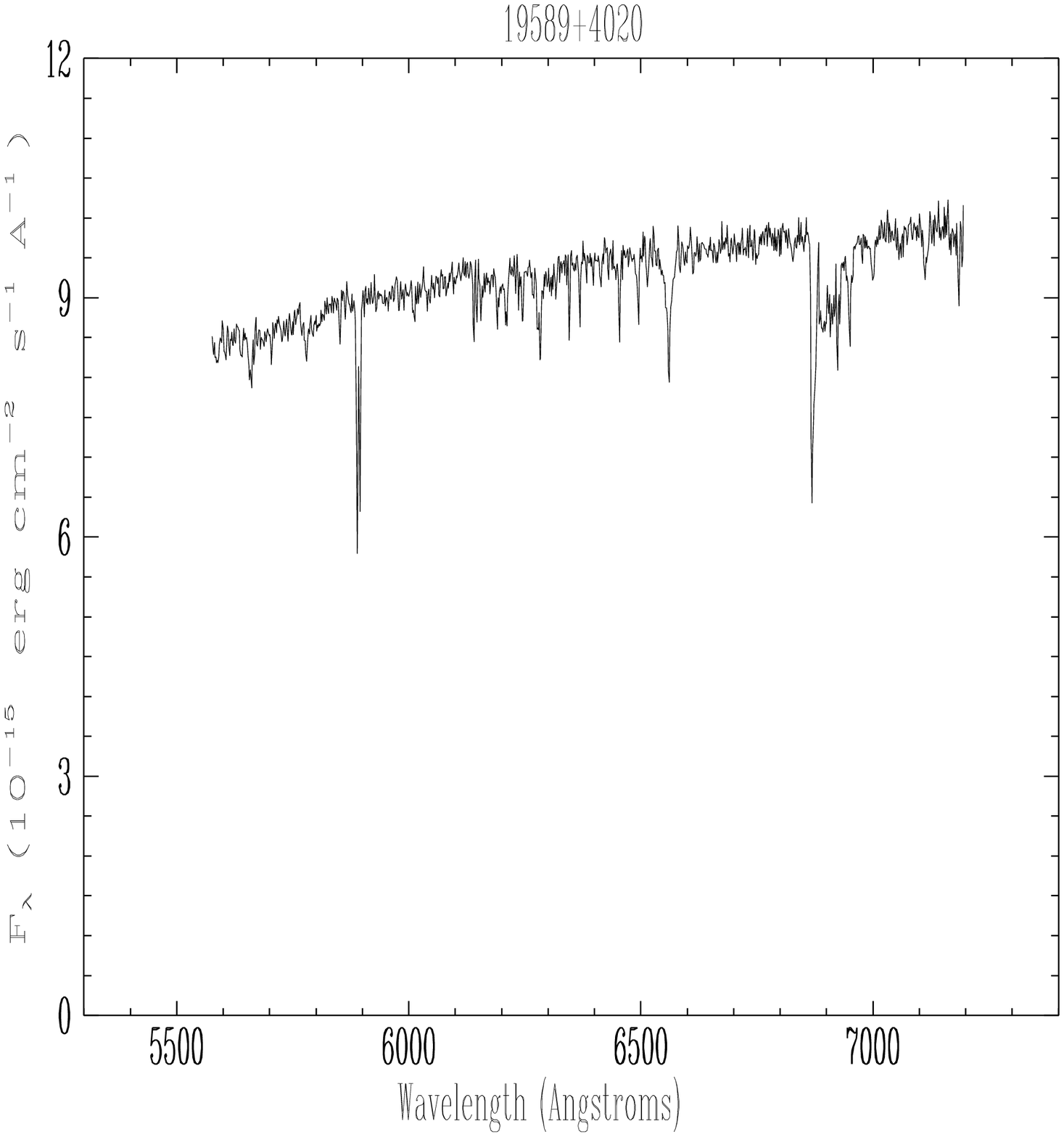}
%\psdraft
\epsfxsize=4cm
\epsfysize=4cm
\epsfbox{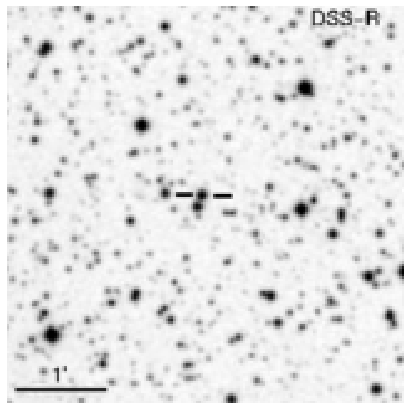}
%\psfull
\end{center}

\begin{center}
\epsfxsize=13.5cm
\epsfysize=4cm
\epsfbox{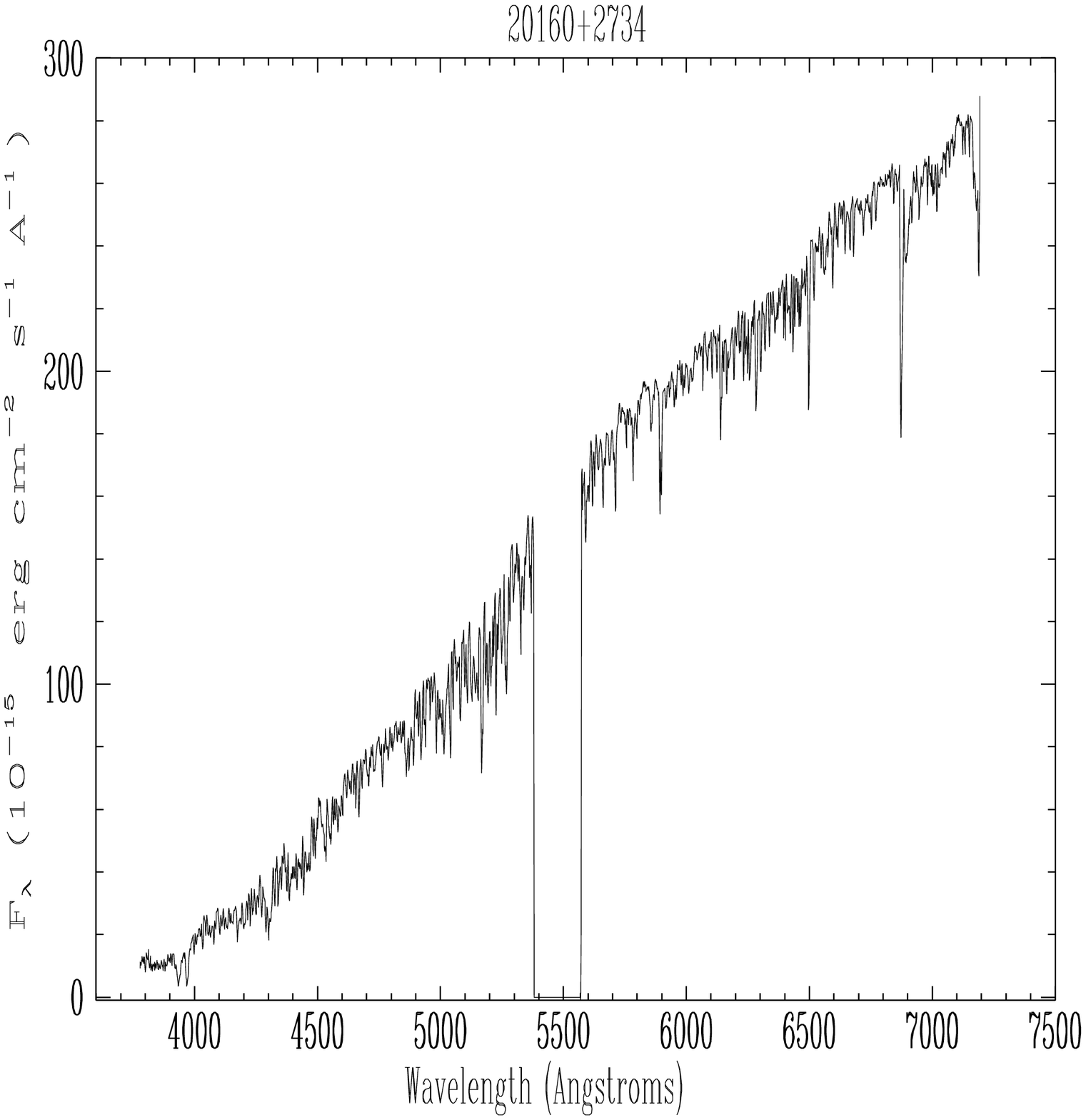}
%\psdraft
\epsfxsize=4cm
\epsfysize=4cm
\epsfbox{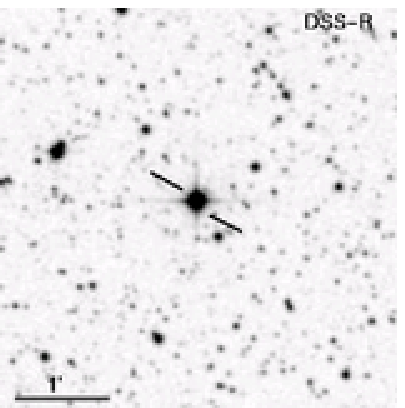}
%\psfull
\end{center}

\begin{center}
\epsfxsize=13.5cm
\epsfysize=4cm
\epsfbox{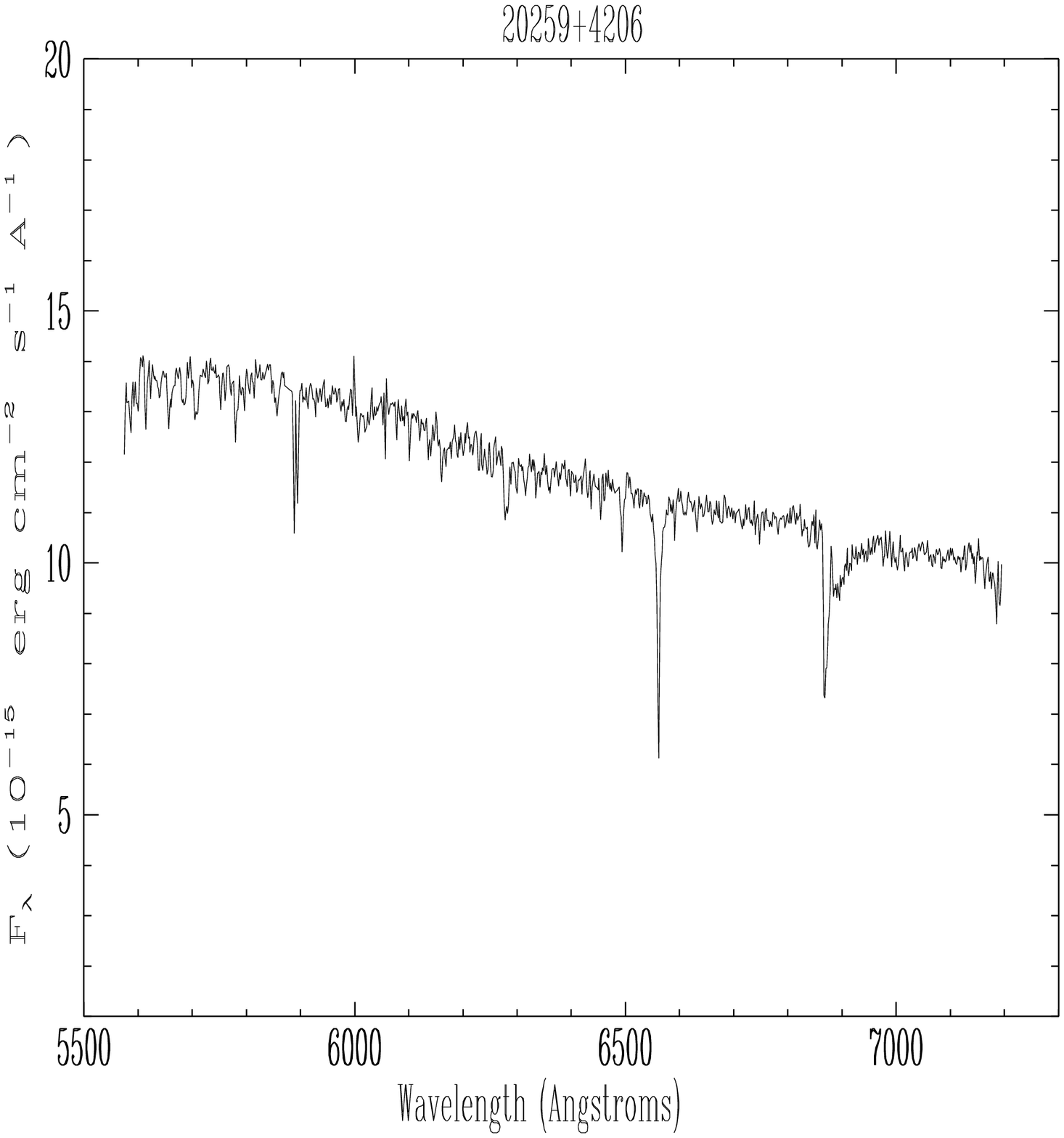}
%\psdraft
\epsfxsize=4cm
\epsfysize=4cm
\epsfbox{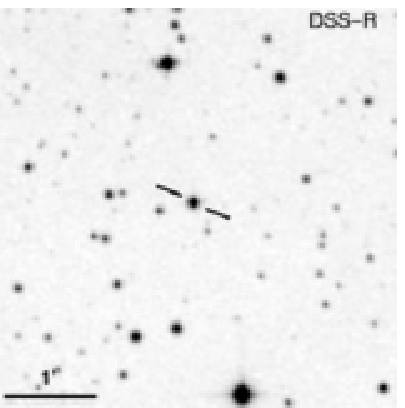}
%\psfull
\end{center}

\begin{center}
\epsfxsize=13.5cm
\epsfysize=4cm
\epsfbox{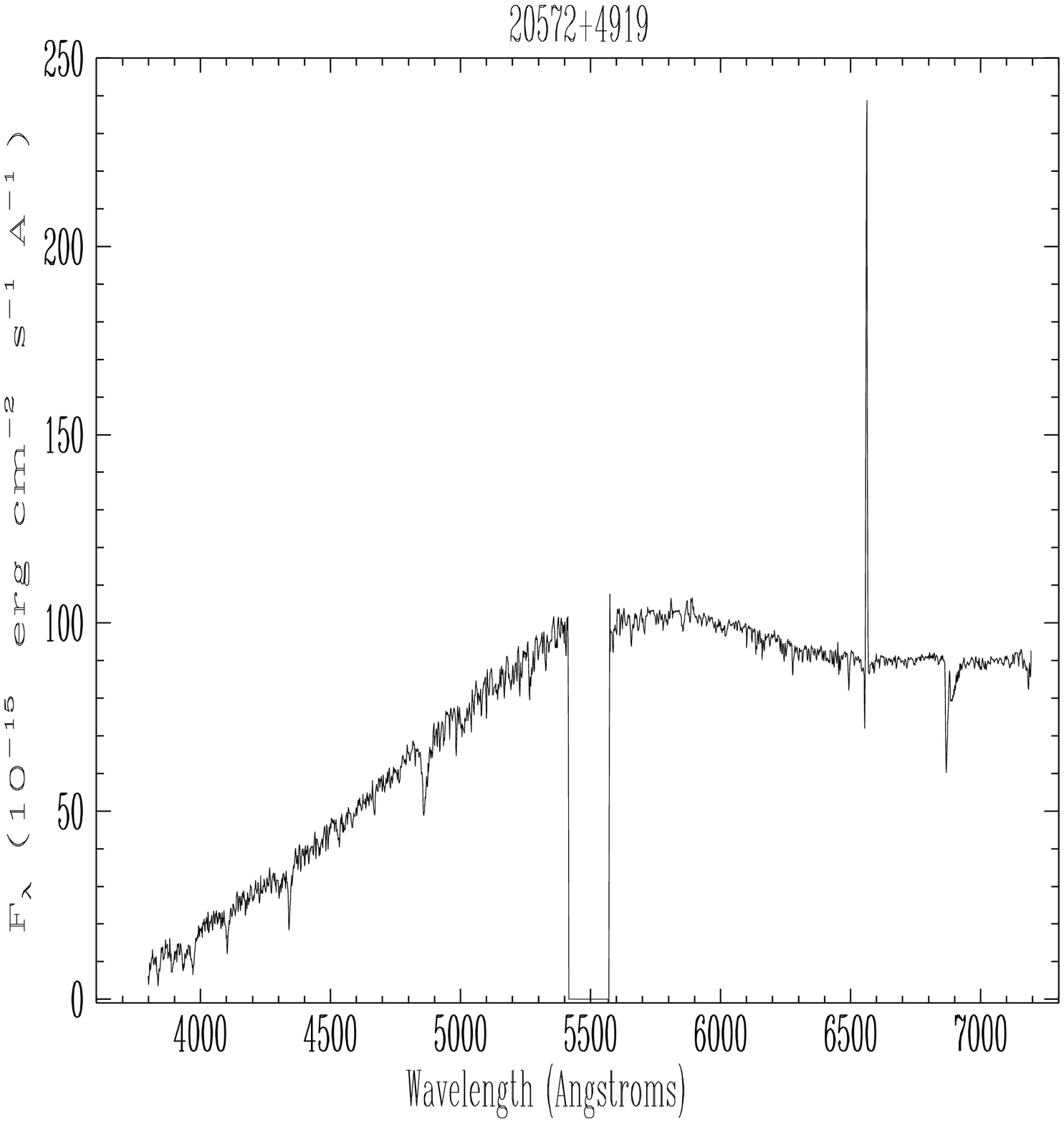}
%\psdraft
\epsfxsize=4cm
\epsfysize=4cm
\epsfbox{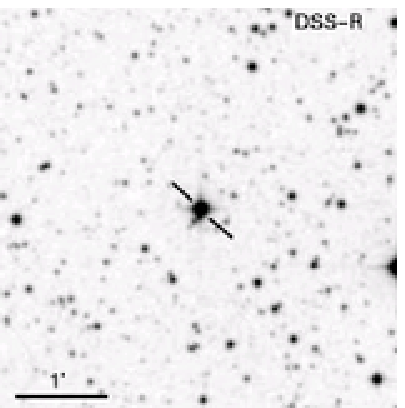}
%\psfull
\end{center}

\caption{Spectra of the objects classified as post-AGB in the sample together with their 
corresponding identification charts (continued). }
\end{figure*}

%-------------------------------------------------------------
%pg22
\setcounter{figure}{0}
\begin{figure*}

\begin{center}
\epsfxsize=13.5cm
\epsfysize=4cm
\epsfbox{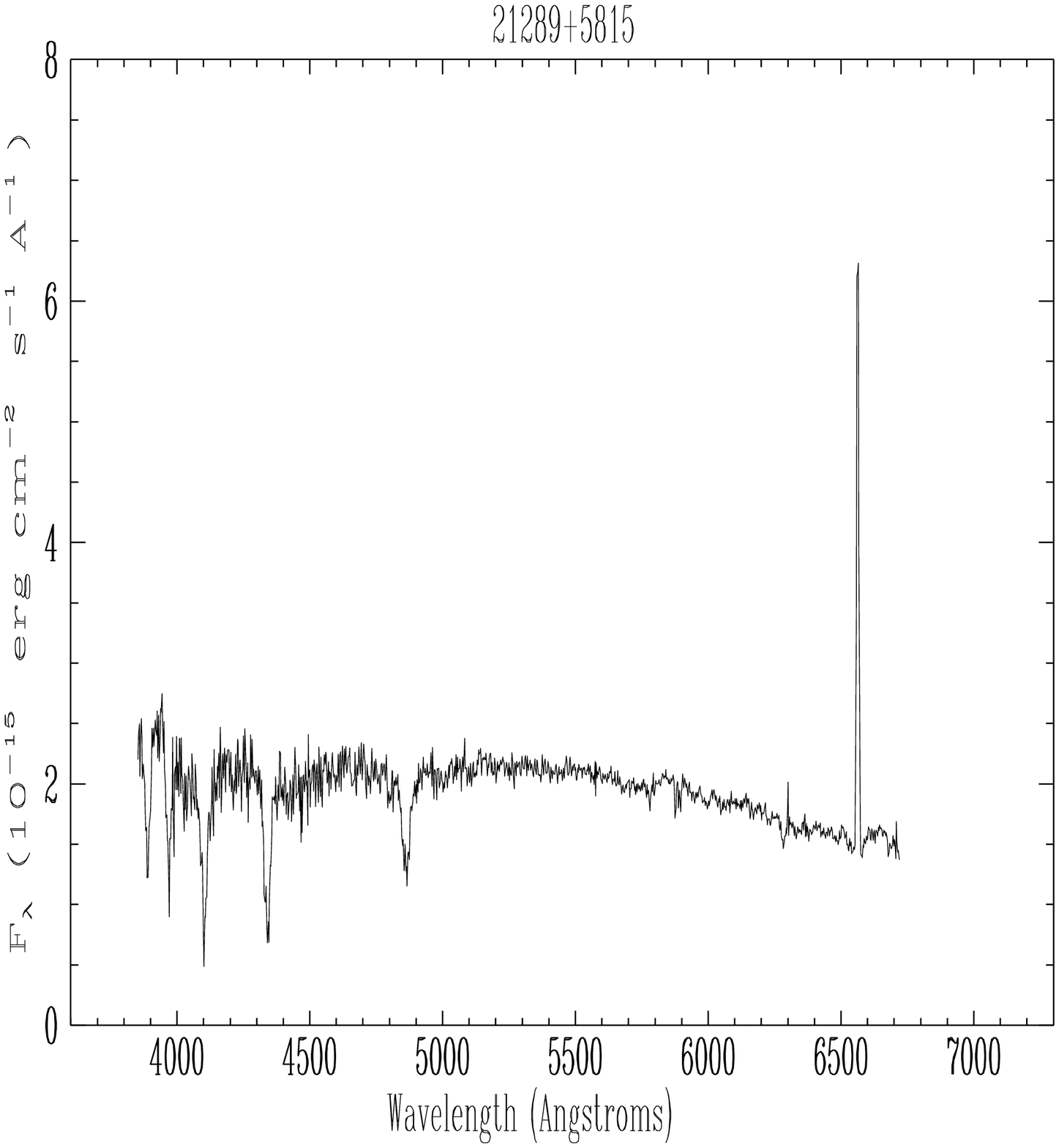}
%\psdraft
\epsfxsize=4cm
\epsfysize=4cm
\epsfbox{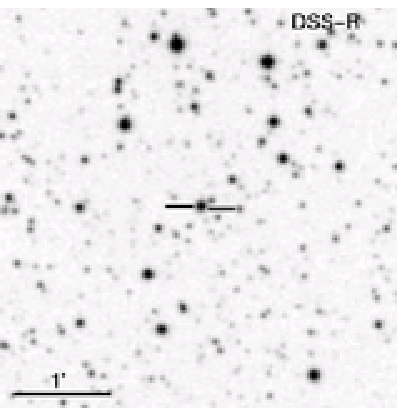}
%\psfull
\end{center}

\begin{center}
\epsfxsize=13.5cm
\epsfysize=4cm
\epsfbox{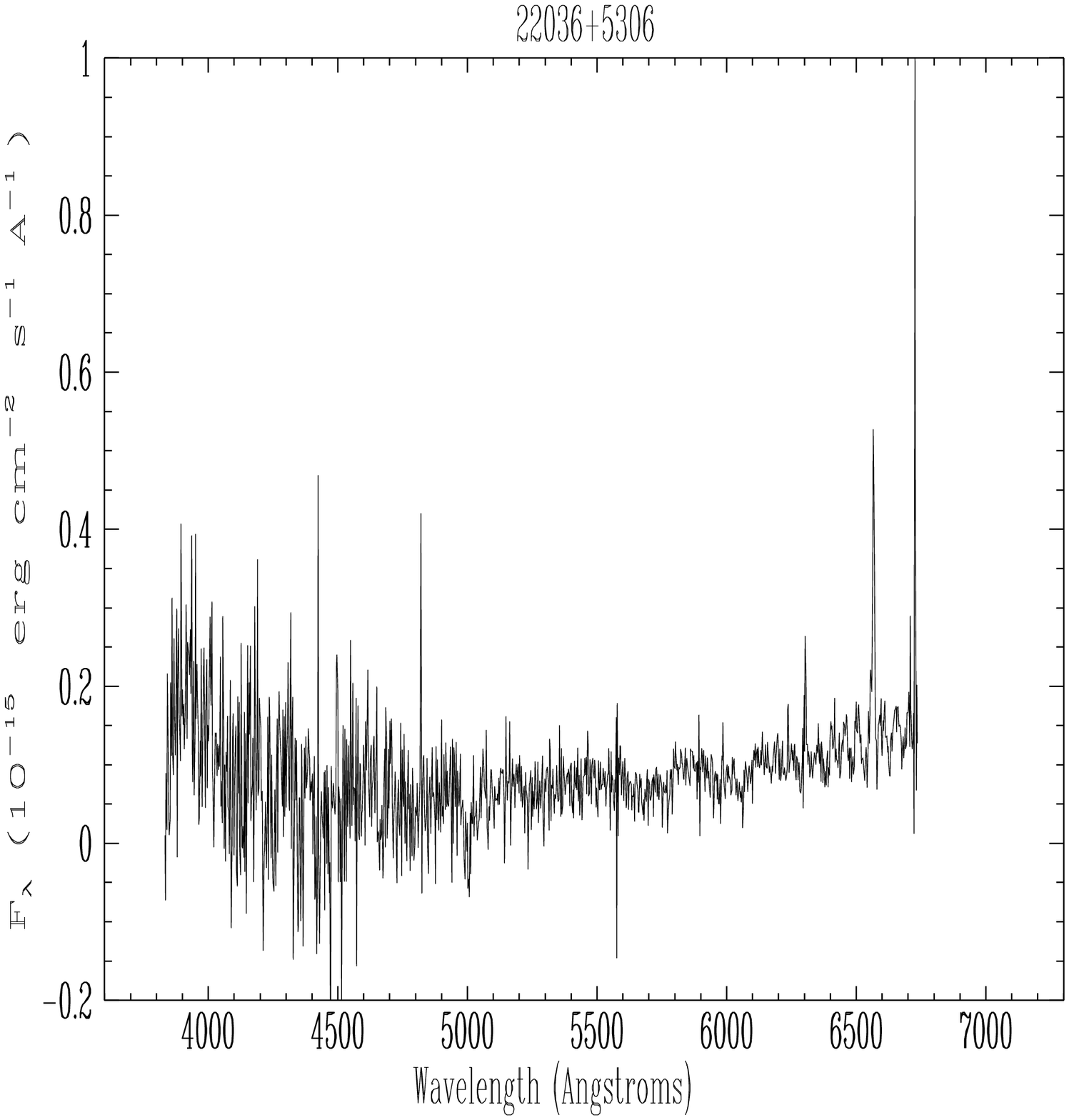}
%\psdraft
\epsfxsize=4cm
\epsfysize=4cm
\epsfbox{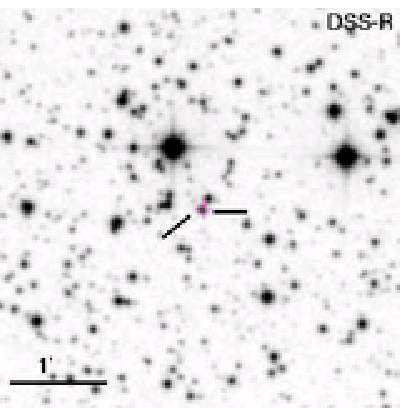}
%\psfull
\end{center}

\begin{center}
\epsfxsize=13.5cm
\epsfysize=4cm
\epsfbox{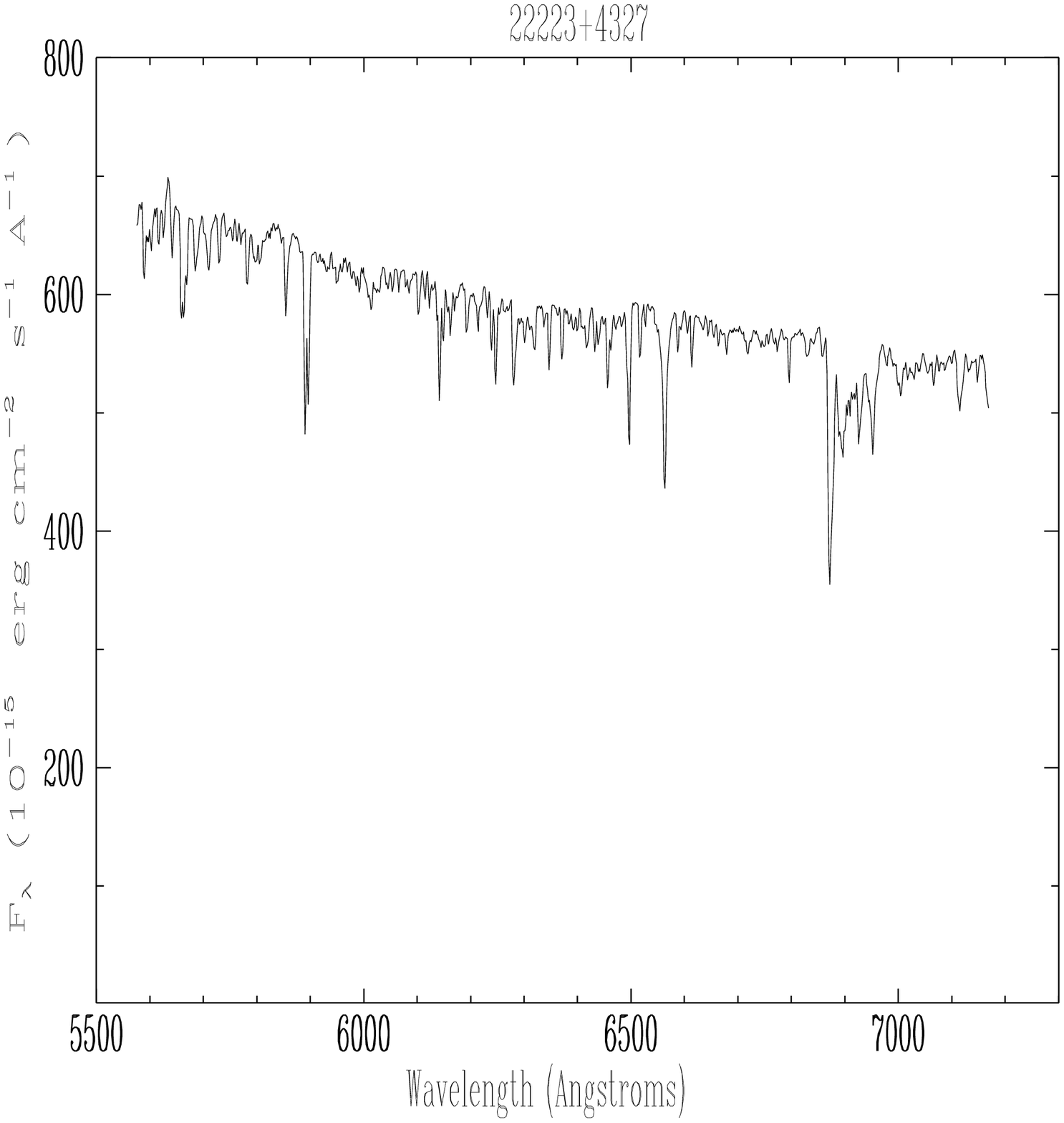}
%\psdraft
\epsfxsize=4cm
\epsfysize=4cm
\epsfbox{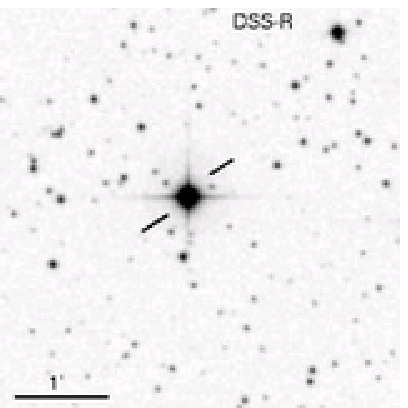}
%\psfull
\end{center}

\caption{Spectra of the objects classified as post-AGB in the sample together with their 
corresponding identification charts (continued). }
\end{figure*}

%%% Local Variables: 
%%% mode: latex
%%% TeX-master: "~/tesis/mitesis/final/tesis"
%%% End: 

\clearpage
\section{Atlas of transition sources}
        \begin{figure*}[h!]

\begin{center}
\epsfxsize=13.5cm
\epsfysize=4cm
\epsfbox{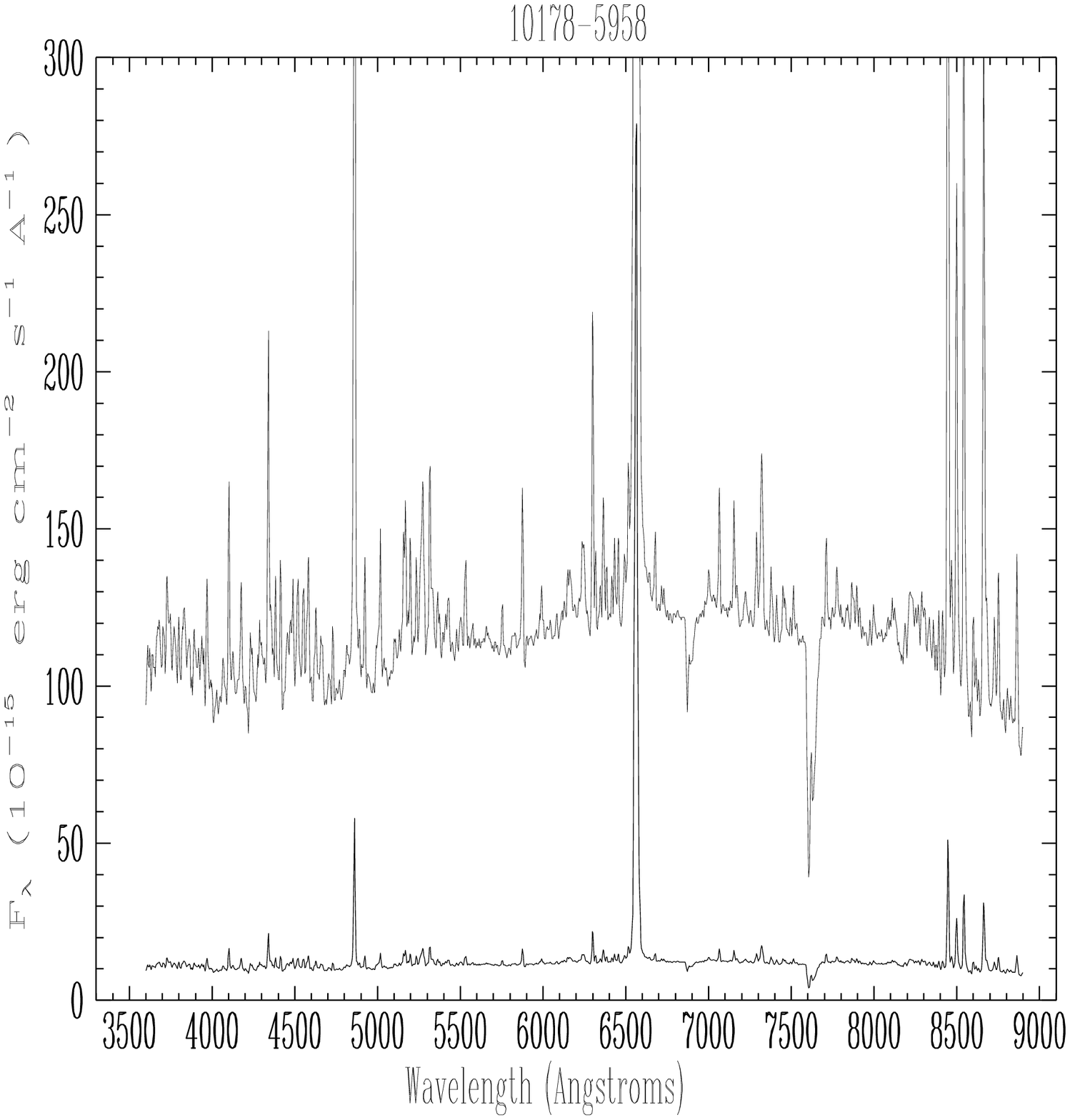}
%\psdraft
\epsfxsize=4cm
\epsfysize=4cm
\epsfbox{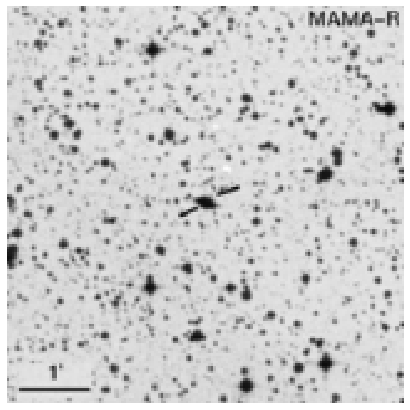}
%\psfull
\end{center}

\begin{center}
\epsfxsize=13.5cm
\epsfysize=4cm
\epsfbox{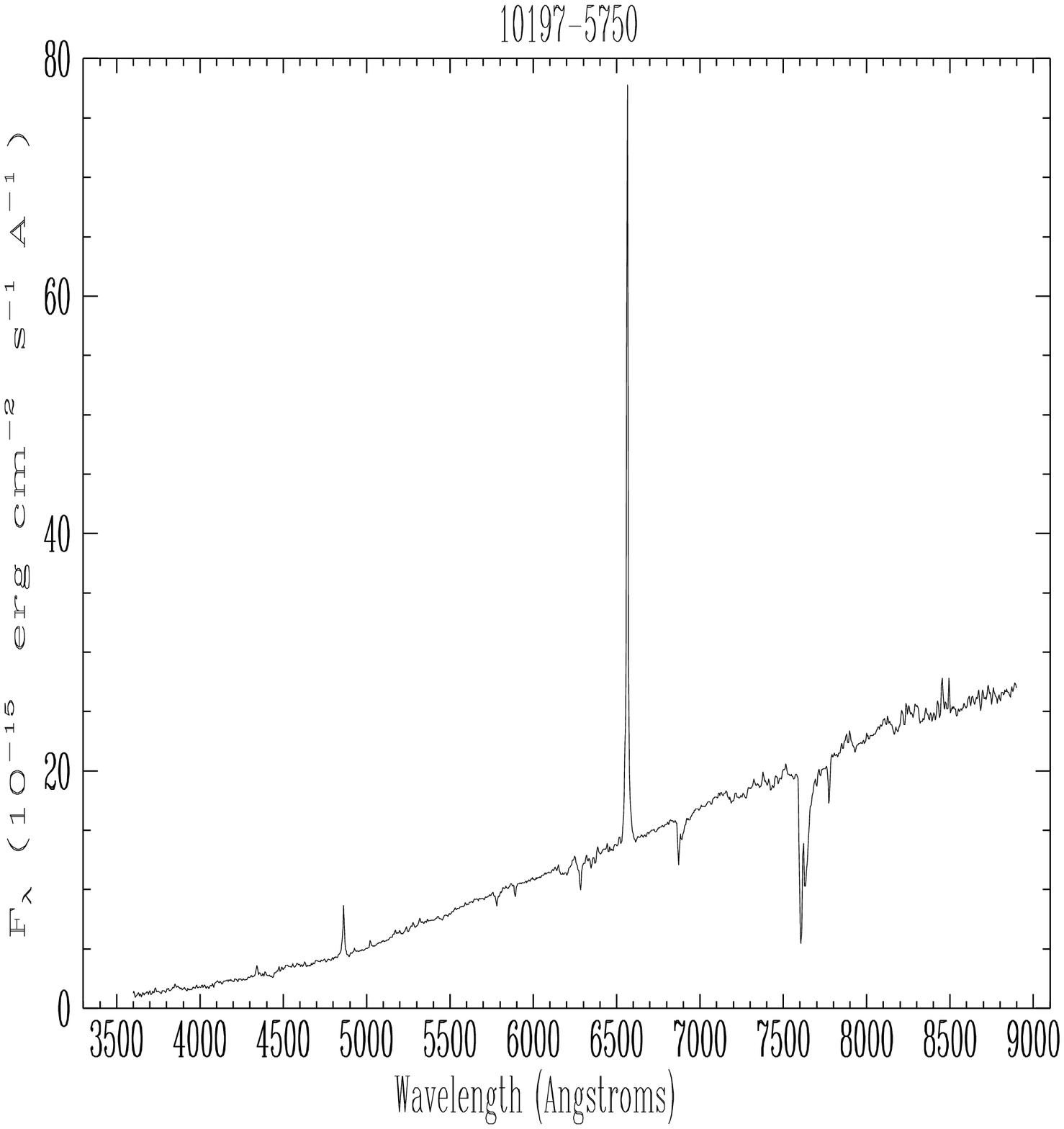}
%\psdraft
\epsfxsize=4cm
\epsfysize=4cm
\epsfbox{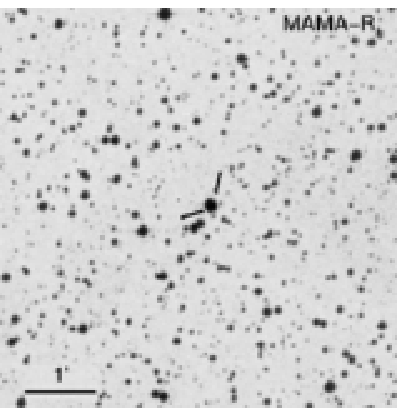}
%\psfull
\end{center}

\begin{center}
\epsfxsize=13.5cm
\epsfysize=4cm
\epsfbox{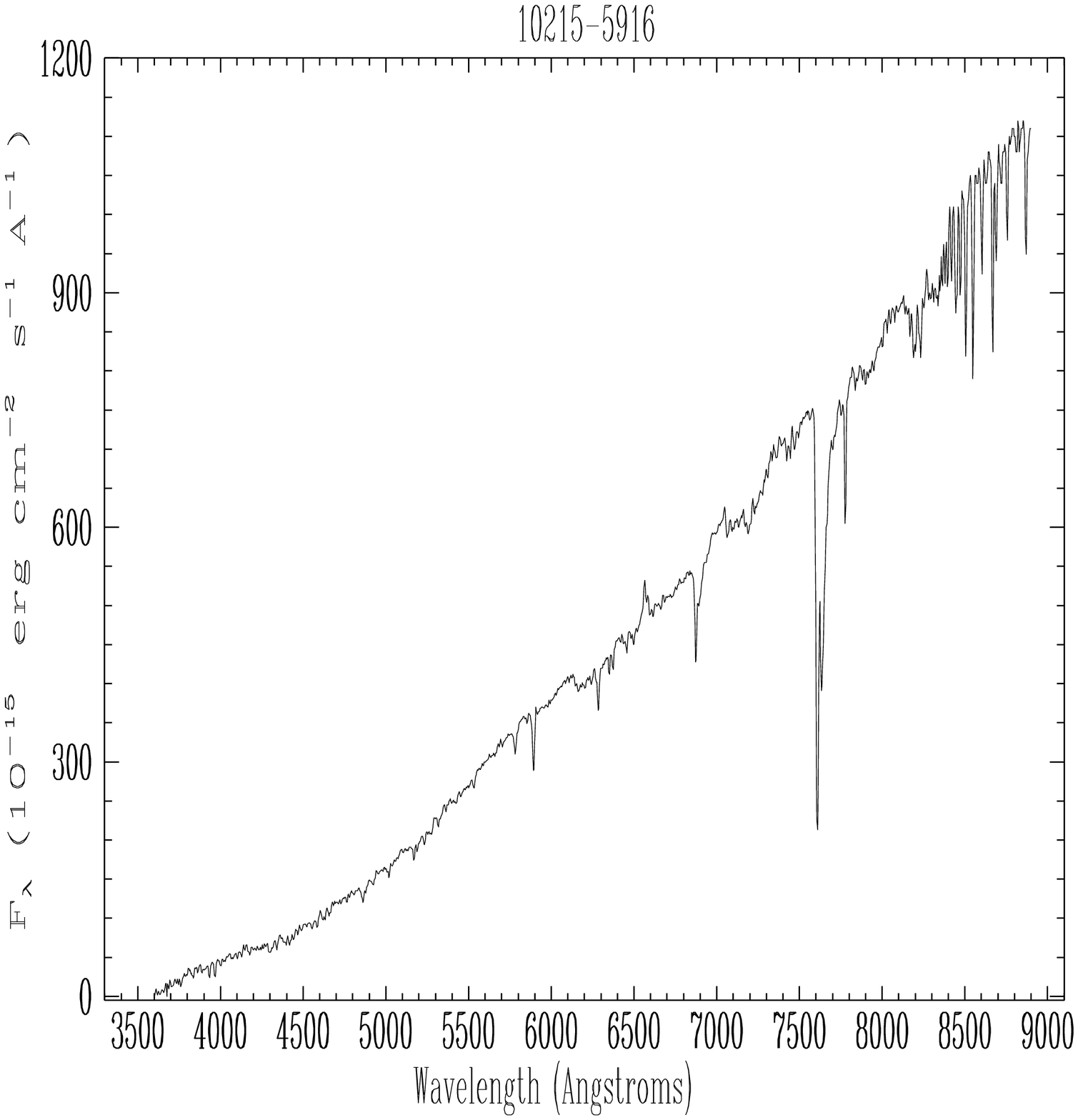}
%\psdraft
\epsfxsize=4cm
\epsfysize=4cm
\epsfbox{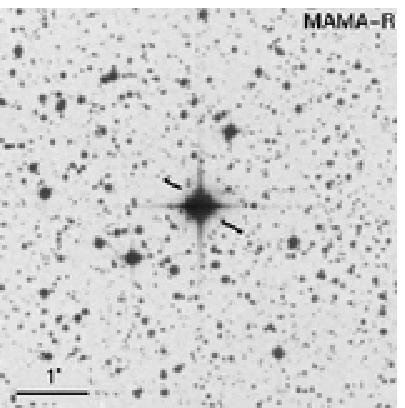}
%\psfull
\end{center}

\begin{center}
\epsfxsize=13.5cm
\epsfysize=4cm
\epsfbox{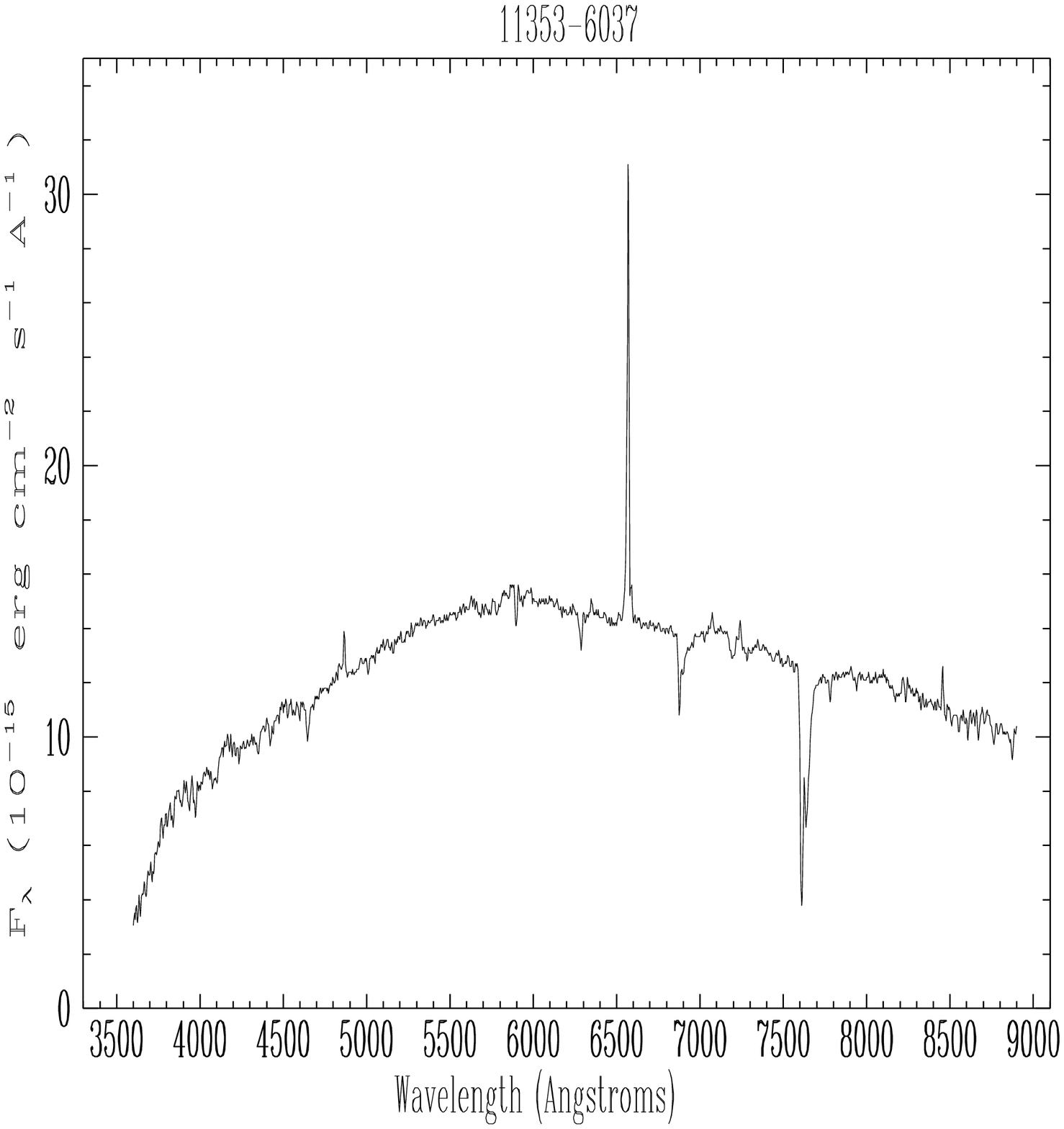}
%\psdraft
\epsfxsize=4cm
\epsfysize=4cm
\epsfbox{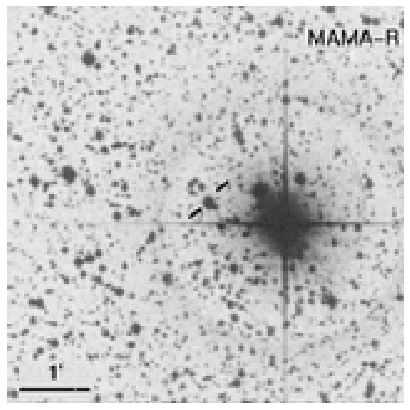}
%\psfull
\end{center}

\caption{Spectra of the transition objects together with their 
corresponding identification charts. }
\end{figure*}

%-------------------------------------------------------------------
%pg2.

\begin{figure*}
\setcounter{figure}{0}

\begin{center}
\epsfxsize=13.5cm
\epsfysize=4cm
\epsfbox{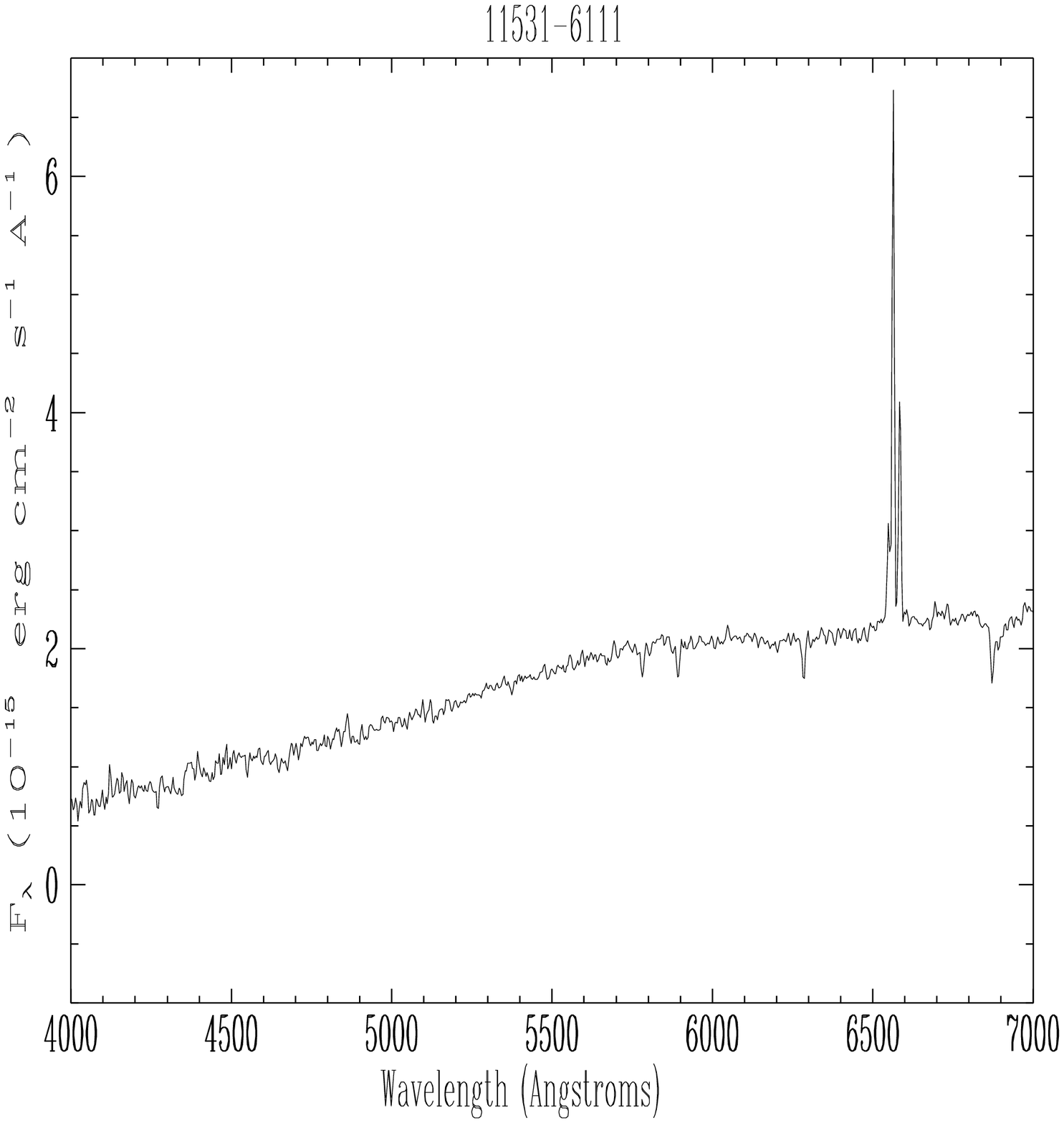}
%\psdraft
\epsfxsize=4cm
\epsfysize=4cm
\epsfbox{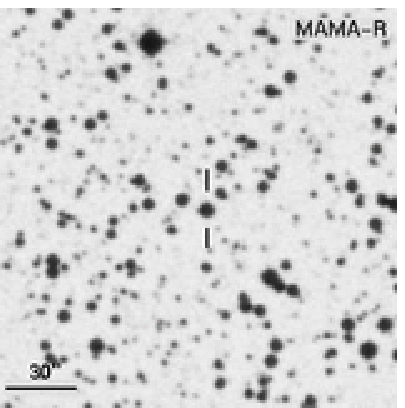}
%\psfull
\end{center}

\begin{center}
\epsfxsize=13.5cm
\epsfysize=4cm
\epsfbox{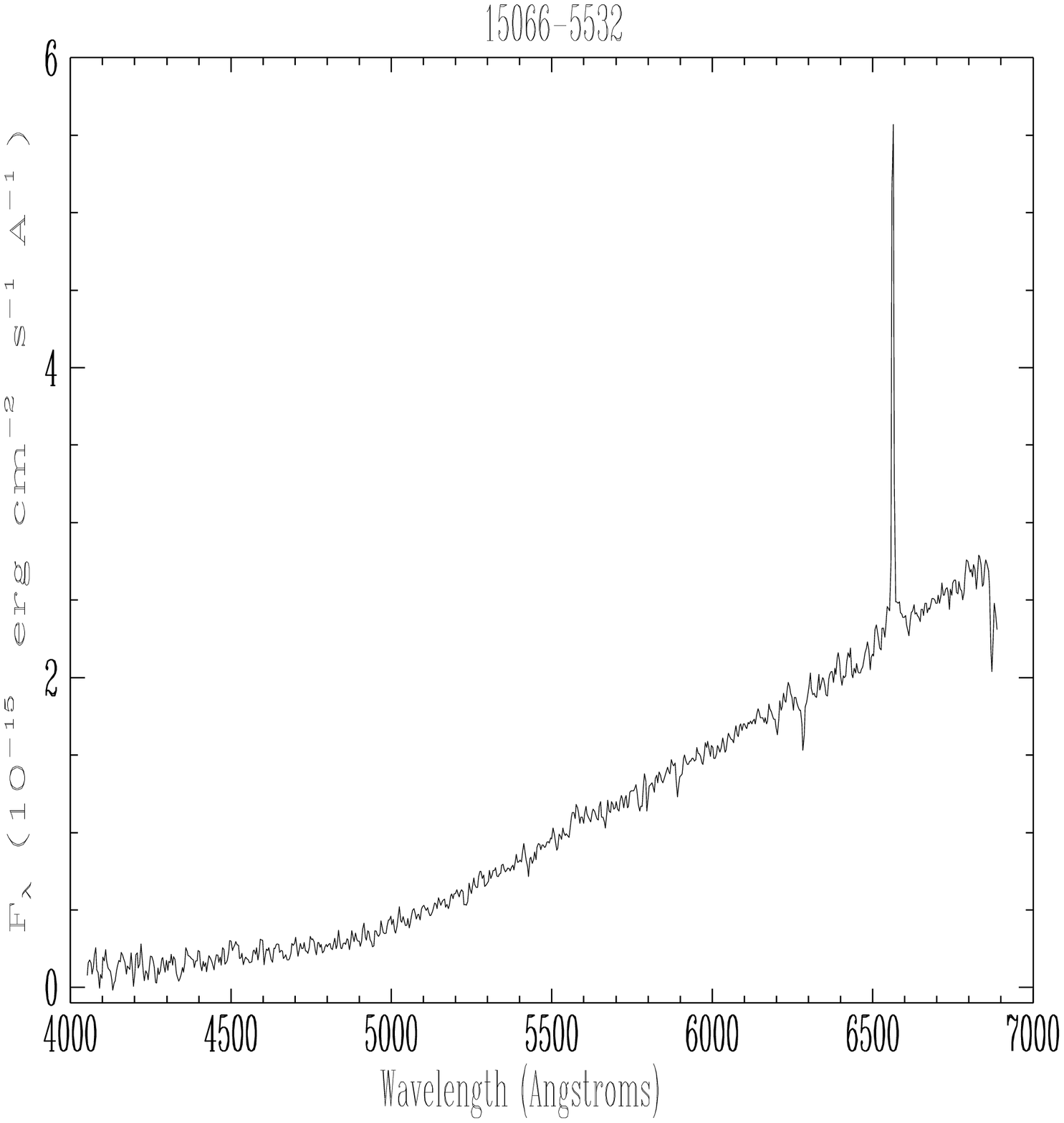}
%\psdraft
\epsfxsize=4cm
\epsfysize=4cm
\epsfbox{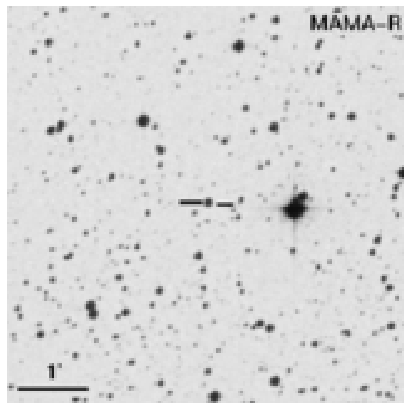}
%\psfull
\end{center}

\begin{center}
\epsfxsize=13.5cm
\epsfysize=4cm
\epsfbox{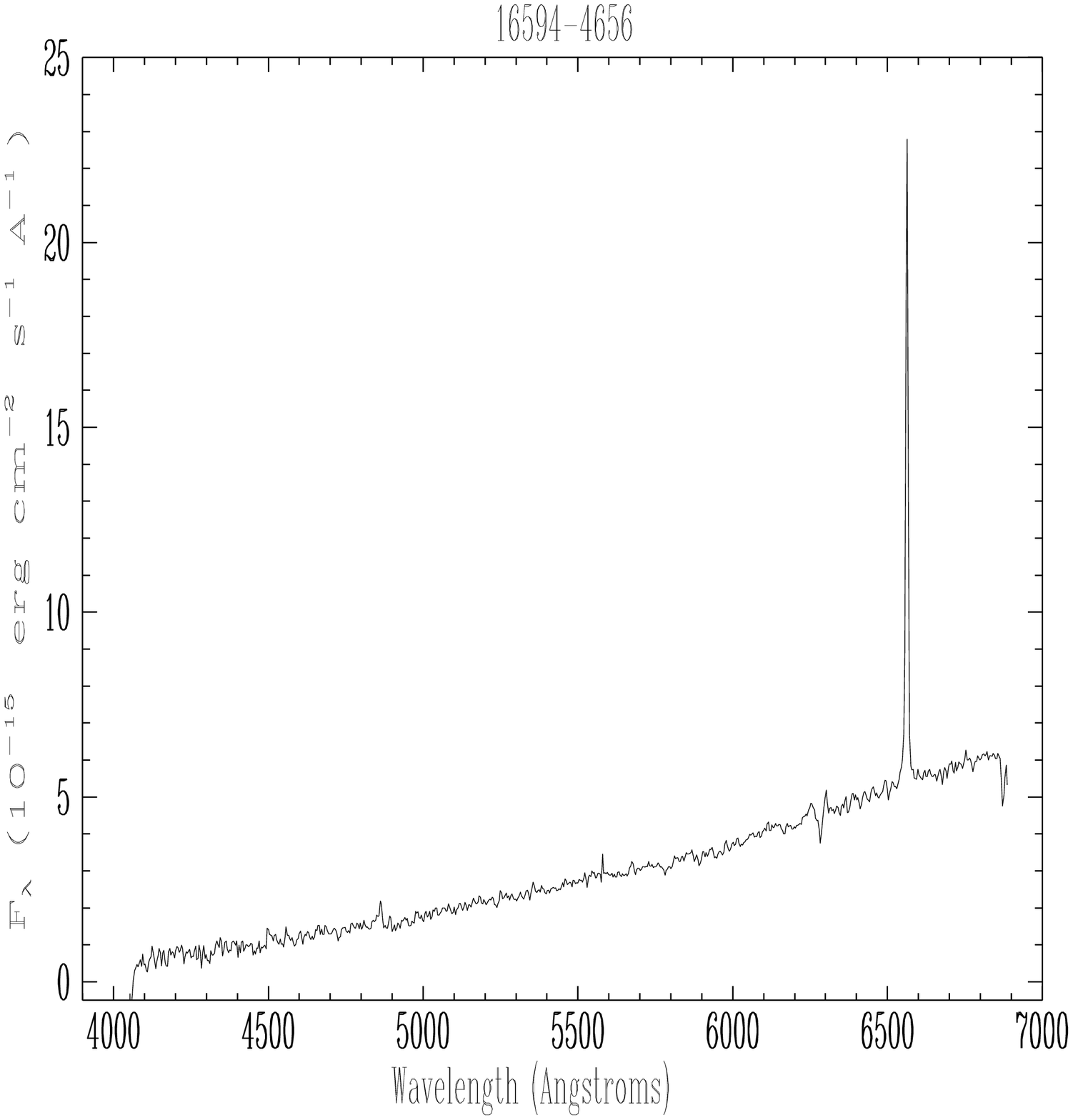}
%\psdraft
\epsfxsize=4cm
\epsfysize=4cm
\epsfbox{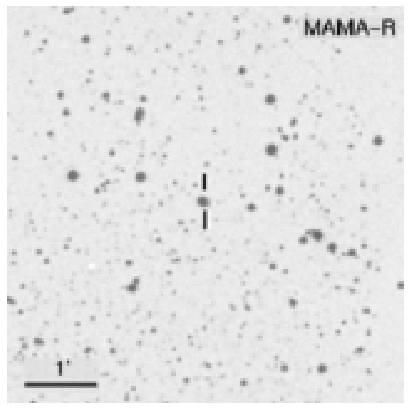}
%\psfull
\end{center}

\begin{center}
\epsfxsize=13.5cm
\epsfysize=4cm
\epsfbox{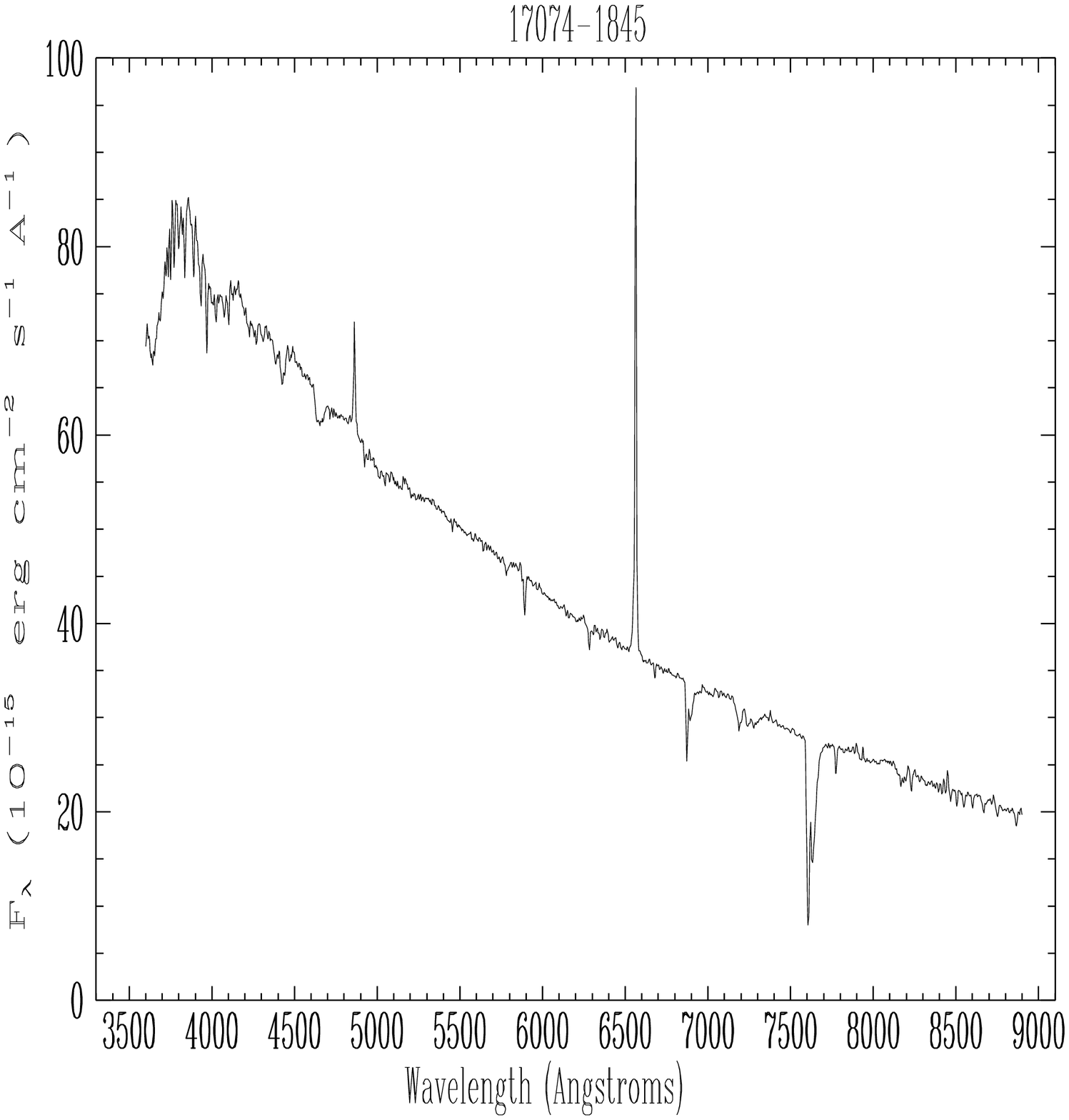}
%\psdraft
\epsfxsize=4cm
\epsfysize=4cm
\epsfbox{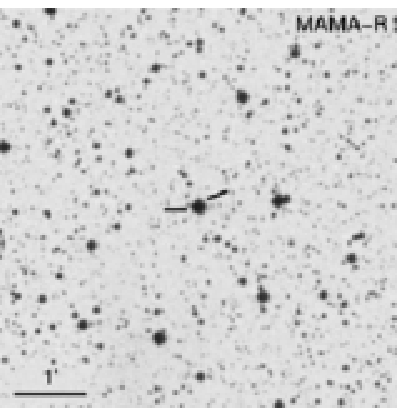}
%\psfull
\end{center}

\begin{center}
\epsfxsize=13.5cm
\epsfysize=4cm
\epsfbox{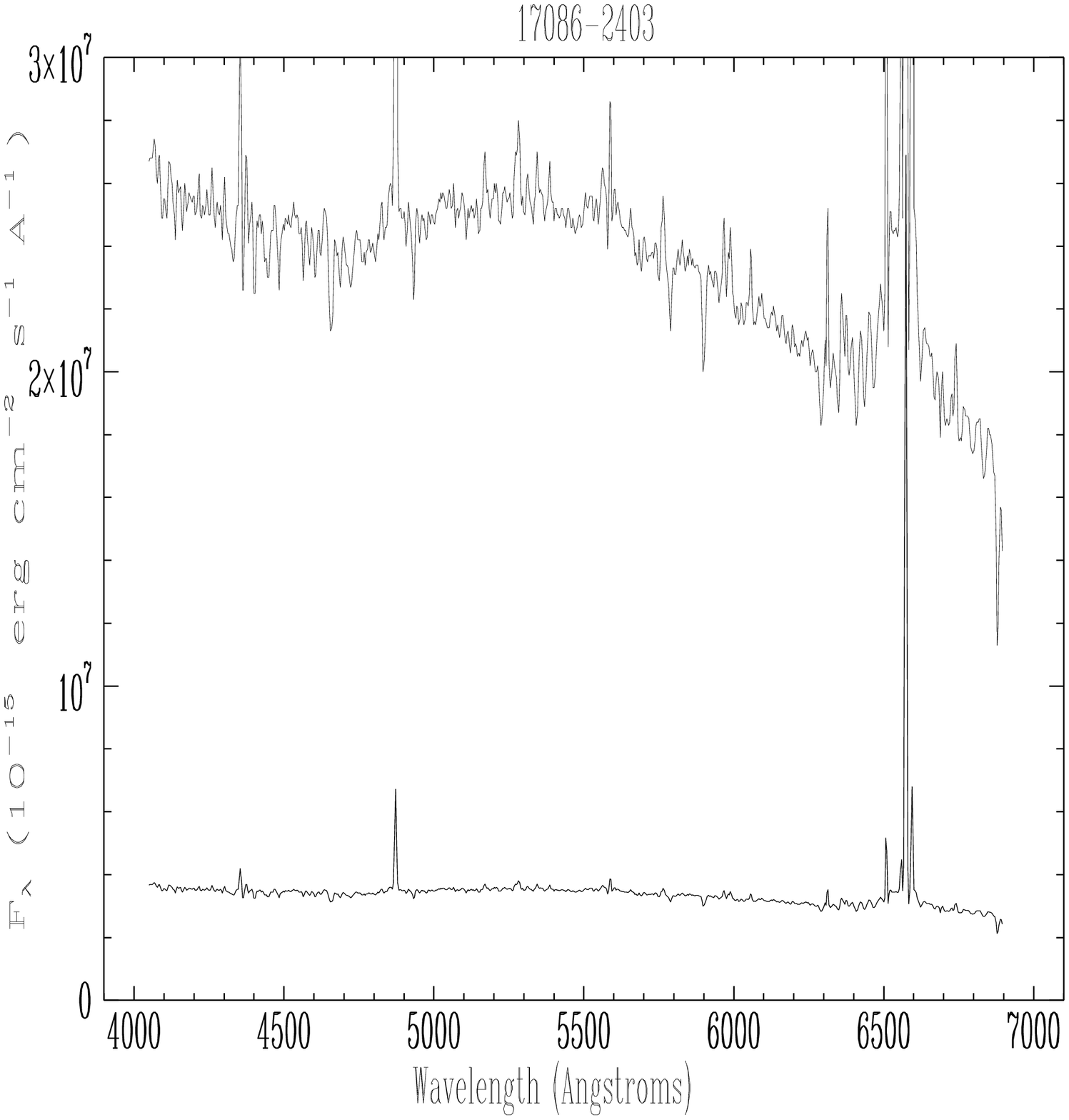}
%\psdraft
\epsfxsize=4cm
\epsfysize=4cm
\epsfbox{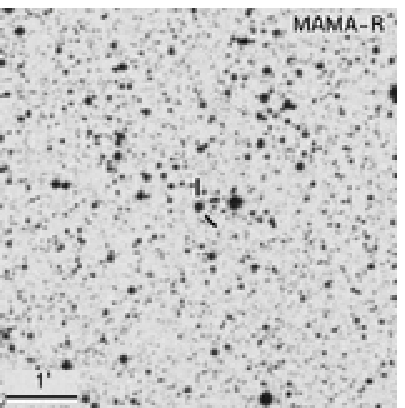}
%\psfull
\end{center}

\caption[]{Spectra of the transition objects together with their 
corresponding identification charts (continued). }
\end{figure*}

%-------------------------------------------------------------------
%pg3.

\begin{figure*}
\setcounter{figure}{0}

\begin{center}
\epsfxsize=13.5cm
\epsfysize=4cm
\epsfbox{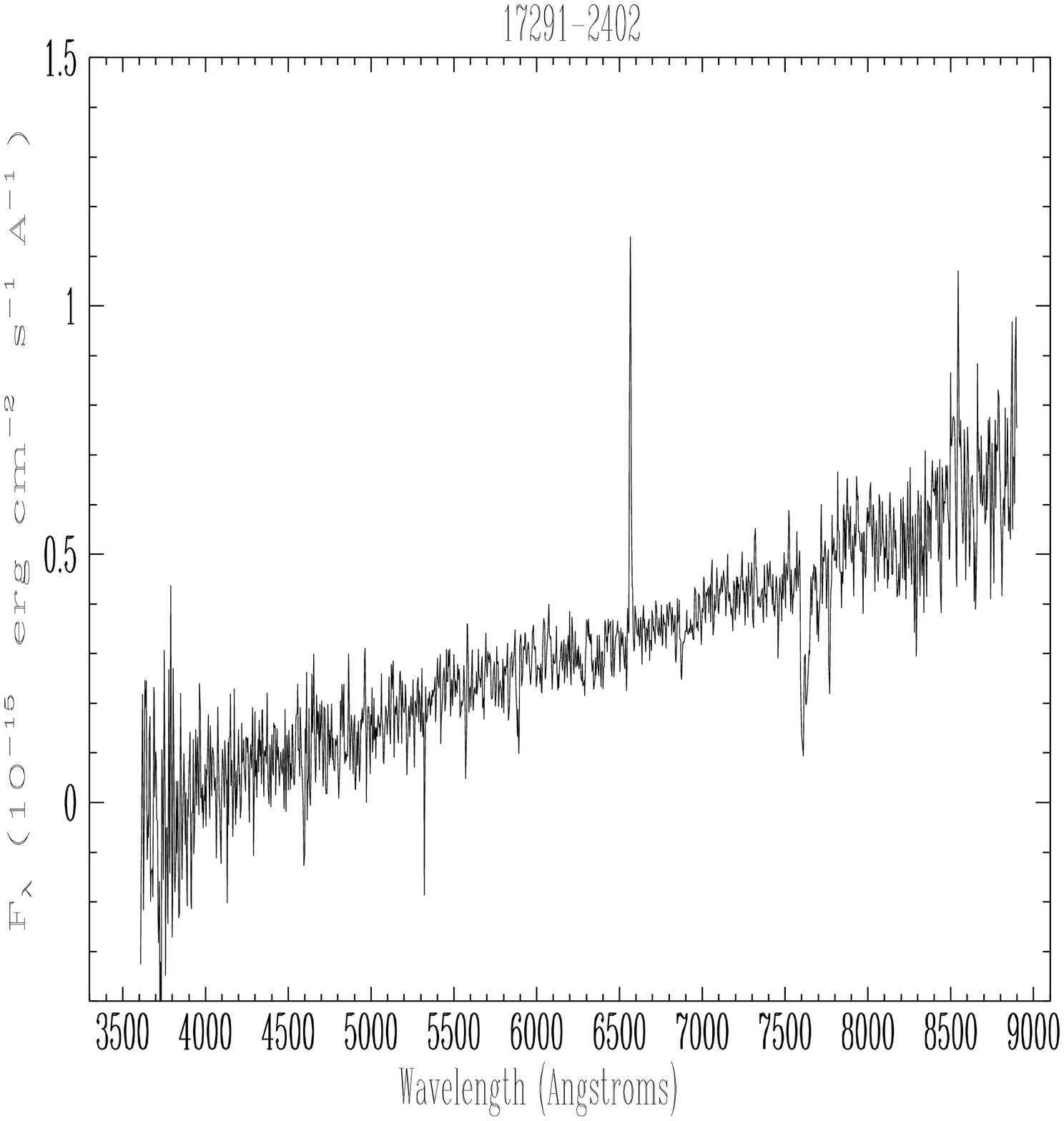}
%\psdraft
\epsfxsize=4cm
\epsfysize=4cm
\epsfbox{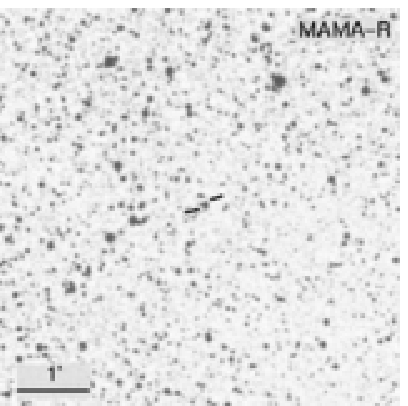}
%\psfull
\end{center}

\begin{center}
\epsfxsize=13.5cm
\epsfysize=4cm
\epsfbox{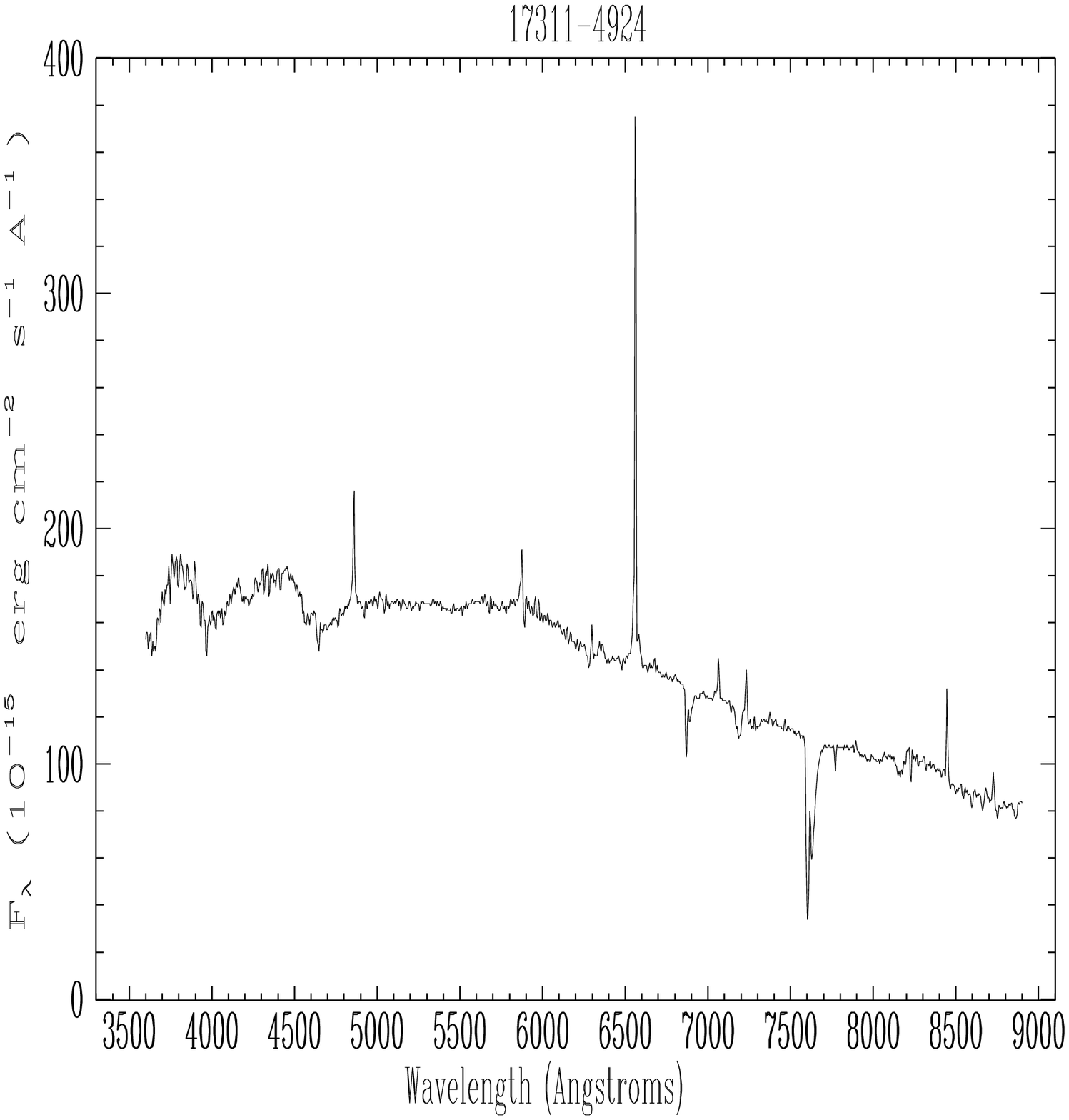}
%\psdraft
\epsfxsize=4cm
\epsfysize=4cm
\epsfbox{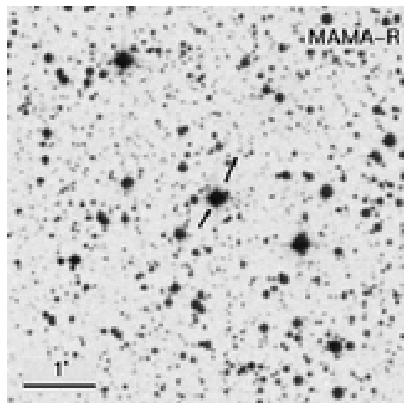}
%\psfull
\end{center}

\begin{center}
\epsfxsize=13.5cm
\epsfysize=4cm
\epsfbox{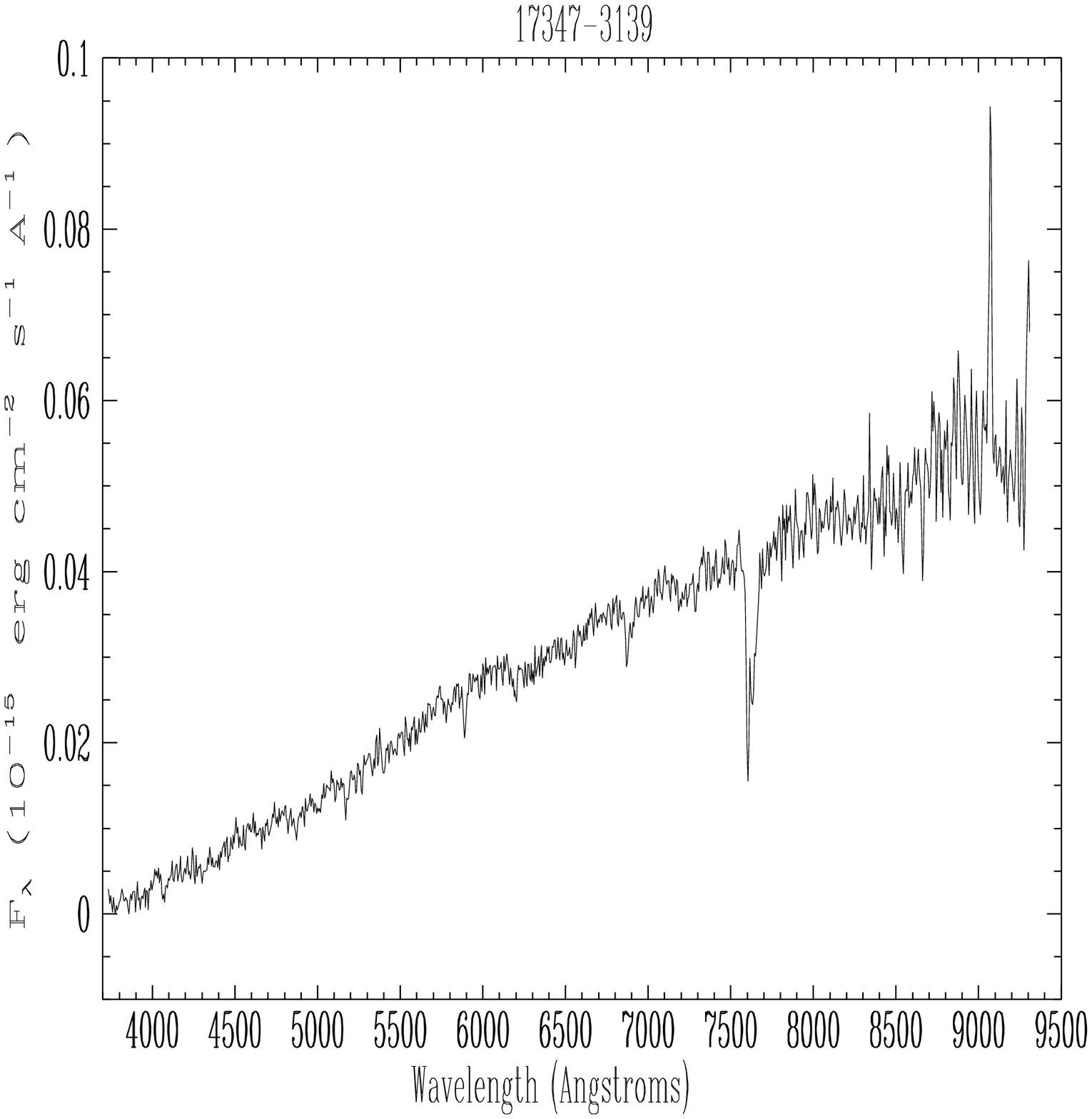}
%\psdraft
\epsfxsize=4cm
\epsfysize=4cm
\epsfbox{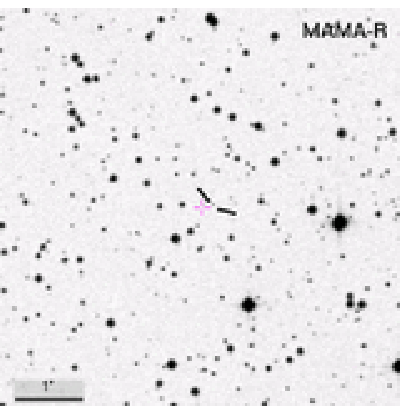}
%\psfull
\end{center}

\begin{center}
\epsfxsize=13.5cm
\epsfysize=4cm
\epsfbox{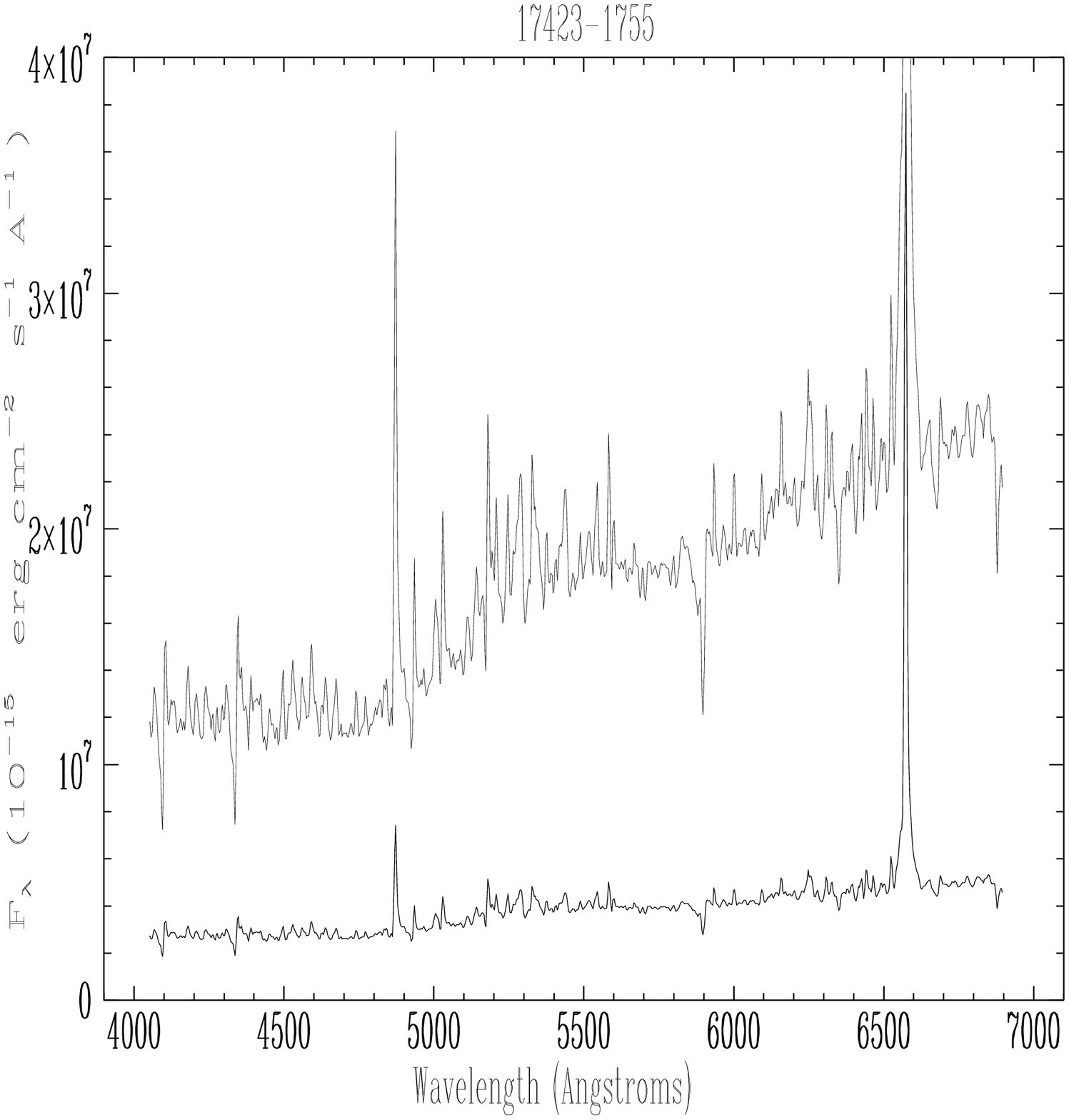}
%\psdraft
\epsfxsize=4cm
\epsfysize=4cm
\epsfbox{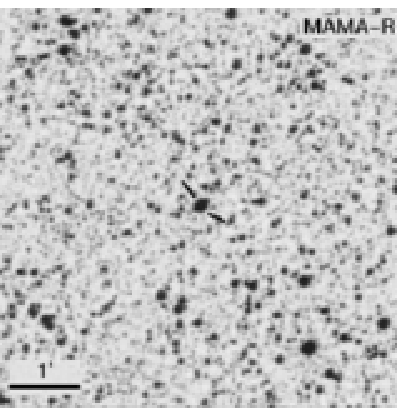}
%\psfull
\end{center}

\begin{center}
\epsfxsize=13.5cm
\epsfysize=4cm
\epsfbox{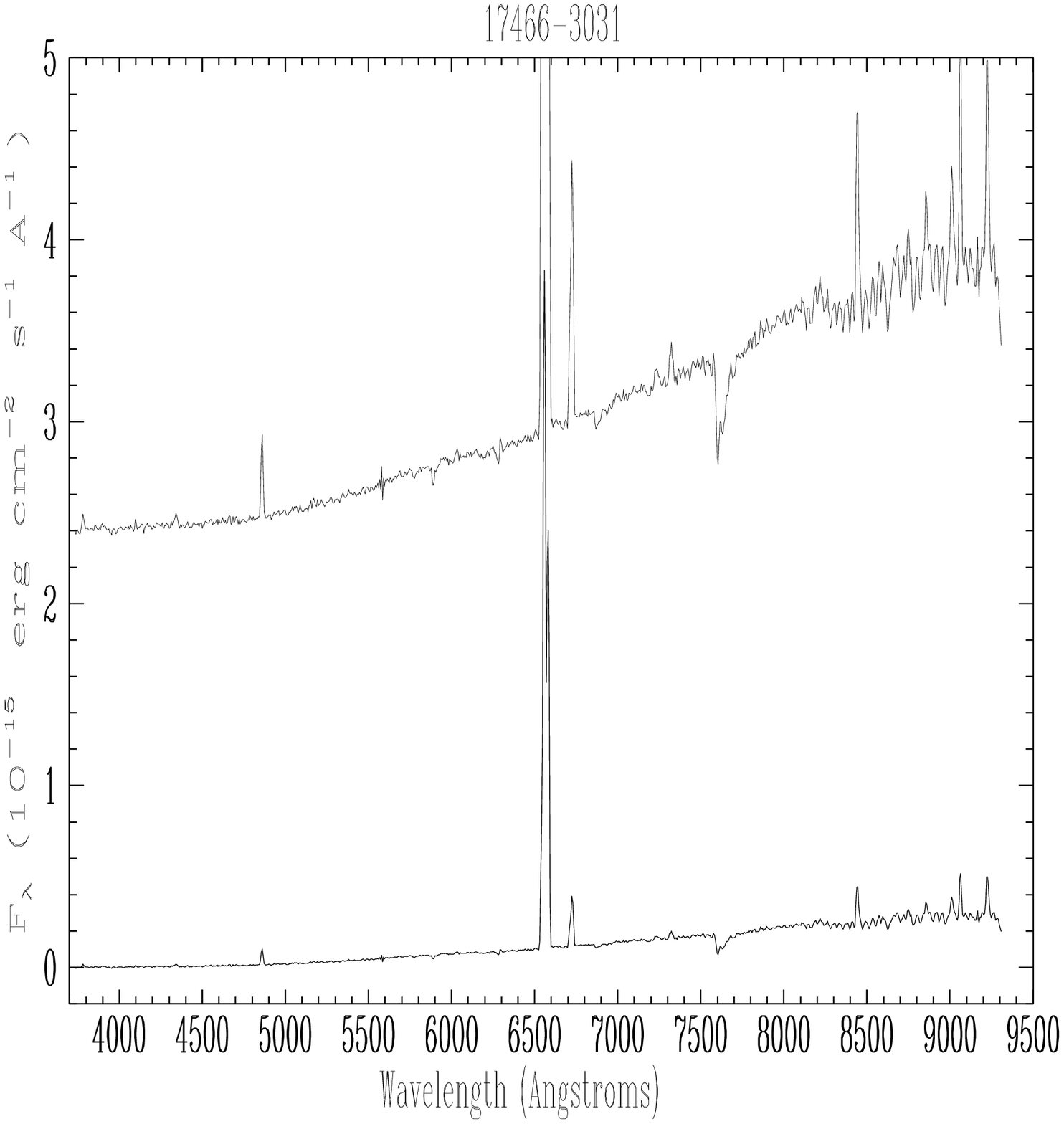}
%\psdraft
\epsfxsize=4cm
\epsfysize=4cm
\epsfbox{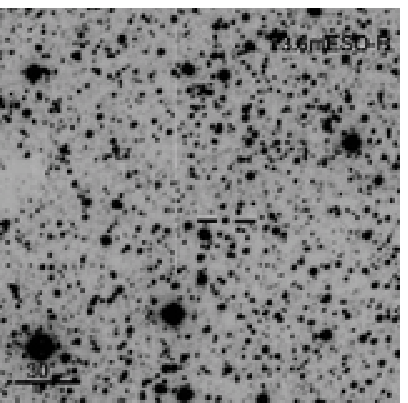}
%\psfull
\end{center}

\caption{Spectra of the transition objects together with their 
corresponding identification charts (continued). }
\end{figure*}

%-------------------------------------------------------------------
%pg4.

\begin{figure*}
\setcounter{figure}{0}

\begin{center}
\epsfxsize=13.5cm
\epsfysize=4cm
\epsfbox{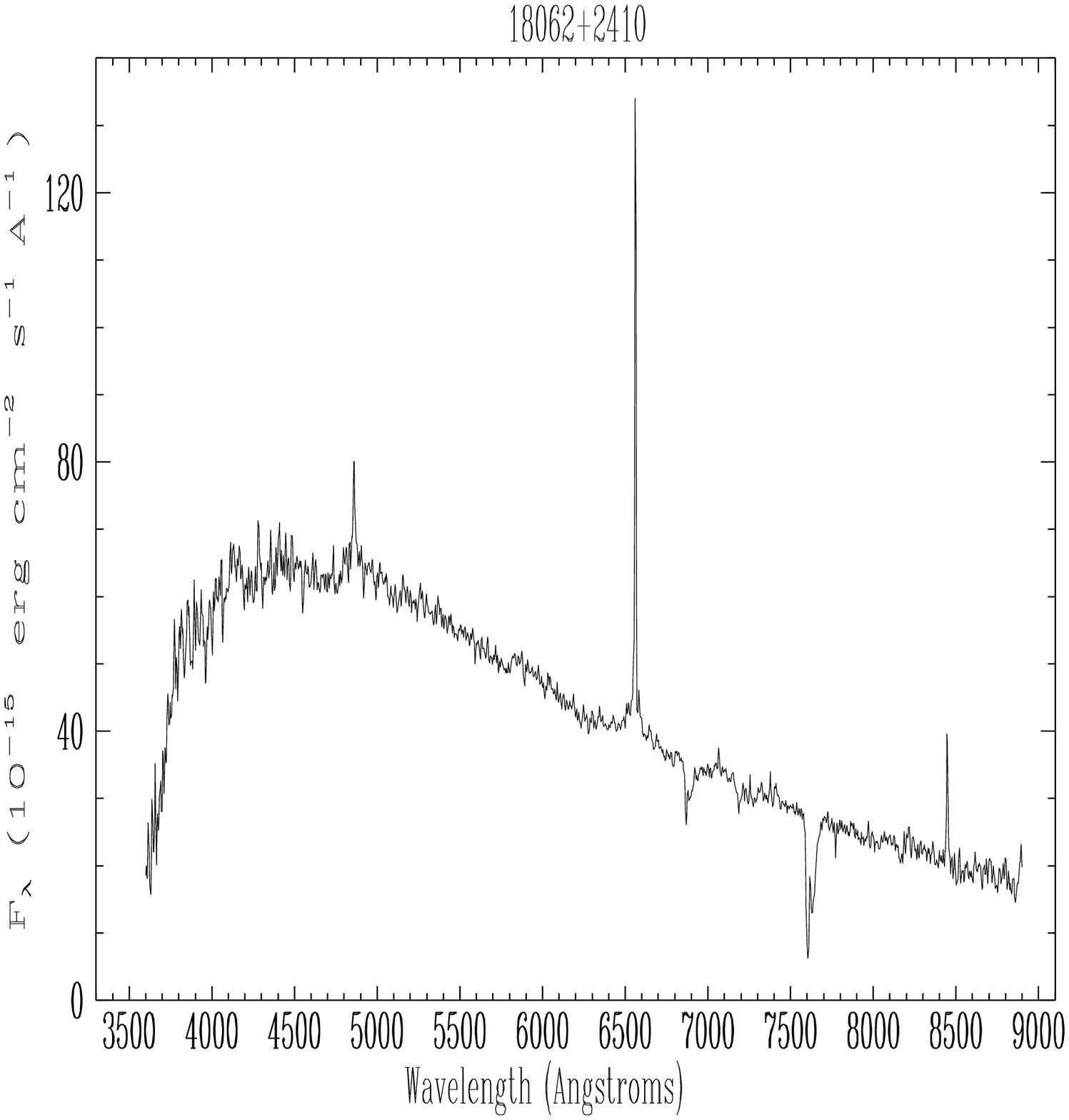}
%\psdraft
\epsfxsize=4cm
\epsfysize=4cm
\epsfbox{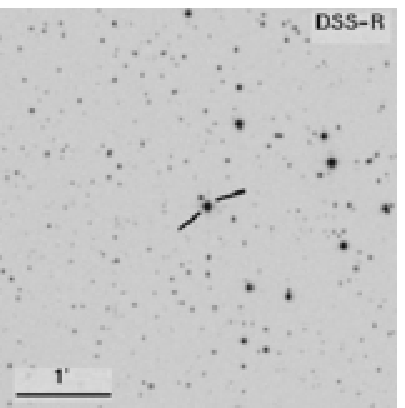}
%\psfull
\end{center}

\begin{center}
\epsfxsize=13.5cm
\epsfysize=4cm
\epsfbox{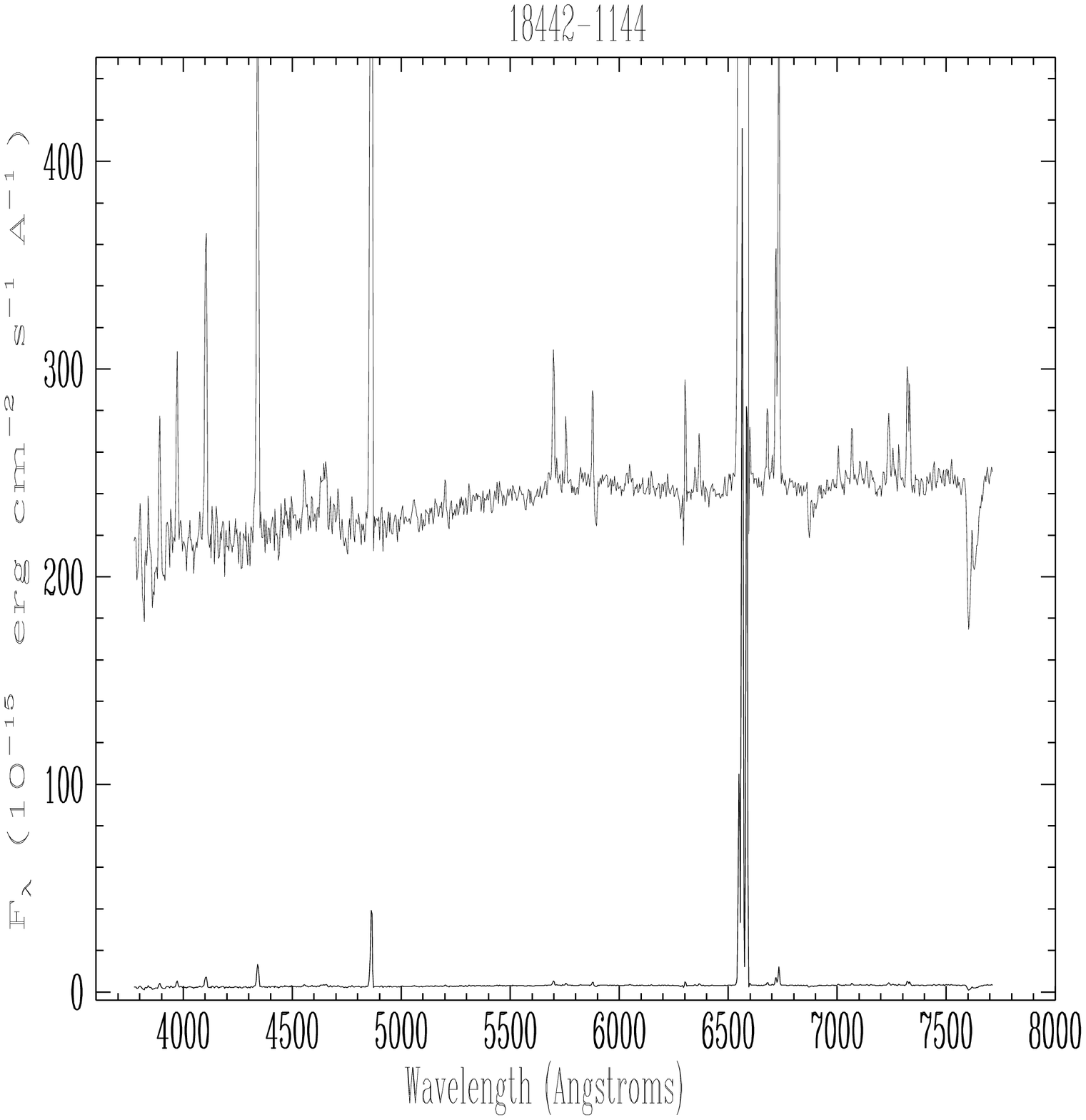}
%\psdraft
\epsfxsize=4cm
\epsfysize=4cm
\epsfbox{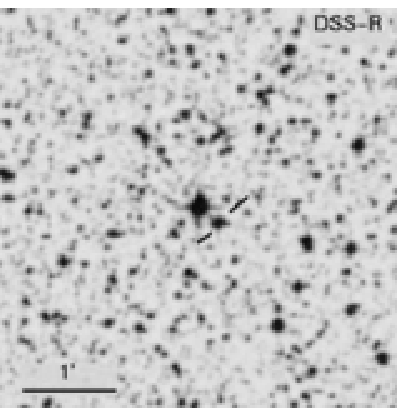}
%\psfull
\end{center}

\begin{center}
\epsfxsize=13.5cm
\epsfysize=4cm
\epsfbox{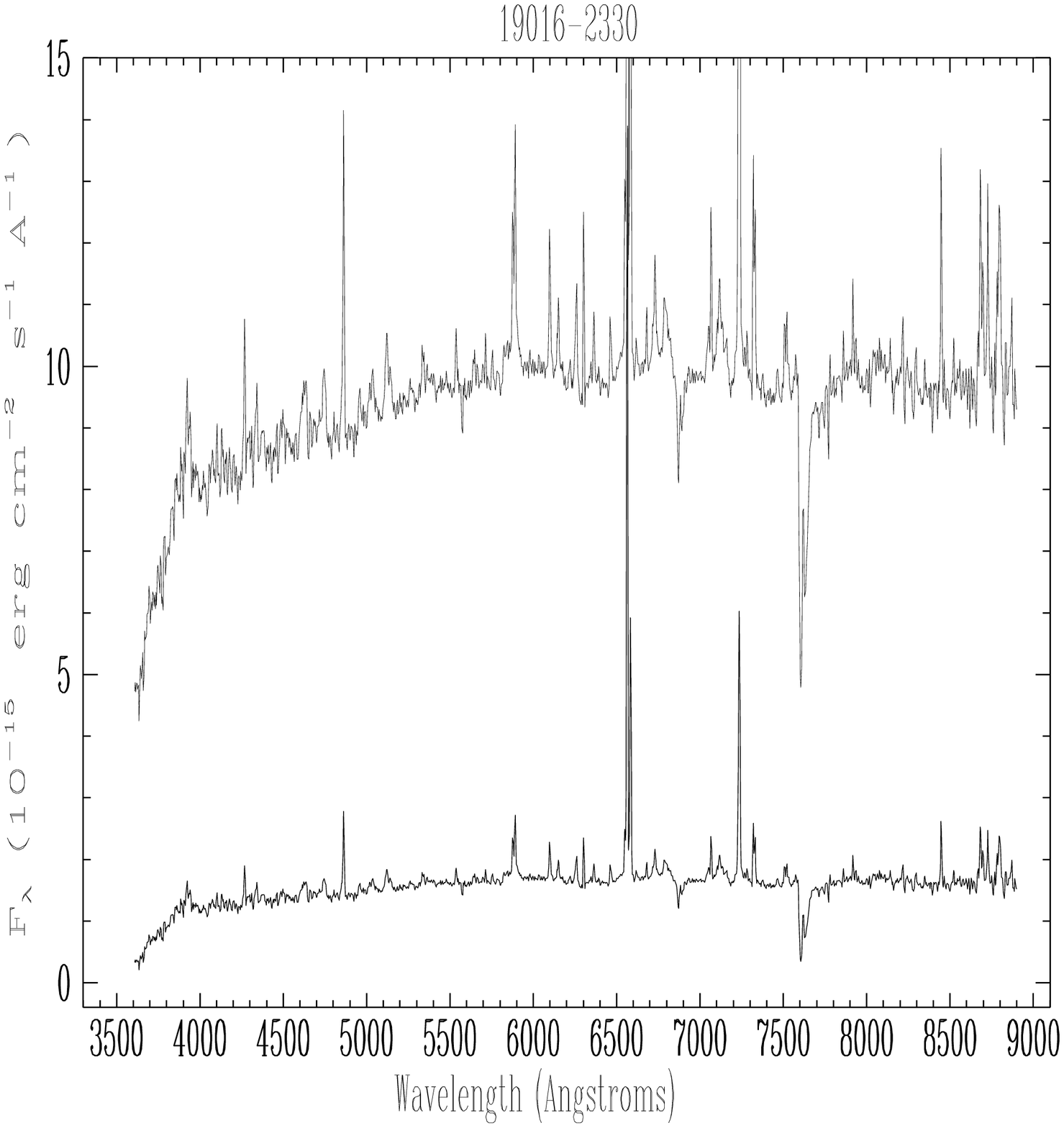}
%\psdraft
\epsfxsize=4cm
\epsfysize=4cm
\epsfbox{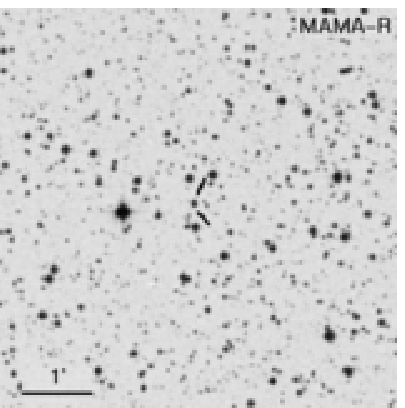}
%\psfull
\end{center}

\begin{center}
\epsfxsize=13.5cm
\epsfysize=4cm
\epsfbox{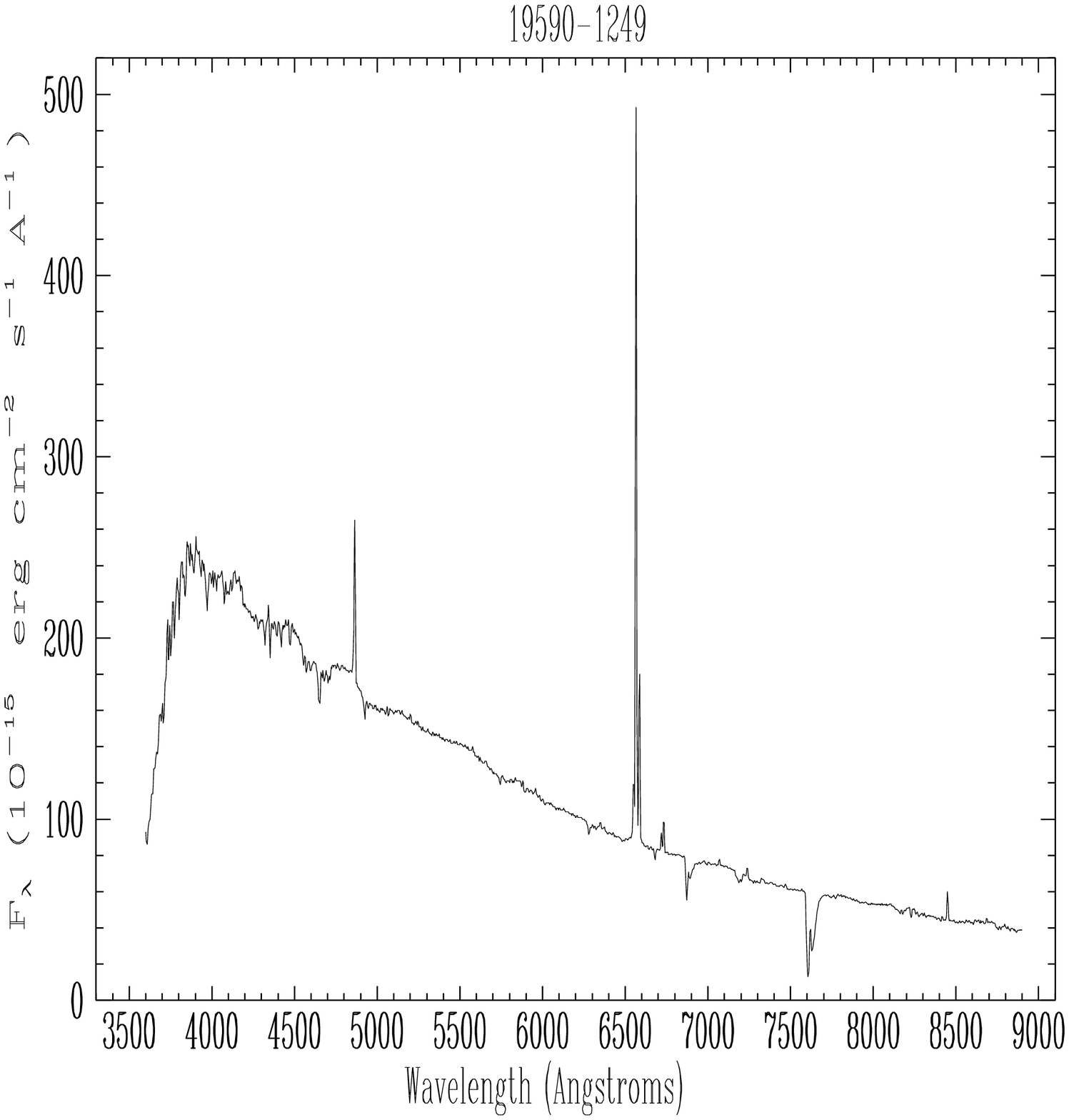}
%\psdraft
\epsfxsize=4cm
\epsfysize=4cm
\epsfbox{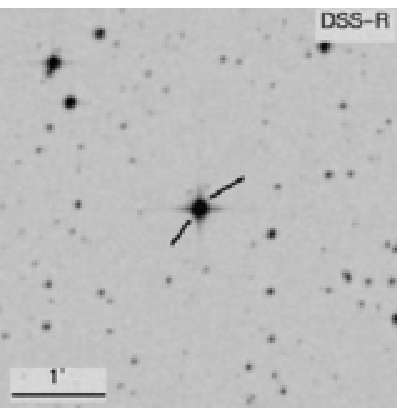}
%\psfull
\end{center}

\begin{center}
\epsfxsize=13.5cm
\epsfysize=4cm
\epsfbox{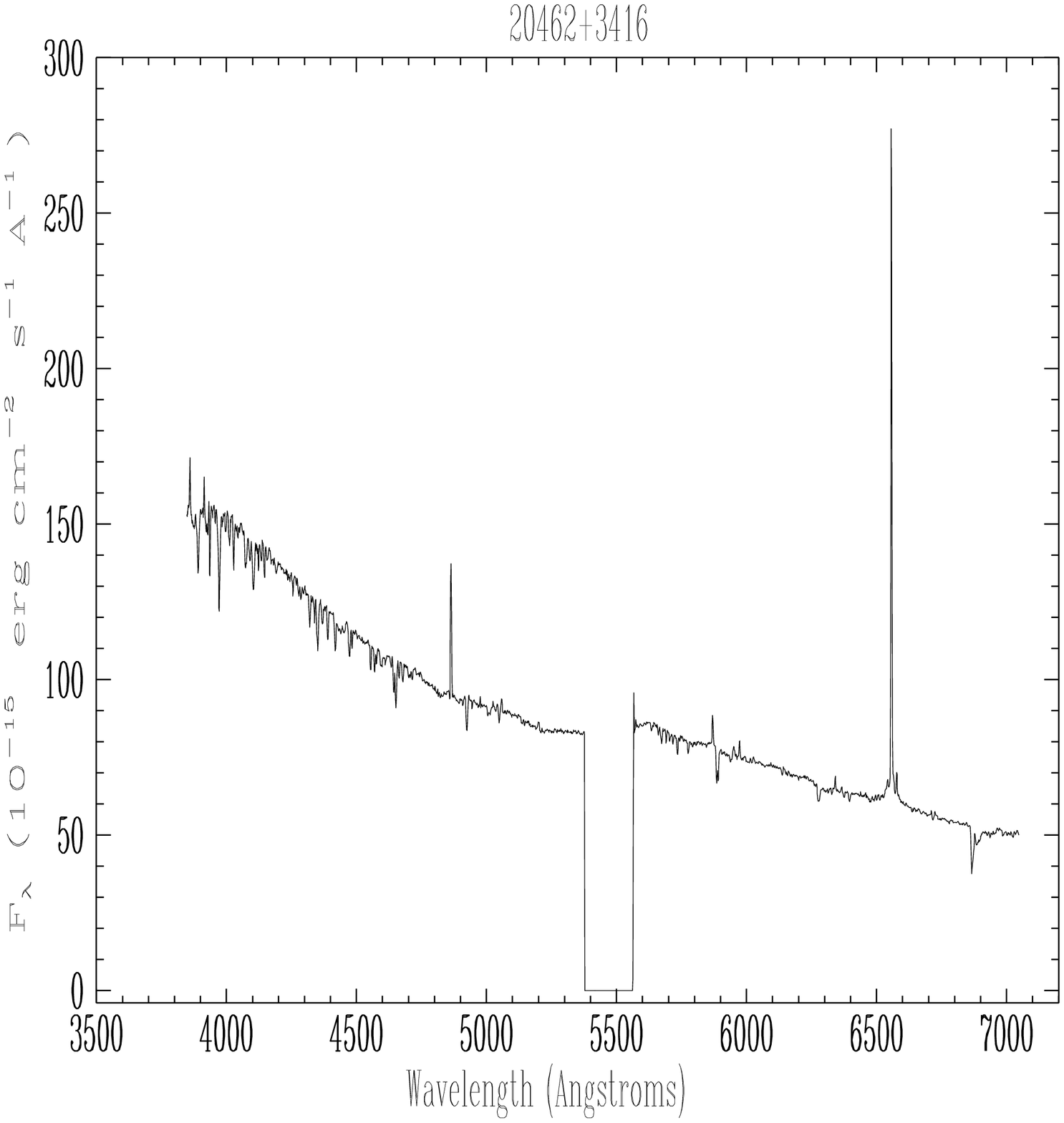}
%\psdraft
\epsfxsize=4cm
\epsfysize=4cm
\epsfbox{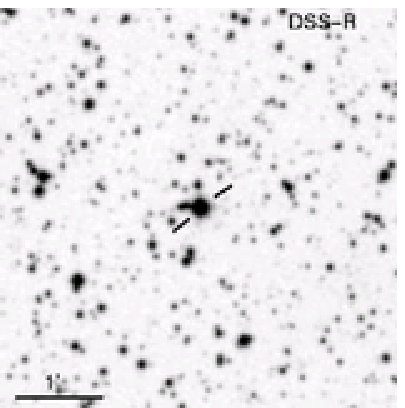}
%\psfull
\end{center}

\caption{Spectra of the transition objects together with their 
corresponding identification charts (continued). }
\end{figure*}

%-------------------------------------------------------------------
%pg4.

\begin{figure*}
\setcounter{figure}{0}

\begin{center}
\epsfxsize=13.5cm
\epsfysize=4cm
\epsfbox{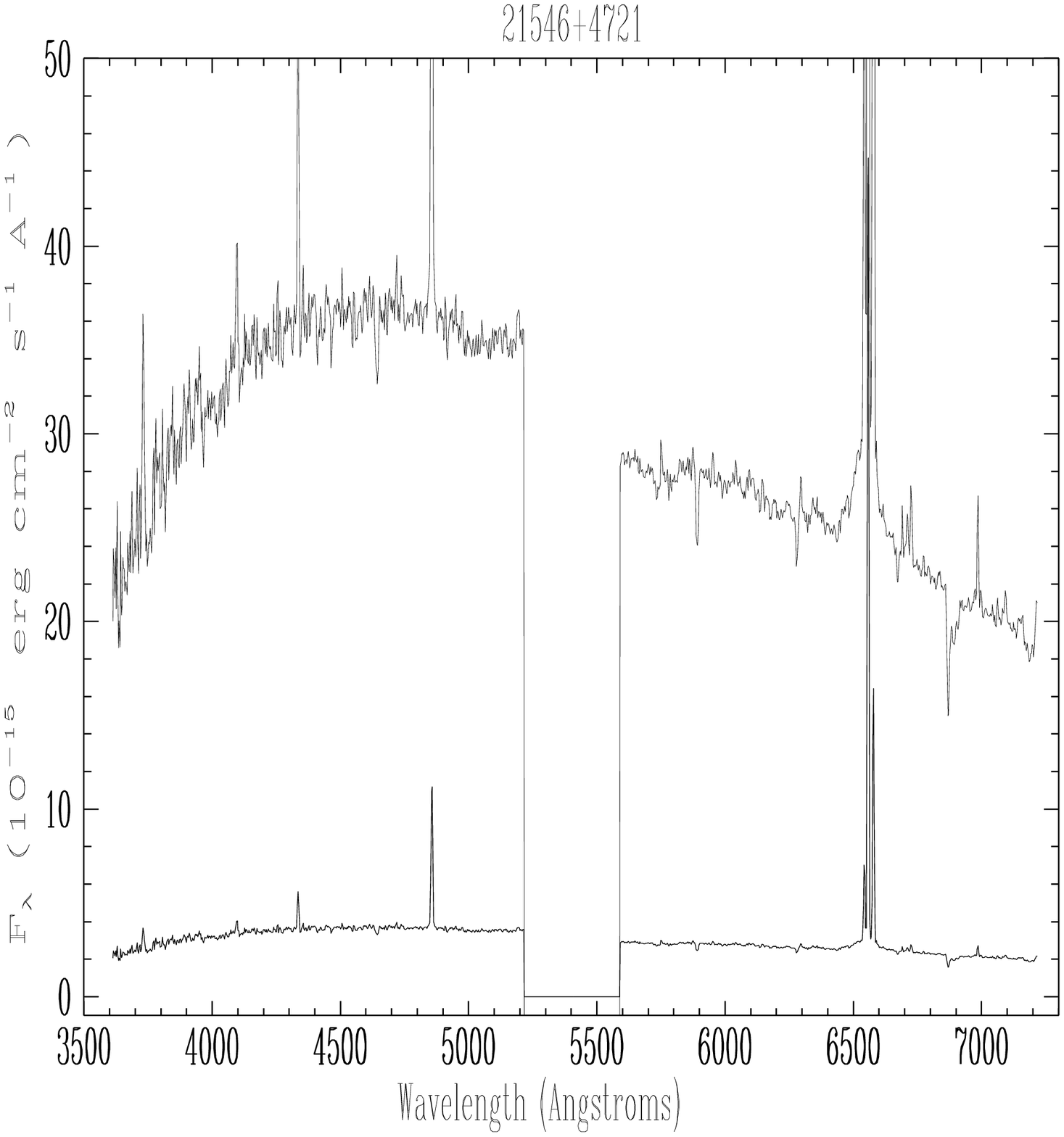}
%\psdraft
\epsfxsize=4cm
\epsfysize=4cm
\epsfbox{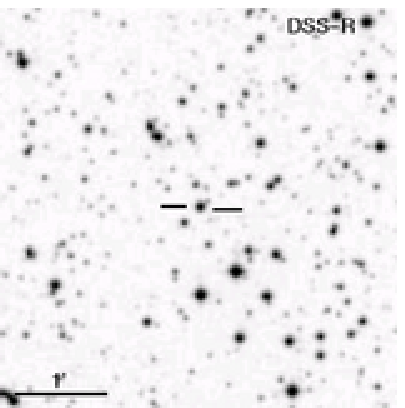}
%\psfull
\end{center}

\begin{center}
\epsfxsize=13.5cm
\epsfysize=4cm
\epsfbox{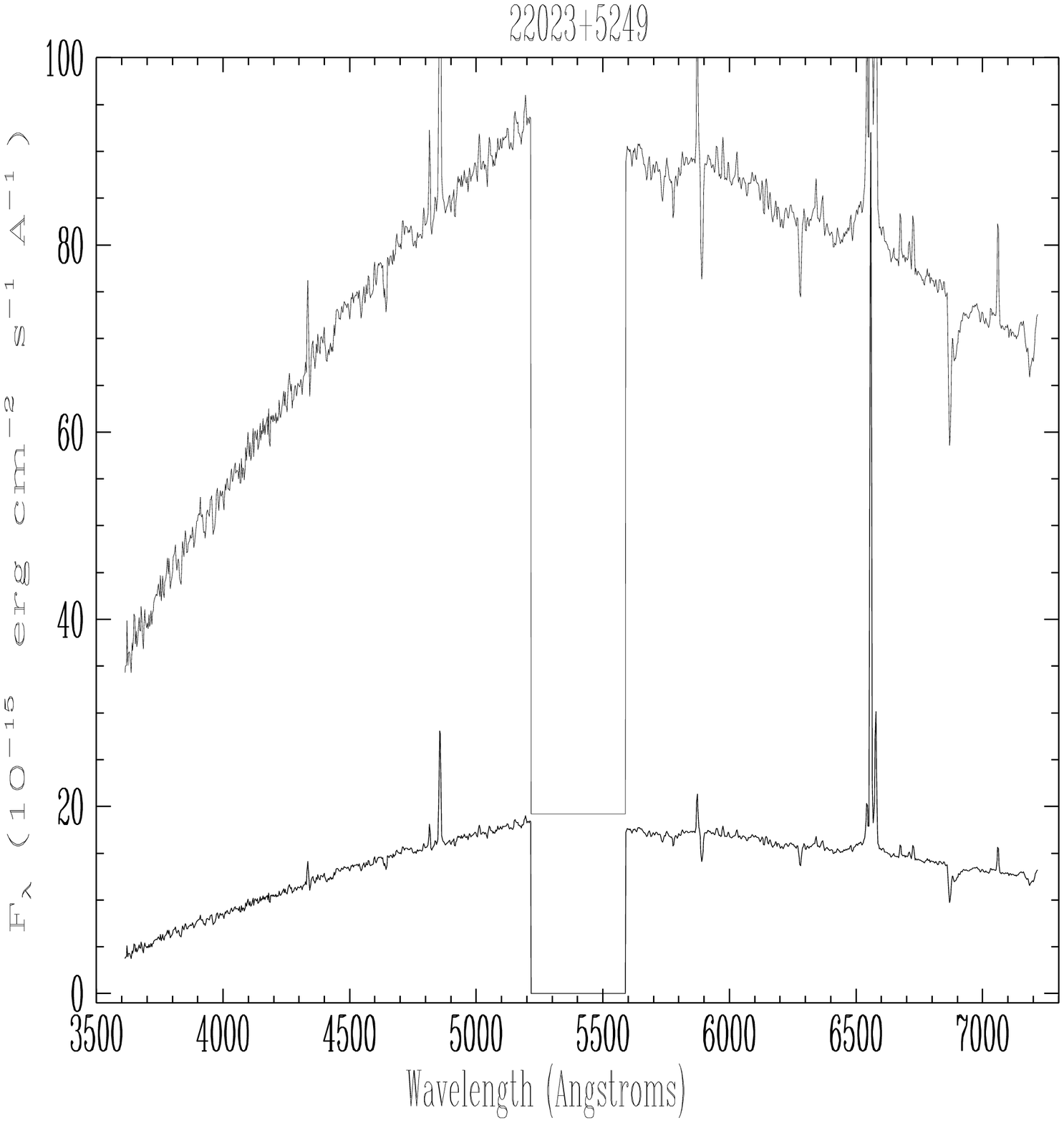}
%\psdraft
\epsfxsize=4cm
\epsfysize=4cm
\epsfbox{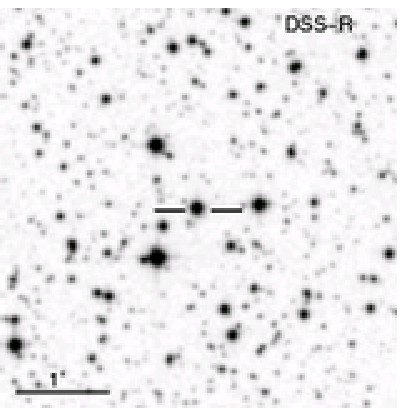}
%\psfull
\end{center}

\caption{Spectra of the transition objects together with their 
corresponding identification charts (continued). }
\end{figure*}

%%% Local Variables: 
%%% mode: latex
%%% TeX-master: "~/tesis/mitesis/final/tesis"
%%% TeX-master: "es"
%%% End: 

\clearpage
\section{Atlas of PNe}
        \begin{figure*}[h!]
\setcounter{figure}{2}

\begin{center}
\epsfxsize=13.5cm
\epsfysize=4cm
\epsfbox{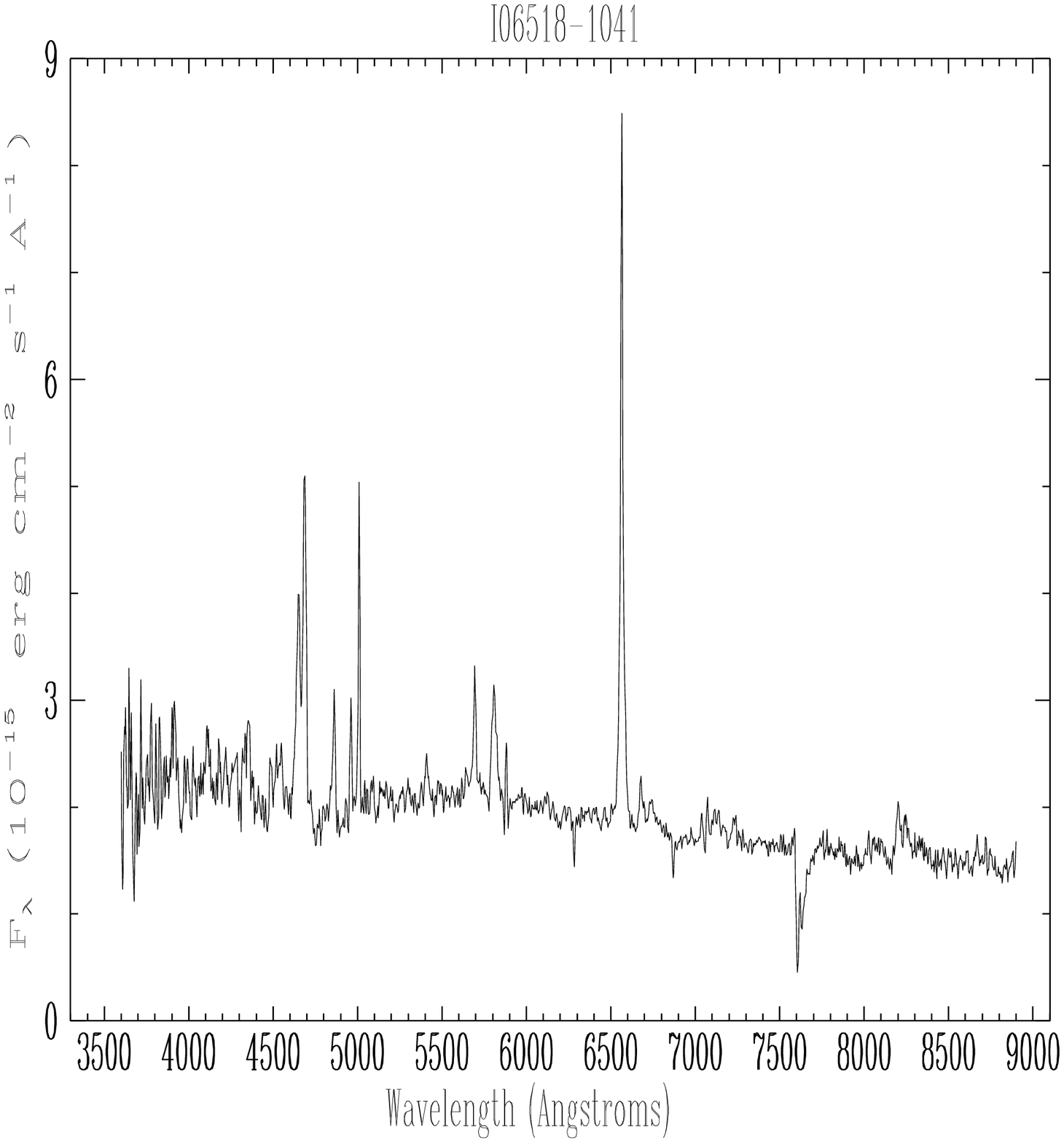}
%\psdraft
\epsfxsize=4cm
\epsfysize=4cm
\epsfbox{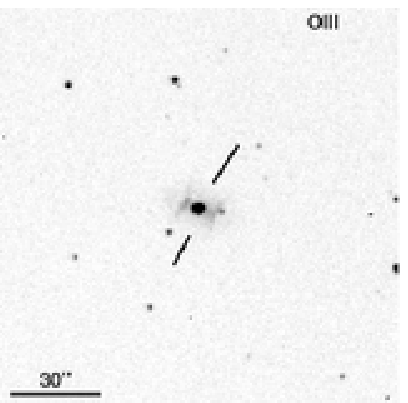}
%\psfull
\end{center}

\begin{center}
\epsfxsize=13.5cm
\epsfysize=4cm
\epsfbox{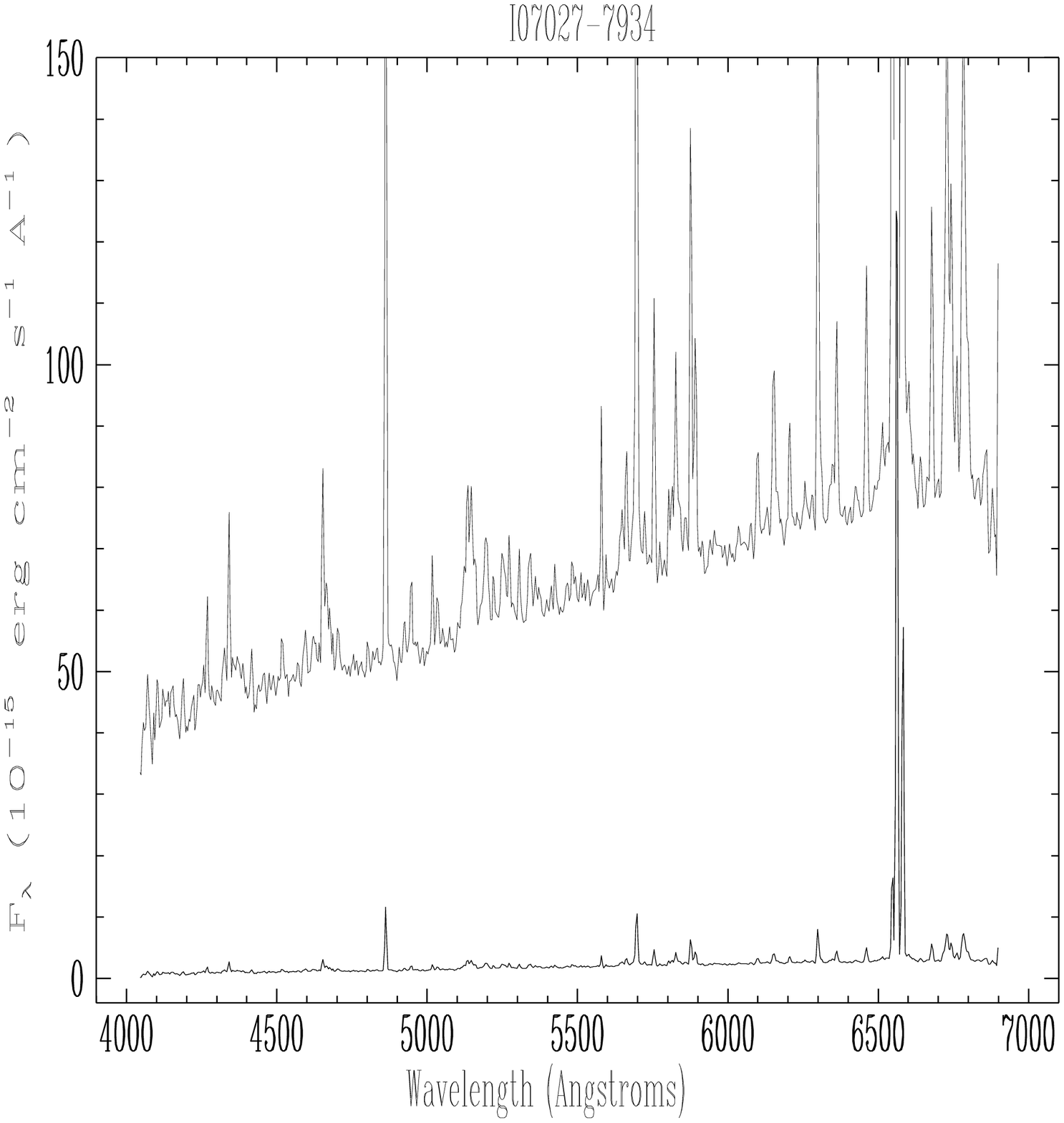}
%\psdraft
\epsfxsize=4cm
\epsfysize=4cm
\epsfbox{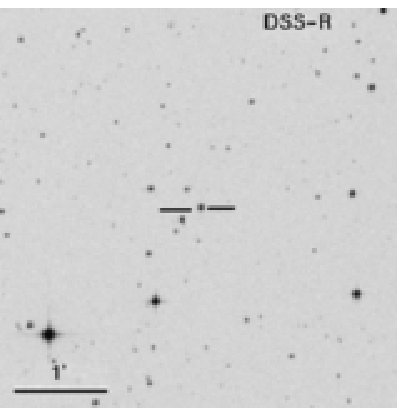}
%\psfull
\end{center}

\begin{center}
\epsfxsize=13.5cm
\epsfysize=4cm
\epsfbox{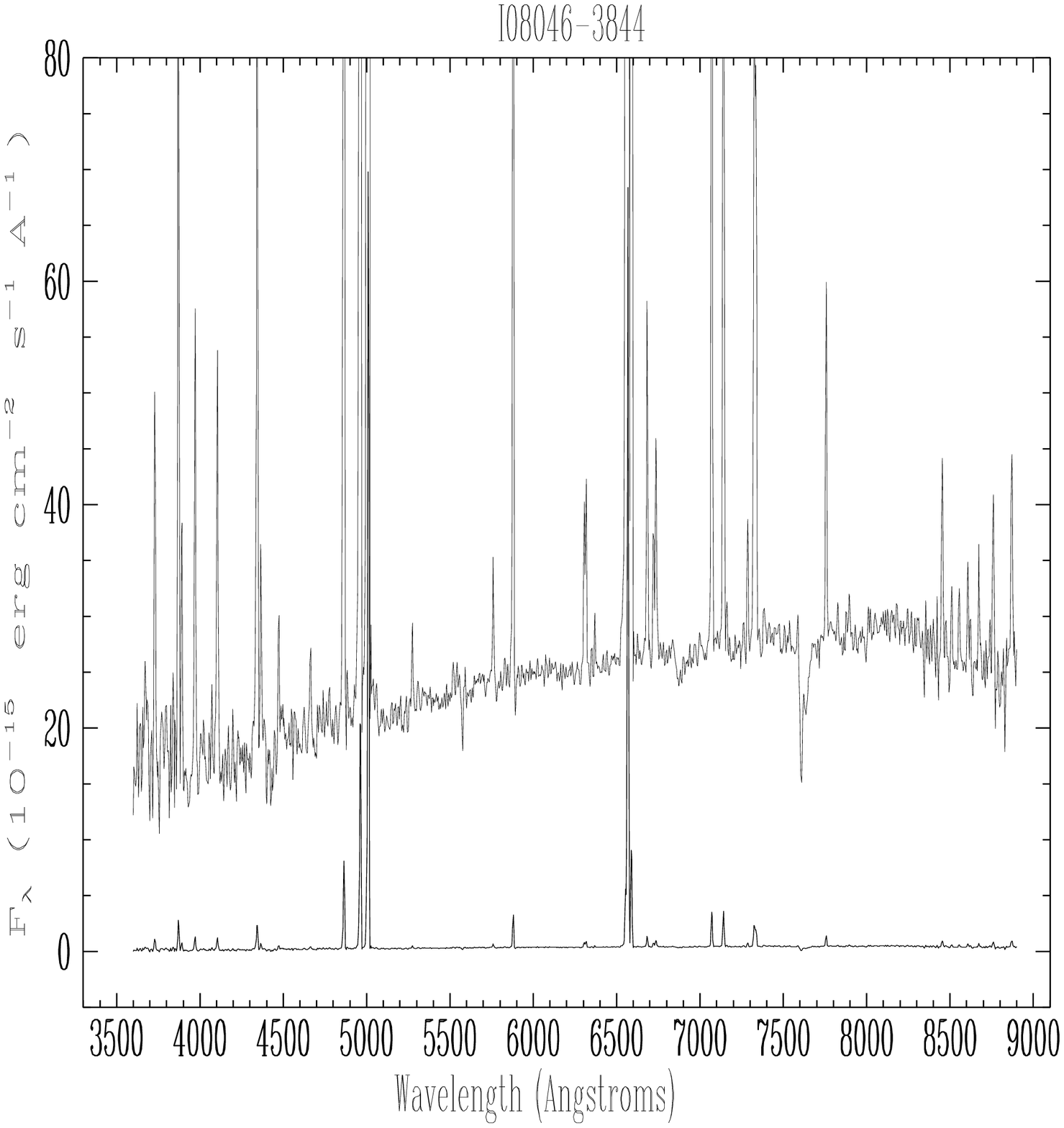}
%\psdraft
\epsfxsize=4cm
\epsfysize=4cm
\epsfbox{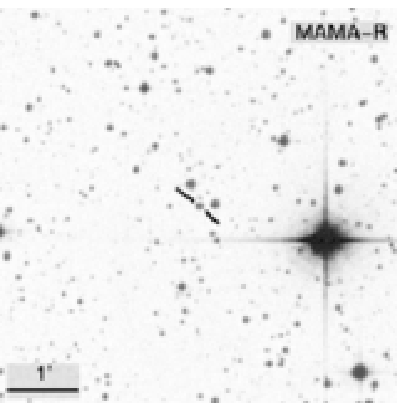}
%\psfull
\end{center}

\begin{center}
\epsfxsize=13.5cm
\epsfysize=4cm
\epsfbox{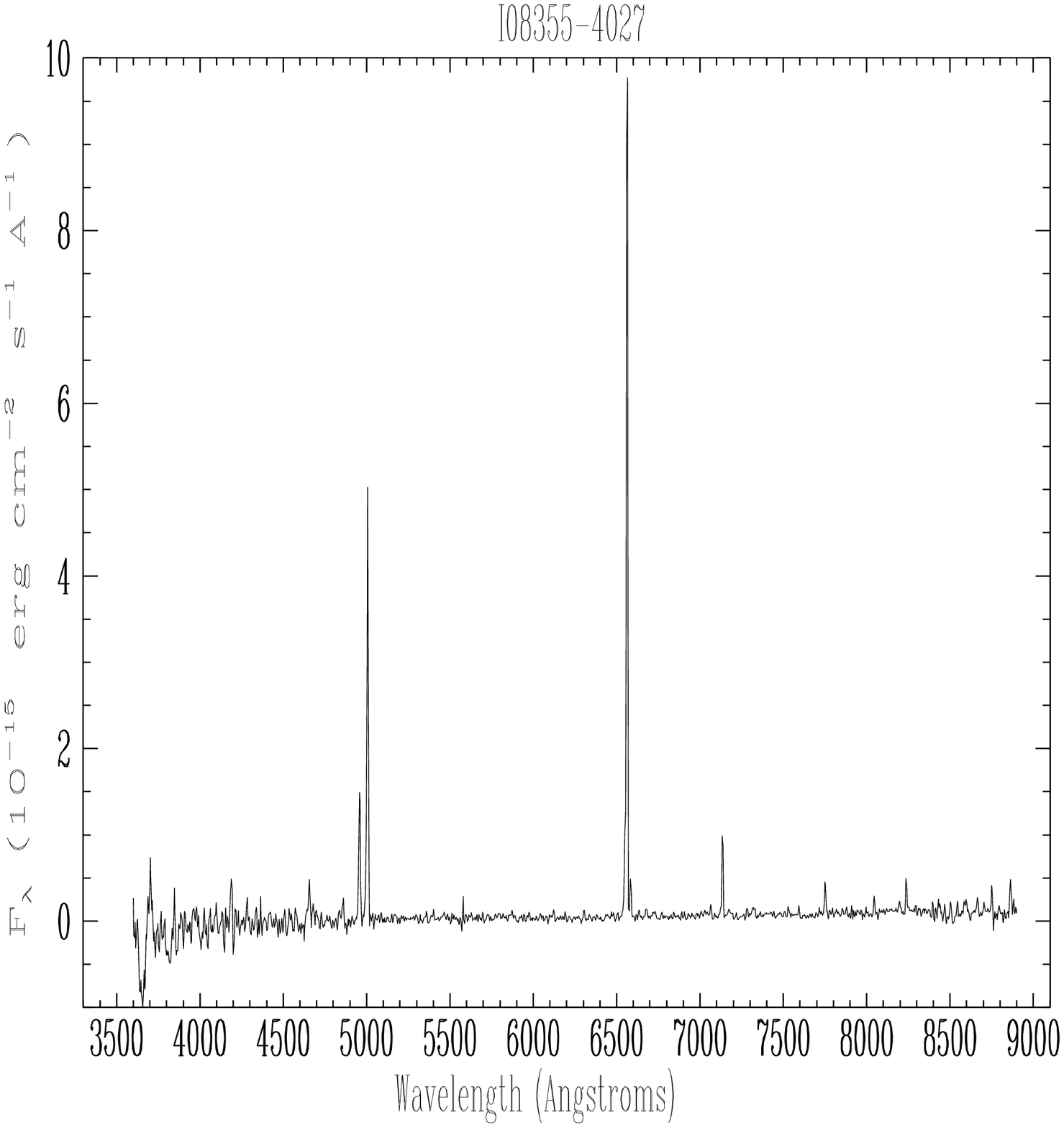}
%\psdraft
\epsfxsize=4cm
\epsfysize=4cm
\epsfbox{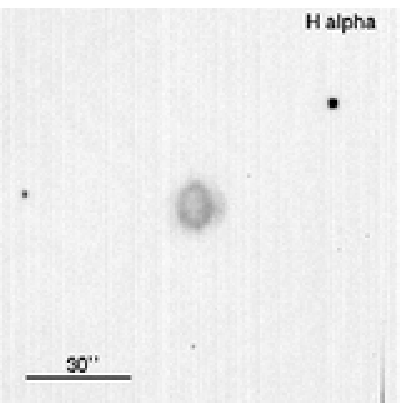}
%\psfull
\end{center}

\caption{Spectra of the PNe in the sample together with their 
corresponding identification charts. }
\end{figure*}

%-------------------------------------------------------------------
%pg2.

\begin{figure*}
\setcounter{figure}{2}

\begin{center}
\epsfxsize=13.5cm
\epsfysize=4cm
\epsfbox{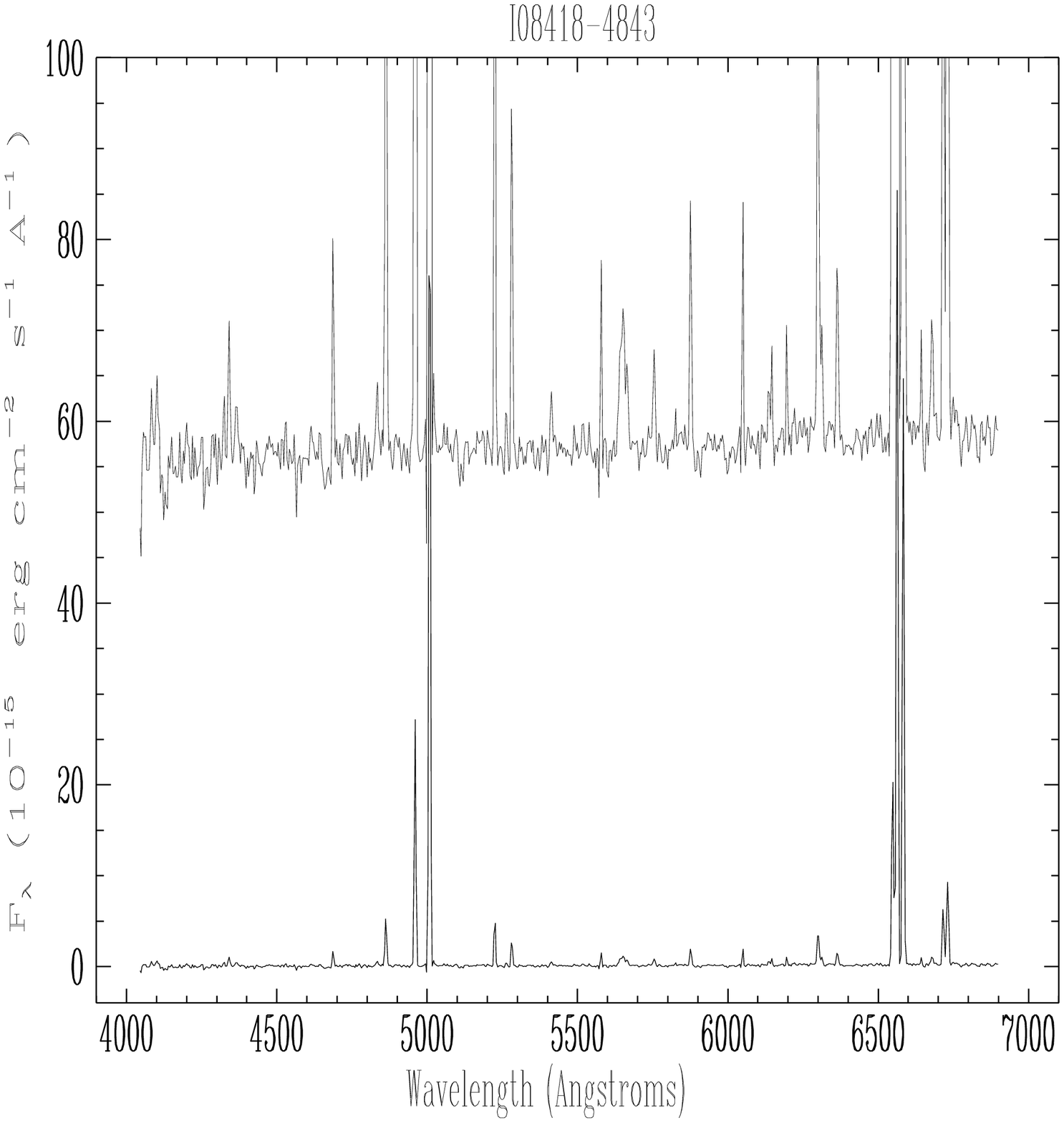}
%\psdraft
\epsfxsize=4cm
\epsfysize=4cm
\epsfbox{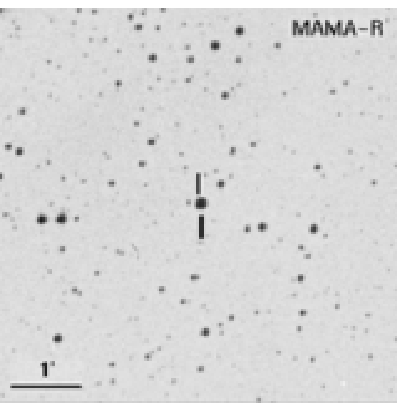}
%\psfull
\end{center}

\begin{center}
\epsfxsize=13.5cm
\epsfysize=4cm
\epsfbox{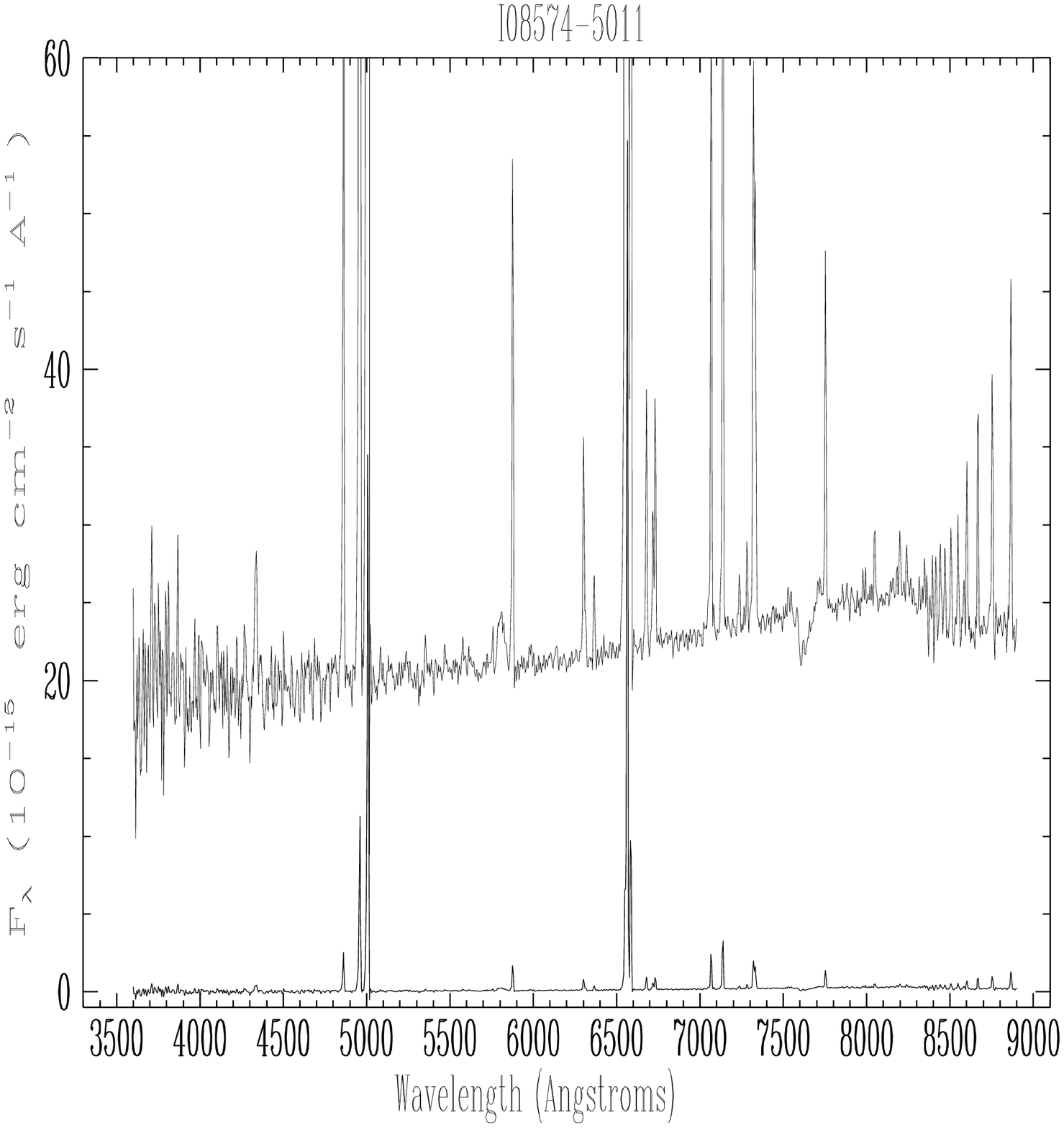}
%\psdraft
\epsfxsize=4cm
\epsfysize=4cm
\epsfbox{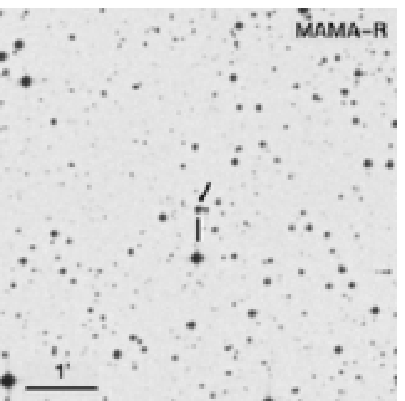}
%\psfull
\end{center}

\begin{center}
\epsfxsize=13.5cm
\epsfysize=4cm
\epsfbox{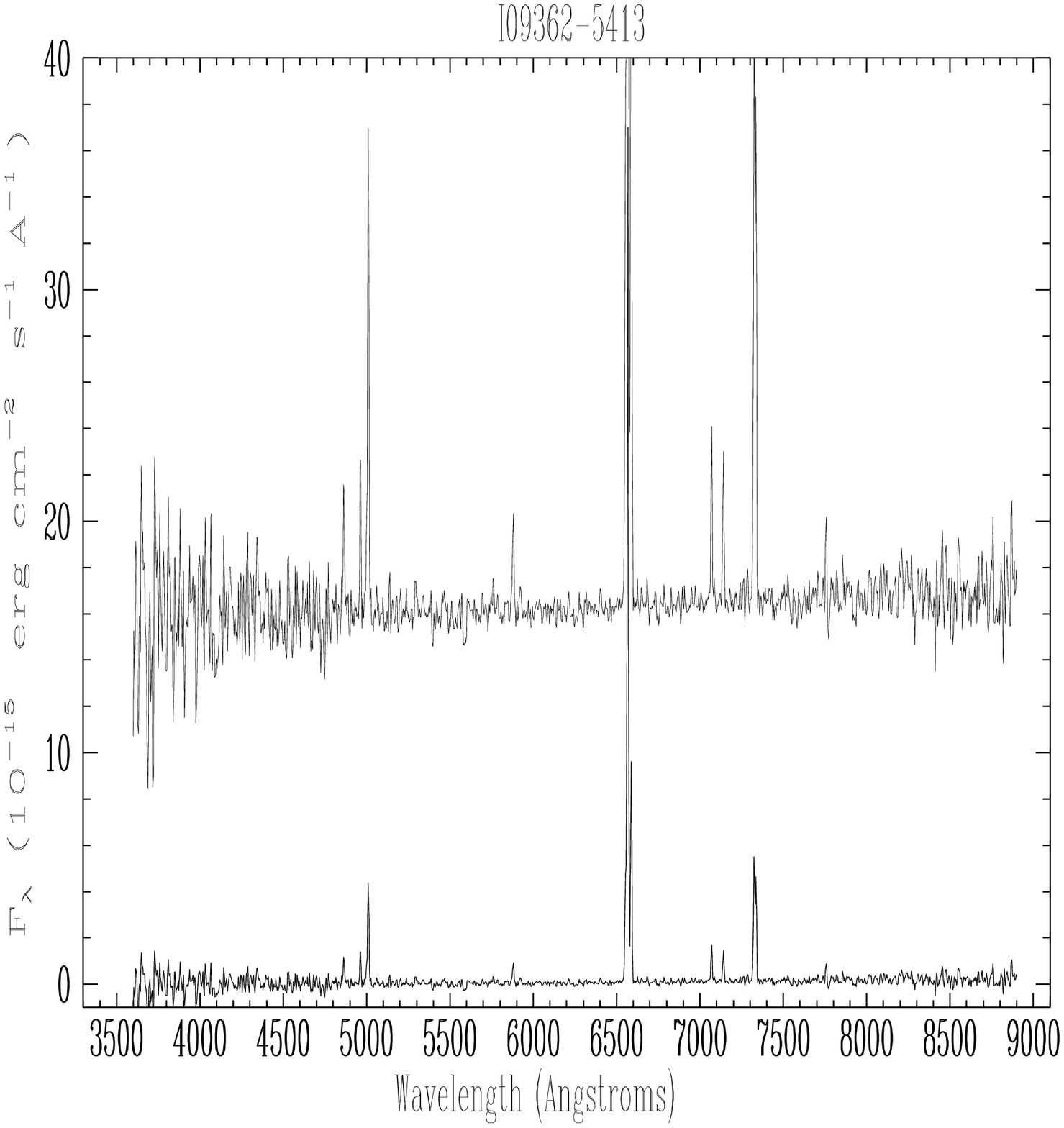}
%\psdraft
\epsfxsize=4cm
\epsfysize=4cm
\epsfbox{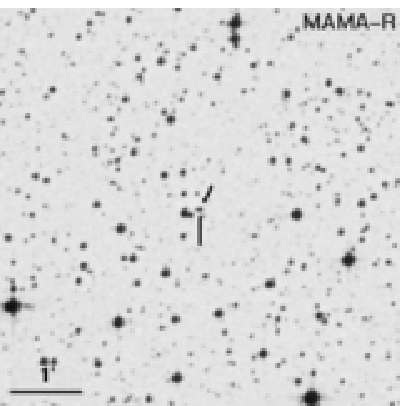}
%\psfull
\end{center}

\begin{center}
\epsfxsize=13.5cm
\epsfysize=4cm
\epsfbox{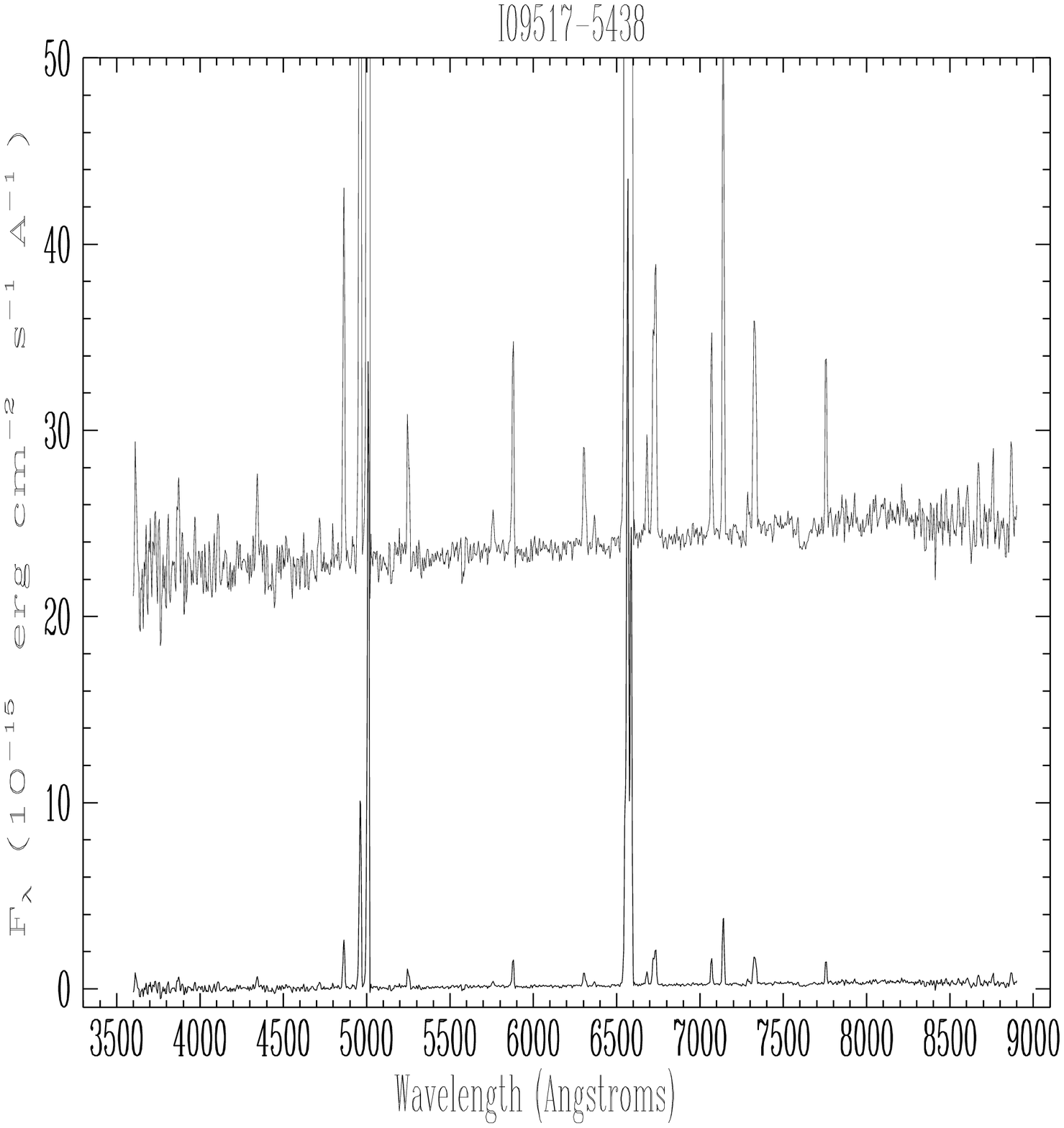}
%\psdraft
\epsfxsize=4cm
\epsfysize=4cm
\epsfbox{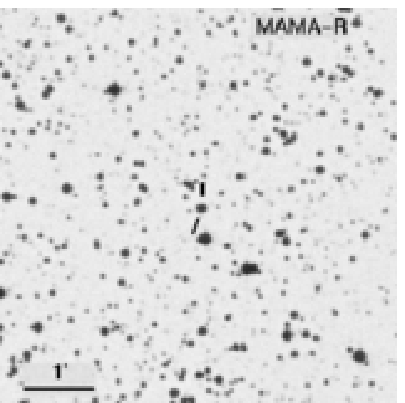}
%\psfull
\end{center}

\begin{center}
\epsfxsize=13.5cm
\epsfysize=4cm
\epsfbox{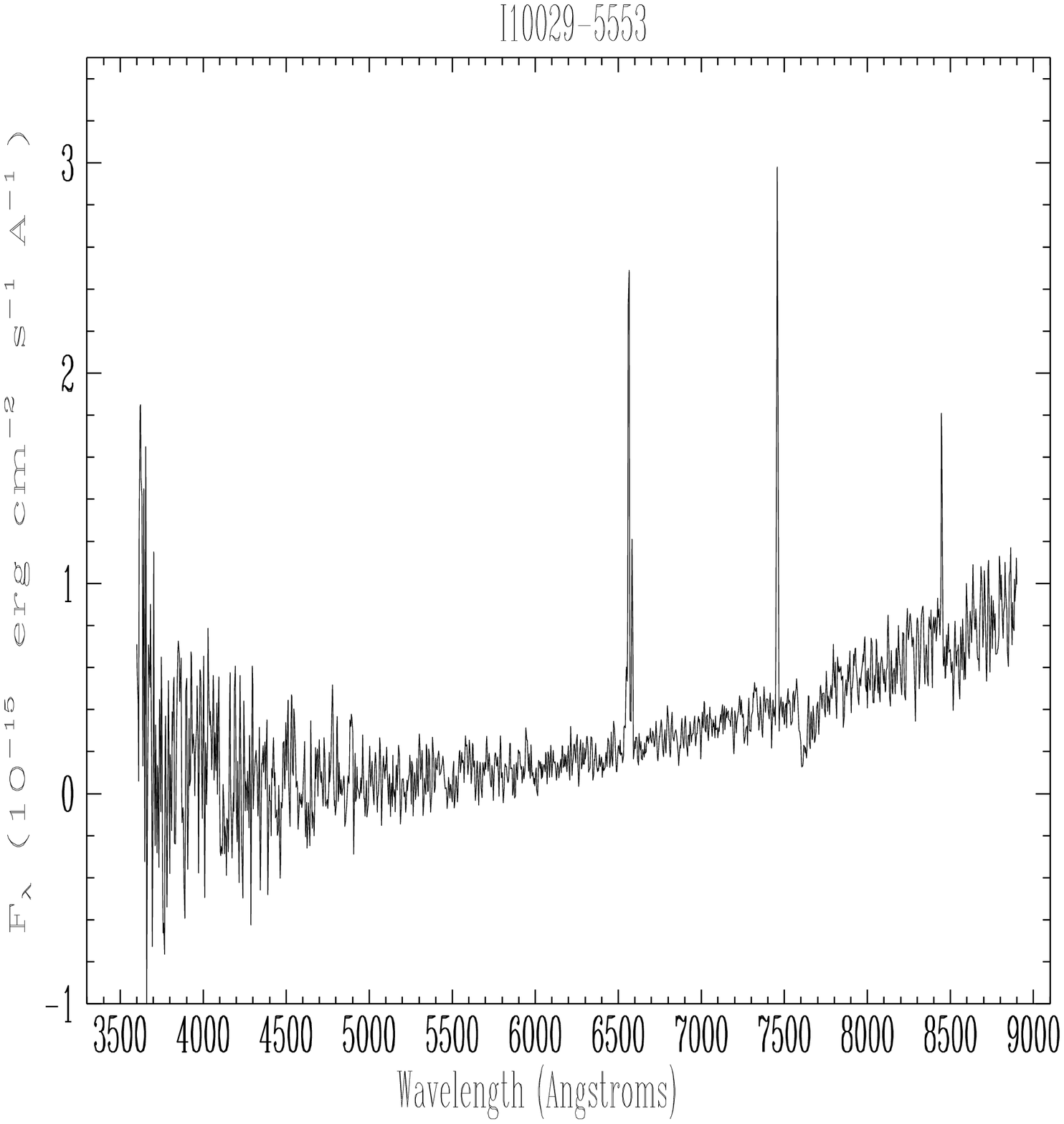}
%\psdraft
\epsfxsize=4cm
\epsfysize=4cm
\epsfbox{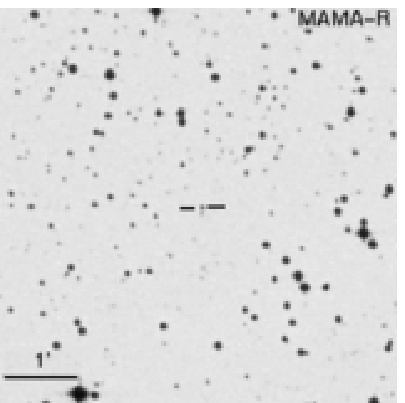}
%\psfull
\end{center}

\caption{Spectra of the PNe in the sample together with their 
corresponding identification charts (continued). }
\end{figure*}

%-------------------------------------------------------------------
%pg3

\begin{figure*}
\setcounter{figure}{2}

\begin{center}
\epsfxsize=13.5cm
\epsfysize=4cm
\epsfbox{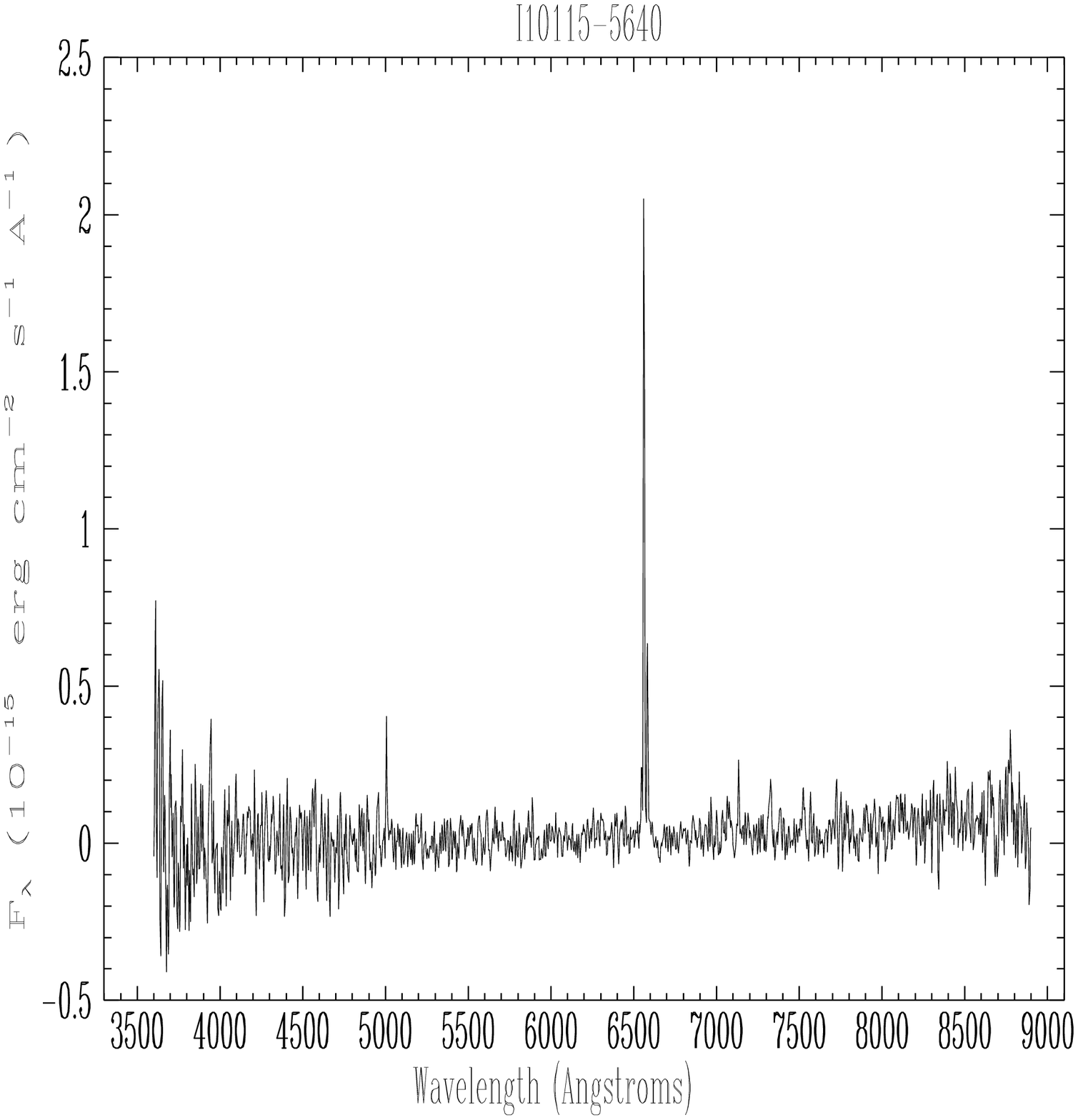}
%\psdraft
\epsfxsize=4cm
\epsfysize=4cm
\epsfbox{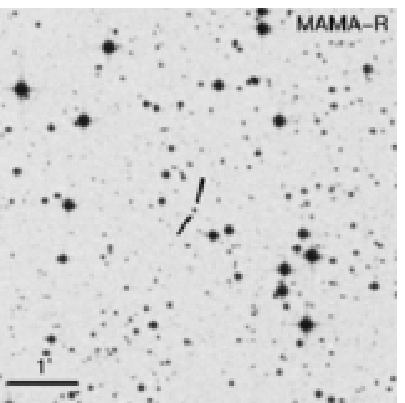}
%\psfull
\end{center}

\begin{center}
\epsfxsize=13.5cm
\epsfysize=4cm
\epsfbox{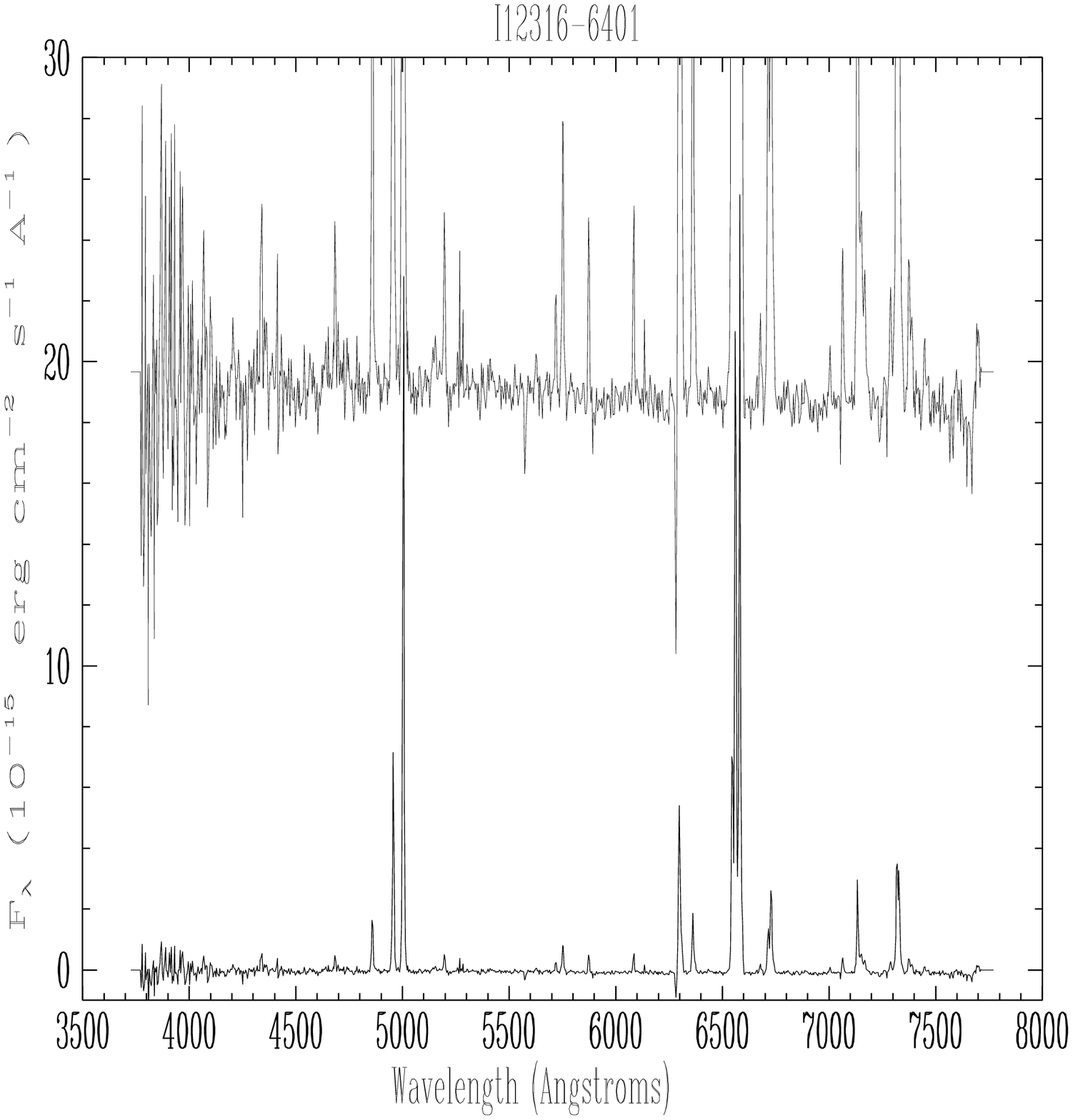}
%\psdraft
\epsfxsize=4cm
\epsfysize=4cm
\epsfbox{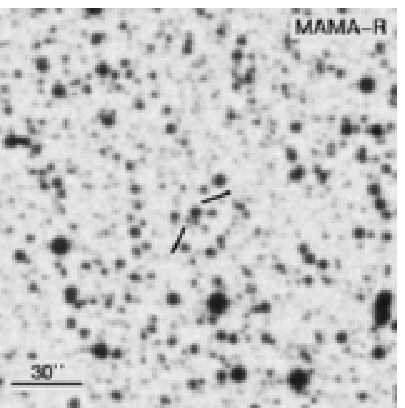}
%\psfull
\end{center}

\begin{center}
\epsfxsize=13.5cm
\epsfysize=4cm
\epsfbox{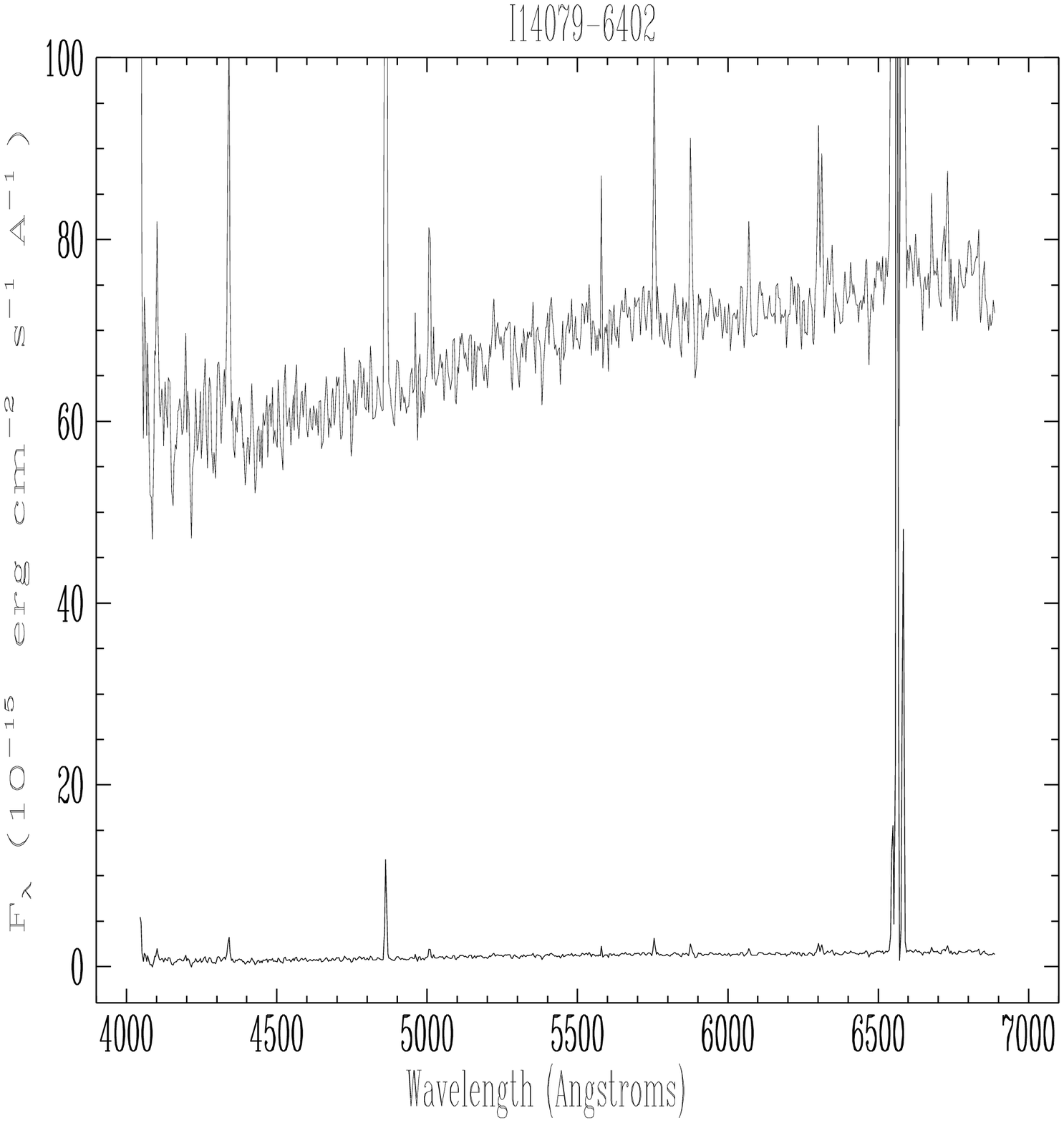}
%\psdraft
\epsfxsize=4cm
\epsfysize=4cm
\epsfbox{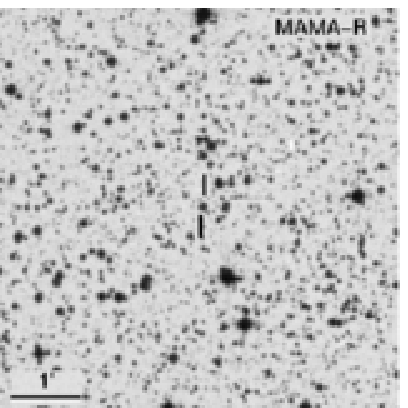}
%\psfull
\end{center}

\begin{center}
\epsfxsize=13.5cm
\epsfysize=4cm
\epsfbox{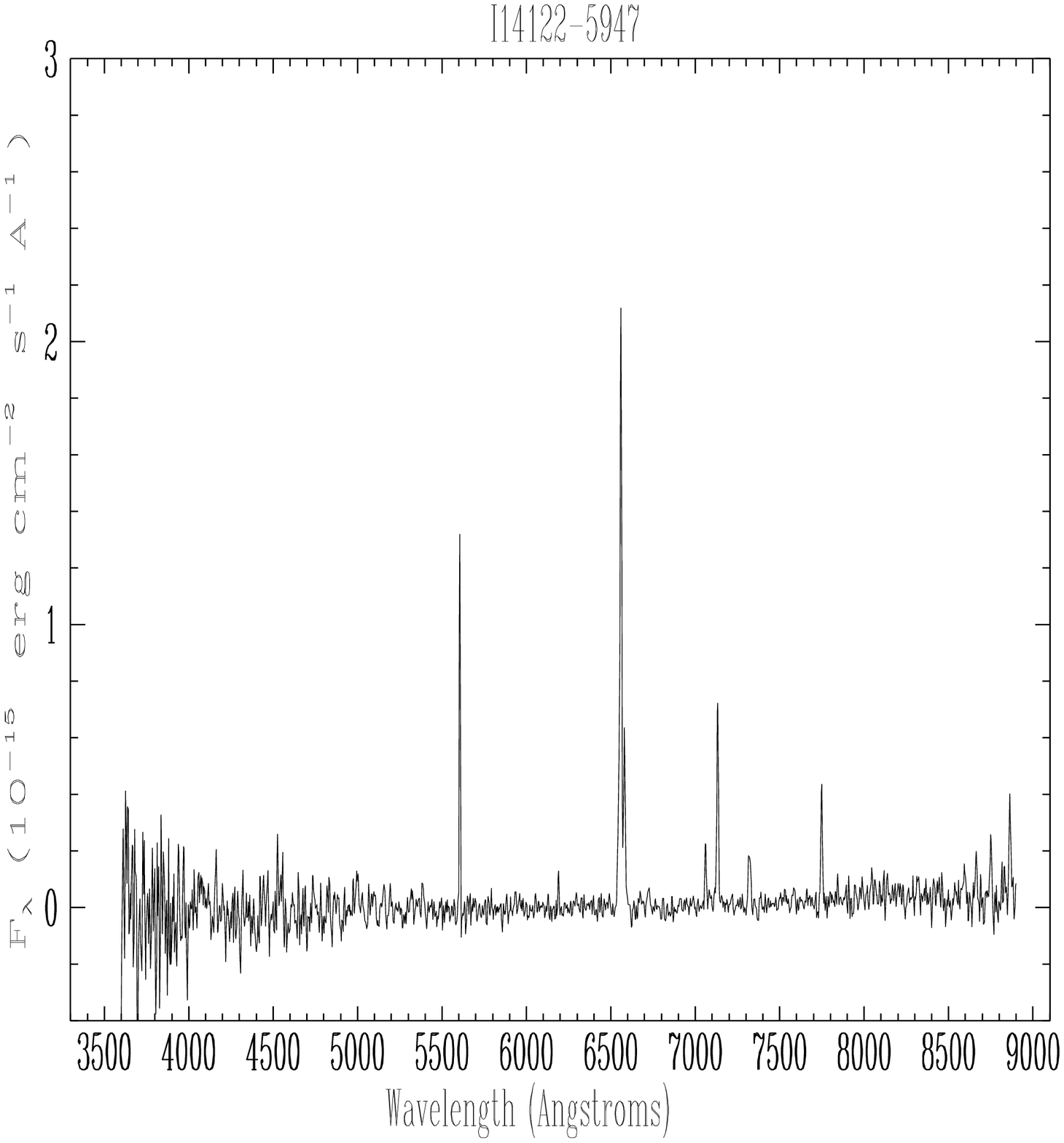}
%\psdraft
\epsfxsize=4cm
\epsfysize=4cm
\epsfbox{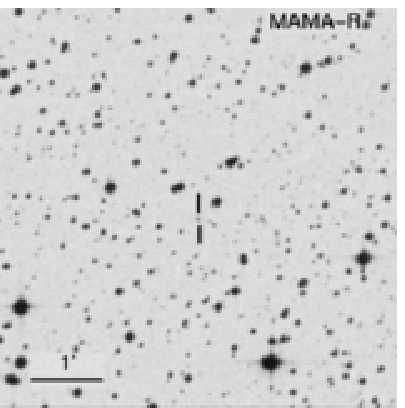}
%\psfull
\end{center}

\begin{center}
\epsfxsize=13.5cm
\epsfysize=4cm
\epsfbox{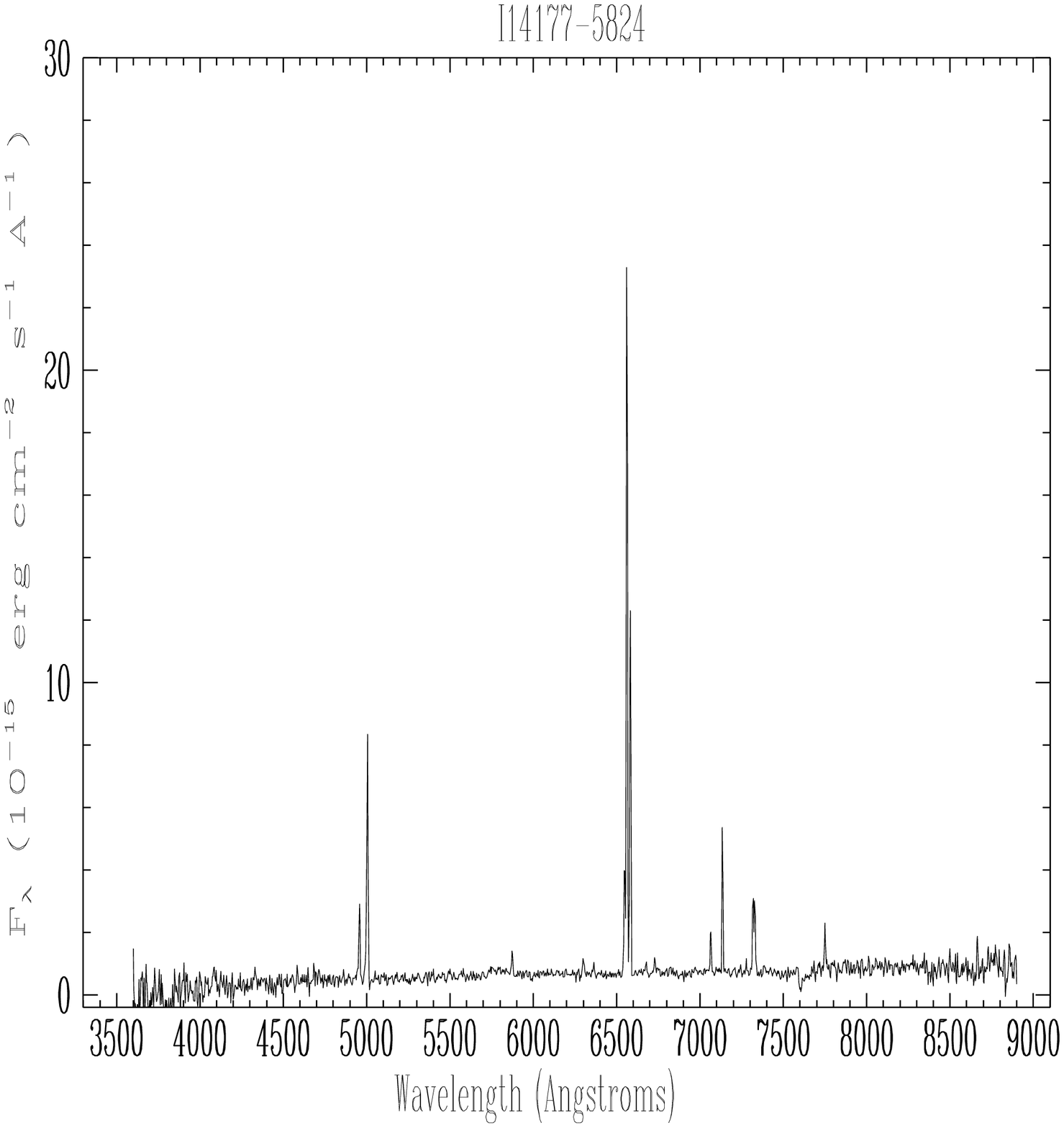}
%\psdraft
\epsfxsize=4cm
\epsfysize=4cm
\epsfbox{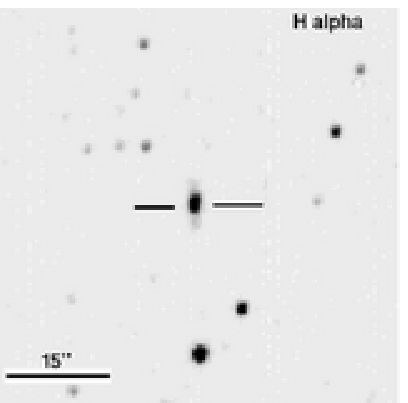}
%\psfull
\end{center}

\caption{Spectra of the PNe in the sample together with their 
corresponding identification charts (continued). }
\end{figure*}

%-------------------------------------------------------------------
%pg4

\begin{figure*}
\setcounter{figure}{2}

\begin{center}
\epsfxsize=13.5cm
\epsfysize=4cm
\epsfbox{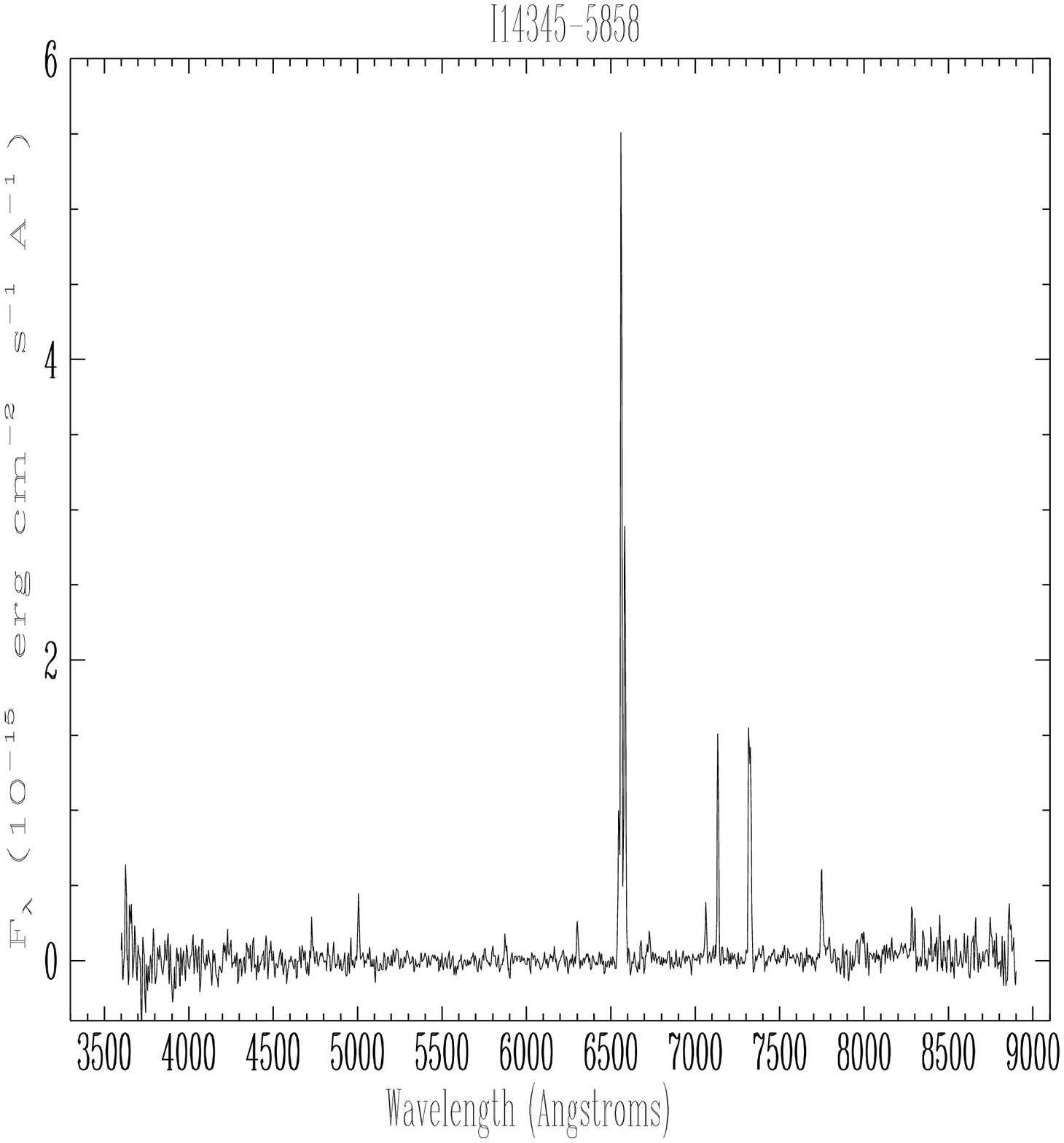}
%\psdraft
\epsfxsize=4cm
\epsfysize=4cm
\epsfbox{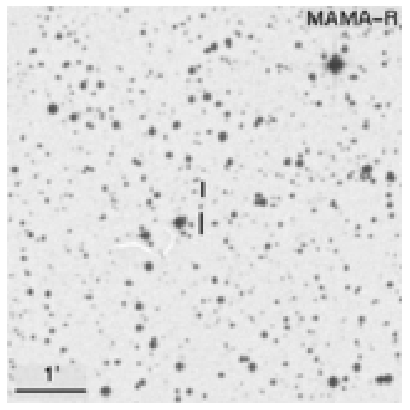}
%\psfull
\end{center}

\begin{center}
\epsfxsize=13.5cm
\epsfysize=4cm
\epsfbox{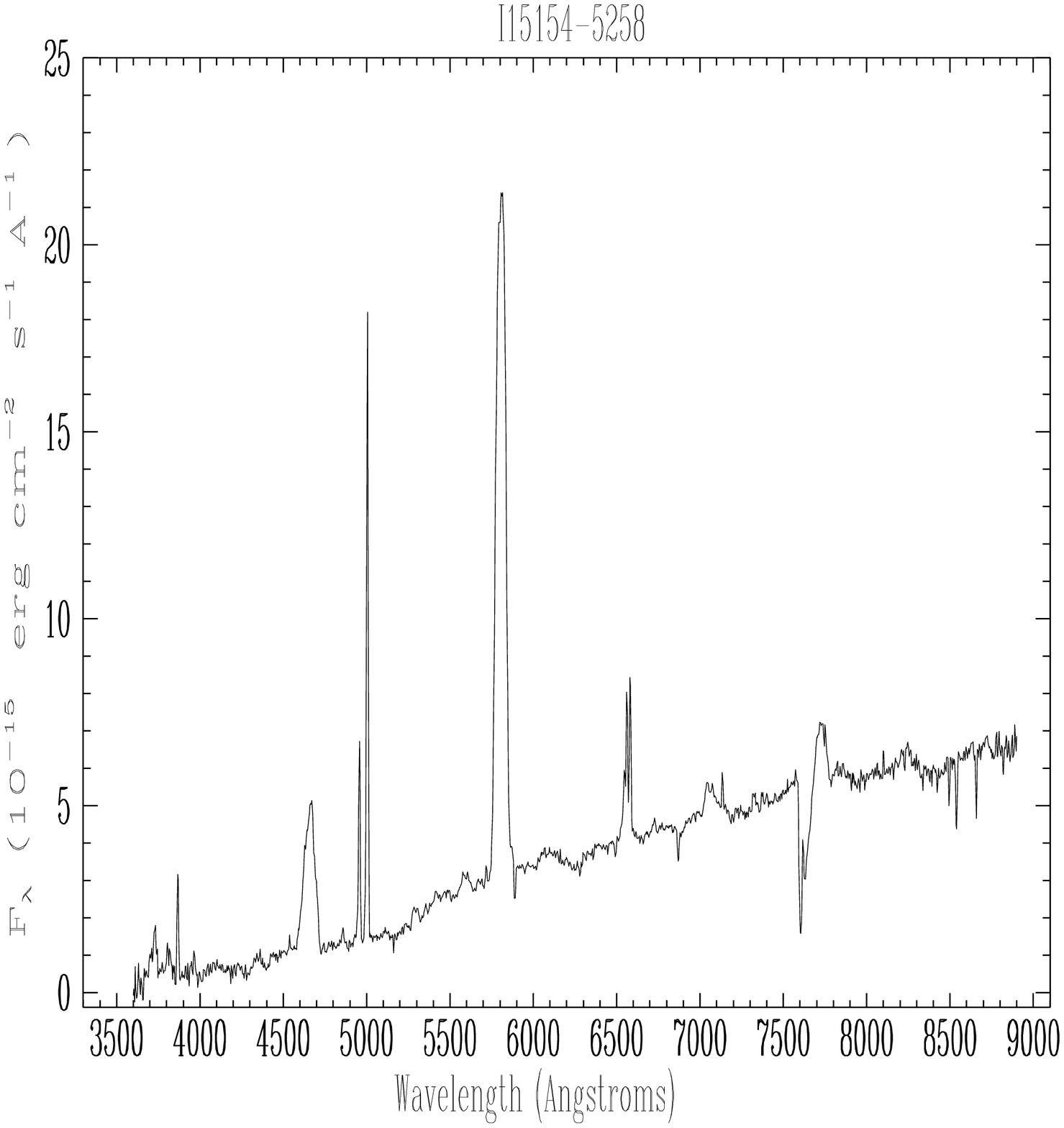}
%\psdraft
\epsfxsize=4cm
\epsfysize=4cm
\epsfbox{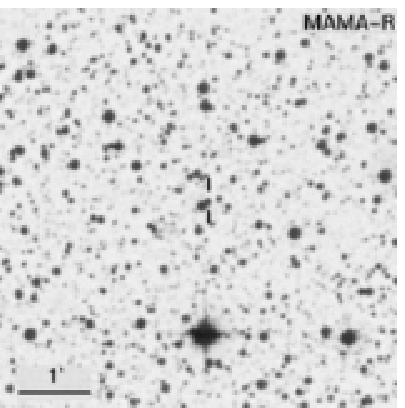}
%\psfull
\end{center}

\begin{center}
\epsfxsize=13.5cm
\epsfysize=4cm
\epsfbox{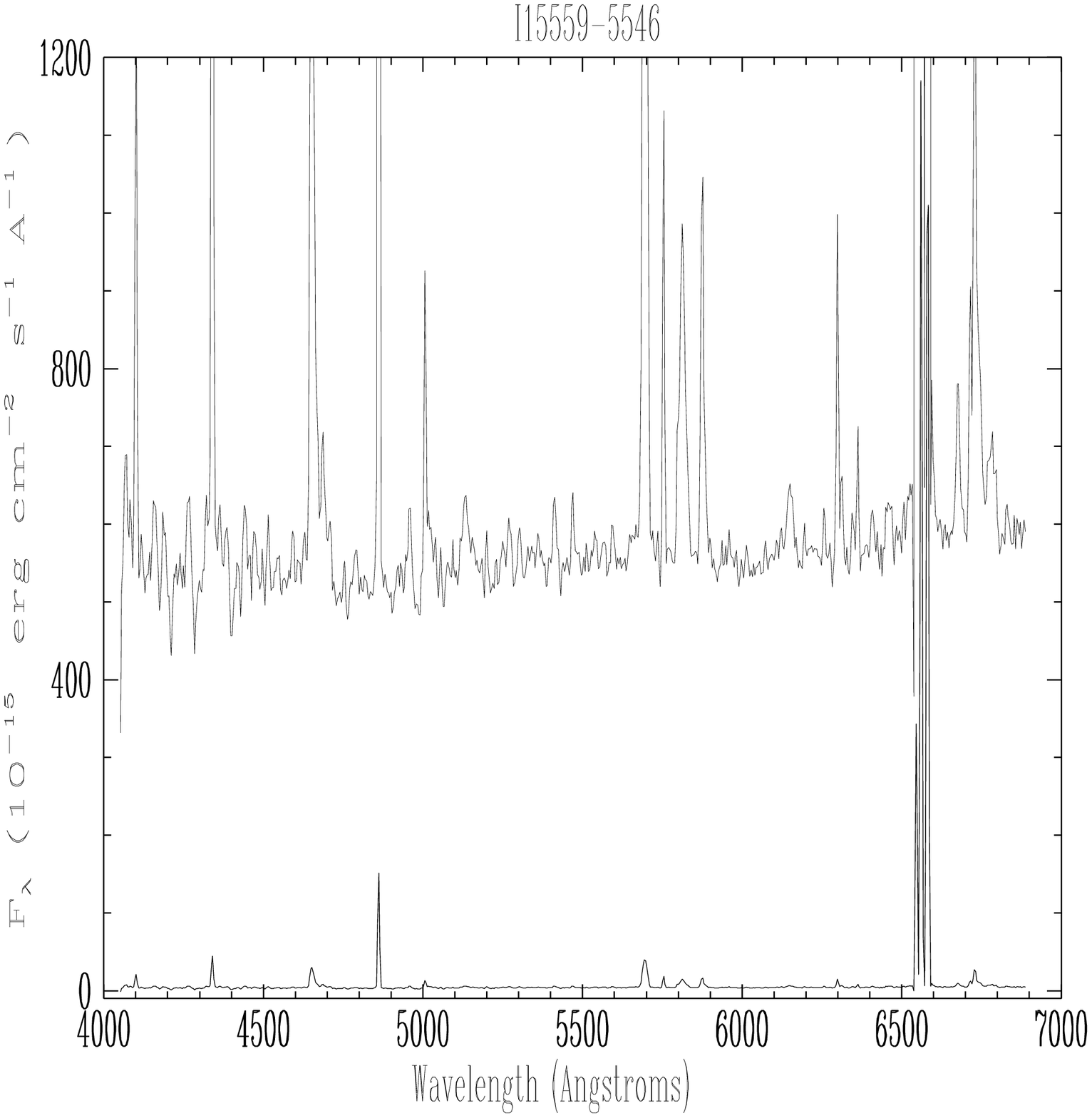}
%\psdraft
\epsfxsize=4cm
\epsfysize=4cm
\epsfbox{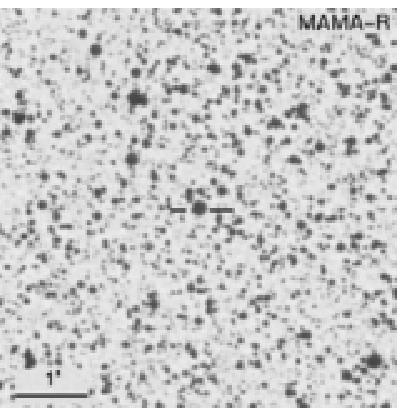}
%\psfull
\end{center}

\begin{center}
\epsfxsize=13.5cm
\epsfysize=4cm
\epsfbox{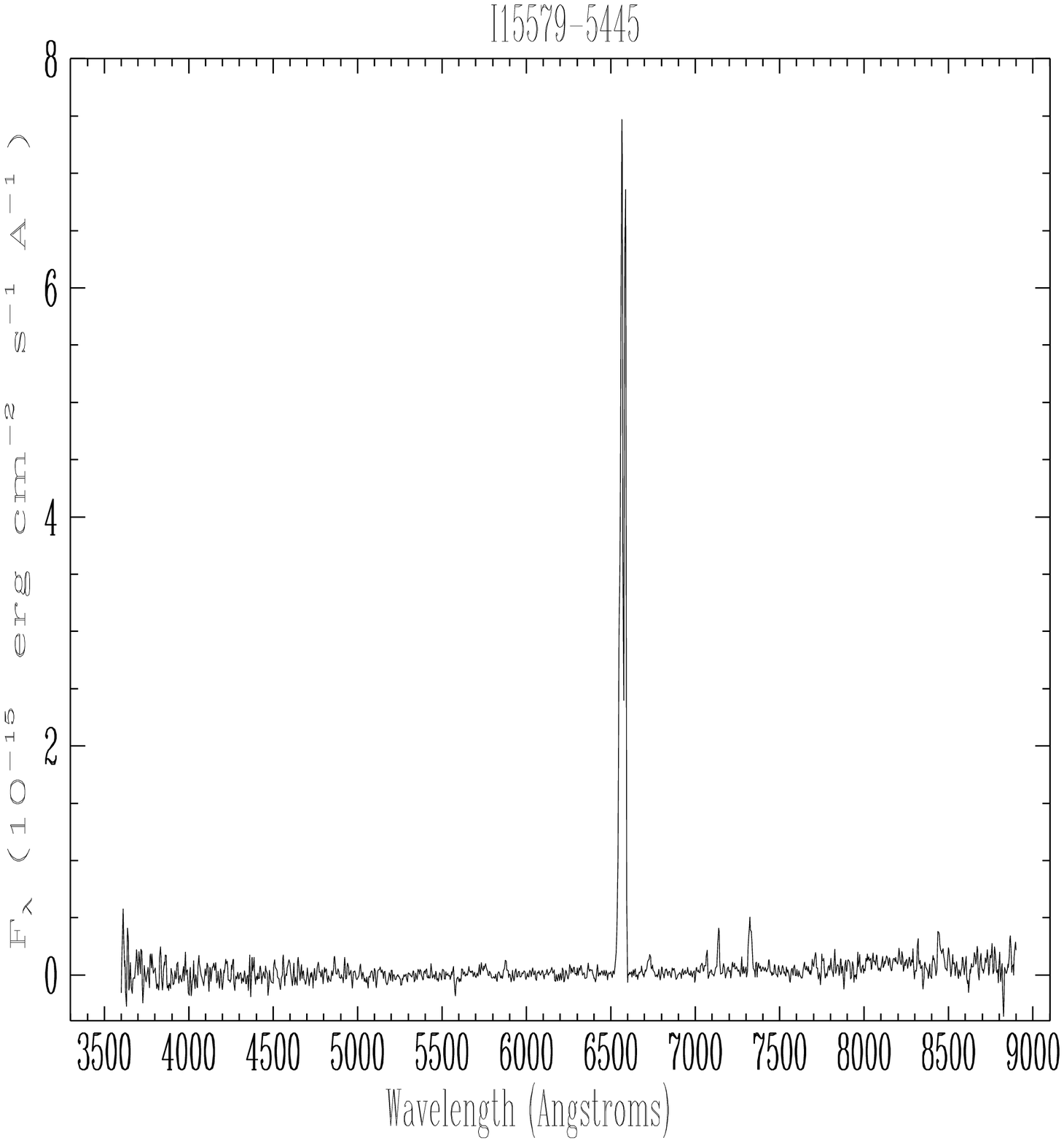}
%\psdraft
\epsfxsize=4cm
\epsfysize=4cm
\epsfbox{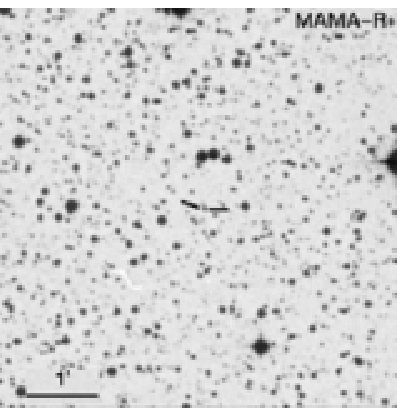}
%\psfull
\end{center}

\begin{center}
\epsfxsize=13.5cm
\epsfysize=4cm
\epsfbox{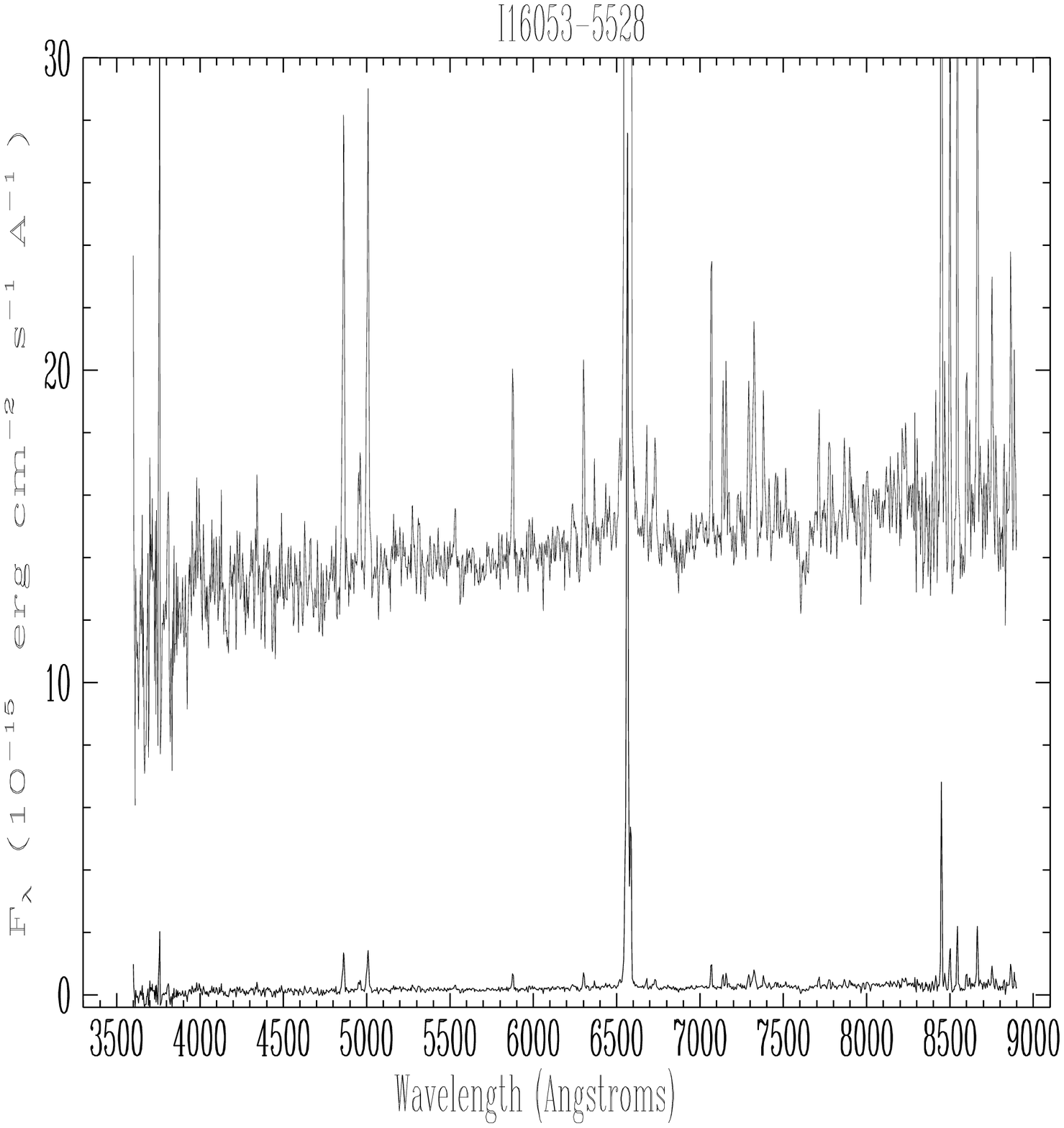}
%\psdraft
\epsfxsize=4cm
\epsfysize=4cm
\epsfbox{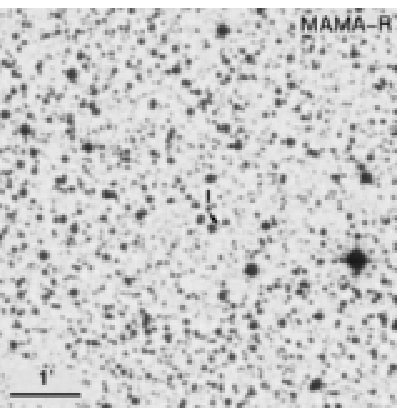}
%\psfull
\end{center}

\caption{Spectra of the PNe in the sample together with their 
corresponding identification charts (continued). }
\end{figure*}

%-------------------------------------------------------------------
%pg5

\begin{figure*}
\setcounter{figure}{2}

\begin{center}
\epsfxsize=13.5cm
\epsfysize=4cm
\epsfbox{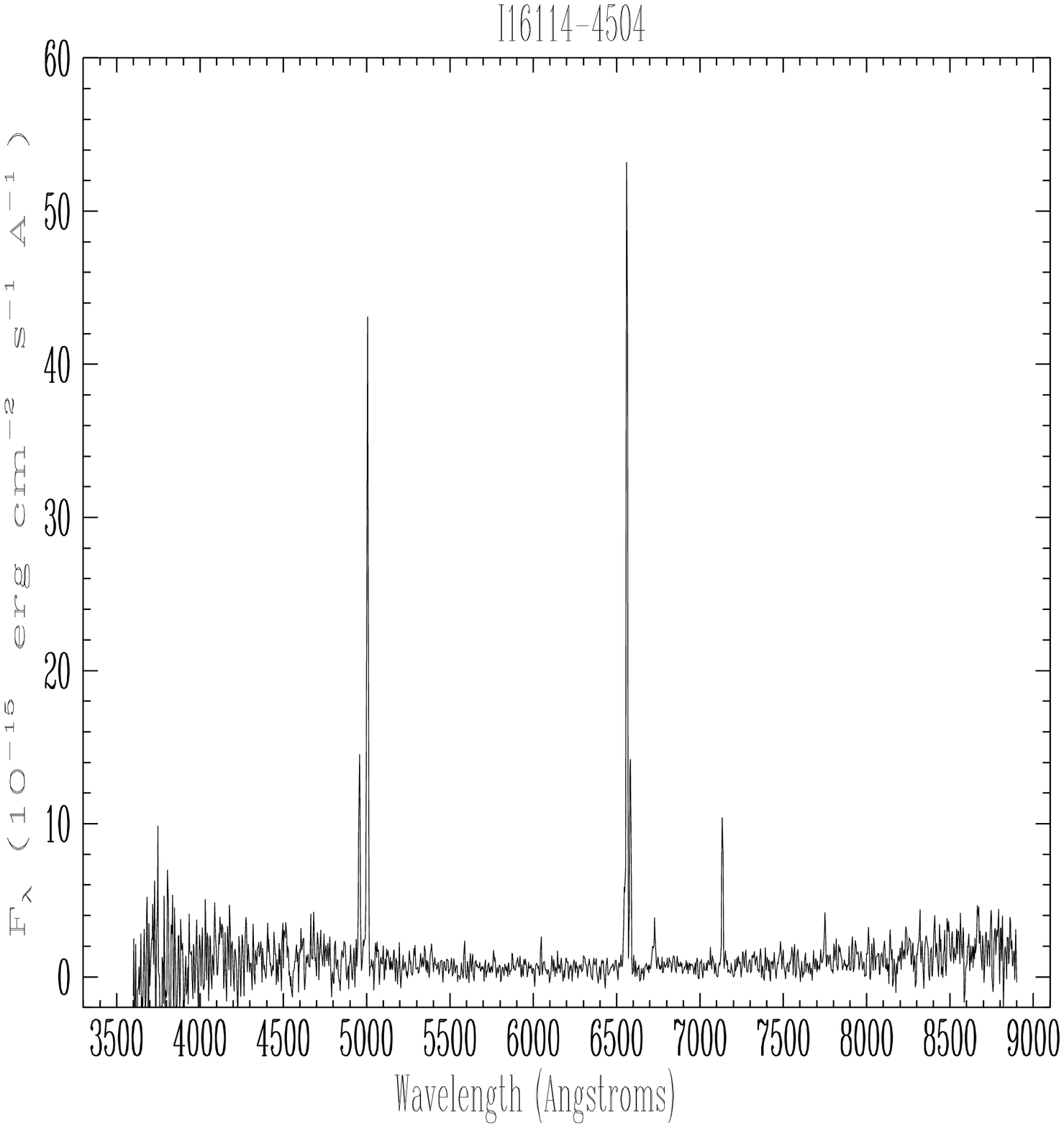}
%\psdraft
\epsfxsize=4cm
\epsfysize=4cm
\epsfbox{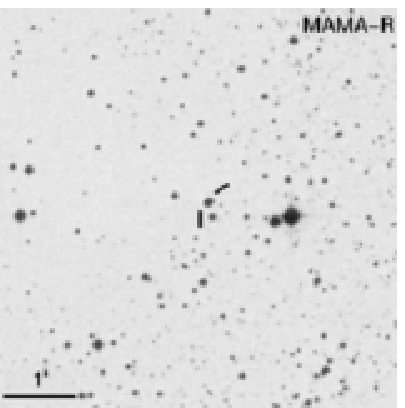}
%\psfull
\end{center}

\begin{center}
\epsfxsize=13.5cm
\epsfysize=4cm
\epsfbox{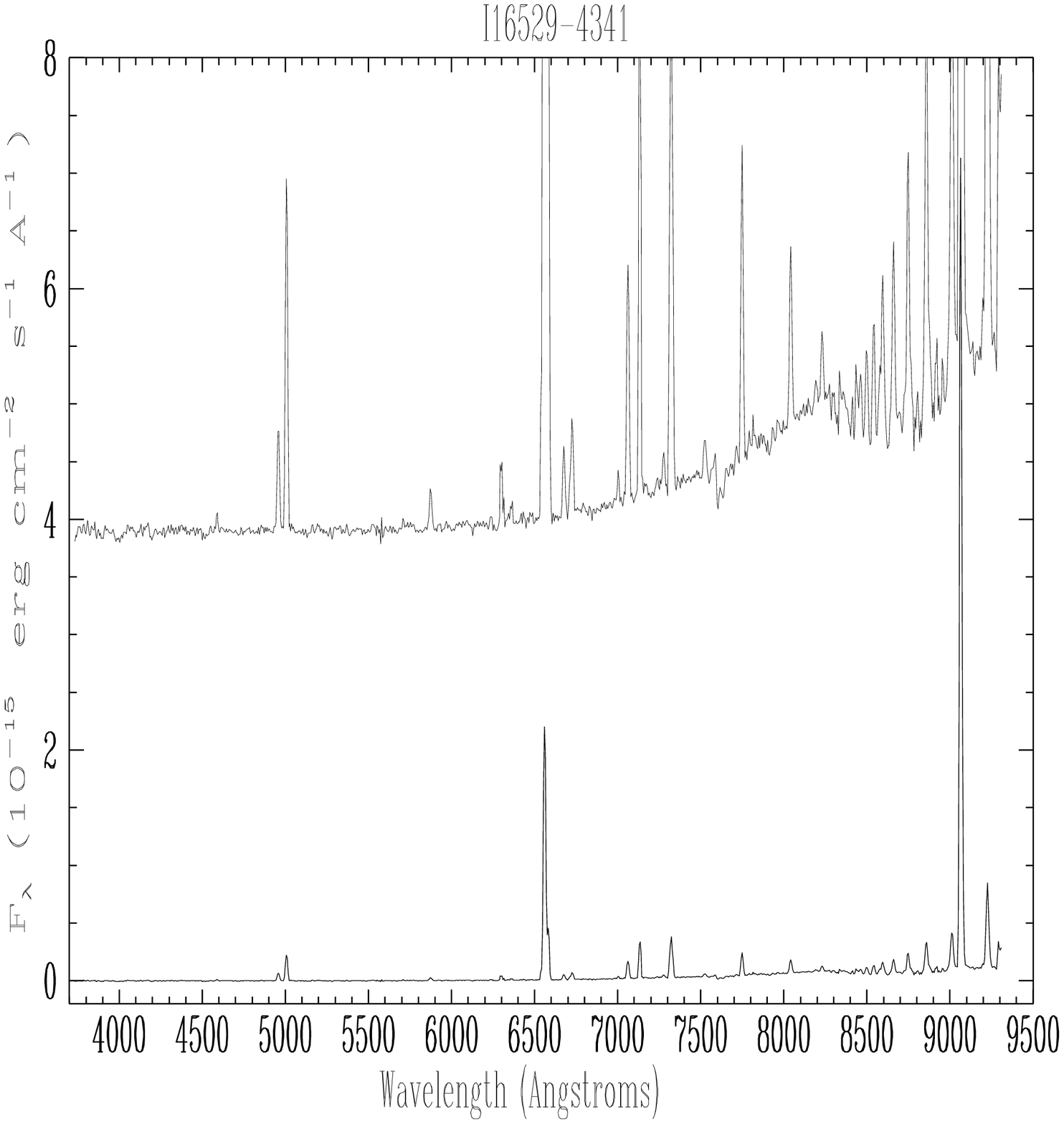}
%\psdraft
\epsfxsize=4cm
\epsfysize=4cm
\epsfbox{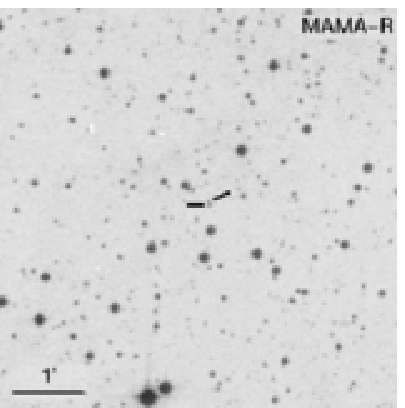}
%\psfull
\end{center}

\begin{center}
\epsfxsize=13.5cm
\epsfysize=4cm
\epsfbox{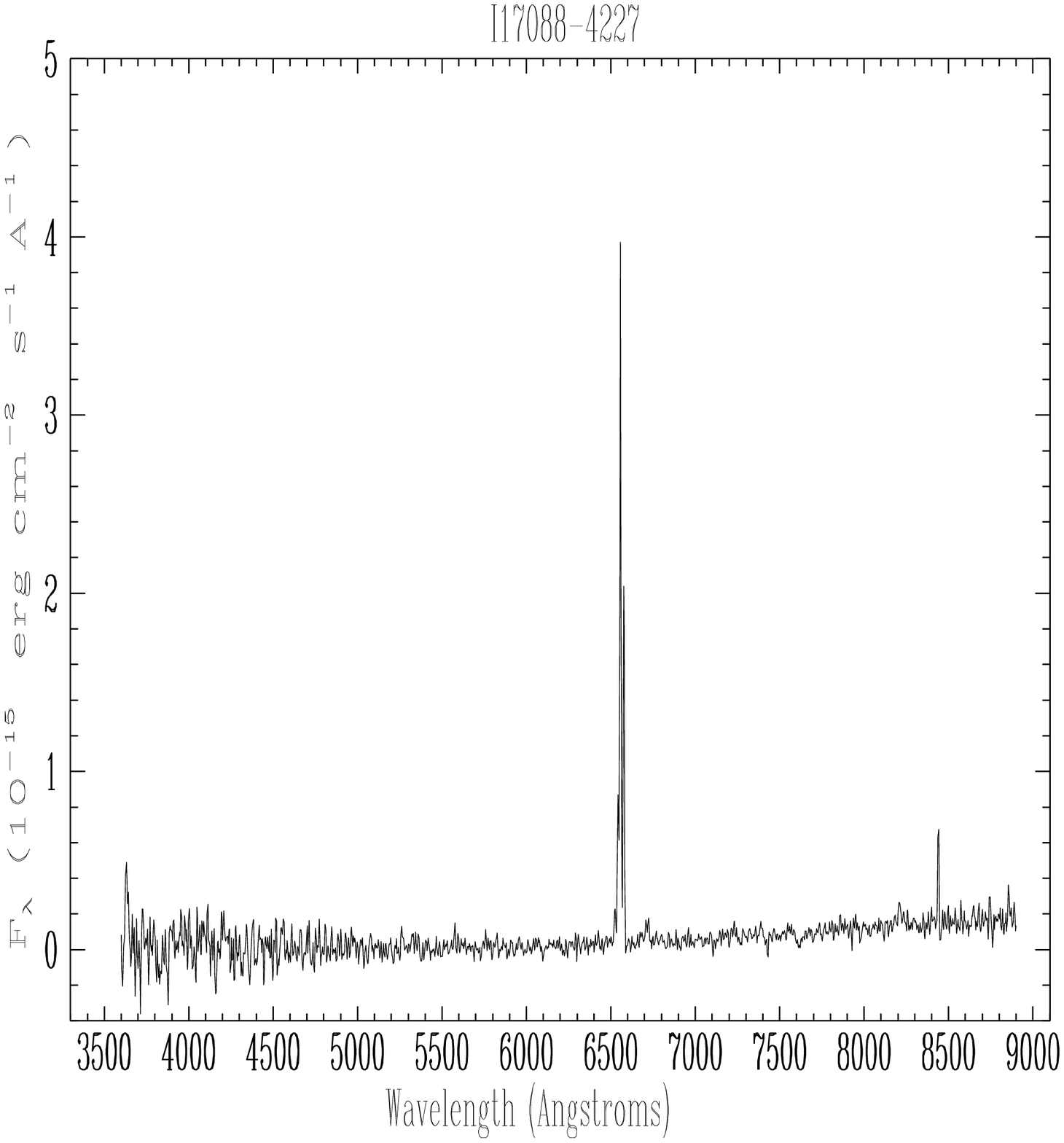}
%\psdraft
\epsfxsize=4cm
\epsfysize=4cm
\epsfbox{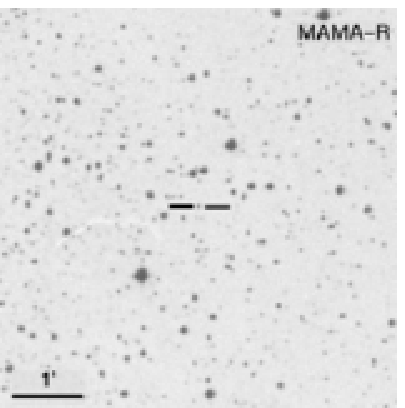}
%\psfull
\end{center}

\begin{center}
\epsfxsize=13.5cm
\epsfysize=4cm
\epsfbox{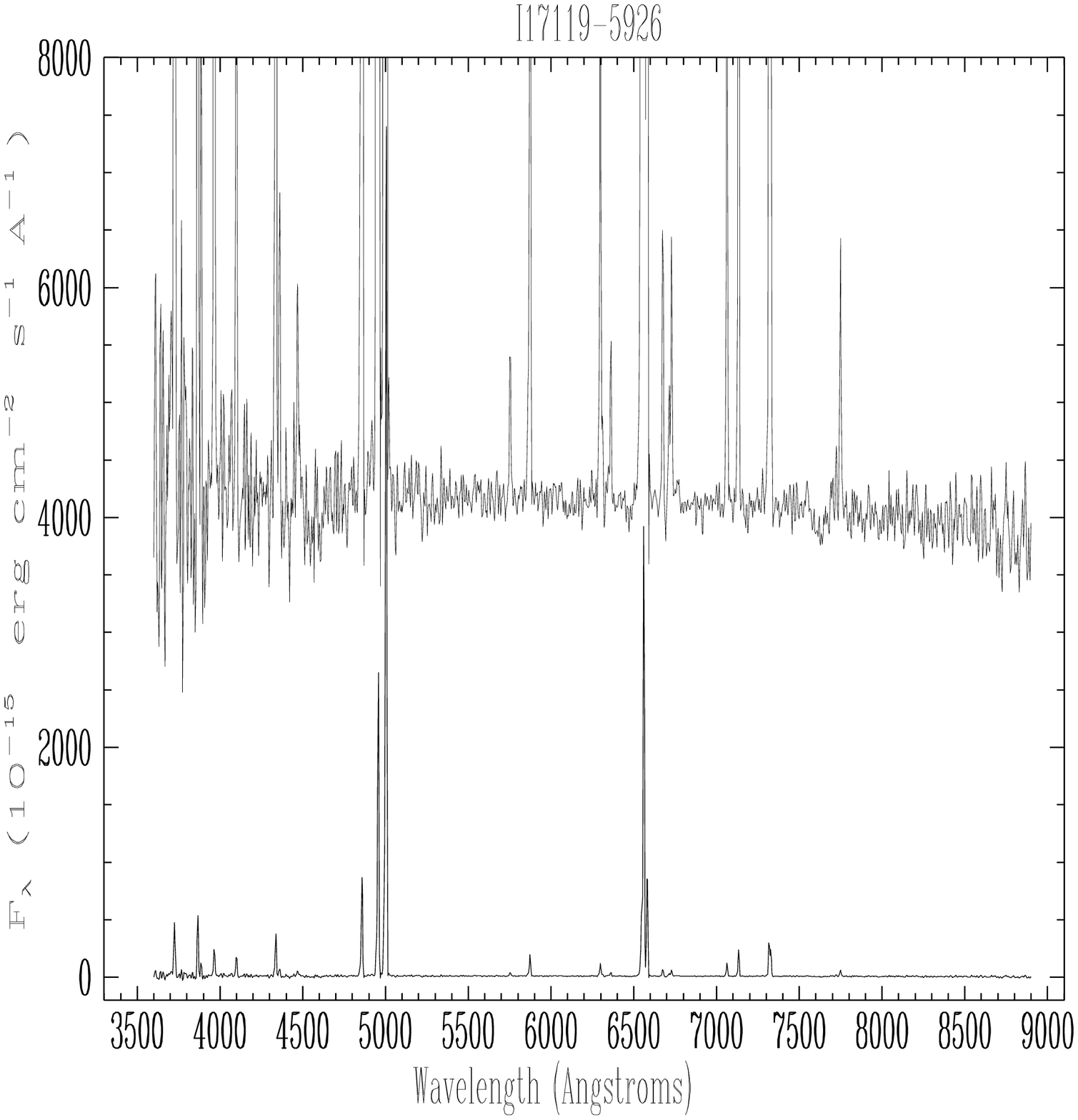}
%\psdraft
\epsfxsize=4cm
\epsfysize=4cm
\epsfbox{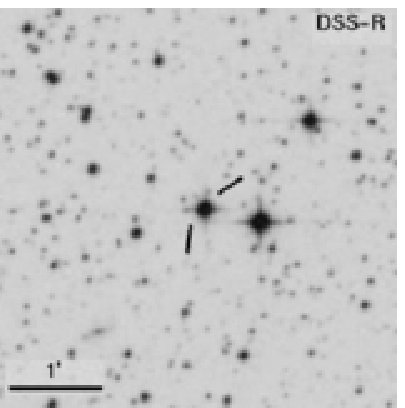}
%\psfull
\end{center}

\begin{center}
\epsfxsize=13.5cm
\epsfysize=4cm
\epsfbox{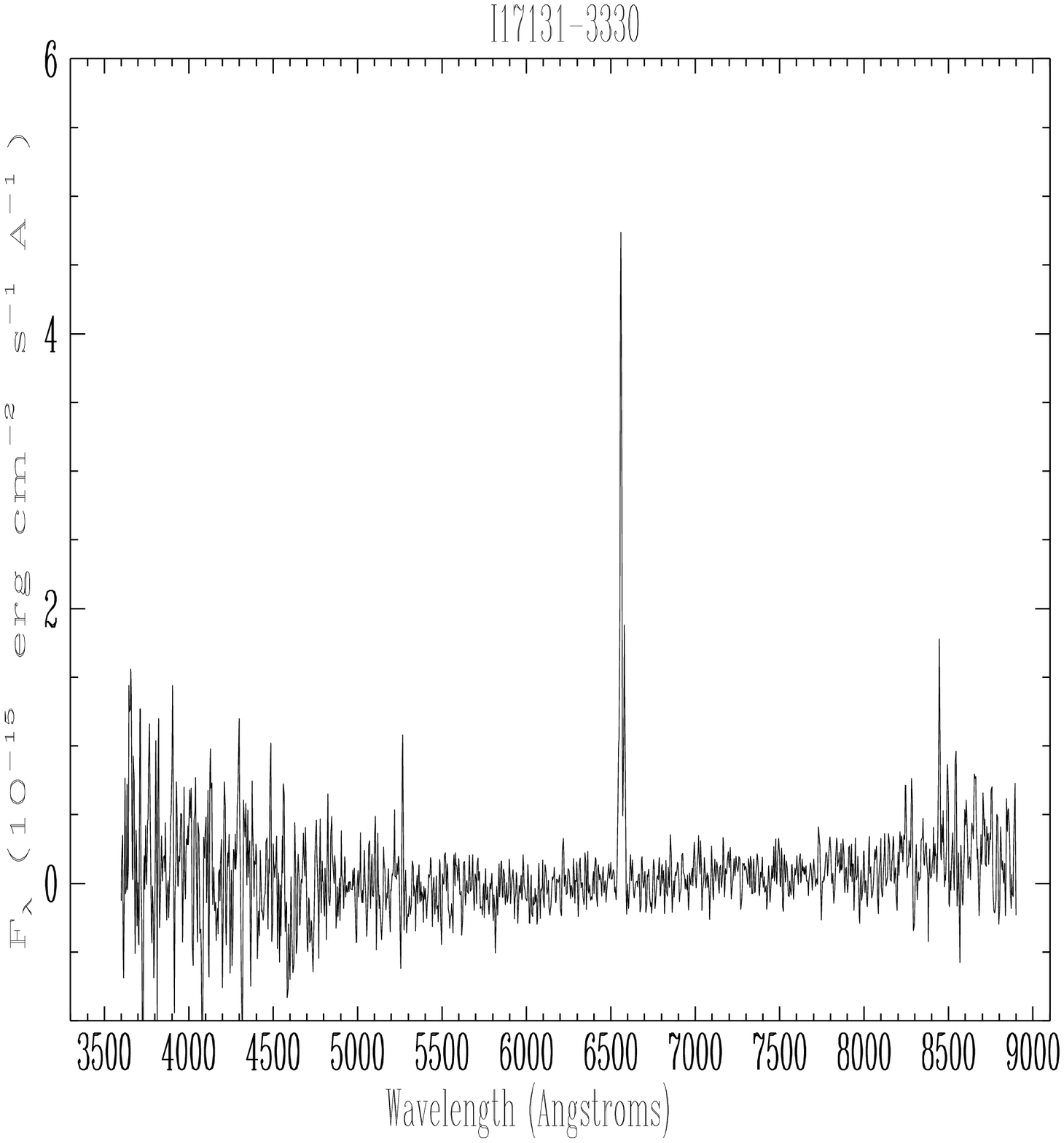}
%\psdraft
\epsfxsize=4cm
\epsfysize=4cm
\epsfbox{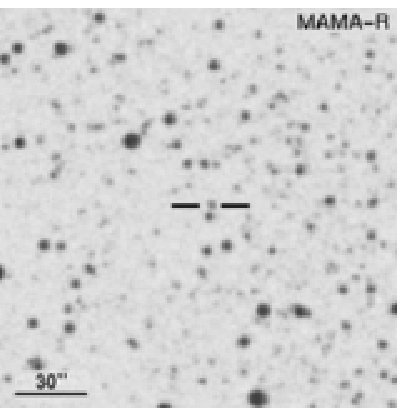}
%\psfull
\end{center}

\caption{Spectra of the PNe in the sample together with their 
corresponding identification charts (continued). }
\end{figure*}

%-------------------------------------------------------------------
%pg6

\begin{figure*}
\setcounter{figure}{2}

\begin{center}
\epsfxsize=13.5cm
\epsfysize=4cm
\epsfbox{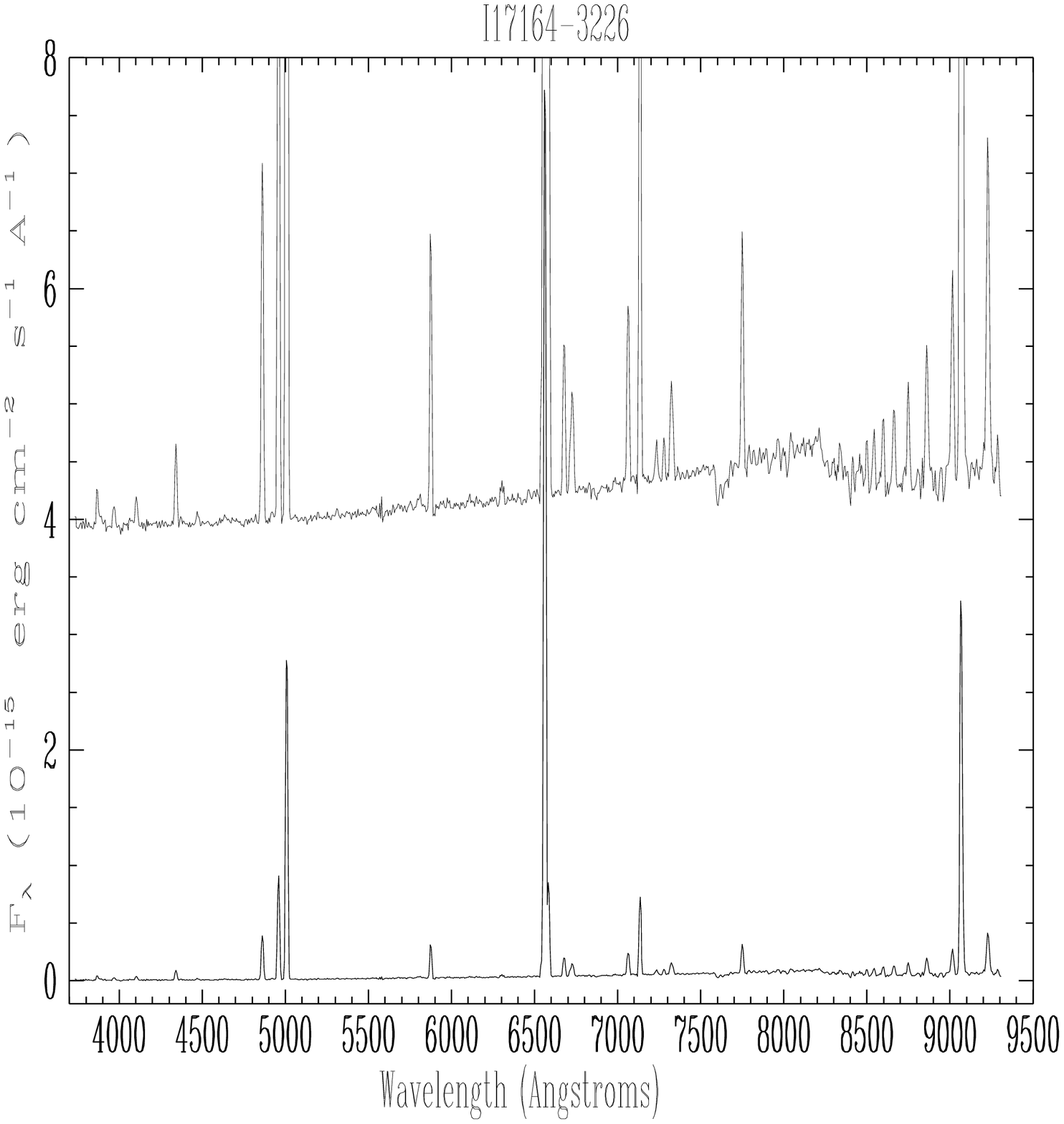}
%\psdraft
\epsfxsize=4cm
\epsfysize=4cm
\epsfbox{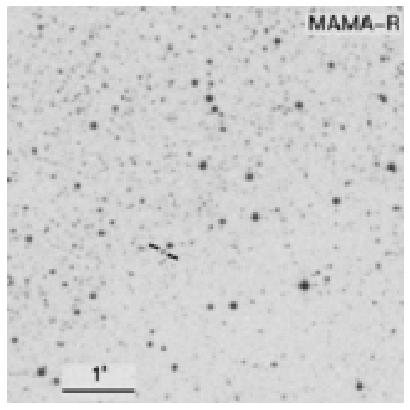}
%\psfull
\end{center}

\begin{center}
\epsfxsize=13.5cm
\epsfysize=4cm
\epsfbox{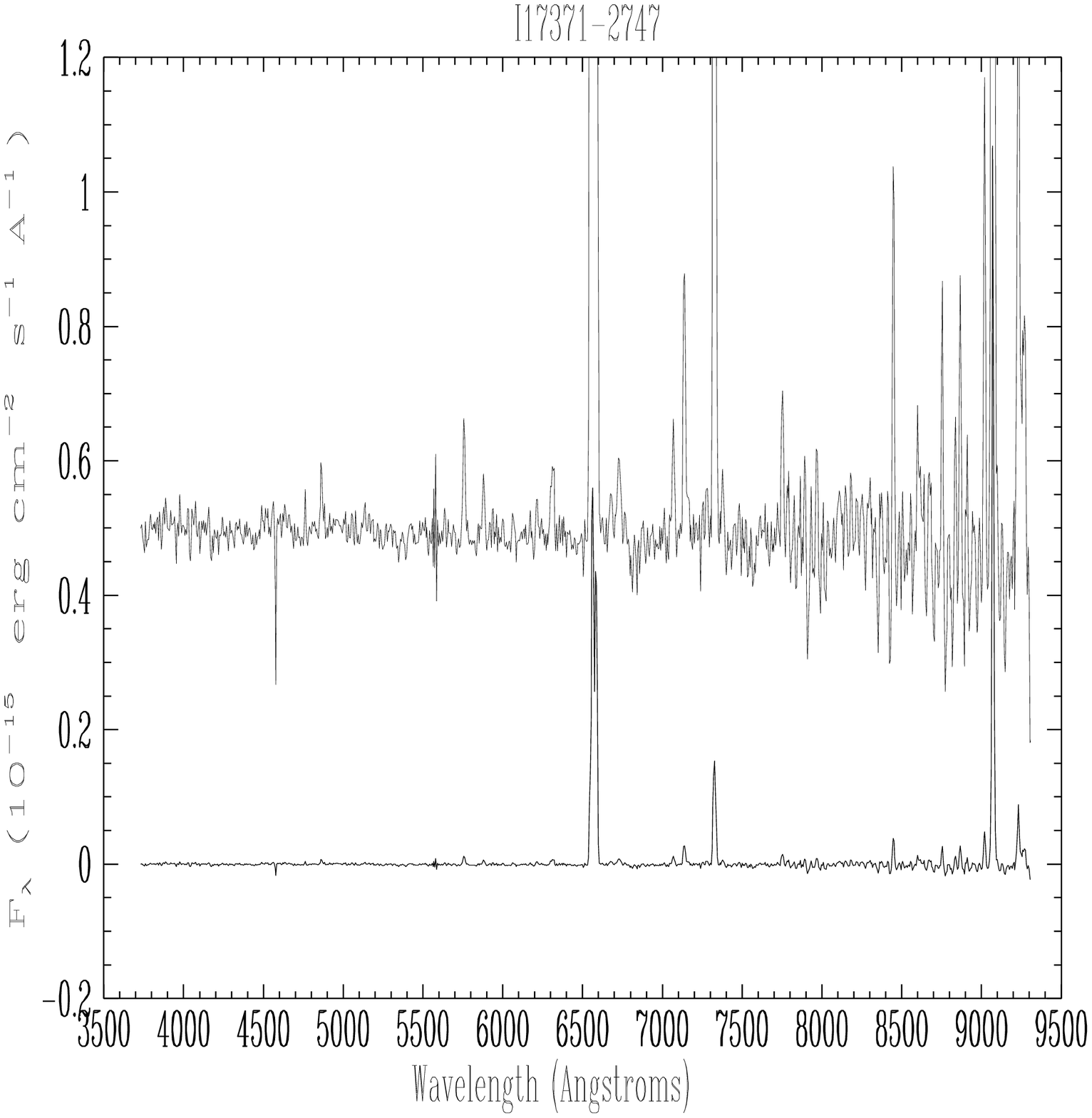}
%\psdraft
\epsfxsize=4cm
\epsfysize=4cm
\epsfbox{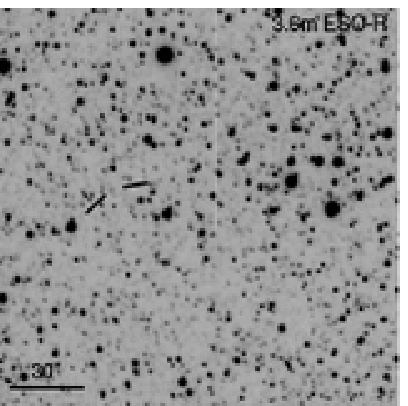}
%\psfull
\end{center}

\begin{center}
\epsfxsize=13.5cm
\epsfysize=4cm
\epsfbox{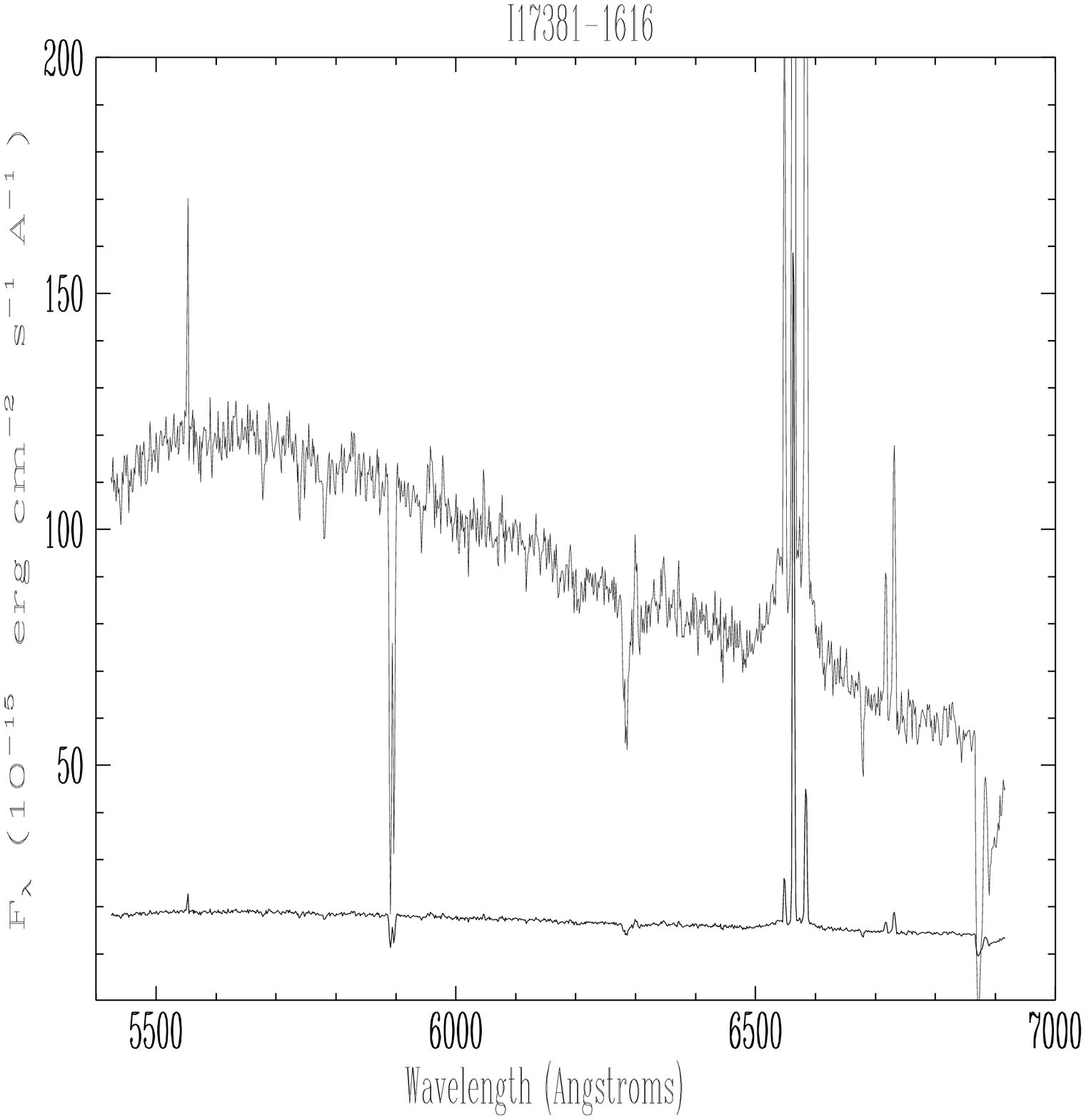}
%\psdraft
\epsfxsize=4cm
\epsfysize=4cm
\epsfbox{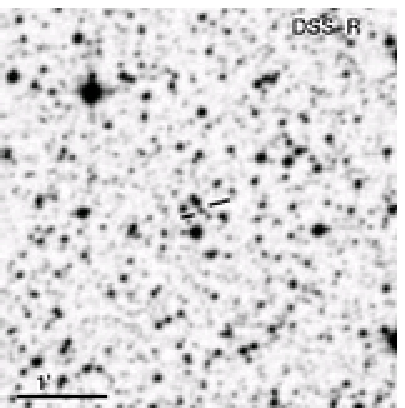}
%\psfull
\end{center}

\begin{center}
\epsfxsize=13.5cm
\epsfysize=4cm
\epsfbox{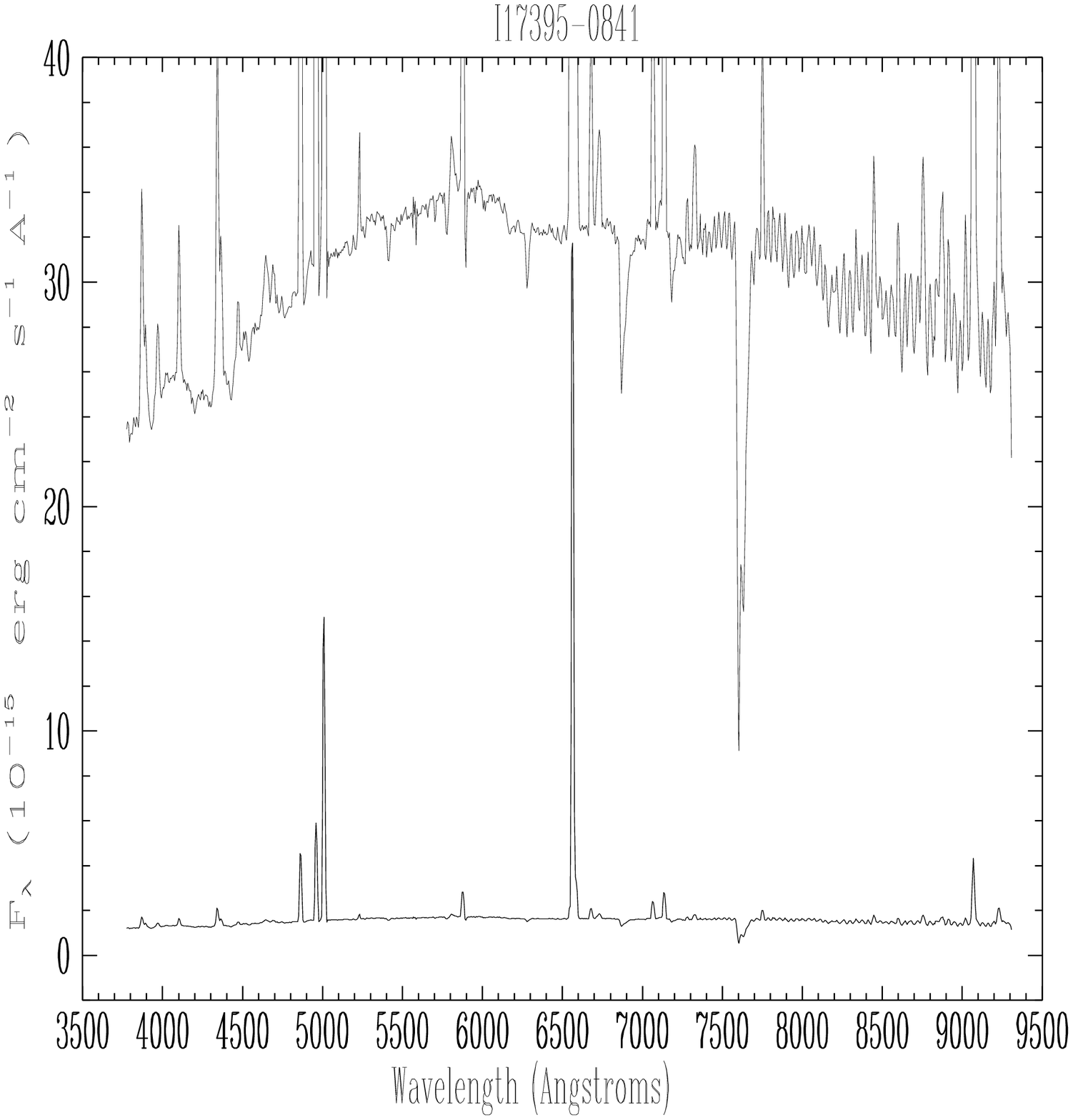}
%\psdraft
\epsfxsize=4cm
\epsfysize=4cm
\epsfbox{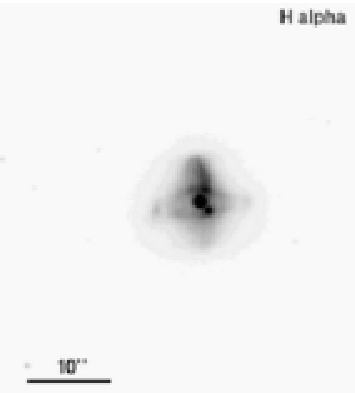}
%\psfull
\end{center}

\begin{center}
\epsfxsize=13.5cm
\epsfysize=4cm
\epsfbox{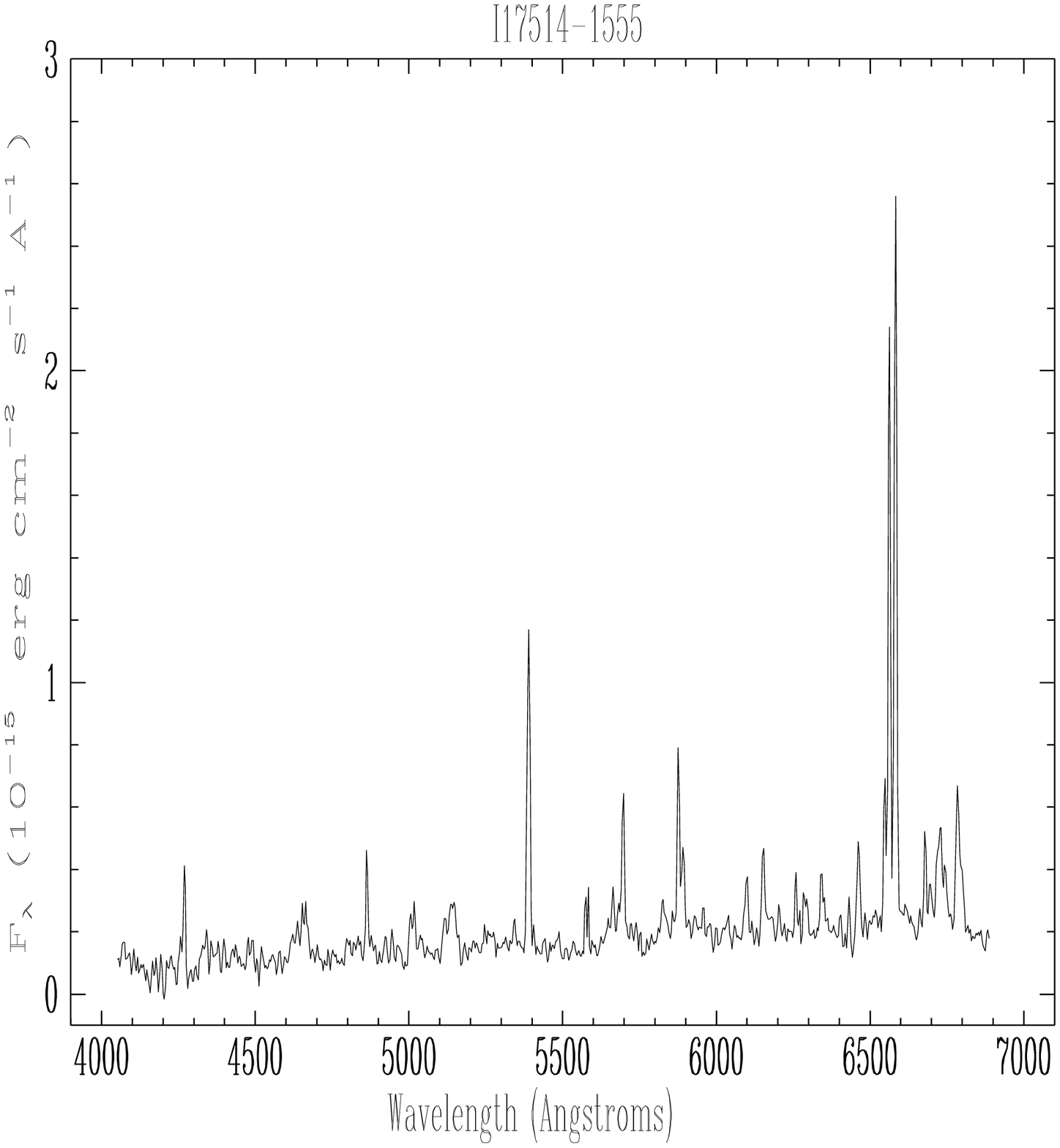}
%\psdraft
\epsfxsize=4cm
\epsfysize=4cm
\epsfbox{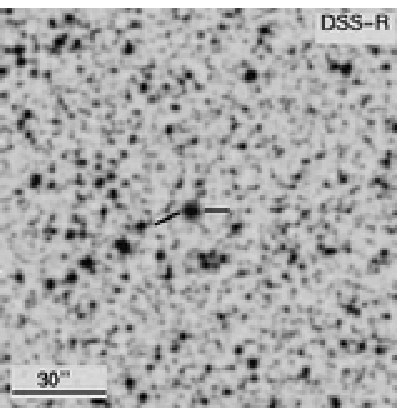}
%\psfull
\end{center}

\caption{Spectra of the PNe in the sample together with their 
corresponding identification charts (continued). }
\end{figure*}

%-------------------------------------------------------------------
%pg7

\begin{figure*}
\setcounter{figure}{2}

\begin{center}
\epsfxsize=13.5cm
\epsfysize=4cm
\epsfbox{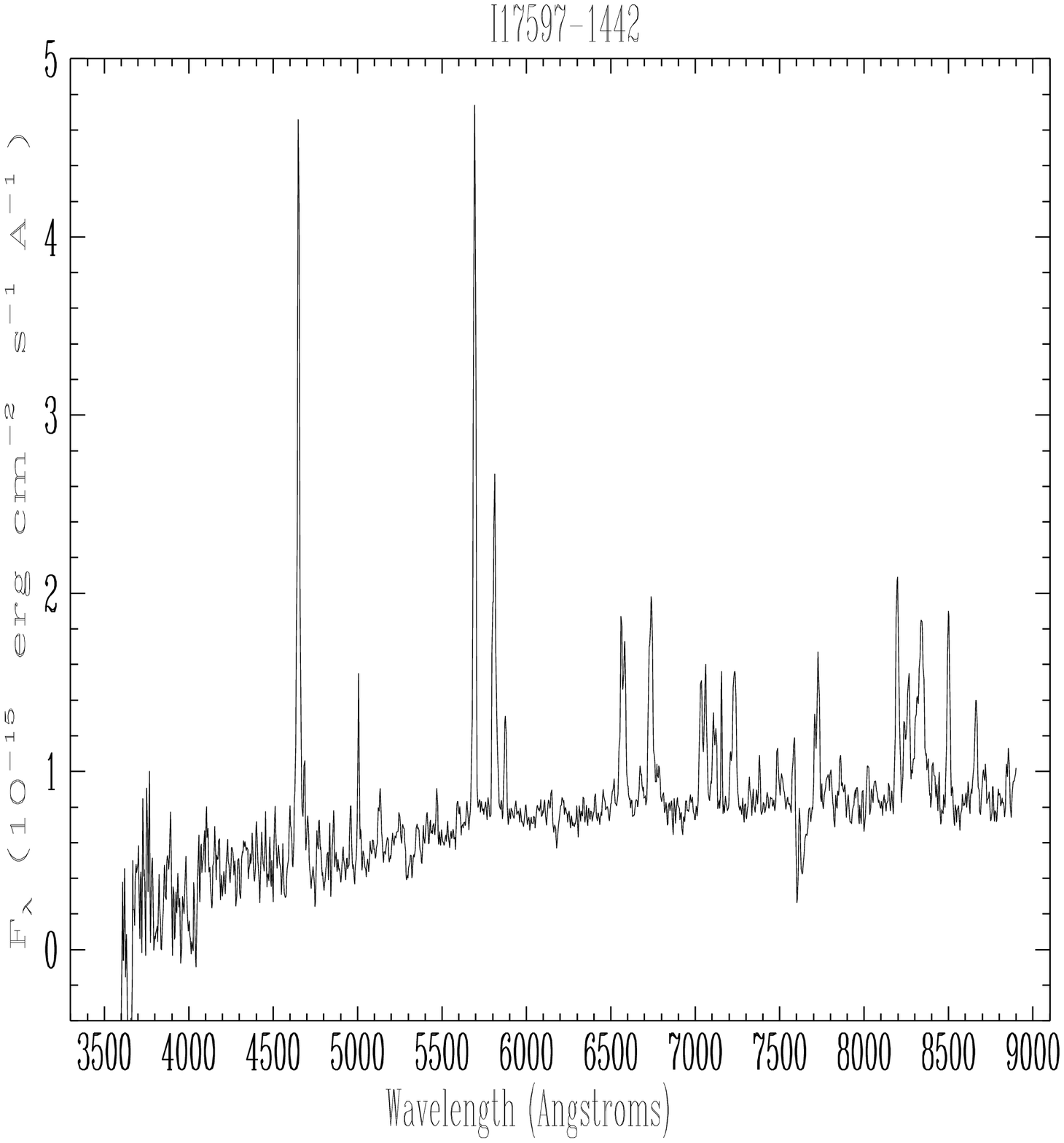}
%\psdraft
\epsfxsize=4cm
\epsfysize=4cm
\epsfbox{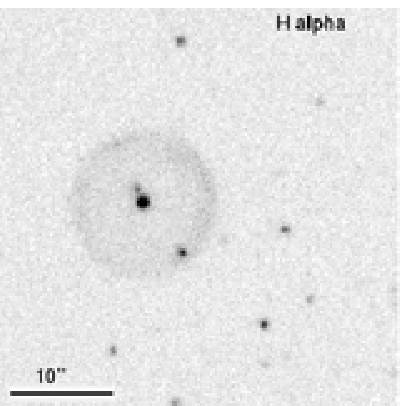}
%\psfull
\end{center}

\begin{center}
\epsfxsize=13.5cm
\epsfysize=4cm
\epsfbox{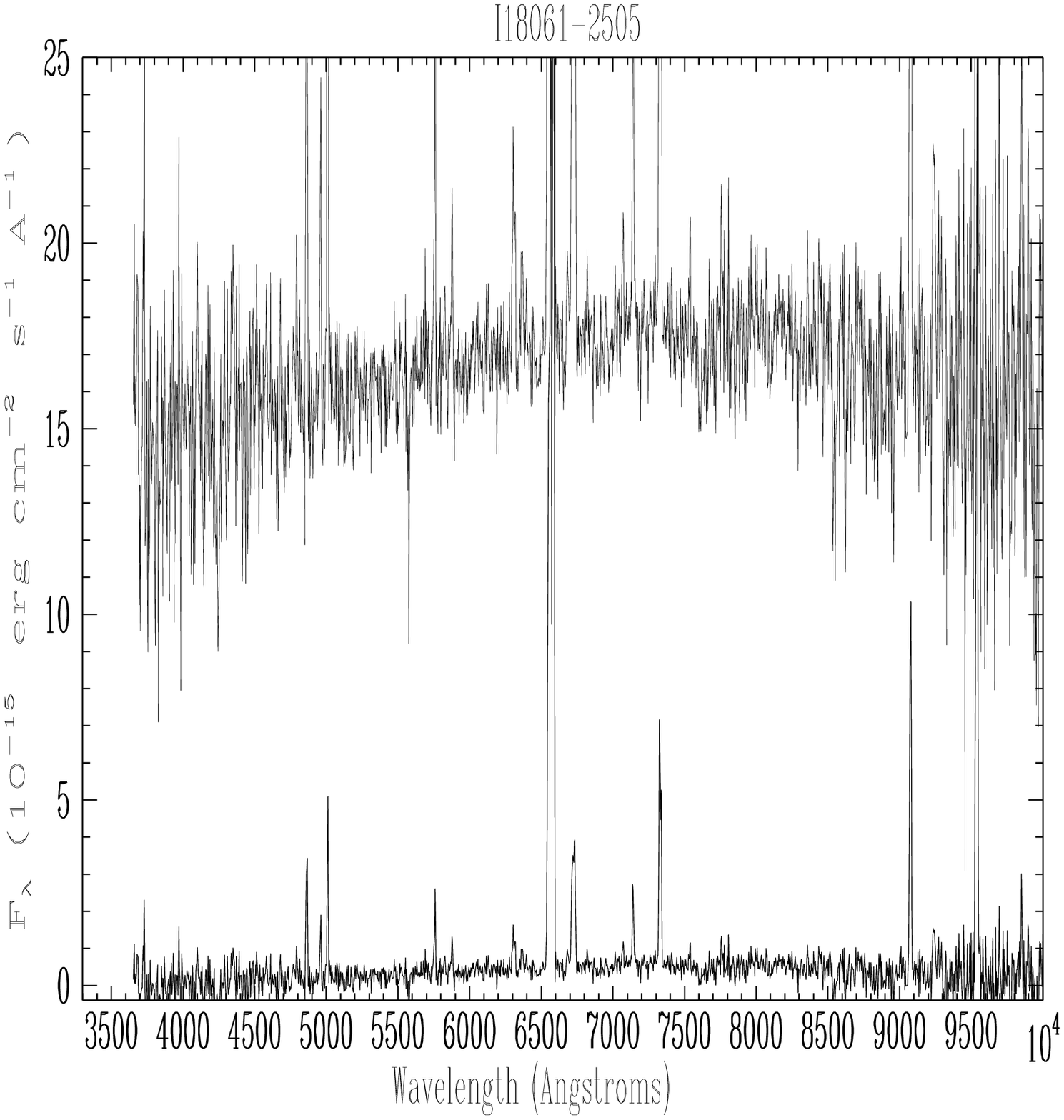}
%\psdraft
\epsfxsize=4cm
\epsfysize=4cm
\epsfbox{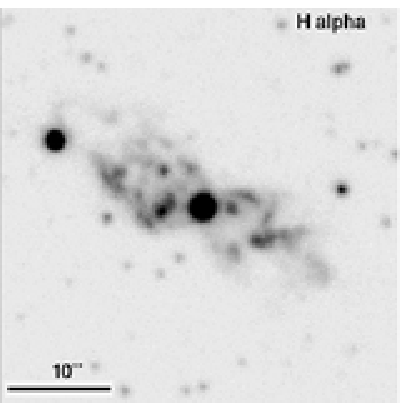}
%\psfull
\end{center}

\begin{center}
\epsfxsize=13.5cm
\epsfysize=4cm
\epsfbox{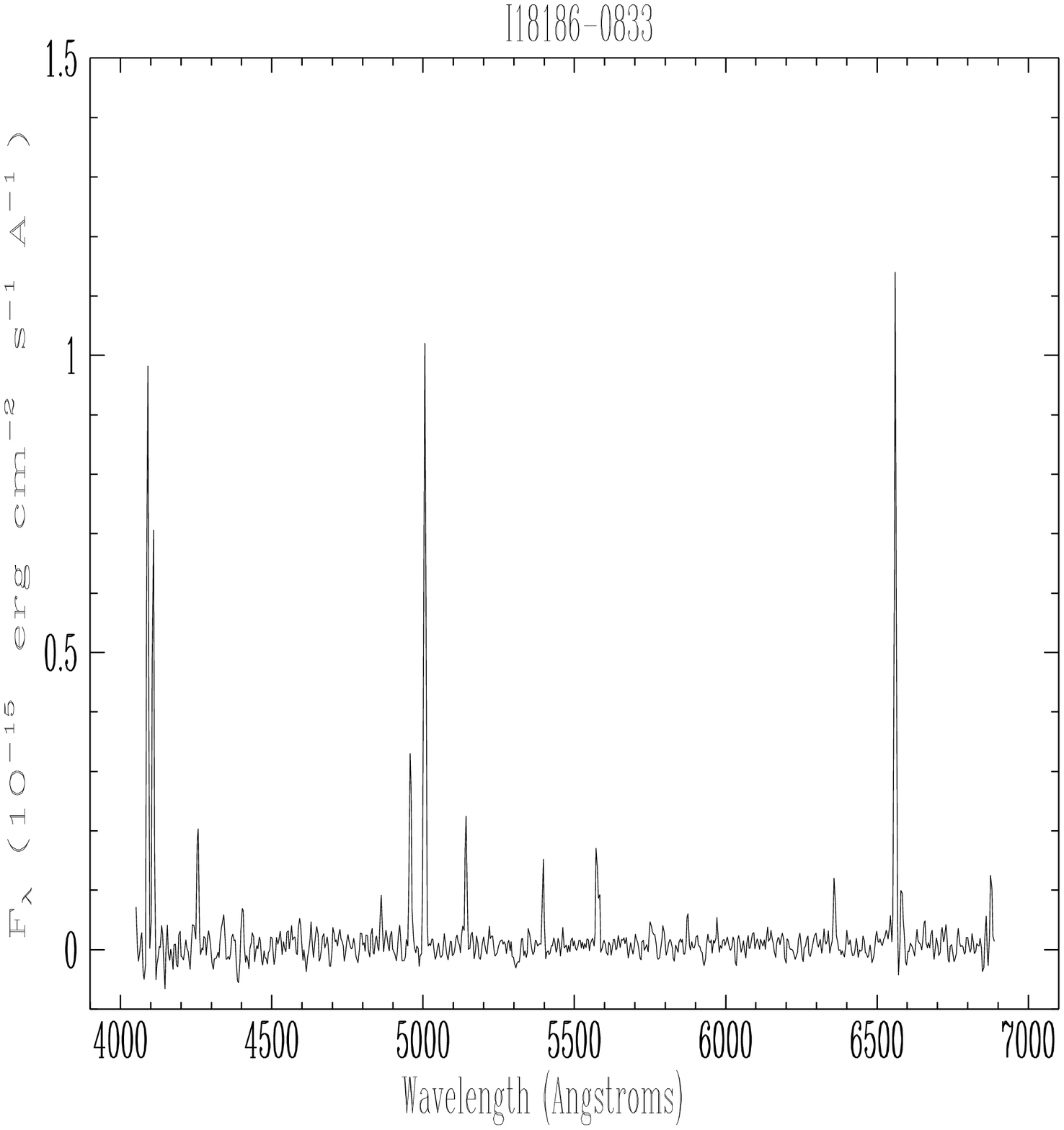}
%\psdraft
\epsfxsize=4cm
\epsfysize=4cm
\epsfbox{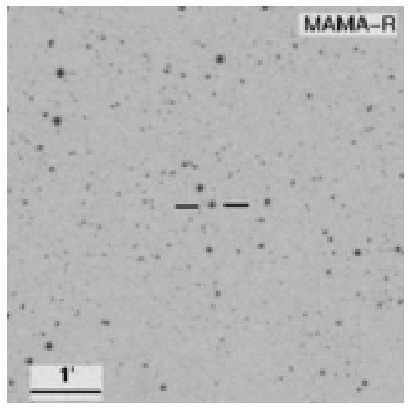}
%\psfull
\end{center}

\begin{center}
\epsfxsize=13.5cm
\epsfysize=4cm
\epsfbox{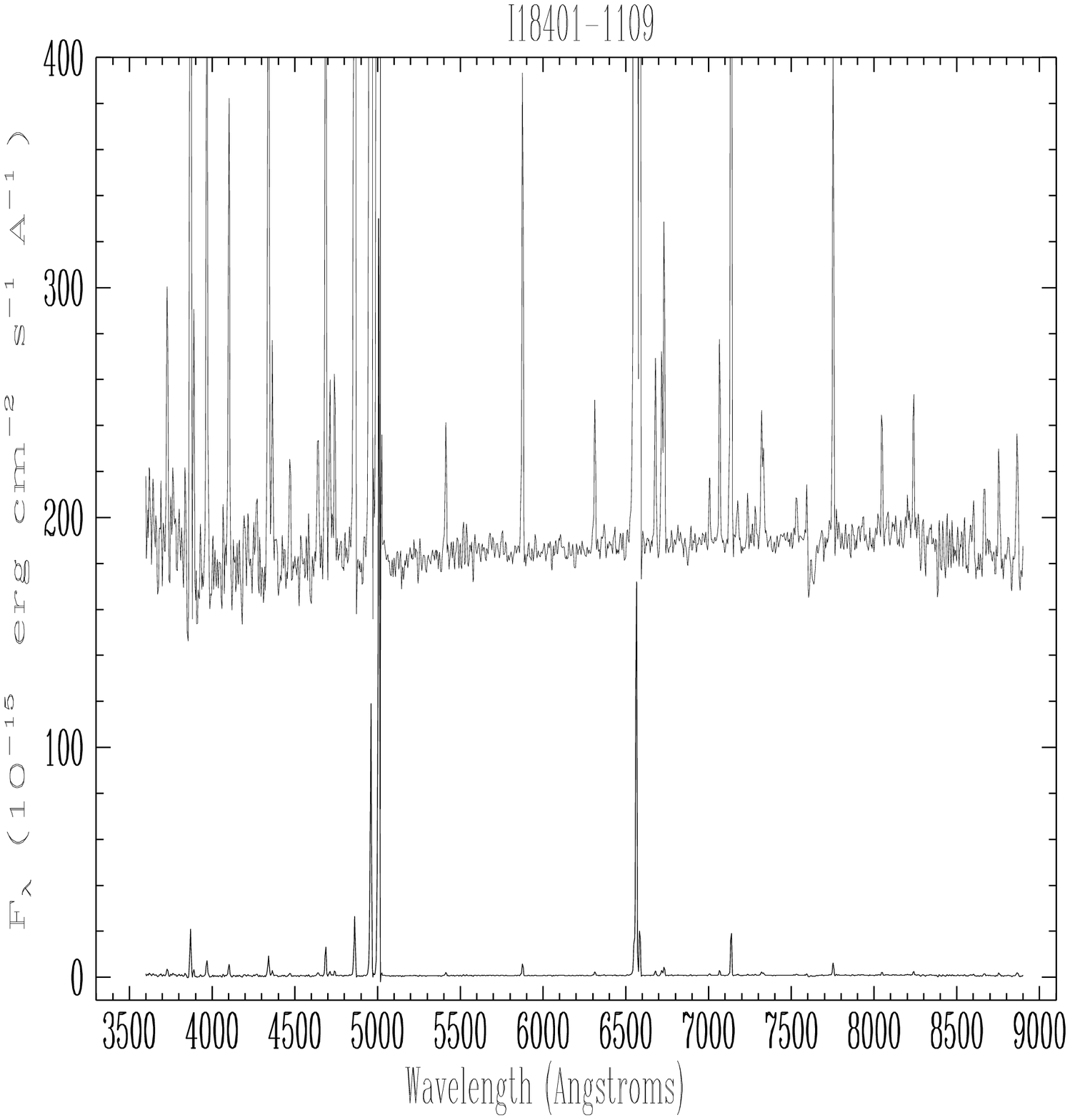}
%\psdraft
\epsfxsize=4cm
\epsfysize=4cm
\epsfbox{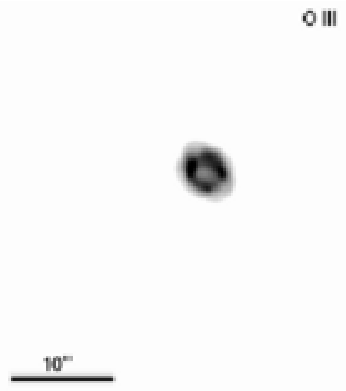}
%\psfull
\end{center}

\begin{center}
\epsfxsize=13.5cm
\epsfysize=4cm
\epsfbox{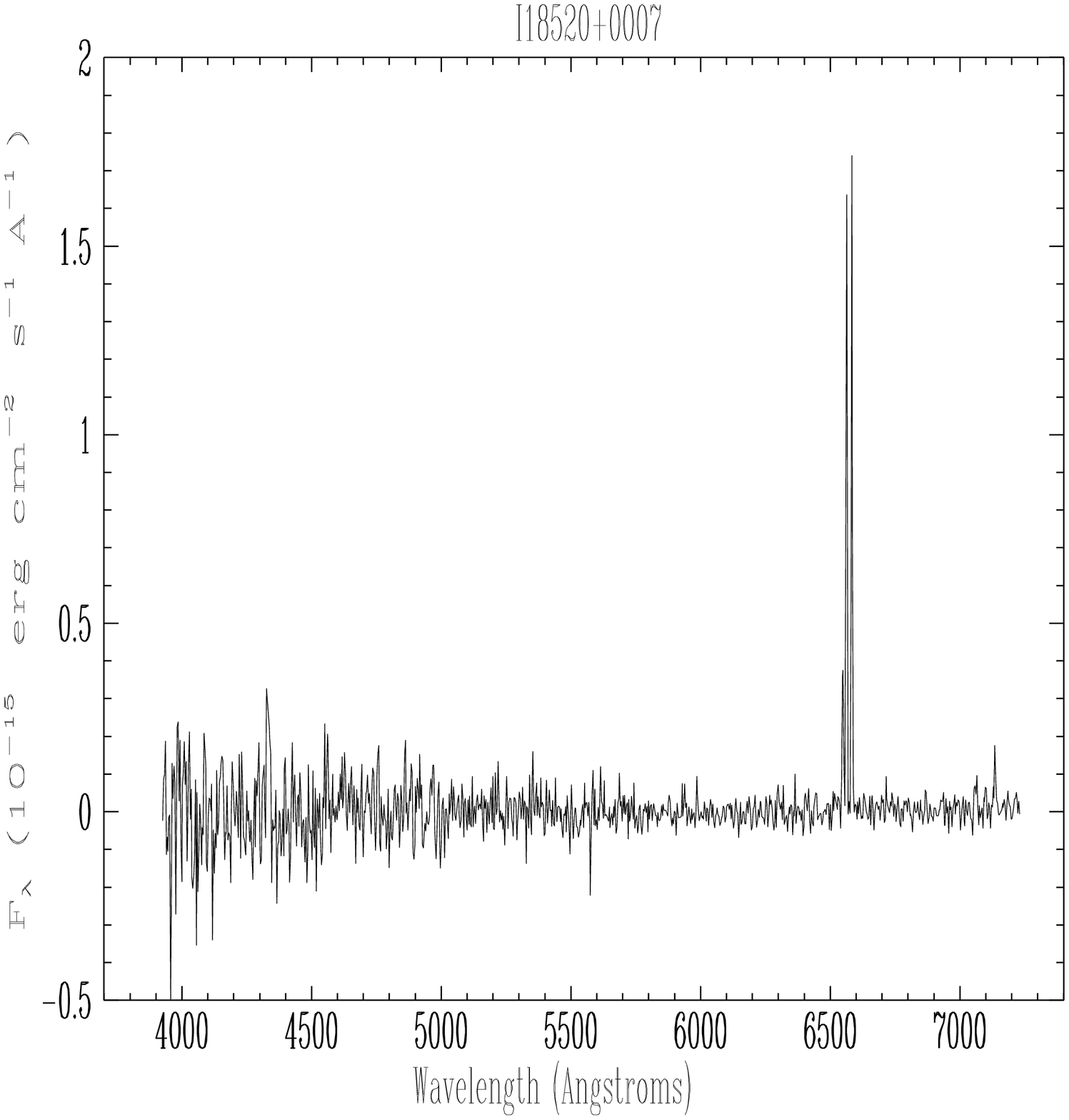}
%\psdraft
\epsfxsize=4cm
\epsfysize=4cm
\epsfbox{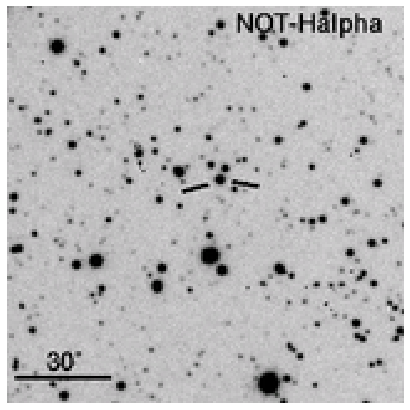}
%\psfull
\end{center}

\caption{Spectra of the PNe in the sample together with their 
corresponding identification charts (continued). }
\end{figure*}

%-------------------------------------------------------------------
%pg8

\begin{figure*}
\setcounter{figure}{2}

\begin{center}
\epsfxsize=13.5cm
\epsfysize=4cm
\epsfbox{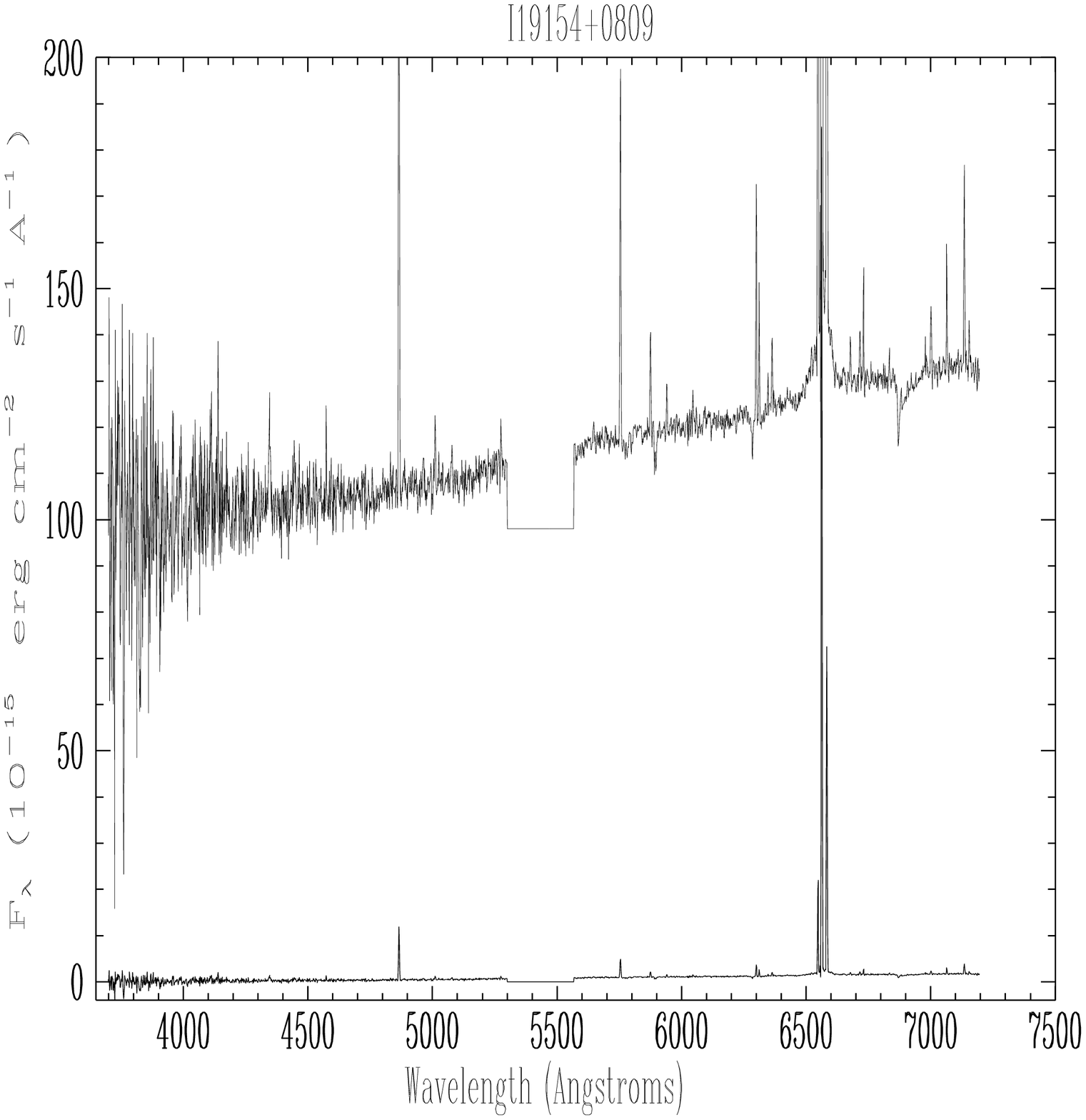}
%\psdraft
\epsfxsize=4cm
\epsfysize=4cm
\epsfbox{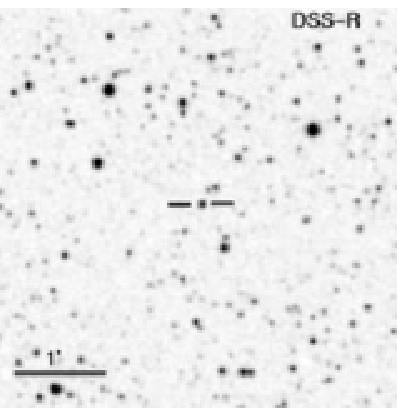}
%\psfull
\end{center}

\begin{center}
\epsfxsize=13.5cm
\epsfysize=4cm
\epsfbox{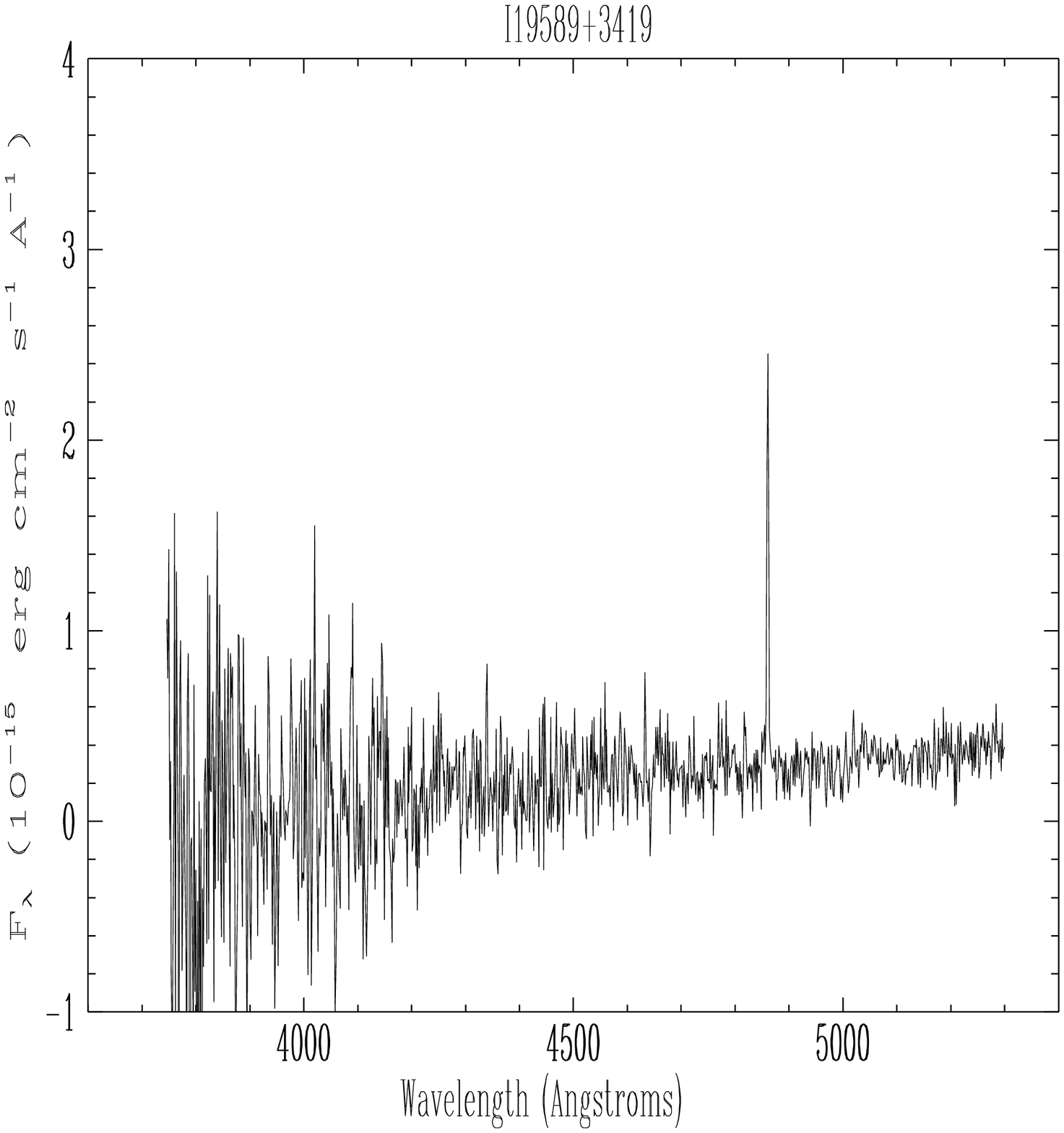}
%\psdraft
\epsfxsize=4cm
\epsfysize=4cm
\epsfbox{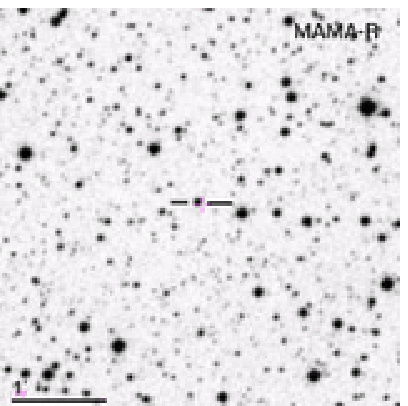}
%\psfull
\end{center}

\caption{Spectra of the PNe in the sample together with their 
corresponding identification charts (continued). }
\end{figure*}

%\end{document}

%%% Local Variables: 
%%% mode: latex
%%% TeX-master: "~/tesis/mitesis/final/tesis"
%%% End: 

\clearpage
\section{Atlas of young stellar objects}
        \begin{figure*}[h!]
\setcounter{figure}{0}

\begin{center}
\epsfxsize=13.5cm
\epsfysize=4cm
\epsfbox{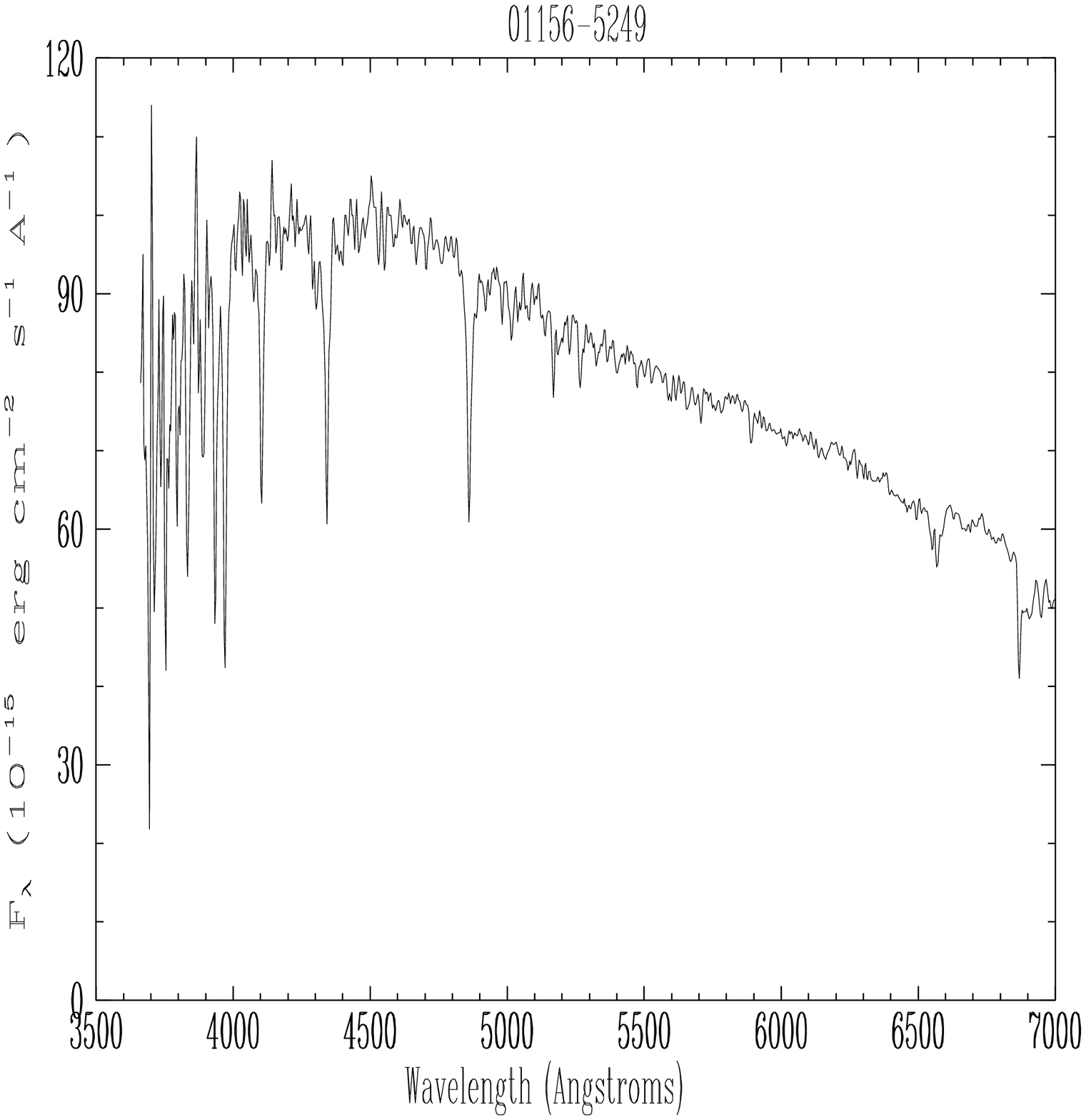}
%\psdraft
\epsfxsize=4cm
\epsfysize=4cm
\epsfbox{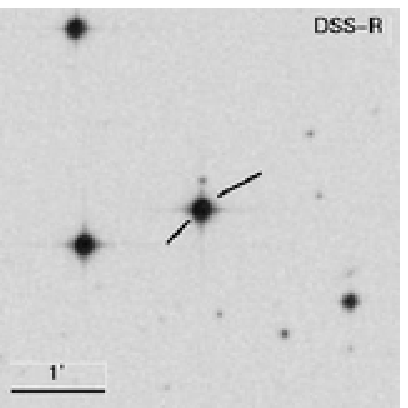}
%\psfull
\end{center}

\begin{center}
\epsfxsize=13.5cm
\epsfysize=4cm
\epsfbox{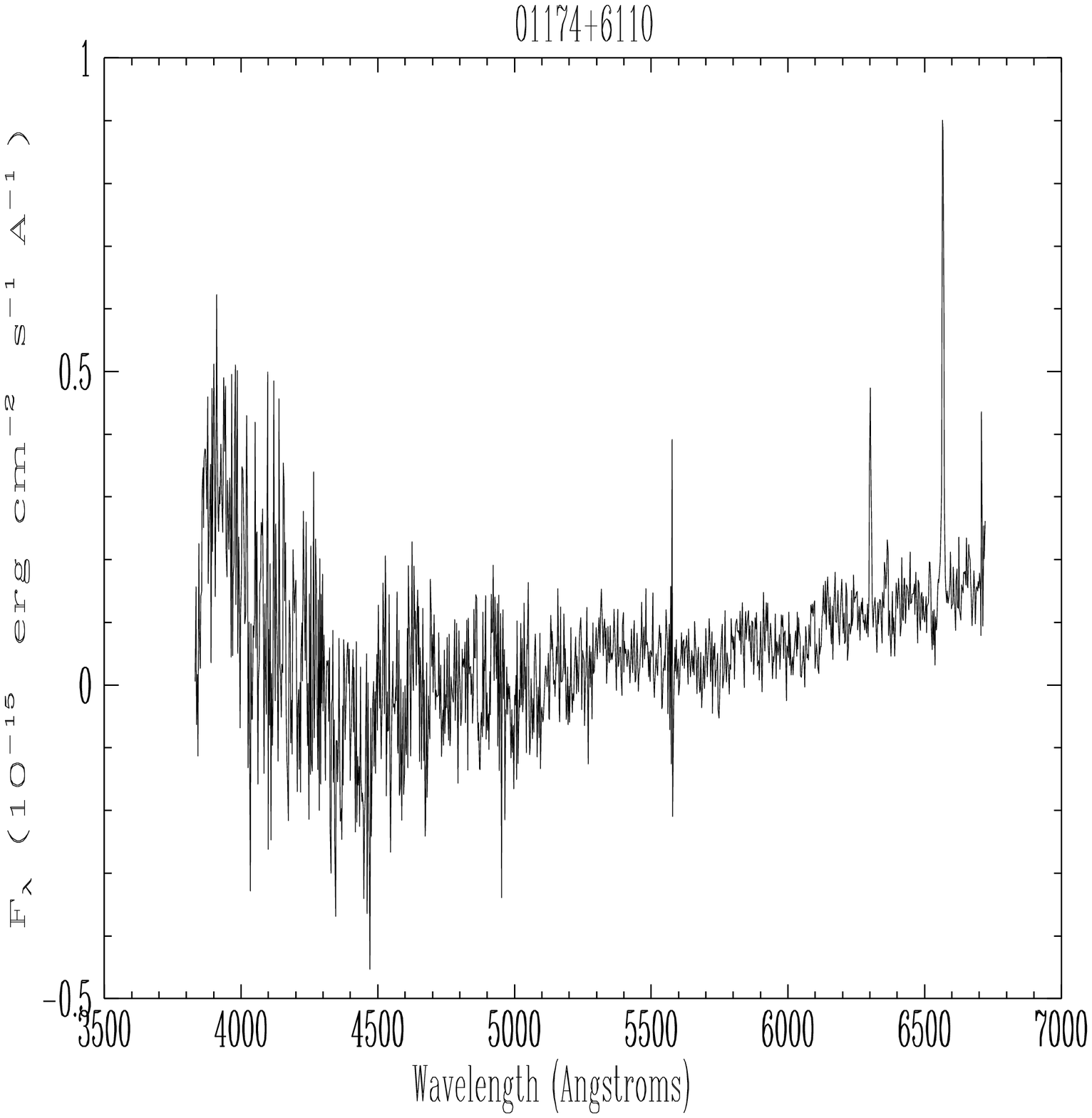}
%\psdraft
\epsfxsize=4cm
\epsfysize=4cm
\epsfbox{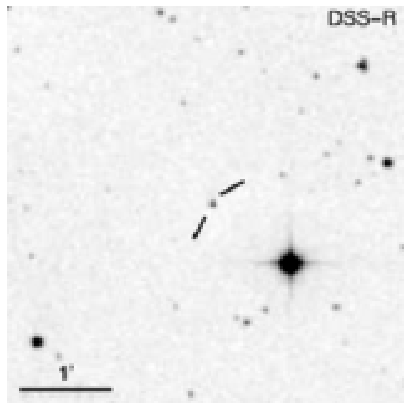}
%\psfull
\end{center}

\begin{center}
\epsfxsize=13.5cm
\epsfysize=4cm
\epsfbox{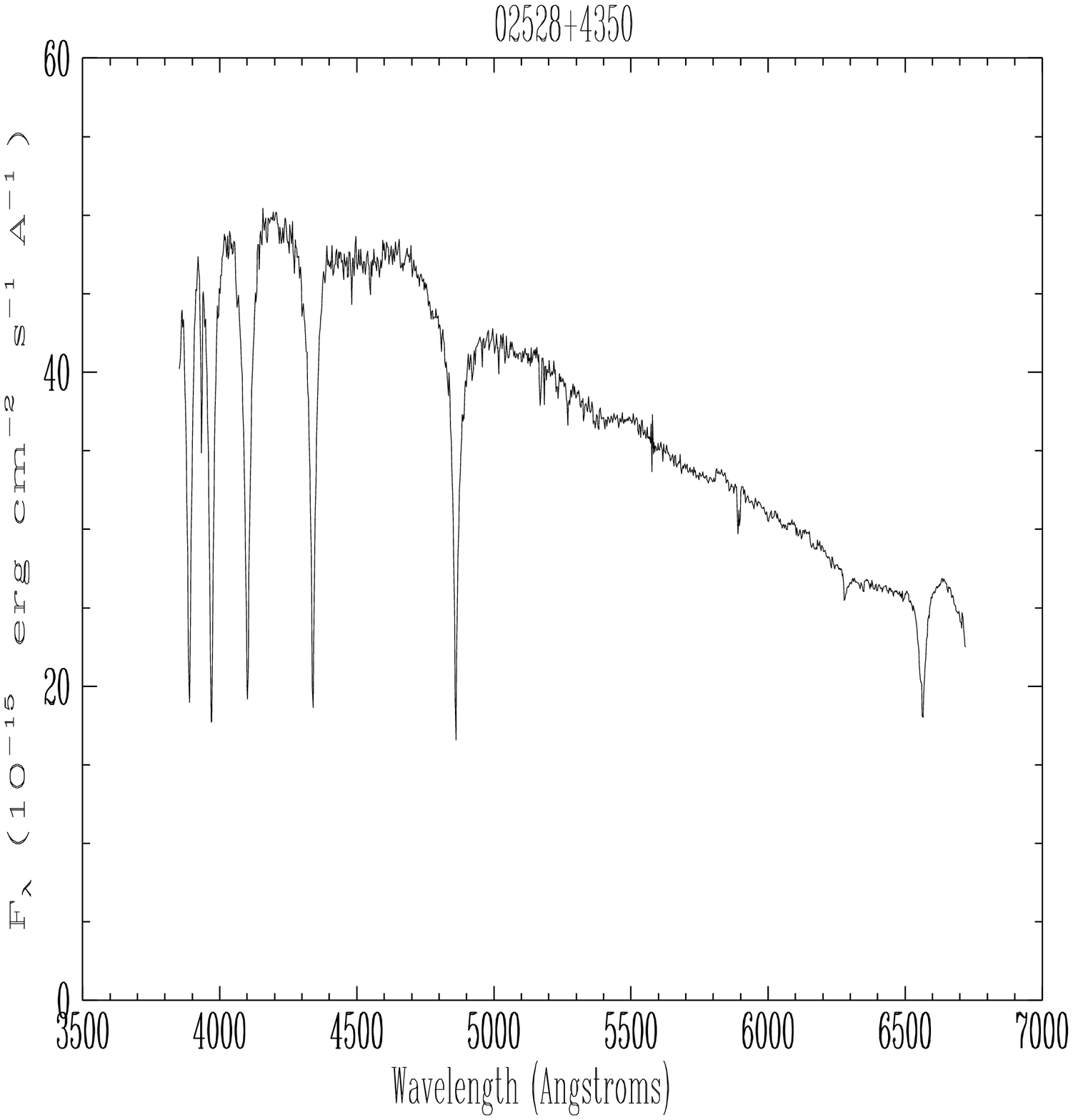}
%\psdraft
\epsfxsize=4cm
\epsfysize=4cm
\epsfbox{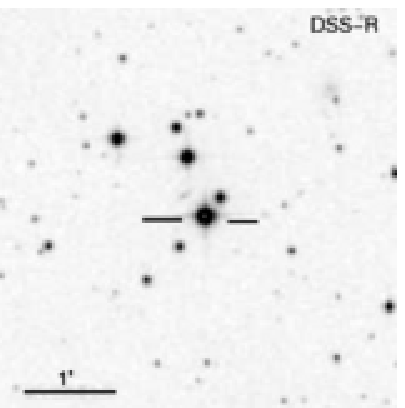}
%\psfull
\end{center}

\begin{center}
\epsfxsize=13.5cm
\epsfysize=4cm
\epsfbox{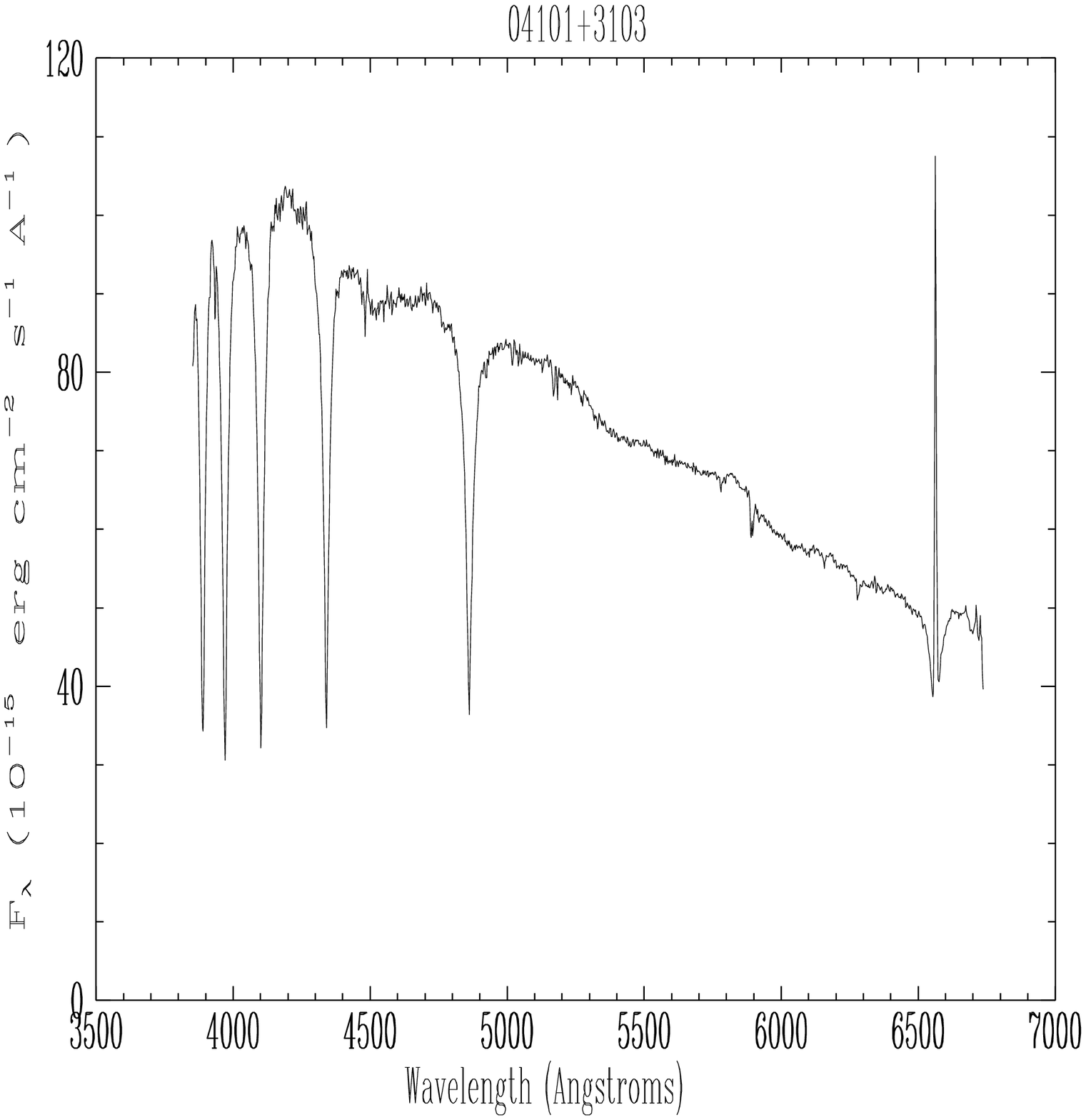}
%\psdraft
\epsfxsize=4cm
\epsfysize=4cm
\epsfbox{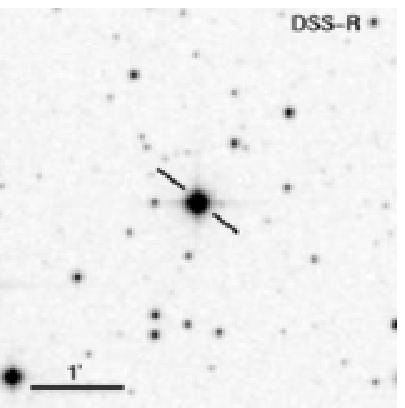}
%\psfull
\end{center}

\caption{Spectra of the objects classified as young stars in the sample together with their 
corresponding identification charts. }
\end{figure*}

%-------------------------------------------------------------------
%pg2.
%\setcounter{figure}{9}
\begin{figure*}
\setcounter{figure}{0}

\begin{center}
\epsfxsize=13.5cm
\epsfysize=4cm
\epsfbox{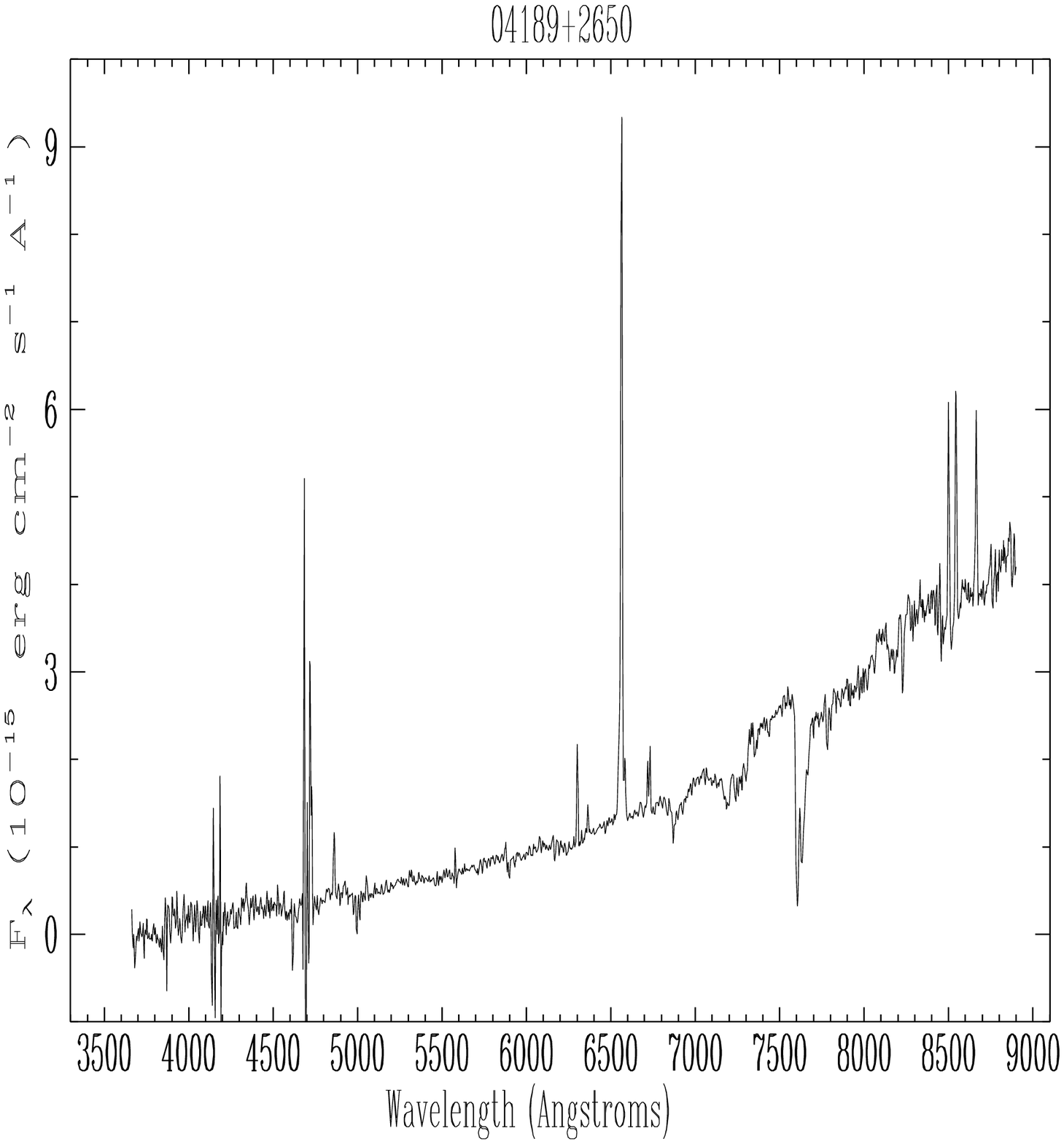}
%\psdraft
\epsfxsize=4cm
\epsfysize=4cm
\epsfbox{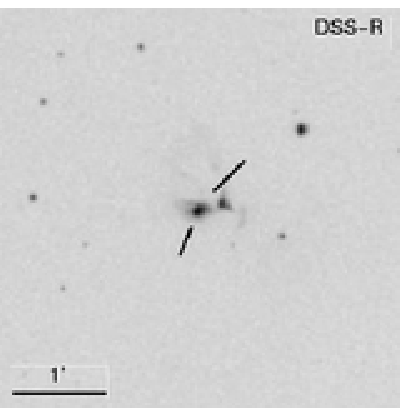}
%\psfull
\end{center}

\begin{center}
\epsfxsize=13.5cm
\epsfysize=4cm
\epsfbox{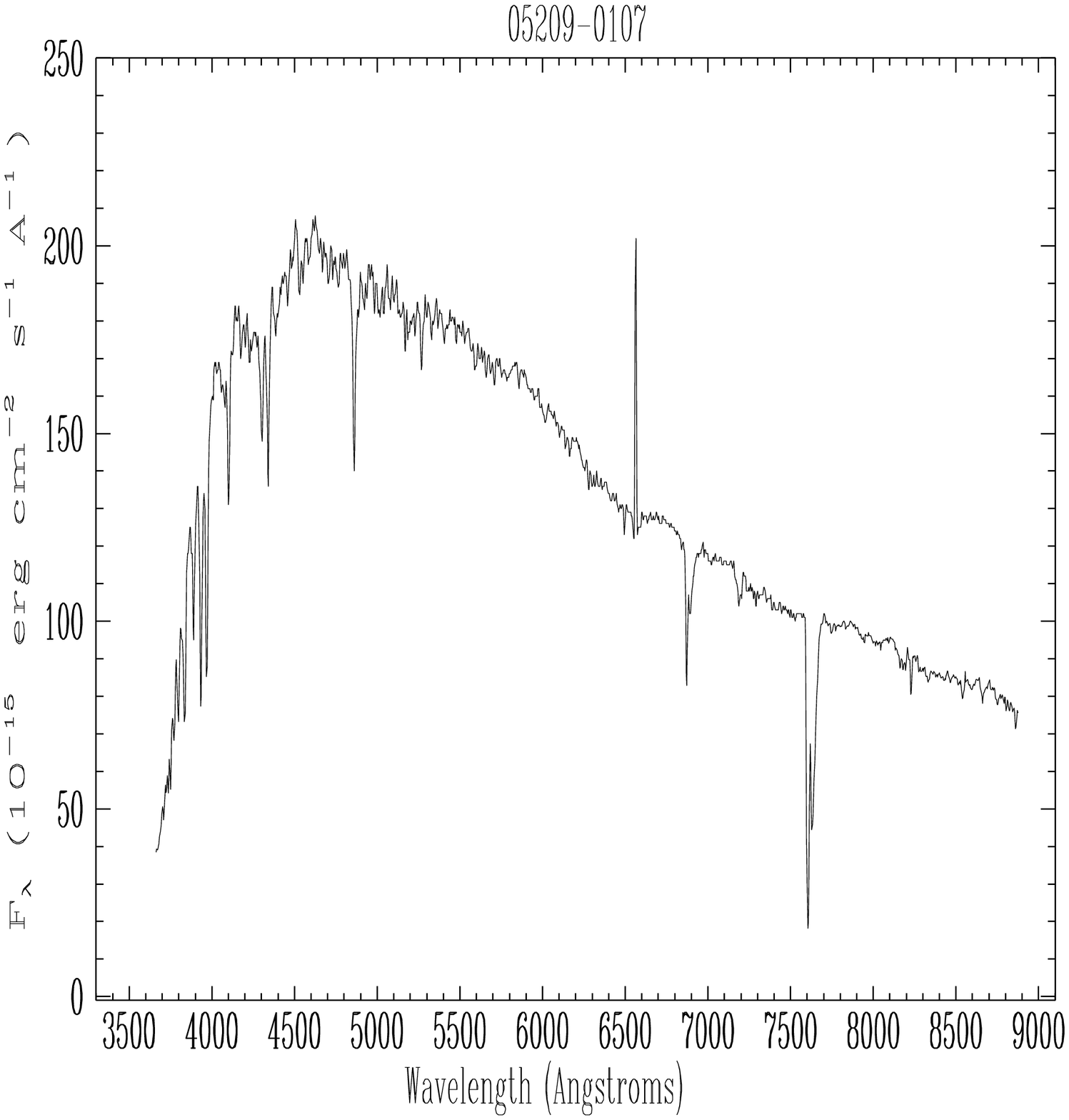}
%\psdraft
\epsfxsize=4cm
\epsfysize=4cm
\epsfbox{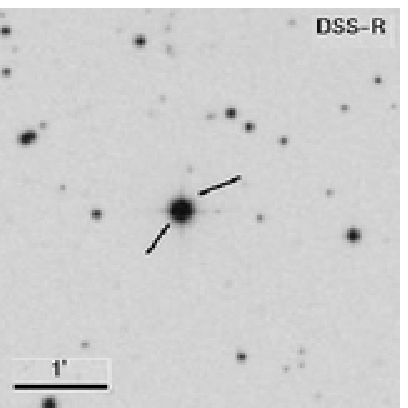}
%\psfull
\end{center}

\begin{center}
\epsfxsize=13.5cm
\epsfysize=4cm
\epsfbox{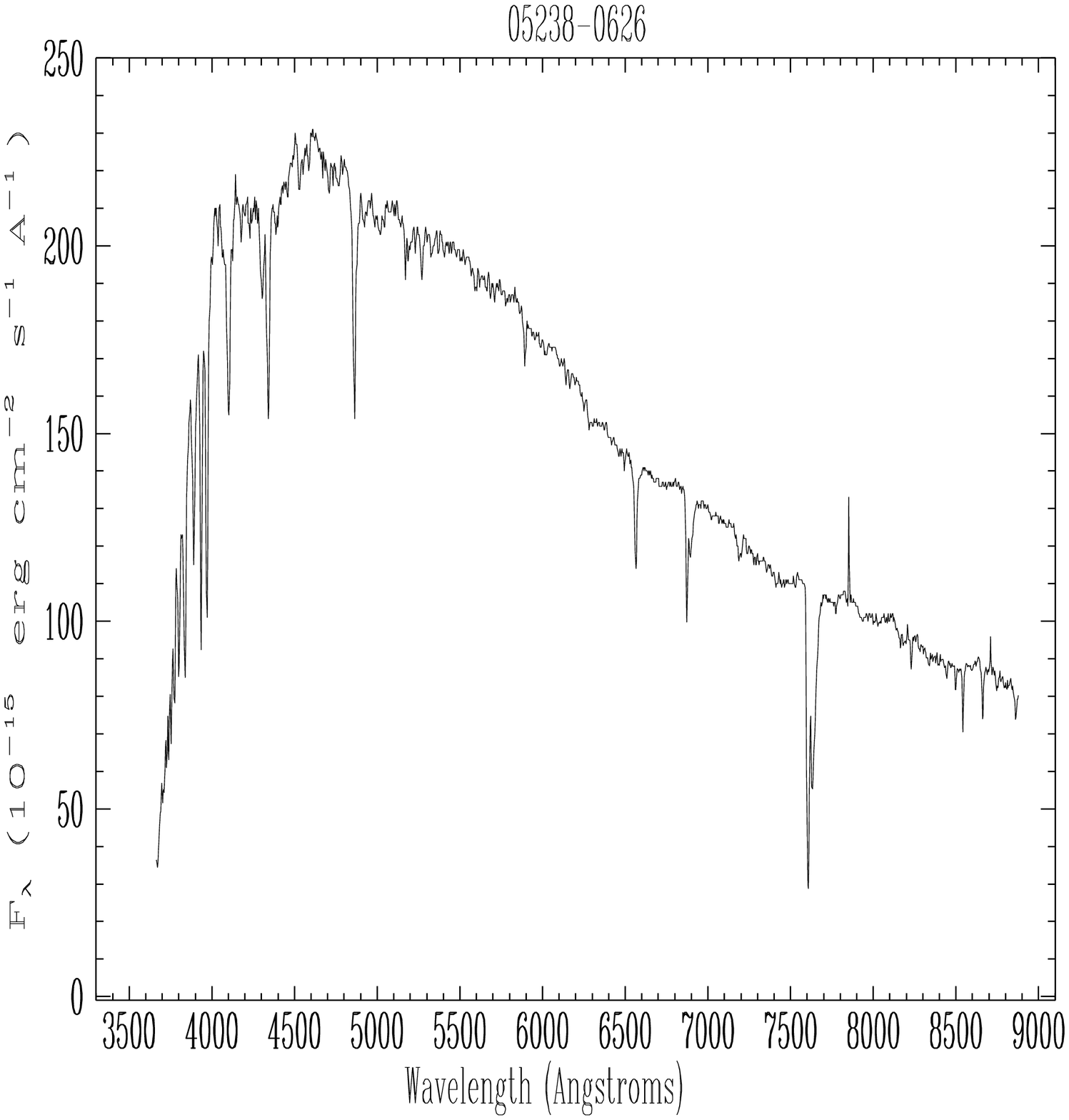}
%\psdraft
\epsfxsize=4cm
\epsfysize=4cm
\epsfbox{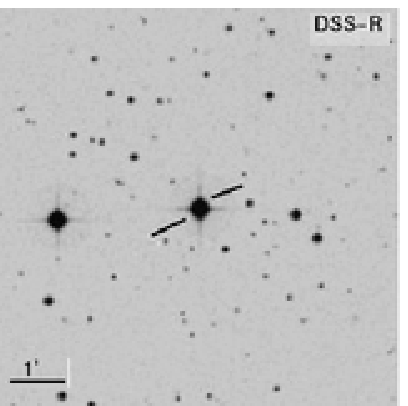}
%\psfull
\end{center}

\begin{center}
\epsfxsize=13.5cm
\epsfysize=4cm
\epsfbox{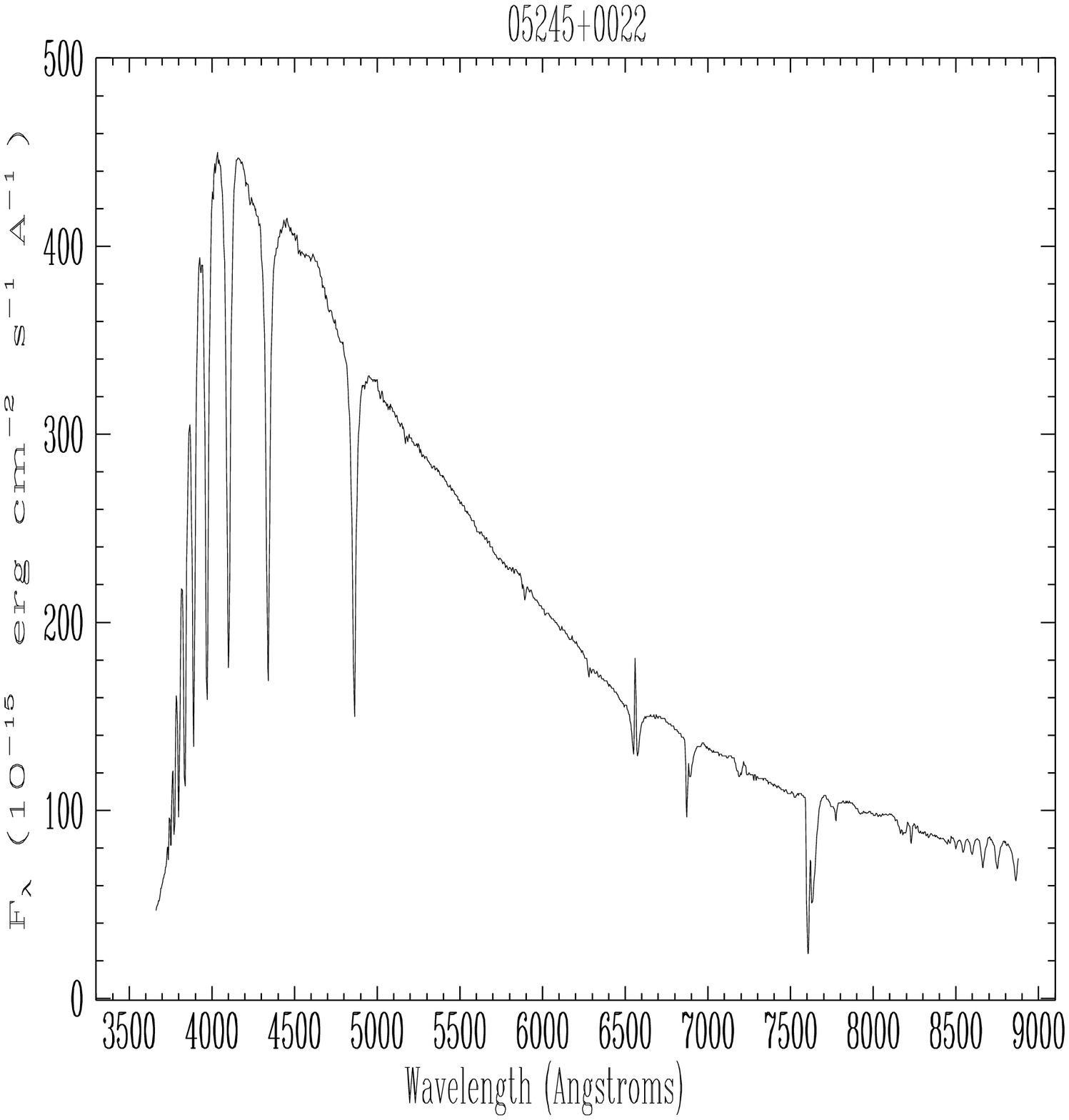}
%\psdraft
\epsfxsize=4cm
\epsfysize=4cm
\epsfbox{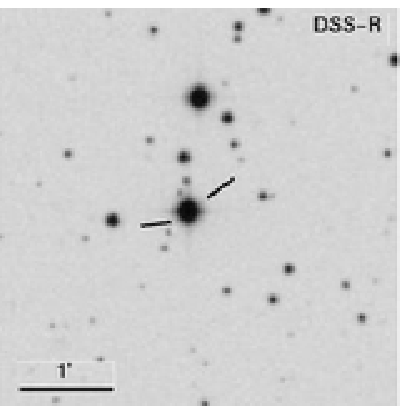}
%\psfull
\end{center}

\begin{center}
\epsfxsize=13.5cm
\epsfysize=4cm
\epsfbox{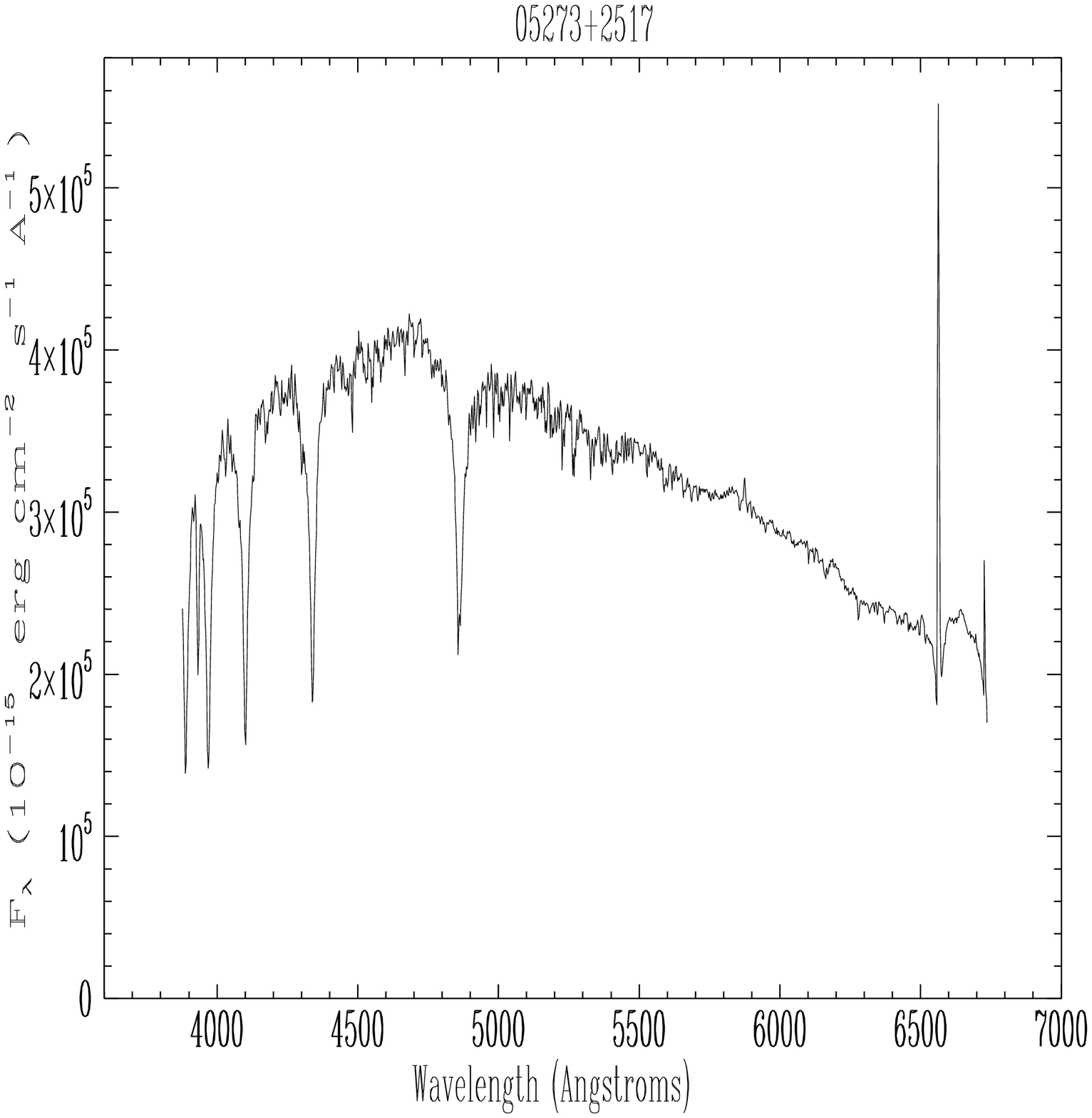}
%\psdraft
\epsfxsize=4cm
\epsfysize=4cm
\epsfbox{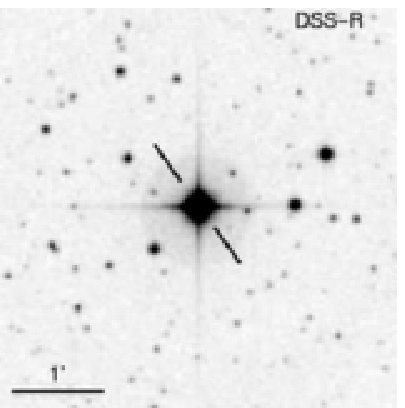}
%\psfull
\end{center}

\caption{Spectra of the objects classified as young stars in the sample together with their 
corresponding identification charts (continued). }
\end{figure*}

%-------------------------------------------------------------------
%pg3.
%\setcounter{figure}{9}
\begin{figure*}
\setcounter{figure}{0}

\begin{center}
\epsfxsize=13.5cm
\epsfysize=4cm
\epsfbox{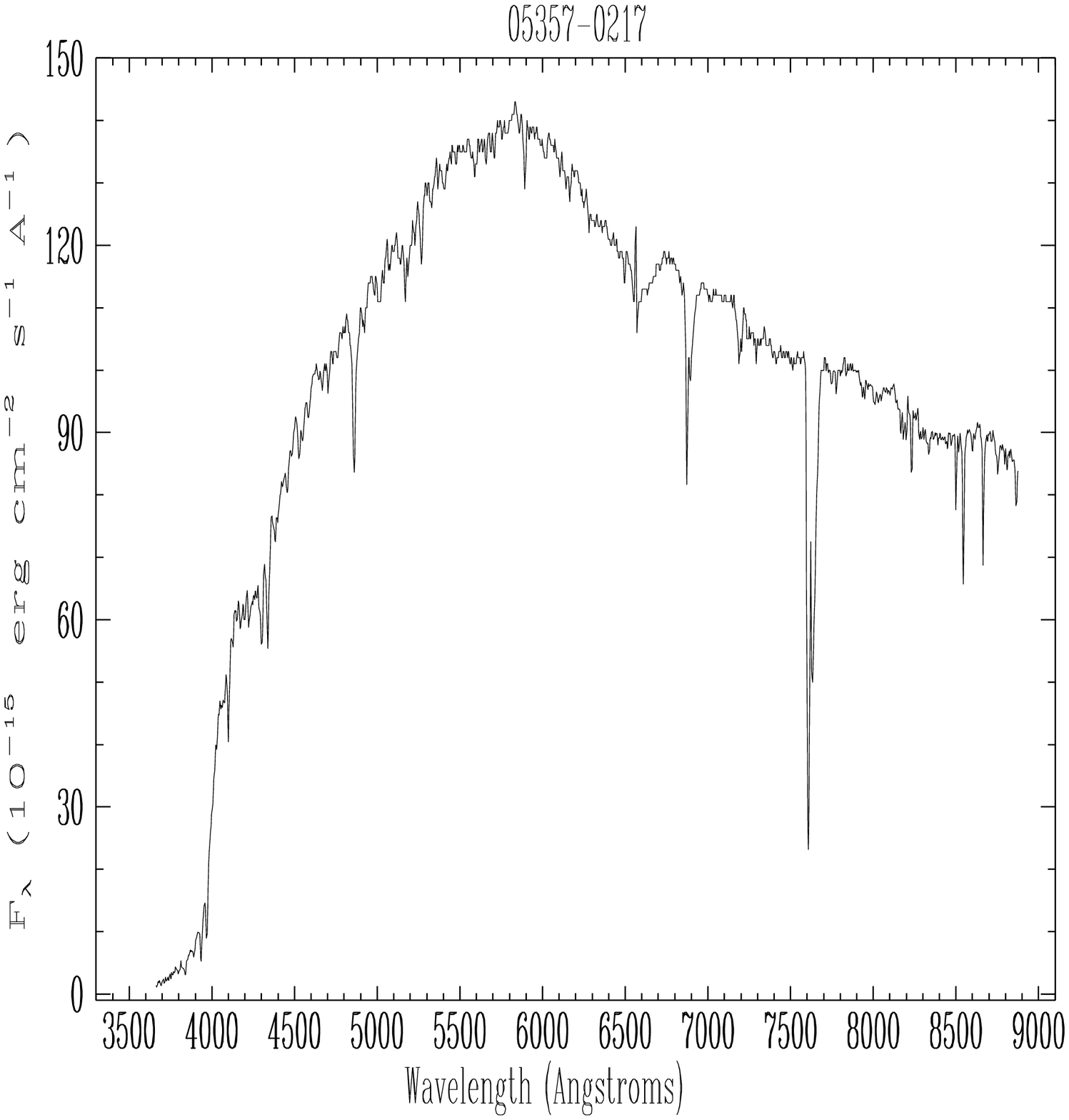}
%\psdraft
\epsfxsize=4cm
\epsfysize=4cm
\epsfbox{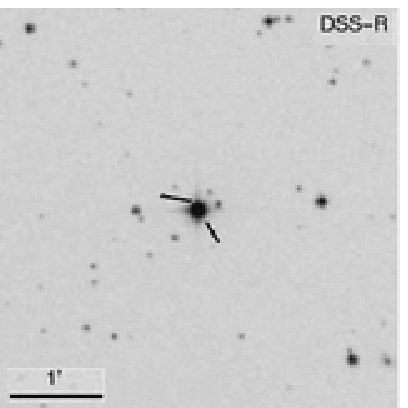}
%\psfull
\end{center}

\begin{center}
\epsfxsize=13.5cm
\epsfysize=4cm
\epsfbox{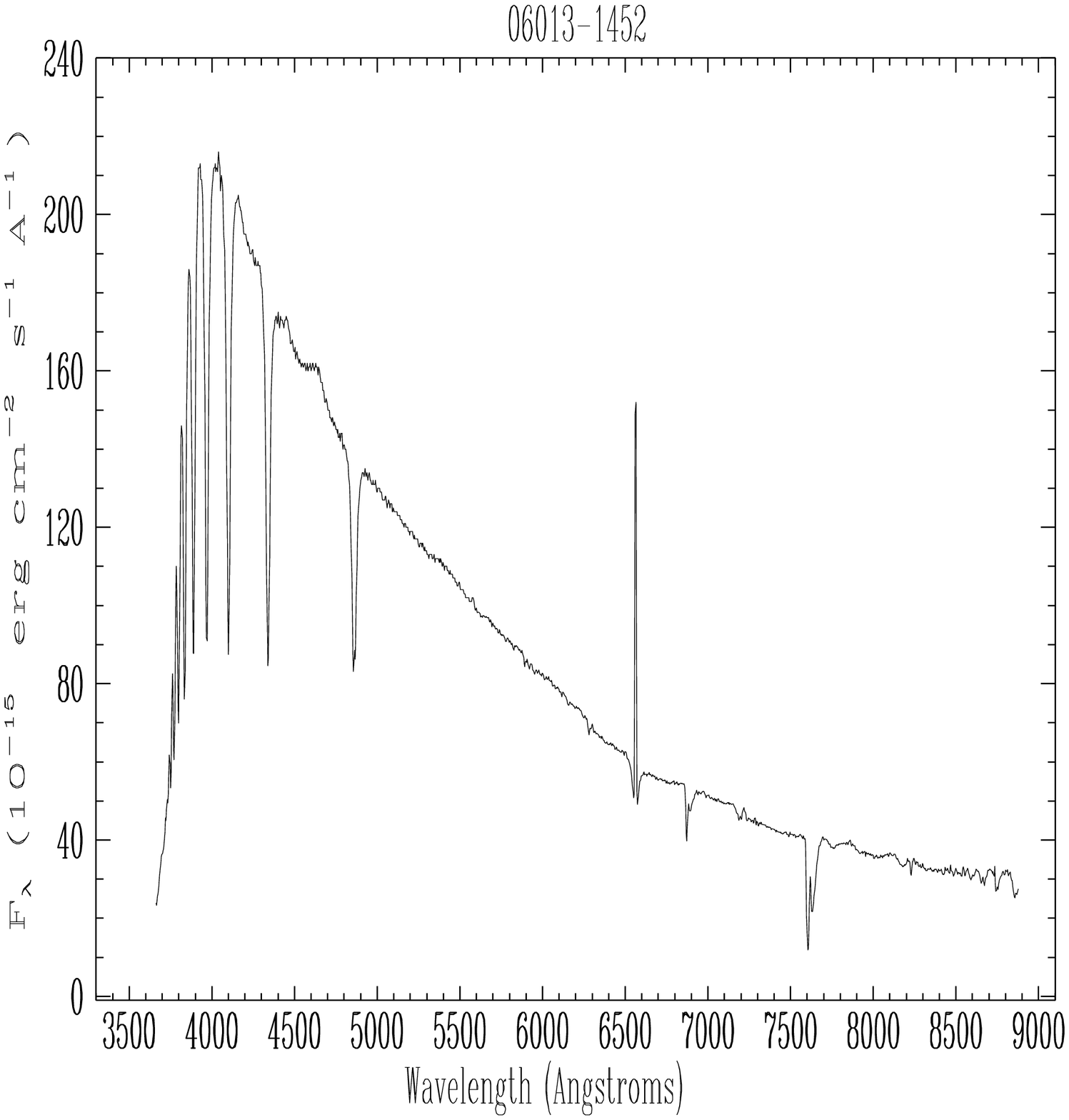}
%\psdraft
\epsfxsize=4cm
\epsfysize=4cm
\epsfbox{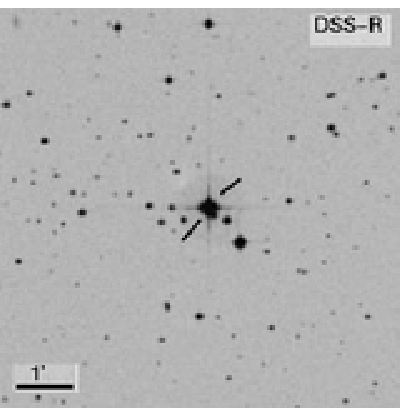}
%\psfull
\end{center}

\begin{center}
\epsfxsize=13.5cm
\epsfysize=4cm
\epsfbox{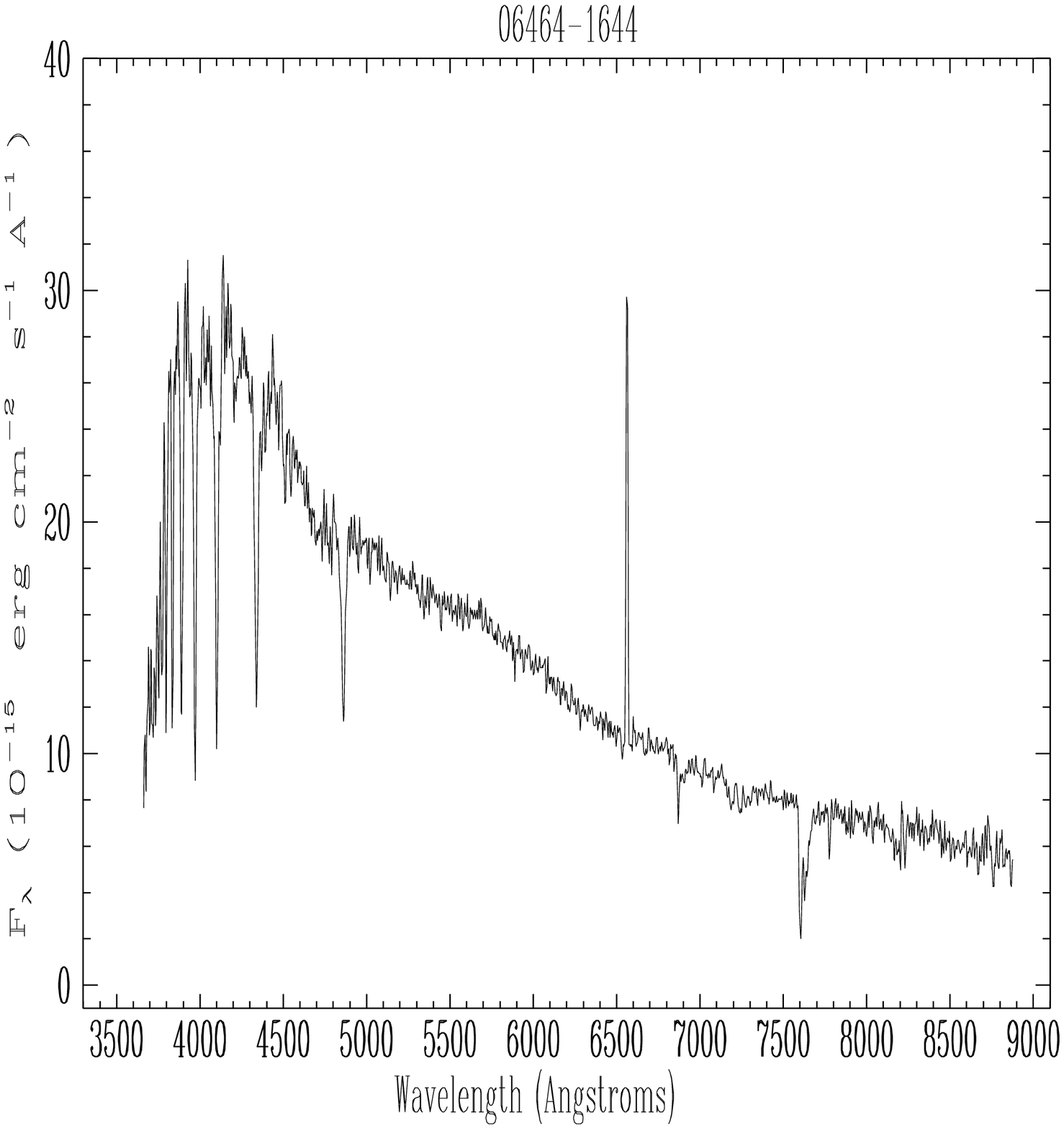}
%\psdraft
\epsfxsize=4cm
\epsfysize=4cm
\epsfbox{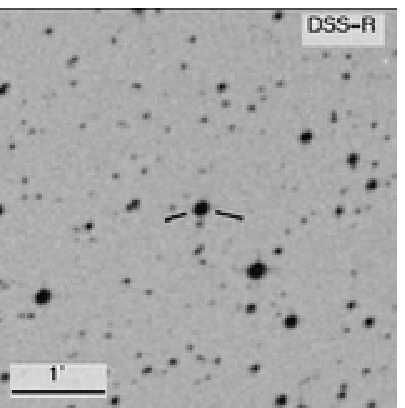}
%\psfull
\end{center}

\begin{center}
\epsfxsize=13.5cm
\epsfysize=4cm
\epsfbox{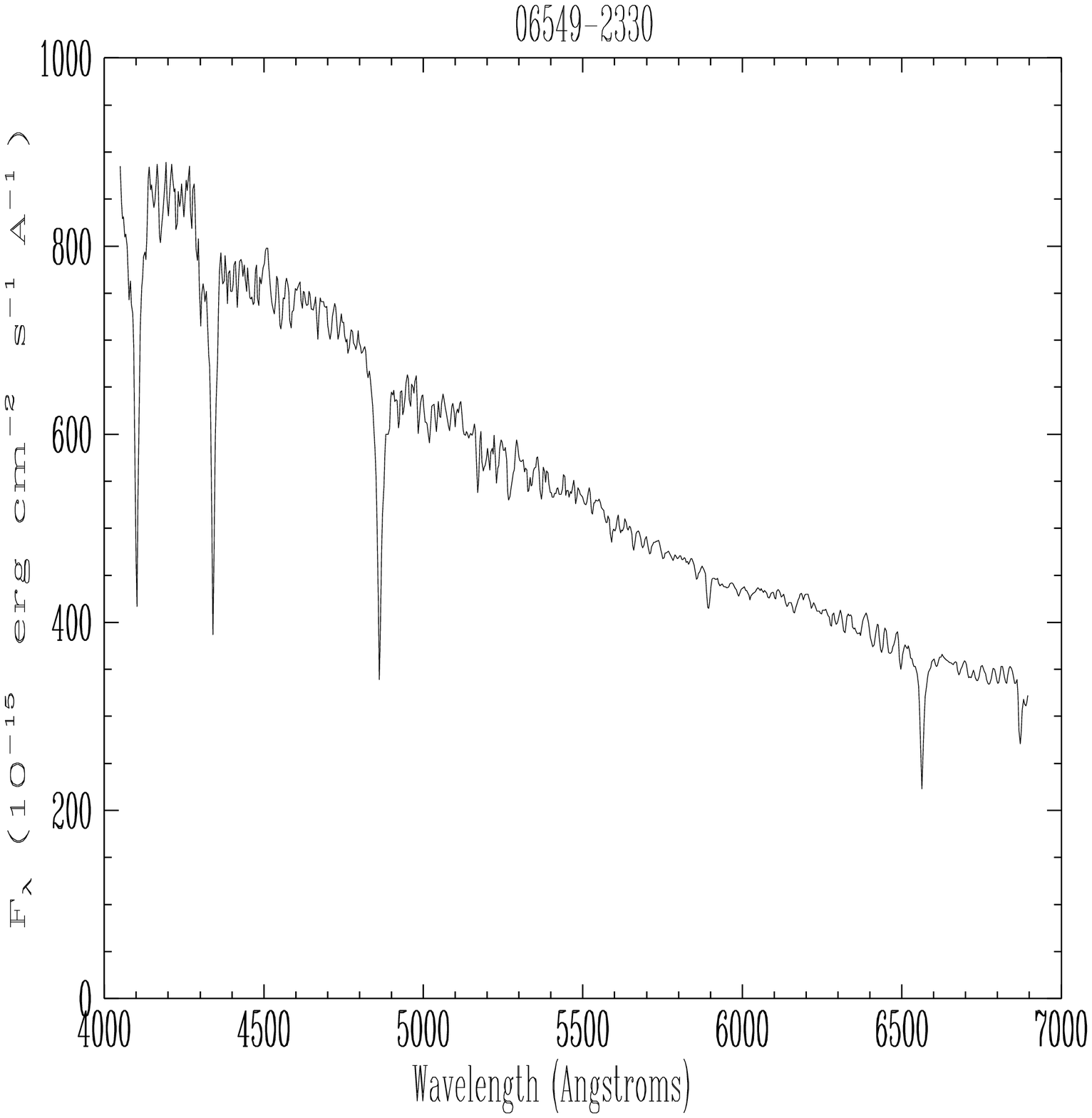}
%\psdraft
\epsfxsize=4cm
\epsfysize=4cm
\epsfbox{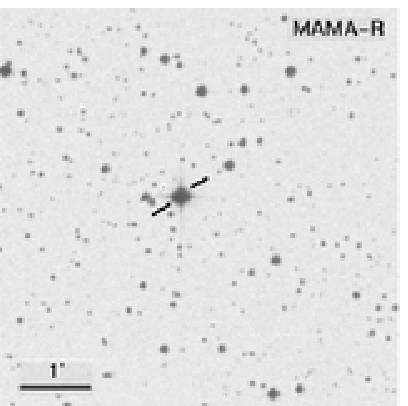}
%\psfull
\end{center}

\begin{center}
\epsfxsize=13.5cm
\epsfysize=4cm
\epsfbox{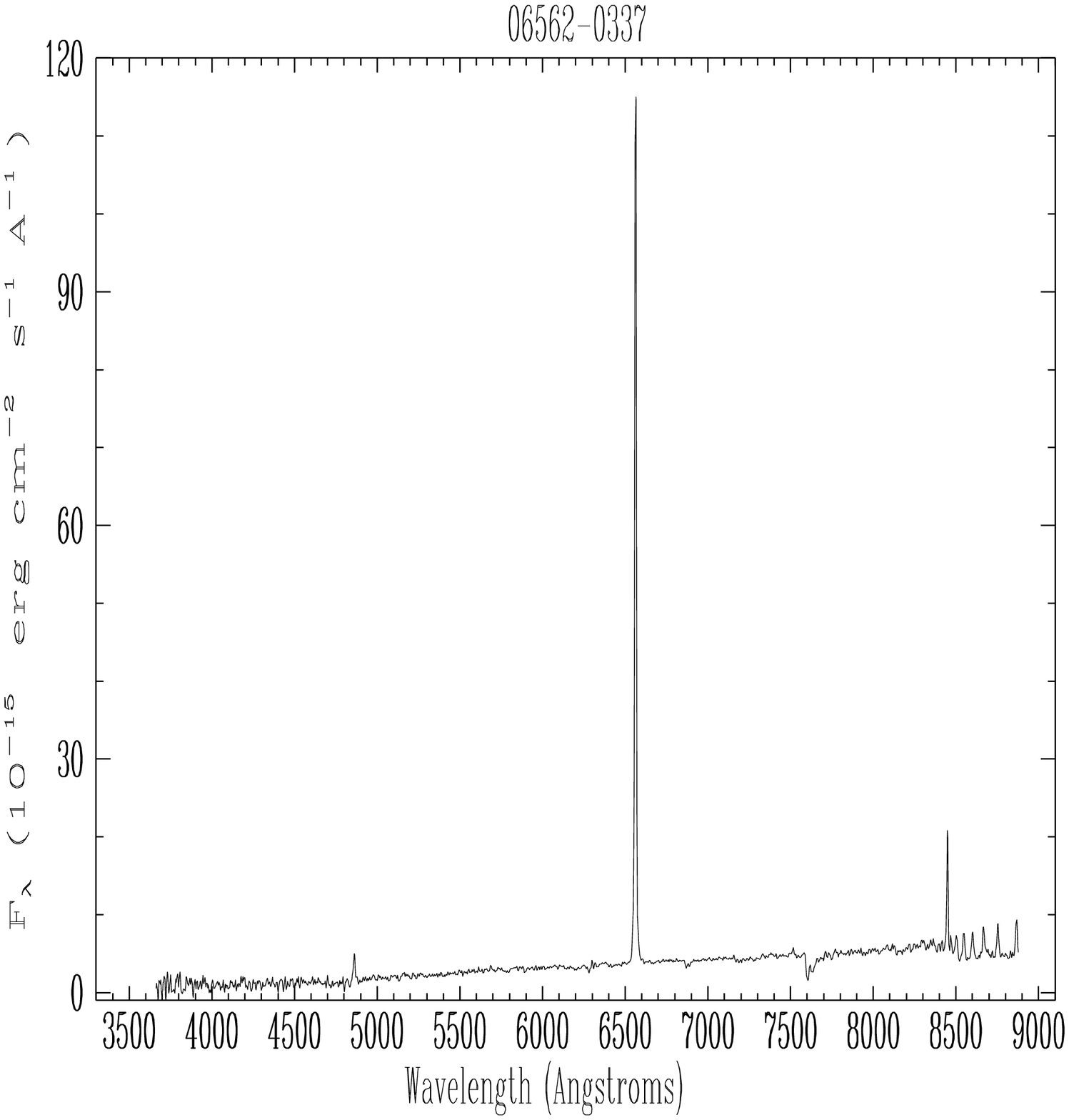}
%\psdraft
\epsfxsize=4cm
\epsfysize=4cm
\epsfbox{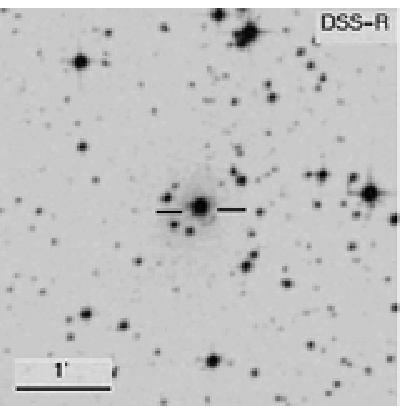}
%\psfull
\end{center}

\caption{Spectra of the objects classified as young stars in the sample together with their 
corresponding identification charts (continued). }
\end{figure*}

%-------------------------------------------------------------------
%pg4
%\setcounter{figure}{9}
\begin{figure*}
\setcounter{figure}{0}

\begin{center}
\epsfxsize=13.5cm
\epsfysize=4cm
\epsfbox{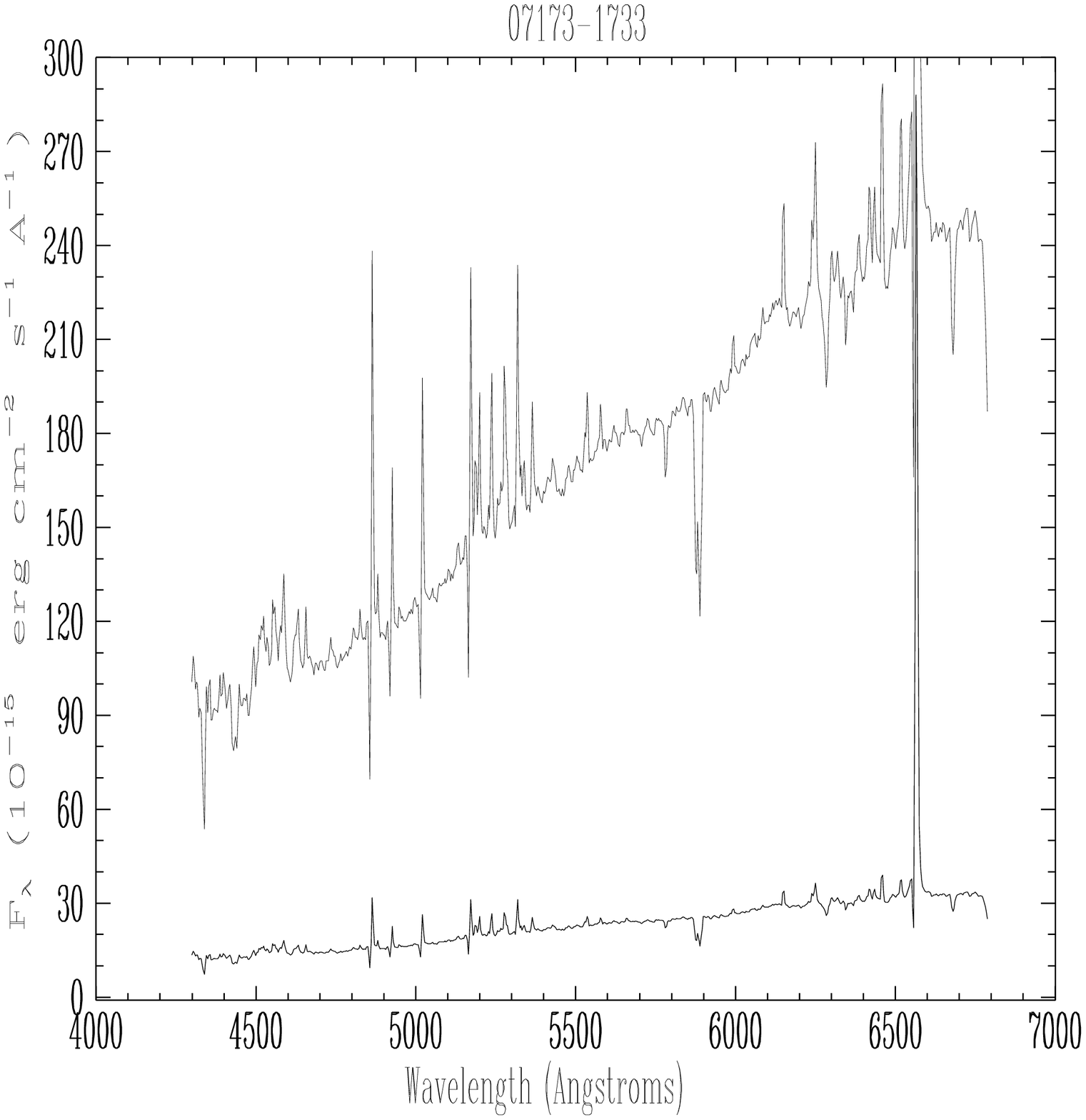}
%\psdraft
\epsfxsize=4cm
\epsfysize=4cm
\epsfbox{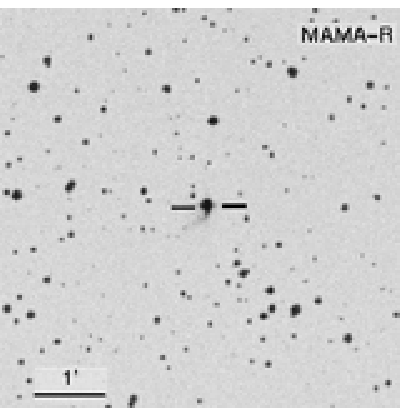}
%\psfull
\end{center}

\begin{center}
\epsfxsize=13.5cm
\epsfysize=4cm
\epsfbox{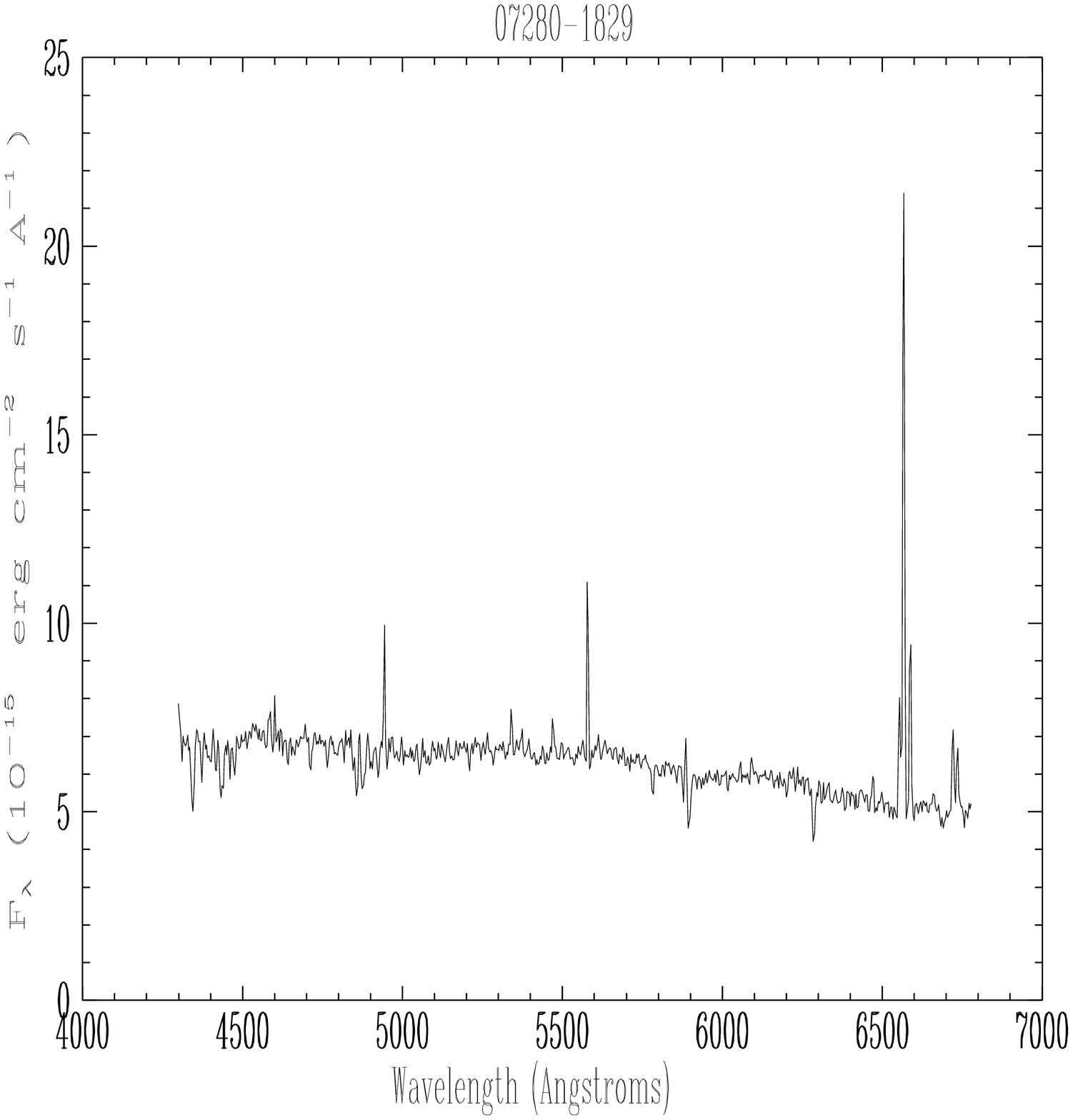}
%\psdraft
\epsfxsize=4cm
\epsfysize=4cm
\epsfbox{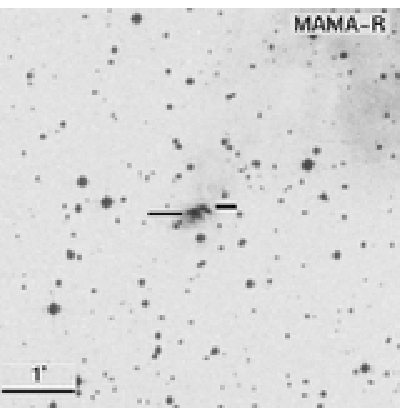}
%\psfull
\end{center}

\begin{center}
\epsfxsize=13.5cm
\epsfysize=4cm
\epsfbox{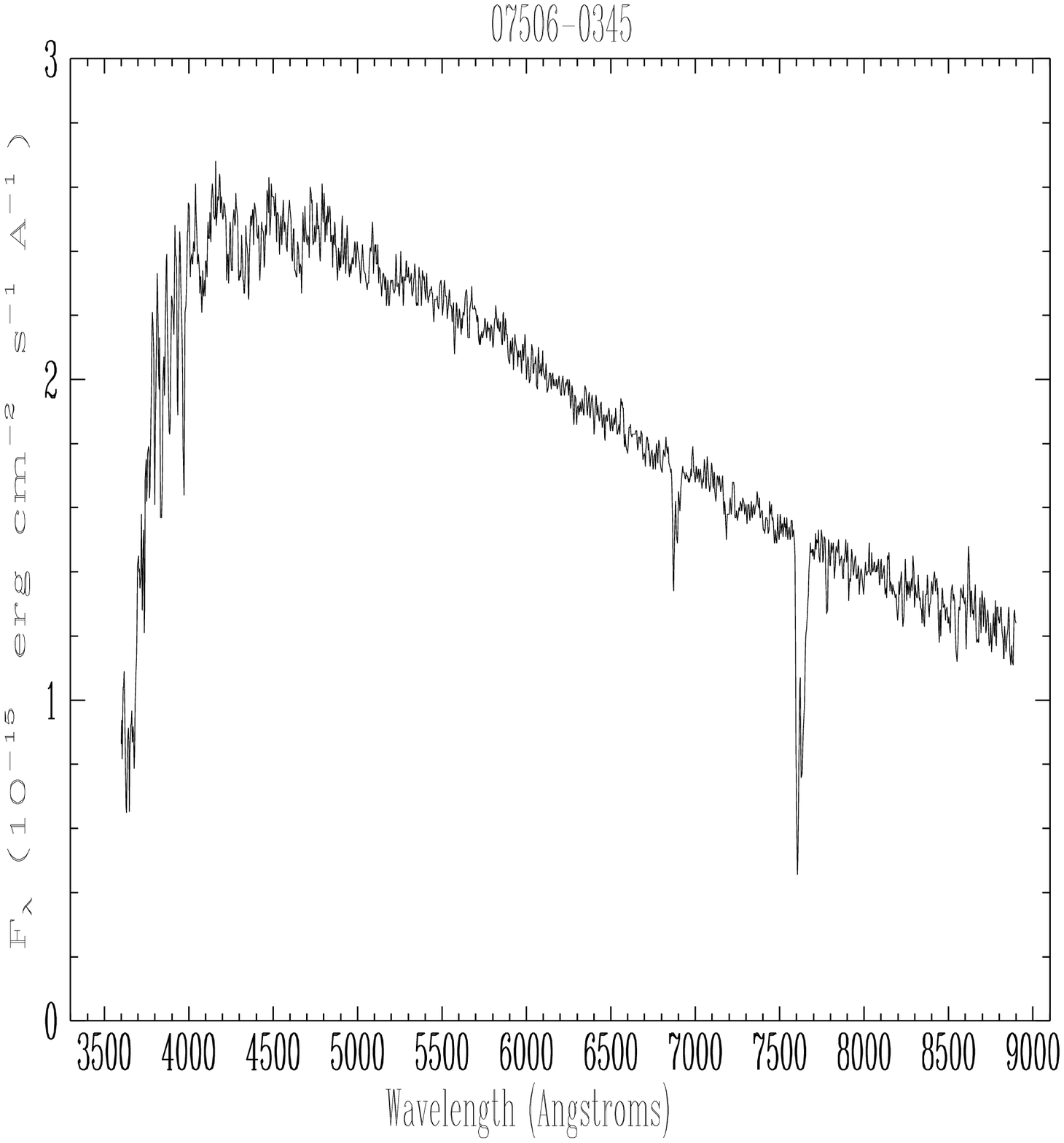}
%\psdraft
\epsfxsize=4cm
\epsfysize=4cm
\epsfbox{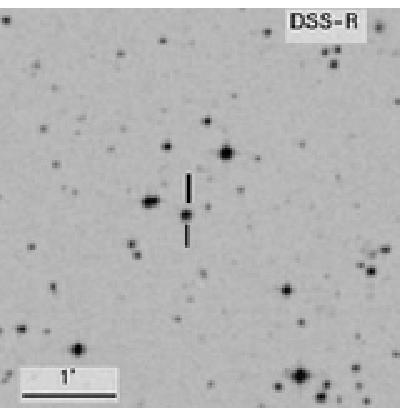}
%\psfull
\end{center}

\begin{center}
\epsfxsize=13.5cm
\epsfysize=4cm
\epsfbox{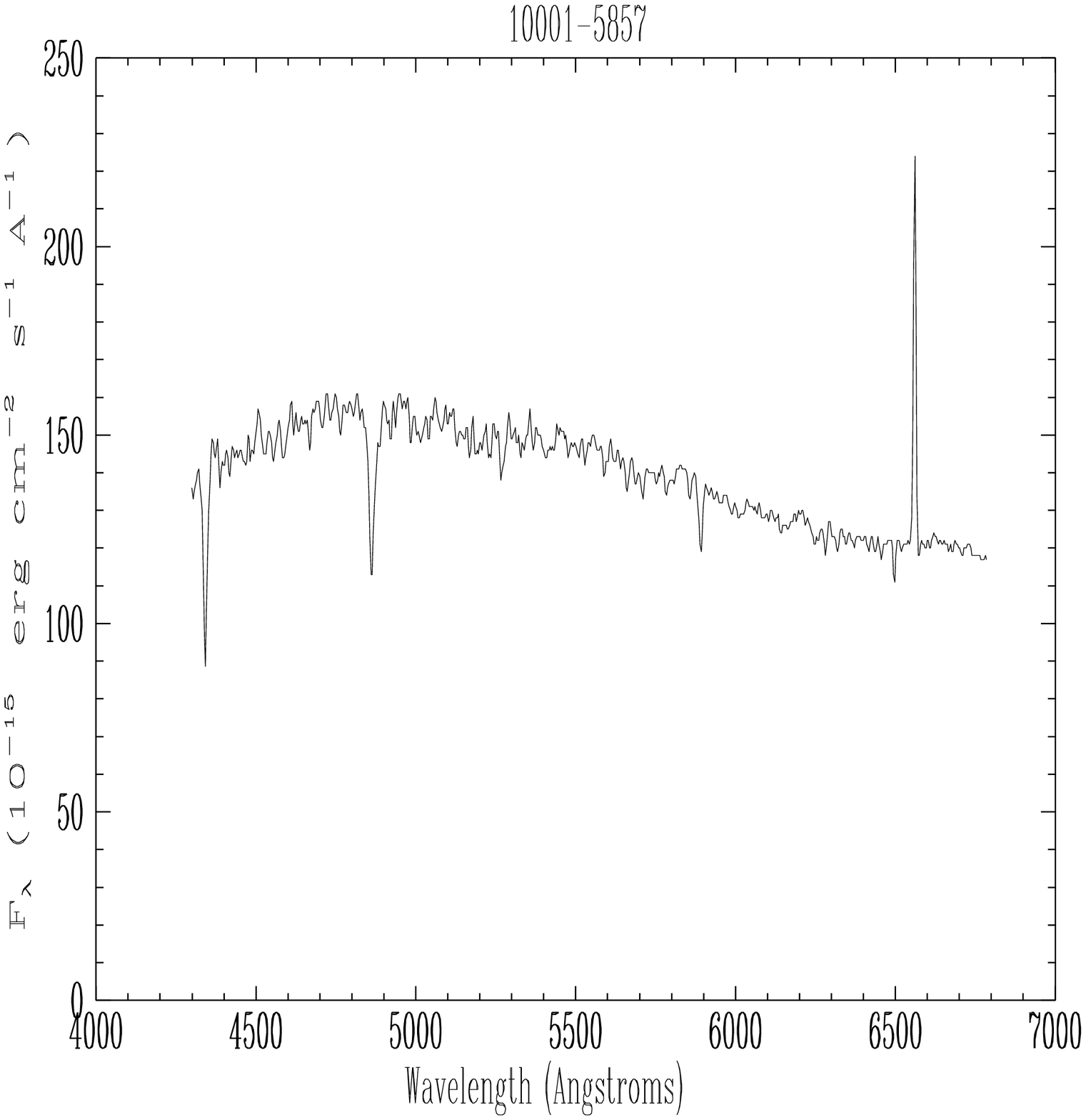}
%\psdraft
\epsfxsize=4cm
\epsfysize=4cm
\epsfbox{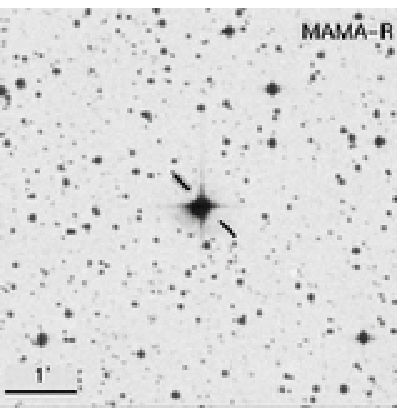}
%\psfull
\end{center}

\begin{center}
\epsfxsize=13.5cm
\epsfysize=4cm
\epsfbox{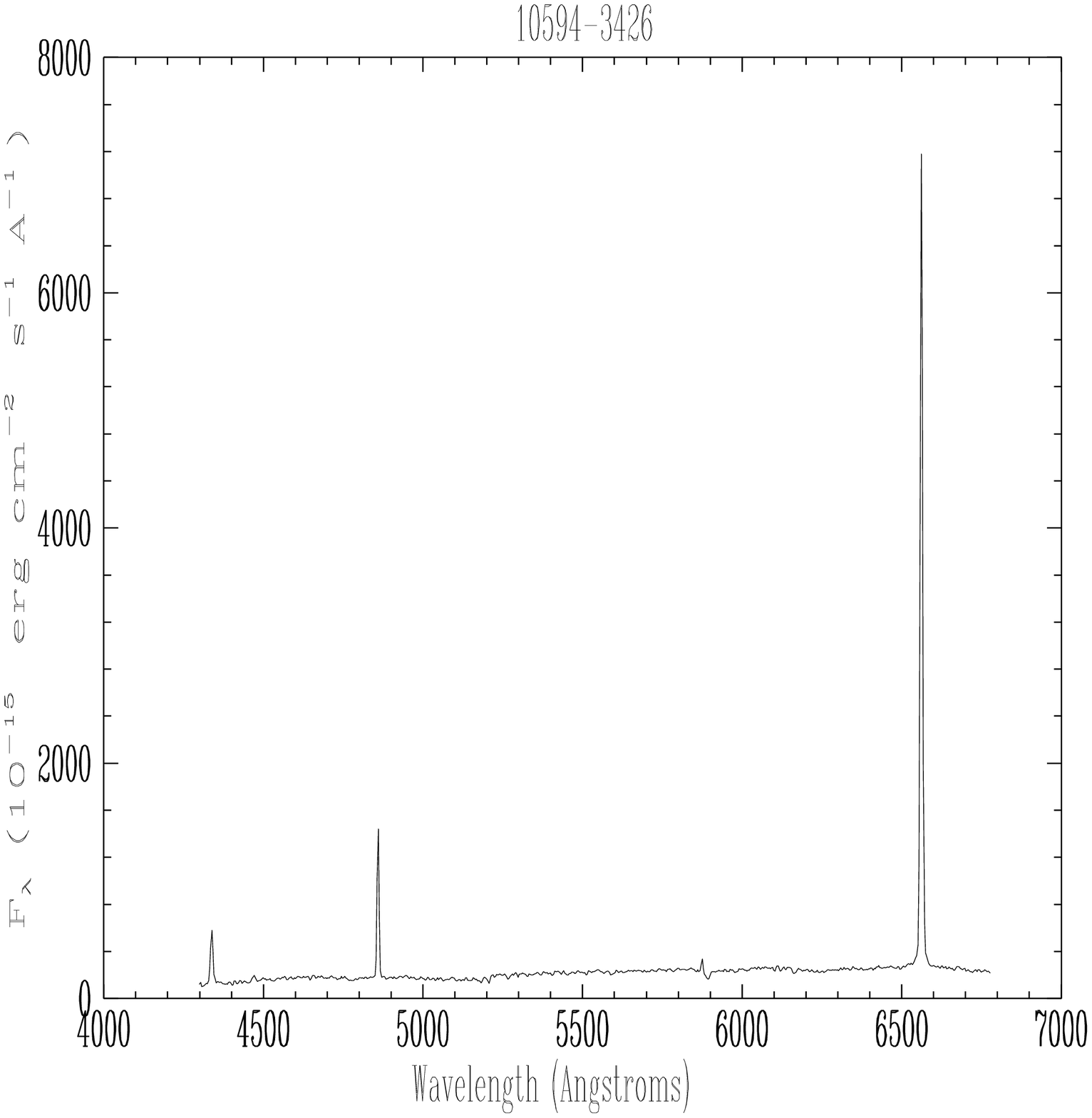}
%\psdraft
\epsfxsize=4cm
\epsfysize=4cm
\epsfbox{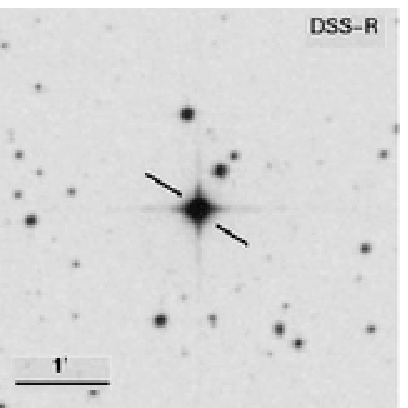}
%\psfull
\end{center}

\caption{Spectra of the objects classified as young stars in the sample together with their 
corresponding identification charts (continued). }
\end{figure*}

%-------------------------------------------------------------------
%pg5
%%\setcounter{figure}{9}
\begin{figure*}
\setcounter{figure}{0}

\begin{center}
\epsfxsize=13.5cm
\epsfysize=4cm
\epsfbox{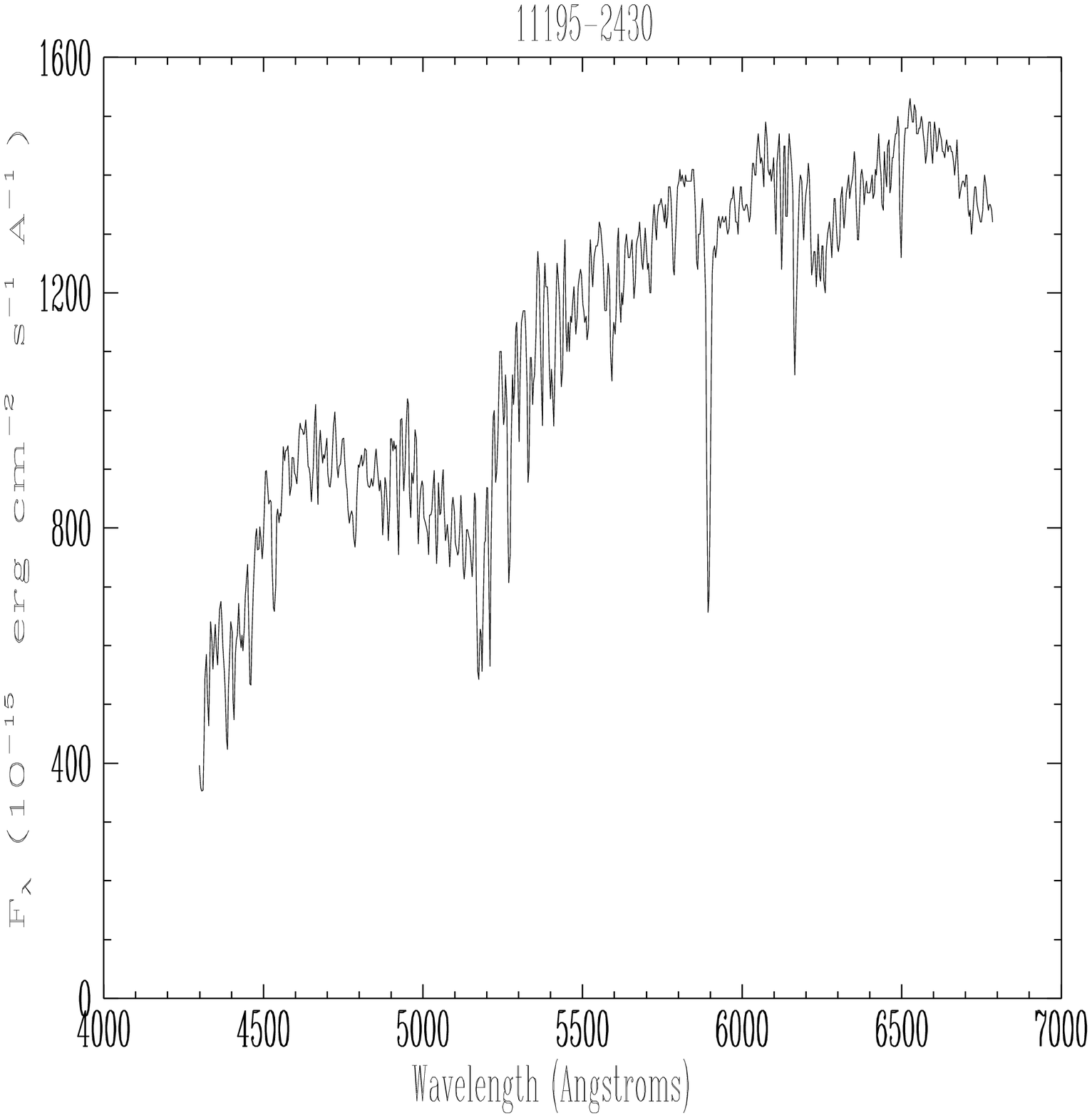}
%\psdraft
\epsfxsize=4cm
\epsfysize=4cm
\epsfbox{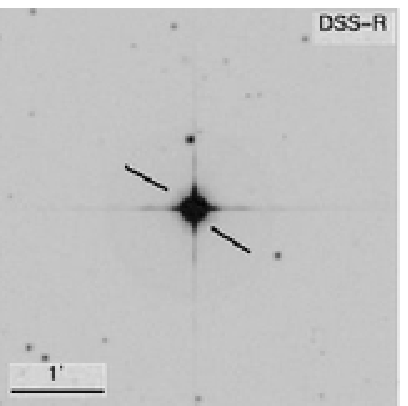}
%\psfull
\end{center}

\begin{center}
\epsfxsize=13.5cm
\epsfysize=4cm
\epsfbox{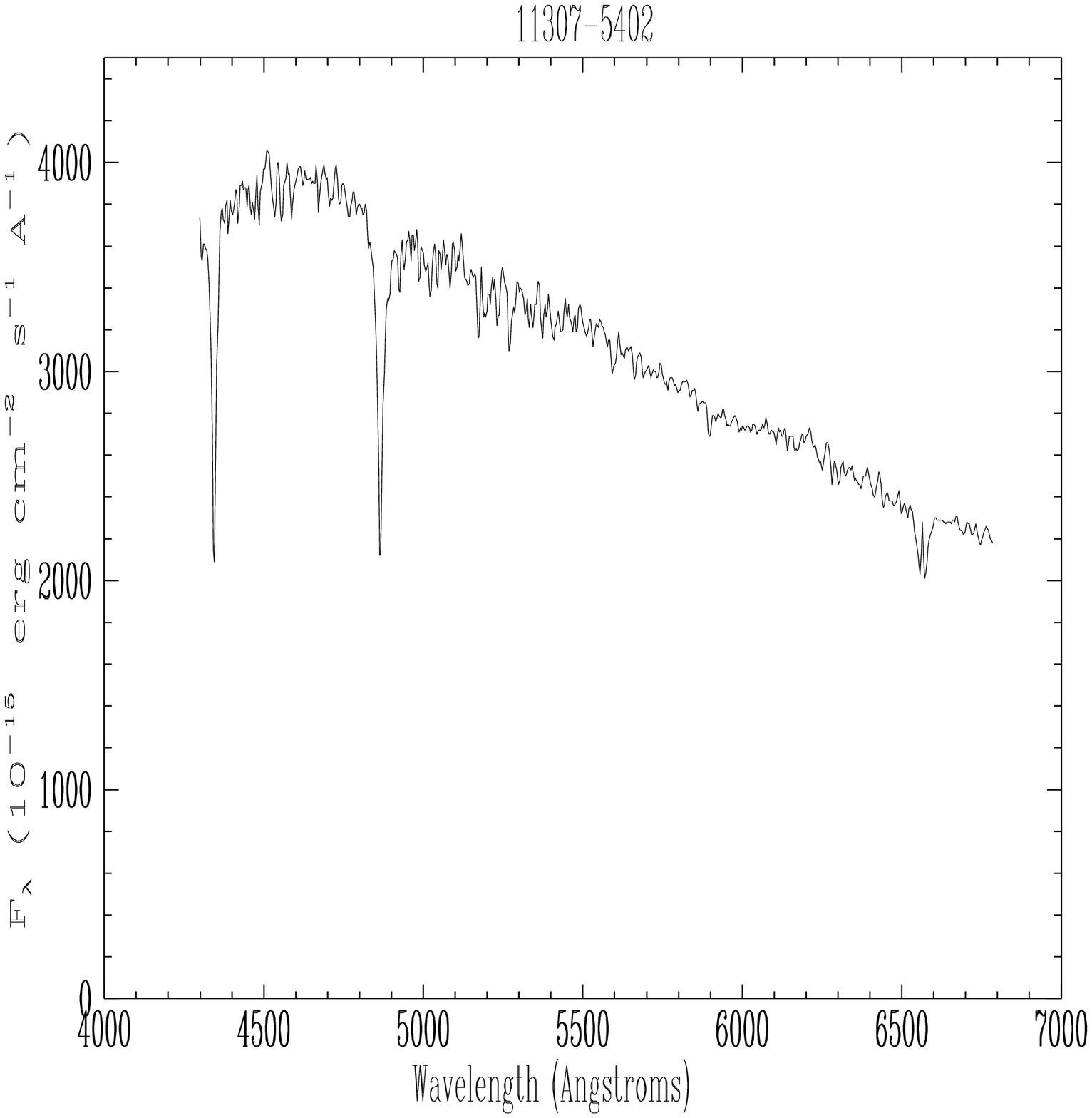}
%\psdraft
\epsfxsize=4cm
\epsfysize=4cm
\epsfbox{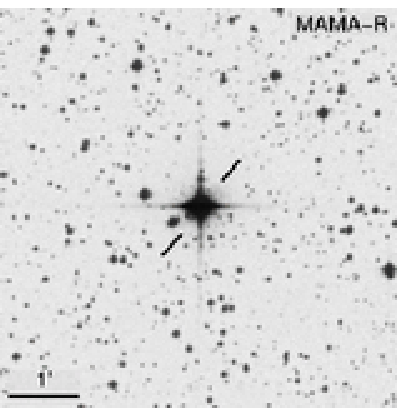}
%\psfull
\end{center}

\begin{center}
\epsfxsize=13.5cm
\epsfysize=4cm
\epsfbox{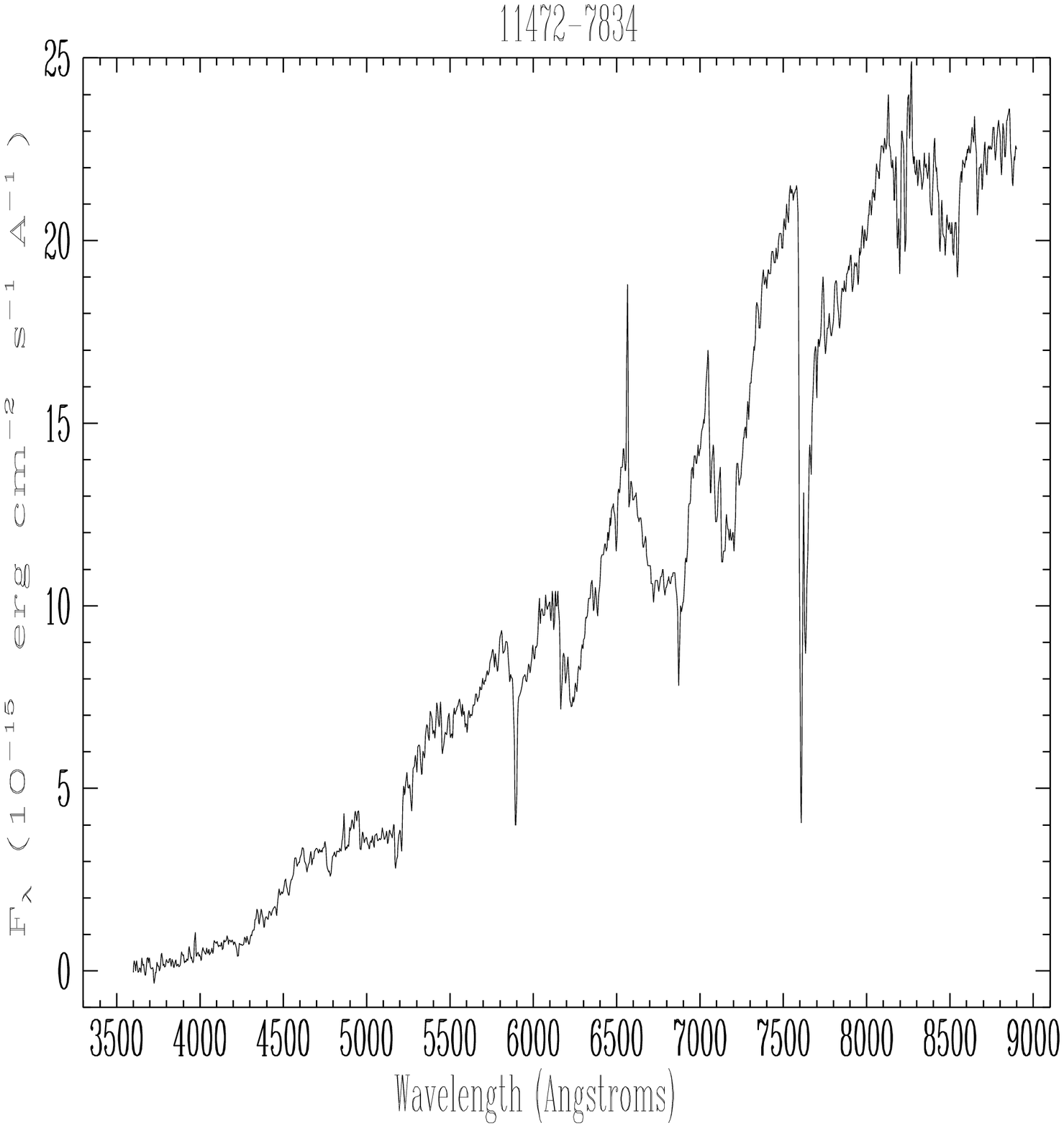}
%\psdraft
\epsfxsize=4cm
\epsfysize=4cm
\epsfbox{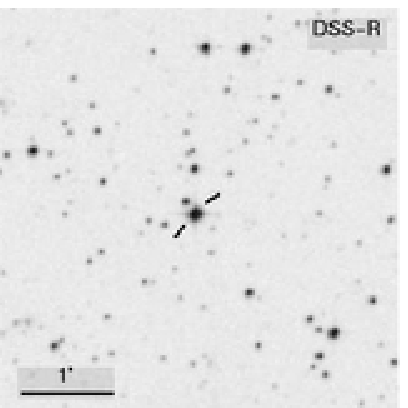}
%\psfull
\end{center}

\begin{center}
\epsfxsize=13.5cm
\epsfysize=4cm
\epsfbox{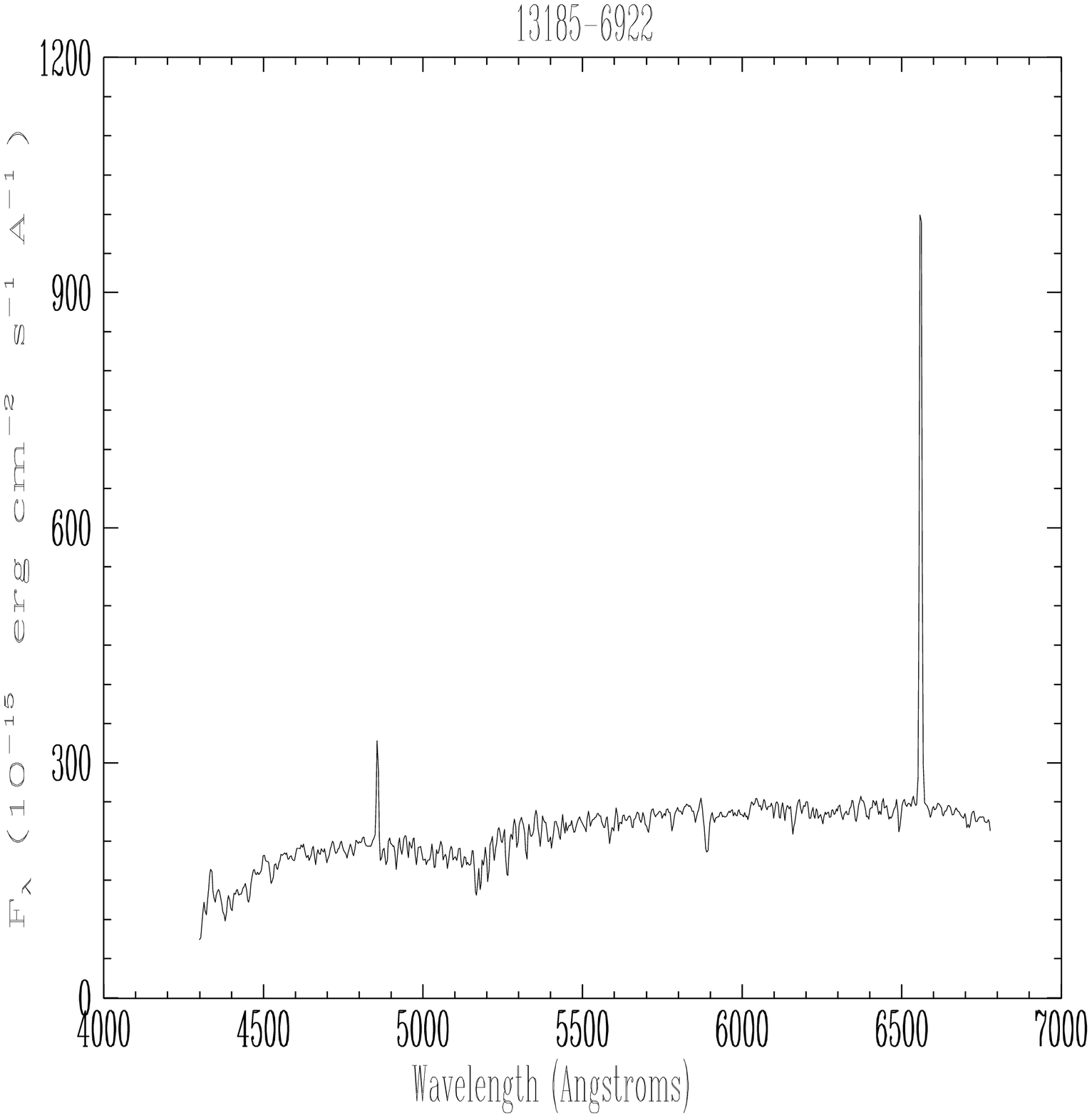}
%\psdraft
\epsfxsize=4cm
\epsfysize=4cm
\epsfbox{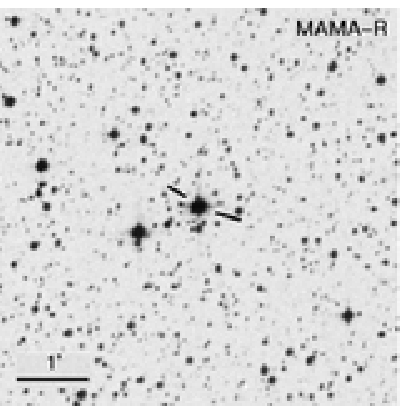}
%\psfull
\end{center}

\begin{center}
\epsfxsize=13.5cm
\epsfysize=4cm
\epsfbox{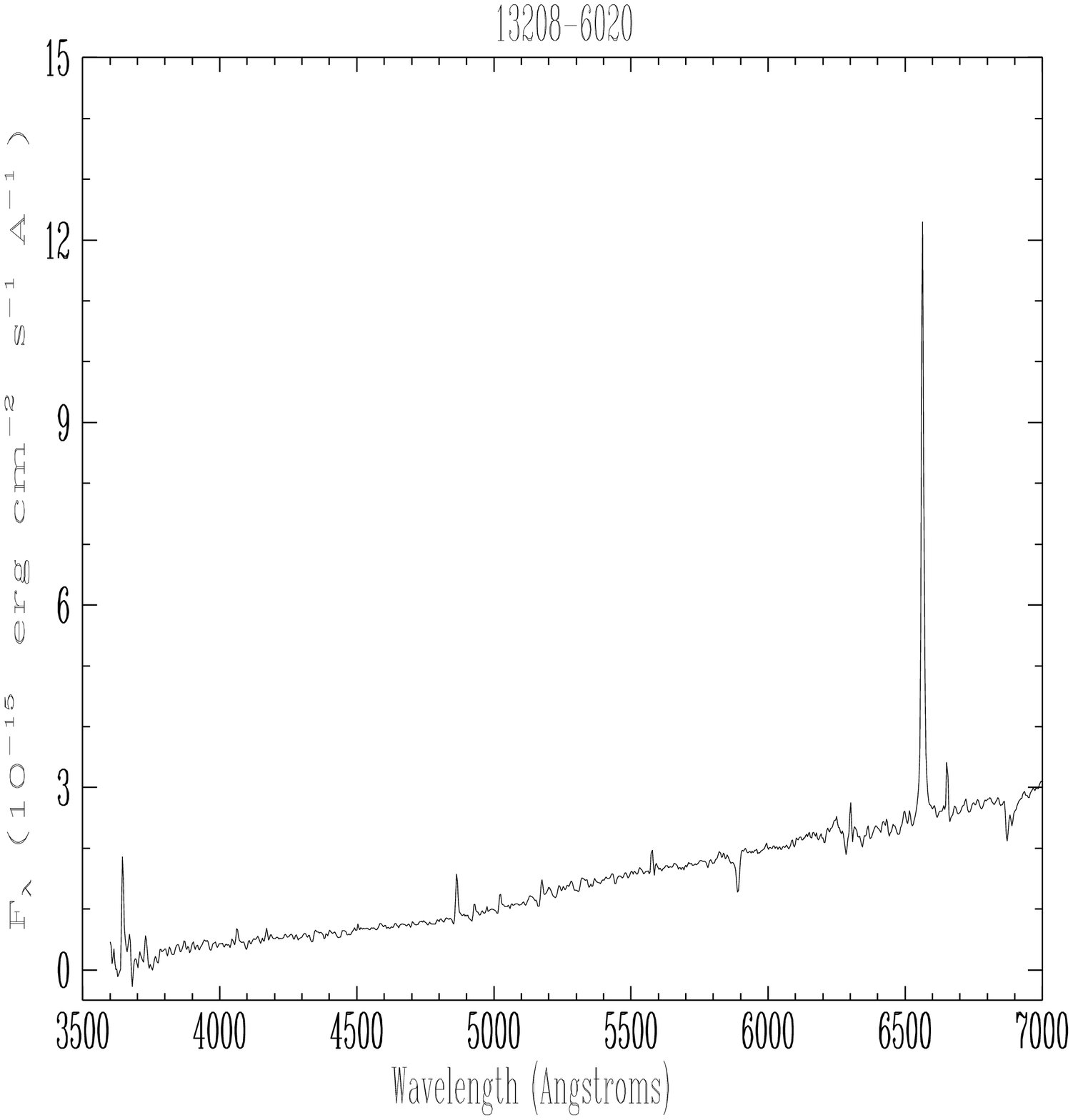}
%\psdraft
\epsfxsize=4cm
\epsfysize=4cm
\epsfbox{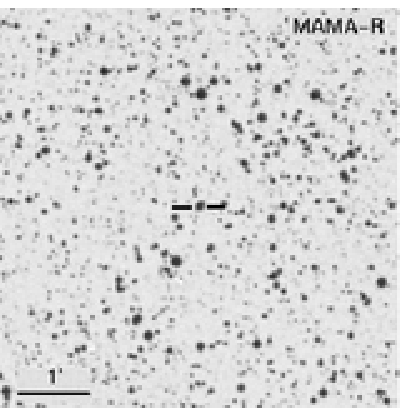}
%\psfull
\end{center}

\caption{Spectra of the objects classified as young stars in the sample together with their 
corresponding identification charts (continued). }
\end{figure*}

%-------------------------------------------------------------------
%pg6
%%\setcounter{figure}{9}
\begin{figure*}
\setcounter{figure}{0}

\begin{center}
\epsfxsize=13.5cm
\epsfysize=4cm
\epsfbox{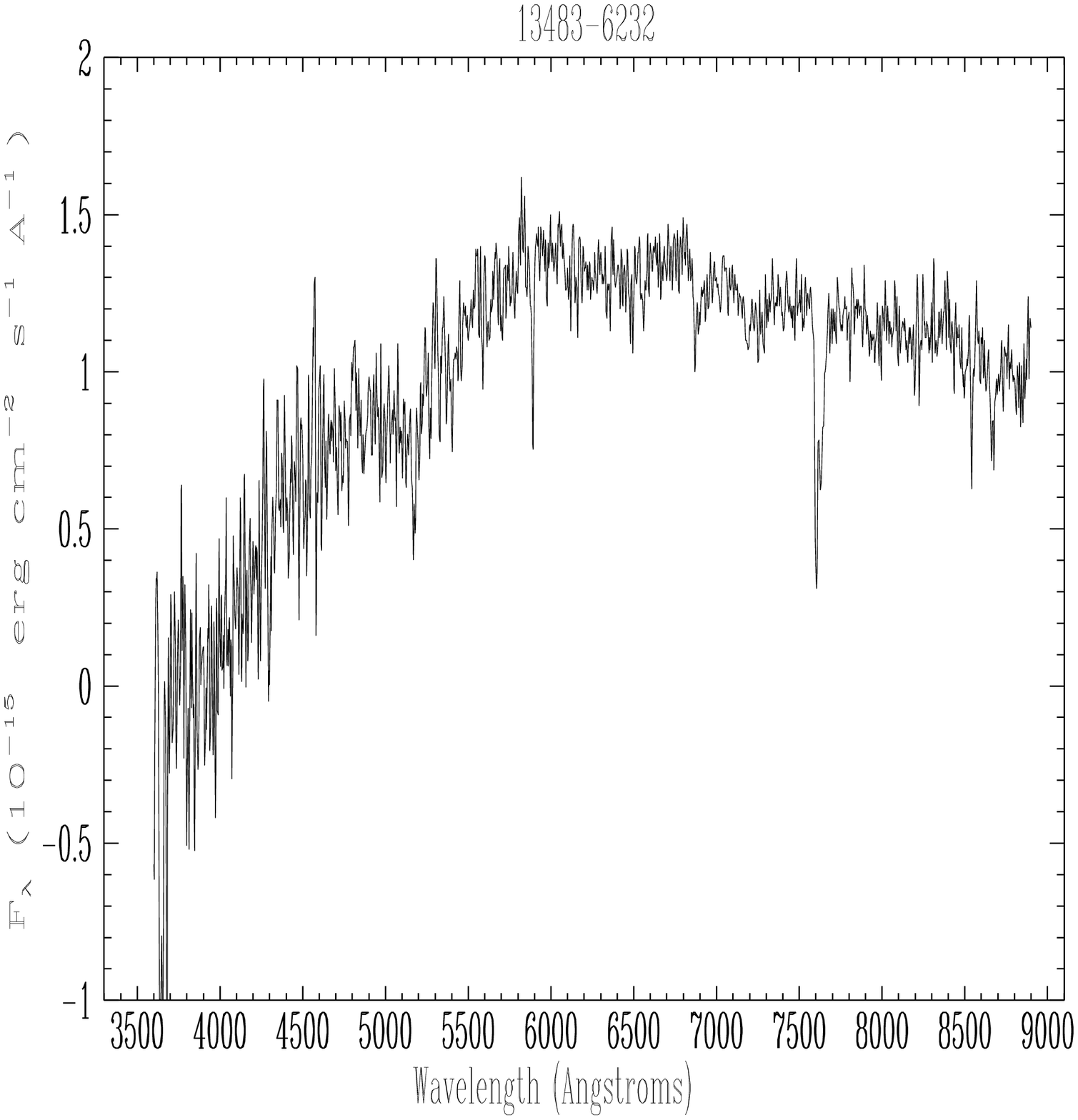}
%\psdraft
\epsfxsize=4cm
\epsfysize=4cm
\epsfbox{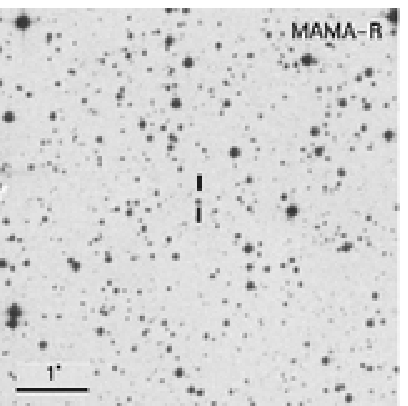}
%\psfull
\end{center}

\begin{center}
\epsfxsize=13.5cm
\epsfysize=4cm
\epsfbox{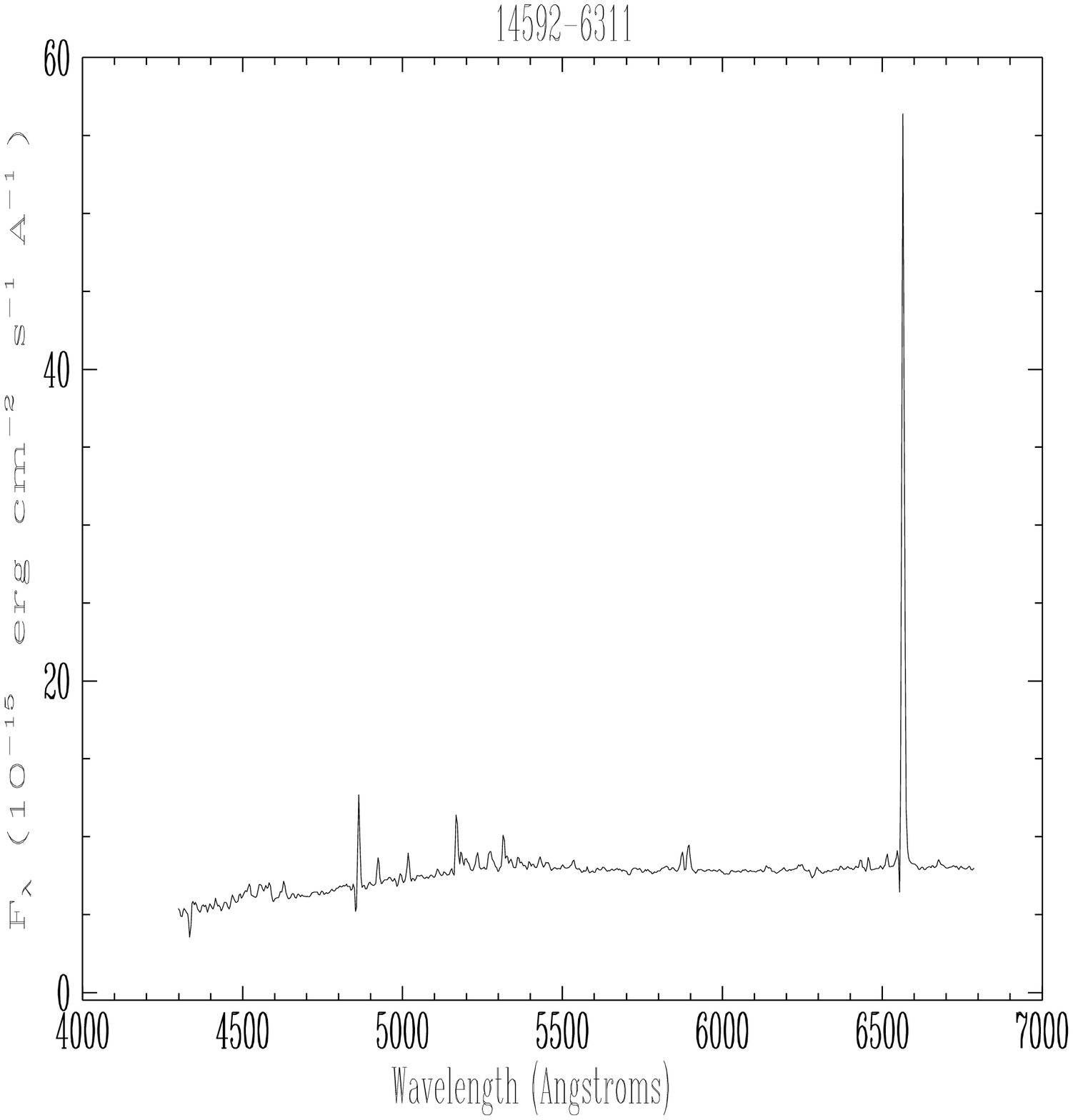}
%\psdraft
\epsfxsize=4cm
\epsfysize=4cm
\epsfbox{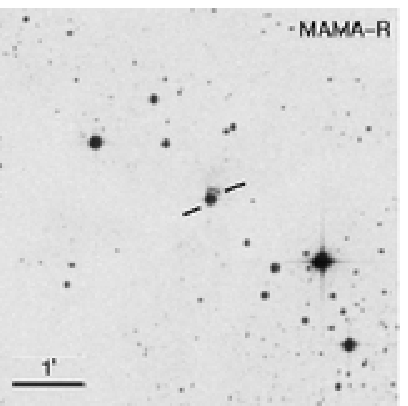}
%\psfull
\end{center}

\begin{center}
\epsfxsize=13.5cm
\epsfysize=4cm
\epsfbox{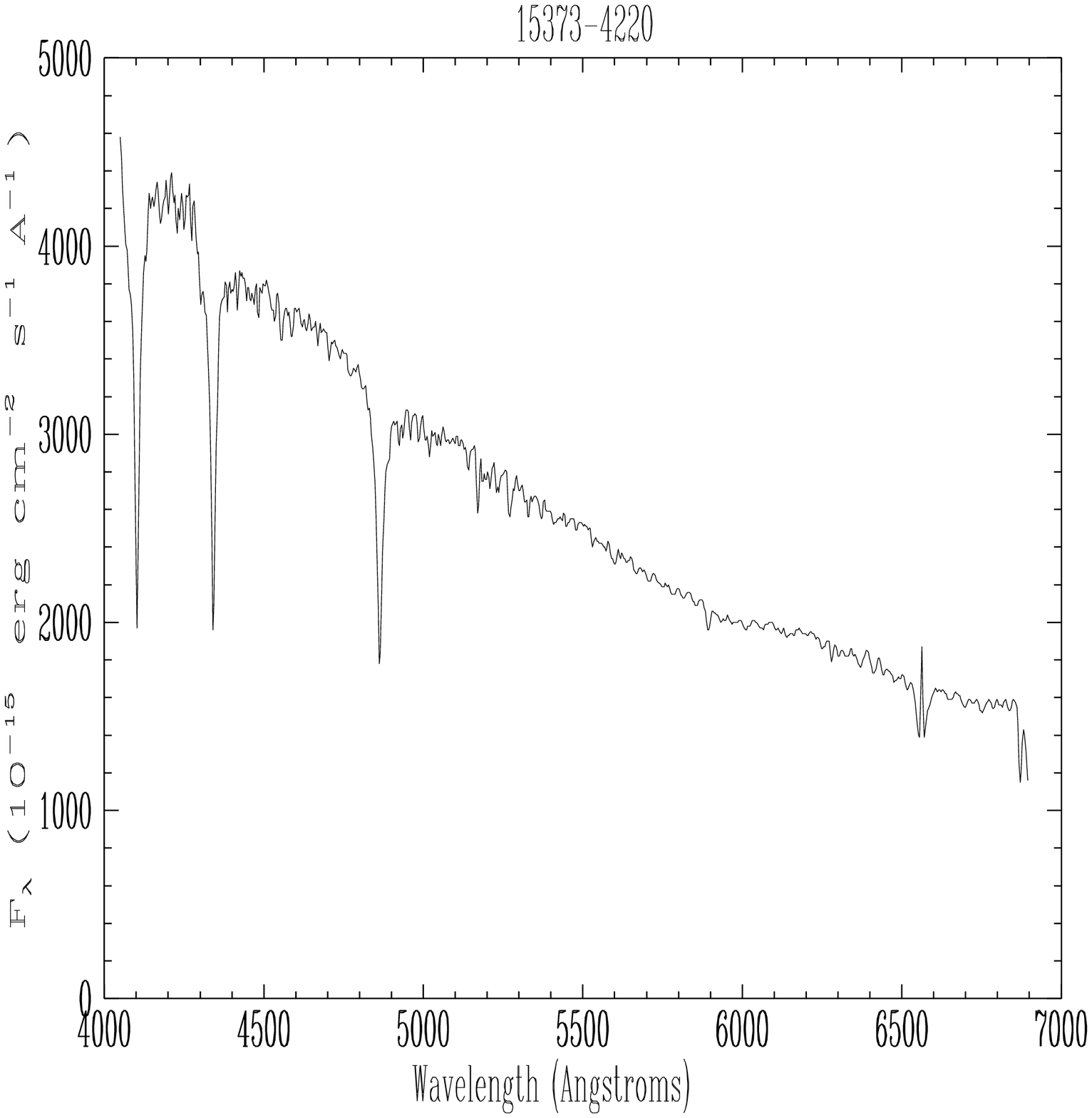}
%\psdraft
\epsfxsize=4cm
\epsfysize=4cm
\epsfbox{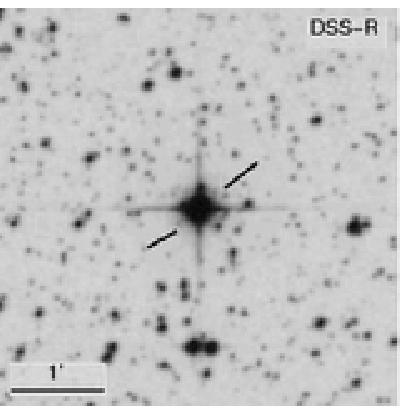}
%\psfull
\end{center}

\begin{center}
\epsfxsize=13.5cm
\epsfysize=4cm
\epsfbox{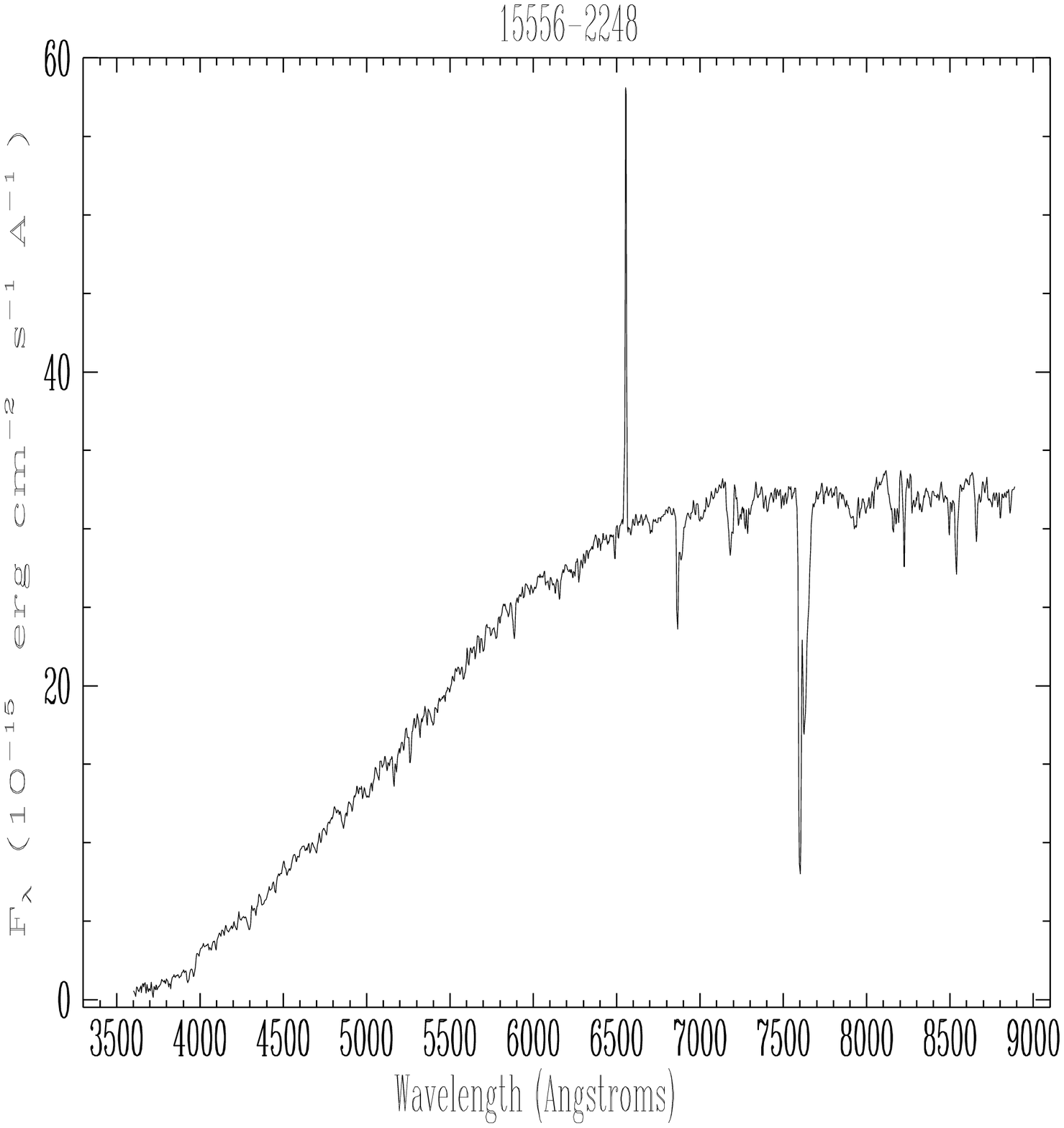}
%\psdraft
\epsfxsize=4cm
\epsfysize=4cm
\epsfbox{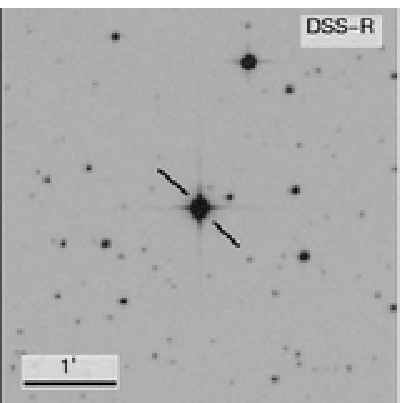}
%\psfull
\end{center}

\begin{center}
\epsfxsize=13.5cm
\epsfysize=4cm
\epsfbox{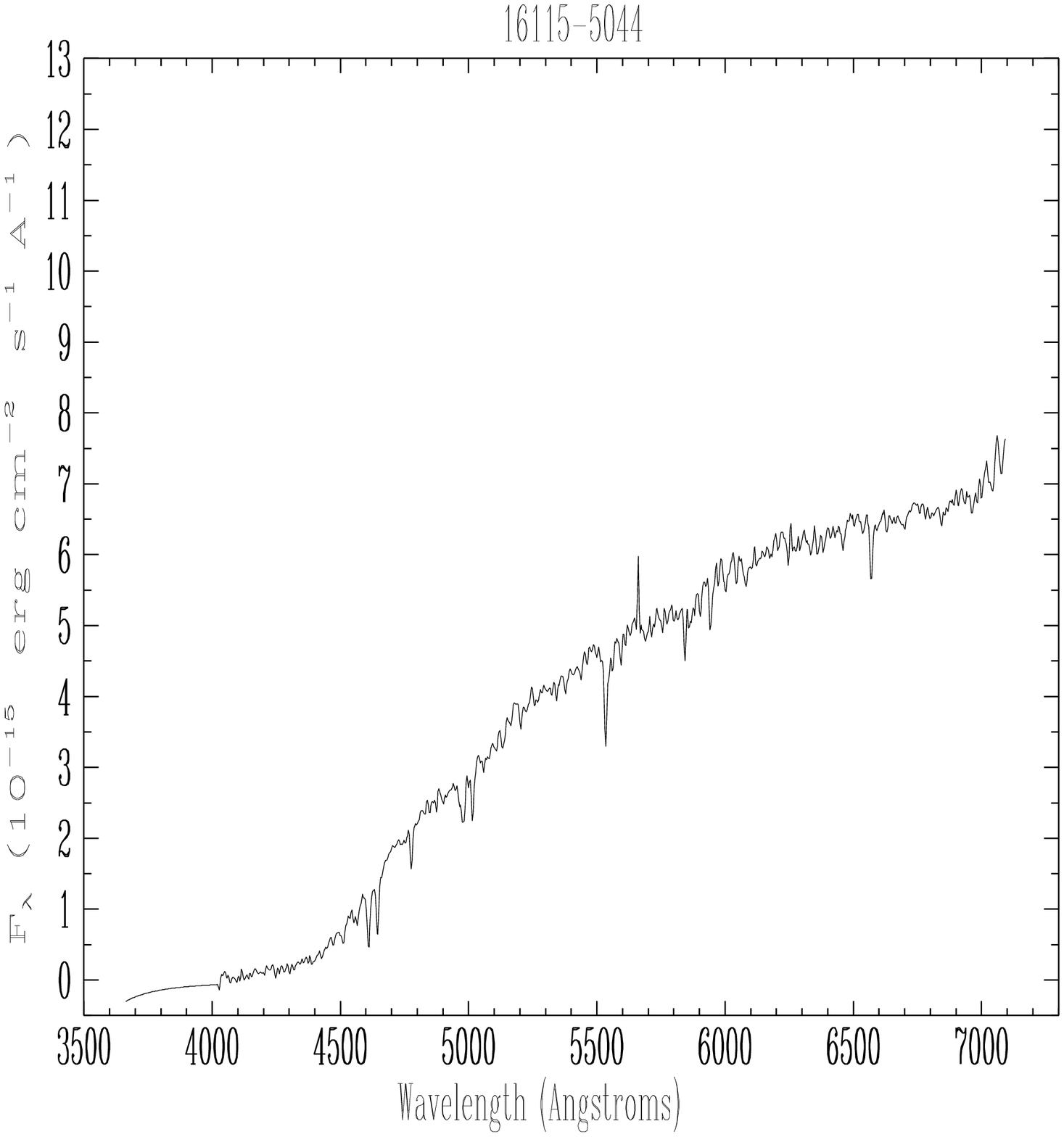}
%\psdraft
\epsfxsize=4cm
\epsfysize=4cm
\epsfbox{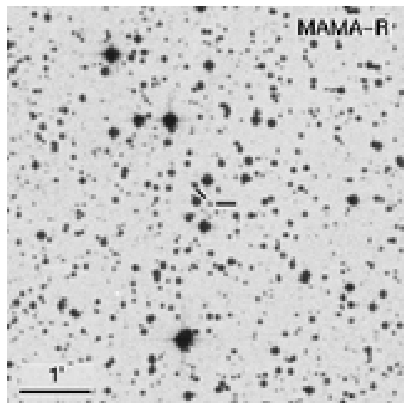}
%\psfull
\end{center}

\caption{Spectra of the objects classified as young stars in the sample together with their 
corresponding identification charts (continued). }
\end{figure*}

%-------------------------------------------------------------------
%pg7
%\setcounter{figure}{9}
\begin{figure*}
\setcounter{figure}{0}

\begin{center}
\epsfxsize=13.5cm
\epsfysize=4cm
\epsfbox{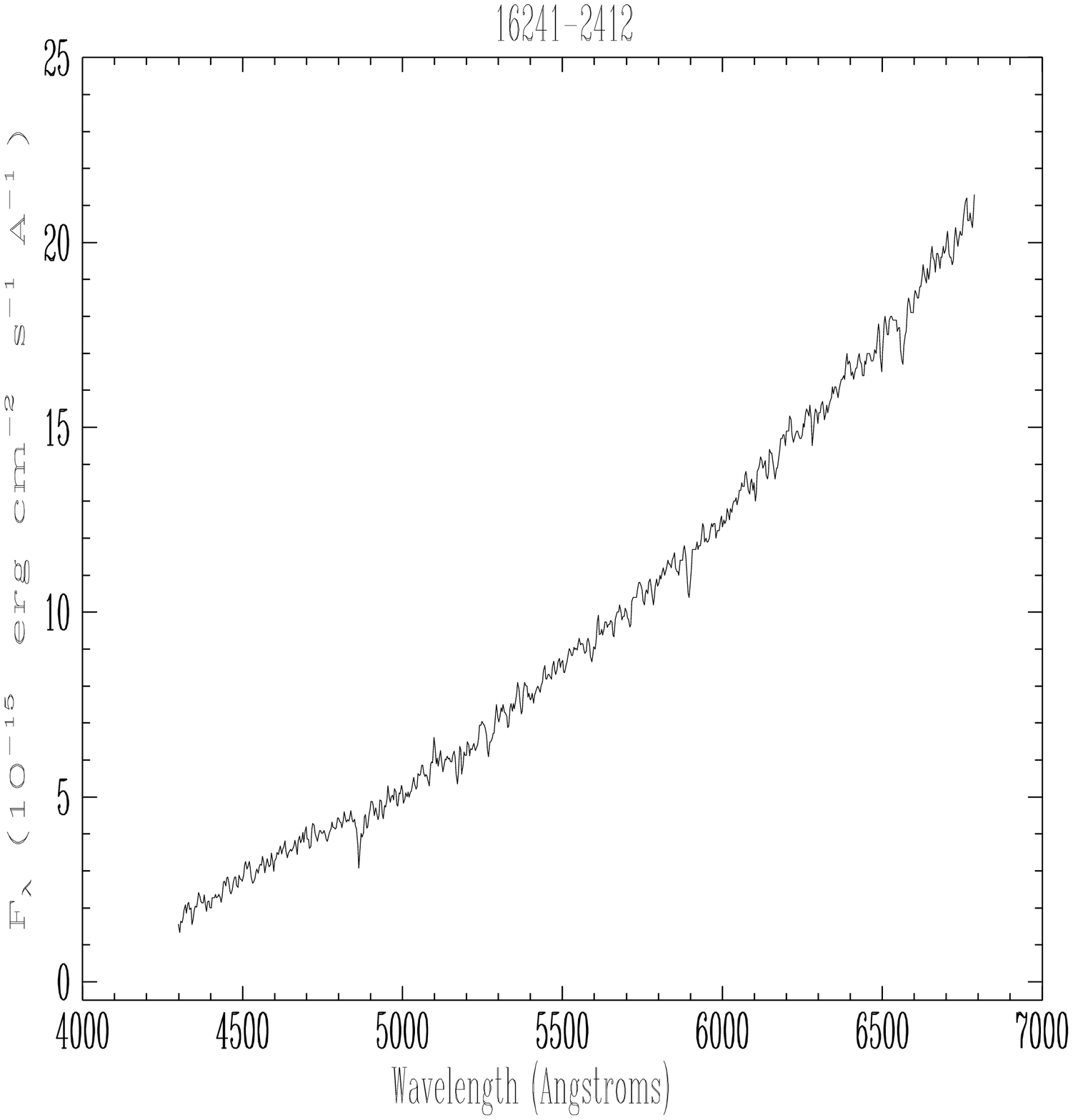}
%\psdraft
\epsfxsize=4cm
\epsfysize=4cm
\epsfbox{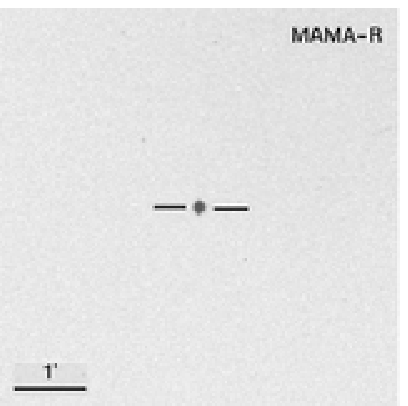}
%\psfull
\end{center}

\begin{center}
\epsfxsize=13.5cm
\epsfysize=4cm
\epsfbox{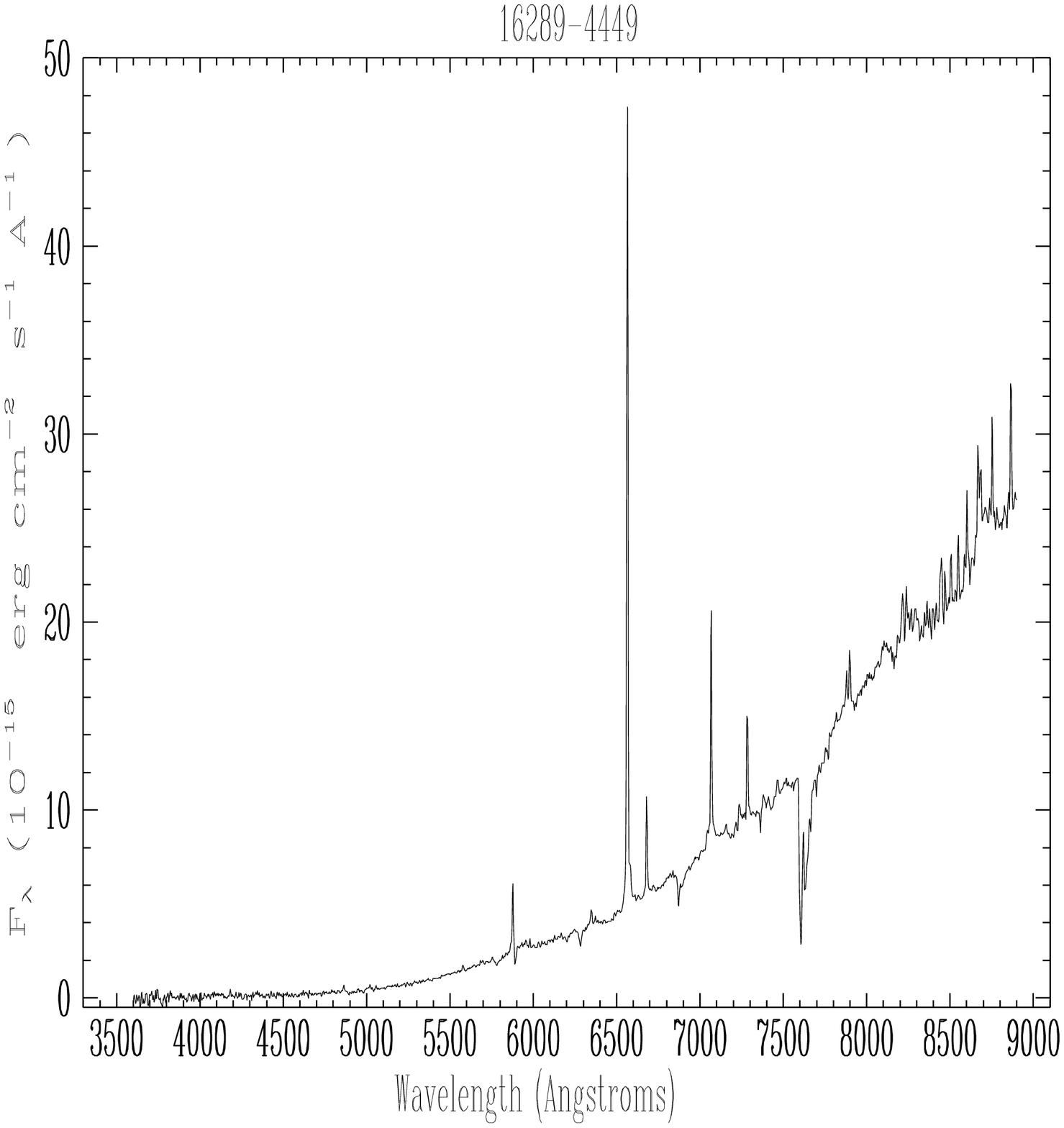}
%\psdraft
\epsfxsize=4cm
\epsfysize=4cm
\epsfbox{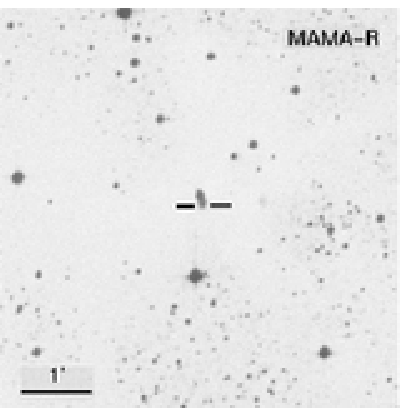}
%\psfull
\end{center}

\begin{center}
\epsfxsize=13.5cm
\epsfysize=4cm
\epsfbox{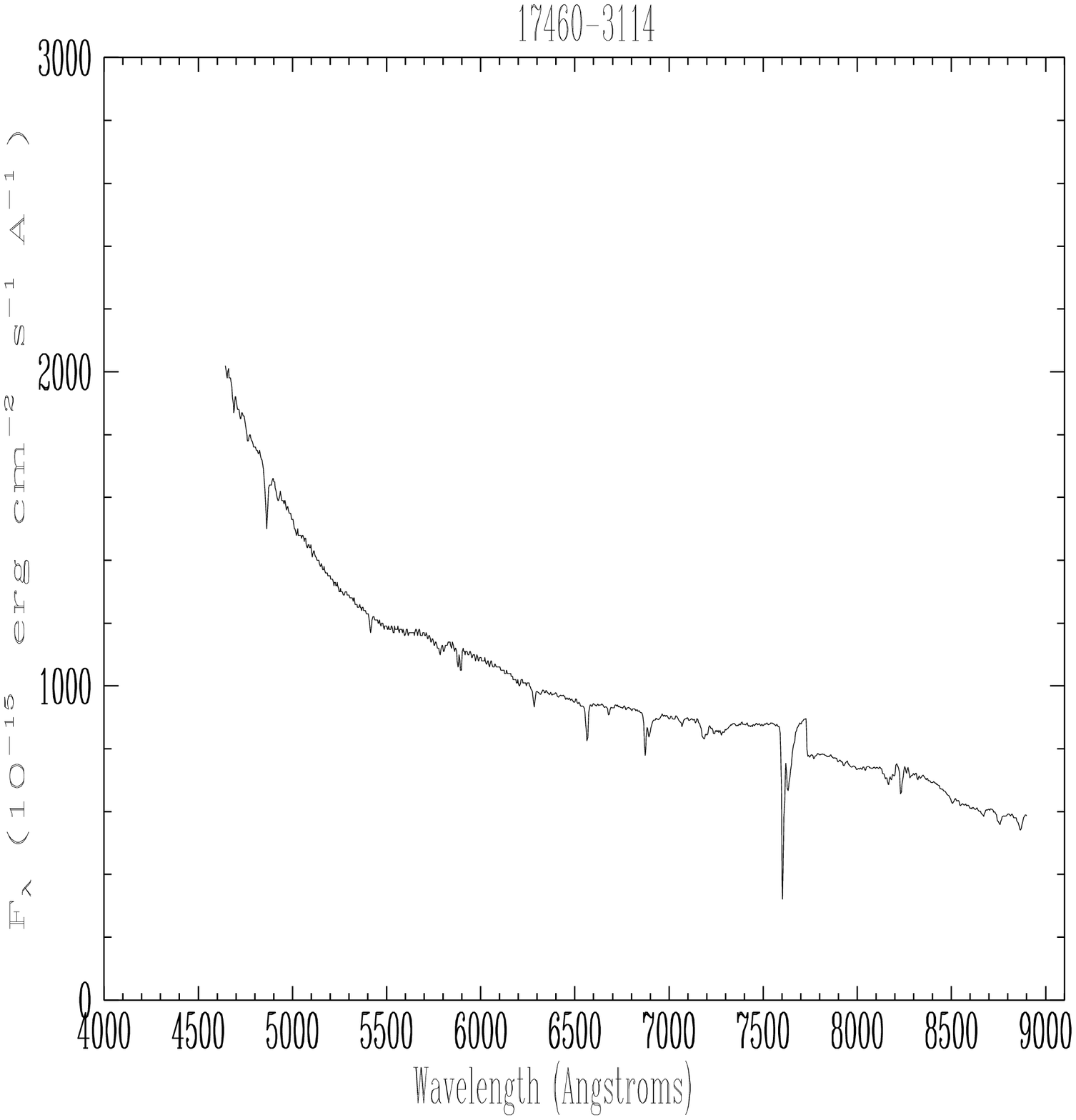}
%\psdraft
\epsfxsize=4cm
\epsfysize=4cm
\epsfbox{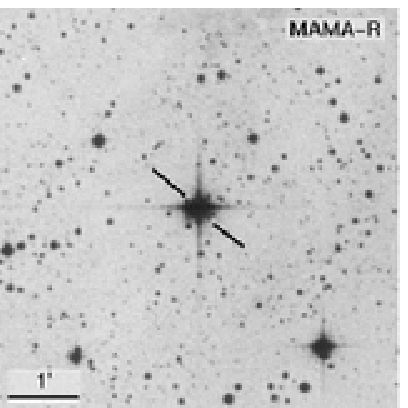}
%\psfull
\end{center}

\begin{center}
\epsfxsize=13.5cm
\epsfysize=4cm
\epsfbox{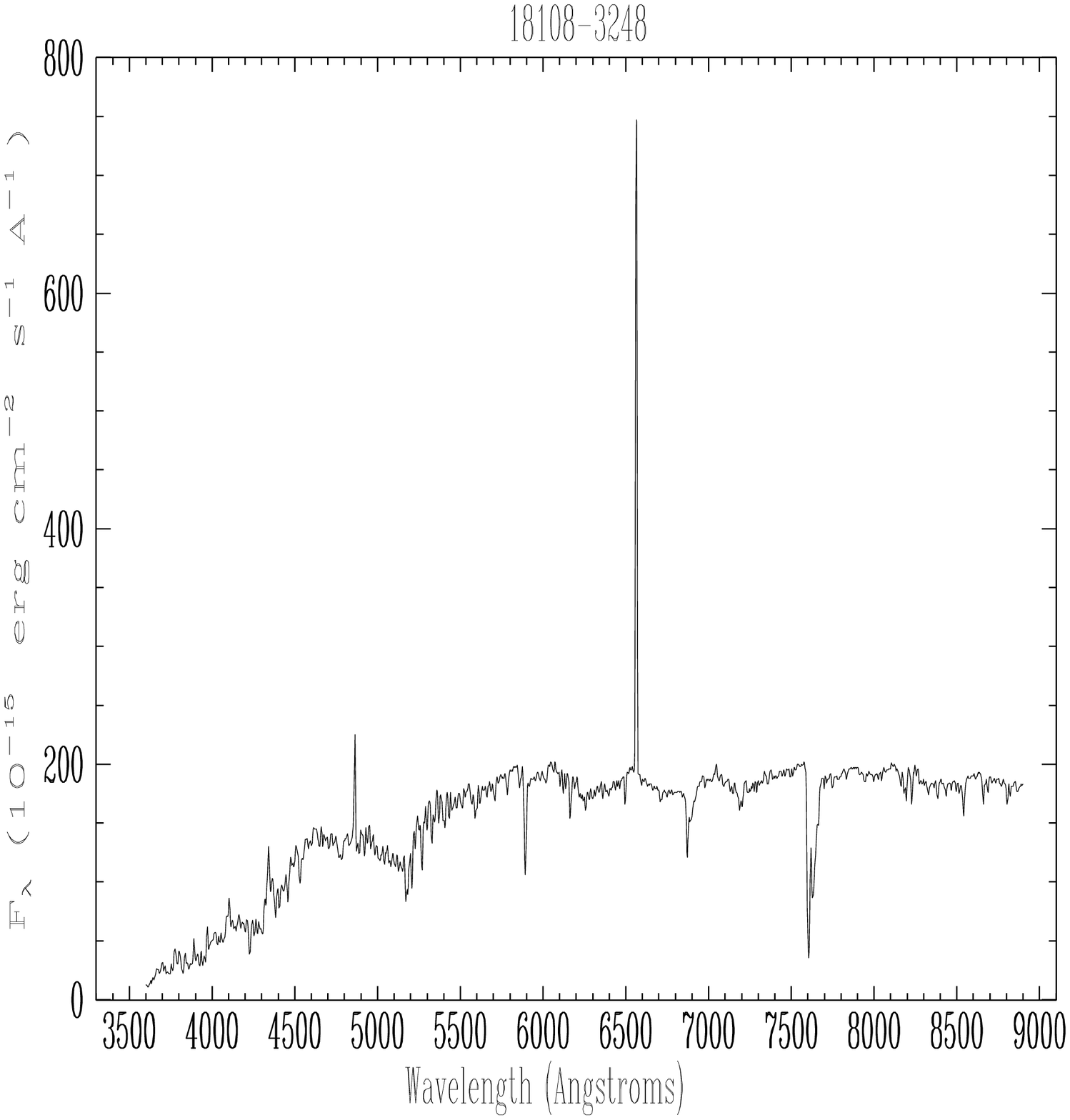}
%\psdraft
\epsfxsize=4cm
\epsfysize=4cm
\epsfbox{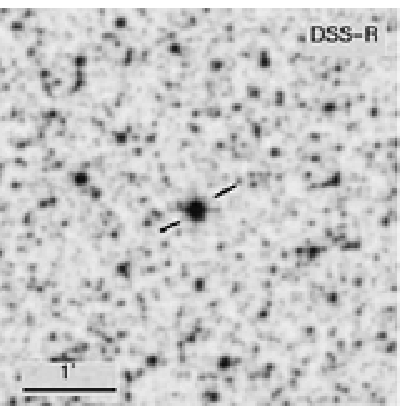}
%\psfull
\end{center}

\begin{center}
\epsfxsize=13.5cm
\epsfysize=4cm
\epsfbox{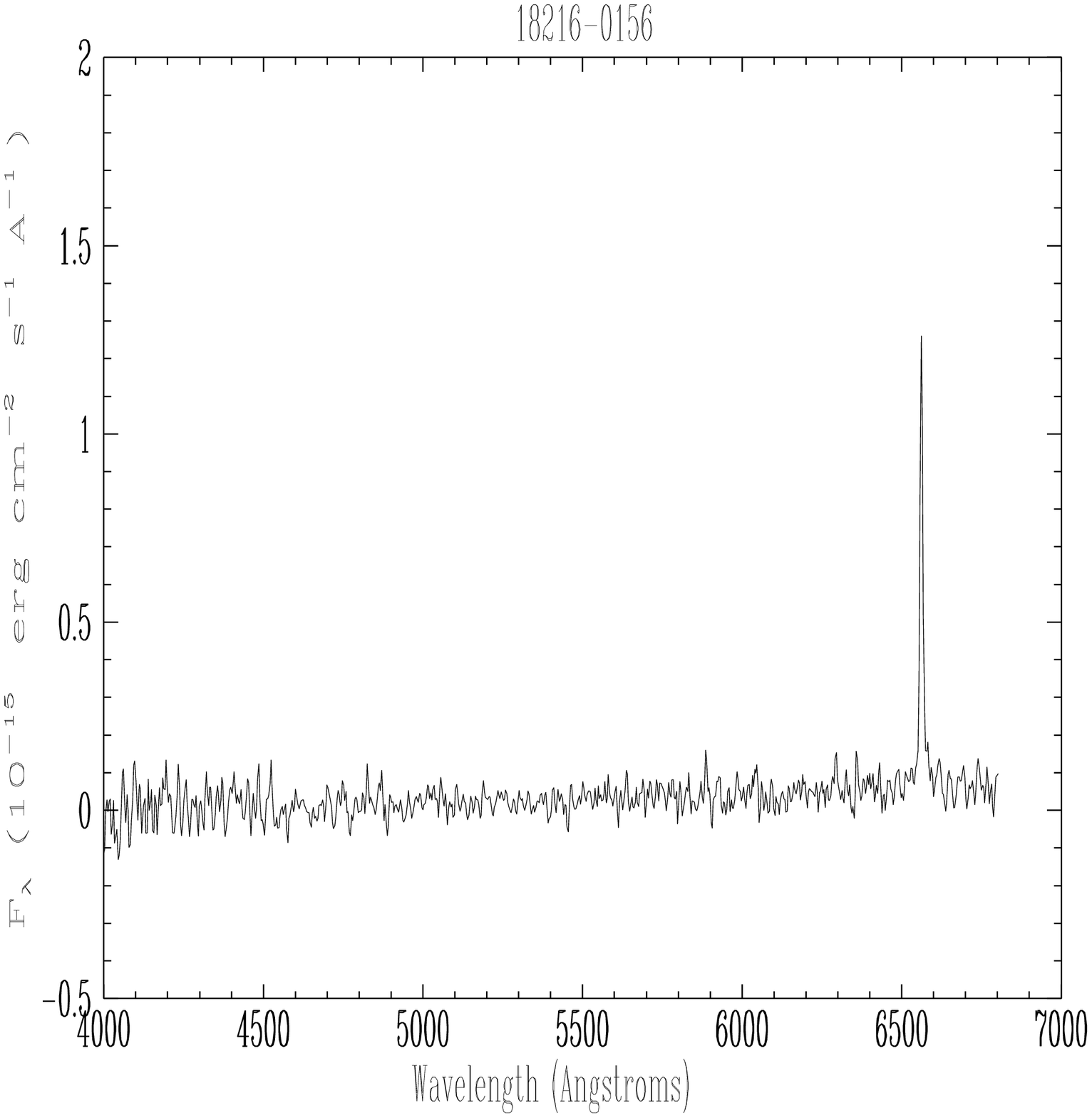}
%\psdraft
\epsfxsize=4cm
\epsfysize=4cm
\epsfbox{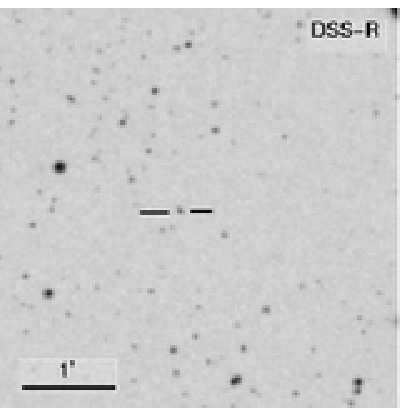}
%\psfull
\end{center}

\caption{Spectra of the objects classified as young stars in the sample together with their 
corresponding identification charts (continued). }
\end{figure*}

%-------------------------------------------------------------------
%pg8
%%\setcounter{figure}{9}
\begin{figure*}
\setcounter{figure}{0}

\begin{center}
\epsfxsize=13.5cm
\epsfysize=4cm
\epsfbox{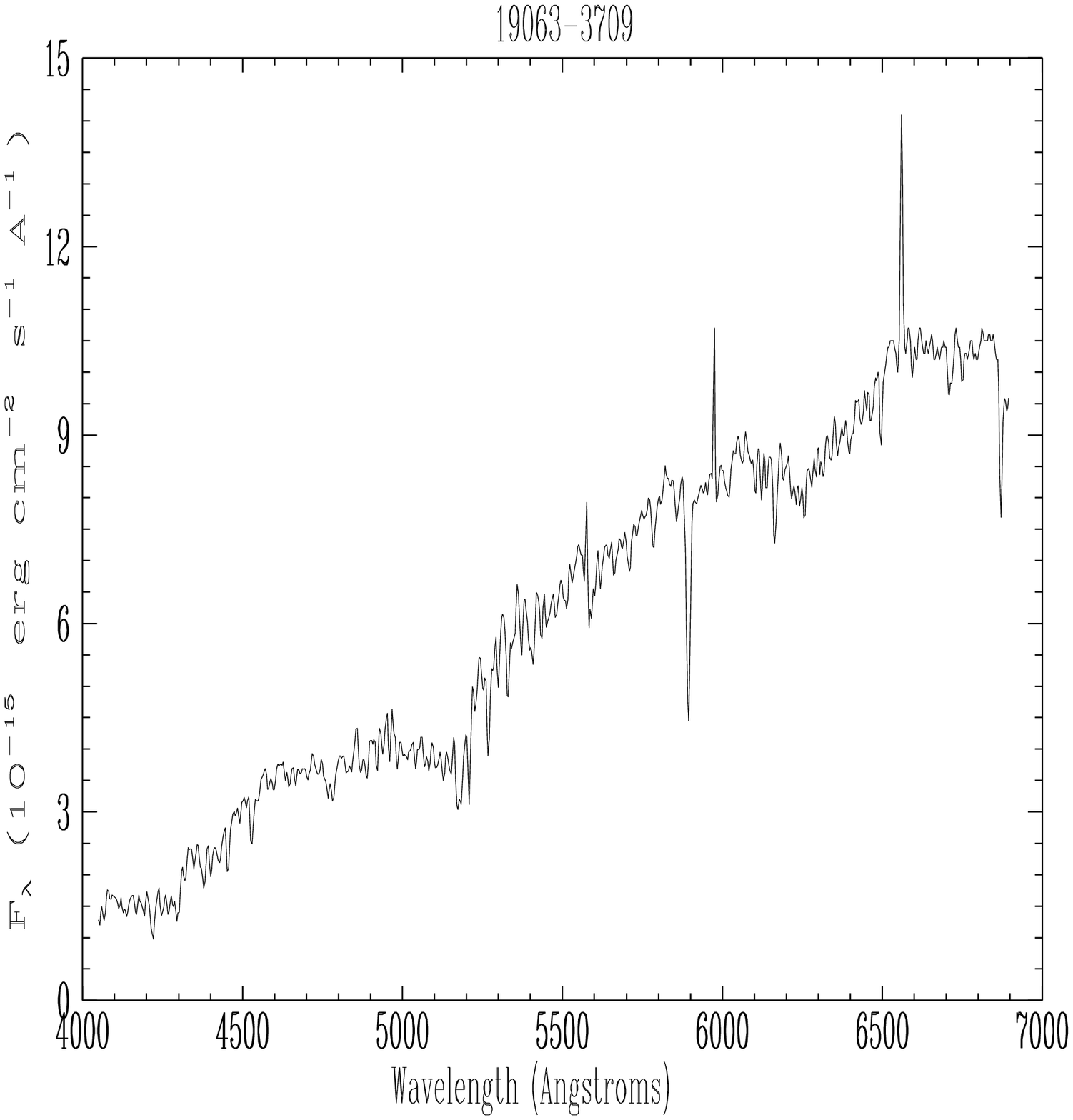}
%\psdraft
\epsfxsize=4cm
\epsfysize=4cm
\epsfbox{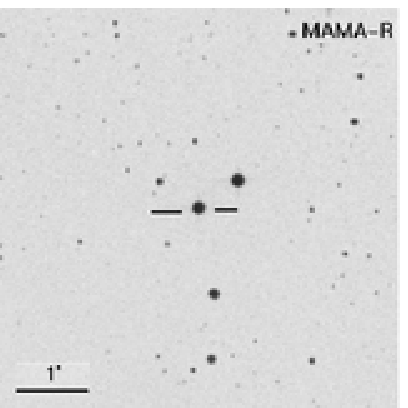}
%\psfull
\end{center}

\begin{center}
\epsfxsize=13.5cm
\epsfysize=4cm
\epsfbox{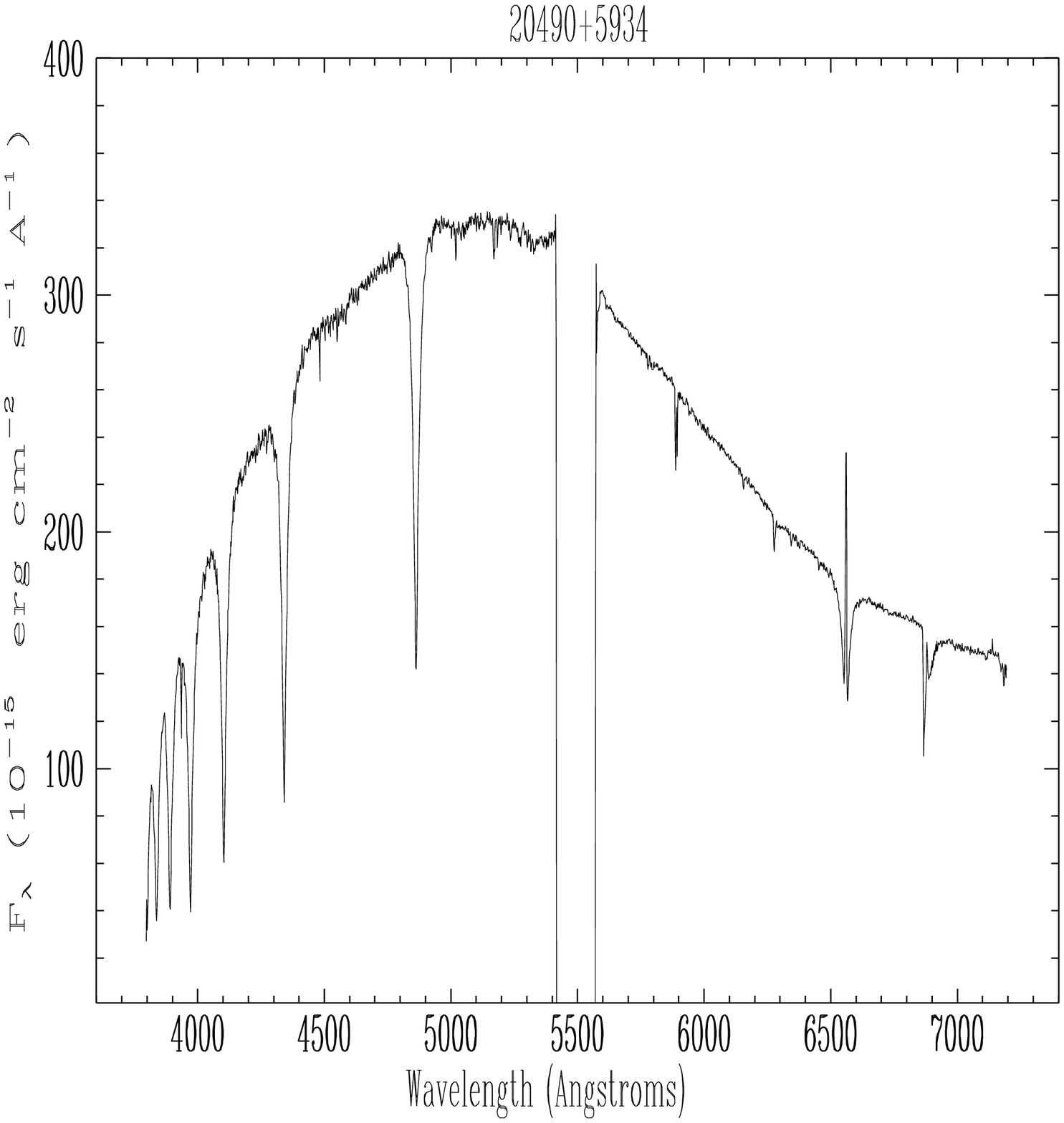}
%\psdraft
\epsfxsize=4cm
\epsfysize=4cm
\epsfbox{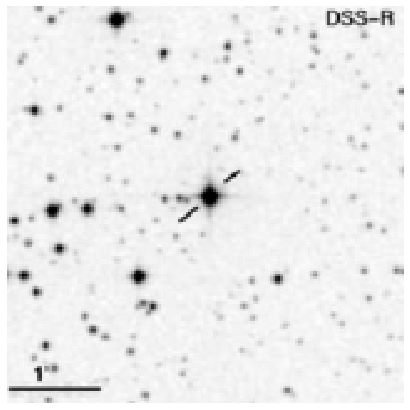}
%\psfull
\end{center}

\begin{center}
\epsfxsize=13.5cm
\epsfysize=4cm
\epsfbox{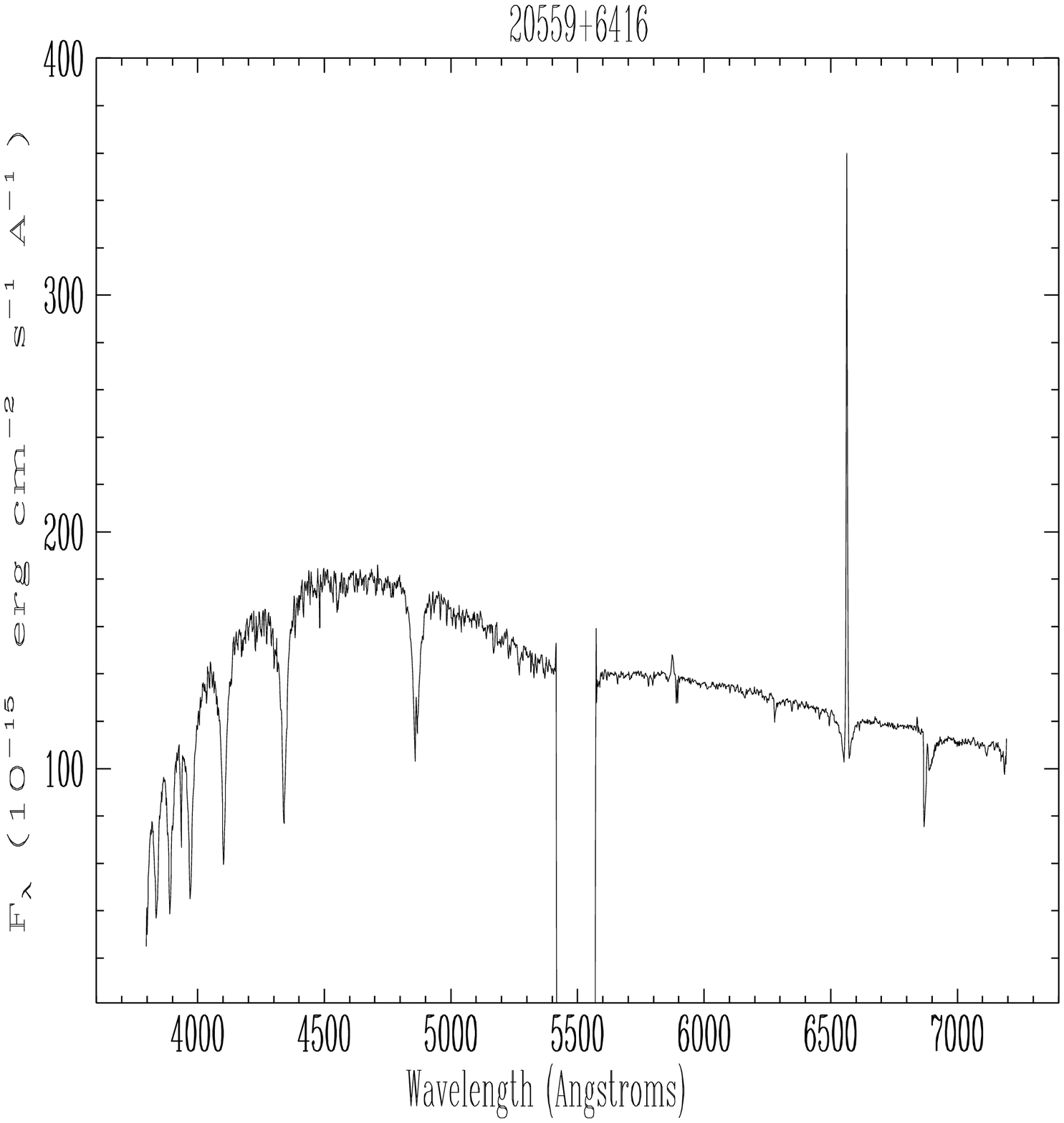}
%\psdraft
\epsfxsize=4cm
\epsfysize=4cm
\epsfbox{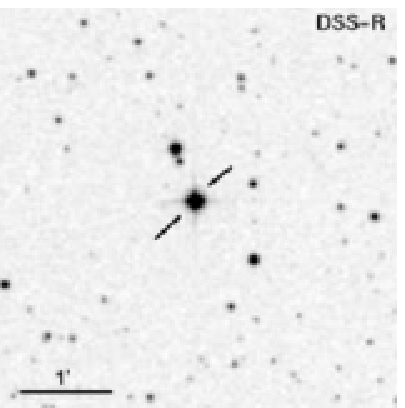}
%\psfull
\end{center}

\begin{center}
\epsfxsize=13.5cm
\epsfysize=4cm
\epsfbox{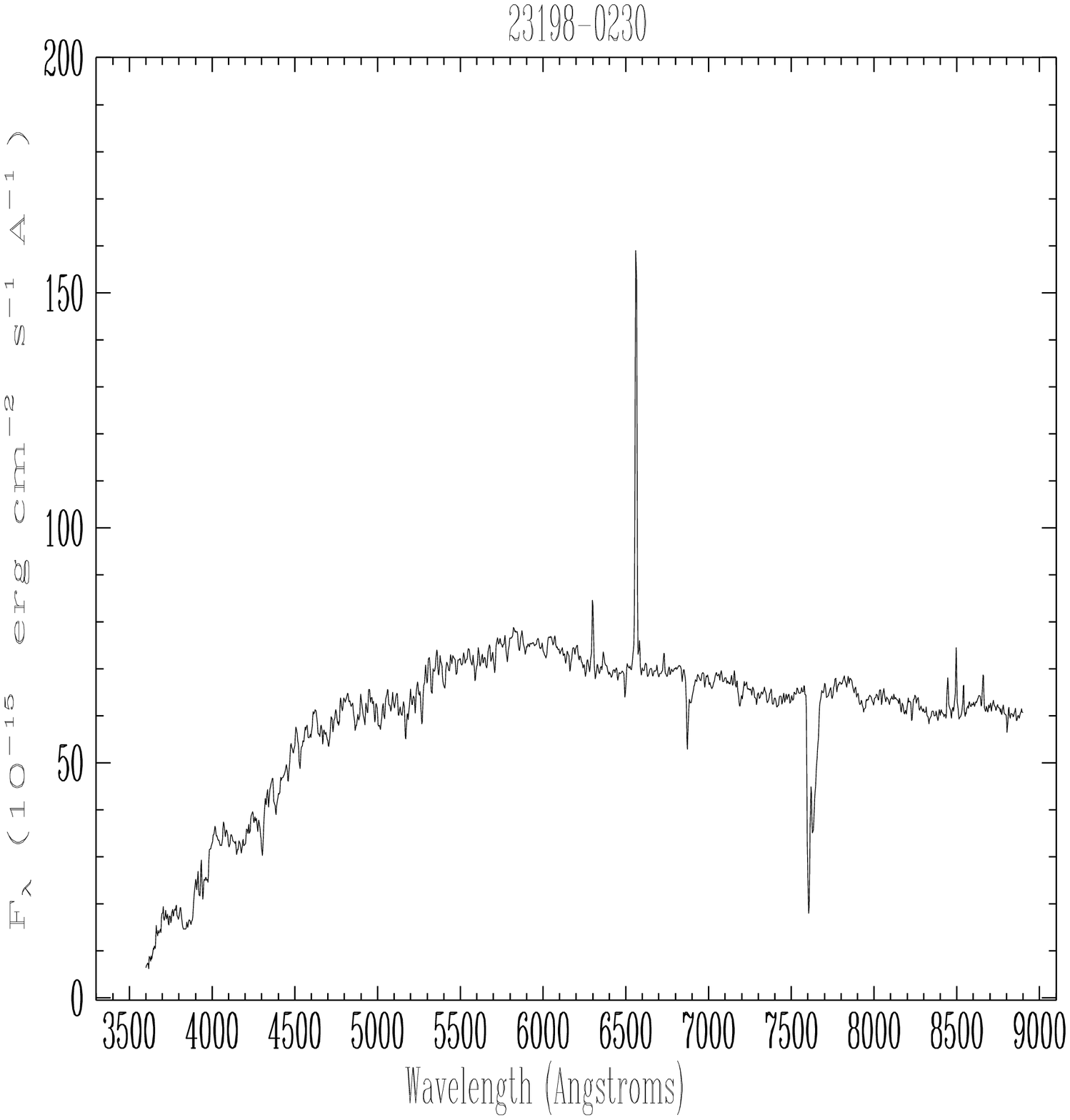}
%\psdraft
\epsfxsize=4cm
\epsfysize=4cm
\epsfbox{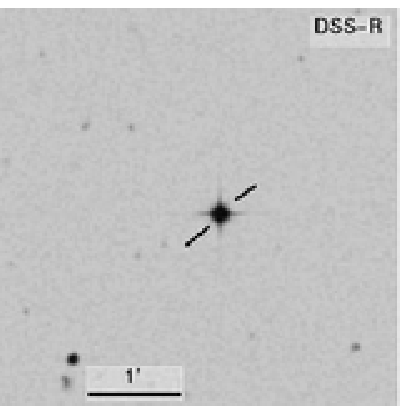}
%\psfull
\end{center}

\caption{Spectra of the objects classified as young stars in the sample together with their 
corresponding identification charts (continued). }
\end{figure*}

%%% Local Variables: 
%%% mode: latex
%%% TeX-master: "~/tesis/mitesis/final/tesis"
%%% End: 

\clearpage
\section{Atlas of galaxies}
        \begin{figure*}[h!]

\begin{center}
\epsfxsize=13.5cm
\epsfysize=4cm
\epsfbox{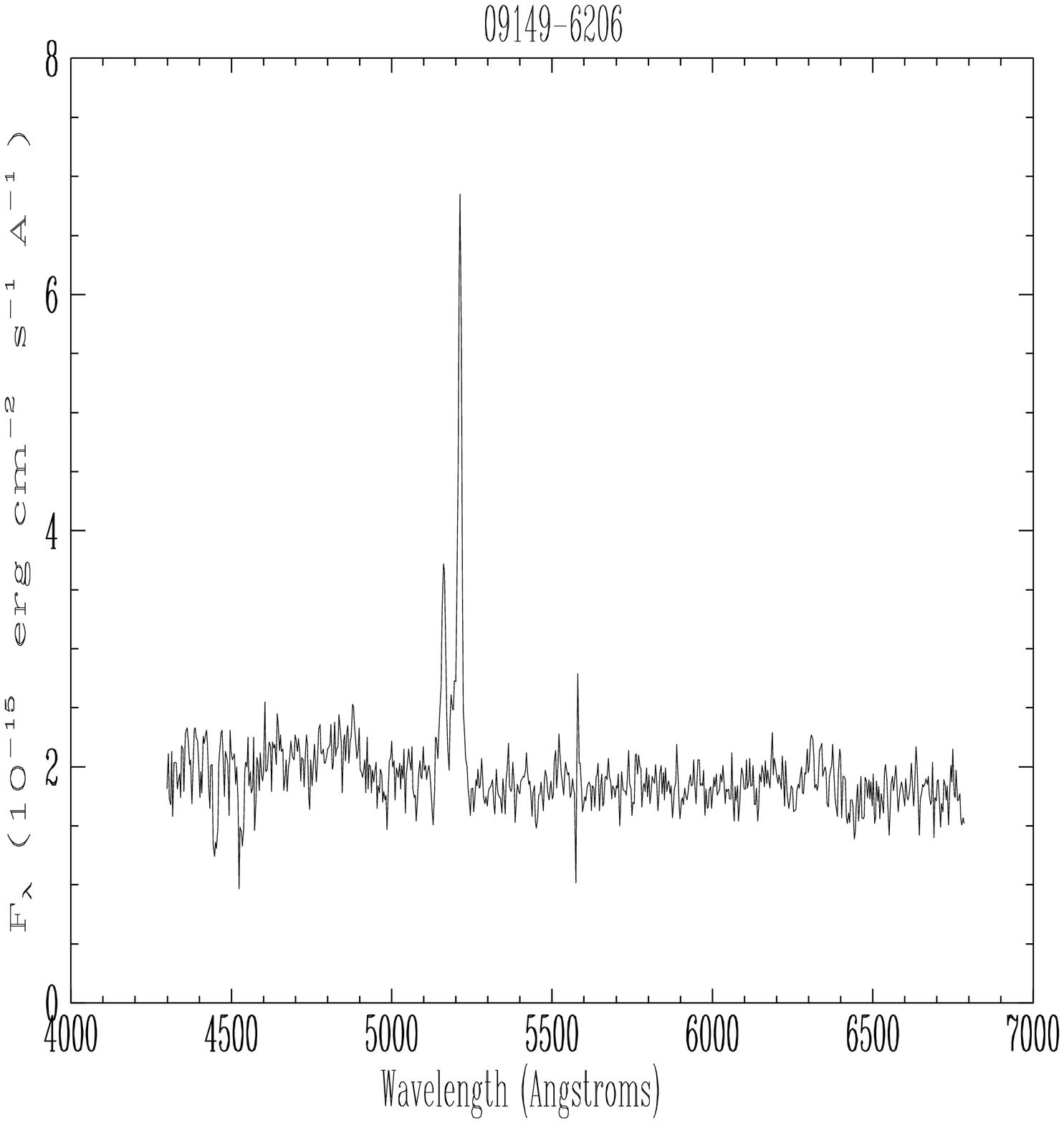}
%\psdraft
\epsfxsize=4cm
\epsfysize=4cm
\epsfbox{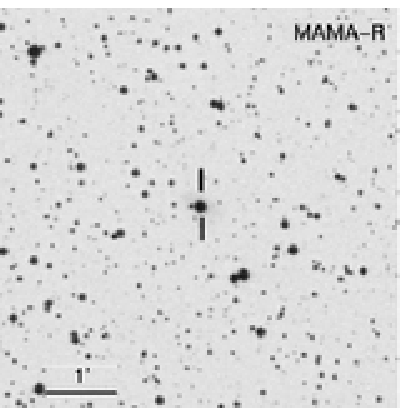}
%\psfull
\end{center}

\begin{center}
\epsfxsize=13.5cm
\epsfysize=4cm
\epsfbox{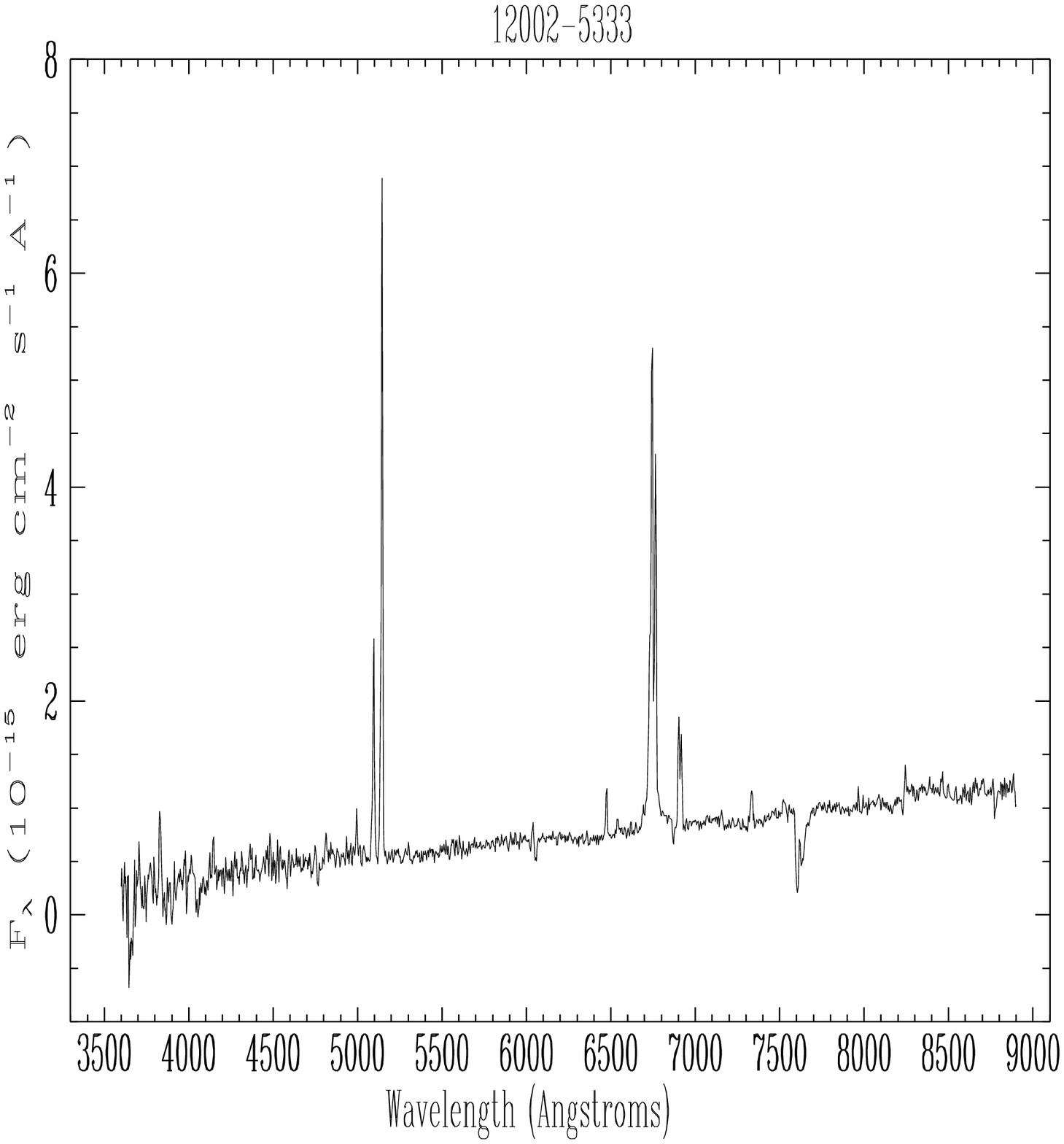}
%\psdraft
\epsfxsize=4cm
\epsfysize=4cm
\epsfbox{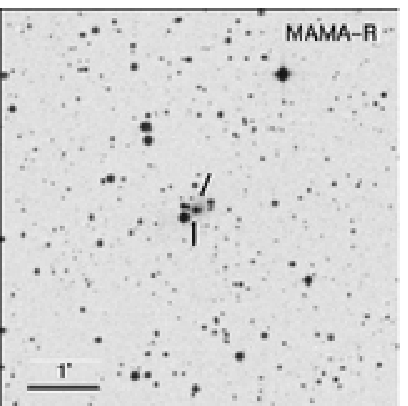}
%\psfull
\end{center}

\caption{Spectra of the sources classified as galaxies in the sample 
together with their corresponding identification charts.}
\end{figure*}

%%% Local Variables: 
%%% mode: latex
%%% TeX-master: "~/tesis/mitesis/final/tesis"
%%% End: 

\clearpage
\section{Atlas of peculiar sources}
        s\begin{figure*}[h!]

\begin{center}
\epsfxsize=13.5cm
\epsfysize=4cm
\epsfbox{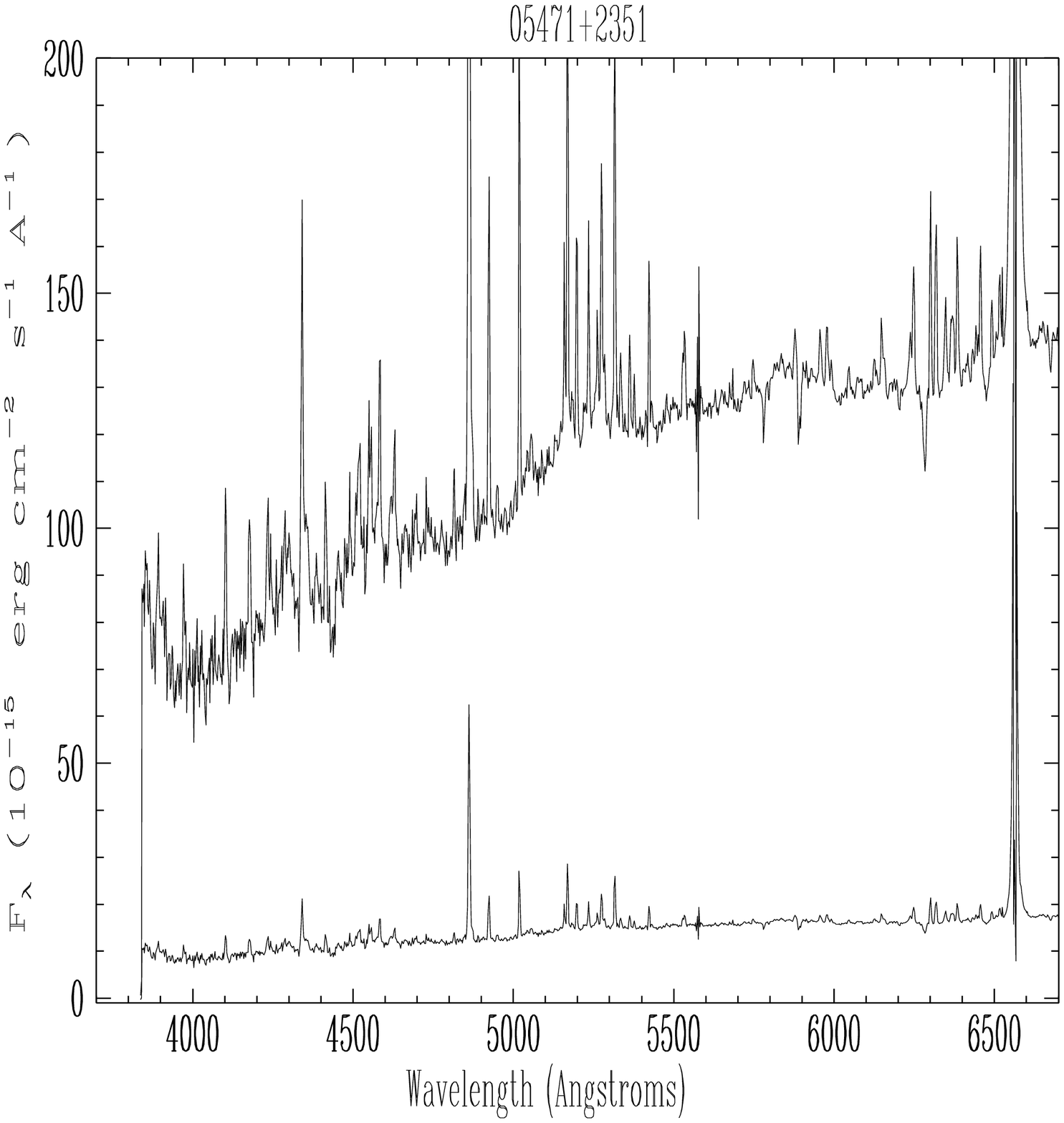}
%\psdraft
\epsfxsize=4cm
\epsfysize=4cm
\epsfbox{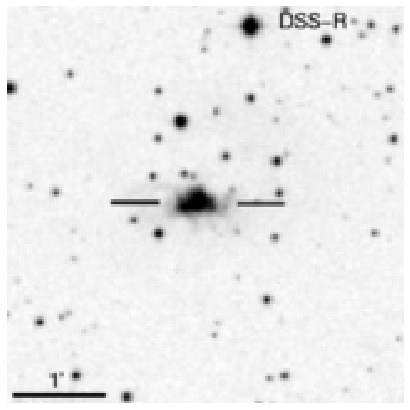}
%\psfull
\end{center}

\begin{center}
\epsfxsize=13.5cm
\epsfysize=4cm
\epsfbox{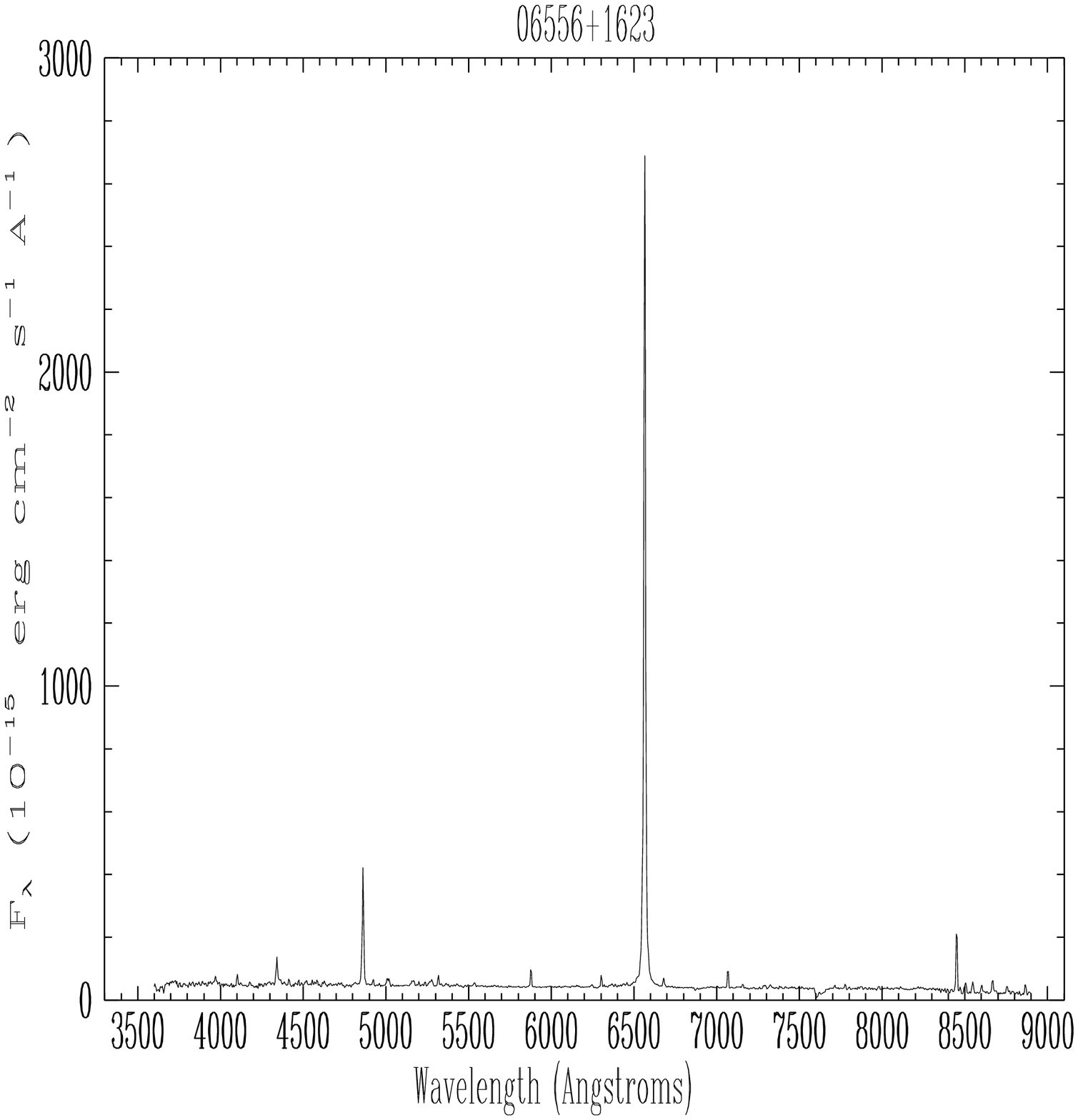}
%\psdraft
\epsfxsize=4cm
\epsfysize=4cm
\epsfbox{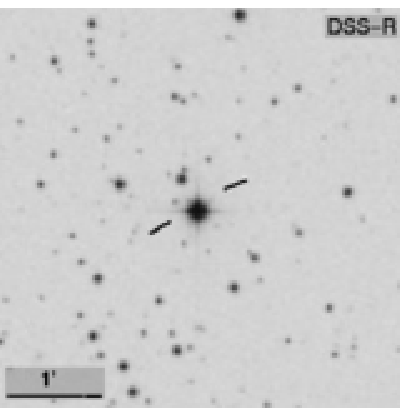}
%\psfull
\end{center}

\begin{center}
\epsfxsize=13.5cm
\epsfysize=4cm
\epsfbox{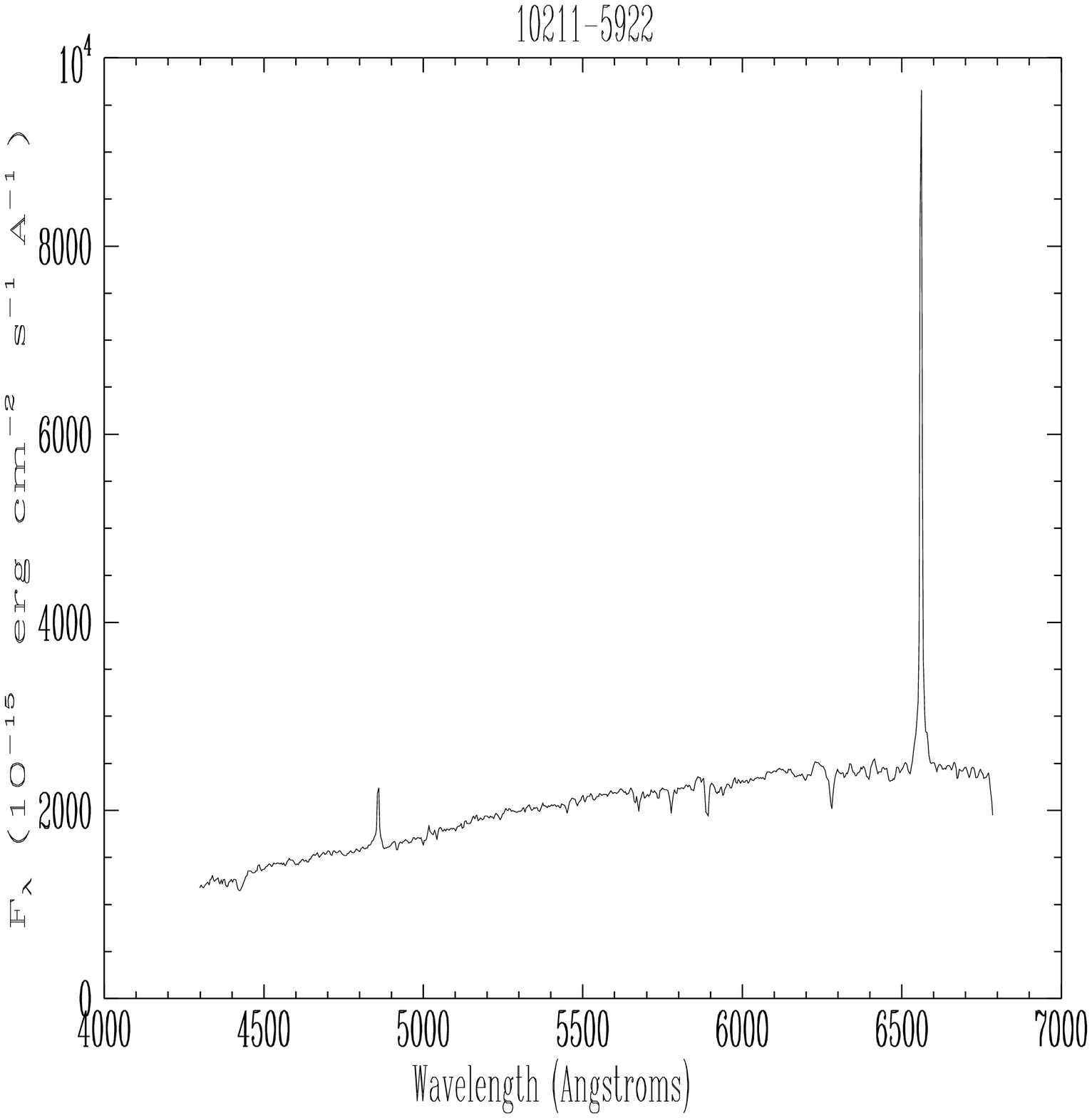}
%\psdraft
\epsfxsize=4cm
\epsfysize=4cm
\epsfbox{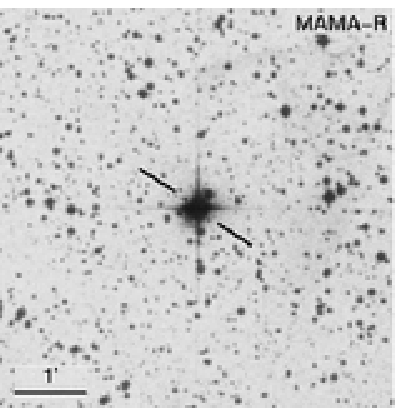}
%\psfull
\end{center}

\begin{center}
\epsfxsize=13.5cm
\epsfysize=4cm
\epsfbox{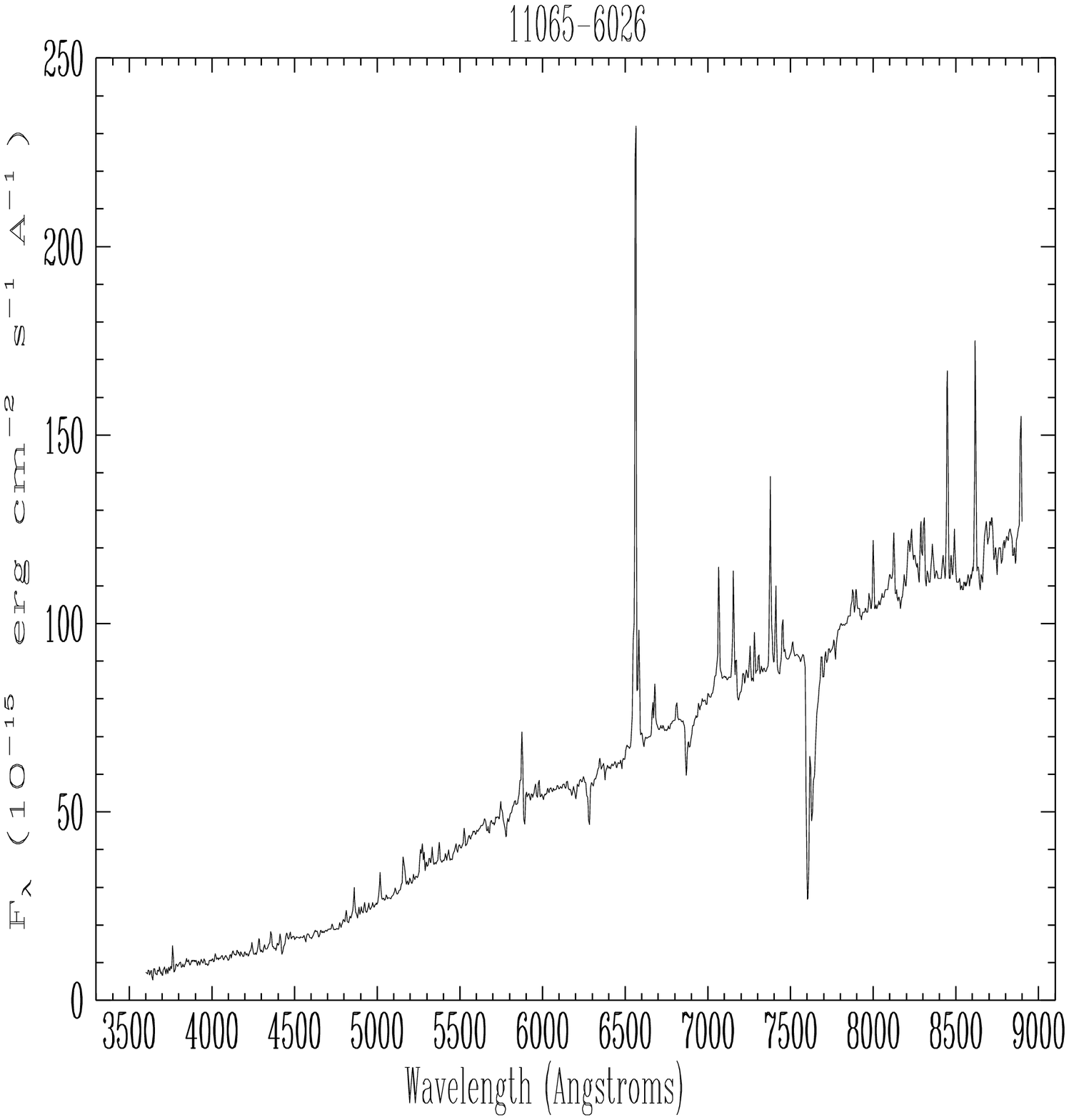}
%\psdraft
\epsfxsize=4cm
\epsfysize=4cm
\epsfbox{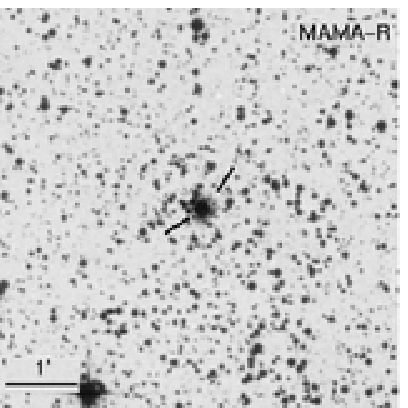}
%\psfull
\end{center}

\begin{center}
\epsfxsize=13.5cm
\epsfysize=4cm
\epsfbox{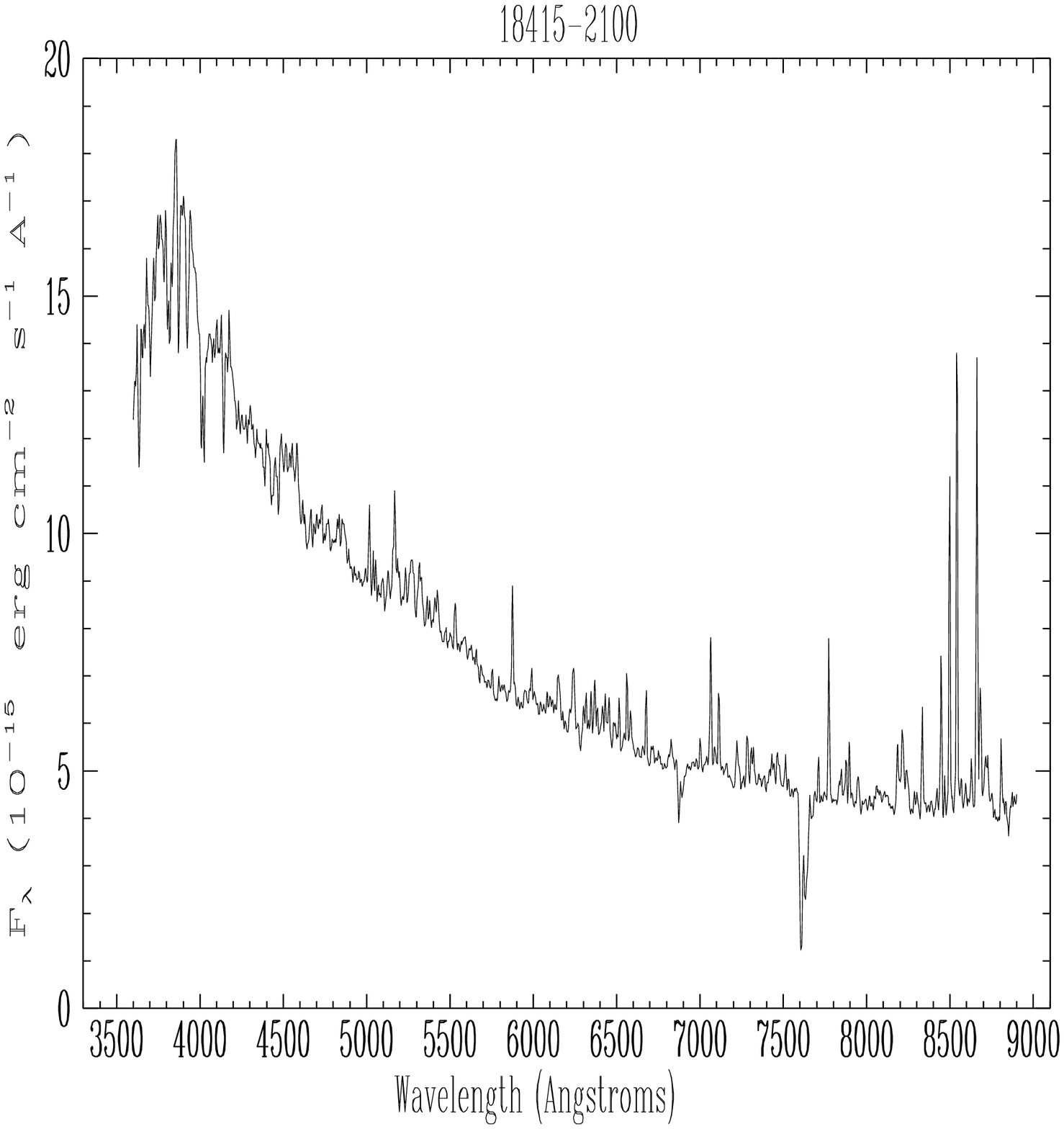}
%\psdraft
\epsfxsize=4cm
\epsfysize=4cm
\epsfbox{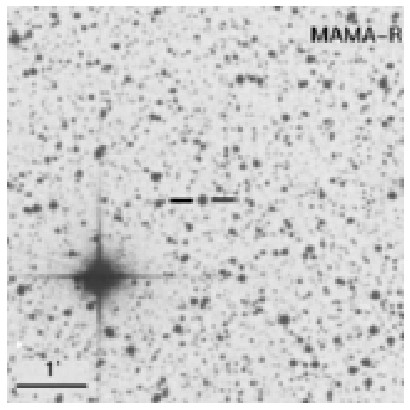}
%\psfull
\end{center}

\caption{Spectra of the peculiar sources in the sample together with their 
corresponding identification charts. }
\end{figure*}

%%% Local Variables: 
%%% mode: latex
%%% TeX-master: "~/tesis/mitesis/final/tesis"
%%% End: 

\end{document}